\documentclass[11pt]{book}
\usepackage{graphicx}
\usepackage{subcaption}
\usepackage{url}
\usepackage[total={6.5in,8.75in}, top=1.2in, left=1.0in, includefoot]{geometry}
\usepackage{eurosym}

\usepackage{amsmath}
\usepackage[figuresright]{rotating}
\usepackage{pdflscape}
\usepackage{amsfonts}
\usepackage{graphbox}
\usepackage[colorlinks = true,
            linkcolor = blue,
            urlcolor  = blue,
            citecolor = blue,
            anchorcolor = blue]{hyperref}
\usepackage{amsmath}
\usepackage{siunitx}
\usepackage{hepunits}
\usepackage{color}

\usepackage[switch,modulo]{lineno}
\newcommand\pubnumber{
    DESY-22-045, IFT--UAM/CSIC--22-028, \\ 
    KEK Preprint 2021-61, PNNL-SA-160884,\\ SLAC-PUB-17662 }
\newcommand\pubdate{January 2023}

\def\Title#1{\begin{center} {\Large{\bf  #1} } \end{center}}
\def\Author#1{\begin{center}{ \sc #1} \end{center}}

\newcommand\pubblock{\rightline{\begin{tabular}{l} \pubnumber\\
         \pubdate \end{tabular}}}
\newenvironment{Abstract}{\begin{quotation} \begin{center}
                       ABSTRACT
     \end{center}\bigskip  }{\end{quotation}}

\def\Acknowledgements{\bigskip  \bigskip \begin{center} \begin{large}
             \bf ACKNOWLEDGEMENTS \end{large}\end{center}}

\def\beq{\begin{equation}}
\def\eeq#1{\label{#1}\end{equation}}
\def\eeqn{\end{equation}}

\def\beqa{\begin{eqnarray}}
\def\eeqa#1{\label{#1}\end{eqnarray}}
\def\eeqan{\end{eqnarray}}
\def\CR{\nonumber \\ }

\def\leqn#1{(\ref{#1})}

\def\overbar#1{\overline{#1}}

\def\etal{{\it et al.}}
\def\ie{{\it i.e.}}
\def\eg{{\it e.g.}}
\def\etc{{\it etc.}}

\def\lsim{\mathrel{\raise.3ex\hbox{$<$\kern-.75em\lower1ex\hbox{$\sim$}}}}
\def\gsim{\mathrel{\raise.3ex\hbox{$>$\kern-.75em\lower1ex\hbox{$\sim$}}}}

\def\L{{\cal L}}

\def\L{{\cal L}}

\def\half{\frac{1}{2}}

\def\del{\partial}
\def\Dslash{\not{\hbox{\kern-4pt $D$}}}
\def\dslash{\not{\hbox{\kern-2pt $\del$}}}
\def\missET{\not{\hbox{\kern-4pt $E$}_T}}

\def\Dlr{\mathrel{\raise1.5ex\hbox{$\leftrightarrow$\kern-1em\lower1.5ex\hbox{$D$}}}}

\def\ee{e^+e^-}
\def\sstw{\sin^2\theta_w}
\def\cstw{\cos^2\theta_w}
\def\mz{m_Z}
\def\gz{\Gamma_Z}

\def\mw{m_W}

\def\MSB{{\bar{M \kern -2pt S}}}
\def\msb{{\bar{\scriptsize M \kern -1pt S}}}

\def\drb{{\bar{\scriptsize D \kern -1pt R}}}
\def\ELER{e^-_Le^+_R}
\def\EREL{e^-_Re^+_L}

\def\eps{\epsilon}
\def\matheuro{{ \mbox{\euro}}}

\def\neu#1{\widetilde\chi^0_{#1}}
\def\s#1{\widetilde{#1}}

\newcommand {\sub}[1]{\ensuremath{_{\mathrm{#1}}}}
\newcommand {\siunit}[2]{\ensuremath{#1\,{\mathrm{#2}}}}
\newcommand{\ccbar}{c\overline{c}}
\newcommand{\bbbar}{b\overline{b}}

\newcommand{\ffbar}{f\overline{f}}
\providecommand{\ppl}  {{\cal P}_{\rm{e}^+}}
\providecommand{\pmi}  {{\cal P}_{\rm{e}^-}}

\providecommand{\peff}  {{\cal P}_{\rm{eff}}}

\newcommand{\afb}{A_{FB}}

\newcommand{\afbnull}{(A_{FB})_0}
\newcommand{\alrfb}{A_{LRFB}}
\newcommand{\alr}{A_{LR}}

\def\sid{{SiD}}
\def\pT{p_T}
\def\micron{ \mu \hbox{m}}
\def\ECal{{ECal}}

\newcommand{\toprule}{\hline}
\newcommand{\midrule}{\hline}
\newcommand{\bottomrule}{\hline}
\def\MXN#1{\mbox{$ M_{\tilde{\chi}^0_#1}                                $}}
\def\MXC#1{\mbox{$ M_{\tilde{\chi}^{\pm}_#1}                            $}}
\def\XP#1{\mbox{$ \tilde{\chi}^+_#1                                     $}}

\def\XPM#1{\mbox{$ \tilde{\chi}^{\pm}_#1                                $}}

\def\XN#1{\mbox{$ \tilde{\chi}^0_#1                                     $}}

\def\p#1{\mbox{$ \mbox{\bf p}_1                                         $}}

\newcommand{\smu}     {\mbox{$ \tilde{\mu}                                 $}}

\newcommand{\sel}     {\mbox{$ \tilde{\mathrm e}                           $}}

\newcommand{\snu}     {\mbox{$ \tilde\nu                                   $}}

\newcommand{\stau}    {\mbox{$ \tilde{\tau}                                $}}
\newcommand{\stone}   {\mbox{$ \tilde{\tau}_1                              $}}

\newcommand{\eeto}    {\mbox{$ {\, \mathrm e}^+ {\mathrm e}^- \to             $}}

\definecolor{Orange}{rgb}{1.0,0.7,0.2}

\def\kek{KEK, Tsukuba, JAPAN}
\def\kekno{{1}}
\def\desy{Deutsches Elektronen-Synchrotron DESY, GERMANY}
\def\desyno{{2}}
\def\oregon{University of Oregon, Eugene, OR USA}
\def\oregonno{{3}}
\def\ucsb{University of California, Santa Barbara, CA USA}
\def\ucsbno{{4}}
\def\pitt{University of Pittsburgh, Pittsburgh, PA USA}
\def\pittno{{5}}
\def\slac{SLAC National Accelerator Laboratory, Menlo Park, CA USA}
\def\slacno{{6}}
\def\ucsc{University of California, Santa Cruz CA USA}
\def\ucscno{{7}}
\def\humboldt{Humbolt University, Berlin, GERMANY}
\def\humboldtno{{8}}
\def\madrid{Universidad Aut\'onoma de Madrid, SPAIN}
\def\madridno{{9}}
\def\minn{University of Minnesota, Minneapolis, MN USA}
\def\minnno{{10}}
\def\chicago{University of Chicago, Chicago, IL USA}
\def\chicagono{{11}}
\def\yale{Yale University, New Haven, CT USA}
\def\yaleno{{12}}
\def\berkeley{University of California, Berkeley, CA USA}
\def\berkeleyno{{13}}
\def\lbnl{Lawrence Berkeley National Laboratory, Berkeley, CA USA}
\def\lbnlno{{14}}
\def\ipmu{Kavli IPMU, University of Tokyo, Kashiwa,  JAPAN}
\def\ipmuno{{15}}
\def\lausanne{EPFL, Lausanne SWITZERLAND}
\def\lausanneno{{16}}
\def\cornell{Cornell Unversity, Ithaca NY USA}
\def\cornellno{{17}}
\def\lal{IJCLab, Universit\'e Paris-Saclay, Orsay FRANCE}
\def\lalno{{18}}
\def\fermilab{Fermi National Accelerator Laboratory, Batavia, IL USA}
\def\fermilabno{{19}}
\def\glasgow{University of Glasgow, Glasgow UK}
\def\glasgowno{{20}}
\def\pnnl{Pacific Northwest National Laboratory, Richland, WA USA}
\def\pnnlno{{21}}
\def\kyushu{Kyushu University, Fukuoka JAPAN}
\def\kyushuno{{22}}
\def\tokyo{University of Tokyo, Tokyo, JAPAN}
\def\tokyono{{23}}
\def\saclay{CEA Saclay, Gif sur Yvette, FRANCE}
\def\saclayno{{24}}
\def\valencia{IFIC, CSIC-Unversity of Valencia, Valencia, SPAIN}
\def\valenciano{{25}}
\def\uta{University of Texas, Arlington, TX USA}
\def\utano{{26}}
\def\kansas{University of Kansas, Lawrence, KS}
\def\kansasno{{27}}
\def\warsaw{University of Warsaw, Warsaw POLAND}
\def\warsawno{{28}}
\def\maryland{University of Maryland, College Park, MD USA}
\def\marylandno{{29}}
\def\belgrade{University of Belgrade, Belgrade, SERBIA}
\def\belgradeno{{30}}
\def\lanl{Los Alamos National Laboratory, Los Alamos, NM USA}
\def\lanlno{{31}}
\def\kanazawa{Kanazawa University, Kanazawa, JAPAN}
\def\kanazawano{{32}}
\def\sokendai{Sokendai, KEK, Tsukuba, JAPAN}
\def\sokendaino{{33}}
\def\nagoya{Nagoya University, Nagoya,  JAPAN}
\def\nagoyano{{34}}
\def\saitama{Saitama University, Saitama, JAPAN}
\def\saitamano{{35}}
\def\oklahoma{University of Oklahoma, Norman, OK USA}
\def\oklahomano{{36}}
\def\hopkins{John Hopkins University, Baltimore, MD USA}
\def\hopkinsno{{37}}
\def\wisconsin{University of Wisconsin, Madison, WI USA}
\def\wisconsinno{{38}}
\def\lancaster{Lancaster University, Lancaster, UK}
\def\lancasterno{{39}}
\def\liplisbon{LIP Laboratorio, Lisbon, PORTUGAL}
\def\liplisbonno{{40}}
\def\toronto{University of Toronto, Toronto, ON CANADA}
\def\torontono{{41}}
\def\stanford{Stanford University, Stanford, CA USA}
\def\stanfordno{{42}}
\def\carleton{Carleton University, Ottawa, ON CANADA}
\def\carletonno{{43}}
\def\cordoba{University of Cordoba, Cordoba, SPAIN}
\def\cordobano{{44}}
\def\beykent{Beykent University, Istanbul, TURKEY}
\def\beykentno{{45}}
\def\iowa{University of Iowa, Iowa City, IA USA}
\def\iowano{{46}}
\def\nebraska{University of Nebraska, Lincoln NE USA}
\def\nebraskano{{47}}
\def\argonne{Argonne National Laboratory, Lemont, IL  USA}
\def\argonneno{{48}}
\def\royalholloway{Royal Holloway University, London, UK}
\def\royalhollowayno{{49}}
\def\cucurova{Cucurova University, Adana, TURKEY}
\def\cucurovano{{50}}
\def\instpoly{Institut Polytechnique de Paris, Palaiseau, FRANCE}
\def\instpolyno{{51}}

\def\ohiostateno{{52}}
\def\karazin{Karazin National University, Kharkiv, UKRAINE}
\def\karazinno{{53}}
\def\oxford{Oxford University, Oxford, UK}
\def\oxfordno{{54}}
\def\cern{CERN, Geneva, SWITZERLAND}
\def\cernno{{55}}
\def\brookhaven{Brookhaven National Laboratory, Upton, NY USA}
\def\brookhavenno{{56}}
\def\ectstar{ECT$^*$, Trento, ITALY}
\def\ectstarno{{57}}
\def\infntrento{INFN-TIFPA Trento, Trento, ITALY}
\def\infntrentono{{58}}
\def\columbia{Columbia University, New York, NY USA} 
\def\columbiano{{59}}
\def\davis{University of California, Davis, CA  USA}
\def\davisno{{60}}
\def\ochanomizu{Ochanomizu University, Tokyo, JAPAN}
\def\ochanomizuno{{61}}
\def\delhi{University of Delhi, New Delhi, INDIA}
\def\delhino{{62}}
\def\daresbury{Daresbury Laboratory, Daresbury, UK}
\def\daresburyno{{63}}
\def\mcgill{McGill University, Montreal, QC CANADA}
\def\mcgillno{{64}}
\def\rutherford{Rutherford Appleton Laboratory, Chilton, UK}
\def\rutherfordno{{65}}
\def\hokkaido{Hokkaido University, Sapporo, JAPAN}
\def\hokkaidono{{66}}
\def\granada{Universidad de Granada, Granada, SPAIN}
\def\granadano{{67}}
\def\lyon{IP2I Lyon, Villeurbanne, FRANCE}
\def\lyonno{{68}}
\def\stonybrook{SUNY Stony Brook, Stony Brook, NY USA}
\def\stonybrookno{{69}}
\def\olddominion{Old Dominion University, Norfolk, VA USA}
\def\olddominionno{{70}}
\def\oakridge{Oak Ridge National Laboratory, Oak Ridge TN USA}
\def\oakridgeno{{71}}
\def\indiana{Indiana University, Bloomington, IN USA}
\def\indianano{{72}}
\def\barcelona{University of Barcelona, Barcelona, SPAIN}
\def\barcelonano{{73}}
\def\cas{Chinese Academy of Sciences, Beijing CHINA}
\def\casno{{74}}
\def\tamu{Texas A\&M University, College Station, TX USA}
\def\tamuno{{75}}
\def\hamburg{University of Hamburg, Hamburg, GERMANY}
\def\hamburgno{{76}}
\def\mainz{University of Mainz, Mainz, GERMANY}
\def\mainzno{{77}}
\def\cantabria{IFCA, CSIC-University of Cantabria, Santander, SPAIN }
\def\cantabriano{{78}}
\def\ciemat{CIEMAT, Madrid, SPAIN}
\def\ciematno{{79}}
\def\saga{Saga University, Saga, JAPAN}
\def\sagano{{80}}
\def\peking{Peking University, Beijing, CHINA}
\def\pekingno{{81}}
\def\harvard{Harvard University, Cambridge, MA  USA}
\def\harvardno{{82}}
\def\melbourne{University of Melbourne, Melbourne, AUSTRALIA}
\def\melbourneno{{83}}
\def\bristol{University of Bristol, Bristol, UK}
\def\bristolno{{84}}
\def\oklahomastate{Oklahoma State University, Stillwater, OK USA} 
\def\oklahomastateno{{85}}
\def\aachen{Aachen University, Aachen GERMANY} 
\def\aachenno{{86}}
\def\jefferson{Thomas Jefferson National Accelerator Facility, Newport News, VA USA}
\def\jeffersonno{{87}}
\def\fudan{Fudan University, Shanghai, CHINA}
\def\fudanno{{88}}
\def\konya{Konya Technical University, Konya, TURKEY}
\def\konyano{{89}}
\def\uclondon{University College London, London, UK}
\def\uclondonno{{90}}
\def\tsukuba{University of Tsukuba, Tsukuba, JAPAN}
\def\tsukubano{{91}}
\def\boskovic{Rudjer Boskovic Institute, Zagreb, CROATIA}
\def\boskovicno{{92}}
\def\osaka{Osaka University, Osaka JAPAN}
\def\osakano{{93}}
\def\osakainst{Osaka Institute of Technology, Osaka JAPAN}
\def\osakainstno{{94}}
\def\karlsruhe{Karlsruhe Institute of Technology, Karlsruhe, GERMANY}
\def\karlsruheno{{95}}
\def\gakugei{Tokyo Gakugei University, Tokyo, JAPAN}
\def\gakugeino{{96}}
\def\tohoku{Tohoku University, Sendai, JAPAN}
\def\tohokuno{{97}}
\def\toyama{Toyama University, Toyama, JAPAN}
\def\toyamano{{98}}
\def\mpimunich{Max Planck Institute, Munich, GERMANY}
\def\mpimunichno{{99}}
\def\helsinki{Helsinki Institute of Physics, Helsinki, FINLAND}
\def\helsinkino{{100}}
\def\ibaraki{Ibaraki University, Mito, JAPAN}
\def\ibarakino{{101}}
\def\edinburgh{University of Edinburgh, Edinburgh, UK}
\def\edinburghno{{102}}
\def\sunyatsen{Sun Yat Sen University, Zhuhai, CHINA}
\def\sunyatsenno{{103}}
\def\eotvos{E\"ov\"os Lor\'and University, Budapest, HUNGARY}
\def\eotvosno{{104}}
\def\osakacity{Osaka City University, Osaka, JAPAN}
\def\osakacityno{{105}}
\def\kyoto{Kyoto University, Kyoto, JAPAN}
\def\kyotono{{106}}
\def\snu{Seoul National University, Seoul, SOUTH KOREA}
\def\snuno{{107}}
\def\bonn{University of Bonn, Bonn, GERMANY}
\def\bonnno{{108}}
\def\kindai{Kindai University, Higashiosaka, JAPAN}
\def\kindaino{{109}}
\def\durham{Durham University, Durham, UK}
\def\durhamno{{110}}
\def\waseda{Waseda University, Tokyo, JAPAN}
\def\wasedano{{111}}
\def\birla{Birla Institute of Technology and Science, Pilani, INDIA}
\def\birlano{{112}}
\def\hiroshima{Hiroshima University, Hiroshima, JAPAN}
\def\hiroshimano{{113}}
\def\ucla{University of California, Los Angeles, CA USA}
\def\uclano{{114}}
\def\witwatersrand{University of Witwatersrand, Johannesburg, SOUTH AFRICA}
\def\witwatersrandno{{115}}
\def\irvine{University of California, Irvine CA USA}
\def\irvineno{{116}}
\def\montenegro{University of Montenegro, Podgorica, MONTENEGRO}
\def\montenegrono{{117}}
\def\northwestern{Northwestern University, Evanston, IL USA}
\def\northwesternno{{118}}
\def\lebanese{Lebanese University, Beirut, LEBANON}
\def\lebaneseno{{119}}
\def\nihon{Nihon University, Tokyo, JAPAN} 
\def\nihonno{{120}}
\def\warwick{University of Warwick, Coventry, UK}
\def\warwickno{{121}}
\def\iwate{Iwate University, Morioka, JAPAN}
\def\iwateno{{122}}
\def\infnrome{INFN, Rome, ITALY}
\def\infnromeno{{123}}
\def\cambridge{University of Cambridge, Cambridge, UK}
\def\cambridgeno{{124}}
\def\nara{Nara Women’s University, Nara, JAPAN}
\def\narano{{125}}
\def\tata{Tata Institute of Fundemental Research, Mumbai, INDIA}
\def\tatano{{126}}
\def\infnmilan{INFN, Milan, ITALY}
\def\infnmilanno{{127}}
\def\unam{Universidad Nacional Aut\'onoma Mexico, Mexico City, MEXICO}
\def\unamno{{128}}
\def\southampton{Southampton University, Southampton, UK}
\def\southamptonno{{129}}
\def\kobe{Kobe University, Kobe, JAPAN}
\def\kobeno{{130}}
\def\triumf{TRIUMF, Vancouver, BC CANADA}
\def\triumfno{{131}}
\def\northeastern{Northeastern University, Boston, MA USA}
\def\northeasternno{{132}}
\def\keio{Keio University, Tokyo, JAPAN}
\def\keiono{{133}}
\def\tohokupro{Tohoku ILC Promotion Council, Sendai, JAPAN}
\def\tohokuprono{{134}}
\def\tokyoinst{Tokyo Institute of Technology, Tokyo, JAPAN}
\def\tokyoinstno{{135}}
\def\alabama{University of Alabama,  Tuscaloosa, AL USA}
\def\alabamano{{136}}
\def\victoria{University of Victoria, Victoria, BC CANADA}
\def\victoriano{{137}}
\def\floridastate{Florida State University, Tallahassie, FL USA}  
\def\floridastateno{{138}}
\def\nippondental{Nippon Dental University, Niigata, JAPAN}
\def\nippondentalno{{139}}
\def\calabria{Universit\'a della Calabria,  Cosenza, ITALY}
\def\calabriano{{140}}
\def\infncalabria{INFN Cosenza, Cosenza, ITALY}
\def\infncalabriano{{141}}
\def\infnlasa{INFN-LASA, Milan, ITALY}
\def\infnlasano{{142}}
\def\weizmann{Weizmann Institute, Rehovot, ISRAEL}
\def\weizmannno{{143}}
\def\cinvestav{CINVESTAV, Mexico City, MEXICO}
\def\cinvestavno{{144}}
\def\manchester{University of Manchester, Manchester, UK}
\def\manchesterno{{145}}
\def\wurzburg{University of W\"urzburg, W\"urzburg, GERMANY}
\def\wurzburgno{{146}}
\def\siegen{University of Siegen, Siegen, CERMANY} 
\def\siegenno{{147}}
\def\desyz{Deutsches Elektronen-Synchrotron DESY, Zeuthen, GERMANY}
\def\desyzno{{148}}
\def\sussex{University of Sussex, Brighton, UK}
\def\sussexno{{149}}
\def\utah{University of Utah, Salt Lake City, UT USA}
\def\utahno{{150}}
\def\heidelberg{University of Heidelberg, Heidelberg, GERMANY}
\def\heidelbergno{{151}}
\def\msu{Michigan State University, Lansing, MI USA}
\def\msuno{{152}}
\def\kokushikan{Kokushikan University,  Tokyo, JAPAN}
\def\kokushikanno{{153}}
\def\niigata{Niigata University, Niigata, JAPAN}  
\def\niigatano{{154}}
\def\yamagata{Yamagata University, Yamagata, JAPAN}
\def\yamagatano{{155}}
\def\istanbul{Istanbul University, Istanbul, TURKEY}
\def\istanbulno{{156}}
\def\delaware{University of Delaware, Newark, DE USA}
\def\delawareno{{157}}
\def\iitmadras{Indian Institute of Technology, Madras, INDIA}
\def\iitmadrasno{{158}}
\def\illinois{University of Illinois, Champaign, IL USA}
\def\illinoisno{{159}}
\def\miyazaki{Miyazaki University, Miyazaki, JAPAN}
\def\miyazakino{{160}}
\def\kogakuin{Kogakuin University, Tokyo, JAPAN} 
\def\kogakuinno{{161}}
\def\hebrew{Hebrew University, Jerusalem, ISRAEL} 
\def\hebrewno{{162}}
\def\krakowinst{Institute of Nuclear Physics, Krakow, POLAND}
\def\krakowinstno{{163}}
\def\technion{Technion - Israel Institue of Technology, Haifa, ISRAEL}
\def\technionno{{164}}
\def\arizona{University of Arizona, Tucson, AZ USA}
\def\arizonano{{165}}
\def\ritsumeikon{Ritsumeikon University, Kyoto, JAPAN}
\def\ritsumeikonno{{166}}
\def\toyamapref{Toyama Prefectural University, Toyama, JAPAN}
\def\toyamaprefno{{167}}
\def\iwatepref{Iwate Prefectural University, Takizawa, JAPAN}
\def\iwateprefno{{168}}
\def\shinsu{Shinsu University, Nagano, JAPAN}
\def\shinsuno{{169}}
\def\riverside{University of California, Riverside, CA USA}
\def\riversideno{{170}}
\def\hawaii{University of Hawaii, Honolulu, HI USA}
\def\hawaiino{{171}}
\def\ibs{Institute for Basic Science, Daejeon, SOUTH KOREA}
\def\ibsno{{172}}
\def\mit{Massachusetts Institute of Technology, Cambridge, MA USA}
\def\mitno{{173}}
\def\nikhef{NIKHEF, Amsterdam, NETHERLANDS}
\def\nikhefno{{174}}
\def\marseille{Aix Marseille Univ, CNRS/IN2P3, CPPM, Marseille, FRANCE}
\def\marseilleno{{175}}
\def\SMUuniv{Southern Methodist University, Dallas, TX USA}
\def\SMUno{{176}}
\def\sogang{CQUEST, Sogang University, Seoul, SOUTH KOREA}
\def\sogangno{{177}}
\def\liverpool{University of Liverpool, Liverpool, UK}
\def\liverpoolno{{178}}
\def\iitbombay{Indian Institute of Technology Bombay, Mumbia, INDIA}
\def\iitbombayno{{179}}
\def\buffalo{SUNY Buffalo, Buffalo, NY USA}
\def\buffalono{{180}}
\def\birmingham{University of Birmingham, Birmingham, UK}
\def\birminghamno{{181}}
\def\washington{University of Washington, Seattle, WA USA}
\def\washingtonno{{182}}
\def\michigan{University of Michigan, Ann Arbor, MI USA}
\def\michiganno{{183}}
\def\umass{University of Massachusetts, Amherst, MA USA}
\def\umassno{{184}}
\def\jaxa{Institute of Space and Astronautical Science, Sagamihara, JAPAN}
\def\jaxano{{185}}
\def\iowastate{Iowa State University Ames, IA USA}
\def\iowastateno{{186}}
\def\zhejiang{Zhejiang University, Zhejiang, CHINA}
\def\zhejiangno{{187}}
\def\infnfrascati{INFN, Frascati, ITALY}
\def\infnfrascatino{{188}}

\begin{document}
\begin{titlepage}
\pubblock

\vskip 1.0in

\Title{The International Linear Collider: Report to Snowmass 2021}

\Author{the ILC International Development Team and the ILC community}

\bigskip

\bigskip

\begin{Abstract}
The International Linear Collider (ILC) is on the table now as a new global energy-frontier accelerator laboratory taking data in the 2030's. The ILC  addresses key questions for our current understanding of particle physics.  It is based on a proven accelerator technology. Its experiments will challenge the Standard Model of particle physics and will provide a new window to look beyond it.    This
document brings the story of the ILC up to date, emphasizing its
strong physics motivation, its readiness for construction, and the
opportunity
 it presents to the US and the global particle physics community.
\end{Abstract}

\end{titlepage}
 
\newpage

\hbox{ \null}

\newpage

\begin{center}

\noindent
Alexander~Aryshev$^\kekno$, 
Ties~Behnke$^\desyno$,
Mikael~Berggren$^\desyno$,
James~Brau$^\oregonno$,
Nathaniel~Craig$^\ucsbno$,	
Ayres~Freitas$^\pittno$,
Frank~Gaede$^\desyno$,
Spencer~Gessner$^\slacno$,
Stefania~Gori$^\ucscno$,
Christophe~Grojean$^{\desyno,\humboldtno}$
% ...

\end{center}

\bigskip

\bigskip

While many of the authors above contributed substantially to the writing of this report, authorship here  mainly represents an endorsement of the goals that this report puts forward. This endorsement is not exclusive of other Higgs factory proposals.  If you would like to add your name in support, please visit \url{https://agenda.linearcollider.org/event/9135/}.  

\newpage

\hbox{\null}

\bigskip

{\footnotesize

\noindent
$^\kekno$\kek\\
$^\desyno$\desy\\
$^\oregonno$\oregon\\
$^\ucsbno$\ucsb\\
$^\pittno$\pitt\\
$^\slacno$\slac\\
$^\ucscno$\ucsc\\
$^\humboldtno$\humboldt\\
$^\madridno$\madrid\\
$^\minnno$\minn\\
$^\chicagono$\chicago\\
$^\yaleno$\yale\\
$^\berkeleyno$\berkeley\\ 
$^\lbnlno$\lbnl\\ 
$^\ipmuno$\ipmu\\
$^\lausanneno$\lausanne\\
$^\cornellno$\cornell\\
$^\lalno$\lal\\
$^\fermilabno$\fermilab\\
$^\glasgowno$\glasgow\\
$^\pnnlno$\pnnl\\
$^\kyushuno$\kyushu\\
$^\tokyono$\tokyo\\ 
$^\saclayno$\saclay\\ 
$^\valenciano$\valencia\\
$^\utano$\uta\\ 
$^\kansasno$\kansas\\ 
$^\warsawno$\warsaw\\ 
$^\marylandno$\maryland\\ 
$^\belgradeno$\belgrade\\
$^\lanlno$\lanl\\
$^\kanazawano$\kanazawa\\
$^\sokendaino$\sokendai\\ 
$^\nagoyano$\nagoya\\ 
$^\saitamano$\saitama\\ 
$^\oklahomano$\oklahoma\\
$^\hopkinsno$\hopkins\\
$^\wisconsinno$\wisconsin\\ 
$^\lancasterno$\lancaster\\
$^\liplisbonno$\liplisbon\\
$^\torontono$\toronto\\
$^\stanfordno$\stanford\\
$^\carletonno$\carleton\\
$^\cordobano$\cordoba\\ 
$^\beykentno$\beykent\\
$^\iowano$\iowa\\ 
$^\nebraskano$\nebraska\\ 
$^\argonneno$\argonne\\ 
$^\royalhollowayno$\royalholloway\\
$^\cucurovano$\cucurova\\ 
$^\instpolyno$\instpoly\\ 
$^\karazinno$\karazin\\ 
$^\oxfordno$\oxford\\ 
$^\lancasterno$\lancaster\\
$^\cernno$\cern\\
$^\brookhavenno$\brookhaven\\
$^\ectstarno$\ectstar\\
$^\infntrentono$\infntrento\\ 
%$^\lebedevno$\lebedev\\
$^\columbiano$\columbia\\ 
$^\davisno$\davis\\ 
$^\ochanomizuno$\ochanomizu\\ 
$^\delhino$\delhi\\ 
$^\daresburyno$\daresbury\\ 
$^\mcgillno$\mcgill\\ 
$^\rutherfordno$\rutherford\\ 
$^\hokkaidono$\hokkaido\\ 
$^\granadano$\granada\\ 
$^\lyonno$\lyon\\
$^\stonybrookno$\stonybrook\\
$^\olddominionno$\olddominion\\
$^\oakridgeno$\oakridge\\
$^\indianano$\indiana\\ 
$^\barcelonano$\barcelona\\ 
$^\casno$\cas\\  
$^\tamuno$\tamu\\ 
$^\hamburgno$\hamburg\\ 
$^\mainzno$\mainz\\ 
$^\cantabriano$\cantabria\\ 
$^\ciematno$\ciemat\\ 
$^\sagano$\saga\\ 
$^\pekingno$\peking\\ 
$^\harvardno$\harvard\\
$^\melbourneno$\melbourne\\
$^\bristolno$\bristol\\ 
$^\oklahomastateno$\oklahomastate\\
$^\aachenno$\aachen\\
$^\jeffersonno$\jefferson\\ 
$^\fudanno$\fudan\\ 
$^\konyano$\konya\\ 
$^\uclondonno$\uclondon\\ 
$^\tsukubano$\tsukuba\\ 
$^\boskovicno$\boskovic\\ 
$^\osakano$\osaka\\ 
$^\osakainstno$\osakainst\\ 
$^\karlsruheno$\karlsruhe\\ 
$^\gakugeino$\gakugei\\ 
$^\tohokuno$\tohoku\\ 
$^\toyamano$\toyama\\
$^\mpimunichno$\mpimunich\\ 
$^\helsinkino$\helsinki\\
$^\ibarakino$\ibaraki\\ 
$^\edinburghno$\edinburgh\\ 
$^\sunyatsenno$\sunyatsen\\ 
$^\eotvosno$\eotvos\\ 
$^\osakacityno$\osakacity\\ 
$^\kyotono$\kyoto\\
$^\snuno$\snu\\ 
$^\bonnno$\bonn\\ 
$^\kindaino$\kindai\\ 
$^\durhamno$\durham\\ 
$^\wasedano$\waseda\\ 
$^\birlano$\birla\\ 
$^\hiroshimano$\hiroshima\\ 
$^\uclano$\ucla\\ 
$^\witwatersrandno$\witwatersrand\\ 
$^\irvineno$\irvine\\ 
$^\montenegrono$\montenegro\\ 
$^\northwesternno$\northwestern\\
$^\lebaneseno$\lebanese\\ 
$^\nihonno$\nihon\\ 
$^\warwickno$\warwick\\
$^\iwateno$\iwate\\ 
$^\infnromeno$\infnrome\\ 
$^\cambridgeno$\cambridge\\ 
$^\narano$\nara\\
$^\tatano$\tata\\ 
$^\infnmilanno$\infnmilan\\ 
$^\unamno$\unam\\ 
$^\southamptonno$\southampton\\ 
$^\kobeno$\kobe\\
$^\triumfno$\triumf\\ 
$^\northeasternno$\northeastern\\
$^\keiono$\keio\\ 
$^\tohokuprono$\tohokupro\\ 
$^\tokyoinstno$\tokyoinst\\ 
$^\alabamano$\alabama\\ 
$^\victoriano$\victoria\\ 
$^\nippondentalno$\nippondental\\ 
$^\calabriano$\calabria\\
$^\infncalabriano$\infncalabria\\
$^\infnlasano$\infnlasa\\ 
$^\weizmannno$\weizmann\\
$^\cinvestavno$\cinvestav\\ 
$^\manchesterno$\manchester\\
$^\wurzburgno$\wurzburg\\ 
$^\siegenno$\siegen\\
$^\desyzno$\desyz\\ 
$^\sussexno$\sussex\\
$^\utahno$\utah\\
$ ^\heidelbergno$\heidelberg\\ 
$^\msuno$\msu\\
$^\kokushikanno$\kokushikan\\ 
$^\niigatano$\niigata\\ 
$^\yamagatano$\yamagata\\ 
$^\istanbulno$\istanbul\\
$^\delawareno$\delaware\\ 
$^\iitmadrasno$\iitmadras\\ 
$^\illinoisno$\illinois\\  
$^\miyazakino$\miyazaki\\ 
$^\kogakuinno$\kogakuin\\ 
$^\hebrewno$\hebrew\\ 
$^\krakowinstno$\krakowinst\\ 
$^\technionno$\technion\\ 
$^\arizonano$\arizona\\ 
$^\ritsumeikonno$\ritsumeikon\\
$^\toyamaprefno$\toyamapref\\ 
$^\iwateprefno$\iwatepref\\ 
$^\shinsuno$\shinsu\\ 
$^\riversideno$\riverside\\ 
$^\hawaiino$\hawaii\\ 
%$^\budkerno$\budker\\
$^\ibsno$\ibs\\ 
$^\mitno$\mit\\ 
$^\nikhefno$\nikhef\\
$^\floridastateno$\floridastate\\ 
$^\marseilleno$\marseille\\ 
$^\SMUno$\SMUuniv\\
$^\sogangno$\sogang\\
$^\liverpoolno$\liverpool\\
$^\iitbombayno$\iitbombay\\ 
$^\buffalono$\buffalo\\
$^\birminghamno$\birmingham\\ 
$^\washingtonno$\washington\\
$^\michiganno$\michigan\\ 
$^\umassno$\umass\\ 
$^\jaxano$\jaxa\\
$^\iowastateno$\iowastate\\ 
$^\zhejiangno$\zhejiang\\ 
$^\infnfrascatino$\infnfrascati

}

\newpage
\hbox{ \null}

\tableofcontents
\newpage

\hbox{ \null}

\newpage

\begin{center}

{\bf {\Large Summary of the Report by Snowmass 2021 Topical Group}  }  
    
\end{center}

This report is a contributed paper written for the Snowmass 2021 study of the future of US particle physics.  It is intended to be a reference document on all aspects of the proposed International Linear Collider (ILC), an electron-positron collider spanning the range of center of mass energies  from the $Z$ pole to 1~TeV.  Although the report is written specifically from the viewpoint of the ILC project, much of the information we have gathered applies equally well to other Higgs factory proposals. Connections to other Snowmass Frontiers are discussed.   To make this information more useful, we reference it here according to the Snowmass 2021 organization.

\bigskip

\bigskip

\noindent{\bf General}

\begin{itemize}
\item {\bf All: } A summary of the report and of the ILC physics case is presented in Chapters 1 and 2.  The current status of the ILC and its potential realization in Japan is presented in Chapter~3.  A general orientation to ILC physics and experimentation is presented in Chapter 5. 

\end{itemize}

\noindent{\bf Energy Frontier}

\begin{itemize} 

  \item {\bf EF01}: Material on the ILC study of the Higgs boson is presented  in Chapters 8 and 10, particularly in Secs. 8.1, 8.2, and 10.2.   The ILC expectations for the precision of Higgs boson couplings are explained in Chapter 12.

\item {\bf EF02}: Material on the implications of the study of the Higgs boson and tests of Beyond-Standard-Model scenarios is presented in Secs. 8.1 and 8.2, and in Chapter 14.

\item {\bf EF03}: Material on study of heavy quarks at the ILC is presented in Secs. 9.3 and 10.1.

\item {\bf EF04}: Material on precision electroweak measurements at the ILC is presented in Chapter~9 
and material on precision theory for the ILC and the interpretation of ILC data using Standard Model Effective Field theory is presented in Chapter 12.

\item {\bf EF05}: Material on precision QCD at the ILC is presented in Sec. 8.4.

\item {\bf EF08}: Material on searches for supersymmetric particles and extended Higgs sectors at the ILC is presented in 
   Sec. 10.5 and 14.2.

\item {\bf EF09}: Material on ILC searches for a wide variety of Beyond-Standard-Model theories, including searches for new particles and decays and precision probes, is presented in Secs.~8.2, 10.1, 10.4, 10.5, 10.6, and Chapters 11 and 14.

\item {\bf EF10}: Material on the ILC searches for particles of dark matter and dark sections is presented in Secs. 10.5, 10.6, 11.3, and 14.1.

\end{itemize}

\newpage
\noindent{\bf Neutrino Physics Frontier}

\begin{itemize} 

  \item {\bf NF01}: Material on searches for TeV-mass particles appearing in models of neutrino mass is presented in Sec. 10.5. 

\end{itemize} 

\noindent{\bf Rare Processes and Precision Measurements}

\begin{itemize} 

  \item {\bf RF06}:  Material on searches for dark sector particles in the ILC fixed target program is presented in Chapter 11.

\end{itemize} 

\noindent{\bf Cosmic Frontier}

\begin{itemize} 

  \item {\bf RF06}:  Material on the ILC searches for particles of dark matter and dark sections is presented in Secs. 10.5, 10.6, 11.3, and 14.1.

\end{itemize} 

\noindent{\bf Theory Frontier}

\begin{itemize} 

   \item {\bf TF02:}  Material on use of Effective Field Theory in the interpretation of ILC data is presented in Chapter  12.

  \item {\bf TF06:}  Material on the precision theory for ILC is presented in Secs. 8.4 andd 12.1.

 \item {\bf TF07:}  Material on the theoretical interpretation of ILC results is presented in Chapters 13 and 14.

 \item {\bf TF06:}  Material on precision theory for ILC is presented in Secs. 8.4 and 12.1.

\item {\bf TF11:}  Material on searches for TeV-mass particles appearing in models of neutrino mass is presented in Sec. 10.5. 
  
\end{itemize}

\noindent{\bf Accelerator Frontier}

\begin{itemize} 

   \item {\bf AF03:}  Material on the ILC accelerator design and R\&D on ILC accelerator technologies is presented in Chapter 4.

  \item {\bf AF04:}  Material on extensions of the ILC to multi-TeV energies is presented in Chapter 15.

 \item {\bf AF06:}  Material on advanced accelerator technologies for the   ILC laboratory is presented in Secs.~15.4 and 15.5. 

 \item {\bf AF07:}  Material on many aspects of ILC accelerator technology is presented in Chapters 4 and 15.  Material on the measurement of energy, luminosity, and beam polarization at the ILC is presented in Section 5.4.

\end{itemize} 

\newpage
\noindent{\bf Instrumentation Frontier}

\begin{itemize} 

\item {\bf IF02 - IF07, IF09:}   The ILD and SiD detectors proposed for the ILC are described as integrated concepts in Chapter 6.  Material on new proposed technologies is presented in Section 6.4.   The material here cuts across the various Instrumentation topical groups.

\end{itemize}

\noindent{\bf Community Engagement Frontier}

\begin{itemize} 

\item {\bf CommF07:}  Material on sustainable accelerator laboratory design and ``Green ILC" is presented in Sec. 4.1.6.

\end{itemize}

\newpage

\chapter{Introduction}   
\label{chap:intro}

The ILC is a proposed next-generation $e^{+} e^{-}$ collider. It starts with $\sqrt{s}=250$~GeV as the Higgs factory.  The precision study of the Higgs boson is the next major goal in collider physics; the ILC will reach important benchmarks in the measurement of the Higgs boson couplings.   Such high precision measurements will provide guidance to the next energy scale for future facilities. At the same time, the ILC provides numerous searches for new physics with monophoton or invisible and exotic Higgs decays, for example into a light dark sector. It can host ancillary experiments with beam dump and/or near IP detectors to search for long-lived and invisible particles. It is technologically mature with a well-understood cost that is about the same as the LHC. The linear design allows further phases at higher as well as lower energies. The ILC can  have a dedicated run at the $Z$ resonance, improving the measurement of the precision electroweak observables by an order of magnitude.  Its length can be  extended to reach the  the $t\bar{t}$ threshold and open $t\bar{t}$ as well as $t\bar{t}H$ production at 500--550~GeV. The site was specifically chosen to allow for an upgrade up to 1~TeV with the same technology, for the Higgs self-coupling measurement and many new physics searches. Superconducting RF cavity technology has  ample room for improvements, allowing for even a 3--4~TeV collider in the same tunnel. Future technologies such as plasma wakefield or dielectric laser accelerators could reach the tens of TeV energy range.

This report is intended to be a comprehensive sourcebook on the ILC, discussing plans for the  accelerator, the experimentation, and the physics analyses and also the physics context and theoretical implications of the ILC measurements.  We hope that it will be useful to those who would like to better understand or evaluate the ILC proposal.  Also, since the physics programs of all proposed Higgs factories are closely aligned, most of our physics discussion will also be helpful in understanding the physics prospects for all facilities of this type. 

\section{Context for the ILC}

We first describe the context for the ILC as it has evolved over half a century of development in particle physics.

%The ILC is a large-scale scientific facility for research in particle physics under development. It aims for collision of high-energy beams of electrons and positrons with the center-of-momentum energy of 250~GeV. It will provide a well characterized environment to study interaction of elementary particles at energies typical of the environment only a trillionth of a second after the Big Bang. The initial focus is to understand the properties of the newly discovered Higgs boson in great precision, which is believed to point to missing pieces in the Standard Model of particle physics. At the same time, it will search for dark matter, study the stability of the Universe, look for clues of unification of forces and matter, and address many other scientific questions. The linear design allows for extension in the future to reach higher energies of collisions. It can also host ancillary experiments at the beam dump, with extracted beams, and near the collision point. 

 The need for a linear collider was recognized already in the 1960's given the energy loss due to unavoidable synchrotron radiation from beams in circular colliders. To achieve power-efficient acceleration, the development of superconducting radio frequency (SCRF) cavities started in earnest in the 1980's. Over four decades, intensive research and development achieved much higher acceleration gradients and reduced the costs of SCRF by more than an order of magnitude. SCRF  provides better tolerances compared to room-temperature klystron-based designs, and was chosen as the ILC  technology in 2005 by the International Technology Recommendation Panel chaired by Barry Barish. The International Committee for Future Accelerators, a body created by the International Union of Pure and Applied Physics in 1976 to facilitate international collaboration in the construction and use of accelerators for high energy physics, recommended the launch of the Global Design Effort (GDE) to produce a Technical Design Report (TDR) for the ILC as an international project. The GDE successfully produced the TDR in 2013 with a purposely site-independent design~\cite{Behnke:2013xla,Baer:2013cma,Adolphsen:2013jya,Adolphsen:2013kya,Behnke:2013lya}.

There is also a long history of discussions on the scientific merit for the ILC. The energy scale of the weak interaction, which makes the Sun burn and allows the synthesis of the chemical elements, was pointed out to be around 250~GeV in 1933 by Enrico Fermi. The need to reach this energy scale has been obvious since then, though the precise target energy was not clear. Early discussions for linear colliders called for 1000~GeV as a safe choice for guaranteed science output. The  GDE focused on 500~GeV for the study of the Higgs boson based on the precision electroweak data of early 2000's. It was only in 2012 that the Higgs boson was discovered at the Large Hadron Collider (LHC) at CERN.  This provided a clear target energy for the ILC at 250~GeV. In the same year, the Japanese Association of High-Energy Physicists (JAHEP) issued a report expressing interest in hosting the ILC in Japan with 250~GeV center-of-momentum energy as its first phase. The European Strategy for Particle Physics updated in 2013 highlighted ``{\it the ILC, based on superconducting technology, will provide a unique scientific opportunity at the precision
frontier.}''  This  was followed by the report of the US Prioritization Panel for Particle Physics Projects (P5) that listed ``{\it Use the Higgs boson as a new tool for discovery}'' as the first among the science drivers for particle physics and stated ``{\it As the physics case is extremely strong, all (funding) Scenarios include ILC support}''.

Intense discussions ensued worldwide on how to realize the ILC. The Japanese government instituted a multitude of committees looking into the scientific and societal merit of hosting the ILC in Japan as well as its technological feasibility and costs. The US government encouraged Japan to host the ILC, with letters from the Secretary of Energy and the Deputy Secretary of State to Japanese Ministers. The 2020 update of the European Strategy for Particle Physics stated ``{\it An electron-positron Higgs factory is the highest-priority next collider}'' and added ``{\it The timely realisation of the electron-positron International Linear Collider (ILC) in Japan would be compatible with this strategy and, in that case, the European particle physics community would wish to collaborate.}'' Following this  update, ICFA created the International Development Team (IDT) in August 2020 to prepare for the creation of prelab towards the realization of the ILC.  The  IDT is hosted by KEK, the national laboratory for high-energy accelerators in Japan.

Since its launch, the IDT has collected information, worked with ICFA, interacted with the community, consulted the funding agencies, to formulate what is required of the ILC Pre-Lab. The Pre-Lab is envisioned to be a four-year process, finalizing the Engineering Design Report for the ILC in a site-specific fashion for the Kitakami mountain range in northern Japan, forging agreements among international partners, and recommending specific experiments for the ILC. 

\section{Outline}

This report will update the information contained in the documents prepared by the ILC for the European Strategy for Particle Physics.  Those documents include a comprehensive review of the ILC up to 2019~\cite{Bambade:2019fyw} and a review of the ILC capabilities for precision measurement~\cite{Fujii:2019zll}.  A comprehensive bibliography for the ILC, up to mid-2020, can be found in \cite{Fujii:2020pxe}.

The outline of this report is the following:  Chapter 2 will
present the most important points of the physics case for the  ILC.
In Chapter 3, we will present the status of the current plan to   realize the
ILC in Japan.

Chapter 4 will present the current state of the ILC accelerator
design, including details of the various ILC energy stages up to a CM
energy of 1~TeV.   This chapter will also discuss the prospects for
extension of the ILC to even higher energies and other issues for ILC
accelerator R\&D.   It will conclude with a discussion of the
opportunities and tentative plans for US contributions to the ILC
accelerator.

Chapter 5 will review the basic aspects of the physics
environment of the ILC---the major physics processes, the plan for
stage-by-stage improvement in the energy and luminosity, and the key
role played in the experimental program by electron and positron beam
polarization.

Chapter 6 will describe the ILC detectors.  We will begin with
descriptions of the two current proposed detectors ILD and SiD,
including the expected measurement capabilities and issues for which
further R\&D is needed.   The chapter will conclude with a survey of
new technologies that offer the promise of further improvements in the
detector capabilities.  Chapter 7 will describe the simulation
framework used in studying the detector capabilites and projecting the
measurement accuracy of physical observables. 

Chapter 8 will describe the planned physics measurements at a CM
energy of 250~GeV.  These include measurements on the Higgs boson and
the $W$ boson, measurements of 2-fermion production, the ILC program
in precision QCD, and descriptions of a number of relevant new
particle searches.

Chapter 9 will describe the ILC program in precision electroweak
measurements.  This includes improvements of the precision electroweak
parameters of the $Z$ boson, both at 250~GeV through the radiative
return reaction $\ee\to \gamma Z$  and through a dedicated program of
running at the $Z$ resonance.  It also includes high-precision
measurements of the $W$ boson mass and width and improved measurements of these
properties for the $Z$ boson.

Chapter 10 will describe the planned physics measurements at CM
energies of 350~GeV and above, up to 1~TeV.   The topics here include
the ILC program of precision measurements of the top quark, the
completion of the measurement of the Higgs boson profile, including the
measurements of the Higgs self-coupling and the top quark Yukawa
coupling, and the ultimate capabilities of the ILC in triple gauge boson
couplings and new particle searches.

Chapter 11 will describe the fixed-target program that the intense,
high-energy electron and positron beams of the ILC will make
available.

Chapters 12-14 will address the interpretation of the ILC
measurements.  Chapter 12 will begin with a review of the status of
precision SM theory for ILC processes.  It will then discuss the
network of tests of the SM available at the ILC. This chapter will
present a unified description of these tests using Standard Model
Effective Field Theory (SMEFT), reviewing the conceptual basis of this
approach and demonstrating its power in providing a unified
interpretation of the full set of ILC experimental measurements.
Chapter 13 will present a theoretical context for the expectation that
the ILC will discover deviations from the SM predictions  and the
relation of such deviations to the most important question now being
asked in particle physics.  Chapter 14 will bring these two lines of
analysis together, quantifying the ability of the ILC to overturn the
SM and to provide evidence of  the more correct underlying
model for particle physics.

Finally, Chapter 15 will lay out some possible futures for the ILC
laboratory with accelerators at still higher energies offering multi-TeV
and multi-10-TeV electron, positron, and photon collisions.

\chapter{Outline of the ILC Physics Case} 
\label{chap:case}

The physics motivation for constructing the ILC is very strong.  The
flagship program of the ILC is the study of the Higgs boson at a much higher
level of precision than will be possible at the LHC. 
The ILC will also carry out precision measurements of the other heavy and
still-mysterious particle in the Standard Model (SM), the top quark.
It will carry out a program of specific searches for postulated new
particles in regions that are very difficult for the LHC to access.
Beyond these specific targets,  the ILC will greatly
improve the level of our understanding of the full set of electroweak
processes in the region up to its final CM energy. In the context of
Standard Model Effective Field Theory (SMEFT), these measurements will work
together to strongly challenge the Standard Model.
   In this chapter, we will introduce each of these
points and prepare the ground for a more detailed discussion later in
this report.

We begin with the 125~GeV Higgs boson. This particle is the
centerpiece of the SM, yet still we know little about it.  From the LHC experiments, we now
know that the couplings  of the Higgs boson agree with
those predicted in the SM, at the level of 20\% accuracy for the major
decay modes. However, this is not nearly sufficient to distinguish the
minimal SM description of the Higgs boson from those of competing
models. According to SMEFT, the deviations of Higgs couplings from SM
predictions are parametrically of the order of $m_h^2/M^2$, where $M$
is the mass scale of additional new particles.   Given the constraints
from particle searches at the LHC, these deviations are expected to be
at most of order 5-10\%, and, to claim discovery of new physics, the
deviations must be measured with high significance. This calls for a
dedicated program to measure the full suite of couplings of the Higgs
boson, and to push the precision of those measurements to the 1\%
level and below.   This requires an $\ee$ collider such as the ILC.

The ILC is well-positioned to carry out this program of
measuring 
the complete profile of the Higgs boson couplings.   At 250~GeV, the
ILC accesses the reaction $\ee\to Zh$, producing  about half a million Higgs bosons, each tagged by a
recoil $Z$ boson at the lab energy of 110 GeV.  Looking in the opposite
hemisphere, we will measure all of the branching ratios of the Higgs
boson down to values of order $10^{-4}$.   These include 10 different
modes of Higgs decay predicted in the SM, and also, possibly, 
invisible, partially-invisible, flavor changing, and other exotic
modes of Higgs decay. By counting recoil $Z$ bosons, we
will obtain an absolute measurement of the cross section for $\ee\to
Zh$, which can then be translated into absolute normalizations of
the various partial widths.

At the second stage of the ILC at 500~GeV, the $W$ fusion reaction
$\ee\to \nu\bar\nu h$ opens up.   This reaction offers an event
sample of about 1 million Higgs boson events in which the only visible
signals in the event are from Higgs decay.  This will not only allow
new measurements to complement the
250~GeV data but also improved understanding of such issues as
$h \to b\bar b/gg/c\bar c$ separation, angular distributions in $h\to
WW^*$, and CP violation tests in $h\to \tau^+\tau^-$.  The combination
of the 250 and 500~GeV programs will give high confidence in any
deviations from the SM detected in the Higgs boson data.

Running at 500~GeV and above also gives access to two important Higgs
couplings that cannot be probed directly in Higgs decays, the Higgs
coupling to $t\bar t$ and the Higgs self-coupling.  Our studies of the
ILC capabilities at 1~TeV predict truly archival measurements of these quantities,
with errors below 2\% and 10\%, respectively.

Different models of new physics beyond the SM affect the various Higgs
couplings differently.  Since the ILC program can determine
each Higgs coupling of the large set available, individually and without
ambiguity, it will provide a pattern of  deviations from the
predictions of the SM that can distinguish different hypotheses about
the underlying model. 

The ILC program of experimental measurements on the Higgs boson will
be described in Chapters 8 and 10 of this report, and the interpretation
of these measurements will be discussed in some detail in Chapters 12
and 14.

The 500~GeV ILC will also give an excellent opportunity for the
measurement of the mass and properties of the top quark.  The mass of
the top quark will be determined
from the position of the sharp threshold in $\ee\to t\bar t$.  The
threshold shape is determined by the short-distance top quark mass, so
that the mass defined in this way, which is needed for high-precision
predictions in and beyond the SM, is determined from the data without
ambiguity.   At $\ee$ colliders, the electroweak form factors of the top quark, which
contain crucial information  about the role of the top quark in
electroweak symmetry breaking, determine the primary top quark pair production cross
section.  Thus, very high precision measurements of these form factors are possible.
The ILC program of measurements on the top quark will be discussed in
Chapter 10 of this report. 

Beyond these SM particles, the ILC has the potential to access
new particles predicted in models beyond the SM.  The LHC experiments
have given powerful access to proposed new particles with couplings to
QCD, but their capability to discover particles with only electroweak
couplings is limited.  All  LHC searches come with caveats
concerning the sizes of electroweak cross sections, the expected decay
patterns, the amount of missing energy, and other features.  Searches
at the ILC will allow these caveats to be eliminated, giving access
to systems with large missing energy and other challenging features, in
particular, to supersymmetry partners of the Higgs boson and to dark
matter in models with compressed spectra.  These issues will be
described in Sections 8 and 10 of this report.

The ILC will dramatically improve the precision of our understanding
of electroweak reactions. For example, the reaction $\ee\to W^+W^-$ has
strong
dependence on both initial- and final-state polarizations. At the ILC,
we will have beam polarization in the initial state and complete
reconstruction of the final state, allowing us
to dissect the structure of the triple-gauge-boson coupling.   The
reactions $\ee\to f\bar f$ allow searches for additional electroweak
resonances that access the 10-TeV mass range and are flavor- and
helicity-specific.  The study of radiative-return events ($\ee\to
\gamma Z$) at 250 GeV will already improve the our precision knowledge
of $Z$-fermion couplings beyond that obtained at LEP.  A dedicated ILC
``Giga-Z'' run at the $Z$ resonance ($5\times 10^9$ $Z$s) will improve
the precision of most electroweak observables by more than an order of
magnitude.   These measurements and others are described in Chapters
8, 9, and 10.

The simplicity of hadronic final states in $\ee$ annihilation also
allows not only higher precision tests of QCD but also new observables
that give insight 
into features such as jet substructure that have come to light at the
LHC.   This new program of QCD measurements will be described in
Chapter 8.

The ILC will also make available the most intense and highest-energy
electron and positron beams for beam dump and dedicated fixed-target
experiments to search for light weakly-interacting particles.   This
program will be described in Chapter 11.

These measurements are very powerful already when they are considered
separately, but they take on increased power when they are combined 
in a coherent way to
stress-test the SM.   This becomes particularly clear
when the full set of SM tests is analyzed using SMEFT.   In this
approach, corrections to the SM are described by contributions to an
effective Lagrangian  from operators of dimension 6 and higher
invariant under the well-tested SM gauge symmetries.  There is only
one Lagrangian; its higher-dimension operators generally contribute to
many electroweak reactions and so receive an array of experimental
constraints. We will describe this method in detail and give examples
of its powerful use in Chapter 12.

There is one more important point that we should make concerning the
program of measurements of the ILC. The goal of testing the SM is not
simply to improve the error bars. It is widely appreciated that the Standard
Model of particle physics, though it is very successful in
describing the results of experiments, is not adequate as a complete
theory of elementary particles.  The goal of the ILC experiments
must be to {\it  prove that the SM is  incomplete}, and, even more, to show the path to
a better understanding of nature.

One way to demonstrate the inadequacy of the SM is to discover a new
resonance that the SM does not account for.  This was the primary goal
of the LHC experiments.  So far, no such resonance has appeared.
There is still considerable room to discover a new resonance at
the HL-LHC, but that window is closing. It is important to open
a new, complementary window, and this is what the ILC's capability for
precision tests of the SM will make available.

 It is not straightforward, though, to demonstrate a deviation
 from the SM through
 precision measurements. First, of all, the deviation must be
 observed with high statistical significance.  Second, the possible
 systematic uncertainties that could mimic the deviation must be under
 complete control.  This calls for multiple cross-checks on the
 sources of uncertainty and, if possible, measurements with different
 sources of systematic uncertainty that can be compared.  Finally, the view
 provided by precision measurements cannot be one-dimensional; rather, it
 should be part of a collective program that has the power to show a
 pattern of discrepancies.   In the best case, a pattern of
 well-established deviations from the SM can point to a common origin
 and thus indicate the nature of the true underlying theory.

The experimental program of the ILC is well-equipped to address these
points.  The general simplicity and cleanliness of $\ee$ annihilation
provides an excellent starting point in the quest for precision.
This environment allows the construction of detectors with high
segmentation and very low material budget, allowing collider event
measurements of unprecedented quality.  In the energy region of the
ILC, electroweak cross sections have a large and well-understood
dependence on beam polarization. With the two signs for each of the
electron and positron beam polarizations, the ILC will provide four 
distinct event samples, each with a distinct combination of physics 
process.  By comparing these samples, we can determine detector
performance and measure important backgrounds from data.  As we
have noted above for the Higgs boson program, changes in the center of
mass energy can also bring in new physics processes that access and
cross-check the variables targetted in precision measurements. The
enabling features of the ILC experimental environment will be
discussed in Chapter 5.   The
capabilities of detectors for the ILC and strategies for further
improvement will be discussed in Chapter 6.  Throughout the succeeding
chapters,we will show these elements at work to ensure the high
quality of the  ILC measurements.

The ILC thus offers a new approach to the discovery of physics beyond the
SM, one of great capability and robustness.   These experiments must be
carried out.  They have the power to lead us to a new stage in our
understanding of fundamental physics.

\chapter{Route to the ILC}  
\label{chap:ILCorg}

This chapter will describe the  organization, schedule, and prospects for the ILC as we currently understand these as of March 1, 2022.  The future plans for the ILC organization are subject to decisions by ICFA in the coming years. In future versions of this report, we will update this section as required.

\bigskip

The worldwide community of particle physicists pursued the dream of realizing a high-energy $e^+ e^-$ linear collider since 1960s. By the end of 20th century, it was clear that such a machine can be built only as an international project because of its scale. The International Committee for Future Accelerators (ICFA) launched the serious effort to come up with a worldwide proposal in November 2003~\cite{initialICFA},  first by creating International Technology Recommendation Panel (ITRP). Under the chairmanship of Barry Barish from Caltech, the ITRP recommended~\cite{ITRP} that the L-band superconducting RF cavity based on niobium is favored over the warm X-band copper-based cavity. ICFA unanimously endorsed this choice at its meeting in August 2004 in Beijing. This marked the beginning of the International Linear Collider (ILC) project.

ICFA launched the Global Design Effort (GDE)~\cite{GDE} in March 2005, with Barry Barish as the director, to produce a technical design for the ILC following the technology decision. Barish was assisted by three regional directors, Michael Harrison (Americas), Kaoru Yokoya (Asia), and Brian Foster (Europe). The design effort was specifically site unspecific, and the GDE was truly an international effort with more than 2000 scientists from more than 300 institutions in 49 countries. It aimed for a center-of-momentum energy of $\sqrt{s}=500$~GeV, with expandability up to 1~TeV. It concluded its mission with the publication of the Technical Design Report (TDR) in June 2013~\cite{ILCTDR}. The site was left for a bid from interested countries.

It was a fortunate coincidence that the discovery of the Higgs boson was announced on July 4, 2012, a year before the publication of the TDR. Given its mass of 125~GeV, it became clear that an $e^+ e^-$ linear collider would be a perfect machine for the precision study of the Higgs boson. ICFA decided to launch a new organization named Linear Collider Collaboration (LCC), with a Linear Collider Board (LCB) as an oversight body, to follow the GDE and coordinate
coordinate global research and development efforts for two next-generation particle physics colliders: the ILC, and the Compact Linear Collider (CLIC) that published its Conceptual Design Report in 2012. The mission of the LCB and LCC was to promote constructing a linear collider as a global project. Members of the collaboration included approximately 2000 accelerator and particle physicists, engineers and other scientists from around the world. ICFA appointed Sachio Komamiya as the chair of the LCB and Lyn Evans as the director of the LCC. Eavns was joined by three associate directors, Mike Harrison for the ILC, Steinar Stapnes for CLIC, and Hitoshi Yamamoto for Physics and Detectors, the deputy director Hitoshi Murayama, and 
three regional directors, Akira Yamamoto (Asia), Brian Foster (Europe), Harry Weerts (Americas), officially starting the LCC in March 2013 with a mandate for three years. The mandate was extended in December 2016, with Harrison replaced by replaced by Shichiro Michizono and Yamamoto by Jim Brau. The chair was succeeded by Tatsuya Nakada (EPFL). In October 2017, the LCC published a  report~\cite{stagingreport} describing the machine parameters and cost for a 250 GeV machine as the first stage.

Before the launch of the LCC, the Japanese Association of High-Energy Physicists (JAHEP) issued a report in October 2012~\cite{JAHEP}, expressing interest in hosting the ILC in Japan with 250~GeV center-of-momentum energy as its first phase, followed by an upgrade to 500~GeV, maintaining the extendability to 1~TeV. This report marked the beginning of an international discussion to build the ILC with Japan as its potential host in mind, and the LCC put an emphasis on adopting the TDR to a site in Japan. 

In parallel, the European Strategy for Particle Physics updated in 2013 highlighted ``{\it the ILC, based on superconducting technology, will provide a unique scientific opportunity at the precision
frontier.}''  This  was followed by the report of the US Prioritization Panel for Particle Physics Projects (P5) that listed ``{\it Use the Higgs boson as a new tool for discovery}'' as the first among the science drivers for particle physics and stated ``{\it As the physics case is extremely strong, all (funding) Scenarios include ILC support}''.

To implement such vision for the ILC, KEK organized the International Working Group that published its report in October 2019~\cite{IWG}. It laid out a framework for cost sharing. Civil engineering will be a responsibility of the Host State. Accelerator components will be provided by all Member States. Construction of conventional facilities will be managed by the ILC Laboratory, and the Host State will provide a significant part of the conventional facilities. The operational cost should be shared among Member States, and the way of sharing should be agreed upon before the construction begins. In addition, it proposed a preparatory laboratory (Pre-Lab) would be established based on a mutual understanding of the laboratories around the world and with the consent of their respective governmental authorities. The Pre-Lab would coordinate the preparatory tasks needed before the construction of the ILC. The Pre-Lab would also assist the inter-governmental negotiations, which are expected to take place in parallel. KEK will play a central role as the host laboratory of the Pre-Lab. After an inter-governmental agreement on the ILC, the Pre-Lab is expected to transition to a full ILC Laboratory. The ILC Laboratory will be responsible for the construction and operation of the ILC accelerator complex.

The Japanese government officially expressed interest in the ILC project at the meeting of the
LCB, with the participation of members of the International Committee for Future Accelerators (ICFA), in March 2019 held at the University of Tokyo~\cite{ICFAmeeting2020}. However, it stayed short of expressing interest
in hosting the ILC.  In February 2020, the Japanese government repeated its position in the joint
LCB-ICFA meeting held at SLAC in February 2020.   In the same year, the 2020 update of the European Strategy for Particle Physics stated ``{\it An electron-positron Higgs factory is the highest-priority next collider}'' and added ``{\it The timely realisation of the electron-positron International Linear Collider (ILC) in Japan would be compatible with this strategy and, in that case, the European particle physics community would wish to collaborate.}'' It should also be noted that more than a hundred Diet members express interest in hosting the ILC in Japan, as well as the local politicians in the area of the proposed site. 

Given all these developments, ICFA  in August 2020,  launched the International Development Team (IDT)~\cite{ICFA2020,IDT2}, replacing the LCC and the LCB, 
 ``{\it as the first step towards the preparatory phase of the ILC project, with a mandate to make preparations for the ILC Pre-Lab in Japan}'' by the end of 2021.

\section{International Development Team}

The mission of the IDT~\cite{IDT} is to ``{\it make preparations for the ILC Pre-Lab in Japan, as the first step of the preparation phase of the ILC project.}'' ICFA appointed Tatsuya Nakada as the chair of the IDT hosted by KEK. The Executive Board (EB) also includes three regional representatives Steinar Stapnes (Europe), Andy Lankford (Americas), Geoffrey Taylor (Asia-Pacific), in addition to chairs of working groups. Each working group has a large number of scientists involved from around the world as can be seen from the websites linked below.

Nakada chairs both the EB and the Working Group 1 (WG1)~\cite{WG1} whose mission is to carry out, together with the Executive Board, the key tasks of developing the function and organizational structure for the ILC Pre-Lab and to support the preparation of scenarios for contributions with national and regional partners. Current members are Paul Collier (CERN), Bruce Dunham (SLAC), Eckhard Elsen (DESY), Brian Foster (Oxford), Juan Fuster (Valencia), Stuart Henderson (Jefferson Lab), Reiner Kruecken (TRIUMF), Joe Lykken (Fermilab), Maksym Titov (Saclay), and Satoru Yamashita (UTokyo).

The Working Group 2 (WG2)~\cite{WG2} is responsible for the accelerator design chaired by Shinichiro Michizono. WG2 is responsible for the preparation of the work plan of the ILC Pre-Lab. There are four subgroups: (1) Superconducting RF Technology (SRF), (2) Damping Rings (DR) / Beam Delivery System (BDS) / Dump, (3) Sources, (4) Civil Engineering. The four subgroups of WG2 are charged to discuss the technical preparation plan and possible schedule and international sharing at the Pre-Lab. 

\begin{figure}[t]
\centering
\includegraphics[width=0.7\textwidth]{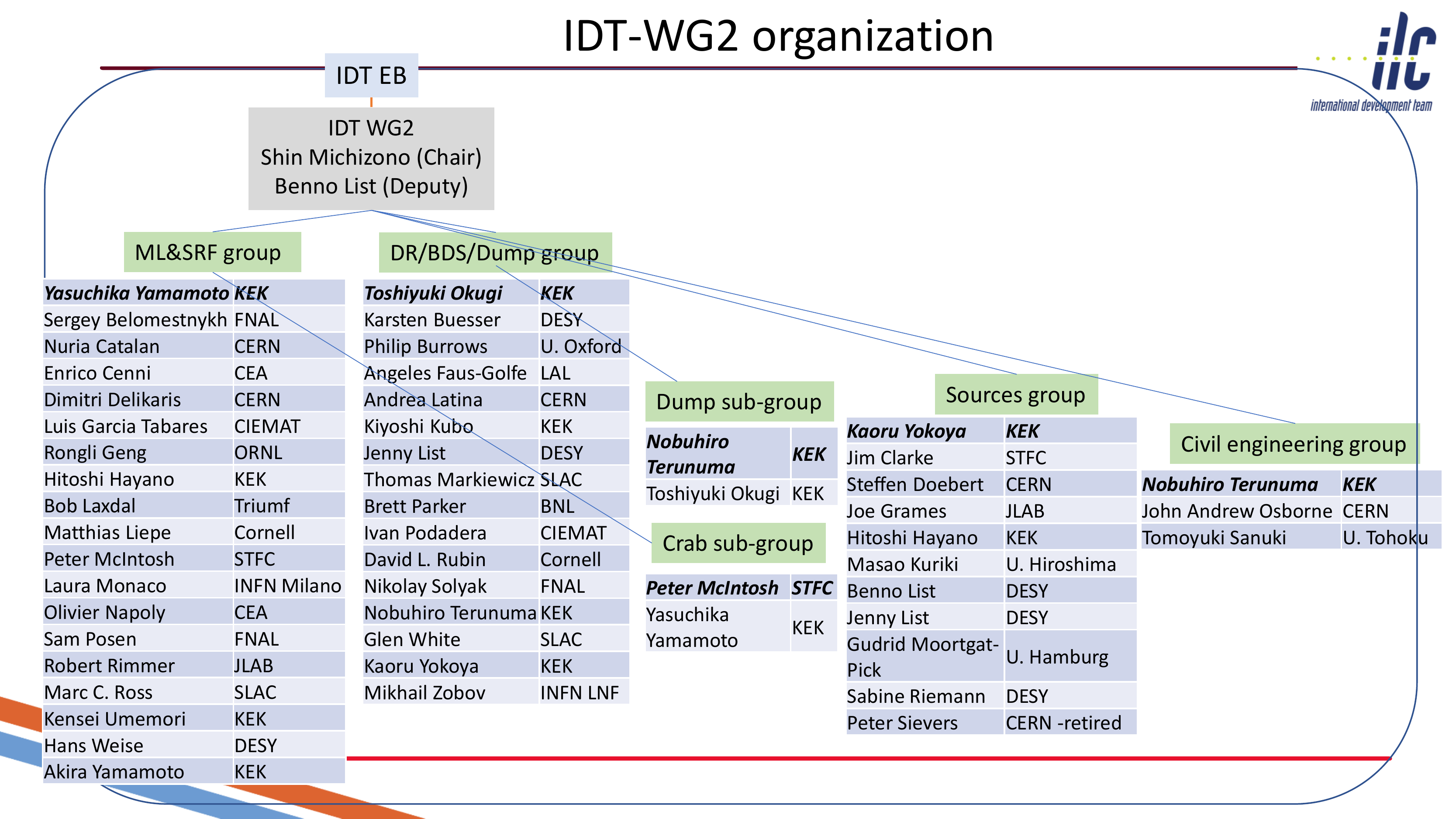}
\caption{The current organization of the IDT WG2.}
\end{figure}

The Working Group 3 (WG3)~\cite{WG3}  is responsible for the physics and detector activities chaired by Hitoshi Murayama. WG3 aims to raise awareness and interest in the ILC development and expand the community, support newcomers to get involved in physics and detector studies, encourage new ideas for experimentations at the ILC. The WG3 Steering Group consists of the coordinator (WG3 Chair), two deputy coordinators, subgroup conveners, and additional members of the Steering Group. The Physics Potential and Opportunities Subgroup~\cite{WG3physics} has many conveners for specific subjects. 

\begin{figure}[t]
\centering
\includegraphics[width=0.7\textwidth]{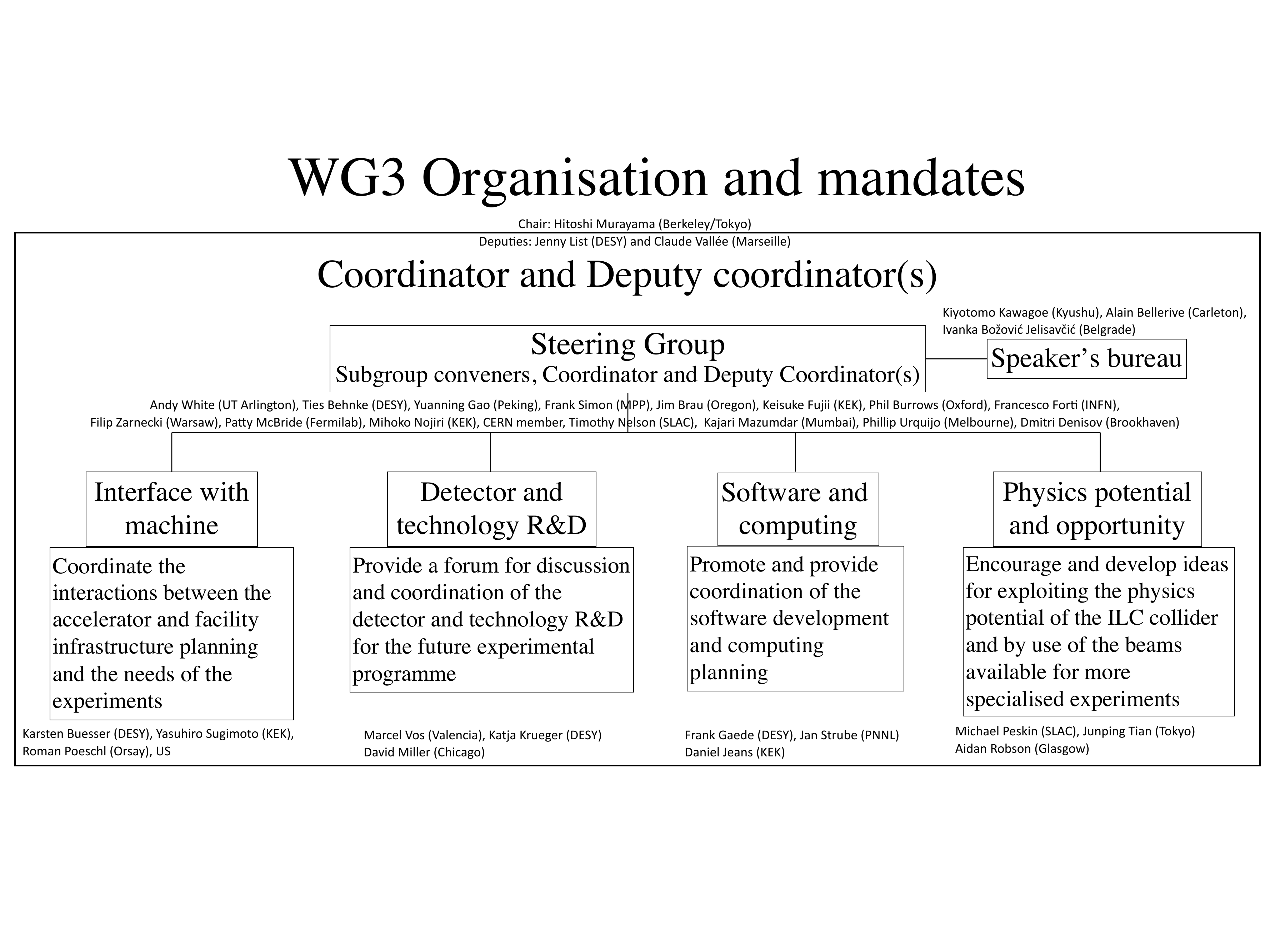}
\caption{The current organization of the IDT WG3.}
\end{figure}

The IDT organized the ILC workshop on Potential Experiments (ILCX2021). At this workshop, it discussed expansion of the scope of the ILC facility beyond the collider experiments for the precision Higgs physics to include~\cite{outsidethetunnel}:
\begin{itemize}
\item potential beam-dump experiments, fixed-target experiments, forward and off-axis detectors, to search for dark matter, long-lived particles, axion, etc,
\item simulating Hawking radiation with strong QED that combines the ILC beam with powerful laser,
\item nuclear physics applications for studies of pentaquarks, tetraquarks, electron-nucleus scattering,
\item industrial applications with neutrons and muons from the beam dump such as studying soft error in self-driving automobiles, archeology, and non-destructive inspection of cargos,
\item hard X-ray free electron laser for biological, medical, and material science,
\item the Green ILC concept to recover spent energy of the beams for other purposes.
\end{itemize}

\section{ILC Pre-Lab}

The IDT put out a \href{https://arxiv.org/abs/2106.00602}{proposal} for the ILC Pre-Lab~\cite{InternationalLinearColliderInternationalDevelopmentTeam:2021guz} on June 1, 2021, to fulfill its mandate.  It proposes a four-year Pre-Lab phase of the ILC for five purposes:
\begin{itemize}
\item Completion of technical preparations and production of engineering design
documents for the accelerator complex.
\item Compilation of design studies and documentation of the civil engineering and
site infrastructure work, and of the environmental impact assessment.
\item Community guidance to develop the ILC physics programme that will fully
exploit its potential.
\item Provision of information to national authorities and to Japanese regional authorities to facilitate development of the ILC Laboratory.
\item Coordination of outreach and communication work.
\end{itemize}
The proposed framework consists of mostly in-kind contribution from various laboratories around the world. The Pre-Lab is envisioned to be a legal entity in Japan to coordinate the activities with support from KEK and Japanese universities. 

\begin{figure}[t]
\centering
\includegraphics[width=0.7\textwidth]{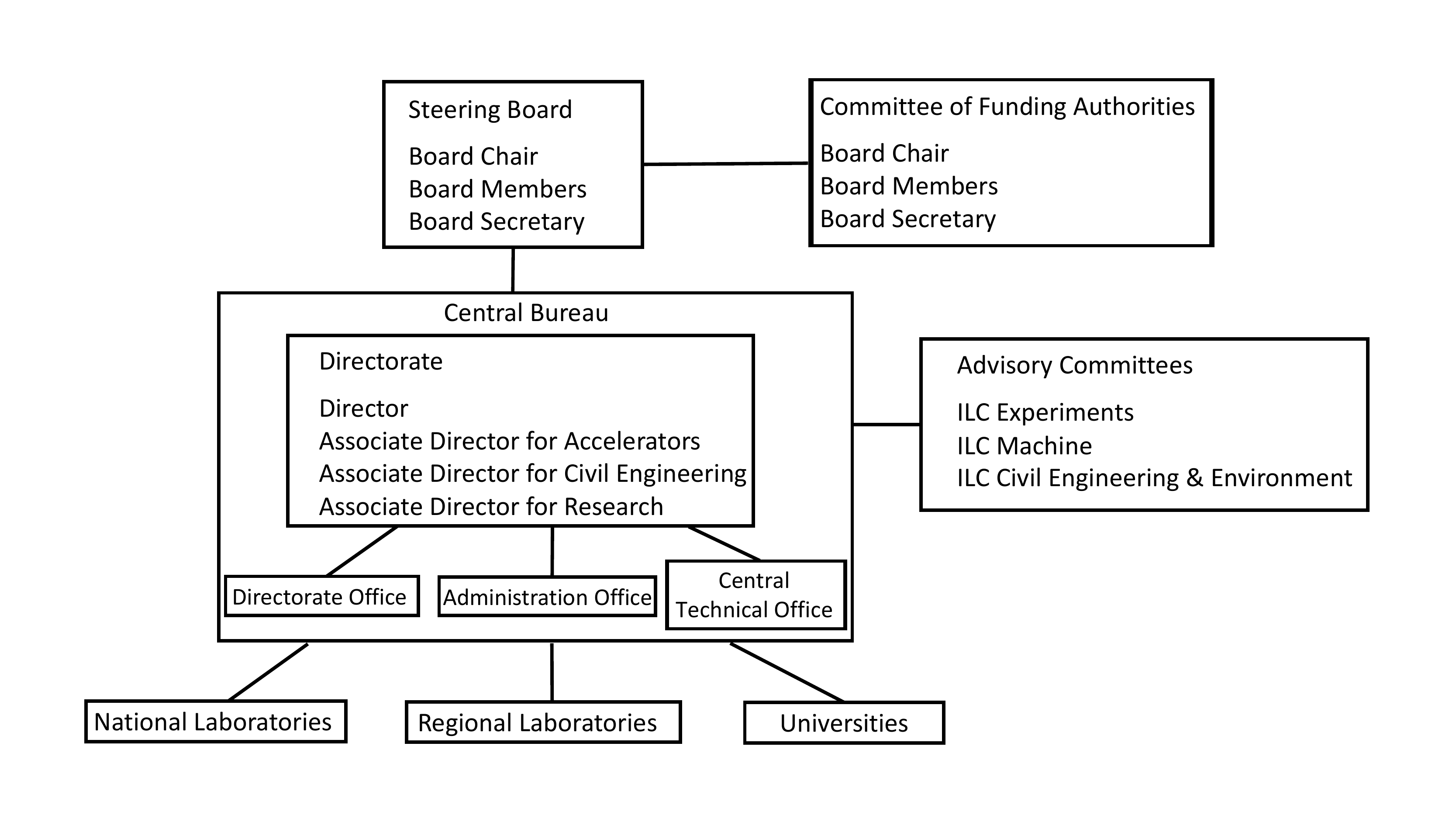}
\caption{The proposed organization of the ILC Pre-Lab.}
\end{figure}
\begin{figure}[t]
\centering
\includegraphics[width=0.7\textwidth]{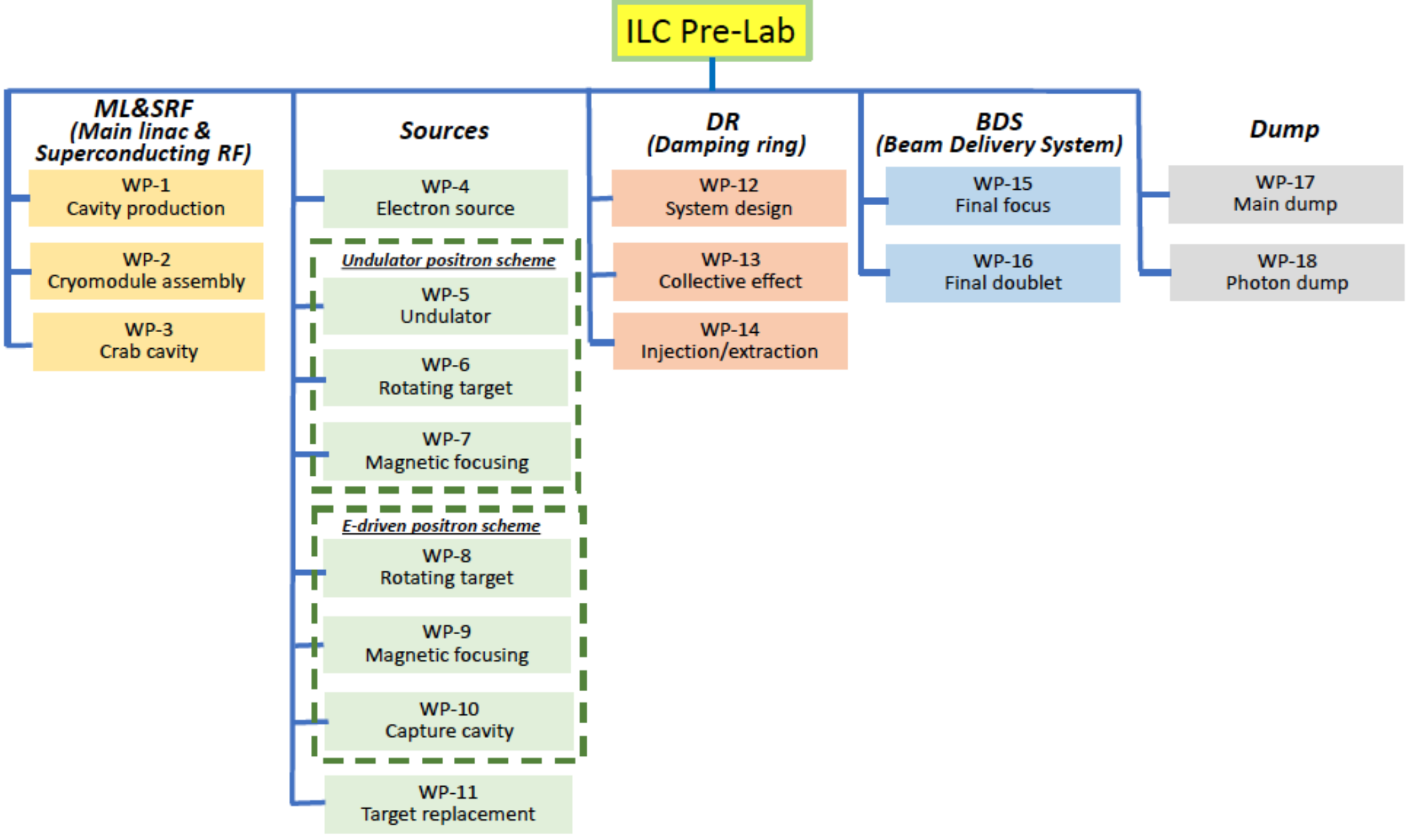}
\caption{The proposed Work Packages of the ILC Pre-Lab.}
\end{figure}

The MEXT Minister Koichi Hagiuda responded to a question during a session of the Diet budget committee concerning the ILC on February 25, 2021. A possible translation of his remark is
\begin{quote} {\it
I am all in favor of building this facility in Japan, but it would require an international cooperation. If the proposal is to spend approximately two hundred million dollars for the preparatory phase, without a clear outlook (to fund the whole ILC project based on international cost sharing), I find it difficult to see how Japanese public would support such a spending. I believe it is imperative to obtain a broad support from both domestic and international communities as a prerequisite.}
\end{quote}

This remark derailed the push to launch the ILC Pre-Lab in Japan. KEK consulted MEXT to prepare a budget request for the ILC Pre-Lab in June 2021 but did not receive an encouragement. KEK in the end decided not to submit a budget request.

On the other hand, MEXT decided to constitute a second phase of the advisory panel to review the progress towards the realization of the ILC since the panel met three years earlier. The panel started its activity in July 2021, and concluded the process in February 2022. 

The final report from the panel, also available in English~\cite{panelreport}, is summarised by KEK~\cite{MEXT}:
\begin{enumerate}
\item The panel recognizes the academic significance of particle physics and the importance of the research activities, including that of a Higgs factory, and understands the value of international collaborative research. However, the panel found that it is still premature to proceed into the ILC Pre-Lab phase, which is coupled with an expression of interest to host the ILC by Japan as desired by the research community proposing the project.
 
\item Given the increasing strain in the financial situation of the related countries, the panel recommends the ILC proponents to reflect upon this fact and to reevaluate the plan. They should reexamine the approach towards a Higgs factory in a global manner taking into account the progress in the various studies such as the Future Circular Collider (FCC) and ILC.
 
\item The panel recommends that the development work in the key technological issues for the next-generation accelerator should be carried out by further strengthening the international collaboration among institutes and laboratories, shelving the question of hosting the ILC.
 
\item For realizing a very large project such as the ILC, cultivating a framework where the related countries can exchange information on their situations and discuss required steps would be important.
 
\item The panel recommends that the research community should continue efforts to expand the broad support from various stakeholders in Japan and abroad by building up trust and mutual understanding through bi-directional communication with the people concerned.
\end{enumerate}

The panel recognizes clearly the importance of particle physics, in particular a Higgs factory. Although the launch of the ILC Pre-Lab is judged to be premature, the report  recommends the development work on the key technological issues, and points out the importance of building up an environment for discussion among governments on the ILC project.

In March 2022, the IDT EB is submitted a proposal to ICFA to continue its effort towards the ILC Pre-Lab under certain conditions. There has to be a substantial increase of funding from MEXT for ``the development work in the key technological issues" to form an international collaboration based on MoUs among the laboratories. Since the Japanese government has not expressed interest in hosting the ILC in Japan, site-specific studies are excluded at this moment. Yet an expanded work on technology development based on international collaboration would advance a major part of the work envisaged for the Pre-Lab. In parallel, international discussions need to be developed in such a way that the Pre-Lab as originally envisioned will start in the 2024--2025 time frame. The site-specific study can commence only after the launch of the Pre-Lab. 

At its meeting in March 2022, ICFA decided to prolong the mandate of the IDT by one year with a statement:  ``In particular, the IDT will work to further strengthen international collaboration among institutes and laboratories, and to expand the broad support from various stakeholders.  ICFA will monitor developments over the next year to assess the availability of resources and progress in international discussions."

Following the statement by ICFA, the IDT has identified time-critical issues in the work packages from the Pre-Lab proposal. Collaborative efforts between KEK and laboratories 
world-wide are being prepared to address them, and these will be formalized by MoUs. In Japan, discussion is advancing between KEK and MEXT for a substantial budget increase, for the Japanese Fiscal Year 2023 starting in April 2023, for the development of ILC-related technologies. Once this is approved, it is expected that the support will continue. The IDT is also preparing to launch an international expert panel to start a general discussion on a global project for a large accelerator facility such as the ILC. Although the panel members are from the particle physics community, the discussion will proceed in close contact with government authorities. It is also planed to have occasional extended panel meetings inviting the government authorities to attend. This is to ensure that the conclusions will be commonly understood by the government authorities. Once this is done for the general discussion on the global project, the panel will proceed to the ILC-specific issues. The second step should lead to the starting of the Pre-Lab and governmental negotiation for the ILC construction. It is planed to have a substantial progress for the first step by the end of 2022. These two  IDT activities are seen positively by MEXT and are supported by the Federation of Diet Members in Japan.  It is hoped that the P5 process will observe these  developments during its deliberations and take them into account in its final recommendations in early 2023.

\section{ILC Laboratory}

Once the Pre-Lab is launched and finishes the Engineering Design Report, and secures an overall agreement to fund the ILC project as a whole, the ILC Laboratory will be launched. Some ideas for the ILC Laboratory have been developed by the GDE/LCB~\cite{GDELCBWG} and the KEK International Working Group~\cite{IWG}. The overall framework for the ILC laboratory will be revisited by the IDT international expert panel in the second phase.   The final decision on the laboratory structure and governance will be decided in the negotiation  among the governments participating in the ILC construction. 

It is envisioned that the ILC construction would take about nine years with an additional year of commissioning. This  would require a stable organization to maintain steady funding and human resources from all participating countries.

\section{Timeline for ILC Detectors}

\begin{figure}[t]
\centering
\includegraphics[width=\textwidth]{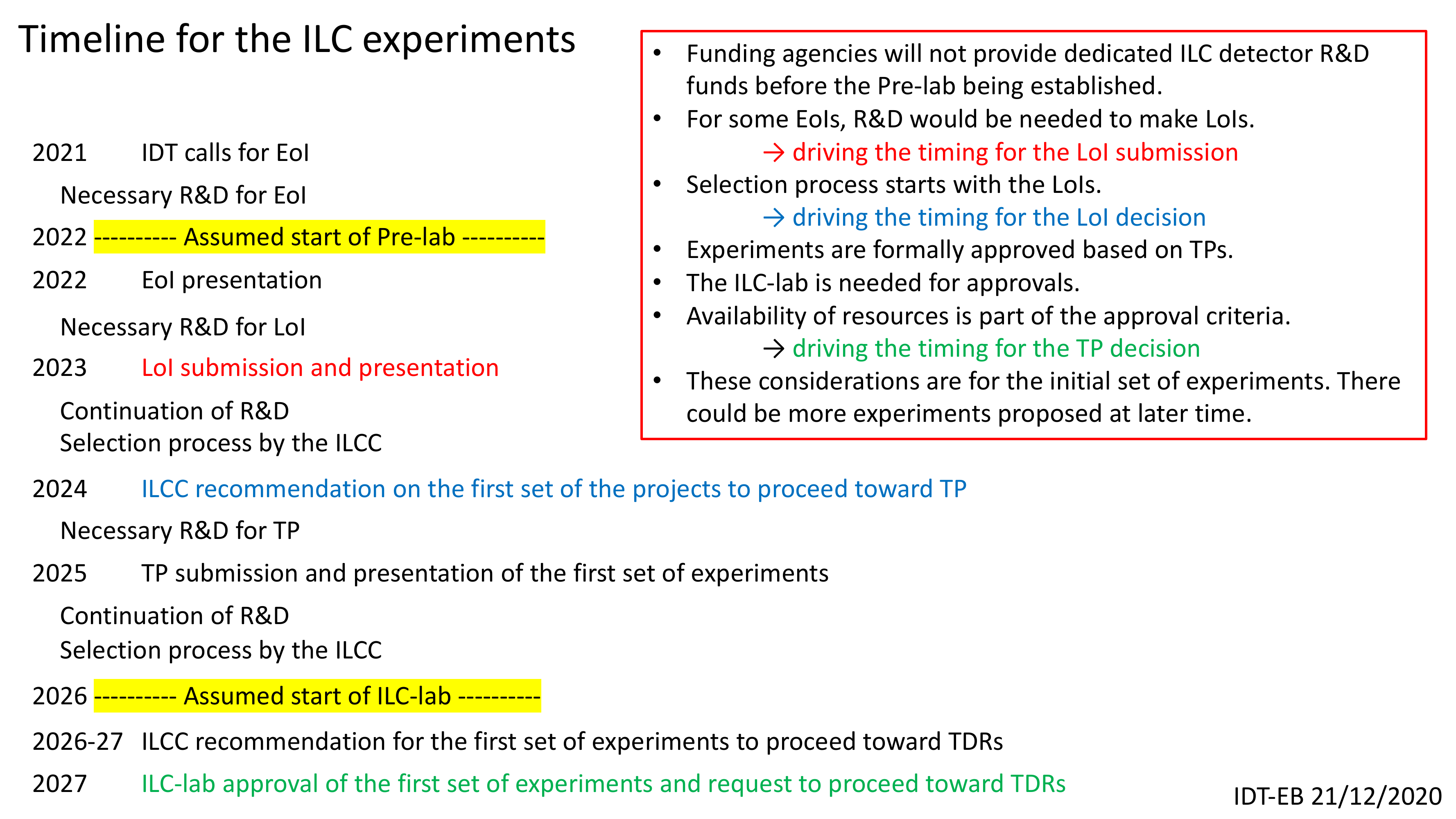}
\caption{The originally envisioned timeline for the Pre-Lab and ILC detectors.}
\label{fig:timeline}
\end{figure}

The originally envisioned timeline for the IDT and Pre-Lab is shown in Fig.~\ref{fig:timeline}.

We do not know when the process will begin to create the Pre-Lab, and unfortunately the schedule is uncertain and delayed. But once it does begin, it leave rather little time for the standard process of Expression of Interest (EoI), Letters of Interest (LoI), and the actual proposal for experiments at the ILC. 

Clearly, it is crucial for all potential experimental proposals to stay up-to-date with the developing technology and science case, to be ready when the opportunity arises.

\chapter{ILC Accelerator} 
\label{chap:accelerator}

\section{ILC accelerator design} 
\label{sec:acc-design}

% CHAPTER ON ILC MACHINE

%{\it Status: Feb 12, 2019}  - revised by MEP
% CHANGELOG
% 13.10.21 BL Updated references to SHIN
%             Removed discussion on gradient limitations, refer to section on R&D instead
%             Removed subsection on upgrades
%             Made figures smaller
%             Fixed luminosity units (Graham)
%  4 March 2022 substantial editing of the Green ILC section (M. Peskin)

The International Linear Collider (ILC) is a $250\,{\mathrm{GeV}}$ (extendable up to $1\,{\mathrm{TeV}}$) linear $e^+e^-$ collider, based on $1.3\,{\mathrm{GHz}}$ superconducting radio-frequency (SCRF) cavities.
It is designed to  achieve a luminosity of $1.35\cdot 10^{34}~{\mathrm{cm}}^{-2}{\mathrm{s}}^{-1}$ and provide an integrated luminosity of $400\,{\mathrm{fb}}^{-1}$ in the first four years of running.
The electron beam will be polarized to $80\,\%$, and positrons with $30\,\%$ polarization will be provided if the undulator based positron source concept is employed. 

Its parameters have been set by physics requirements first outlined in 2003,
updated in 2006, and thoroughly discussed over many years with the physics user community. 
After the discovery of the Higgs boson it was decided that an initial energy of $250\,{\mathrm{GeV}}$ provides the opportunity for a precision Standard Model and Higgs physics programme at a reduced initial cost~\cite{Evans:2017rvt}.
Some relevant parameters are given in Table~\ref{tab:ilc-params}.
This design evolved from two decades of R\&D, described in Sec.~\ref{chap:intro}, 
an international effort coordinated first by the GDE under ICFA mandate and since 2013 by the Linear Collider Collaboration (LCC). 

%% Alex wrote: We have 3 options here: either we use "landscape" or "turn" or "sidewaystable"
%% in my opinion, "sidewaystable" looks better.

%\begin{landscape} 
\begin{sidewaystable*} %inserted by Alex Aryshev 19/02/22
\centering 
%\begin{table*}[tbhp] % Original 
%\begin{turn}{90} %inserted by Alex Aryshev 19/02/22
\begin{tabular}{lcccccccc}
Quantity & Symbol & Unit & Initial & ${\mathcal{L}}$ Upgrade & Z pole & \multicolumn{3}{c}{Upgrades} \\
\hline
Centre of mass energy & $\sqrt{s}$ & ${\mathrm{GeV}}$ & $250$ & $250$ & $91.2$ & $500$ & $250$ & $1000$ \\
Luminosity & \multicolumn{2}{c}{${\mathcal{L}}$ ~~~~$10^{34}{\mathrm{cm^{-2}s^{-1}}}$} & $1.35$ & $2.7$ &  $0.21/0.41$ & $1.8 / 3.6$ & $5.4$ & $5.1$ \\
Polarization for $e^-/e^+$ & $P\sub{-}(P\sub{+})$ & \% & $80(30)$ &  $80(30)$ & $80(30)$ & $80(30)$ & $80(30)$ & $80(20)$ \\
Repetition frequency &$f\sub{{rep}}$ & ${\mathrm{Hz}}$  & $5$ & $5$ &  $3.7$ & $5$ & $10$ & $4$ \\
Bunches per pulse  &$n\sub{{bunch}}$ & $1$  & $1312$ & $2625$ &  $1312/2625$  & $1312 / 2625$ & $2625$ & $2450$ \\
Bunch population  &$N\sub{{e}}$ & $10^{10}$ & $2$ &  $2$ &  $2$ & $2$ & $2$ & $1.74$ \\
Linac bunch interval & $\Delta t\sub{{b}}$ & ${\mathrm{ns}}$ & $554$ & $366$ &  $554/366$ & $554 / 366$ &  $366$ &$366$ \\
Beam current in pulse & $I\sub{{pulse}}$ & ${\mathrm{mA}}$& $5.8$ & $8.8$ &   $5.8/8.8$ & $5.8/8.8$ & $8.8$ & $7.6$  \\
Beam pulse duration  & $t\sub{{pulse}}$ & ${\mathrm{\mu s}}$ & $727$ & $961$ & $727/961$ & $727 / 961$ & $961$ & $897$ \\
Average beam power  & $P\sub{{ave}}$   & ${\mathrm{MW}}$ & $5.3$ & $10.5$ &   $1.42/2.84^{*)}$ &  $10.5 / 21$  & $21$ & $27.2$ \\
RMS bunch length  & $\sigma^*\sub{{z}}$  & ${\mathrm{mm}}$ & $0.3$ & $0.3$ & $0.41$ & $0.3$ & $0.3$ &  $0.225$ \\
Norm. hor. emitt. at IP & $\gamma\epsilon\sub{{x}}$ & ${\mathrm{\mu m}}$& $5$ & $5$ & $6.2$ &  $5$ & $5$ & $5$  \\ 
Norm. vert. emitt. at IP & $\gamma\epsilon\sub{{y}}$ & ${\mathrm{nm}}$ & $35$ & $35$ &  $48.5$ &  $35$ & $35$ & $30$ \\ 
RMS hor. beam size at IP  & $\sigma^*\sub{{x}}$ & ${\mathrm{nm}}$  & $516$ & $516$ &  $1120$ & $474$ & $516$ &  $335$ \\
RMS vert. beam size at IP &$\sigma^*\sub{{y}}$ & ${\mathrm{nm}}$ & $7.7$  & $7.7$  &   $14.6$ & $5.9$ & $7.7$  & $2.7$ \\
Luminosity in top $1\,\%$ & ${\mathcal{L}}\sub{0.01} / {\mathcal{L}}$ &  & $73\,\%$  &  $73\,\%$ &  $99\,\%$ & $58.3\,\%$ & $73\,\%$ & $44.5\,\%$\\
Beamstrahlung energy loss & $\delta\sub{BS}$ &  & $2.6\,\%$  & $2.6\,\%$  &   $0.16\,\%$ & $4.5\,\%$ &$2.6\,\%$  & $10.5\,\%$ \\
Site AC power  & $P\sub{{site}}$ &  ${\mathrm{MW}}$ & $111$ & $128$ &   $94/115$ & $173 / 215$ & $198$ & $300$ \\
Site length & $L\sub{{site}}$ &  ${\mathrm{km}}$ & $20.5$ & $20.5$  &  $20.5$ & $31$ & $31$ & $40$ \\
\end{tabular}
%\end{turn} %inserted by Alex Aryshev 19/02/22
\caption{Summary table of the ILC accelerator parameters in the initial $250\,{\mathrm{GeV}}$ staged configuration  and possible upgrades.
A \siunit{500}{GeV} machine could also be operated at \siunit{250}{GeV} with \siunit{10}{Hz} repetition rate, bringing the maximum luminosity to \siunit{5.4\cdot 10^{34}}{cm^{-2} s^{-1}}~\cite{Harrison:2013nvaxxx}. 
*): For operation at the $Z$-pole additional beam power  of \siunit{1.94/3.88}{MW} is necessary for positron production.
%{\it COMPLETE NUMBERS}
\label{tab:ilc-params}}
%\end{table*}
\end{sidewaystable*}
%\end{landscape} %inserted by Alex Aryshev 19/02/22

The design of the ILC accelerator is governed by the goal of high power-efficiency.
The overall power consumption of the accelerator complex during operation is $111\,{\mathrm{MW}}$ at  $250\,{\mathrm{GeV}}$ and is limited to $300\,{\mathrm{MW}}$ at  $1\,{\mathrm{TeV}}$, which is about $70\,\%$ more than today's peak power consumption of CERN \cite{CERN:2020aa}.
% Stapnes at ALCW2018: 1.35TWh in 2012 -> 154MW on average in 2012
This is achieved by the use of SCRF technology for the main accelerator, which offers a high RF-to-beam efficiency through the use of superconducting cavities, operating at $1.3\,{\mathrm{GHz}}$, where high-efficiency klystrons are commercially available.
At accelerating gradients of $31.5$ to $35\,{\mathrm{MV/m}}$ this technology offers high overall efficiency and reasonable investment costs, even considering the cryogenic infrastructure needed for the operation at $2\,{\mathrm{K}}$.

The underlying TESLA technology is mature, with a broad industrial base throughout the world, and is in use at a number of free electron laser facilities that are in operation (FLASH~\cite{schreiber_faatz_2015,Vogt:2018wvy} and European XFEL~\cite{bib:xfel}) at DESY, Hamburg), under construction (LCLS-II~\cite{bib:lcls-ii} at SLAC, Stanford) or in preparation (SHINE~\cite{Zhao:2018lcl,Huang:2021a} in Shanghai) in the three regions Asia, Americas, and Europe that contribute to the ILC project.
In preparation for the ILC, Japan and the U.S. have founded a collaboration for further cost optimisation of the TESLA technology.
In recent years, new surface treatment technologies utilising nitrogen during the cavity preparation process, such as the so-called   nitrogen infusion technique, have been developed at Fermilab, with the prospect of  achieving higher gradients and lower loss rates with a less expensive surface preparation scheme than assumed in the TDR (see Sec.~\ref{sec:acc-beyond}).

When the Higgs boson was discovered in 2012, the Japan Association of High Energy Physicists (JAHEP) made a proposal to host the ILC in Japan~\cite{JAHEP:2012a,JAHEP:2012b}. 
Subsequently, the Japanese ILC Strategy Council conducted a survey of possible sites for the ILC in Japan, looking for  suitable geological conditions for a tunnel up to $50\,{\mathrm{km}}$ in length (as required for a $1\,{\mathrm{TeV}}$  machine), and the possibility to establish a laboratory where several thousand international scientists can work and live. 
As a result, the candidate site in the Kitakami region in northern Japan, close to the larger cities of Sendai and Morioka, was found to be the best option. 
The site offers a large, uniform granite formation with no currently active faults and a geology that is well suited for tunnelling.
Even in the great Tohoku earthquake in 2011, underground installations in this rock formation were essentially unaffected~\cite{bib:sanuki:desy2017}, which underlines the suitability of this candidate site. 

This section starts with a short overview of the changes of the ILC design between the publication of the TDR in $2013$ and today, followed by a description of the SCRF technology, and a description of the overall accelerator design and its subsystems. 
Thereafter, possible upgrade options are laid out, the Japanese candidate site in the Kitakami region is presented, and costs and schedule of the accelerator construction project are shown.

%===============================================================================
 \begin{figure*}[tbhp]
 \begin{center}
 \includegraphics[width=\hsize]{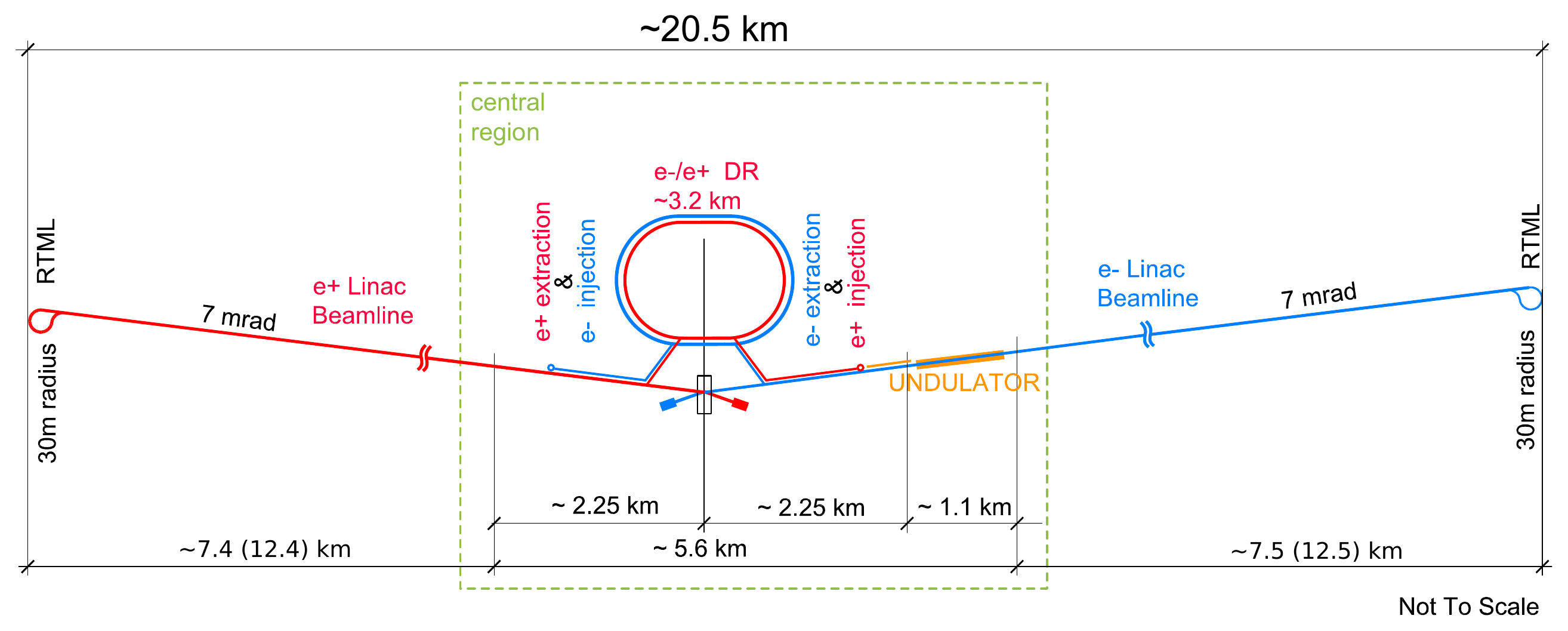}
\caption{Schematic layout of the ILC in the $250\,{\mathrm{GeV}}$ staged configuration.
\label{fig:ilc-schematic}}
 \end{center}
 \end{figure*}

\subsection{Design evolution since the TDR}
\label{sec:design_evo}

Soon  after the discovery of the Higgs boson, the TDR for the ILC accelerator was published in 2013~\cite{Adolphsen:2013jya,Adolphsen:2013kya} after 8 years of work by the Global Design Effort (GDE).
The TDR design was based on the requirements set forth by the ICFA mandated parameters committee~\cite{Heuer:2006}:
\begin{itemize}
\item a centre-of-mass energy of up to \siunit{500}{GeV},
\item tunability of the centre-of-mass energy between $\sqrt{s} = \siunit{200}{GeV}$
 and \siunit{500}{GeV},
\item  a luminosity sufficient to collect \siunit{500}{fb^{-1}} within four years of operation, taking into account a three-year a ramp up.  This corresponds to a final luminosity of \siunit{250}{fb^{-1}} per year and an instantaneous luminosity of ${\mathcal{L}} = \siunit{2 \cdot 10^{34}}{cm^{-2}\,s^{-1}}$,
\item an electron polarization of at least $80\,\%$,
\item  the option for a later upgrade to energies  up to \siunit{1}{TeV}.
\end{itemize}

The accelerator design presented in the TDR met these requirements at an estimated construction cost of \siunit{7,982}{MILCU} for a Japanese site, plus \siunit{22.9}{Mh} (million hours) of labor in participating institutes~\cite[Sec. 15.8.4]{Adolphsen:2013kya}. 
Costs were expressed in ILC Currency Units, ILCU, where \siunit{1}{ILCU} corresponds to \siunit{1}{US\$} at 2012 prices.

In the wake of the Higgs discovery and the JAHEP proposal to host the ILC in Japan\cite{JAHEP:2012a,JAHEP:2012b}, plans were made for a lower cost facility operating at $\sqrt{s} = \siunit{250}{GeV}$ near the maximum of the $e^+e^- \to Zh$  cross section.
A revised plan based on the TDR~\cite[Sect. 12.5]{Adolphsen:2013kya} and  subsequent analyses~\cite{Dugan:2014} was made for a machine with \siunit{125}{GeV} polarized beams and a luminosity of ${\mathcal{L}} = \siunit{1.35 \cdot 10^{34}}{cm^{-2}\,s^{-1}}$, capable of delivering about \siunit{200}{fb^{-1}} per year, or \siunit{400}{fb^{-1}} within the first four years of operation.

Several other changes of the accelerator design have been approved by the ILC Change Management Board since 2013, in particular:
\begin{itemize}
\item The free space between the interaction point and the edge of the final focus quadrupoles ($L^*$) was unified between the ILD and SiD detectors~\cite{bib:cr-0002}, facilitating a machine layout with the best possible luminosity for both detectors.
\item A vertical access shaft to the experimental cavern was foreseen~\cite{bib:cr-0003}, allowing a CMS-style assembly concept for the detectors, where large detector parts are built in an above-ground hall while the underground cavern is still being prepared. 
\item The shield wall thickness in the Main Linac tunnel was reduced from $3.5$ to \siunit{1.5}{m}~\cite{bib:cr-0012}, leading to a significant cost reduction. This was made possible by dropping the requirement for personnel access during beam operation of the main linac.
\item Power ratings for the main beam dumps, and intermediate beam dumps for beam aborts and machine tuning, were reduced to save costs~\cite{bib:cr-0013}.
\item A revision of the expected horizontal beam emittance at the interaction point at \siunit{125}{GeV} beam energy, based on improved performance expectations for the damping rings and a more thorough scrutiny of beam transport effects at lower beam energies, lead to an increase of the luminosity expectation from $0.82$ to \siunit{1.35 \cdot 10^{34}}{cm^{-2}\,s^{-1}}~\cite{bib:cr-0016}.
\item The active length of the positron source undulator has been increased from $147$ to \siunit{231}{m} to provide sufficient intensity at \siunit{125}{GeV} beam energy~\cite{PWG:2018a}.
\end{itemize}

These changes contributed to an overall cost reduction, risk mitigation, and improved performance expectation.

Several possibilities were evaluated for the length of the initial tunnel. 
Options that include building tunnels with the length required for a machine with $\sqrt{s} = \siunit{350}{GeV}$ or \siunit{500}{GeV}, were considered.
In these scenarios, an energy upgrade would require the installation of additional cryomodules (with RF and cryogenic supplies), but little or no civil engineering activities.
In order to be as cost effective as possible the final proposal endorsed by ICFA~\cite{ICFA:2017} does not include these empty tunnel options. 

While the length of the main linac tunnel was reduced, the beam delivery system and the main dumps are still designed to allow for an energy upgrade up to  $\sqrt{s} = \siunit{1}{TeV}$.

\begin{figure}[thbp]
\begin{center}
   \includegraphics[width=0.5\hsize]{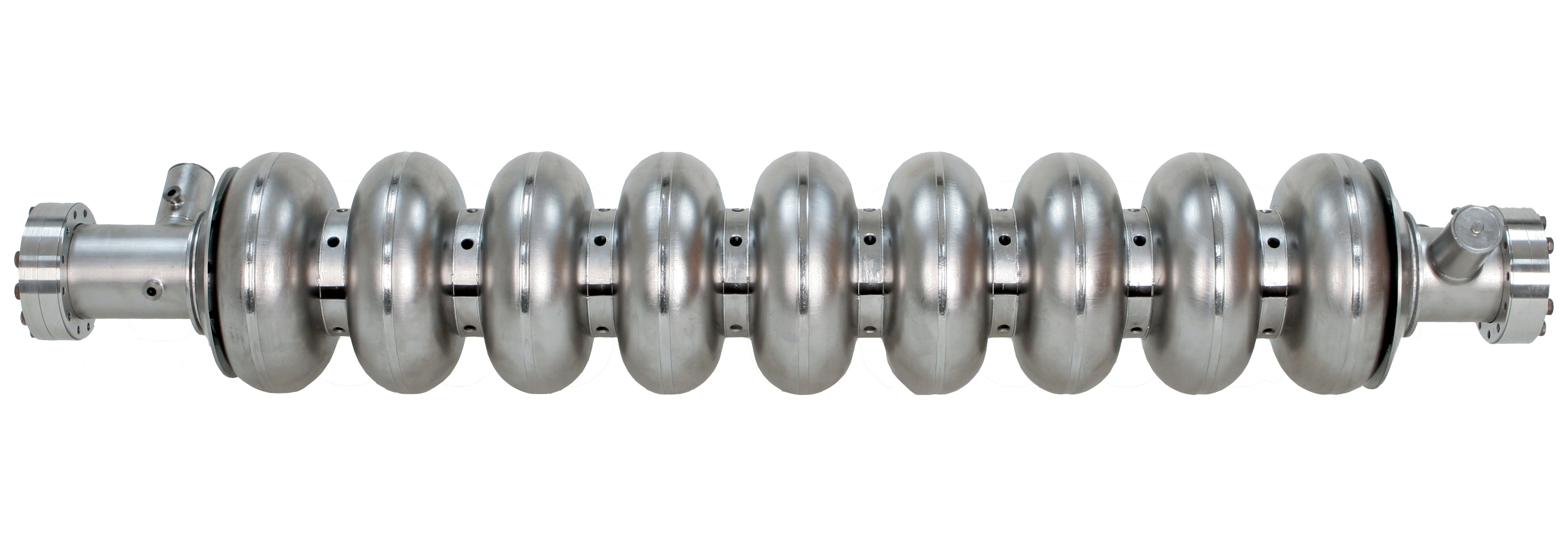}
\end{center}
\caption{A $1.3\,{\mathrm{GHz}}$ superconducting niobium nine-cell cavity.
}
% Figure from TDR Executive Summary
% https://svnsrv.desy.de/k5websvn/wsvn/General.ilctdr/tags/FormOne-Print-Release/tdres/accel/figs/tesla9cell-cavity-2.jpg
\label{fig:tesla-cavity}
\end{figure}

\subsection{Superconducting RF Technology}
\label{sec:SRFtech}

%{\it Description of the TESLA SCRF technology - 4 pages
%
%Stresses long experience - FLASH, STF, XFEL, broad industrial base - LCLS-II, SHINE:
%addresses concerns mentioned in Nomura report and others (yield / gradient, MARX modulator)
%
%Nomura issue: MARX modulator
%
%Figures: Cavity, Cryomodule (Rey Hory) 
%
%}

\begin{figure}[bthp]
\begin{center}
   \includegraphics[width=0.5\hsize]{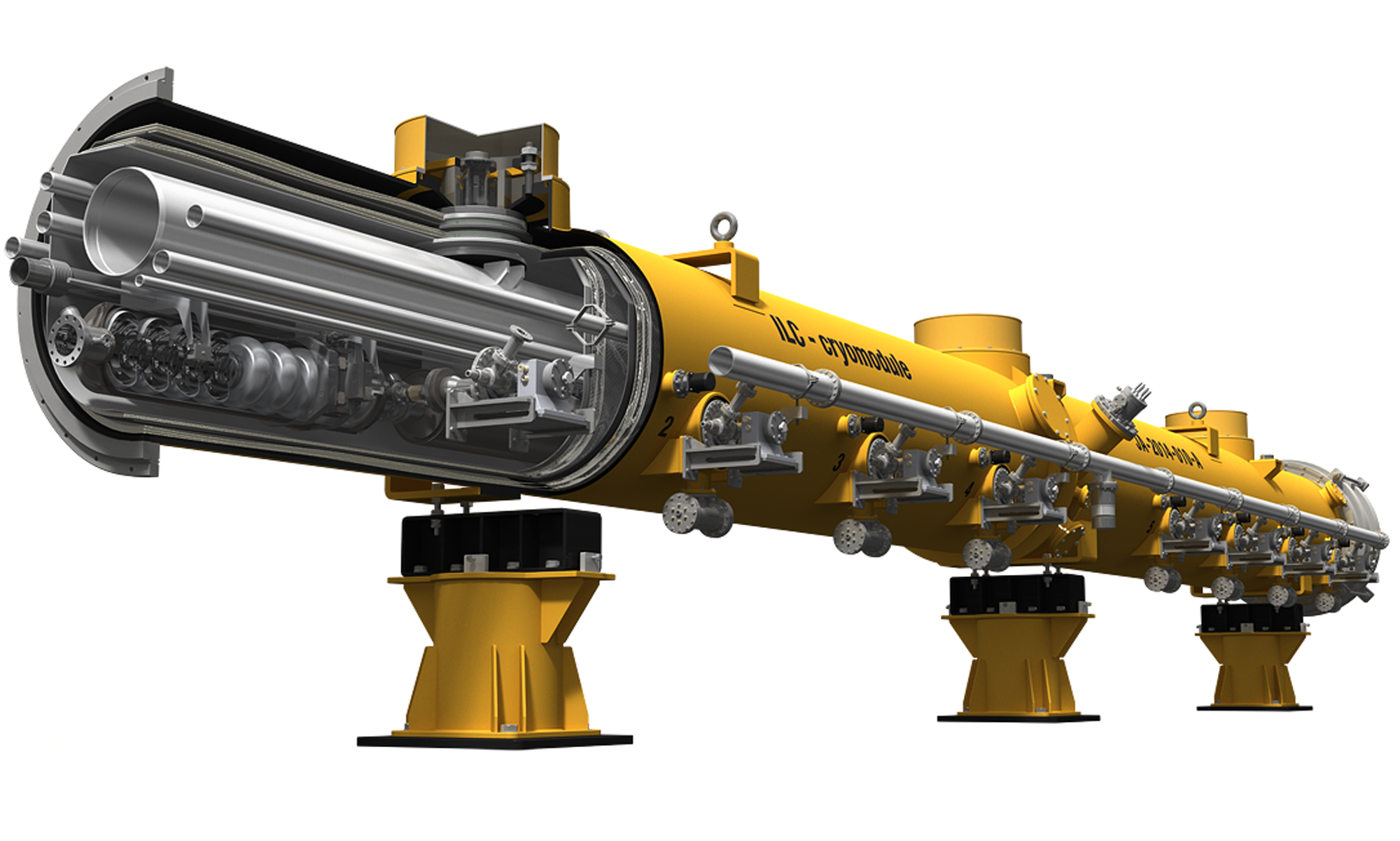}
\end{center}
\caption{An ILC type cryomodule. \copyright Rey.Hori/KEK.}
% Figure from Rey.Hori - check copyright!
% http://www.linearcollider.org/images/pid/1000890/gallery/10_ILC_cryomodule.jpg
\label{fig:cryomodule}
\end{figure}

The heart of the ILC accelerator consists of the two superconducting Main Linacs that accelerate both beams from \num{5} to \siunit{125}{GeV}.
These linacs are based on the TESLA technology:
beams are accelerated in \siunit{1.3}{GHz} nine-cell superconducting cavities made of niobium and operated at \siunit{2}{K} (Fig.~\ref{fig:tesla-cavity}). These 
 are assembled into cryomodules comprising nine cavities or eight cavities plus a quadrupole/corrector/beam position monitor unit, and all necessary cryogenic supply lines (Fig.~\ref{fig:cryomodule}). 
Pulsed klystrons supply the necessary radio frequency power (High-Level RF HLRF) to the cavities by means of a waveguide power distribution system and one input coupler per cavity.

This technology was primarily developed at DESY for the TESLA accelerator project that was proposed in 2001.
Since then, the TESLA technology collaboration~\cite{bib:ttc} has been improving this technology, which is now widely used around the world.
As discussed in Section~\ref{sec:acc-beyond}, the TESLA technology is based on a history of superconducting accelerator projects of more than 50 years, starting in the 1970s with the Ilinois Microtron Superconducting Linac and the Stanford Superconductiong Accelerator.
Today, a large number of superconducting accelerators such as CEBAF at  Jefferson Lab, SNS at Oak Ridge Laborratory, or FRIB at Michigan State University, to name just a few U.S. facilities, are in operation, demonstrating the success of this approach. 

%\subsubsection{The Economics of Superconducting Linacs}
%
%The total energy dissipation $P\sub{d}$ in a linac cavity of length $L$ and a total voltage $V = g L$ at a the accelerating gradient $g$ is given by
%$$
%  P\sub{d} = g V R\sub{s} \lambda / C
%$$
%where $R\sub{s}$ is the material's surface resistance, $\lambda$ is the RF frequency, and $C$ (in ${\mathrm{\Omega^2}}$) depends only on the cavity shape, not its size.
%The dissipated power in a single cavity grows with the square of the gradient, 
%and the overall power grows linearly with the gradient for a given accelerating voltage.

% BL 13.10.21 Figure removed
%\begin{figure*}[tbhp]
%   \includegraphics[width=0.8 \hsize]{chapters/accelerator/figures/LBandSRFNbCavityGradientEvolution_ProjectImpact_rev28dec2018}
%\caption{Development of the gradient of SRF cavities since 1970
%\cite[updated]{Geng:2015glc}.
%}
% Paper is published under CC-BY-3.0
%\label{fig:gradients}
%\end{figure*}

\subsubsection{The quest for high gradients}
\label{subsubsec:highgrad}

The single most important parameter for the cost and performance of the ILC is the accelerating gradient $g$.
The TDR baseline value is an average gradient $g = \siunit{31.5}{MV/m}$ for beam operation, with a $\pm 20\,\%$ gradient spread between individual cavities.
Recent progress in R\&D for high gradient cavities raises the hope to increase
the gradient by  $10\,\%$  to  $g = \siunit{35}{MV/m}$, which would reduce the total cost of the \siunit{250}{GeV} accelerator by about  $6\,\%$.
%% 6% = 50BY / 803BY
% Copied from TDR Vol 3.II p. 23
To achieve the desired gradient in beam operation, the gradient achieved in the low-power vertical test (mass production acceptance test) is specified $10\,\%$ higher to allow for operational gradient overhead for low-level
RF (LLRF) controls, as well as some degradation during cryomodule assembly (few ${\mathrm{MV/m}}$).
%Figure~\ref{fig:gradients} shows how the achievable gradients have evolved over the past 50 years~\cite{Geng:2015glc}.
%Figure~\ref{fig:SRFprogress} shows how the achievable gradients have evolved over the past 50 years.
Section~\ref{sec:acc-beyond} discusses the evolution of achievable gradients have evolved over the past 50 years, and the prospects for further improvements.

\paragraph{Gradient impact on costs:}
To the extent that the cost of cavities, cryomodules and tunnel infrastructure is independent of the achievable gradient, the investment cost per GeV of beam energy is inversely proportional to the average gradient achieved. This is the reason for the enormous cost saving potential from higher gradients.
This effect is partially offset by two factors:  the energy stored in the electromagnetic field of the cavity, and the dynamic heat load to the cavity from the electromagnetic field.  These grow quadratically with the gradient for one cavity, and therefore linearly for a given beam energy.
The electromagnetic energy stored in the cavity must be replenished by the RF source during the filling time that precedes the time when the RF is used to accelerate the beam passing through the cavity; this energy is lost after each pulse and thus reduces the overall efficiency and requires more or more powerful modulators and klystrons.
The overall cryogenic load is dominated by the dynamic heat load from the cavities, and thus operation at higher gradient requires larger cryogenic capacity.
Cost models that parametrise these effects indicate that the minimum of the investment cost per GeV beam energy lies at \num{50} or more GeV, depending on the relative costs of tunnel, SCRF infrastructure and cryo plants, and depending on the achievable $Q_0$~\cite{Adolphsen:2011a}. 
Thus,  the optimal gradient is significantly higher than the value of approximately \siunit{35}{MV/m} that is currently realistic; this emphasises  the relevance of achieving higher gradients.

It should be noted that in contrast to the initial investment, 
the operating costs rise when the gradient is increased, and this must be factored into the cost model. 
The reason for this is that the energy stored in the cavity, which is lost after each pulse, as well as the heat generated in the cavity walls rise with the square of the gradient, thus leading to a rise of electricity need with gradient that is linear to first order.

% 13.10.21 BL paragraph on gradient limitations removed 
% \paragraph{Gradient limitations:}

\paragraph{Results from European XFEL cavity production:}

\begin{figure}[htbp]
\begin{center}
   \includegraphics[width=0.7\hsize]{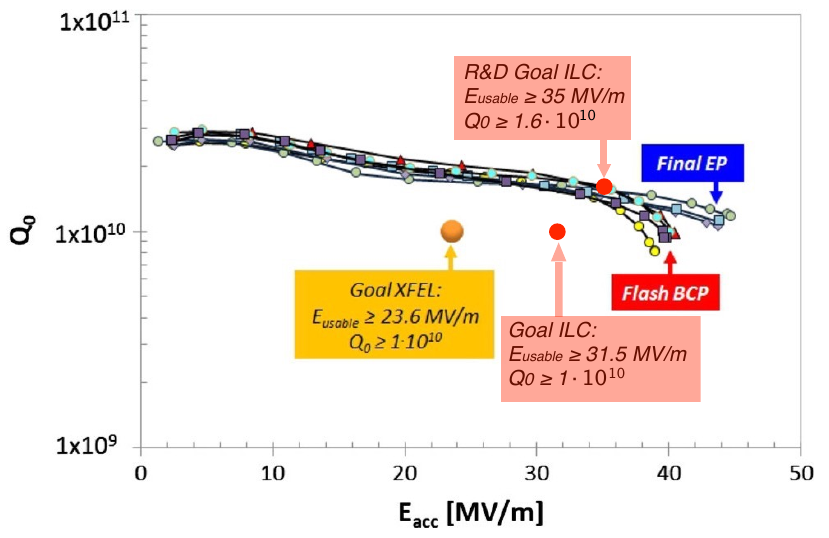}
\end{center}
\caption{Examples of the $Q_0\,(E\sub{acc})$ curves of some of the best
cavities, either treated at RI using ``EP final'', or at EZ using
``BCP flash.''
\cite[Fig. 19]{Singer:2016fbf}. 
RI employs a production process that closely follows the ILC specifications, with a final electropolishing step.
The ILC gradient / $Q_0$ goals are overlaid.}
% Paper published under CC-BY-3.0
\label{fig:cavity-gradient}
\end{figure}

The production and testing of $831$ cavities for the European XFEL~\cite{Singer:2016fbf,Reschke:2017gjp} provides the biggest sample of cavity production data so far. 
Cavities were acquired from two different vendors, Research Instruments (RI) and Zanon Research (EZ).
RI employed a production process with a final surface treatment closely following the ILC specifications, including a final electropolishing (EP) step,
while EZ used buffered chemical polishing (BCP).
The European XFEL specifications asked for a usable gradient of \siunit{23.6}{MV/m} with a $Q_0 \ge 1  \cdot 10^{10}$ for operation in the cryomodule;
with a $10\,\%$ margin this corresponds to a target value of \siunit{26}{MV/m} for the performance in the vertical test stand for single cavities.
Figure~\ref{fig:cavity-gradient} shows the $Q_0$ data versus accelerating gradient of the best cavities received, with several cavities reaching more than \siunit{40}{MV/m}, significantly beyond the ILC goal, already with $Q_0$ values that approach the target value $1.6\cdot10^{10}$ that is the goal of future high-gradient R\&D.

\begin{figure}[tbhp]
\begin{center}
   \includegraphics[width=0.5\hsize]{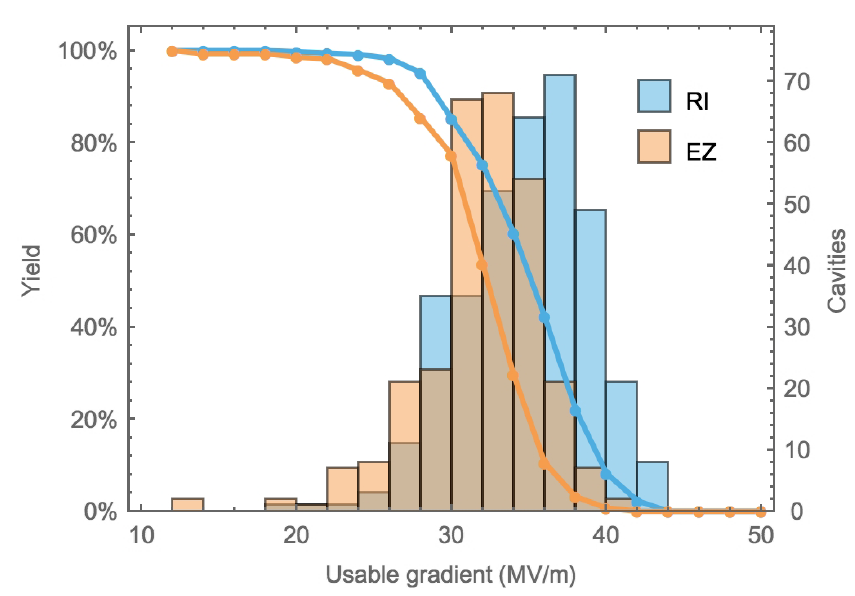}
\end{center}
\caption{Distribution and yield of the ``as received'' maximum
gradient of cavities produced for the European XFEL, separated by vendor \cite[Fig. 33]{Reschke:2017gjp}. 
Vendor RI employs a production process that closely follows the ILC specifications, with a final electro polishing step.}
% Paper published under CC-BY-4.0
\label{fig:cavity-yield}
\end{figure}

European XFEL production data, in particular from vendor RI, provide excellent statistics for the cavity performance as received from the vendors, as shown in Fig.~\ref{fig:cavity-yield}.
For vendor RI, the yield for cavities with a maximum gradient above \siunit{28}{MV/m} is $85\,\%$, with an average of \siunit{35.2}{MV/m} for the cavities that pass the cut.

Since the European XFEL performance goal was substantially lower than the ILC specifications, cavities with gradient below \siunit{28}{MV/m}, which would not meet ILC specifications, were not generally re-treated for higher gradients, limiting our knowledge of the effectiveness of re-treatment for large gradients.
Still, with some extrapolation it is possible to extract yield numbers applicable to the ILC specifications ~\cite{bib:Walker:2017.lcws}.

The European XFEL data indicate that after re-treating cavities with gradients outside the ILC specification of $\siunit{35}{MV/m} \pm 20\,\%$, \ie, below \siunit{28}{MV/m}, a yield of $94\,\%$ for a maximum gradient above \siunit{28}{MV/m} can be achieved, with an average value of \siunit{35}{MV/m}, meeting the ILC specification.
Taking into account limitations from $Q_0$ and the onset of field emission, the usable gradient is lower.  
This gives a $82\,(91)\,\%$ yield and an average usable gradient of \siunit{33.4}{MV/m} after up to one (two) re-treatments.
The re-treatment and testing rate is significantly higher than assumed in the TDR, but the European XFEL experience shows that re-treatment can mostly be limited to a simple high-pressure rinse (HPR) rather than an expensive electropolishing step.  

Overall, the European XFEL cavity production data demonstrate that it is possible to mass-produce cavities meeting the ILC specifications as laid out in the TDR with the required performance and yield.

% PARAGRAPHS ON R&D omitted here from European Stretagy, as they are covered later

\subsubsection{Choice of RF frequency}

% Copied from TDR Vol 3.II p. 23
The choice of operating frequency is a balance between the higher cost of larger, lower-frequency cavities and the increased cost at higher frequency associated with the lower sustainable gradient from the increased surface resistivity. 
The optimum frequency is in the region of \siunit{1.5}{GHz}, but during the early R\&D on the technology, \siunit{1.3}{GHz} was chosen due to the commercial availability of high-power klystrons at that frequency.

\subsubsection{Cavities}

The superconducting accelerating cavities for the ILC are nine-cell structures made out of high-purity niobium (Fig.~\ref{fig:tesla-cavity}), with an overall length of \siunit{1.25}{m} and an active length of \siunit{1.038}{m}.
Cavity production starts from niobium ingots which are forged and rolled into \siunit{2.8}{mm} thick niobium sheets that are individually checked for defects by an eddy current scan and optical inspection~\cite{Adolphsen:2013jya}.
Cavity cells are produced by deep-drawing the sheets into half cells, \num{18} of which are joined by electron beam welding with two end groups to form the whole structure.
This welding process is one of the most critical and cost-intensive steps of the cavity manufacturing procedure. 
Utmost care must be taken to avoid irregularities, impurities and inclusions in the weld itself, and deposition of molten material at the inner surface of the cavity that can lead to field emission.

After welding, the inner surface of the cavity must be prepared.
The process is designed to remove material damage incurred by chemical procedures during the fabrication process, chemical residues from earlier production steps, hydrogen in the bulk niobium from earlier chemical processing, and contamination from particles.
In a last step, the cavity is closed to form a hermetically sealed structure ready for transport.
The treatment steps involve a series of rinses with ethanol or high pressure water, annealing in a high purity vacuum furnace at \siunit{800^\circ}{C} and \siunit{120^\circ}{C}, and electropolishing or buffered chemical polishing.
The recipe for the surface preparation has been developed over a long time.  Still, it remains subject to optimisation, since it is a major cost driver for the cavity production and largely determines the overall performance and yield of the cavities.
In particular the electropolishing steps are complicated and costly, as they require complex infrastructure and highly toxic chemicals.
An important advantage of nitrogen infusion and other novel surface treatment processes (see Sec.~\ref{subsec:highgradient}) is that the final electropolishing step is omitted.

Careful quality control during the production process is of high importance.
At the European XFEL, several quality controls were conducted by the manufacturer during production, with nonconformities reported to the institute responsible for the procurement, where a decision was made whether to accept or reject a part~\cite{Singer:2016fbf}. 
With this ``build to print'' approach, in which the manufacturer guarantees that a precise production process will be followed but does not guarantee a specific performance, procurement costs are reduced, because the manufacturer does not carry, and does not charge for,  the performance risk.

Upon reception from the manufacturer, cavities are tested in a vertical cryostat (``vertical test''), where $Q_0$ is measured as a function of the gradient.
Cavities that fall below the specified gradient goal are re-treated by an additional (expensive) electropolishing step or a comparatively simple high-pressure rinse. 
After retreatment, the vertical test is repeated.

Re-treatment and tests constitute a major cost driver in cavity production. 
For the ILC TDR, it was assumed that $25\,\%$ of the cavities would fall below the \siunit{28}{MV/m} gradient threshold and undergo re-treatment and a second vertical test.
European XFEL data from RI that followed the ILC production recipe indicate that $15\,\%$ to $37\,\%$ of the cavities fall below \siunit{28}{MV/m}, depending  on whether the maximum or the ``usable'' achieved gradient is considered~\cite{bib:Walker:2017.lcws}.
% However, European XFEL experience also shows that, in most of the cases, a high-pressure rinse is sufficient as re-treatment to remove surface defects, which is a cost saving compared to the electropolishing assumed in the TDR.

After successful testing, prior to installation in the cryomodule, cavities are equipped with a magnetic shield and the frequency tuner, which exerts mechanical force on the cavity to adjust the resonant frequency to the frequency of the external RF field~\cite[Sect. 3.3]{Adolphsen:2013kya}.

\subsubsection{Power coupler}

The power coupler transfers the radio frequency (RF) power from the waveguide system to the cavity. 
In the ILC, a coupler with a variable coupling coefficient is employed; this is realised using  a movable antenna.  Another role of the coupler is to 
separate the cavity vacuum from the atmospheric pressure in the waveguide, and to  insulate the cavity at \siunit{2}{K} from the surrounding room temperature.
Thus,  the coupler has to fulfill a number of demanding requirements: transmission of high RF power with minimal losses and no sparking, vacuum tightness and robustness against window breaking, and minimal heat conductivity.  
As a consequence, the coupler design is highly complex, with a large number of components and several critical high-tech manufacturing steps.

\begin{figure}[htbp]
\begin{center}
   \includegraphics[width=0.5\hsize]{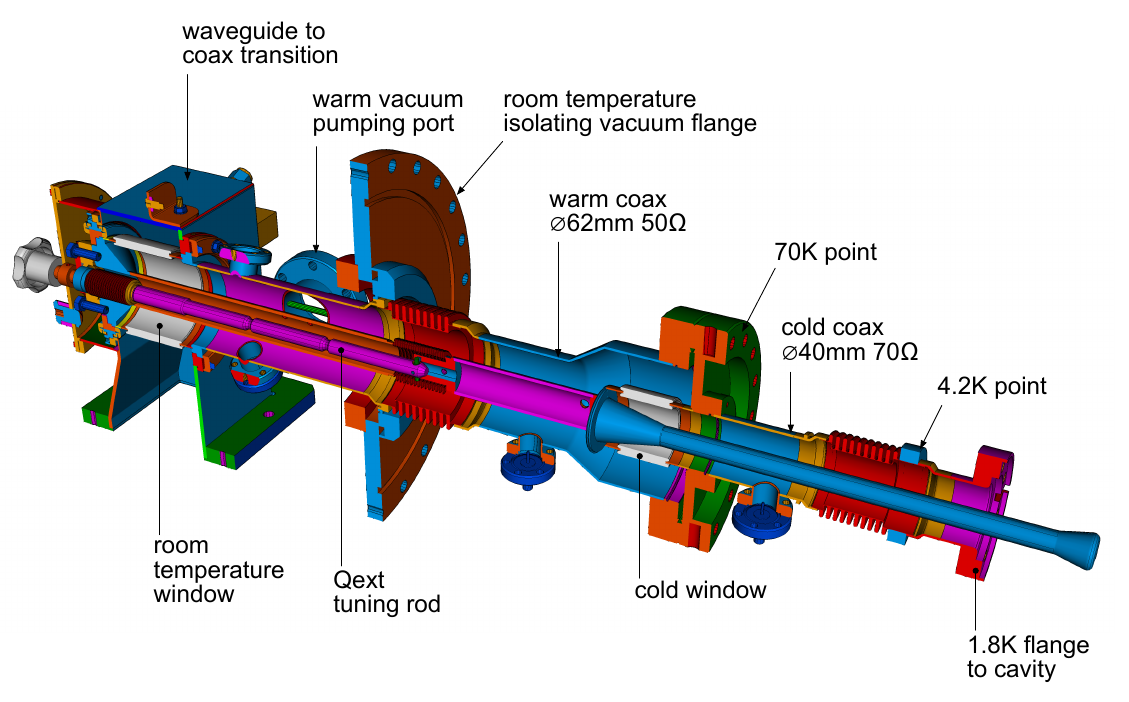}
\end{center}
\caption{An European XFEL type coupler.
  % Fig from ILC TDR, should be OK to use
}
\label{fig:xfelcoupler}
\end{figure}

The baseline coupler design was originally developed in the 1990s for the TESLA Test Facility (TTF, now FLASH) at DESY,
and has since been modified by a collaboration of LAL and DESY for use in the European XFEL.
About 840 of these couplers (depicted in Fig. \ref{fig:xfelcoupler}) were fabricated by three different companies for the  European XFEL~\cite{Kaabi:2013wna},  where 800 are now in operation.
A lot of experience has been gained from this production~\cite{Sierra:2017wyc}.

\subsubsection{Cryomodules}

To facilitate transportation, installation and operation, 8 or 9 cavities are integrated into a \siunit{12.6}{m} long cryomodule~(Fig.~\ref{fig:cryomodule}), which houses the cavities, thermal insulation, and all necessary supply tubes for liquid and gaseous helium at \siunit{2-80}{K} temperature.

\begin{figure}[htbp]
\begin{center}
   \includegraphics[width=0.5\hsize]{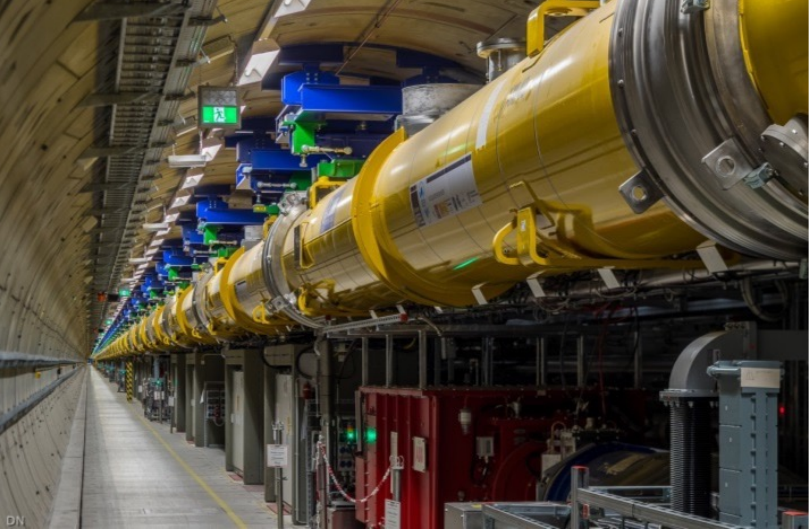}
\end{center}
\caption{View of installed cryomodules in the tunnel of the European XFEL~\cite{Reschke:2018ywk}.
}
% Paper is published under CC-BY-3.0
\label{fig:xfel-tunnel}
\end{figure}

Nine of these cryomodules are connected in the tunnel to form a cryostring with a common liquid helium supply.  RF for one such string is provided by two klystrons.
No separate helium transfer line is necessary, as all helium transport lines are integrated within the modules.  A  quadrupole / beam position monitor / corrector magnet unit  is mounted instead of the 9th cavity in every third module.
Figure~\ref{fig:xfel-tunnel} shows installed cryomodules in the tunnel of the  European XFEL~\cite{Reschke:2018ywk}.

Cryomodule assembly requires a dedicated facility with large clean rooms, especially trained, experienced personnel, and thorough quality control~\cite{Berry:2017gpt}.
The cryomodules are certified for liquid helium pressure of up to \siunit{2}{bar}.  Thus they must  conform to the applicable pressure vessel codes, which brings with it very stringent documentation requirements for all pressure bearing parts~\cite{Peterson:2011zz}.

For the European XFEL project, 103 cryomodules were produced at production rate is close to the rate envisaged for a possible European contribution of 300 cryomodules to a \siunit{250}{GeV} ILC in Japan.

While the design gradient for European XFEL accelerator modules of \siunit{23.6}{MV/m} is significantly lower than the aim of \siunit{31.5-35}{MV/m} for the ILC, a number of cryomodules have been built around the world that come close to or reach the ILC TDR specification of \siunit{31.5}{MV/m}: An European XFEL prototype module at DESY reached \siunit{30}{MV/m}~\cite{Kostin:2009a}, Fermilab has demonstrated cryomodule operation at the ILC specification of \siunit{31.5}{MV/m}~\cite{Broemmelsiek:2018iqr}, and KEK has reported stable pulsed operation of a cryomodule at \siunit{36}{MV/m}~\cite{Yamamoto:2018kml}.

\begin{figure}[htbp]
\begin{center}
   \includegraphics[width=0.5\hsize]{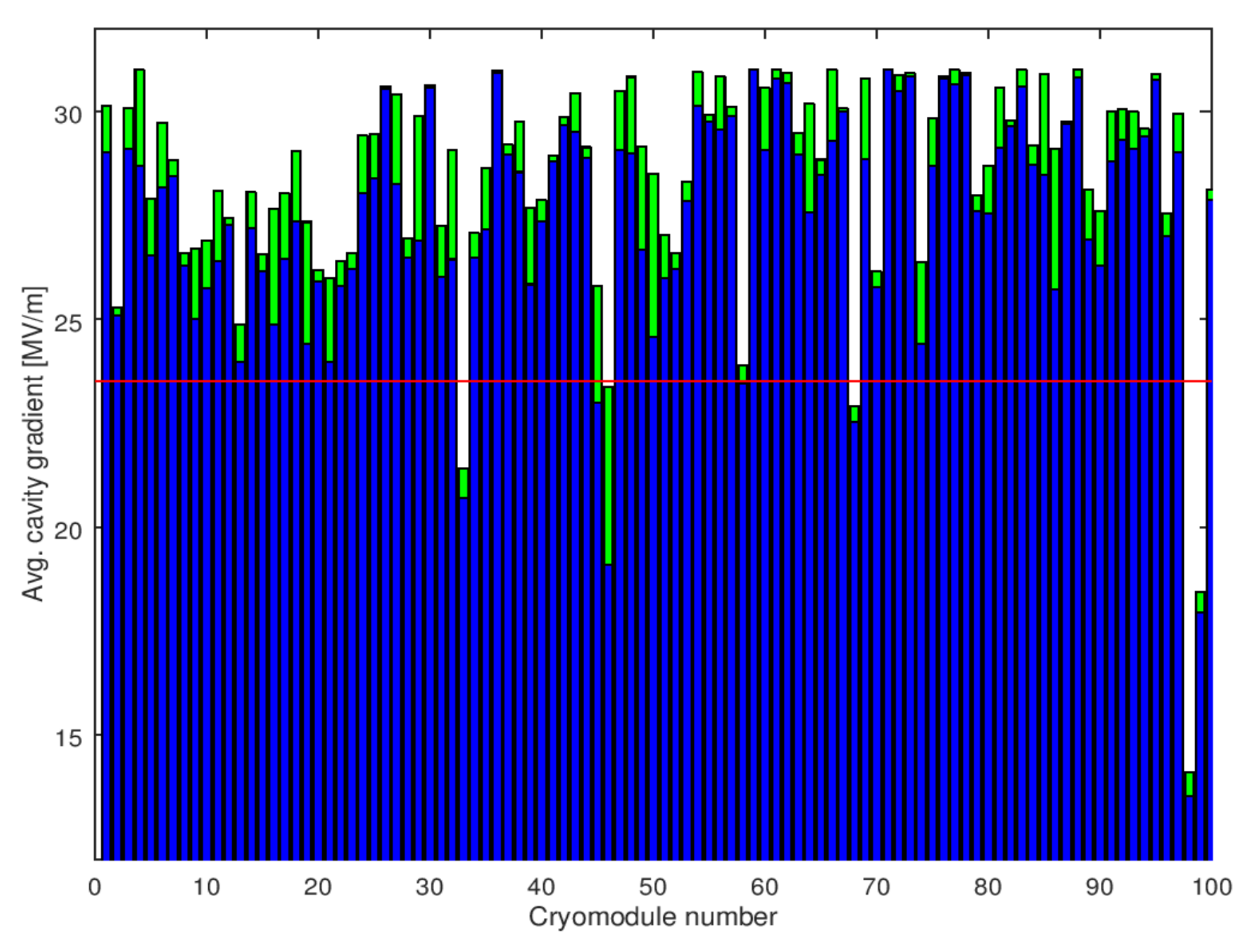}
\end{center}
\caption{Average of the operating (blue) and maximum
(green) gradient for cavities in each European XFEL serial-production cryomodule.
The specification of \siunit{23.6}{MV/m} is marked by a red line
\cite{Kasprzak:2018kkr}.  Modules 98 and 99 were assembled from the lowest-performing cavities.}
% Paper is published under CC-BY-3.0
\label{fig:cryomodules-performance}
\end{figure}

Figure~\ref{fig:cryomodules-performance} shows the average cavity gradients per cryomodule for the European XFEL serial-production cryomodules~\cite{Kasprzak:2018kkr}. 
In the tests, the gradients were limited administratively to \siunit{31}{MV/m}; the true maxima might be higher.
For almost all of the modules, the cavity gradients are significantly above the European XFEL specification of \siunit{23.6}{MV/m}.

\subsubsection{Plug-compatible design}

In order to allow various designs of sub-components from different countries and vendors to work together in the same cryomodule, a set of interface definitions has been internationally agreed upon.
This ``plug-compatible'' design ensures that components are interchangeable between modules from different regions and thus reduces the cost risk.
Corresponding interface definitions exist for the cavity, the fundamental-mode power coupler, the mechanical tuner and the helium tank.
The ``S1Global'' project~\cite{bib:s1g} has successfully built a single cryomodule from several cavities equipped with different couplers and tuners, demonstrating the viability of this concept.

\subsubsection{High-level radio-frequency}

The high-level radio-frequency (HLRF) system provides the RF power that drives the accelerating cavities.
The system comprises modulators, pulsed klystrons, and a waveguide power distribution system.

\paragraph{Modulators:}
The modulators provide the short, high-power electrical pulses required by the pulsed klystrons from a continuous supply of electricity. 
The ILC design foresees the use of novel, solid state Marx modulators.
These modulators are based on a solid-state switched capacitor network, where capacitors are charged in parallel over the long time between pulses, and discharged in series during the short pulse duration,
transforming continuous low-current, low voltage electricity into short high-power pulses of the required high voltage of \siunit{120}{kV} at a current of \siunit{140}{A}, over \siunit{1.65}{ms}.
Such Marx modulators have been developed at SLAC~\cite{Kemp:2011zz} 
and successfully tested at KEK~\cite{Gaudreau:2014pza}.
However, long-term data about the required large mean time between failures (MTFB) are not yet available.

\paragraph{Klystrons:}
The RF power to drive the accelerating cavities is provided by \siunit{10}{MW} L-band multi-beam klystrons. 
Devices meeting the ILC specifications were initially developed for the TESLA project, and later for the European XFEL.
They are now commercially available from two vendors (Thales and Toshiba), both of which provided klystrons for the European XFEL.
The ILC specifications ask for a $65\,\%$ efficiency (drive beam to output RF power), which are met by the existing devices.

\begin{figure}[htbp]
\begin{center}
   \includegraphics[width=0.7\hsize]{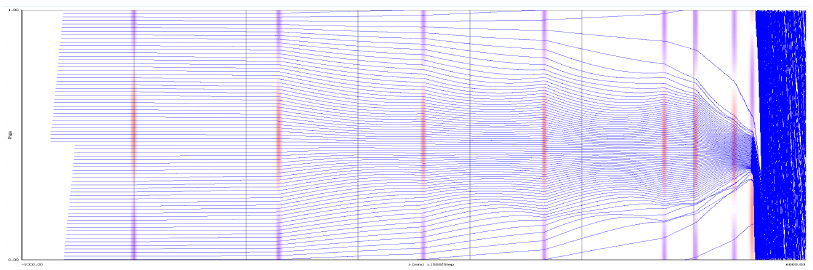}
\end{center}
\caption{Electron phase  profile of an \siunit{800}{MHz} klystron employing the Core Oscillation Method (COM)~\cite{Constable:2017hha}.
}
% Paper is published under CC-BY-3.0
\label{fig:com}
\end{figure}

Recently, the High Efficiency International Klystron Activity (HEIKA) collaboration~\cite{Syratchev:2015a, Gerigk:2018ebm} has been formed that investigates novel techniques for high--efficiency klystrons.
Taking advantage of modern beam dynamic tools, methods such as the Bunching, Alignment and Collecting (BAC) method~\cite{Guzilov:2014a} and the Core Oscillation Method (COM)~\cite{Constable:2017hha} (Fig.~\ref{fig:com})
 have been developed that promise increased efficiencies up to $90\,\%$~\cite{Baikov:2015bif}.  
One advantage of these methods is that it is possible to increase the efficiency of existing klystrons by equipping them with a new electron optics, as was demonstrated retrofitting an existing tube from VDBT, Moscow. 
This increased the output power by almost 50\,\% and its efficiency from 42\,\% to 66\,\%~\cite{Jensen:2016a}.

To operate the ILC at an increased gradient of \siunit{35}{MV/m} would require that the maximum klystron output power is increased from $10$ to \siunit{11}{MW}. 
It is assumed that this will be possible by applying the results from this R\&D effort to high-efficiency klystrons.
 
\paragraph{Local Power--Distribution System (LPDS):}

In the baseline design, a single RF station with one modulator and klystron supplies RF to $39$ cavities, which corresponds\footnote{Out of three cryomodules, two have nine cavities and one has eight plus a quadrupole, which amounts to $26$ cavities for a three cryomodule unit. Three such units share a cryogenic supply and are connected to two klystrons, see Sect.~\ref{sec:acceldesign}.} to $4.5$ cryomodules~\cite[Sec. 3.6.4]{Adolphsen:2013kya}. Then  $2$ klystrons drive a $9$ cryomodule cryo-string unit.
The power is distributed by the LPDS, a system of waveguides, power dividers and loads. 
All cavities from a $9$-cavity module and half of a $8$--cavity module are connected in one LPDS, and three such LPDS units are connected to one klystron.
This arrangement allows an easy refurbishment such that a third klystron can be added to a cryo-string, increasing the available power per cavity by $50\,\%$ for a luminosity upgrade (cf.\ Sec.~\ref{sec:lumi-upgrade}).

The LPDS design must provide a cost--effective solution for the  distribution of 
the RF power with minimal losses, and at the same time provide the 
flexibility to adjust the power delivered to each cavity by at least $\pm20\,\%$ to allow for 
the specified spread in maximum gradient. 
The LPDS design therefore contains remotely controlled, motor-driven Variable Power Dividers (VPD), phase shifters, and H--hybrids that can distribute the power with the required flexibility.
This design allows one to optimise the power distribution during operation, based on the cavity performance in the installed cryomodule, and thus to get the optimum performance out of the system.
It does not require a measurement of the individual cavity gradients after the module assembly, and is thus compatible with the ILC production scheme, where only a fraction of the cryomodules are tested.
This is a notable difference from  the scheme employed at the European XFEL, where $100\,\%$ of the modules were tested, and the the power distribution for each module was tailored to the measured cavity gradients, saving investment costs for the LPDS but making the system less flexible.

\subsubsection{Cryogenics}

The operation of the large number of superconducting cryomodules for the main linacs and the linacs associated with the sources requires a large--scale supply of liquid helium.
The cyomodules operate at \siunit{2}{K} and are cooled with superfluid helium, which at \siunit{2}{K} has a vapor pressure of about \siunit{32}{mbar}.

The accelerator is supplied with liquid helium by several cryogenic plants~\cite[Sec. 3.5]{Adolphsen:2013kya} of a size similar to those in operation at CERN for the LHC, at Fermilab, and DESY,
with a cooling capacity equivalent to about \siunit{19}{kW} at \siunit{4.5}{K}.
The \siunit{2}{K} and \siunit{4.5}{K} helium refrigerators are located in an underground access hall~\cite{bib:cr-0014} that is connected to the surface, where the helium compressors, gas tanks and further cryogenic infrastructure are located.
The total helium inventory is approximately $310,000$ liquid liters or about $41$ metric tonnes, about one third of the LHC's helium inventory.  A factor 2 more helium is needed for 
500~GeV operation.

%\subsubsection{Existing and Future Facilities}

% XXXXX MORE HERE  XXXXXXXX

\subsubsection{Series production and industrialisation}

Due to the construction of the European XFEL, the industrial basis for the key SCRF components is broad and mature.
In all three regions (Europe, America, Asia), several vendors for cavities have been qualified for ILC type cavities, and provided cost estimates in the past.
RF couplers have also been successfully produced by European and  American vendors for the European XFEL and LCLS-II projects.

ILC/TESLA type cryomodules have been built in laboratories around the world (DESY, CEA in Europe, FNAL and JLAB in America, KEK in Asia).
Series production has been established in America at Fermilab and JLAB for LCLS-II.
The largest series production was conducted by CEA in France~\cite{Weise:2014zqa,Berry:2017gpt}, again for the European XFEL, with the assembly of \num{103} cryomodules in total by an industrial partner under the supervision of CEA personnel, with a final throughput of one cryomodule produced every four working days.

ILC type pulsed \siunit{10}{MW} klystrons are commercially available from two vendors in Japan and Europe.

%===============================================================================

\subsection{Accelerator design}
\label{sec:acceldesign}

\subsubsection{Electron and positron sources}
\label{par:beampol}

The electron and positron sources are designed to produce \siunit{5}{GeV} beam pulses with a bunch charge that is $50\,\%$ higher than the design bunch charge of \siunit{3.2}{nC} ($\siunit{2\cdot 10^{10}}{e}$), in order to have sufficient reserve to compensate  any unforeseen inefficiencies in the beam transport.
In the baseline design, both sources produce polarized beams with the same time structure as the main beam, \ie, $1312$ bunches in a $\siunit{727}{\mu s}$ long pulse.

The electron source design~\cite{Adolphsen:2013kya} is based on the SLC polarized electron source, which has demonstarted that the bunch charge, polarization and cathode lifetime parameters are feasible.
The long bunch trains of the ILC do require a newly developed laser system and powerful preaccelerator structures, for which preliminary designs are available.
The design calls for  a Ti:sapphire laser impinging on a photocathode based on a strained GaAs/GaAsP superlattice
structure, which will produce electron bunches with an expected polarization of \siunit{85}{\%},
sufficient for \siunit{80}{\%} beam polarization at the interaction point, as demonstrated at SLAC~\cite{Alley:1995ia}.

The positron source poses a larger challenge. 

In the baseline design, hard circularly polarized photons are produced in a helical undulator driven by the main electron beam.  
These are converted to $e^+ e^-$ pairs in a target of \siunit{1}{m} diameter rotating at \siunit{100}{m/s}.
Positrons are captured in a flux concentrator or a quarter wave transformer, accelerated to \siunit{400}{MeV} in two normal conducting prea-ccelerators followed by a superconducting accelerator very similar to the main linac, before they are injected into the damping rings at \siunit{5}{GeV}.
The positrons inherit a longitudinal polarization of $30\,\%$ from the circularly polarized photons.
The positron polarization thus achieved is $30\,\%$.
The E-166 experiment at SLAC has successfully demonstrated this concept  \cite{Alexander:2009nb}, albeit at intensities much lower than foreseen for the ILC. 
Technological challenges of the undulator source concept are the target heat load, the radiation load in the flux concentrator device, and the dumping of 
the high intensity photon beam remnant.

As an alternative, an electron-driven positron source concept has been developed.
In the electron-driven scheme, a \siunit{3}{GeV} electron beam from a dedicated normal conducting linac produces positrons in a rotating target.
The electron drive beam, being independent from the main linac, has a completely different time structure. 
Positrons are produced in $20$ pulses at \siunit{300}{Hz} with $66$ bunches each.  With this scheme, it takes about \siunit{67}{ms} to produce the  positrons needed for a single Main Linac pulse with its $1312$ bunches, compared to \siunit{0.8}{ms} for the undulator source.
This different time structure spreads the heat load on the target over a longer time, allowing a target rotation speed of only 5~m/s rather than 100~m/s, which reduces the engineering complexity of the target design, in particular the vacuum seals of the rotating parts.
Although not free from its own engineering challenges, such as the high beam loading in the normal conducting cavities, the electron driven design is currently considered to be a low risk design that is sure to work.

Aside from the low technical risk, the main advantage of the electron driven design is the independence of positron production and electron main linac operation, which is an advantage for accelerator commissioning and operation in general.
In particular, electron beam energies below \siunit{120}{GeV} for operation at the $Z$ resonance or the $WW$ threshold would be no problem.
The undulator source, on the other hand, offers the possibility to provide beams at the maximum repetition rate of \siunit{10}{Hz} given by the damping time in the damping rings of \siunit{100}{ms}, whereas the electron driven scheme is limited to \siunit{6}{Hz} due to the additional \siunit{66}{ms} for positron production.
The main difference between the concepts is the positron polarization offered by the undulator source, which adds significantly to the physics capabilities of the machine.  
The physics implications of positron polarization is discussed later in the report, in Sec.~\ref{sec:polarization}. 

Both concepts have been reviewed recently \cite{PWG:2018a} inside the ILC community, with the result that both source concepts appear viable, with no known show stoppers, but they
require some more engineering work. 
The decision on the choice will be taken once the project has been approved, based on the physics requirements, operational aspects, and technological maturity and risks. 

\paragraph{Beam polarization and spin reversal}

At the ILC, the electron beam and potentially the positron beam are longitudinally polarized at the source, \ie,  the polarization vector is oriented parallel or antiparallel 
to the beam direction.
Whenever a longitudinally polarized beam of energy $E\sub{beam}$ is deflected by an angle $\theta\sub{bend}$, the polarization vector undergoes a precession through an angle $\theta\sub{pol} =  \gamma a \theta\sub{bend}$~\cite{Moffeit:2005pb}, 
with the Lorentz factor $\gamma = E\sub{beam}/m\sub{e}$ and the electron's anomalous magnetic moment $a = (g-2)/2$. 
To preserve the longitudinal beam polarization during the long transport from the source through the damping rings to the start of the main linac, which involves many horizontal bends, the beam polarization vector is rotated into the transverse plane, perpendicular to the damping ring plane, before the beam is transferred to the damping rings, and rotated back to a longitudinal direction by a set of spin rotators at the end of the RTML (see Sec.~\ref{sec:rtml}).
Through the use of two rotators, it is possible to bring the polarization vector into any desired direction, and compensate any remaining net precession between these spin rotators and the interaction point, so that any desired longitudinal or transverse polarization at the IP can be provided.

To control systematic effects, fast helicity reversal is required.  This is helicity reversal of each beam independently, on a pulse to pulse basis, which must be achieved without a change of the magnetic fields of the spin rotator magnets.
For the electron beam, a fast helicity reversal is possible through a flip of the cathode laser polarization.  For the undulator-based positron source, the photon polarization is given by the undulator field.  Two parallel sets of spin rotators in front of the damping rings are used that rotate the polarization vector either to the $+y$ or $-y$ direction.  With this scheme,
fast kickers can select a path through either of the two spin rotators and thus provide a fast spin reversal capability~\cite{Moffeit:2005pb,Malysheva:2016jdr}.

\subsubsection{Damping rings}

The ILC includes two oval damping rings of \siunit{3.2}{km} circumference, sharing a common tunnel in the central accelerator complex.
The damping rings reduce the horizontal and vertical emittance of the beams by almost six orders of magnitude\footnote{The normalized vertical emittance of the positrons is reduced from $\gamma \epsilon_{\mathrm{y}} \approx 8\,{\mathrm{mm}}$ to $20\,{\mathrm{nm}}$.} within a time span of only \siunit{100}{ms}, to provide the low emittance beams required at the interaction point. 
Both damping rings operate at an energy of \siunit{5}{GeV}.

The damping rings' main objectives are
\begin{itemize} 
\item to accept electron and positron beams at large emittance and produce the low-emittance beams required for high-luminosity production.
\item to dampen the incoming beam jitter to provide highly stable beams.
\item to 
delay bunches from the source and allow feed-forward systems to compensate for pulse-to-pulse variations in parameters such as the bunch charge.
\end{itemize}

Compared to today's fourth generation light sources, the target value for the normalized beam emittance ($\siunit{4}{\mu m}$/\siunit{20}{nm} for the normalised horizontal / vertical beam emittance) is low, but not a record value, and it is thus considered to be a realistic goal.

The main challenges for the damping ring design are to provide
\begin{itemize} 
\item a sufficient dynamic aperture to cope with the large injected emittance of the positrons.
\item a low equilibrium emittance in the horizontal plane.
\item a very low emittance in the vertical plane.
\item a small damping time constant.
\item damping of instabilities from electron clouds (for the positron DR) and fast ions (for the electron DR).
\item a small (\siunit{3.2-6.4}{ns}) bunch spacing, requiring very fast kickers for injection and ejection.
\end{itemize}

Careful optimization has resulted in a TME (Theoretical Minimum Emittance) style lattice for the arcs that balances a low horizontal emittance with the required large dynamic aperture~\cite[Chap. 6]{Adolphsen:2013kya}. 
Recently, the horizontal emittance has been reduced further by lowering the dispersion in the arcs through the use of longer dipoles~\cite{bib:cr-0016}.
The emittance in the vertical plane is minimised by careful alignment of the magnets and tuning of the closed orbit to compensate for misalignments and field errors, as demonstrated at the CESR-TA facility~\cite{Billing:2011zc}.

The required small damping time constant requires large synchrotron radiation damping, which is provided by the insertion of $54$ wigglers in each ring.
This results in an energy loss of up to $7.7\,{\mathrm{MV}}$ per turn and up to $3.3\,{\mathrm{MW}}$ RF power to 
store the positron beam at the design current of $390\,{\mathrm{mA}}$.  This
actually exceeds the average beam power of the accelerated positron beam, $2.6\,{\mathrm{MW}}$ at 
a $250\,{\mathrm{GeV}}$.

Electron cloud (EC) and fast ion (FI) instabilities limit the overall current in the damping rings to about \siunit{400-800}{mA}, where the EC limit that affects the positrons is assumed to be more stringent. 
These instabilities arise from electrons and ions being attracted by the circulating beam towards the beam axis. 
A low base vacuum pressure of \siunit{10^{-7}}{Pa} is required to limit these effects to the required level.
In addition, gaps between bunch trains of around $50$ bunches are required in the DR filling pattern, which permits the use of clearing electrodes to mitigate EC formation.
These techniques have been developed and tested at the CESR-TA facility~\cite{Conway:2012zza}

In the damping rings, the bunch separation is only \siunit{6.4}{ns} (\siunit{3.2}{ns} for a luminosity upgrade to $2625$ bunches). 
Extracting individual bunches without affecting their emittance requires kickers with rise/fall times of \siunit{3}{ns} or less.
Such systems have been tested at ATF~\cite{Naito:2010zzb}.

The damping ring RF system will employ superconducting cavities operating at half the Main Linac frequency (\siunit{650}{MHz}).
Klystrons and accelerator modules can be scaled from existing \siunit{500}{MHz} units in operation at CESR and KEK~\cite[Sec. 6.6]{Adolphsen:2013kya}.

\subsubsection{Low emittance beam transport: ring to Main Linac (RTML)}
\label{sec:rtml}

The Ring to Main Linac (RTML) system~\cite[Chap. 7]{Adolphsen:2013kya} is responsible for transporting and matching the beam from the Damping Ring to the entrance of the Main Linac.
Its main objectives are
\begin{itemize} 
\item transport of the beams from the Damping Rings at the center of the accelerator complex to the upstream ends of the Main Linacs,
\item collimation of the beam halo generated in the Damping Rings,
\item rotation of the spin polarization vector from the vertical to the desired angle at the IP (typically, in longitudinal direction).
\end{itemize}

The RTML consists of two arms for the positrons and the electrons. 
Each arm comprises a damping ring extraction line transferring the beams from the damping ring extraction into the main linac tunnel, a long low emittance transfer line (LTL), the turnaround section at the upstream end of each accelerator arm, and a spin rotation and diagnostics section.

The long transport line is the largest, most costly part of the RTML.
The main challenge is to transport the low emittance beam at \siunit{5}{GeV} with minimal emittance increase, and in a cost-effective manner, considering that the total length of both arms is about \siunit{14}{km} for the \siunit{250}{GeV} machine.

In order to preserve the polarization of the particles generated in the sources, their spins are rotated into a vertical direction (perpendicular to the Damping Ring plane) before injection into the Damping Rings. 
A set of two rotators~\cite{Emma:1995kf} employing superconducting solenoids allows to rotate the spin into any direction required.

At the end of the RTML, after the spin rotation section and before injection into the bunch compressors (which are considered part of the Main Linac, not the RTML~\cite{bib:cr-0010}), a diagnostics section allows measurement of  the emittance and the  coupling between the  horizontal and vertical plane.
A skew quadrupole system is included to correct for any such coupling.

A number of circular fixed-aperture and rectangular variable-aperture collimators in the RTML provide betatron collimation at the beginning of the LTL, in the turn around and before the bunch compressors.

\subsubsection{Bunch compressors and Main Linac}

\begin{figure}[htbp]
\begin{center}
   \includegraphics[width=0.7\hsize]{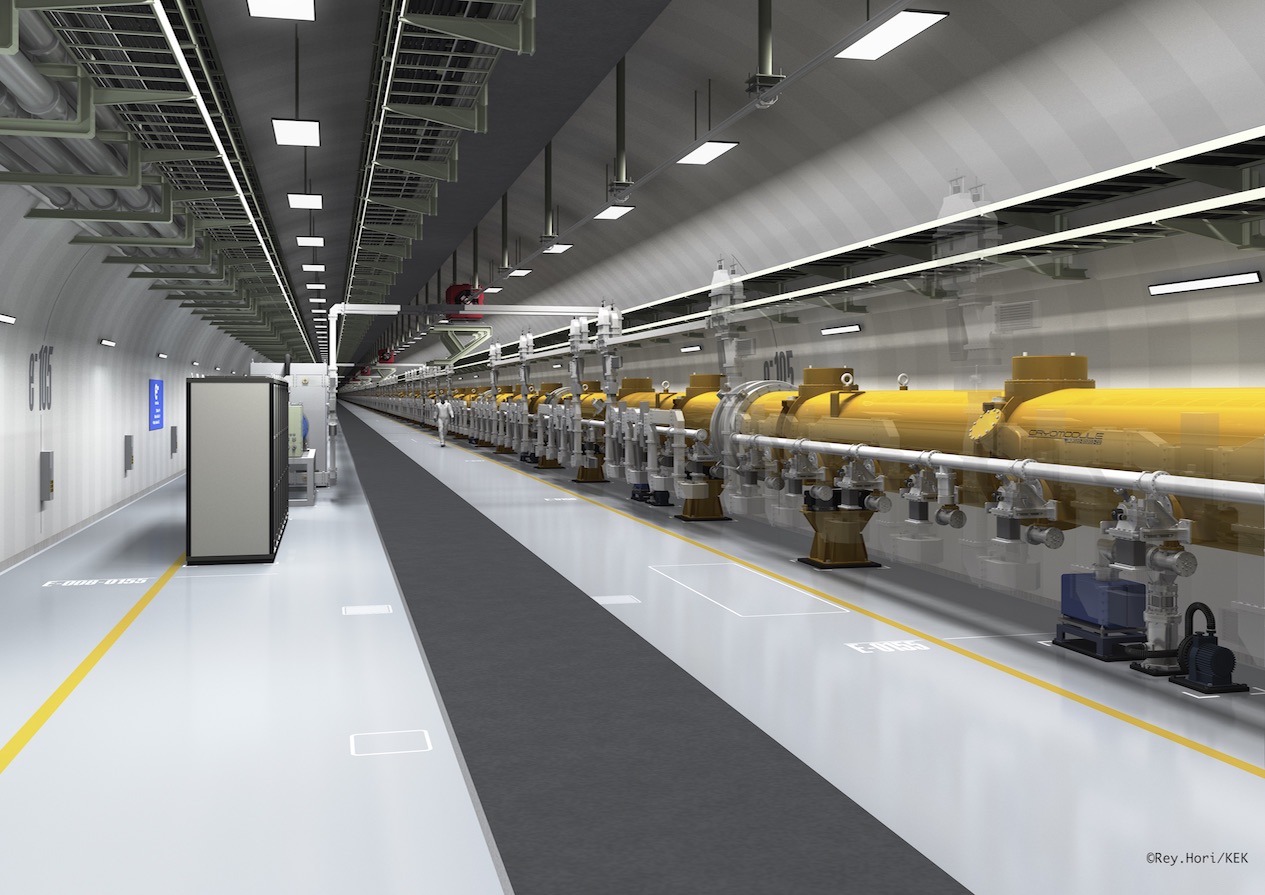}
\end{center}
\caption{Artist's rendition of the ILC Main Linac tunnel. The shield wall in the middle has been removed.
\copyright Rey.Hori/KEK.}
\label{fig:ilc-tunnel}
\end{figure}

\begin{figure}[htbp]
\begin{center}
   \includegraphics[width=0.7\hsize]{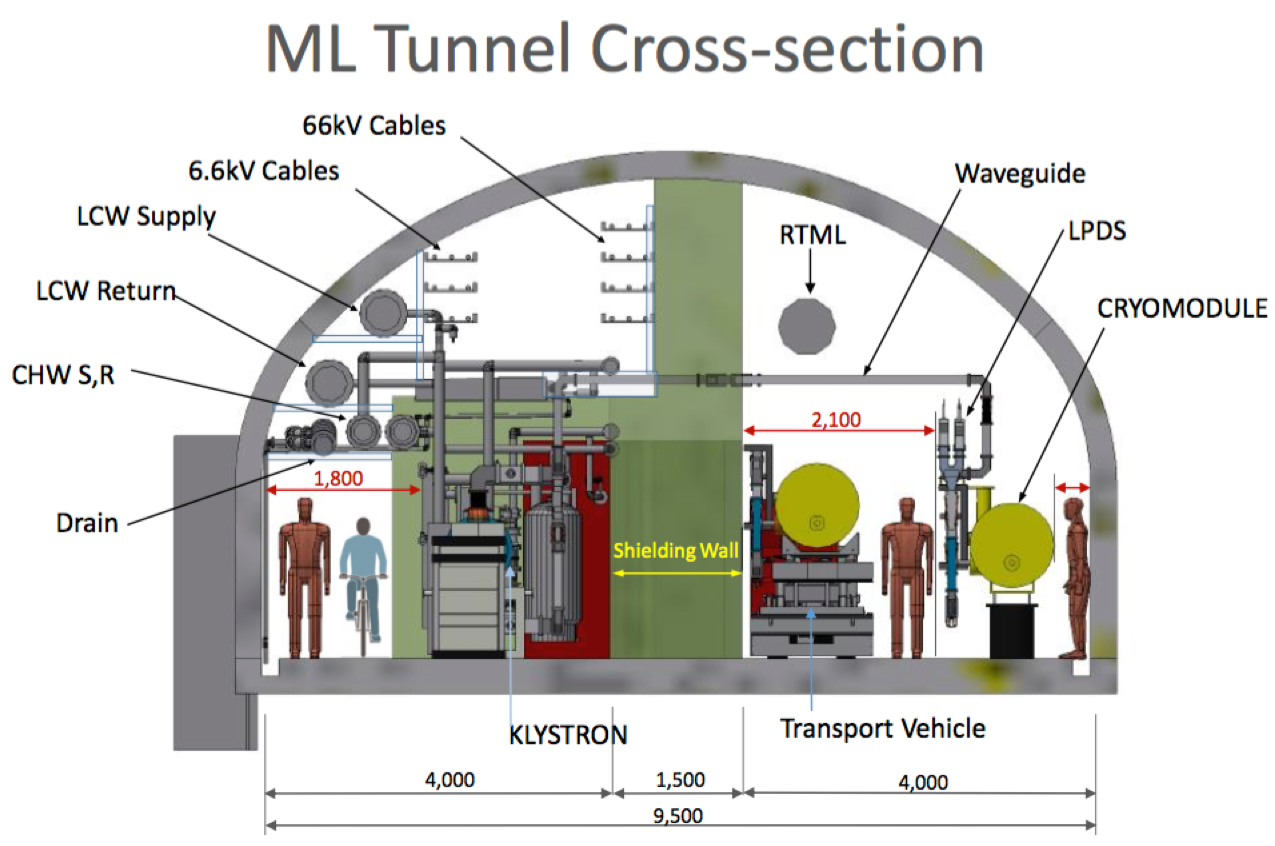}
\end{center}
\caption{Cross section through the Main Linac tunnel.}
\label{fig:ml-tunnel}
\end{figure}

At the heart of the ILC are the two Main Linacs, which accelerate the beams from $5$ to \siunit{125}{GeV}.
The linac tunnel, as depicted in Figs.~\ref{fig:ilc-tunnel} and \ref{fig:ml-tunnel}, has two parts, separated by a shield wall. 
One side (on the right in Fig.~\ref{fig:ilc-tunnel}) houses the beamline with the accelerating cryomodules as well as the RTML beamline hanging on the ceiling.
The other side contains power supplies, control electronics, and the modulators and klystrons of the High-Level RF system.
The concrete shield wall (indicated as a dark-grey strip in in Fig.~\ref{fig:ilc-tunnel}) has a thickness of \siunit{1.5}{m}~\cite{bib:cr-0012}.
The shield wall allows access to the electronics, klystrons and modulators during operation of the klystrons with cold cryomodules, protecting personnel from X-ray radiation emanating from the cavities caused by dark currents.
Access during beam operation, which would require a wall thickness of \siunit{3.5}{m}, is not possible.

The first part of the Main Linac is a two-stage bunch compressor system~\cite[Sec. 7.3.3.5]{Adolphsen:2013kya}, each consisting of an accelerating section followed by a wiggler. 
The first stage operates at \siunit{5}{GeV}, with no net acceleration, the second stage accelerates the beam to \siunit{15}{GeV}.
The bunch compressors reduce the bunch length from $6$ to \siunit{0.3}{mm}.

After the bunch compressors, the Main Linac continues for about \siunit{6}{km} with a long section consisting entirely of cryomodules, bringing the beam to \siunit{125}{GeV}. 

\paragraph{RF distribution:}

Each cryomodule contains $9$ cavities, or for every third module, $8$ cavities and a package with a superconducting quadrupole, corrector magnets, and beam position monitor.
Nine such modules, with a total of 78 cavities, are powered by $2$ klystrons and provide \siunit{2.54 (2.82)}{GeV} at a gradient of \siunit{31.5 (35)}{MV/m}.
Table~\ref{tab:ml-units} gives an overview over the units that form the linacs.
The waveguide distribution system allows an easy refurbishment to connect a third klystron for a luminosity upgrade.
The $50\,\%$ RF power increase would allow $50\,\%$ higher current through smaller bunch separation, and longer beam pulses because of a reduced filling time, so that the number of bunches per pulse and hence the luminosity can be doubled, while the RF pulse duration of \siunit{1.65}{ms} stays constant.

\paragraph{Cryogenic supply:}

A $9$ module unit forms a cryo string, which is connected to the helium supply line with a Joule-Thomson valve.
All helium lines are part of the cryomodule, obviating the need for a separate helium transfer line. 
Up to $21$ strings with $189$ modules and \siunit{2.4}{km} total length can be connected to a single plant; 
this is limited by practical plant sizes and the gas--return header pressure drop.  

\begin{table}[htbp]
\begin{center}
\begin{tabular}{llrr}
Unit & Comprises & Length & Voltage \\
\hline
Cavity & \siunit{1.038}{m} active length & \siunit{1.25}{m} & \siunit{32.6~/~36.2}{MV} \\
Cryomodule & $8\,^2/_3$ cavities & \siunit{12.65}{m} & \siunit{282~/~314}{MV} \\
RF Unit & $4.5$ cryomodules & \siunit{58.2}{m} & \siunit{1.27~/~1.41}{GV} \\
Cryostring & 2 RF units & \siunit{116.4}{m} & \siunit{2.54~/~2.82}{GV} \\
Cryounit & up to 21 cryostrings & \siunit{2454}{m} & \siunit{53.4~/~39.3}{GV} \\
\end{tabular}
\end{center}
\caption{
  \label{tab:ml-units}
  Units that make up the main linacs. 
  The voltage takes into account that the beam is $5^\circ$ shifted in phase (``off crest'') for longitudinal stability, and is given for an average gradient of \siunit{31.5 / 35}{MV/m}.
  A RF unit is powered by one klystron, each cryostring is connected by a valve box to the liquid helium supply, and a cryounit is supplied by one cryogenic plant.
  Total lengths include additional space between components.  
  Cryomodules comprise $9$ or $8$ cavities, in a $2:1$ mixture, resulting in $8\,^2/_3$ cavities per cryomodule on average.
}
\end{table}

\paragraph{Cost reduction from larger gradients:}

\begin{figure*}[htbp]
\begin{center}
   \includegraphics[width=\hsize]{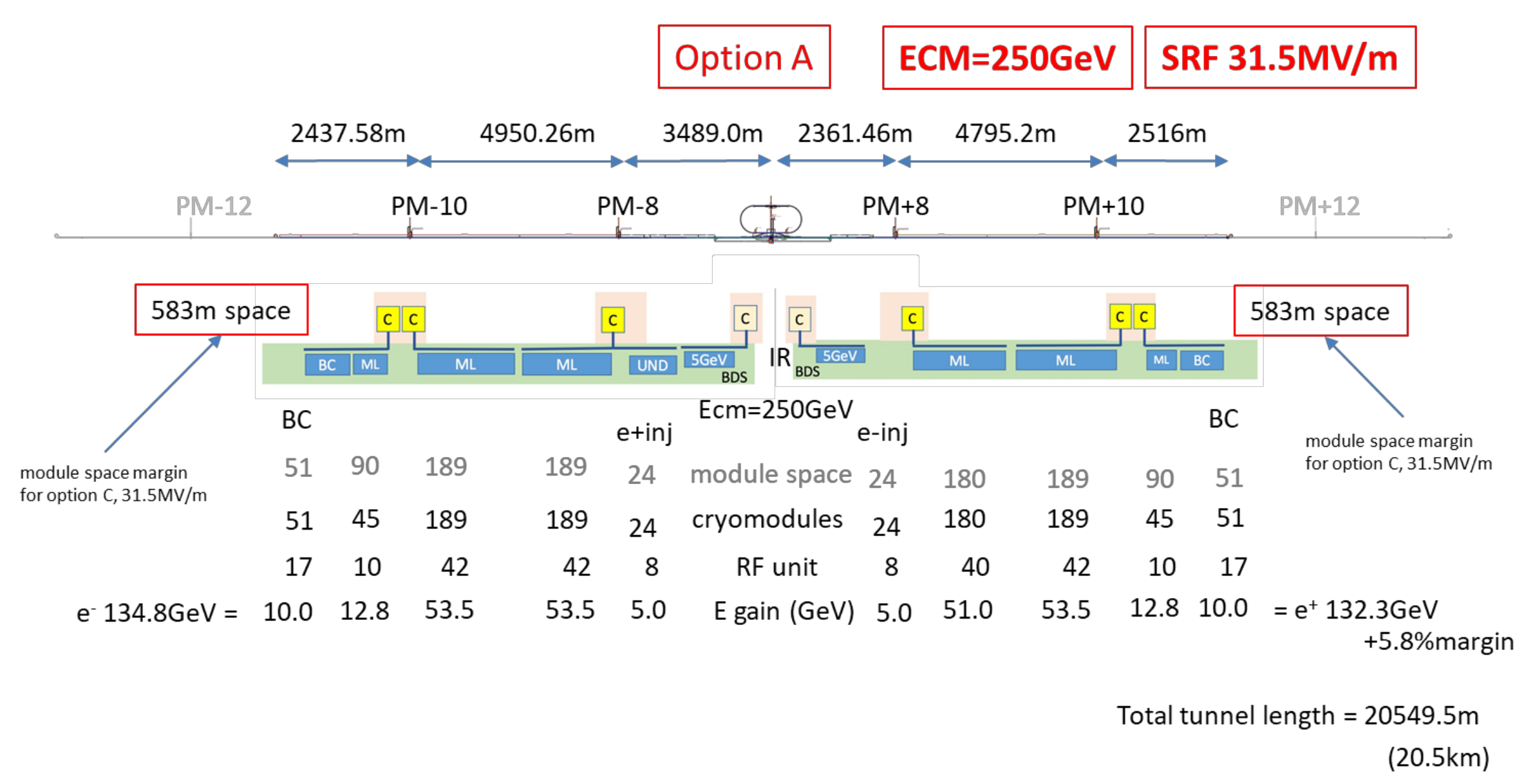}
   \includegraphics[width=\hsize]{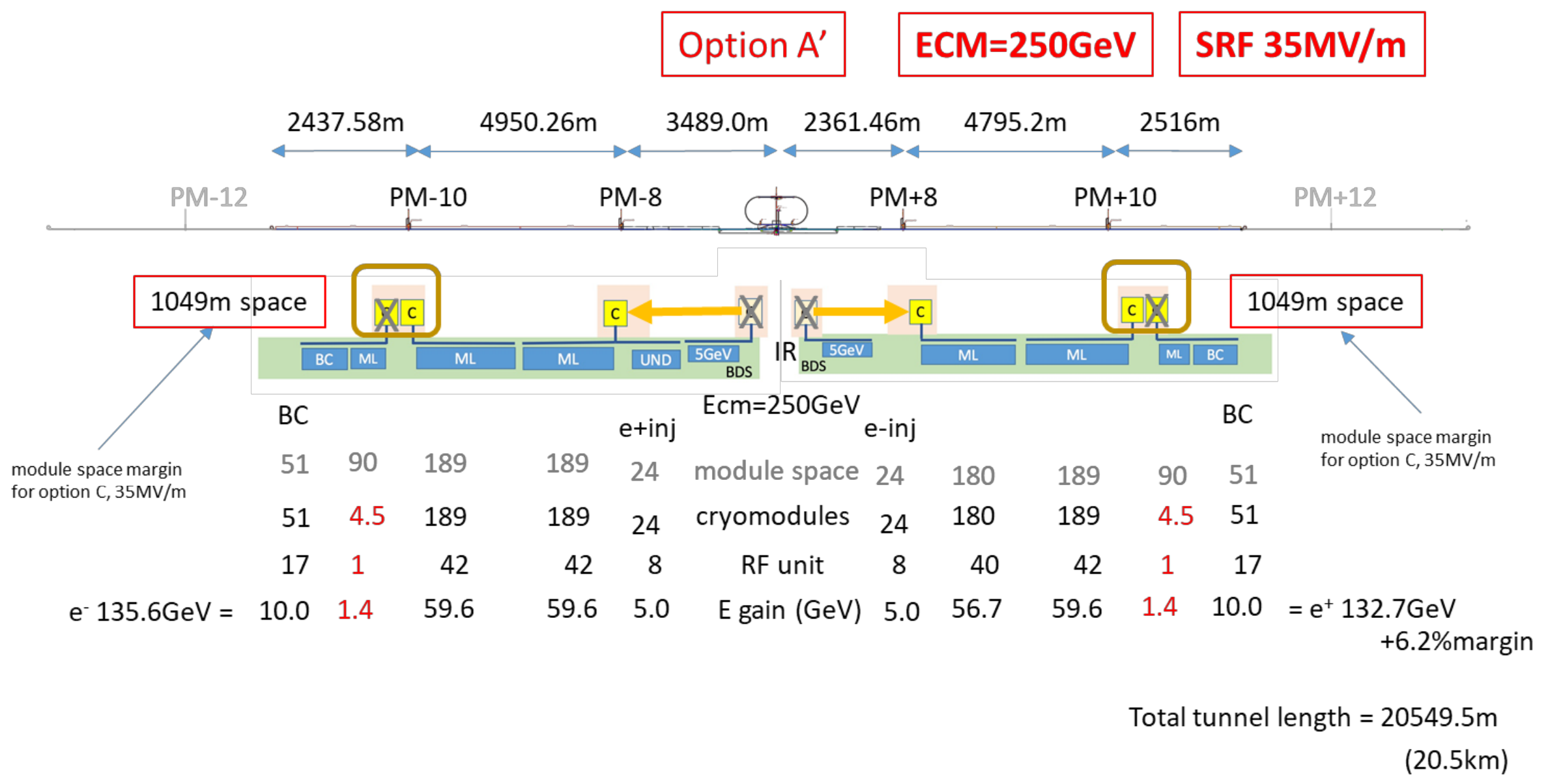}
\end{center}
\caption{Cryogenic layout for a gradient of \siunit{31.5}{MV/m} (top) and \siunit{35}{MV/m} (bottom)~\cite{Evans:2017rvt}.
``Module space'' indicates how many cryomodules can be physically installed, ``cryomodules'' and ``RF unit'' indicates the number of actually installed modules and klystrons (one klystron per 4.5 cryomodules). ``E gain'' indicates the energy gain in GeV. ``BC'', ``ML'', ``e+ inj'', ``e- inj'' and ``UND'' refer to the sections with need for liquid helium: bunch compressor, main linac, 5GeV boosters in the positron and electron source, and the positron source undulator section, respectively. PM$\pm8, 10, 12$ refer to access hall locations, ``C'' to cryo plants; meter numbers on top indicate the length of the corresponding section.}
\label{fig:ml-cryo-opta}
\end{figure*}

Figure~\ref{fig:ml-cryo-opta} shows the layout of the cryogenic supply system for the \siunit{250}{GeV} machine.
At the top, the situation is depicted for the gradient of \siunit{31.5}{MV/m} with a quality factor of $Q\sub{0}=1.0\cdot 10^{10}$, as assumed in the TDR~\cite{Adolphsen:2013kya}. 
In this case, the access points PM$\pm 10$ would house two cryogenic plants, each supplying up to $189$ cryomodules or an equivalent cryogenic load.   In this configuration $6$ large plants in the access halls plus $2$ smaller plants in the central region would be needed.
The bottom picture shows the situation for a gradient of \siunit{35}{MV/m} with $Q\sub{0}=1.6\cdot 10^{10}$, as could be expected from successful R\&D. 
The increased gradient would  allow reduction of the total number of cryomodules by roughly $10\,\%$ from $987$ to $906$. The increased quality factor would reduce the dynamic losses such that $4$ cryo plants would provide sufficient helium.

In general, the accelerator is designed to make good use of any anticipated performance gain from continued high gradient R\&D,  in the case that raising the gradient is seen to be  beneficial from an economical point of view, without incurring unwanted technology risk.

\subsubsection{Beam delivery system and machine detector interface}
\label{subsubsec:bds_mdi}

The Beam Delivery System (BDS) transports the $e^+/e^-$ beams from the end of the main linacs, focuses them to the required small beam spot at the Interaction Point (IP), brings them into collision, and transports the spent beams to the main dumps~\cite[Chap. 8]{Adolphsen:2013kya}.
The main functions of the BDS are
\begin{itemize}
\item measuring the main linac beam parameters and matching it into the final focus.
\item protecting beamline and detector from mis-steered beams~\footnote{On the electron side, the protective fast beam abort system is actually located upstream of the positron source undulator.}.
\item removing large amplitude (beam--halo) and off--momentum particles from the beam to minimize background in the detector.
\item accurately measuring the key parameters energy and polarization before and after the collisions.
\end{itemize}
The BDS must provide sufficient diagnostic and feedback systems to achieve these goals.

The BDS is designed such that it can be upgraded to a maximum center-of-mass energy of \siunit{1}{TeV}; components such as the beam dumps, that are not cost drivers for the overall project but would be cumbersome to replace later, are dimensioned for the maximum beam energy from the beginning.
In other places, such as the energy collimation dogleg, those components necessary for \siunit{125}{GeV} beam operation are installed and space for a later upgrade is reserved.

Overall, the BDS is \siunit{2254}{m} long from the end of the main linac (or the undulator and target bypass insert of the positron source on the electron side, respectively) to the IP.

\paragraph{Diagnostics and collimation section:}
The BDS starts with a diagnostics section, where emittance, energy and polarization are measured and any coupling between the vertical and horizontal planes is corrected by a set of skew quadrupoles.
The energy measurement is incorporated into the machine protection system and can, \eg,  extract off-momentum bunches caused by a klystron failure in the main linac that would otherwise damage the machine or detector.
An emergency dump~\cite{bib:cr-0013} is dimensioned such that it can absorb a full beam pulse at \siunit{500}{GeV}, sufficient for \siunit{1}{TeV} operation.

The diagnostics section is followed by a collimation system, which first removes beam halo particles (betatron collimation). 
Then, off-momentum particles are removed.
In this energy collimation section, sufficient dispersion must be generated by bending the beam in a dogleg, while avoiding excessive synchrotron radiation generation in dispersive regions that leads to an increase of the horizontal emittance.
This emittance dilution effect grows as $E\sub{beam}^6$ at constant bending radius for the normalised emittance, and determines the overall length of the energy collimation section for a maximum \siunit{500}{GeV} beam energy to about \siunit{400}{m}.

\paragraph {Final focus with feedback system and crab cavities:}

The final focus system demagnifies the beam to the required spot size of \siunit{516 \times 7.7}{nm^2} by means of a final quadrupole doublet.
Even the relatively small energy spread of $\approx 0.1\,\%$ leads to a significant spread of the focal length of the doublet and requires a correction to achieve the desired beam size, which is realised by a local chromaticity correction scheme~\cite{Raimondi:2000cx}.

To bring the beams to collision with the neccessary nanometer accuracy requires a continuous compensation of drift and vibration effects.
Along the ILC, the pulse length and bunch separation (\siunit{727}{\mu s} and \siunit{554}{ns}, respectively) are large enough to allow corrections between pulses as well as within a bunch train (intratrain feedback).
Beam-beam offsets of a fraction of the beam size lead to a measurable deflection of the outgoing beams,and these measurements are used to feed fast stripline kickers that stabilize the beam. 
Finally, the \siunit{3.9}{GHz} crab cavities close to the interaction point are incorporated that rotate the bunches to compensate for the \siunit{14}{mrad} beam crossing angle~\cite[Sect. 8.9]{Adolphsen:2013kya}.

\begin{figure}[htbp]
\begin{center}
   \includegraphics[width=0.8\hsize]{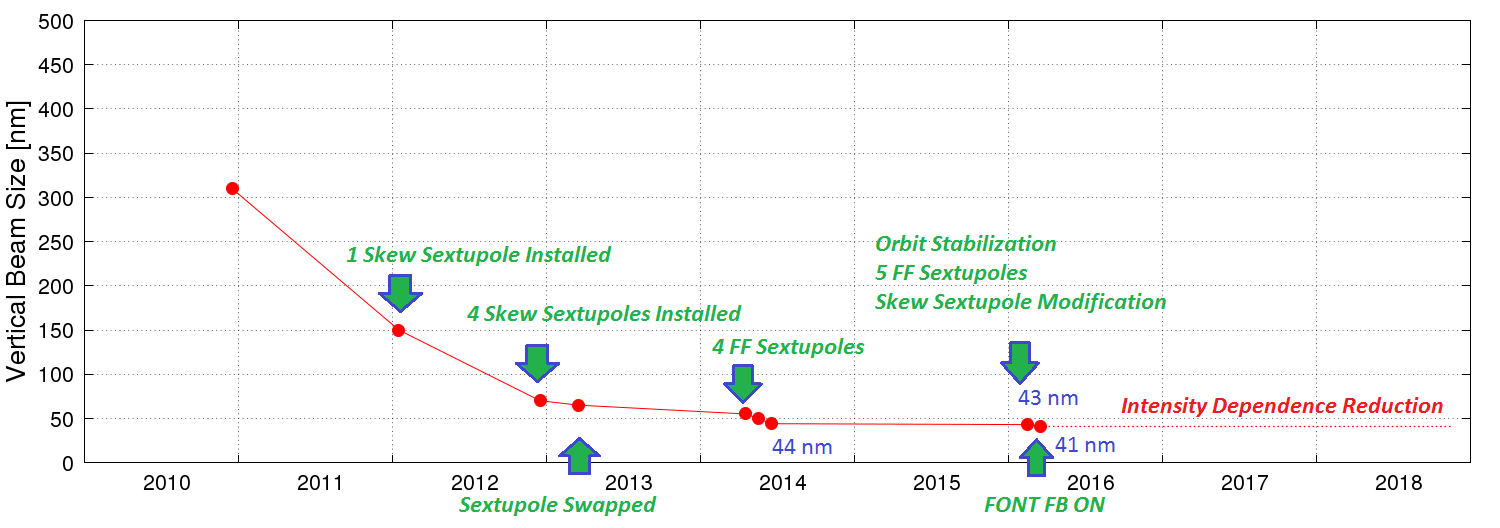}
\end{center}
\caption{Beamsizes achieved at the Accelerator Test Facility 2 (ATF2) as a function of time~\cite{bib:atf2esu}. The latest result (\siunit{41}{nm}~\cite{Okugi:2017jji}) is within $10\,\%$ of the goal beam size of \siunit{37}{nm}.}
\label{fig:atf-results}
\end{figure}

\paragraph {Test results from ATF2:}
The Accelerator Test Facility 2 (ATF2) was built at KEK in 2008 as a test bench for the ILC final focus scheme~\cite[Sec. 3.6]{Adolphsen:2013jya}.
Its primary goals were to achieve a \siunit{37}{nm} vertical beam size at the interaction point (IP), and to demonstrate beam stabilisation at the nanometer level~\cite{Grishanov:2005ek,Grishanov:2006kx}.
After scaling for the different beam energies (ATF2 operates at $E\sub{beam}=\siunit{1.3}{GeV}$), the \siunit{37}{nm} beam size corresponds to the TDR design value of $\sigma\sub{y}^* = \siunit{5.7}{nm}$ at \siunit{250}{GeV} beam energy.
As Fig.~\ref{fig:atf-results} shows, this goal has been reached within $10\,\%$~\cite{Okugi:2017jji} by the successive application of various correction and stabilisation techniques, 
validating the final focus design, in particular the local chromaticity correction~\cite{White:2014vwa}.

The fifth generation FONT5 feedback system~\cite{Apsimon:2018bpq} for the ILC and CLIC has also been tested at the ATF2, where a beam stabilisation to \siunit{41}{nm} has been demonstrated, in excellent agreement with the predicted one given the incoming bunch jitter and bunch-to-bunch correlation~\cite{Ramjiawan:2018egu}.

Since November 2016, intensity-dependence effects on the ATF2 beam size have been studied extensively.  
They show a degradation of the beam size with increasing intensity that is compatible with the effect of  wakefields.
Simulations and experiments in ATF2 show that the effect is not important when scaled to ILC, 
and could be mitigated by including a dedicated ``wakefield knob'' in the routine tuning procedure.

\paragraph {Machine detector interface (MDI):}

% describe experimental hall, final focus

% XXXXXXXXXXXXX BEGIN NEW 20190219 XXXXXXXXXXXXXX

The ILC is configured to have two detectors that share one interaction point, with one detector in data taking position at any time, in a so--called ``push--pull'' operation~\cite[Sec. 8.4]{Adolphsen:2013jya}.
Both detectors are mounted on movable platforms that allow an exchange of the detectors within approximately \siunit{24}{hours}.

In the push--pull scheme, the innermost final focus quadrupole ``QD0'', a slim, superconducting magnet package combined with a sextupole for local chromaticity correction, is installed within the detectors. 
The other part of the final focus doublet (``QF1'') is located outside the detector on a bridge, and does not move with the detector.
Since the TDR, the free space $L^*$ between interaction point and the QD0 edge has been harmonised to a common value of $L^*=\siunit{4.1}{m}$~\cite{bib:cr-0002}, which facilitates the design of a final focus optics that delivers optimal and equal performance to both detectors.

The detectors are located in an underground cavern. 
In contrast to the TDR design, it is foreseen to have a large vertical access shaft~\cite{bib:cr-0003}, which permits a CMS--style detector installation concept, in which the detectors are assembled in large modules in a surface hall and lowered into the hall by means of a gantry crane capable of lowering pieces that weigh up to \siunit{4000}{t}.
As the CMS experience shows, this concept significantly reduces the schedule risk associated with the experimental hall, since the cavern needs to be available for detector installation only one or two years prior to commissioning.

% \paragraph {Extraction line}

% XXXXXXXXXXXXX END NEW 20190219 XXXXXXXXXXXXXX

\paragraph {Main dump:}

The main beam dumps~\cite[Sect. 8.8]{Adolphsen:2013kya} are rated for a maximum beam power of \siunit{17}{MW}~\cite{bib:cr-0013}, enough for a \siunit{1}{TeV} upgrade of the accelerator.
The main dump design is based on the successful SLAC \siunit{2.2}{MW} beam dump~\cite{Walz:1967nz}.
It  utilises water at \siunit{10}{bar} pressure (to prevent boiling) as the absorber medium. 
The main engineering challenges lie in the safe recombination of the produced oxyhydrogen gas and in the safe containment and disposal of radioisotopes, in particular tritium and $^7{\mathrm{Be}}$ produced from spallation processes.
The entry window is another component that must be carefully designed. 

% XXXXXXXXXXXXX BEGIN NEW 20190219 XXXXXXXXXXXXXX

\paragraph {Measurement of beam energy, luminosity, and beam polarization:}
This paragraph gives a brief overview on the BDS components which serve the measurements of beam energy, luminosity, and beam polarization. These measurements and their combination with additional information from $e^+e^-$ collision data will be discussed in more detail in Sec.~\ref{sec:lep_precision}.

Two energy spectrometers, one located \siunit{700}{m} upstream of the IP, the other \siunit{55}{m} downstream, provide independent and complementary measurements of the beam energy with an accuracy of \siunit{100}{ppm}~\cite{Boogert:2009ir}.

The luminosity is measured to $10^{-3}$ accuracy from low angle Bhabha scattering in the luminosity calorimeters of the main collider experiments, typically covering polar angles from $30$ to \siunit{90}{mrad}. Additional calorimeters (BeamCal) in the region $5$ to \siunit{30}{mrad} provide a fast signal that is sensitive to the beam sizes and offsets of the colliding beam, and that can thus be used for their tuning, as part of an intra-beam feedback system.

Beam polarization is measured by means of Compton scattering~\cite{Baier:1969hw}: electrons that scatter off green or infrared light laser photons lose enough energy that they can be detected in a spectrometer; their momentum spectrum is used to fit the beam polarization~\cite{Vormwald:2015hla}.
Two such polarimeters are located \siunit{1800}{m} upstream and \siunit{150}{m} downstream of the IP, which allows to interpolate the precise polarization at the IP and control the systematics, including effects from precession of the polarization vector by transverse fields and depolarizing effects in the interaction, which lead to a sizeable variation of the polarization within the bunch during the collision. Each polarimeter will measure the local polarization up to a relative scale uncertainty of $0.25\%$, whose impact on physics measurements will be largely mitigated by extracting the luminosity-weighted average polarisation values from the $e^+e^-$ collision data themselves. It is expected that uncorrectable point-to-point uncertainties will be one to even two orders of magnitude smaller, dominated by residual variations in the beam conditions.

%===============================================================================
% BL 13.10.21 subsection on upgrade options removed completely, because this is a separate section now
% \subsection{Upgrade options \label{subsec:upg-opt}}
%===============================================================================

% XXXXXXXXXXXXX BEGIN NEW 20220114 by K.Yokoya XXXXXXXXXXXXXX

\subsection{\boldmath Operation at the $Z$-pole}
\label{subsec:Zpole_accelerator}

The TDR described the design of ILC for the energy range between \siunit{200}{GeV} and \siunit{500}{GeV} 
with possible upgrade to  \siunit{1}{TeV}. The project starts with \siunit{250}{GeV} as the Higgs factory. 
However, once the ILC for \siunit{250}{GeV} is built, it is still possible to operate it below the lowest of these energy regions---in particular at the $Z$ pole at a center-of-mass energy of \siunit{91.2}{GeV}.  Of course, the luminosity will be  
lower than at \siunit{250}{GeV}. 

The first issue for the $Z$-pole operation is positron production. Since the electron beam of energy 
\siunit{45.6}{GeV} cannot produce sufficient number of positrons by the undulator scheme, 
another electron beam (\siunit{125}{GeV}) dedicated to positron production is necessary. 
This is achieved by operating the $e^-$ part of the accelerator at twice the repetition frequency, with alternating beams for positron production and for physics collisions.
In this scheme the maximum repetition frequency of collision allowed by the AC power system 
of ILC250 turns out to be \siunit{3.7}{Hz}. 
Thus, the machine operation cycle is as follows:
\begin{enumerate}
\item  Create \siunit{5}{GeV} electron beam (1312 bunches) and store it in DR for 1/(2$\times$\siunit{3.7}{Hz}) = \siunit{135}{ms}.
\item  Extract the electron beam from DR, accelerate it to \siunit{125}{GeV} in the electron main linac, 
   let it go through the undulator, create positron beam, accelerate the positron beam to \siunit{5}{GeV} 
   and store it in the positron DR for \siunit{135}{ms}. 
\item  Create the next \siunit{5}{GeV} electron beam and store it in DR for \siunit{135}{ms} in parallel with step 2.
\item  Extract the electron/positron beams from each DR, accelerate to \siunit{45.6}{GeV} 
   in the electron/positron main linacs and collide them.
\end{enumerate}
This one cycle takes 2$\times$\siunit{135}{ms} = \siunit{270}{ms} = 1/\siunit{3.7}{Hz}.
The spent electron beam after step 2 is transported to the special beam dump 
(designed for up to \siunit{8}{MW}).

In this scenario the positron main linac is operated at \siunit{3.7}{Hz}, whereas the electron main linac 
is at \siunit{7.4}{Hz}, one pulse accelerates electrons to \siunit{125}{GeV} at the full gradient (\siunit{31.5}{MV/m}) and 
the next pulse to \siunit{45.6}{GeV} at a gradient of \siunit{8.76}{MV/m} by adopting reduced klystron power\footnote{For the latter low energy pulse one might imagine 
accelerating the beam to \siunit{45.6}{GeV} at full gradient and 
turning off the power in the rest of the linac. This would have the advantage of 
better beam dynamics in the linac and a consequently smaller emittance at the 
end of the linac. However this 
scheme does not work with the baseline accelerator design because 
it is not possible to detune quickly the cavities in the rest of the 
electron linac as needed to avoid beam loading.
%but detuning by the mechanical tuner cannot be done at \siunit{3.7}{Hz}.)
Implementation of such a full gradient low energy pulse would 
need a dedicated rest-of-linac bypass for the electron beamline.}.
% So there may be the possibility of finding a 
% solution ? (Graham Wilson)

The time for damping in DR is \siunit{135}{ms}, shorter than the \siunit{200}{ms} of standard \siunit{5}{Hz} operation. 
This is feasible because the power system of DR can accept up to \siunit{10}{Hz}. 

There are many issues to be considered in addition to above such as
\begin{itemize}
\item  Required wiggler strength in DR and re-evaluation of the dynamic aperture
\item  Beam dynamics in the low-gradient main linac under alternating gradient (31.5 and \siunit{8.76}{MV/m}) 
          operation mode. (Orbit correction for colliding beam only).
\item  Tight horizontal collimation depth due to large geometric emittance. Momentum band-width 
         of the BDS system (a longer bunch 
         length of \siunit{0.41}{mm} is adopted to reduce the beam energy spread). 
\item  The wakefield effects in BDS due to low energy and long bunch.
\item   Beam-beam interaction with large disruption parameter ($D_y \approx 32$).
\end{itemize}
These issues are discussed in detail in~\cite{yokoya2019}.

The relevant parameters are listed in Tab.~\ref{tab:ilc-params}. The luminosity is estimated to be 
$2.05\times 10^{33} /\mbox{cm}^2/\mbox{s}$  with $1312$ bunches per pulse 
and $4.1\times 10^{33} /\mbox{cm}^2/\mbox{s}$  with twice the number of bunches. 
The expected beam polarization is the same as in the \siunit{250}{GeV} case, i.e. $>80$\% for electrons 
and $\sim$30\% for positrons.

In the case that the e-driven positron source is adopted instead of the undulator source, 
simple \siunit{5}{Hz} operation is possible as at \siunit{250}{GeV} and, therefore, the luminosity would be higher 
by a factor 5/3.7, i.e. $2.8\times 10^{33} /\mbox{cm}^2/\mbox{s}$  with 1312 bunches.  In this case, the  
 positron beam would not be  polarized.

In summary, $Z$-pole operation of the ILC is possible with similar performance for both positron source concepts under discussion.

% XXXXXXXXXXXXX END NEW 20220114 by K.Yokoya XXXXXXXXXXXXXX

\subsection{Civil engineering and site}

\begin{figure}[p]
   \includegraphics[width=\hsize]{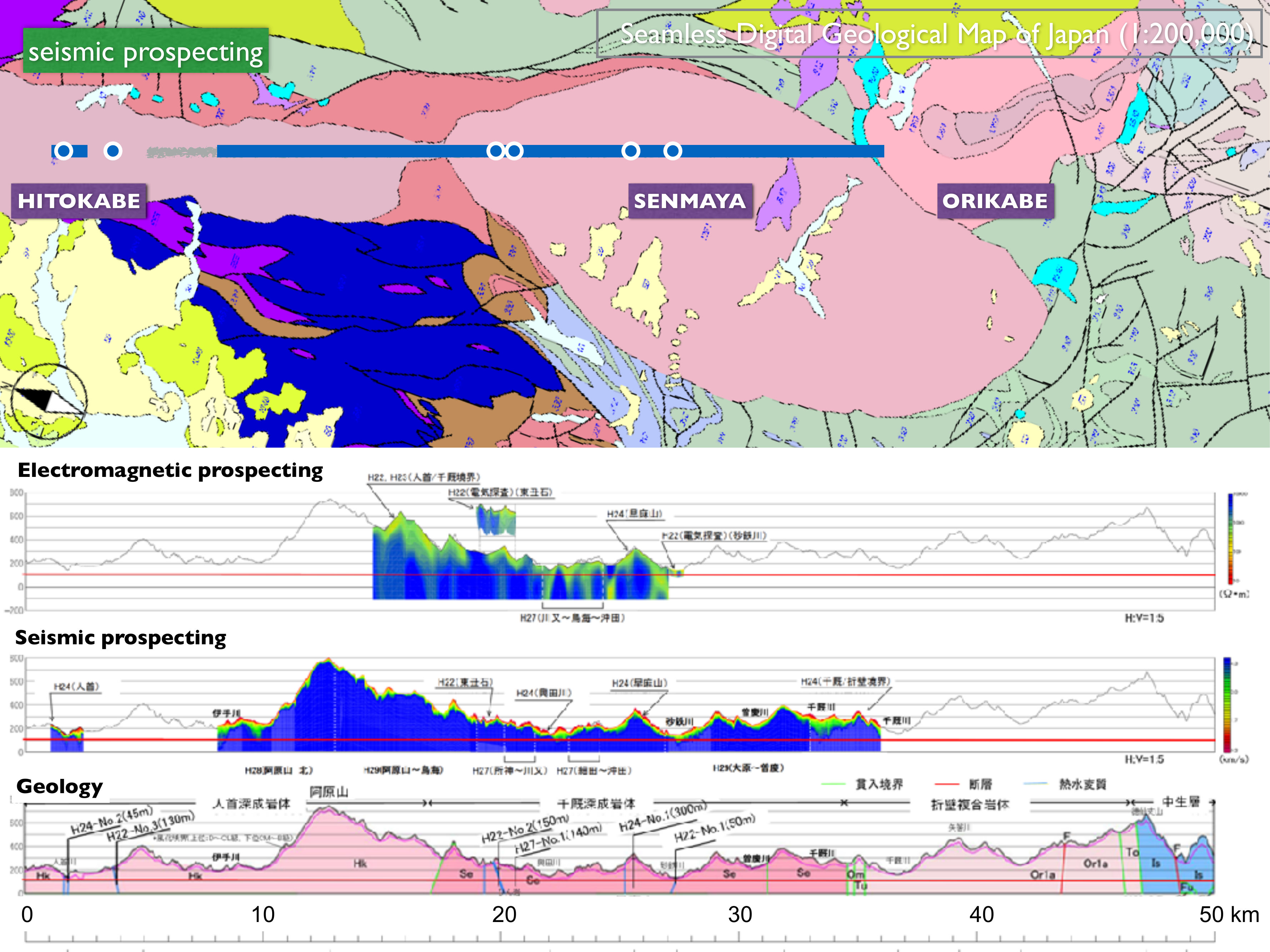}
\caption{Geological situation at the Kitakami site.}
\label{fig:kitakami-geology}
\end{figure}

In 2014, the ILC Strategy Council announced the result of its candidate site evaluation for the best possible ILC site in Japan~\cite{ILCSC:2014a}.
The evaluation was conducted by a number of Japanese experts from universities and industry, and reviewed by an international commitee. 
It considered technical as well as socio-environmental aspects, and concluded that the candidate site in the Kitakami region is best suited for the ILC.

\begin{figure}[htbp]
\begin{center}
   \includegraphics[width=0.7\hsize]{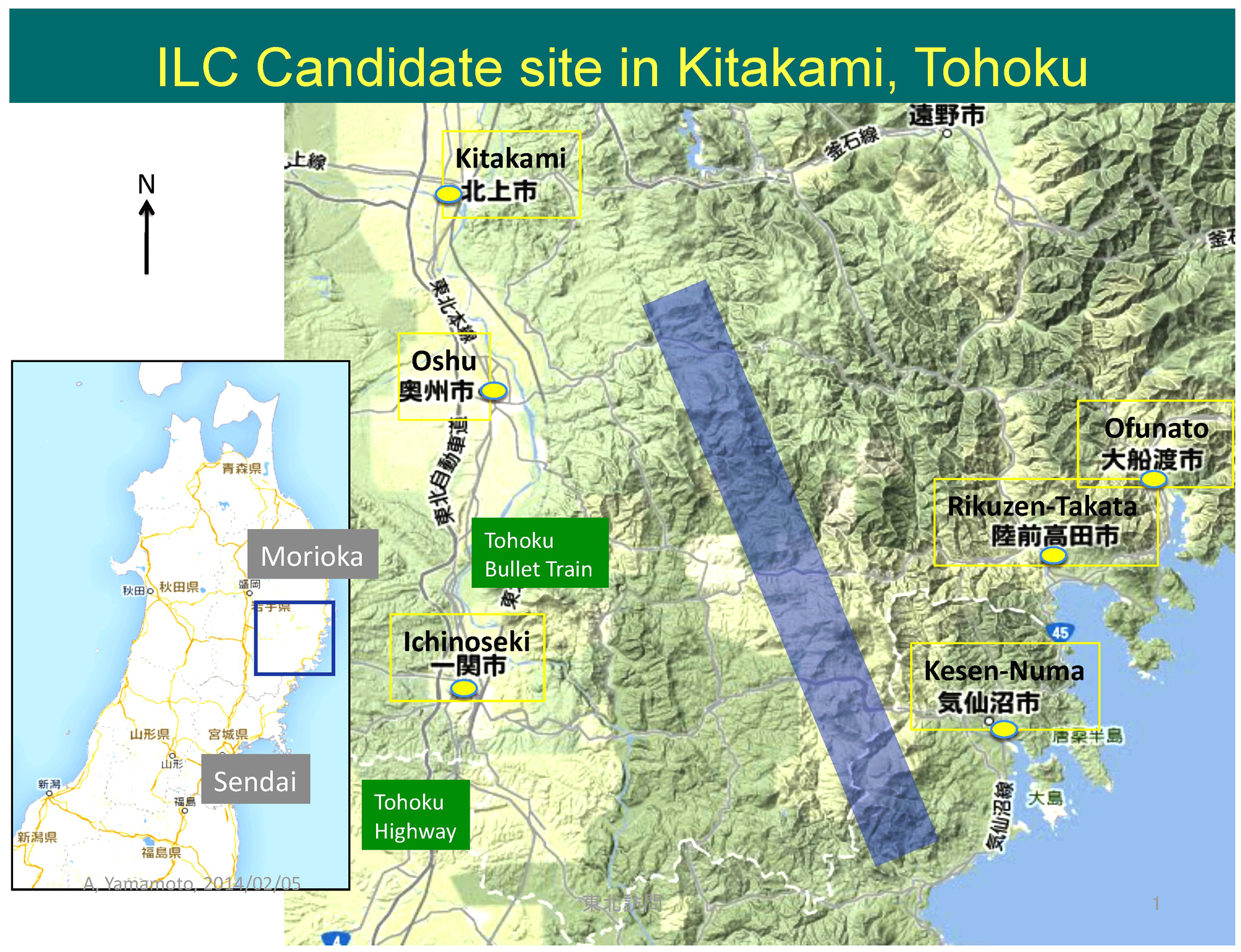}
\end{center}
\caption{The Kitakami candidate site for the ILC~\cite{Warmbein:2014a}.}
\label{fig:kitakami-site}
\end{figure}

The site (Fig.~\ref{fig:kitakami-site}) is located in the Japan's northern Tohoku region, not far from Sendai with its international airport, in the prefectures of Iwate and Miyagi.
The closest cities are Ichinoseki, Oshu, and Kitakami, which all offer Shinkansen (bullet train) access to Sendai and Tokyo.
The closest harbour is in the city of Kesen-Numa.
The coastal region in this area was severely hit by the great Tohoku earthquake in 2011. 
Both prefectures are supportive of the ILC project and view it as an important part of their strategy to recover from the earthquake disaster.

The Kitakami site was largely selected because of its excellent geological condition. 
The proposed ILC trajectory lies in two large, homogeneous granite formations, the Hitokabe granite in the north and Senmaya granite to the south.
The site provides up to \siunit{50}{km} of space, enough for a possible \siunit{1}{TeV} upgrade or more, depending on the achievable accelerating gradient.  
Extensive geological surveys have been conducted in the area, including boring, seismic measurements, and electrical measurements~\cite{Sanuki:2015a}, as shown in Fig.~\ref{fig:kitakami-geology}.
The surveys show that the rock is of good quality, with no active seismic faults in the area.

Earthquakes are frequent throughout Japan, and the accelerator and detectors need  proper supports that isolate them from vibrations during earthquakes and micro tremors~\cite{Sanuki:2018b}. 
Proven technologies exist to cope with all seismic events, including magnitude 9 earthquakes such as the great Tohoku earthquake. 

% XXXXX MENTION TUNNELLING TECHNOLOGY (NAT), Cross section, Tomo's talk on earthquake vibes XXXXXXXX

Vibration measurements taken during the construction of a road tunnel show that accelerator operation would be possible during the excavation of a tunnel for an energy upgrade~\cite{Sanuki:2018a}.

% XXXXXXX CONTINUE HERE XXXXXXXXXX

%===============================================================================

%\newpage

\subsection{Green ILC}

https://www.overleaf.com/project/5feb77d659085f27f0653fa4\label{sec:greenilc}

%{\it Author Takayuki Saeki. Please give comments to takayuki.saeki@kek.jp by email.}
% revised by Michael Peskin.  Please keep him in the loop.

The design of the ILC is based on Superconducting RF  technology, which is more efficient than the normal conducting technology in terms of the energy consumption. However, still the total energy consumption of ILC is 111 MW at 250-GeV initial phase, 163 MW at 500-GeV phase and 300 MW at 1-TeV phase as shown in Table~\ref{tab:ilc-params}.  These values are comparable to the energy consumption of the LHC but still large in absolute terms.  The world is moving to carbon-neutrality as a goal, and this should apply also to the major laboratories of particle physics. 
  This being so,  the reduction of energy consumption and the usage of sustainable energy, and thus the efficient and sustainable design of the ILC, are crucial issues that must be addressed, especially to cooperate with the local community in the regions of the ILC site in Japan. For this purpose, the Advanced Accelerator Association (AAA) in Japan, consisting of members from both industry and academia, organized the  "Green-ILC Working Group (WG)".   The Green-ILC WG  collaborates with the international team of the ILC. Its activities include studies on the efficient design of ILC components, accelerator sub-systems, the overall system design, and even an  ILC city hosting the laboratory campus. The ILC team has been continuously communicating with the local community of the ILC Kitakami site about its carbon-neutrality policy. In addition, Green ILC activities have been contributing to the ICFA panel on sustainable colliders and accelerators.  Because the available resources within ILC group are limited, we are cooperating with industry, the local communicty of ILC Kitakami site, and ICFA for mutually beneficial activities in this area.

\begin{figure}[t]
 \begin{center}
 \includegraphics[width=0.9\hsize]{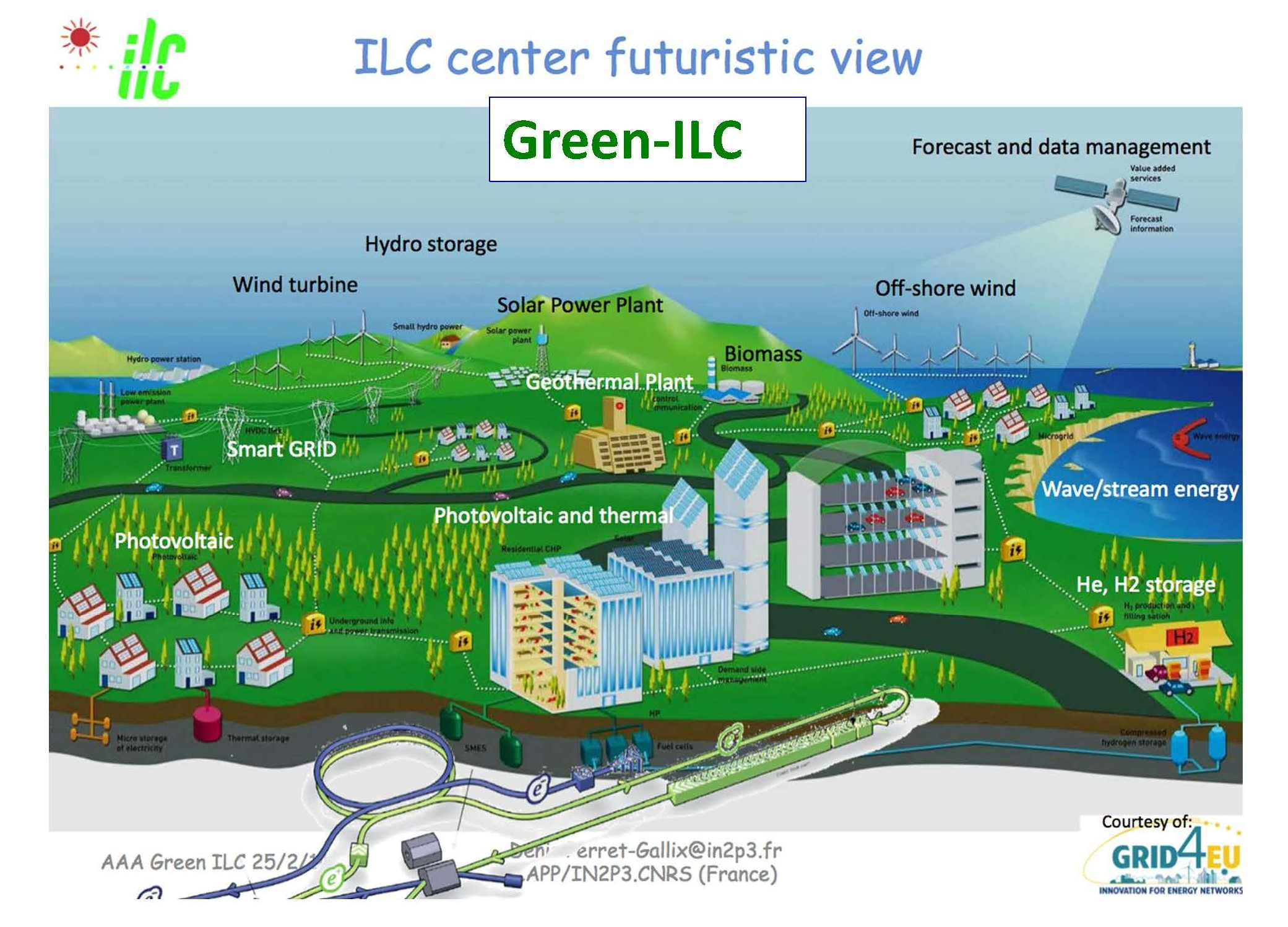}
 \caption{Schematic for the concept of Green ILC.}
 \label{fig:GILC-Concept}
\end{center}
\end{figure}

\subsubsection{History of Green ILC activities}

The Green ILC activity was triggered by three presentations from the ILC group at the  2nd Workshop on Energy Sustainable Science at Research Infrastructure (2nd ESSRI WS) in October 2013, in which the strategy of Green ILC, aiming at the sustainable and efficient design of the ILC, was presented~\cite{Yoshioka2013,DPG2013,Suzuki2013}. At that time, the vision of Green ILC was put forward, as reflected in the schematic illustration shown in Fig.~\ref{fig:GILC-Concept}. Soon after 2nd ESSRI WS (2013), a session including four presentations was organized for Green ILC in the LCWS2013 workshop in Tokyo in November 2013. In February 2014, the Green-ILC WG (WG) was organized under the  Advanced Accelerator Association Promoting Science and Technology (AAA), an association of 102 corporate organizations from industry and 41  institutional organizations from academia~\cite{AAAhomepage}. The proposals and discussions of the Green-ILC WG have been summarized in the Green-ILC WG  Report-2016~\cite{GreenILCpage}, which includes papers on  green accelerators in the world, green components for the ILC, energy recovery and storage for the ILC, and plans for a  Green ILC city. Most recently, the  ILC team has been discussing relevant issues  with the community of the ILC Kitakami-site through the Tohoku ILC Project Development Center~\cite{TohokuILCpage}, which was established in August 2020.   The Green ILC WG has also been contributing to the ICFA panel on sustainable colliders and accelerators since this was initiated in 2015~\cite{ICFA:Panel}.

\subsubsection{R\&D and proposals of components for Green ILC}

\begin{figure}[tb]
 \begin{center}
 \includegraphics[width=0.95\hsize]{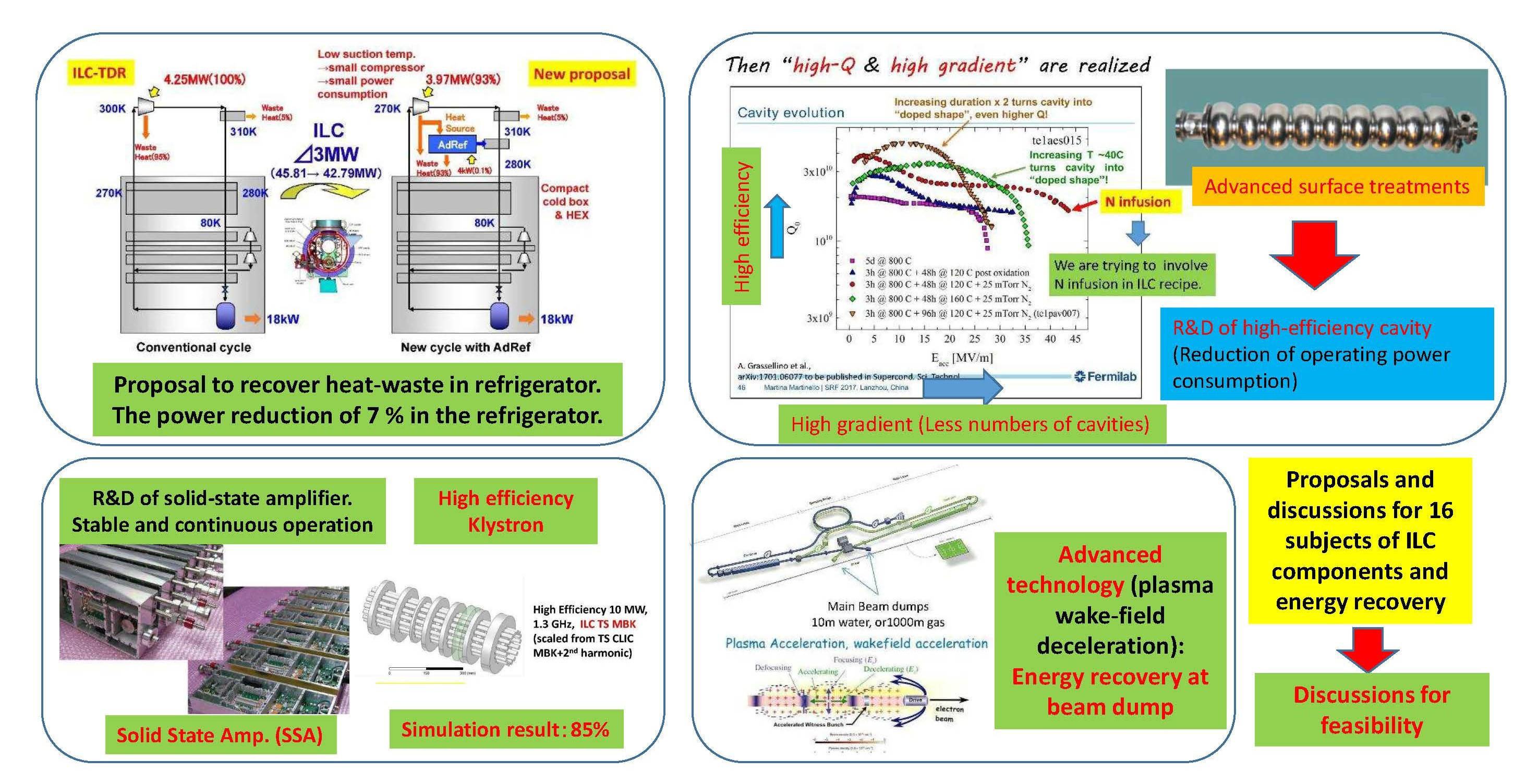}
 \caption{Illustrations of R\&D for high-efficiency components of the ILC. Upper left: a proposal for efficient refrigerator, Lower left: proposals for a high efficiency power source; Upper right: examples of high-Q and high-gradient cavities;  Lower right: a proposal for a beam dump with wake field deceleration.}
 \label{fig:GILC-Components}
\end{center}
\end{figure}

In the Green-ILC WG and in recent Linear Collider workshops, we have  discussed various subjects for the efficient and sustainable design of the ILC. Figure~\ref{fig:GILC-Components} illustrates some of the results that have been presented.

The upper left-hand box of  Fig.~\ref{fig:GILC-Components} shows an examples of an  efficient refrigerator proposed by a company at the Green-ILC meeting of the  AAA~\cite{GreenILCpage}. The waste heat from the refrigerator is recovered in the heat circuit and reused, and the power consumption is reduced by 7\% in total compared to the refrigerator design of described in the ILC TDR.

 The lower left-hand box of  Fig.~\ref{fig:GILC-Components} shows examples of R\&D on efficient power sources for the  ILC,  a Solid State Amplifier (SSA) and high-efficiency klystron~\cite{GILC:ILCX2021}.  The usage of SSA in ILC has historically been difficult because of its high cost, but recently the cost of SSA has been  decreasing rapidly. The power source design with SSA for ILC might become feasible within a few years. If the SSA is used in the ILC, a quick and hot swap of SSA modules is possible. This will lead to a high mean time between failures and a short mean time to repair in the ILC operation. Because the refrigerator of the ILC would operate continuously even in the stand-by mode during repair, this will contribute to the reduction of the power consumption of the  ILC. Recent development of a new klystron technology~\cite{Cai2020} and the availability of modern computer tools will allow us to boost the efficiency of the L-band klystron from around 65\% in existing ILC commercial tubes to almost 85\% in the new design. The fabrication of prototype klystrons to realize this new technology is under study now.

The upper right-hand box of  Fig.~\ref{fig:GILC-Components} shows plots of Q vs. $E_{acc}$ for SRF cavities after  applying various advanced surface treatments~\cite{Grass2017SUST,Martina2017}. Some advanced surface treatments are found to provide high Q and high $E_{acc}$ at the same time.  The LCLS-II project~\cite{bib:lcls-ii} has applied a nitrogen-doping surface treatment to realize high Q  at gradients somewhat lower than those of ILC. The technology of high Q and high gradient might be introduced into  the design of the ILC once the yield rate for these surface treatments in mass production has been studied systematically.  The ILC design using such high-gradient cavities will reduce the length of the linac, thus reducing both the construction cost and the energy consumption in the ILC operation.

\begin{figure}[tb]
 \begin{center}
 \includegraphics[width=0.95\hsize]{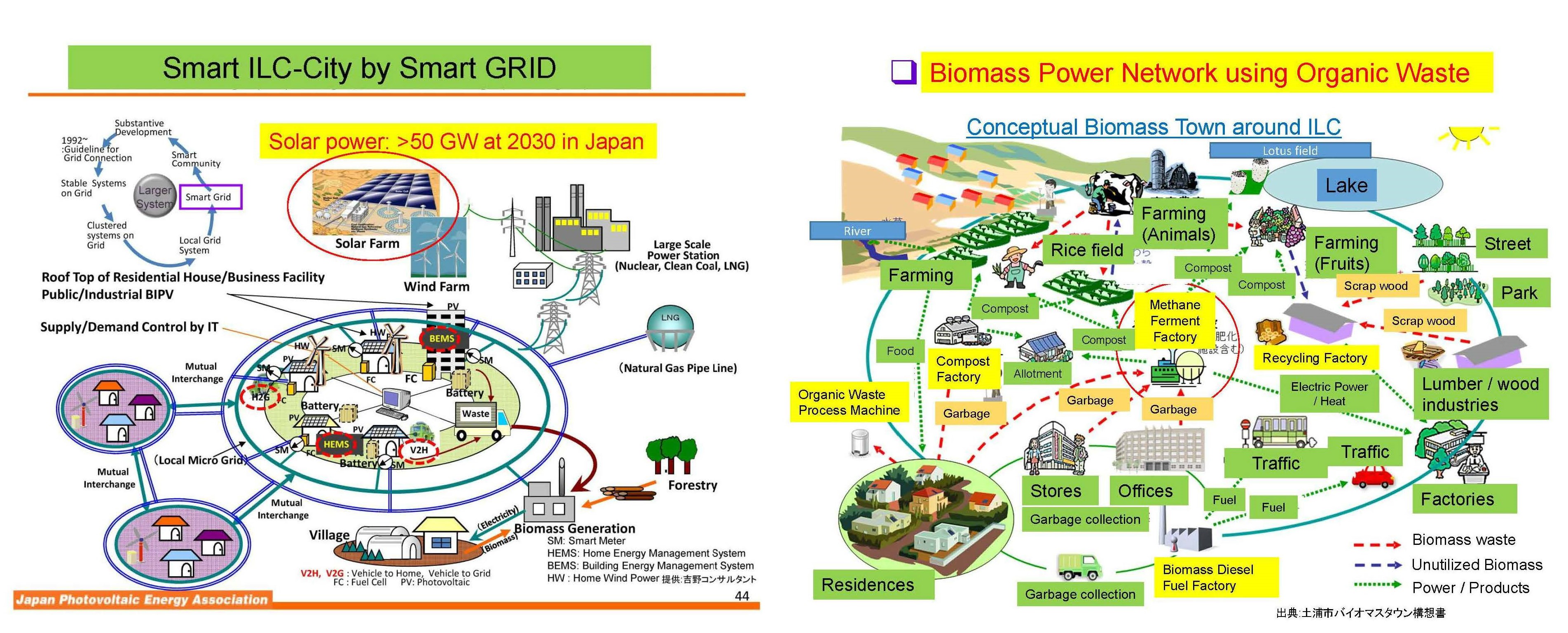}
 \caption{Proposals for a  smart power grid (left) and  biomass power network (right) for the  ILC city.}
 \label{fig:GILC-GridBiomass}
\end{center}
\end{figure}

\begin{figure}[tb]
 \begin{center}
 \includegraphics[width=0.7\hsize]{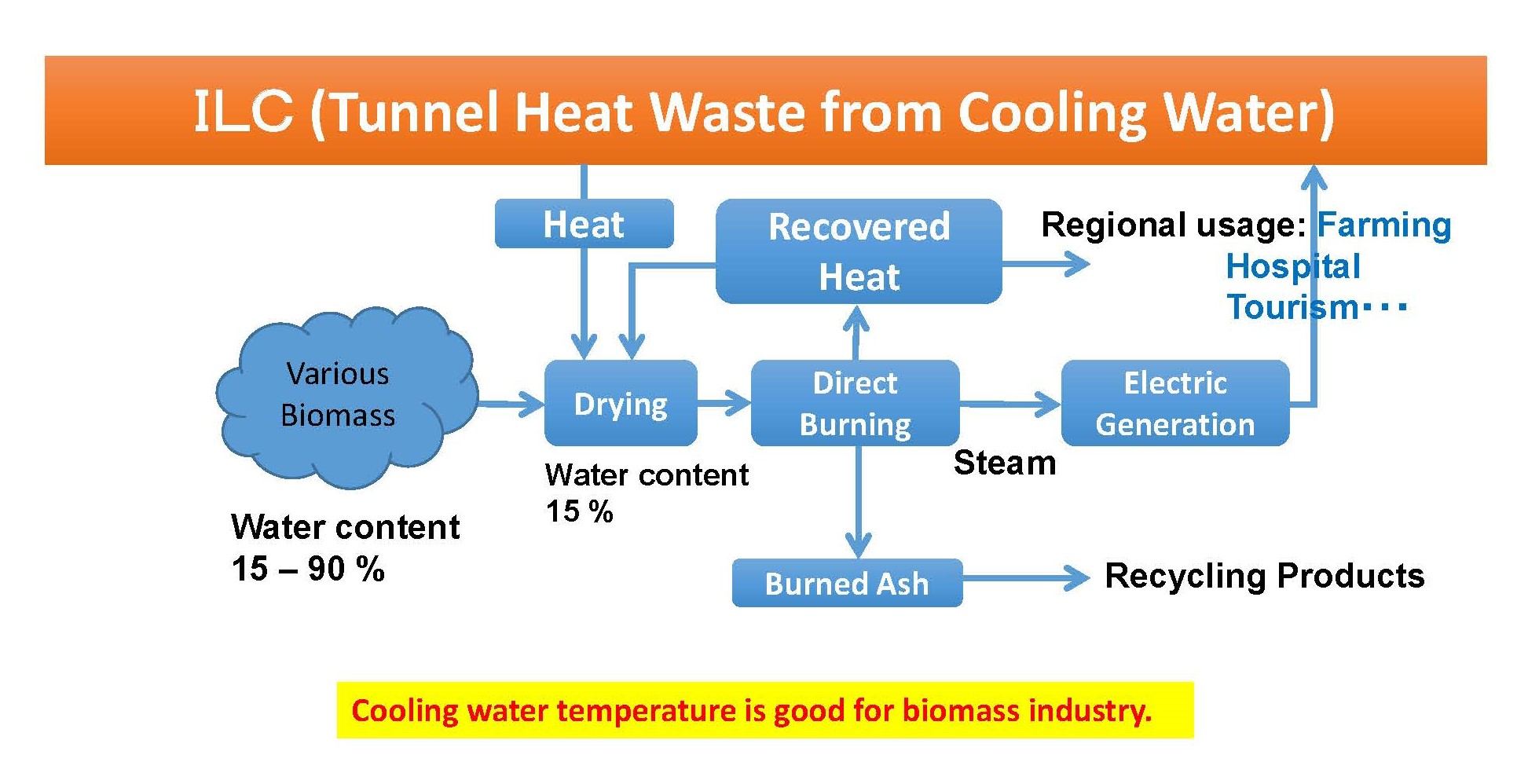}
 \caption{A proposal to use the waste heat of ILC tunnel for drying biomass.}
 \label{fig:GILC-Tunnel-Recovery}
\end{center}
\end{figure}

The lower right-hand box of 
Fig.~\ref{fig:GILC-Components}  shows a proposal for a new beam dump design using the technology of wake-field deceleration~\cite{Saeki:ESS2017,Saeki:IEEE2016}. The beam dump of the  ILC, including its cooling water system, is a  high radiation area and must be heavily shielded.  The concept of wake-field deceleration of the  beam can reduce the  radiation level of the beam dump dramatically,  even enough  that it is feasible to recover the heat energy of beam dump through cooling water. This technology is not confirmed yet, but it might resolve a common problem of the beam dump for all accelerators and contribute to their sustainability.

\subsubsection{Proposal for the operation of Green ILC}

A proposal for the  reduction and modulation of energy consumption by the  ILC by scheduling its operation modes was discussed in the Green ILC session of the ILCX2021 workshop~\cite{GILC:ILCX2021}. The energy consumption of ILC varies depending on the operation mode, which switches among modes such as full beam operation, reduced beam, standby, and stop (repair).   The various modes can be scheduled according to the available regenerative energy resources and demand for  electric power in the region of the  ILC site.  Variations of the demand of electric power with the  season and time of day can also be taken into account. Moreover, the power consumption of ILC can be modulated by the use of pre-chilled water and/or liquid helium. If we consider the optimum scheduling of ILC operations to minimize and modulate the power consumption, this will contribute to both the efficient and sustainable operation of the ILC and to improved use of electric power in the local area that surrounds it.

\subsubsection{R\&D and proposals for a Green ILC city}

The Green-ILC WG discussed  the design of an ILC city that includes the ILC Laboratory campus. If the ILC is realized in Japan, it is likely that the ILC Laboratory and a surrounding new ILC city will be built near the ILC machine. In that case, the city will be newly constructed, and so advanced concepts for an efficient and sustainable city might be introduced.  Fig. ~\ref{fig:GILC-GridBiomass} shows some concepts and proposals for a smart power grid  (left) and a biomass power network (right) for the  ILC city~\cite{GreenILCpage}.

In the smart power grid, solar power farms and biomass power stations are included.  The  ILC machine would be connected to the smart grid  and the operation modes of ILC would be organized  in concert with its daily schedule of energy production.
The biomass power network for the  ILC city would include  methane fermentation, biomass diesel fuel production, and scrap wood recycling factories. Biomass would be  collected through the network, and various kinds of energy would be  produced by biomass and  distributed to residents, offices, buildings, facilities, and factories. The electric power produced by the biomass power network would be provided to the smart grid network. The ILC machine would also contribute to the biomass power network by the use of waste heat from the  ILC tunnel, as shown in Fig.~\ref{fig:GILC-Tunnel-Recovery}. The recycling of waste heat from cooling water in the ILC tunnel has been  proposed by a company that uses such waste heat to  dry biomass.  Then energy can be produced as electricity and heat by burning the biomass~\cite{GreenILCpage}.

\subsubsection{Cooperation with ICFA  international panel on sustainable colliders and accelerators}

\begin{figure}[t]
 \begin{center}
 \includegraphics[width=0.6\hsize]{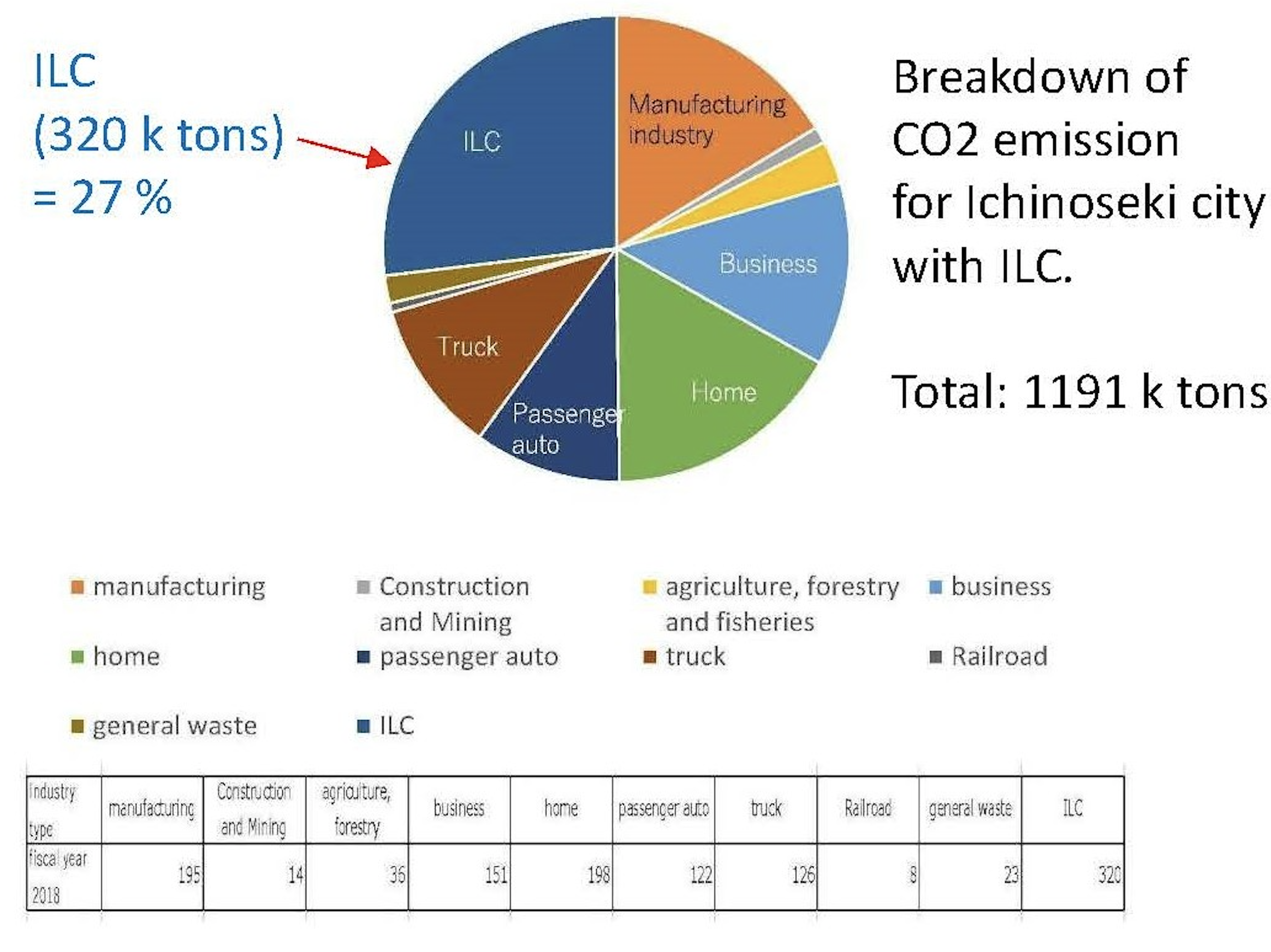}
 \caption{Breakdown of total CO$_2$ emission in Ichinoseki city, including the contribution from the ILC Laboratory.}
 \label{fig:GILC-CO2-Ichinoseki}
\end{center}
\end{figure}

The ICFA international panel on sustainable colliders and accelerators started in 2015~\cite{ICFA:Panel}. The panel started with 20 members, of whom 2 were from the  ILC group. The ICFA panel has been discussing and promoting  strategy and coordination for sustainable colliders and accelerators, energy efficient accelerator concepts, energy efficient and sustainable accelerator technologies, and energy management of large research facilities, with networking across the laboratories. It has also been  providing close and active communications among various projects and regions. Green ILC activities have been contributing to the ICFA panel from its beginning.  Because available resources in the ILC team are limited, we are cooperating with the ICFA panel in mutually beneficial studies of these issues.

\subsubsection{Studies and proposals for organization of the local area around the  ILC Kitakami-site in Japan}

Because the ILC will be a major user of energy and source of atmospheric CO$_2$, we have a responsibility to work with the local community to mitigate this and bring the plan for the ILC Laboratory as close to carbon-neutrality as possible.   As explained above, the ILC team is in close communication on this issue with local authorities through the  Tohoku ILC Project Development Center~\cite{TohokuILCpage}.  Currently, the ILC is expected to emit 320 kilotons of CO$_2$ per year, compared to 871~ kilotons of CO$_2$ emitted in 2018 by Ichinoseki City, the closest city to the  ILC Kitakami site; see Fig.~\ref{fig:GILC-CO2-Ichinoseki}.   Forests in this local area can absorb about 300 kilotons/year.  It  then is feasible for the ILC Laboratory, working with local authorities, to shape its planning to offset these losses~\cite{GILC:ILCX2021,Yoshioka2020PASJ}.  In particular:  (1) The ILC community should develop energy-saving technologies and  not only apply them  to the  ILC, but also  give them back to society. (2) The ILC community should cooperate with the community of the area to increase the percentage of renewable energy in the area. (3) The ILC Laboratory should integrate into its construction plan a program of sound management of the local forestry industry to increase the absorption of CO$_2$.

The ILC Laboratory has much expertise to bring to bear on the wise use of energy resources and sustainable energy policies, and we can also draw on insights from and collaborations with the broader scientific community.  We consider it a priority to work with members of the local community not only to make the ILC more green, but also to assist the transition to greener policies in the region that hosts the ILC Laboratory.

\subsection{ILC cost and schedule}

\label{sec:costandschedule}

%{\it 
%Description of Cost estimate and schedule - 1 page 
%
%include human resources, cost reduction effect by R\&D, operating costs
%}

For the Technical Design Report, the construction cost of the ILC accelerator was carefully evaluated from a detailed, bottom--up, WBS (Work Breakdown Structure)-based cost estimation~\cite[Sect. 15]{Adolphsen:2013kya}.
The TDR estimate distinguishes two cost categories: Value accounts for materials and supplies procured from industry and is given in ILCU (ILC Currency Unit, where $\siunit{1}{ILCU} = \siunit{1}{US\$}$ in 2012 prices), and Labor accounts for work performed in the participating institutions and is given in person--hours or person--years\footnote{One person--year corresponds to $1700$ working hours.}.

The Value of acquired goods reflects its worth in the local currency of the purchasing institution. 
Therefore, conversion of Value between currencies is performed based on Purchasing Power Parities (PPP), which are regularly evaluated and published by the OECD~\cite{OECD:2018,Eurostat:2012}, rather than currency exchange rates. 
The PPP values reflect local price levels and thus depend on the type of goods and the country, but fluctuate significantly less than currency exchange rates.
Therefore, conversions from ILCU to other currencies cannot not be made on the basis of exchange rates to the U.S. dollar, but on PPP values.

The TDR estimate covers the cost of the accelerator construction, assumed to last 9 years plus one year of commissioning. 
It includes the cost for the fabrication, procurement, testing, installation, and commissioning of the whole accelerator, its components, and the tunnels, buildings \etc, and the operation of a central laboratory at the site over the construction period. 
It does not, however, cover costs during the preparation phase preceding the start of construction work (``ground breaking''), such as design work, land acquisition, infrastructure (roads, electricity, water) for the site.

Based on the TDR cost estimate, an updated cost estimate was produced for the \siunit{250}{GeV} accelerator. 
This updated cost estimate includes the cumulative effect of the changes to the design since the TDR (see Sect.~\ref{sec:design_evo}), and evaluates the cost for the reduced machine by applying appropriate scaling factors to the individual cost contributions of the TDR cost estimate.

\begin{figure*}[tb]
 \begin{center}
 \includegraphics[width=0.36\hsize]{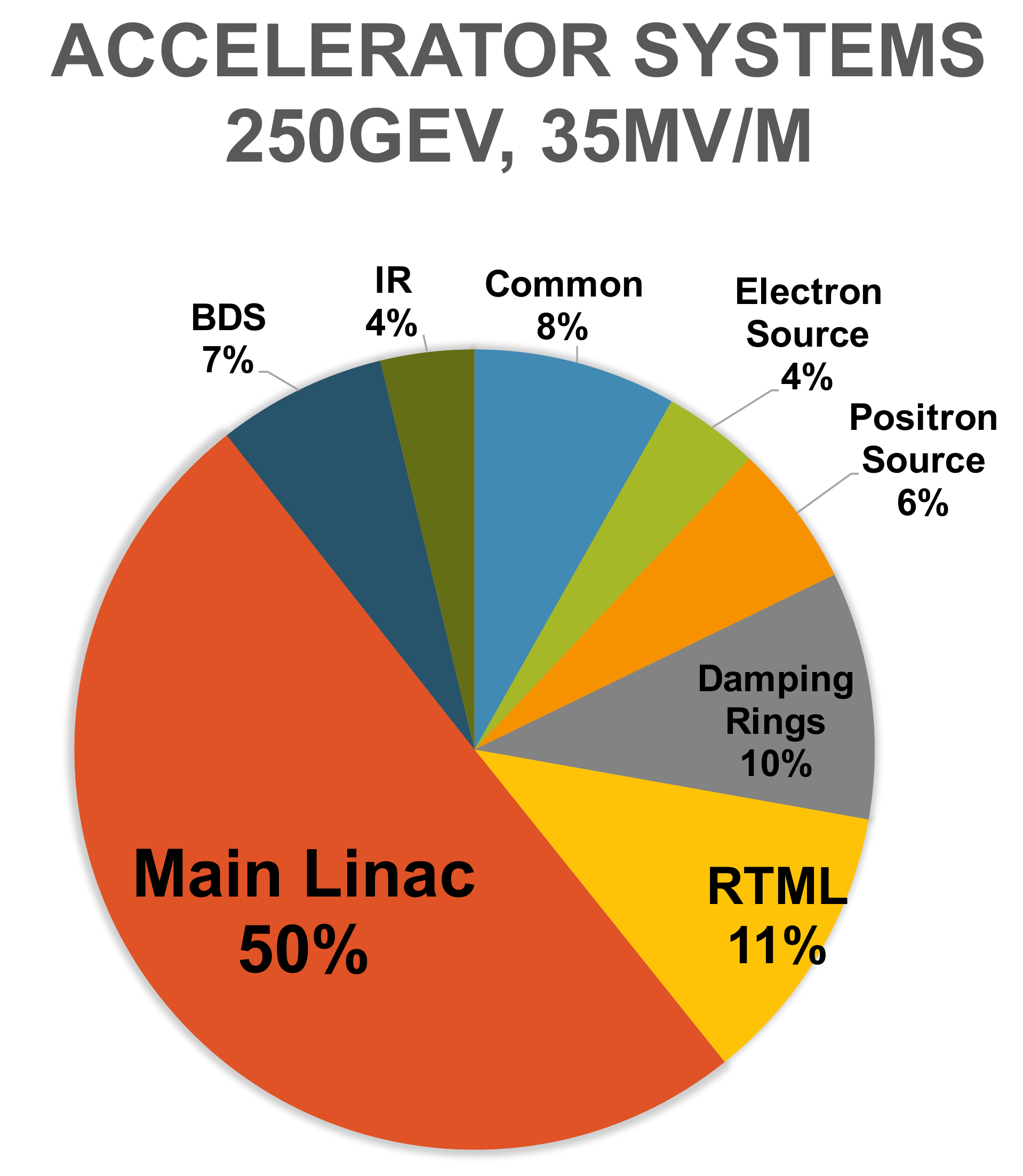}
 \includegraphics[width=0.4\hsize]{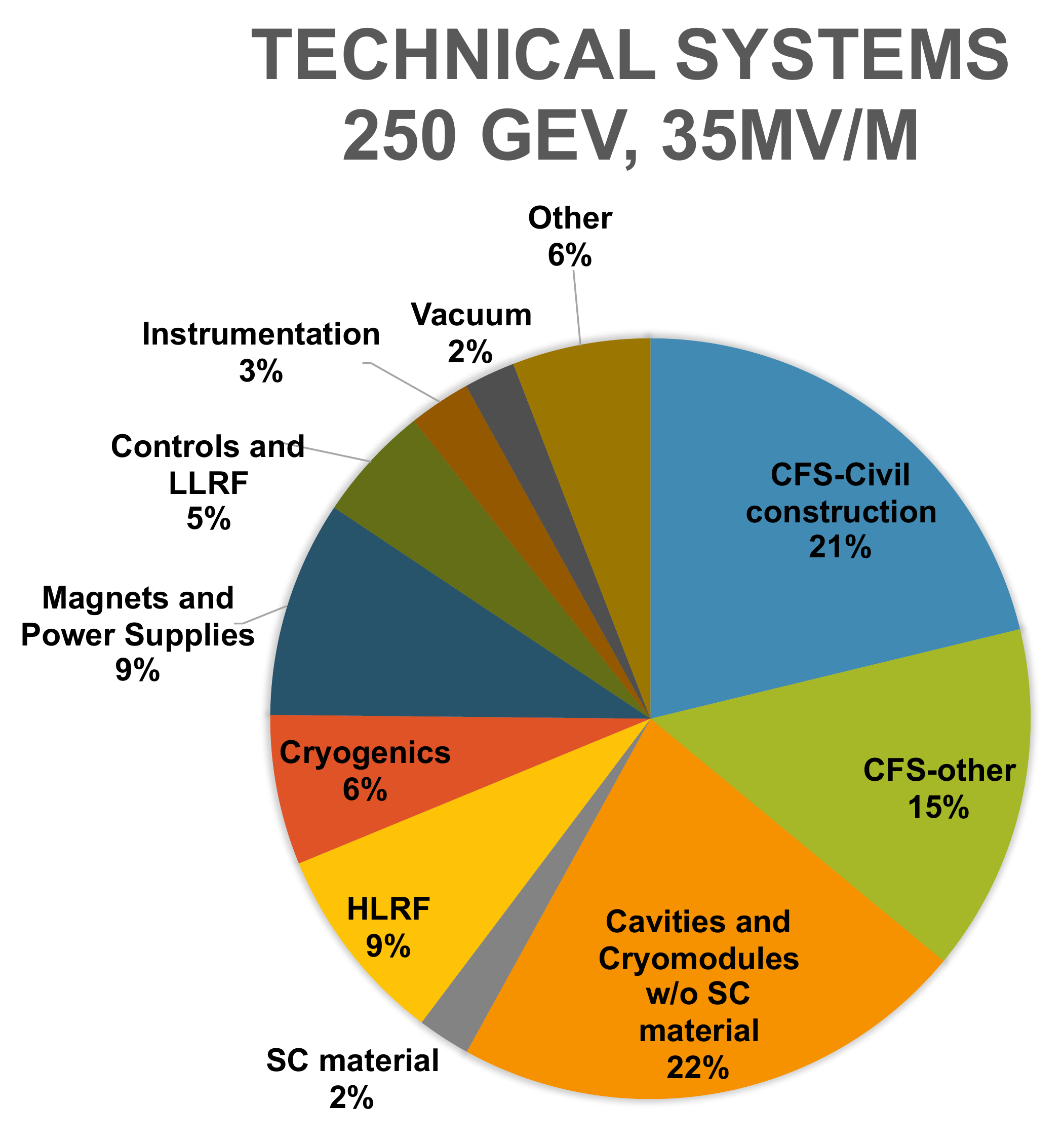}
\caption{Breakdown of Value costs into accelerator systems (left) and technical systems (right) for the \siunit{250}{GeV} ILC accelerator, assuming that cost reduction measures are successful and a gradient of \siunit{35}{MV/m} can be reached.
\label{fig:costs}}
 \end{center}
 \end{figure*}

The resulting Value estimate for the ILC accelerator at \siunit{250}{GeV} is 
\siunit{4,780-5,260}{MILCU}~\cite{Evans:2017rvt} in 2012 prices, where the lower number assumes a cavity gradient of \siunit{35}{MV/m}, while the higher number is based on the TDR number of \siunit{31.5}{MV/m}. 
In addition, \siunit{17,165}{kh} (thousand person-hours) are required of institutional Labor.

In 2018, the ILC Advisory Panel of the Japanese Ministry of Education, Culture, Sports, Science and Technoloy (MEXT) concluded its review of the ILC~\cite{ILCAP:2018}. 
For this review, costs were evaluated in Japanese Yen in 2017 prices, taking into account the local inflation for goods and construction costs.
For the purpose of this estimate, also the Labor costs were converted to Yen to yield \siunit{119.8}{G\yen}, resulting in a total range of the accelerator construction cost of \siunit{635.0 - 702.8}{G\yen}, where the range covers uncertainties in the civil construction costs (\siunit{18}{G\yen}) and of the gradient (\siunit{49.8}{G\yen}).
For the this estimate, conversion rates of $\siunit{1}{US\$} = \siunit{100}{JP\yen}$ and $\siunit{1}{\matheuro} = \siunit{1.15}{US\$}$ were assumed.

Operation costs of the accelerator and the central laboratory are estimated to be \siunit{36.6-39.2}{G\yen} (about \siunit{318-341}{M\matheuro}) per year.

\section{ILC staging up to 1 TeV} 
\label{sec:acc-staging}

%\begin{linenumbers}

\subsection{Introduction}

The requirements for ILC physical characteristics \cite{Bambade:2019fyw} define a continuous range of center-of-mass energy from \siunit{92}{GeV} ($Z$-pole \cite{Yokoya:2019rhx}) to \siunit{500}{GeV} with the possibility of additional upgrading to a center-of-mass energy of \siunit{1}{TeV}. The GDE has focused on providing a reliable design and cost estimation for the \siunit{200-500}{GeV} base machine. The design is a price-performance optimized solution for a given energy range. The center-of-mass energy of \siunit{250}{GeV} can be realized with a straight machine \siunit{20}{km} long, and the energy of \siunit{500}{GeV} can be achieved if it is expanded to \siunit{30}{km}. To be as cost-effective as possible, the final ILC proposal \cite{Asai:2017pwp} approved by ICFA does not include empty tunnel options for future upgrades. Despite the fact that the length of the main tunnel of the linear accelerator has been reduced, the beam delivery system and main dumps are designed to allow for an energy upgrade up to \siunit{1}{TeV}.

The development of accelerator structures with higher acceleration gradients can lead to a significant increase in energy while maintaining a compact infrastructure. To date, significant progress has been made in the development of structures with a gradient well above the $31.5$~MV/m required for the ILC, and even above the $45$~MV/m as required for the \siunit{1}{TeV} ILC \cite{Grassellino:2018tqg,Grassellino:2017bod}. In the longer term, structures with an alternative shape or with a thin-film Nb$_{3}$Sn coating or multilayer coating can significantly improve the performance of linear particle accelerators \cite{gurevich2006,kubo2021superheating}. Newer acceleration schemes can achieve even higher gradients as discussed in Sect. 4.3. Finally, the emergence of acceleration schemes based on plasma wake field acceleration or other advanced concept could open up the ILC energy regime up to \siunit{30}{TeV}. Thus, the ILC laboratory has the potential to support a higher energy electron-positron collider. The ability to increase energy levels makes the Linear Collider a very flexible tool, allowing in response to a new discoveries at the LHC. There are several options for upgrading the ILC in terms of energy, luminosity and beam polarization.

The level of detail of the staging and upgrade scenarios is significantly less mature than the baseline. In particular, the TeV upgrade parameters and associated conceptual design represent a relatively simple and straightforward scaling of the base machine based on assumptions about higher achievable operating parameters for SCRF technology with an average acceleration gradient of $45$~mV/m with $Q_{0} = 2 \times 10^{10}$. Achieving these values requires further research and development beyond the basic technology. It is anticipated that this R\&D will continue in parallel with both construction and operation of the base machine, so that the expansion of the core linear accelerators required to increase particle energy will benefit from improved technology. In addition, accelerator research and development should continue to dramatically increase particle collision energy in preparation for future experimental efforts that may indicate the existence of new particles and new phenomena at higher energy.

Both luminosity enhancement and low-energy staging are based on existing technology and do not require additional research and development. For upgrades to TeV energies, a design approach that has minimal impact on the operation of the ILC is desired. The two sets of parameters presented for upgrading to TeV (the so-called low and high beamstrahlung) were obtained after careful consideration of the physical impact.

It should be emphasized that the flexibility in the choice of beam parameters remains one of the key advantages of the ILC. It can be adjusted whenever new ideas and discoveries either from (HL-) LHC or from the ILC itself set new requirements. 

\subsection{Parameters}

Table~\ref{tab:ilc-params} shows the main ILC parameter sets for center-of-mass energy of \siunit{250}{GeV}  and of \siunit{500}{GeV}, the luminosity upgrade of these, and a possible set of parameters for the energy \siunit{1}{TeV}. The parameters for the first stage of the \siunit{250}{GeV} machine are identical to the baseline parameters set for this energy.

\subsection{Luminosity upgrade}
\label{sec:lumi-upgrade}

The ILC luminosity can be improved by increasing the luminosity per bunch (or by the charge of the bunch) or by increasing the number of bunches per second \cite{Harrison:2013nva}. Increasing the brightness per bunch requires a smaller vertical beam size, which can be achieved by tighter focusing and / or lower beam emittance. However, this approach invariably involves high perturbation of the beam, resulting in the risk of luminosity loss due to improper beam steering. Thus, a very accurate feedback system is required. The ILC design also allows the number of bunches to collide per second to be increased by doubling the number of bunches per pulse and possibly increasing the pulse repetition rate. Doubling the number of bunches per pulse from the base number of $1312$ to $2625$ will require a decrease in the time separation between bunches from $554$~ns to $366$~ns, which will lead to an increase in the beam current from $5.8$~mA to $8.8$~mA, which will require installation of $50\%$ more klystrons and modulators. Since the RF pulse duration of $1.65$~ms will not change, the cryogenic load will also not change. The beam pulse duration increases from $714\:\mu$s to $961\:\mu$s. The choice of the distance between bunches is consistent with both the harmonic number of the damping ring and the duration of the RF pulse of the main linear accelerator. Doubling the number of bunches would double the beam current in the damping rings. For a positron ring, this may exceed the limitations associated with the electron cloud instability. To reduce this risk, the damping ring tunnel is large enough to accommodate a third damping ring so that the positron current can be distributed over the two rings. Basic schematics for electron and positron sources are specified to produce more bunches needed for upgrades. RTML, and in particular the SCRF RF linear accelerator sections for beam compressors - are already compatible with a large number of bunches. 

The pulse repetition rate is \siunit{5}{Hz} in the base configuration. However, when ILC for the center-of-mass energy \siunit{500}{GeV} is built and it is operated at \siunit{250}{GeV}, the repetition rate can be increased up to \siunit{10}{Hz} so as to further double the luminosity. The RF and the wiggler magnet systems of the damping rings can be reinforced so that the beams can be damped within \siunit{100}{ms}. The klystrons in the bunch compressors and the main linacs can be operated at \siunit{10}{Hz}. The wall plug power system and the cryogenics of the main linacs can accept this mode because the accelerating gradient is low. The positron source must be improved for higher target heat load. Thus, the luminosity at \siunit{250}{GeV} can be increased to $5.4 \times 10^{34} \,\mbox{cm}^{-2}\mbox{s}^{-1}$ with 2625 bunches per pulse and \siunit{10}{Hz}.

The invasive nature of the additional cryogenic power installation requires a shutdown, during which all additional RF power supply must be installed. This will also include additional water cooling and the required wall plug power, although pipe sizes are already specified for the additional baseline load and do not need upgrades. In particular, the $25\%$ increase in cryogenic load (mainly due to high power coupler losses and HOM losses due to higher current) is within the base specification. All beam position monitors (and other instruments) are compatible with shorter beam spacing. Beam dynamics problems (multi-bunch effects) are also acceptable, and high power couplers and HOM couplers/absorbers are specified in the baseline for higher beam currents.

\subsection{Energy upgrade}

An obvious advantage of a linear collider is the possibility of an energy upgrade. In principle, the main linear accelerator can be expanded with the cost that is proportional to added length (i.e. added beam energy) with some additional costs of moving the turnarounds and compressors. Additional costs arise if the beam delivery system (BDS), including the beam dumps, has to be expanded to cope with the increased beam energy. The current ILC BDS is designed to be easily modified to operate at center-of-mass energies up to \siunit{1}{TeV} at minimal cost. Depending on the actual gradient achieved during the construction of the ILC, maximum $162$ cryomodules can be installed in previously unoccupied space in the tunnel reserved for the timing constraint in addition to those required to reach $250$~GeV, which will increase the center of mass energy by approximately $50$~GeV to about $300$~GeV, and two additional cryogenic plants may need to be installed. Further increases in energy will require the expansion of the tunnel. As noted above, an accelerator with a total length of at least 50 km can be placed on the Kitakami site, which is more than enough for center-of-mass energy of \siunit{1}{TeV}. Any expansion of the accelerator system can be accomplished by adding new cryomodules at the low energy (upstream) ends of the accelerator without the need to move already installed modules.

The upgrade can take place in two phases: a preparation phase, when the accelerator is still running and producing data, and an installation phase, when the accelerator stops. During the preparation phase, the necessary components will be purchased and manufactured, in particular cryomodules, klystrons and modulators. At the same time, civil engineering will continue to excavate new access tunnels, underground halls and the main tunnel. Recent research shows that the level of vibration caused by tunneling will allow construction of tunnel extensions close to the existing ones without impacting machine operation \cite{bib:sanuki:lcws2018}, minimizing the required shutdown time. During the installation phase, the newly built tunnels will be connected to the existing ones, the beam lines at the turnarounds and wiggler sections of the bunch compressors will be dismantled, and new cryomodules and a new turnaround and bunch compressors will be installed. In parallel, any necessary changes can be made to the positron source and the final focus of the machine. Since the cryomodules would be ready for installation at the beginning of the shutdown period, it is anticipated that the shutdown could be limited to about a year for an energy upgrade

The choice of beam parameters and luminosity increase for the $1$~TeV upgrade is also based on direct scaling from a set of base parameters, but more limited by additional considerations related to higher energy and average beam power:

\begin{enumerate}
\item  The total wall plug power required for the modified machine must be below some realistic limit (assumed to be $300$~MW);
\item  The beam current and pulse duration must be compatible with injectors, damping rings and the main linear accelerator of the basic design;
\item  Energy losses due to beamstrahlung should be acceptable, and the maximum pair-production angle should be limited at the maximum luminosity per bunch crossing.
\end{enumerate}

Limiting the total wall plug power requires reducing the repetition rate from $5$~Hz to $4$~Hz, while the need to maintain the RF pulse length in the original main linear accelerator at approximately $1.6$~ms and the choice of the damping ring harmonic number limits the number of bunches to $2450$. The limits of beamstrahlung depend on physics, therefore, for the study of physical and detector groups, a set of parameters with high beamstrahlung radiation with $\delta_{BS}\sim 10\%$ and, accordingly, a higher luminosity $5.11 \times 10^{34}$~cm$^{-2}$s$^{-1}$ was proposed. 
The parameter set is based on the reduced charge of one bunch 
($1.7 \times 10^{10}$), shorter bunch length ($250\mu$m and $225\mu$m for low and high $\delta_{BS}$, respectively), and increased horizontal beam size for controlling beamstrahlung and pair-production angle, while the vertical beta function at the interaction point (IP) is further reduced to increase the luminosity per bunch crossing \cite{yokoya2001}. The bunch lengths and IP beta functions are within the range of bunch compressor and final focusing systems.  

Increasing the beam energy will require the expansion of the main SCRF linear accelerators to provide an additional $250$~GeV per beam. The beam current for the $1$~TeV upgrade ($7.6$~mA) is higher than the baseline ($5.8$~mA) but less than that for luminosity upgrade ($8.8$~mA) for $500$~GeV design, suggesting some level of modification. Assuming the luminosity upgrade is the first to occur; the injectors (sources and damping rings) will be reused unchanged. Compressor sections along with the RTML will be moved to the beginning of the extended linear accelerators. It is also necessary to lengthen the $5$~GeV long-transfer line from the damping ring to the turn-around. The beam delivery system will require the installation of additional dipoles to provide the required higher integrated field strength. The cost and schedule of the upgrade is entirely dependent on the expansion of the main linear accelerators. One of the key cost considerations is the choice of an accelerating gradient. Ongoing R\&D for high gradient SCRF is expected to continue in parallel with the construction and operation of the base machine. With this in mind, it is assumed that when the linear accelerator technology is upgraded, a higher gradient and quality factor are incorporated. The actual choice of these options will clearly depend on the state-of-the-art at the time of the upgrade. However, for the purposes of this discussion, an average acceleration gradient of $45$~MV/m with $Q_{0} = 2 \times 10^{10}$ will be assumed. Using the existing baseline linear accelerator with bunch compressors described in~\cite[Sec. 7.3.3.5]{Adolphsen:2013kya}, main linac weak quadrupole magnets for the $15$-$25$~GeV energy range, and normal FODO lattice for $25$-$125$~GeV beam energy, there are four key consequences for the upgrade:

\begin{enumerate}
\item  The beam current and pulse length must be compatible with the existing RF installation and cryogenic refrigeration capacity.
\item The extension part of the main linac tunnel and the turn around must be ready for bunch compressors and a weak quadrupole section relocation.
\item The new, higher accelerating gradient SRF cavities should be installed.
\item  The rest of the original linear accelerator will use the FoFoDoDo lattice as opposed to the basic FoDo lattice, which will result in weaker focusing and larger beta function values. Simulation of the beam dynamics showed that the growth of the vertical emittance can be kept within acceptable limits.
\end{enumerate}

\subsection{Positron source}

The undulator-based positron source must be compatible with the initial energy of the electron beam of $500$~GeV. The solution is to replace the baseline helical undulator with a shorter one, with a longer period and a smaller field. The upgraded undulator will provide a photon beam similar to the baseline so that the same target and capture device can be used without modification \cite{evans2017}. One of the important considerations is the opening angle of photons, which is halved for higher beam energy; this makes collimating photons for polarization more challenging. Currently, a conservative estimate of $20\%$ polarization is considered acceptable, but higher values may be possible, provided that a suitable solution is found for collimating photons with a smaller aperture \cite{ushakov2013}.The baseline design geometry of the target-bypass dogleg for the high-energy electron beam already accommodates the $500$~GeV beam transport with a few percent horizontal emittance growth \cite{jones2010}, although additional dipole magnets will need to be installed. The electron-driven positron source is compatible with Energy upgrade as it is.

\subsection{RTML}

The two-stage compressor system will need to be ``relocated'' to a new location upstream. This scenario assumes that a new two-stage compressor will be installed, as well as a new turnaround and an extended transport line. Also, during the shutdown for the final installation of the warm wiggler base sections and cryomodules, the most upstream sections of the main linear accelerator will be updated as discussed in the ``Energy upgrade'' subsection. The original turnaround will be disconnected and bypassed by a new long transport line. It is likely that the space between the original and the upgraded linac will also be used for additional diagnostic and dump systems, including an emergency extraction dump to protect the machine, similar to the one found at the linac exit (BDS entrance).

\subsection{Beam Delivery System (BDS)}

The BDS geometry (length and average bend radius) is already compatible with the transport of a $500$~GeV beam with an acceptable increase in the emittance generated by synchrotron radiation \cite{evans2017}. Additional dipoles are required (as well as appropriate power supplies and cooling) to be installed in drift spaces provided in the base grid. The main high power dumps have already been designed for higher average beam powers to avoid the need to replace them during modernization (dumps will become radioactive after several years of operation).

\subsection{Polarization upgrade}

It is assumed that at center-of-mass energies up to $500$~GeV, ILC beams will have at least $80\%$ of the electron polarization at IP in combination with a positron polarization of $30\%$ for an undulator positron source. At $1$~TeV, the positron polarization will reach at least $20\%$.

 At beam energies above $125$~GeV, the flux of undulator photons increases rapidly. Photon polarization is maximal at zero angle of radiation emission; it reduces and even inverts at large angles. Thus, collimation of the excess photon flux at large radiation angles increases the net polarization. Thus, as an upgrade option, $60\%$ polarization of positrons at IP can be possible at a center-of-mass energy of $500$~GeV with the addition of a photon collimator. 
 
 The design of the accelerator includes sets of spin rotators, which allow to choose any desired direction of the polarization vectors at the IP. The baseline running scenario considers data taking with longitudinal polarization configurations as default, but data sets with transverse polarization could be added as upgrades.

\subsection{Summary}

These chapters examined incremental upgrade and upgrade options other than the $500$~GeV baseline scheme and demonstrates the greater design flexibility and capabilities of the ILC installation. The basic design already contains the possibility to simply increase luminosity by doubling the average beam power ($50\%$ increase in average RF power). The parameters and scope of future upgrades to center-of-mass energy of $1$~TeV were presented, based on the expansion of the main linear accelerators with minimal impact on the existing (baseline) machine. The construction of the extended machine, in principle, could proceed in parallel with the physical launch, with minimal interruption for connecting the baseline and modernized linear accelerators and the subsequent commissioning of the machine. The physical parameters (luminosity) for retrofitting to TeV energies represent a compromise between the physical requirements of the beam-beam (limiting bremsstrahlung and pair-production angle) and the desire to limit the total required wall plug power to about 
$300$~MW.

%\end{linenumbers}

\section{R\&D program on superconducting RF}
\label{sec:acc-beyond}

The technology for the ILC at 250 GeV is "shovel-ready". The TDR was completed some time ago.  The technology has been demonstrated and industrialized.  A large-scale prototype (the European XFEL) has been installed, and in operation.  New large scale facilities---LCLS-II, ESS, PIP-II, and SHINE---are soon to be commissioned or under construction.  Extensive SRF infrastructure exists worldwide for cavity fabrication, surface treatment, clean assembly, cold testing, and cryomodule assembly.  Major SRF facilities are available at DESY, CERN, INFN, Saclay, Orsay, INFN, KEK, JLAB, Cornell, Fermilab, MSU, and at several industries around the world.  New infrastructure is becoming available for upcoming projects such as for ESS in Europe, and PAPS in China.  New industries in S. Korea, China and Japan are rapidly growing familiar with SRF technology.

The decade 2010--2020 has brought enormous progress to the physics, technology and applications of SRF cavities as will be presented in the following Sections. Here we summarize some of the highlights.  Unprecedented high Q values have been attained up to $E_{acc}= 20$--30~MV/m. These advances were achieved by novel surface preparation techniques, such as nitrogen doping, cold-electropolishing, and 300~C baking, along with special cavity cool-down procedures to eliminate the residual resistance contribution from trapped DC magnetic flux. These recent accomplishments have translated into significant increases (by a factor of $>$~2–-3) in the efficiency of CW particle accelerators (e.g., LCLS-II at SLAC) operated at medium accelerating fields of about 20~MV/m.  

On the high gradient frontier, most relevant for ILC, new treatments of Nitrogen-Infusion and  two-step baking (75/120~C) have paved the way for gradients near 50~MV/m.  Applying these advanced treatments to improved shape cavities developed earlier holds the prospect of gradients close to 60~MV/m. A radical step of replacing the standing wave TESLA type structure with a Travelling Wave structure will open the door to gradients of 70~MV/m by lowering the peak surface fields.  But much development work will be needed to reach these exceptional levels. 

In Section~\ref{sec:SRFtech}, we have described the evolution of superconducting RF technology up 
to the
present and explained the robustness of the ILC plan for operation at a  nominal gradient of 31.5~MeV/m.
However, superconducting RF technology 
continues to move forward.  To reduce the cost of the 250~GeV ILC, to reduce the cost of the upgrade to 500~GeV, and to propose affordable designs for the ILC at 1~TeV and beyond, it is 
important to continue to improve this technology to achieve gradients as high as possible in cavities 
that can be produced reliably by industry.   In this section, we will describe the R\&D program to improve the gradient of  superconducting RF cavities.   The improvements that we describe here go beyond
the baseline ILC design, but we expect that they will be brought into play as the ILC evolves to higher 
energy.   The far-future application of extremely high-gradient superconducting RF to take the ILC beyond 1~TeV will be  described in Sec.~\ref{sec:futureSRF}.

\subsection{Gradient status for the ILC baseline 250 GeV}

\begin{figure}
\begin{center}
\includegraphics[width=0.80\hsize]{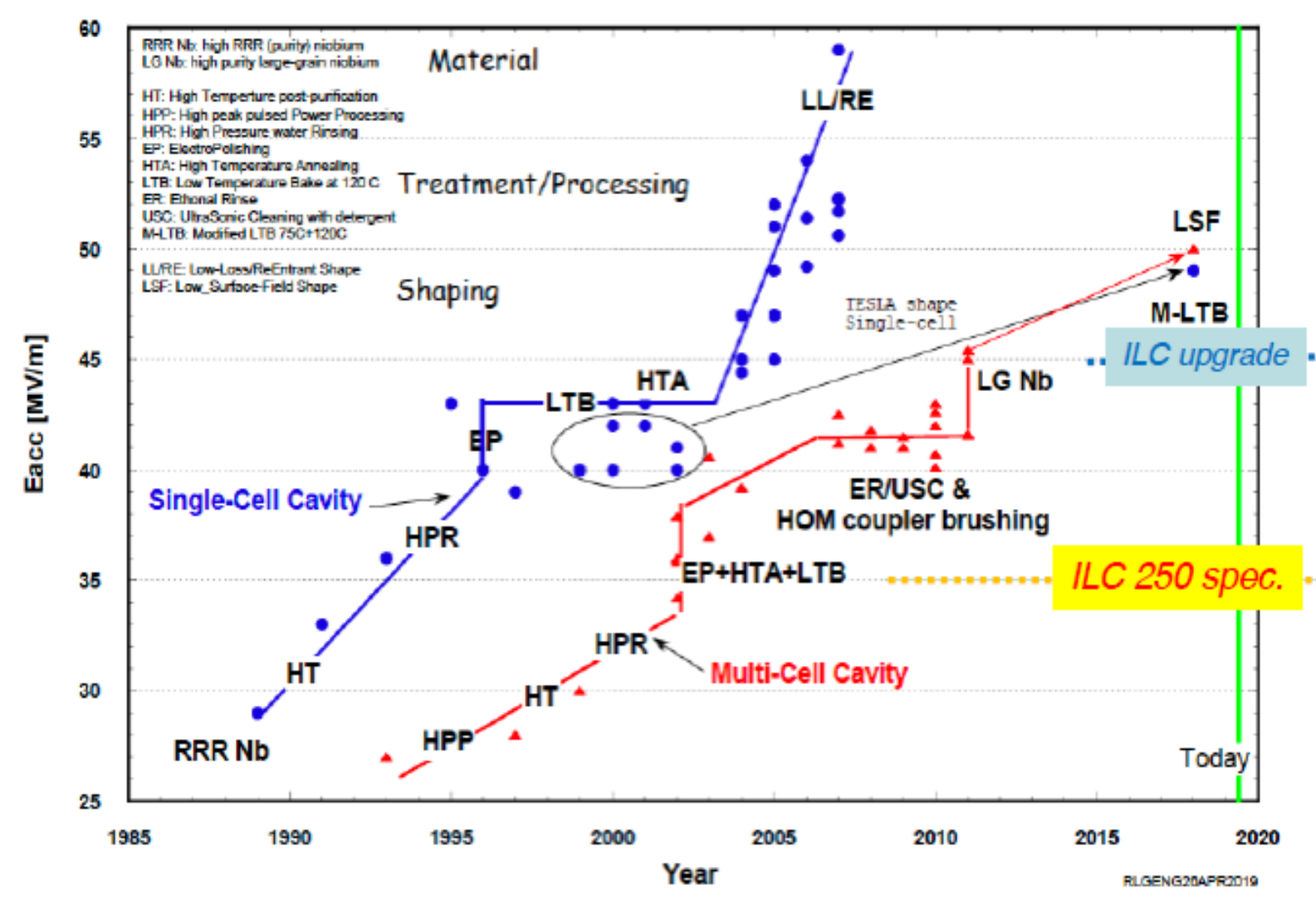}
\end{center}
\caption{Steady progress in single and multi-cell cavity gradients over 3+ decades \cite{Yamamoto:2019}.}
\label{fig:SRFprogress}
\end{figure}

\begin{figure}
\begin{center}
\includegraphics[width=0.60\hsize]{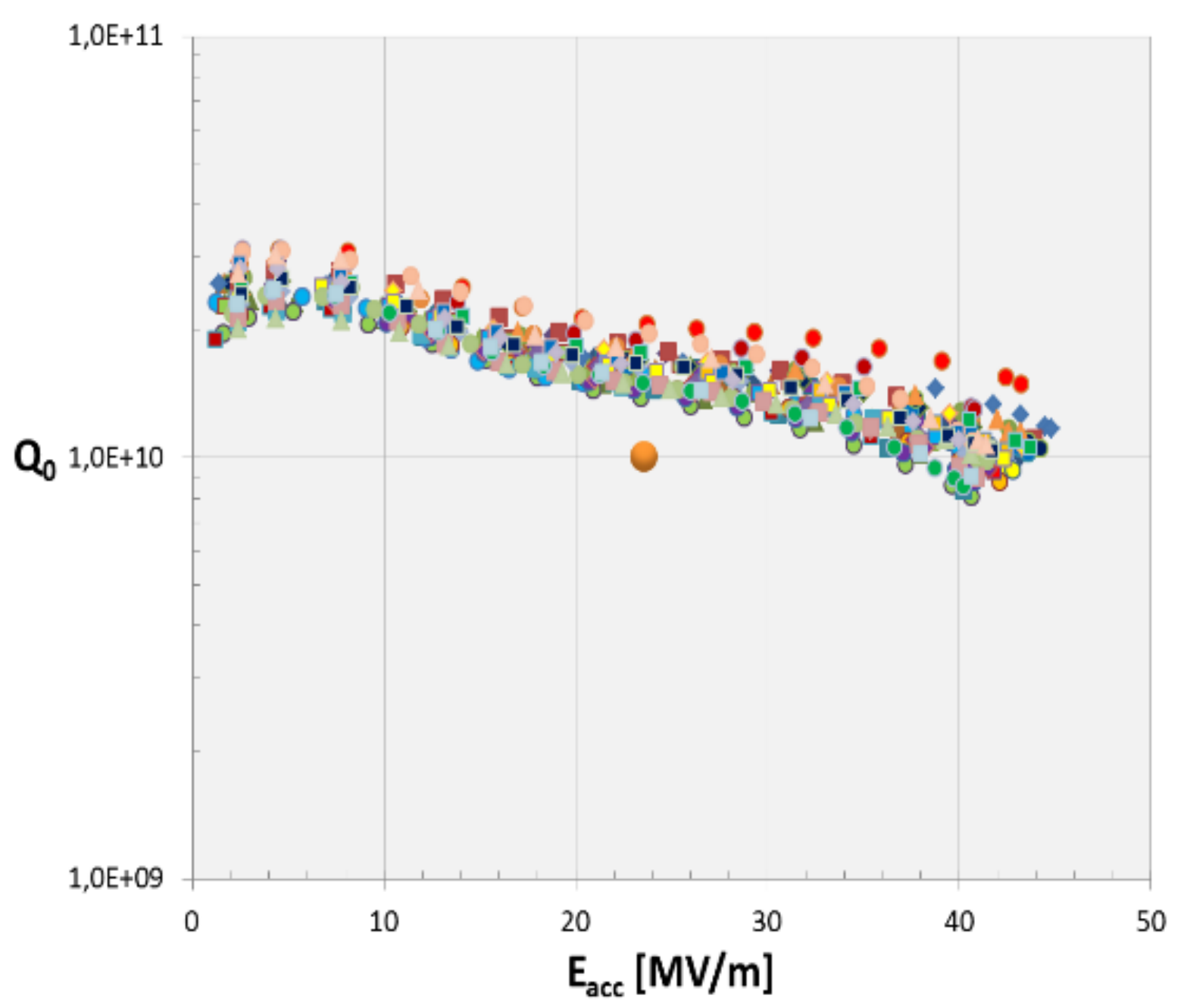}
\end{center}
\caption{9-cell test results from DESY on $> 40$ cavities produced and treated by Research Instruments (RI) \cite{Pekeler:2016}.}
\label{fig:DESYresults}
\end{figure}

Figure~\ref{fig:SRFprogress}  shows the steady progress in single and multicell cavity gradients
~\cite{Yamamoto:2019} over the last 3+ decades coming from high purity, high Residual Resistivity Ration (RRR) Nb, electropolishing, 800~C furnace treatment for H removal, 100`atm. high pressure water rinsing (HPR) for removal of field emission particulates, and final baking at 120`C for removal of the high field Q-slope.  These procedures establish a standard ILC cavity preparation and treatment recipe from which cavity gradients of 35 MV/m are expected, as observed from the European XFEL production run.    More than 40 ``best'' cavities from the European XFEL production run showed 40--45 MV/m \cite{Pekeler:2016}, as shown in Fig.~\ref{fig:DESYresults}.  At DESY, two large grain 9-cell cavities reached 45 MV/m \cite{Singer:2013ha}.

%Figure 1
%%%%%%%%%%%%%%%%
%%%%%%%%%%%%%%%%%%%%%%%%%%%%%%%%%%%%%%%%%%%%%%%%

%Figure 2.  
%%%%%%%%%%%%%%%%
%%%%%%%%%%%%%%%%%%%%%%%%%%%%%%%%%%%%%%%%%%%%%%%%

Key areas of further development over the last 5 years have been for higher Q values at medium gradients (16--22 MV/m) for CW operation with the invention of new techniques of Nitrogen doping \cite{Grassellino:2015bzl,Romanenko:2013wga}.  Nitrogen doping for high Q has already been applied to the construction of a large (4--8 GeV) new accelerator, LCLS-II, and its high energy upgrade LCLS-II-HE. For LCLS-II, more than 300 cavities in more than 35 cryomodules, have been delivered to SLAC, and most of these are already  installed.   For LCLS-II-HE, ten 1.3~GHz 9-cell N-doped cavities have reached average $3.5\times 10^{10}$  at 25.7~MV/m. 

\begin{figure}
\begin{center}
\includegraphics[width=0.80\hsize]{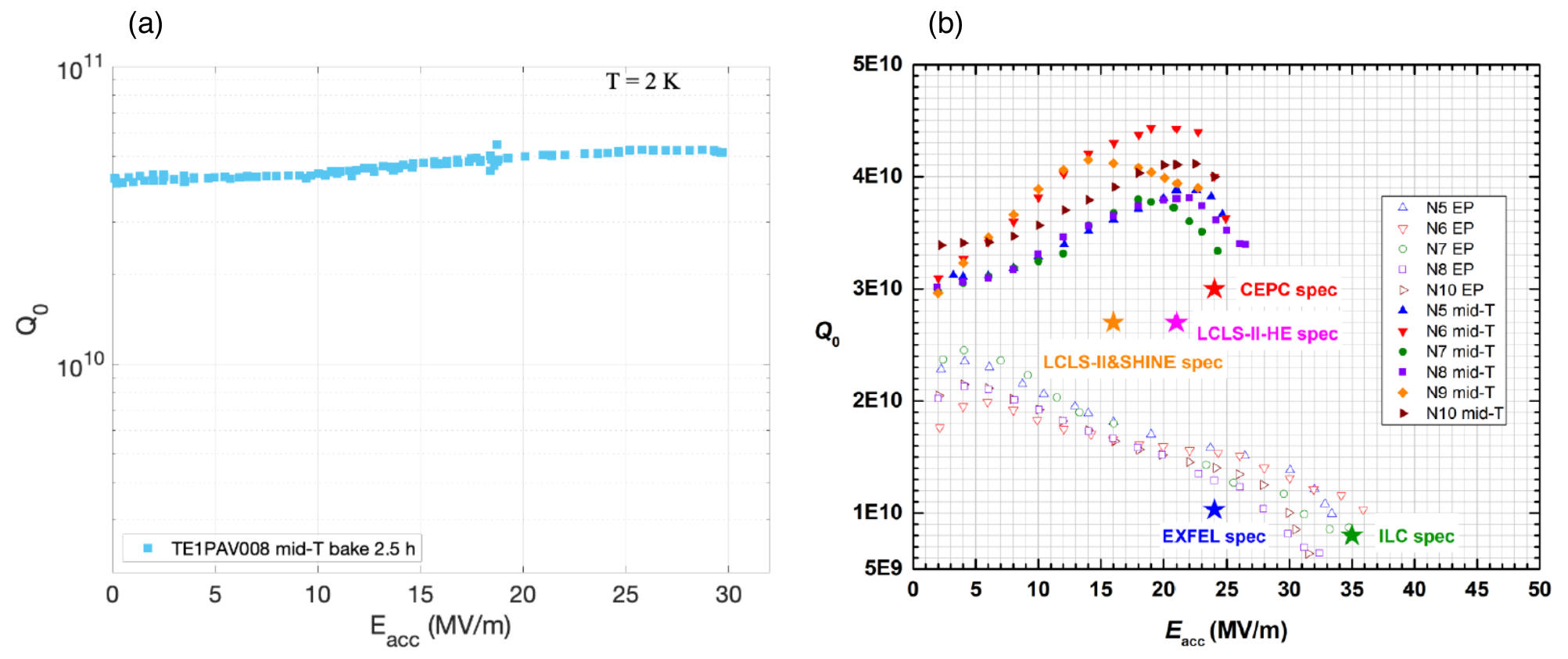}\\
\includegraphics[width=0.40\hsize]{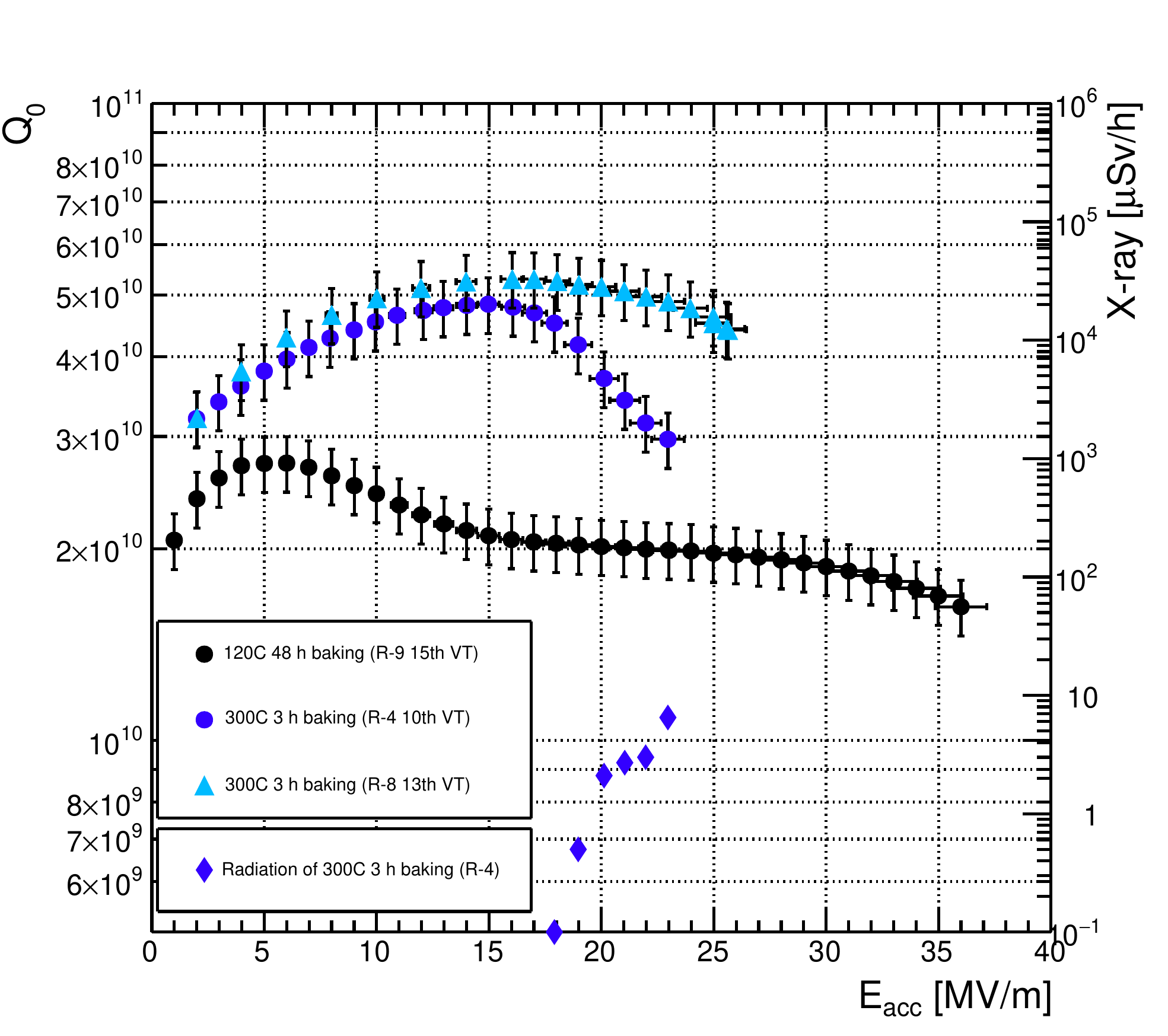}\\
\end{center}
\caption{ (a) Q = $5\times 10^{10}$ at 30 MV/m by baking at 300~C to dissolve the natural oxide (and other surface layers) into the bulk, but not exposing the cavity to air or water before RF measurements; (b) IHEP (China) results on mid-T baking for 9-cell cavities compared to results on the same cavities with the standard ILC treatment~\cite{He:2020ptd}; (c) 300~C baking results from KEK on single cell cavities after high pressure water rinsing.  Standard ILC treatment results are also included.}
\label{fig:Qresults}
\end{figure}

Further improvements can be expected from impressive  developments~\cite{Posen:2019rlz}
 that show Q = $5\times 10^{10}$ at 30~MV/m by baking at 300~C (mid-T baking) to dissolve the natural oxide (and other surface layers) into the bulk, but not exposing the cavity
 to air or water before RF measurements.  It is interesting to note how the Q rises with field, as seen for
 N-doping (Fig.~\ref{fig:Qresults}(a))..  After exposure to air, followed by HPR, the Q dropped to $2\times 10^{10}$ at 30 MV/m.  Surface
 analysis of similarly treated samples show a Nitrogen peak at a few nm below the surface, suggesting that N is present at the surface and has diffused into the Nb to give the doping effect.  IHEP in China followed up on these encouraging results with several 9-cell TESLA cavities with successful  results~\cite{He:2020ptd}, as shown in Fig.~\ref{fig:Qresults}(b).  After mid-T (300~C) furnace bake, and HPR, all the 9-cell cavities 
demonstrate high Q in the range of 3.5--$4.4\times 10^{10}$ at the gradient between 16--24~MV/m, as shown in  Fig.~\ref{fig:Qresults}(b). These cavities have all exceeded the specification of LCLS-II HE ($2.7\times 10^{10} $ at 21MV/m).  KEK is also pursuing the mid-T baking option.  After mid-T baking and high pressure water rinsing, single cell cavities reach Q values of $5\times 10^{10}$ at 16~MV/m and quench fields of 20--25~MV/m~\cite{Ito:2021xpy}.

% Figure 3
%%%%%%%%%%%%%%%%
%%%%%%%%%%%%%%%%%%%%%%%%%%%%%%%%%%%%%%%%%%%%%%%%

\subsection{High Gradient (45 MV/m) SRF for upgrade paths to 1 TeV}
\label{subsec:highgradient}

Section~\ref{sec:acc-staging}  discusses ILC energy upgrade paths from 250 GeV to 380 GeV (Top Factory), 500 GeV and 1000 GeV.  For the 1000 GeV upgrade (Scenario B), the 2013 ILC TDR uses a gradient of 45~MV/m with $Q_0 =2 \times  10^{10}$ for the additional linac from 500 GeV to 1000 GeV.  The SRF parameters are chosen on the forward-looking assumptions of advances in SRF technology derived from R\&D which will continue in parallel to both construction and operation of ILC 250 GeV to 1000 GeV.  Such extrapolations in SRF performance are reasonably based on expectations from proof-of-principle results already in hand.   As discussed further below, single cell cavities with improved treatment reach 49 MV/m, and single cell cavities with improved shapes that reach 52--59~MV/m. 

\subsubsection{Nitrogen Infusion}

\begin{figure}
\begin{center}
\includegraphics[width=0.60\hsize]{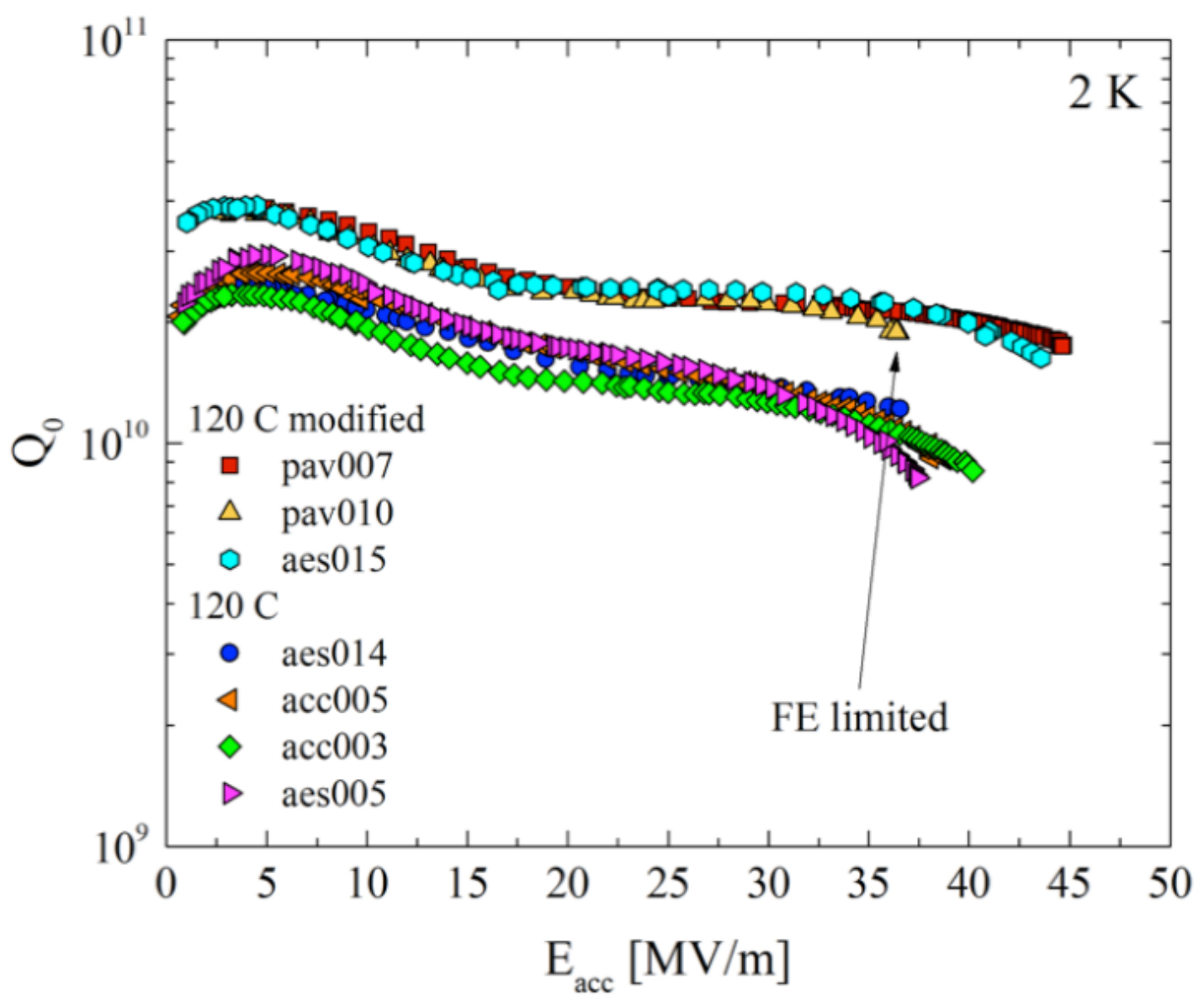}
\end{center}
\caption{Comparison of the performance of several 1-cell cavities from N-infusion with cavities prepared by the standard ILC recipe of EP and 120~C baking.}
\label{fig:Qcomparison}
\end{figure}

On the high gradient frontier (with higher Q’s), the invention of Nitrogen infusion~\cite{Grassellino:2017bod}, stemming from Nitrogen-doping, 
demonstrates gradients of 40--45~MV/m as shown in Fig.~\ref{fig:Qcomparison}, and compared to the performance
 of cavities prepared with the standard ILC recipe.  JLAB has shown success with infusion\cite{Dhakal:2020qhb},  but KEK~\cite{Umemori:2019pup} 
and DESY~\cite{Wenskat:2019hyk} have found the technique to be sensitive to the quality of the infusion furnace, and difficult
 to implement.

\subsubsection{Two-Step Baking and Cold Electropolishing}

% Figure 4
%%%%%%%%%%%%%%%%
%%%%%%%%%%%%%%%%%%%%%%%%%%%%%%%%%%%%%%%%%%%%%%%%

\begin{figure}
\begin{center}
(a) \hskip 4.0in  (b) \\
\includegraphics[width=0.45\hsize]{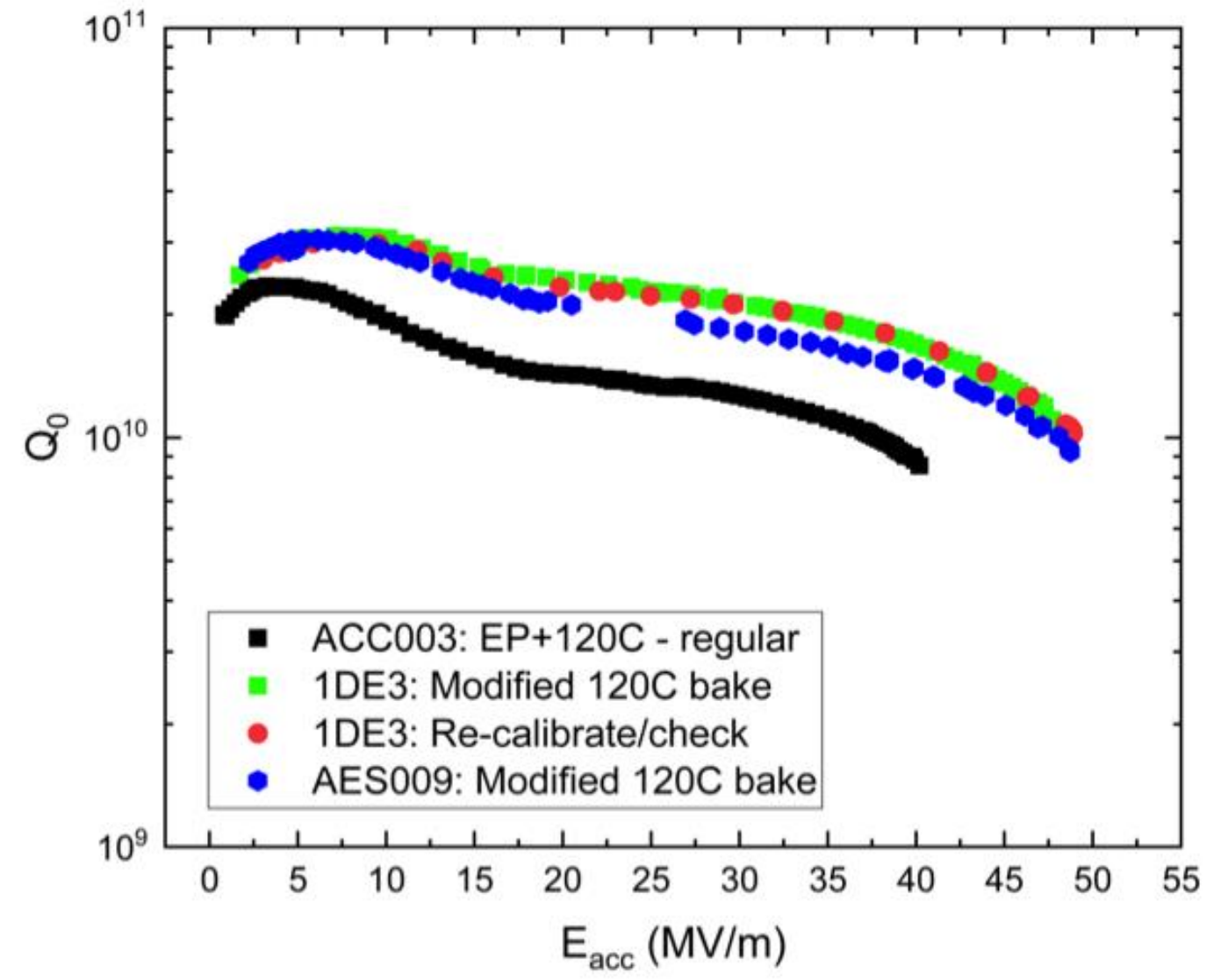}\ \ \raise 25pt \hbox{ \includegraphics[width=0.45\hsize]{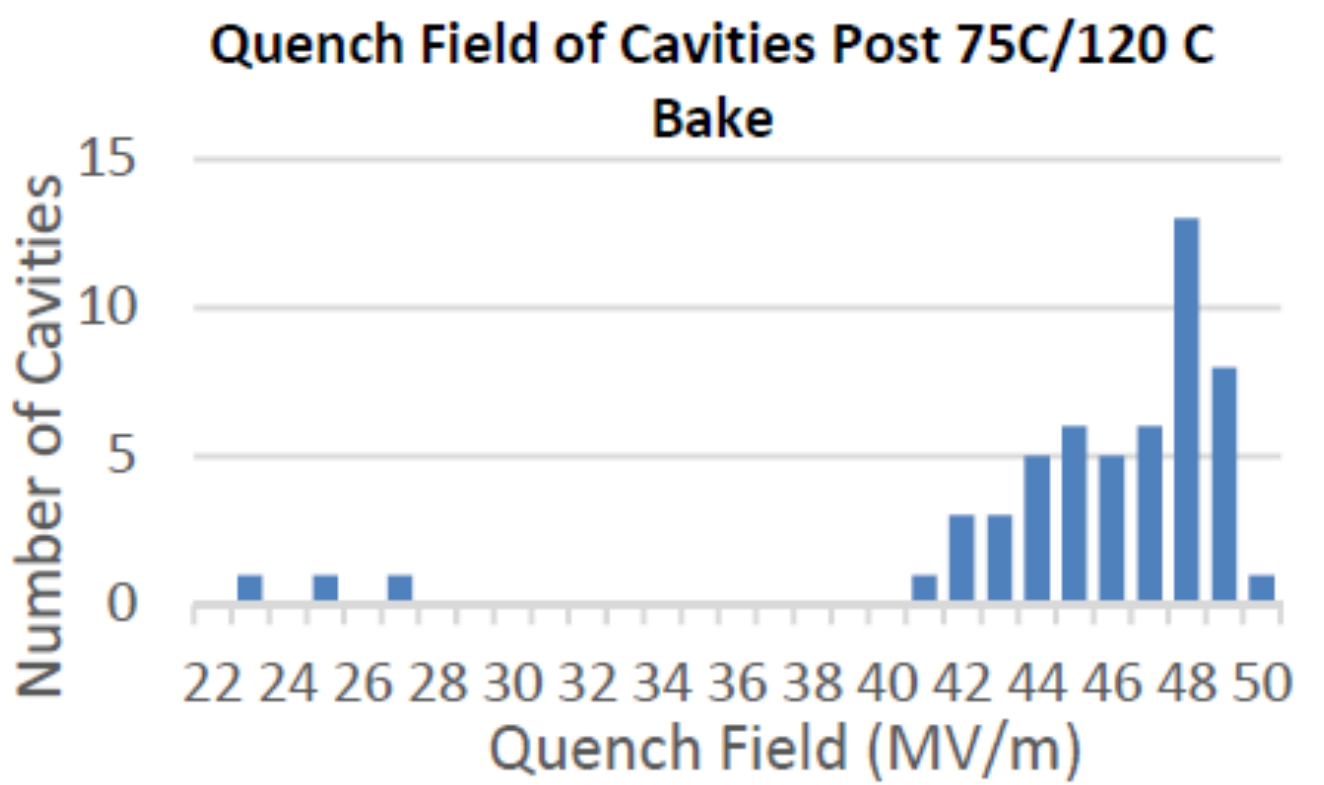}}
\end{center}
\caption{ (a):  $Q$ vs. $E$ curve of 1-cell cavity reaching 49 MV/m from Cold EP/optimized baking (75/120~C) compared to the curve of a cavity prepared by the standard ILC recipe. (b) Histogram of gradients of a large number of single cell cavities prepared by Cold EP/optimized baking (75/120~C). }
\label{fig:Gradients}
\end{figure}

In another new development, extraordinarily high quench fields for 1.3~GHz niobium TESLA-shaped SRF cavities, some near 50~MV/m have been achieved with the 75/120~C bake surface treatment developed at FNAL, as shown in Fig.~\ref{fig:Gradients}(a). Two-Step baking with Cold Electropolishing~\cite{Grassellino:2018tqg} show gradients in the range of 40--50~MV/m (average 45 MV/m), as depicted in Fig.~\ref{fig:Gradients}(b). Note that 3 cavities that quench below 28 MV/m were found to have physical defects that likely limited the performance.

%Figure 5 
%%%%%%%%%%%%%%%%
%%%%%%%%%%%%%%%%%%%%%%%%%%%%%%%%%%%%%%%%%%%%%%%%

 \subsection{Toward 60 MV/m - advanced shape cavities}

\begin{figure}
\begin{center}
\includegraphics[width=0.70\hsize]{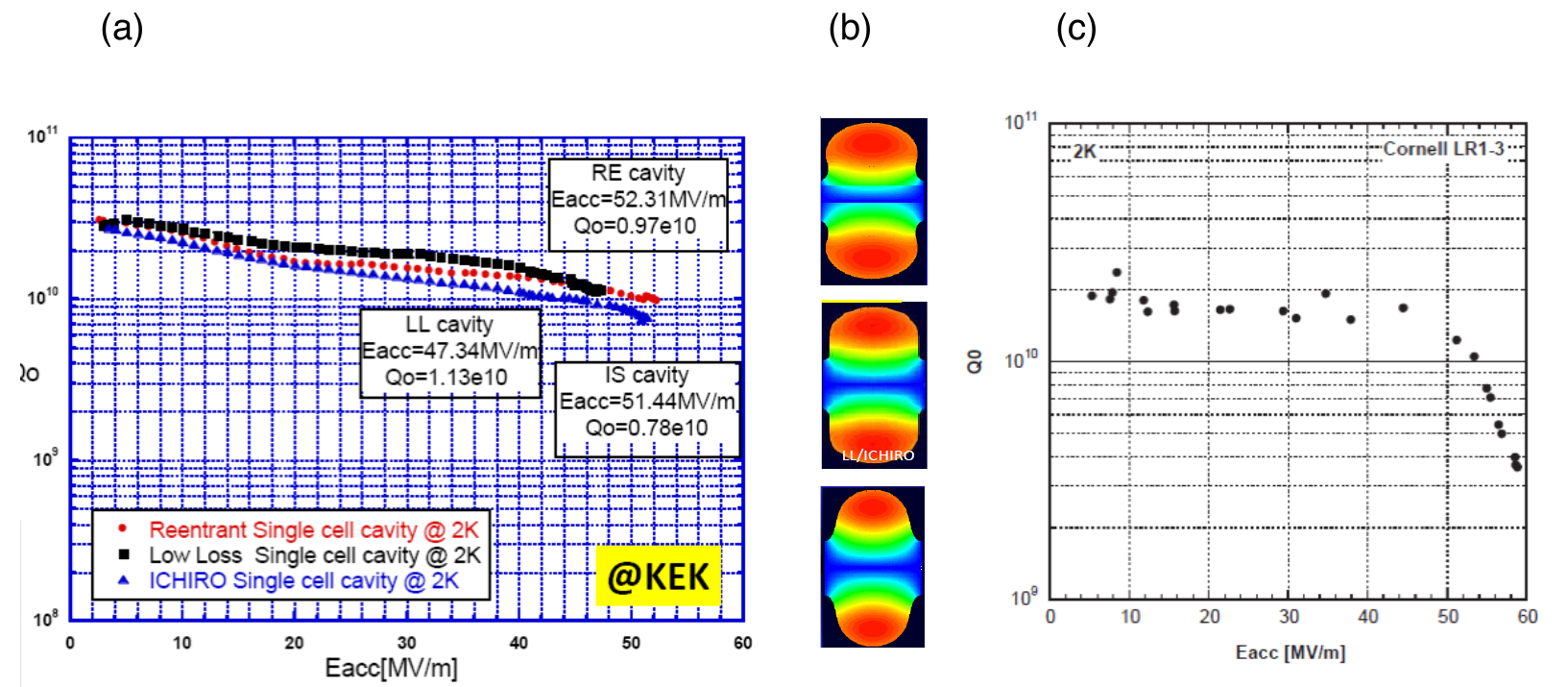}
\end{center}
\caption{ (a) Gradients greater than 50~MV/m demonstrated in single cell cavities of various improved shapes. (b) Comparison of RE (top), LL/ICHIRO (middle) and TESLA (bottom) cavity cell shapes.  Here,  color is magnetic field intensity, red highest, blue lowest.   (c) Record gradient near 59~MV/m demonstrated with the RE shape (60~mm aperture).}
\label{fig:HighGradients}
\end{figure}

Continuing along the gradient frontier, multicell cavities of Re-entrant (RE)~\cite{Shemelin:2002gk,Shemelin:2003fr}, Low-Loss (LL)~\cite{Sekutowicz:2002} and ICHIRO~\cite{Saito:2004} shapes (Fig.~\ref{fig:HighGradients}(b)) have been introduced to lower $H_{pk}/E_{acc}$ 10 - 20\% by rounding the equator to expand the surface area of the high magnetic field region, and by allowing $E_{pk}/E_{acc}$ to rise by about 20\%.  The Re-entrant shape has an $\Omega$-like profile with $H_{pk}/E_{acc}$ = Oe/35.4/(MV/m), $E_{pk}/E_{acc}$ = 2.28 (for 60 mm aperture) as compared to 42.6 Oe/(MV/m) and $E_{pk}/E_{acc}$ = 2.0 for the standard TESLA shape (70 mm aperture).  
The GR/Q value for the re-entrant shape is about 34\% higher than the TESLA shape, which reduces cryogenic losses.  The 20\% increase in $E_{pk}$ makes cavities with the new shapes more susceptible to field emission, but we can expect progress in field emission reduction with cleaner surface preparation developments over the coming decades. 

The motivation in trying the new shape was that quench, governed by $H_{pk}$, is a hard limit, whereas field emission, governed by $E_{pk}$, can be improved by better engineering.   The Low-Loss shape with 60 mm aperture has $H_{pk}/E_{acc}$ = 36.1 Oe/(MV/m), and $E_{pk}/E_{acc}$ = 2.36,  and a 23\% higher GR/Q than the TESLA shape. (Here G stands for the Geometry Factor of the cavity shape, and R/Q is the geometric shunt impedance.) The ICHIRO shape is a variant of the LL  shape.  A relative newcomer to the advanced shape effort is the LSF shape~\cite{Li:2008}, which is a small refinement of the LL shape.  This obtains $H_{pk}/E_{acc}$ = 37.1 Oe/(MV/m)  without raising $E_{pk}/E_{acc}$  (=  1.98). For  comparison, the RE shape with 60 mm aperture has $H_{pk}/E_{acc}$ = 35.4 Oe/MV/m, and $E_{pk}/E_{acc}$ = 2.28. 

Many single cell cavities with the advanced shapes were built, prepared with the standard ILC recipe, and tested to demonstrate gradients of 50 – 54 MV/m with Q0 values above $10^{10}$~\cite{Furuta:2006jj,Eremeev:2007zza}, as shown in Fig.~\ref{fig:HighGradients}(a). A record field of 54 MV/m at Q about $10^{10}$ was set by a single cell Re-entrant cavity with 60 mm aperture, and 59 MV/m at Q about $3\times 10^9$ (see Fig.~\ref{fig:HighGradients}(c)~\cite{Geng:2005xr}) for the same cavity.  However, the best multi-cell cavities of the new shapes have only reached 42~MV/m~\cite{Reece:2013faa}, mostly due to the dominance of field emission.  A 5-cell cavity of the LSF shape recently tested at JLAB showed 50 MV/m gradient in three of the five cells~\cite{Geng:2018} by exciting several modes of the fundamental pass-band.  

As we have seen earlier, the newly developed, two-step bake 
procedure has demonstrated a gradient of 49~MV/m in TESLA shape 1-cell cavities.   Combining the two-step bake with one of the advanced shape cavities has the potential of improving the gradients toward 60~MV/m.  For example, the Low-Loss shape has the potential for 18\% 
improvement from 49 to 58~MV/m.  But no laboratory has attempted such combined efforts as yet.  

%Figure 6 
%%%%%%%%%%%%%%%%
%%%%%%%%%%%%%%%%%%%%%%%%%%%%%%%%%%%%%%%%%%%%%%%%

Depending on the R\&D resources available, we can anticipate that 9-cell accelerator structures, using some of the advanced techniques discussed here, could achieve the single-cell levels of 55~MV/m  in about 10 years.  This estimate is based on the historical time lag seen in Fig~\ref{fig:SRFprogress} between 1-cell results and 9-cell results.  To reach 70~MV/m in Travelling Wave structures could take another 5--10 years, considering that these are  based on the familiar superconductor Nb and considering that first efforts on Travelling Wave cavities have already started. 

\subsubsection{Cost reduction efforts}

The energy upgrades would also benefit from cost saving measures under exploration, such as niobium material cost reduction (15-25\%) for sheet production directly from ingots (with large grains), and/or from seamless cavity manufacturing from tubes using hydroforming, or spinning, instead of the expensive machining and electron beam welding procedures now in practice.  Cost-reducing avenues
 for cryomodules~\cite{Peterson:2016teh} are to connect cryomodules in continuous, long strings similar to cryostats for long strings of superconducting magnets, saving the cost for the expensive ends.  The elimination of the external cryogenic transfer line by placing all cryogenic supply and return services in the cryomodule also reduce costs, not only directly for the cryogenic components, but also by reducing tunnel space required.  Additional cost reductions and efficiency improvements (not included in the TDR 1~TeV estimate) can be also be expected from improved klystron and modulator technology.
In Sec.~\ref{sec:futureSRF},  we discuss the ILC upgrade path from 1~TeV to 2~TeV based on gradients/Q of 55 MV/m/$2\times 10^{10}$ obtained by the best new treatments, such as the two-step bake/Cold EP, applied to advanced shape structures, such as the Low-Loss structure, built from Niobium.    This section also provides tables summarizing the main parameters of the 2~TeV ILC upgrade path to be compared to CLIC 1.5~TeV and the 
70--80 MV/m SRF upgrade paths to 3~TeV
As discussed in Sec.~\ref{sec:futureSRF}, we consider the ILC upgrade path from 1~TeV
 to 3~TeV based on very high gradient SRF opened by R\&D underway on two fronts: 
\begin{enumerate}
\item  Optimized travelling wave (TW) superconducting structures~\cite{Kanareykin:2005wn,Kostin:2015tws,Shemelin:2021} with effective gradients up to 70~MV/m, along with 100\% increase in $R$/Q.  This reduces the dynamic heat load by 40\% due to increase of $GR$/Q, where $G$ is the geometry factor of TW structure.  
\item  80 MV/m/$1\times 10^{10}$ gradient/Q potential for Nb$_3$Sn~\cite{Shemelin:2021}  at 4.2 K, based on extrapolations from high power pulsed measurements on single cell Nb3Sn cavities.     
\end{enumerate}

\subsubsection{Travelling wave structures}
 
\begin{figure}
\begin{center}
\includegraphics[width=0.60\hsize]{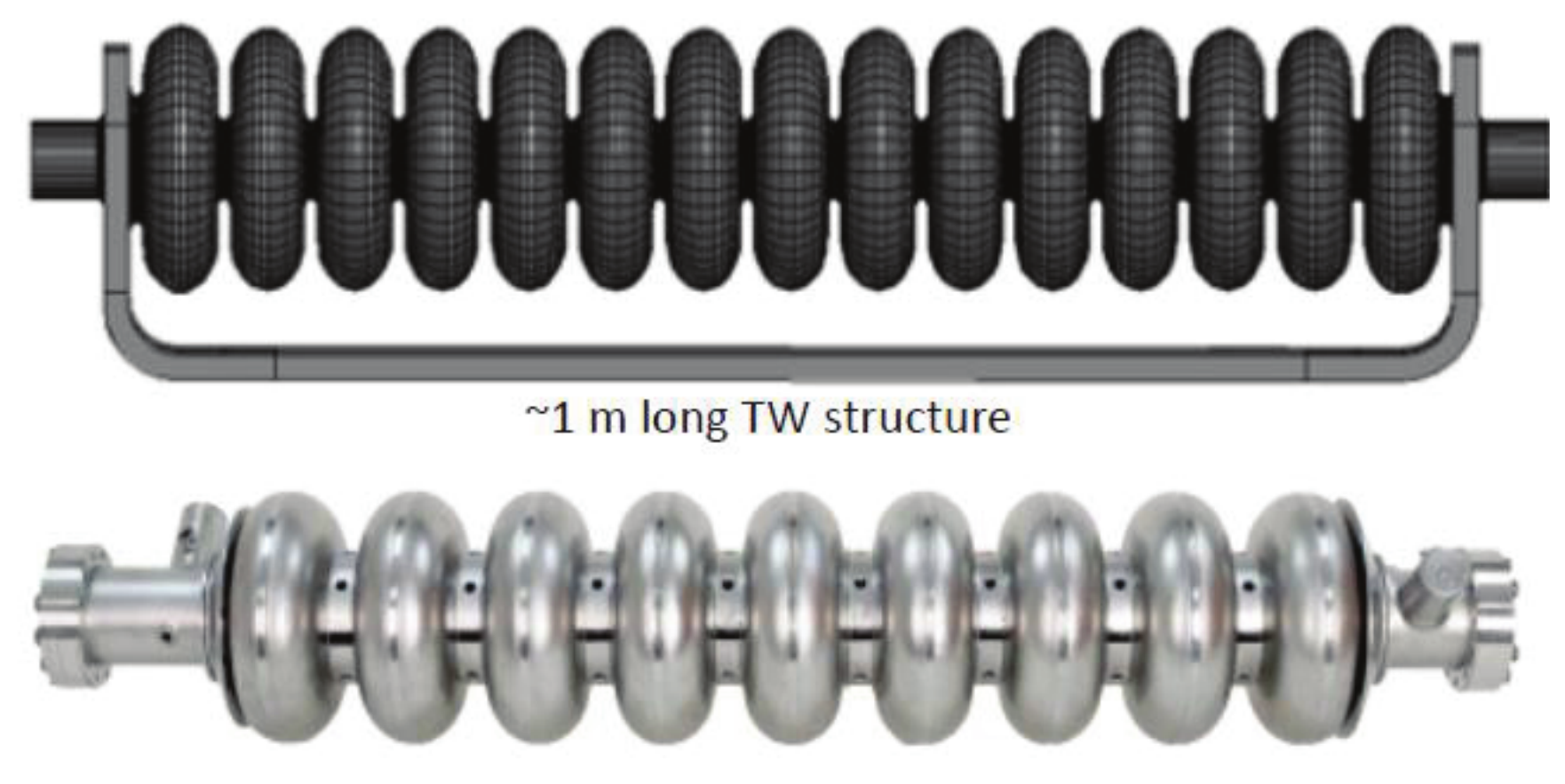}
\end{center}
\caption{The TW structure compared to the one-meter standing-wave ILC structure.}
\label{fig:One-meter}
\end{figure}

Travelling wave (TW) structures offer several main advantages compared to standing wave (SW) structures: substantially lower peak magnetic ($H_{pk}/E_{acc}$), lower peak electric field ($E{pk}/E_{acc}$) ratios, together with substantially higher $R$/Q (for lower cryogenic losses).  The emphasis for future design is to lower  $H_{pk}/E_{acc}$, as much as possible, since $H_{pk}$ presents a hard ultimate limit to the performance of Nb cavities via the critical superheating field.  But, as Fig~\ref{fig:One-meter}  shows, the TW structure requires twice the number of cells per meter as for the SW structure in order to provide the proper phase advance (about 105 degrees), as well as a feedback waveguide for redirecting power from the end of the structure back to the front end of accelerating structure, which avoids high peak surface fields in the accelerating cells.  The feedback requires careful tuning to compensate reflections along the TW ring to obtain a pure traveling wave regime at the desired frequency.

%Figure 7
%%%%%%%%%%%%%%%%
%%%%%%%%%%%%%%%%%%%%%%%%%%%%%%%%%%%%%%%%%%%%%%%%

As discussed in Sec.~\ref{sec:futureSRF}, to obtain a luminosity comparable to CLIC 3~TeV, the beam bunch charge for the 3~TeV upgrade can be 3 x lower than the bunch charge for 0.5~TeV.  Hence it is possible to lower the cavity aperture (from 70 mm to 50 mm) without severe penalty in wake-fields to obtain an overall 48\% reduction in $H_{pk}/E_{acc}$, and factor of 2 gain in $R$/Q over the TESLA standing wave structure.   
Accordingly, we examine the impact of 70~MV/m for the 3 TeV ILC upgrade to obtain a luminosity comparable to CLIC 3 TeV.   Section~\ref{sec:futureSRF} provides tables summarizing the main parameters of the 70 MV/m ILC upgrade path as compared to CLIC 3 TeV, including capital costs, AC powers, energy spreads and backgrounds at the IP.  
Modelling and optimization calculations are underway for TW structure optimization~\cite{Shemelin:2021}.   Table~\ref{tab:Cells} shows one set of optimized parameters for optimized cell shape, phase advance, and 50~mm aperture that yield $H_{pk}/E_{acc}$ = 28.8 Oe/(MV/m) with $E_{pk}/E_{acc}$ = 1.73.  Since $H_{pk}/E_{acc}$ is 42.6~Oe/MV/m and $E_{pk}/E_{acc}$ = 2 for the TESLA structure, the TW structure has reduced the critical parameter $H_{pk}/E_{acc}$ by 48\%!  The geometrical parameters for the cell shape are defined in the inset figure accompanying Table~\ref{tab:Cells}.  If results for the best single cell TESLA shape cavities prepared today ($E_{acc} = 49$~MV/m,  $H_{pk} = 209$~Oe) can be reached in such a TW structure it will be possible to reach $E_{acc} = 72.5$~MV/m.  The 100\% $R$/Q increase lowers the dynamic heat load and cryogenic power needed for high gradients.  

The high group velocity in the TW mode also increases the cell-to-cell coupling from 1.8\% for the TESLA structure to 2.3\%.  Thus TW structures have less sensitivity to cavity detuning errors, making tuning easier, despite the larger number of cells.  Studies~\cite{Shemelin:2021} show that the cell shape can be fine tuned to avoid multipacting, without increasing $H_{pk}$ more than 1\%.  HOM damping is under study.  Preliminary results show that the first 10 monopole modes up to 7 GHz show no trapping.   

Many significant challenges must still be addressed along the TW development path.   High circulating power in the feedback waveguide must be demonstrated.  Cavity fabrication and surface processing procedures and fixtures must deal with (roughly) double the number of cells per structure.  

%Table1 
%%%%%%%%%%%%%%%%
\begin{table}
\begin{center}
\includegraphics[width=0.33\hsize]{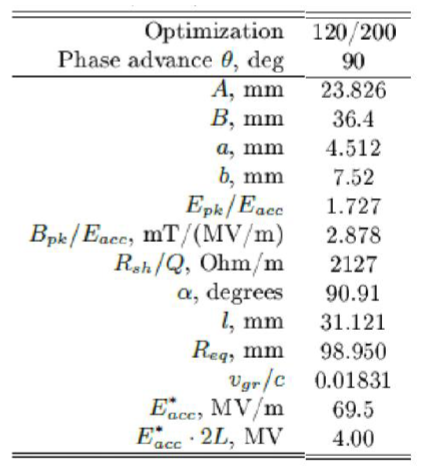}\ \ \ \ \includegraphics[width=0.25\hsize]{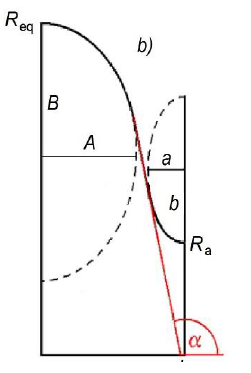}
\end{center}
\caption{Parameters of optimized cells with limiting surface fields: $E_{pk}$ = 120~MV/m and $B_{pk}$ = 200~mT, aperture radius $R_a$ = 25~mm.  $E_{acc}$ is the accelerating rate when the limiting surface fields are achieved.  $2L$ is the cell length = 57.55~mm.  An 18-cell structure (1.036~m) will have the nearly same active length as the TESLA structure (1.061~m).  (from \cite{Shemelin:2003fr}, Table~II, column~2).}
\label{tab:Cells}
\end{table}
%%%%%%%%%%%%%%%%%%%%%%%%%%%%%%%%%%%%%%%%%%%%%%%%

\begin{figure}
\begin{center}
\includegraphics[width=0.70\hsize]{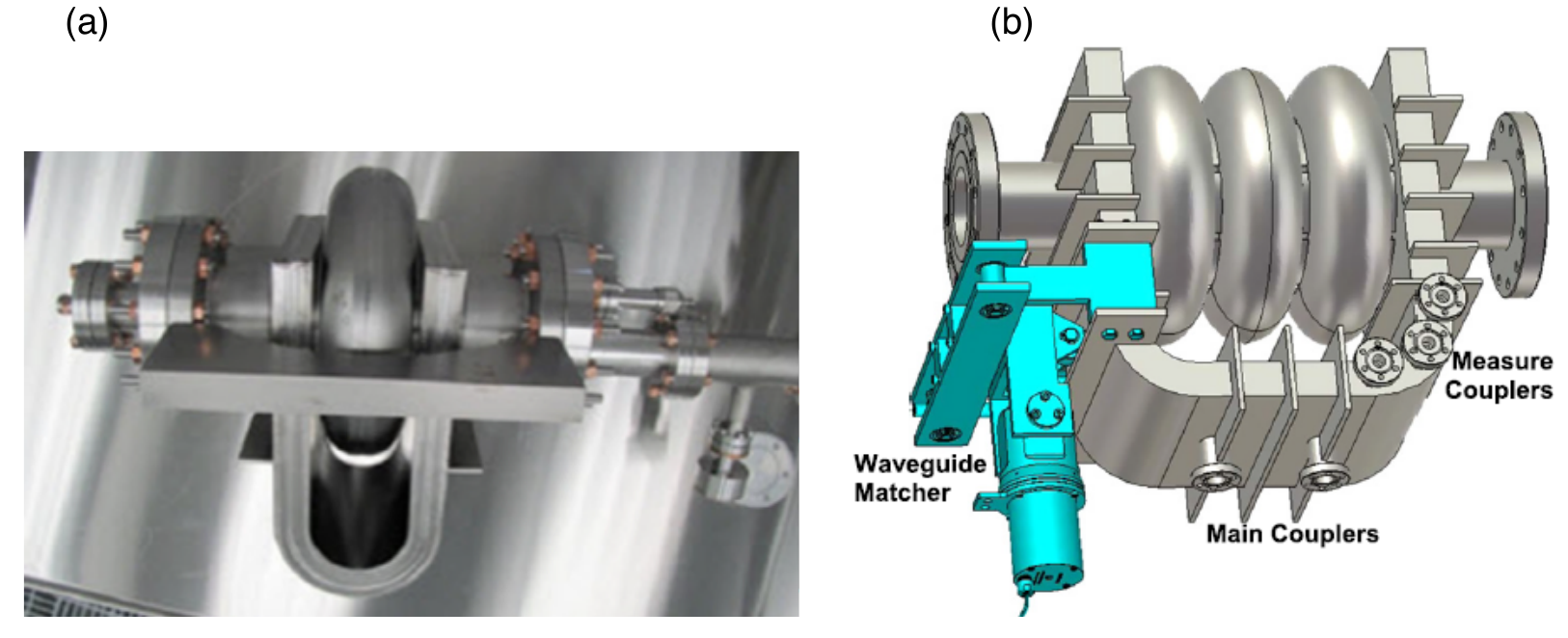}
\end{center}
\caption{ (a) 1-cell TW Niobium structure with return waveguide, treated by BCP and tested to reach 26 MV/m. (b) 3-cell TW structure built but not yet tested.} 
\label{fig:TWstructures}
\end{figure}

First structure fabrication and testing efforts have started for TW cavity development~\cite{Kanareykin:2005wn,Kostin:2015tws}.  With the relatively easier BCP treatment only, the first single cell TW cavity (Fig.~\ref{fig:TWstructures}(a)) with recirculating waveguide achieved 26 MV/m accelerating gradient, limited by the high field Q-slope, as expected for BCP.   This result is very encouraging for a first attempt.   A 3-cell Nb TW structure with recirculating waveguide (Fig.~\ref{fig:TWstructures}(b))was designed and fabricated but has not yet been tested. 

In Sec.~\ref{sec:futureSRF}, we consider the ILC upgrade path from 1~TeV to 3~TeV based on 70~MV/m TW Nb cavities and Q = $3\times 10^{10}$, to be compared to CLIC 3~TeV.  That Section provides tables summarizing the main parameters of the 3~TeV ILC with CLIC 3~TeV.

%Figure 8
%%%%%%%%%%%%%%%%
%%%%%%%%%%%%%%%%%%%%%%%%%%%%%%%%%%%%%%%%%%%%%%%%

\subsection{\texorpdfstring{Nb$_3$Sn}{Nb3Sn}}

A15 compounds (a series of inter metallic compounds with the chemical formula A$_3$B, where A is a transition metal and B can be any element) are intermetallic and brittle in the bulk form, so SRF structures are produced as a thin layer on the inner surface of an already formed structure. Nb$_3$Sn is the most explored
compound, and Nb$_3$Sn films on a Nb structure give  the best results so far~\cite{Posen:2017ltl,Posen:2019kks,Posen:2019}.  Still, as yet, this  does not give as high gradients as pure Nb cavities.  The A15 phase is in the composition range of 18–25\% Sn. The superconducting properties $T_c$, $\Delta$,  and $H_c$, depend strongly on the Sn content~\cite{Godeke:2006}. Perfect ordering in the stoichiometric phase is achieved close to stoichiometry (at 24.5 at\%) 
where $H_{sh}$ is 420~mT as compared to Nb’s $H_{sh}$ of 220~mT at 0~K~\cite{Catalani:2008,Lin:2012,Kubo:2020qwt}.  Accordingly, we can expect the upper limit of the gradient to be 400 mT or near 95 MV/m. 

Nb$_3$Sn films a few microns thick can be deposited on 
the inner surface of Nb cavities by the Sn vapor diffusion process, that is, by exposing  the Nb surface  to Sn vapor ($10^{-3}$~mbar) in an UHV furnace at temperatures between 1050~C and 1250~C.  In general, the Nb$_3$Sn films produced exhibit good material quality with Sn content of about 25\%, $T_c$ of about 18~K, $\Delta$ from 2.7 to 3.2~meV~\cite{Becker:2015}.  The correct stoichimetry is established by a ``phase locking'' feedback process~\cite{Rudman1984}.   The process is diffusion-limited, resulting in good thickness uniformity over large surface areas~\cite{Peininger1988}, and at the same time it produces very clean grain boundaries~\cite{LeeMao2020}.  Coating results are typically reproducible for the same Nb cavity substrate, but have been seen to vary between different cavities. 

\begin{figure}
\begin{center}
\includegraphics[width=0.80\hsize]{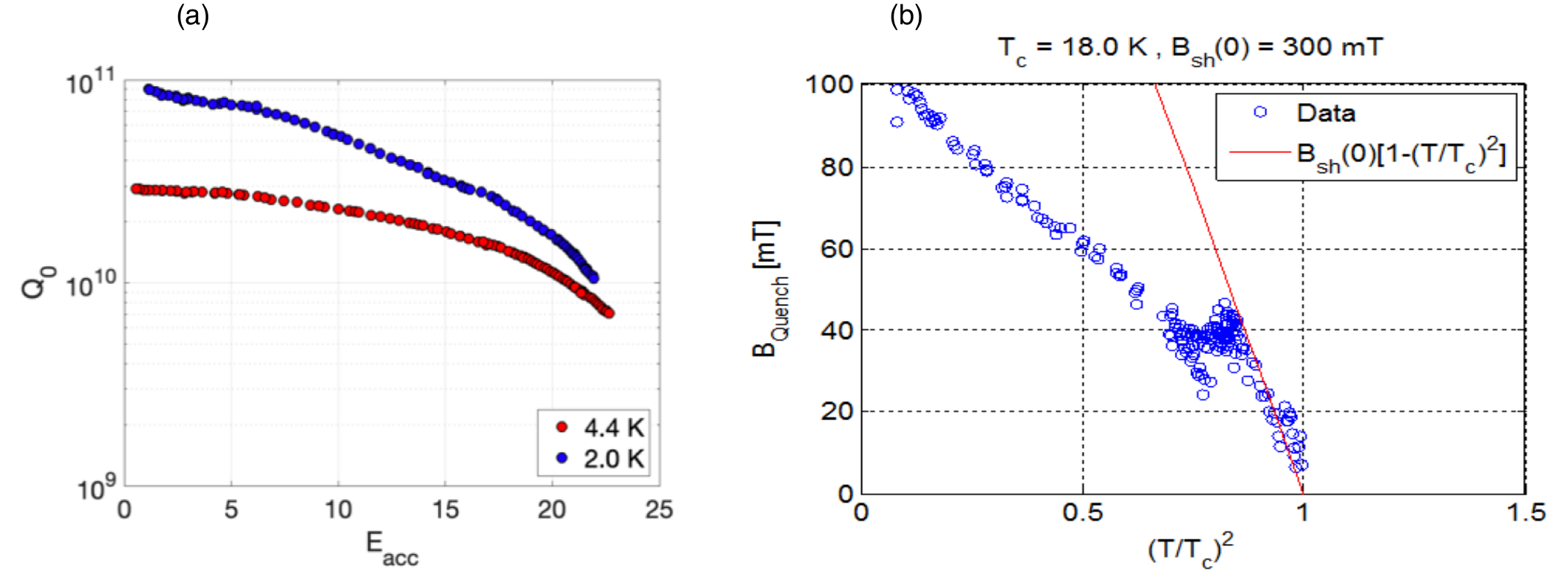}
\end{center}
\caption{(a) Record CW behavior for Nb$_3$Sn coated with the solid state diffusion method at Fermilab showed $E_{acc}$ =  23 MV/m.  Excellent performance was obtained by making a smooth thin film about 1~$\mu$m. (b) Measurements of the critical RF field of Nb$_3$Sn using high power pulsed RF.  The high temperature results extrapolate to a maximum surface magnetic field of 300 mT, which would translate to $E_{acc}$ = 85 MV/m for a Low-Loss shape cavity.}
\label{fig:Nb3Sn}
\end{figure}

Some of the limitations of Nb$_3$Sn films created through this process arise from the sensitivity of the thermodynamic critical field $H_c$ (and therefore the superheating field) to the exact Sn concentration.  For example, a Sn 
depletion of 3\% reduces $H_c$ by 75\%.  Other difficulties are the high surface roughness at Nb$_3$Sn grain boundaries possibly causing local field enhancement.  Somewhat thinner (1 $\mu$m) layers give smoother surfaces and best results (Fig.~\ref{fig:Nb3Sn}).

Since its origin more than 40 years ago, most practitioners of the Sn vapor deposition process have encountered a Q-slope problem and gradient limits~\cite{Hillenbrand1975,Heinrichs1984,Penninger1985,Kneisel1996,Muller1997}.  By the late 1990's, 1.5~GHz single and multi-cell Nb cavities coated with Nb$_3$Sn were investigated up to peak accelerating fields of 15-30~MeV/m.
 The best case of a flat $Q$ vs $E$ curve out to 23 MV/m has been achieved at Fermilab~\cite{Posen:2017ltl,Posen:2019kks,Posen:2019} The performance at 4.2~K is also very attractive showing $Q_0 > 10^{10}$ at gradient of 18 MV/m.  The latest films have smaller surface roughness (by a factor of 2), smaller thickness (1~$\mu$m vs 2--3~$\mu$m) and smaller grain size (0.7~$\mu$m vs 1.2~$\mu$m).   Careful material science is still required to understand and confidently control the Nb$_3$Sn crystal growth dynamics so as to produce low-loss surfaces.

%Figure 9
%%%%%%%%%%%%%%%%
%%%%%%%%%%%%%%%%%%%%%%%%%%%%%%%%%%%%%%%%%%%%%%%%

High power pulsed RF measurements (Fig.~\ref{fig:Nb3Sn}(b)) at Cornell on a Nb$_3$Sn cavity show encouraging trends for very high gradients~\cite{ Posen:2017ltl}.   At high temperature ($T > 15$~K), the results track the high superheating field, extrapolating to 300 mT ($E_{acc}\approx80$~MV/m) at zero temperature.  But at lower temperature, thermal limitations take over to limit the highest field to about 100 mT (24MV/m) which is close to the CW result of 22 MV/m.  

Theoretical studies~\cite{Kubo:2013nra,Kubo:2016dzb,Posen:2015eua,Gurevich:2017vnn}  and DC critical field measurements~\cite{Ito:2019ads} suggest that thin layers of high temperature superconductors such as Nb$_3$Sn or TiN or MgB$_2$ deposited on the RF surface of Nb cavities could lead to substantially higher gradients than possible with Nb.   The phase-space for such new development is quite extensive~\cite{Ito:2019ads}.  We expect that the enthusiasm of the proponents in each area will continue to drive efforts.  The road to an accelerating cavity with gradients higher than possible with Nb is likely to be very long.    

In Sec.~\ref{sec:futureSRF}, which discusses possible  ILC upgrade beyond the  TDR, we consider the ILC upgrade path from 1~TeV to 3~TeV assuming that the promise of this program can be met. We present designs are based on  Nb$_3$Sn cavities with gradients of 80~MV/m and Q of $1\times 10^{10}$.  The discussion there provides tables summarizing the main parameters of the 3~TeV ILC based on this technology, to be compared to CLIC 3~TeV design, and also discusses the potential benefits from 80~MV/m Nb$_3$Sn.  

There is a second approach to the creation of accelerating cavities based on Nb$_3$Sn thin films.  This is to use a substrate made of an inexpensive and thermally efficient material, Cu or bronze, coating the inner surface with an A15 superconductor such as Nb$_3$Sn.  A machine whose RF structures are made of bulk Cu or bronze operating with the properties of Nb$_3$Sn would deliver higher gradient and higher temperature of operation and would give some cost reduction with respect to the use of a Nb substrate.  The use of a Cu substrate would also take advantage of the long experience and recent advances in the fabrication of Cu accelerating structures.

The strategy of Nb$_3$Sn on Cu is currently at a very early stage because most funding in the U.S. has been devoted the the Sn vapor deposition process described above.  However, there are small efforts in the U.S. and abroad for producing Nb$_3$Sn on bronze or Cu. There is as yet no actual tested structure with Q$\sim  10^{10}$ and useful accelerating fields. 

TO deposit Nb$_3$Sn on the Cu or bronze substrate, the Sn vapor diffusion process described above cannot be used, since Cu melts at 1085$^\circ$C. But several alternative methods have been developed to solve this problem.
One possible method is to directly deposit  Nb$_3$Sn superconducting material,  with no heat treatment required. One such method~\cite{Franz:2015lel} was developed a few years ago within an Italian student program~\cite{Barzi:2019pxx}.   A different method has been demonstrated by a group at the Technische Universit\"at Darmstadt, which performs direct deposition of Nb$_3$Sn at 435$^\circ$C using magnetron sputtering in a co-sputtering mode from
 two targets~\cite{Schafer2020}. 

Another, simpler, electroplating technique to coat Nb surfaces with Cu and Sn layers from aqueous solutions and produce Nb$_3$Sn  during a standard heat treatment at ~700$^\circ$C was developed and made reproducible at FNAL in the last few years~\cite{Barzi:2015hpe}. The know-how was then transferred to KEK within an U.S.-Japan Science and Technology Cooperation Program in HEP. Critical temperatures of 17.6 to 17.8 K are routinely achieved. The technology is presently used by Akita Kagaku Co. Ltd, a Japanese electroplating company, to coat 3 GHz Nb cavities. This technique can be implemented on Cu surfaces also, after sputtering them with Nb by using a magnetron system. The main challenge of this method is to achieve the required purity in the Nb$_3$Sn superconducting phase.

Another venue to coat Cu and/or bronze was invented by the Japanese National Institute for Materials Science (NIMS). This process builds upon the A15 superconducting wire technology and also exploits the heat treatment temperature reduction effect of the Cu as the ternary element of the Nb-Sn-Cu phase diagram. In the presence of Cu as the ternary element, the maximum temperature of the heat treatment cycle needed for Nb$_3$Sn  formation is less than 700$^\circ$C. When Cu is not present in the system, as for instance in the Sn vapor diffusion process, the temperature required from the binary phase diagram is much higher. This temperature difference affects the grain size of the Nb$_3$Sn phase, in that the grain size through the bronze process is much finer than that achieved in the Sn vapor diffusion process. It is thought that this method would be suitable to use on SRF cavities fabricated by hydro-forming. 

Most of these coating methods are scalable to standard SRF cavity cells. More details on these various coating methods, and a discussion of their scaling to standard SRF cavity cells, is presented in~\cite{Barzi:2022vtw}.  It seems important to bring one of these methods to the level of development at which the promise of Nb$_3$Sn on Cu or bronze could be investigated in working SRF cavities.  Then it will be possible to attack the issues of surface roughness, purity, and high  Q and perhaps to demonstrate the production of cavities with favorable performance, reproducibility, and cost.

\section{ILC Accelerator technical preparation plan}
\label{sec:acc-RandD}

Although much work has already been done to establish the ILC design
and technical readiness, a number of issues remain to the studied to
prepare the final design of the ILC.  The techical basis for the ILC was
fully documented ten years ago in the ILC Technical Design
Report and its Addendum~~\cite{Adolphsen:2013jya,Adolphsen:2013kya}.
Still, three sets of issues need to be studied anew.  First, it is
necessary to revisit all of  the items to  understand whether any
updates are called based on more recent R\&D results  (including the
past ten years of SRF cost reduction R\&D)
and consistency with the ILC 
staging plan~\cite{Evans:2017rvt}.  Second, because the TDR work was
done without a specific site in view. issues related to the site must
 be addressed again for the specific candidate site in the Tohoku
 region of Japan.  Finally, the MEXT advisory panel and the Science
 Council of Japan have called attention to  some remaining technical issues
 that need to be resolved during the ILC preparation
 period~\cite{MEXTadv,SCJreport}. 

The International Development Team (IDT) was established by the International Committee for the Future Accelerators in August 2020 to prepare for establishing the ILC Pre-lab as the first step toward the construction of the ILC in Japan. IDT-WG2 is now identifying the accelerator-related activities for the ILC Pre-lab necessary before starting the construction of the ILC. The ILC Pre-lab activities are expected to continue about 4 years and the principal accelerator activities of the ILC Pre-lab are technical preparations and engineering design and documentation. The deliverables of the Pre-lab accelerator activities, both technical preparations and engineering design and documentation, will be provided as in-kind contributions by member laboratories of the Pre-lab. Overall management of worldwide Pre-lab accelerator activities will be provided by the Associate Director for Accelerators, assisted by the Central Technical Office. Similarly, each technical preparation and engineering design work package will be led by a manager drawn from one of the member laboratories, guided by the domain and common technology managers. The detailed organization chart for Pre-lab accelerator activities will be defined by the Pre-lab Directorate. The ILC Machine Advisory Committee (ILCMAC), in its advisory role to the Associate Director for Accelerators, will monitor technical progress and review the engineering design and documentation.
A full description of technical preparation is given in the document “Technical Preparation and Work Packages (WPs)
during ILC Pre-lab”~\cite{TPWP2021}. In this section, we will briefly
review this plan. 

%%%%%%%%%%%%%%%%
\begin{figure}
\begin{center}
\includegraphics[width=0.80\hsize]{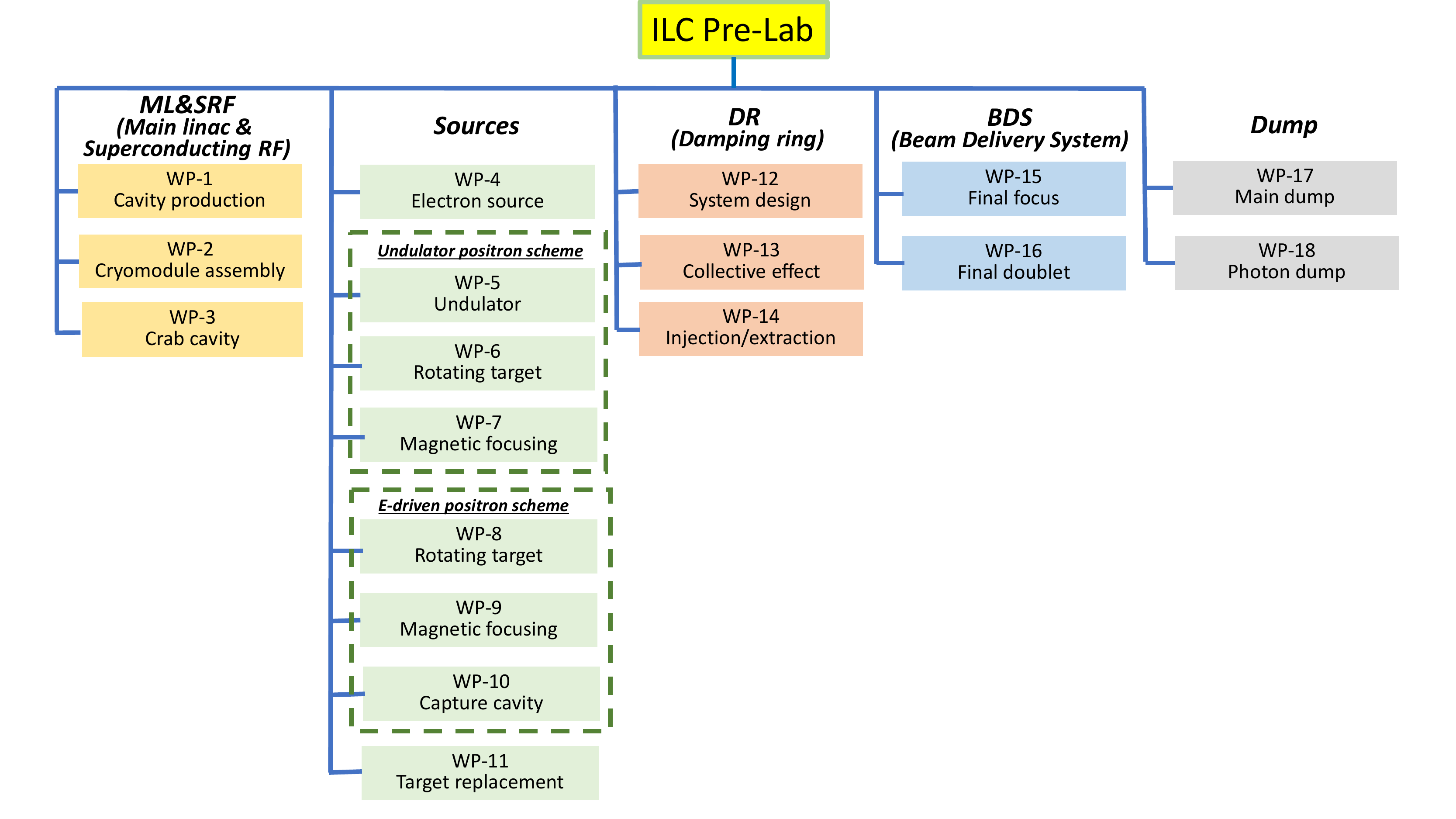}
\end{center}
\caption{Summary of the Work Packages for the technical preparations
that will be carried out during the ILC Pre-Lab period.}
\label{fig:WorkPackages}
\end{figure}
%%%%%%%%%%%%%%%%%%%%%%%%%%%%%%%%%%%%%%%%%%%%%%%%

The Work Packages for the technical  preparation activities cover the following topics:
\begin{itemize}
\item {\bf Main Linac (ML) and SRF production}: 
 Cavity and Cryomodule (CM) global production readiness will be demonstrated
 through the 
 fabrication of  roughly 40  cavities in each of the  3 regions,  the
 requirement of 
 RF performance achieved with $\geq 90$\% success  demonstrated with
 sufficient statistics by using a part (about a half) of the 40
 cavities in each region, and the fabrication of 2 CMs in each of
 the three regions using 40\% of the cavities fabricated. 
\item  {\bf ML global integration}:  The program of
 global CM transfer will be conducted to demonstrate the the 
CM production satisfies  satisfies high-pressure gas safety (HPGS)
 regulations, safe transport across oceans, and the qualification of
 the CM performance after shipping from Europe and the Americas to
 Japan across the oceans. One of the two CMs in each region wil be used
 for this purpose. We plan to accomplish this goal with two
 steps. In the first step, if transport-test CMs (fully constructed
 but not suitable for use in the linac) are available from LCLS-II
 and/or European XFEL, those will be used to test simple
 transportation and to gather important information about stress,
 acceleration, \etc, excluding the HPGS regulation process. In the
 second step, the ILC prototype CM developed during the ILC Pre-lab
 phase will be shipped to Japan, including the HPGS regulation process
 and 
the full CM quality assurance program within the ILC Pre-lab phase period.
\item {\bf Positron source} : The final design will be
 selected from either the 
 an undulator-driven or the electron-driven 
option and its technology readiness will  be demonstrated.
\item {\bf Damping Ring (DR) and Beam Delivery System (BDS)}:
 Readiness of the nanobeam technology for the DR, based on work at the
 ATF3 and related facilities, and the BDS systems will be
 demonstrated, 
particularly including the fast kicker and feedback controls.
\item {\bf Beam dump}: A system design will  be established, including
 beam window handling, cooling water circulation, and safety
 assurance. 
\end{itemize} 
A total of 18 WPs (3 ML\&SRF, 8 Sources, 3 DR, 2 BDS, and 2 Dumps) are
proposed as illustrated in Fig.~\ref{fig:WorkPackages} and summarized
in an extended list below. The classification of some items should be
clarified.  The crab cavity (WP-3) will be installed
in the BDS area, but is classified as ML\&SRF since the crab cavity
uses SRF technology. The photon dump (WP-18) will be used for the
undulator positron source. However, this WP-18 is classified as dump
due to its specialty. The target replacement (WP-11) is a common WP
for undulator and e-driven positron sources. These relationships are
also shown in Fig.~\ref{fig:WorkPackages}.

The explicit tasks of the WPs are as follows:
\begin{itemize}
\item {\bf WP-1 (ML\&SRF)}:  Cavity Industrial Production Readiness
($3 \times 40$ Cavities)
   \begin{itemize}
     \item Cavity industrial production readiness to be demonstrated, including cavities with He tank + magnetic shield for cavity, high-pressure-gas regulation, surface-preparation/heat treatment (HT)/Clean-room work, partly including the 2nd pass, vertical test (VT)
 \item Plug compatibility, Nb material, and recipe for surface
 treatment to be reconfirmed/decided 
\item Cavity Production Success yield to be confirmed (before He tank
 jacketing)
\item Tuner baseline design to be established
\end{itemize}
Note: Infrastructure for surface treatment, HT, VT, pre-tuning, \etc,
is the responsibility of each region.
\item {\bf WP-2 (ML\&SRF)}: Cryomodule (CM) Assembly, 
Global Transfer and Performance Assurance ($3\times 2$ CMs)
\begin{itemize} 
\item Coupler production readiness to be demonstrated, including
preparation/RF processing ($3\times 20$ Couplers)
\item  Tuner production readiness to be demonstrated, including reliability verification
($3\times 20$ Tuners)
\item Superconducting Magnet (SCM: Q+D combined) production readiness
to be demonstrated ($3\times 3$ SCMs, 1 prototype + 2 in each region)
\item CM production readiness to be demonstrated including
high-pressure-gas, vacuum vessel (VV), cold-mass, and assembly
(cavity-string, coupler, tuner, SCM, \etc)
\item CM test including degradation mitigation (in 2-CM joint work, etc.) at assembly site before ready for CM transportation
\item CM Transportation cage and shock damper to be established
\item Ground transportation practice, using mockup-CM
\item Ground transportation test, using production-CM longer than European XFEL
\item Global transport of CM by sea shipment (requiring longer container)
\item Performance assurance test after CM global transport (at KEK)
\item Returning transport of CM back to home country (by sea shipment)
\end{itemize}
Note: Infrastructure for coupler conditioning: klystron, baking
furnace, and associated environment is the responsibility of each
region.  Also, hub-lab infrastructure for the CM production, assembly, and test 
 is the responsibility of each
region.
\item {\bf WP-3 (ML\&SRF)}:  Crab Cavity (CC) for BDS  (2 CCs + 1 for
SRF validation)
\begin{itemize}
\item Decision of installation location with cryogenics/RF location
accelerator tunnel
\item Confirmation of the complete CC system specifications
\item Development of CC cavity/coupler/tuner integrated design (ahead of Preliminary CC technology Down-selection)
\item Preliminary CC technology down-selection (2 cavity options)
\item CC Model-work and Prototype production and high-power validation of CC cavity/coupler/tuner integrated system for two primary candidates (ahead of final CC technology Down-selection) 
\item Harmonized operation of the two prototype cavities in a vertical
test to verify ILC synchronization performance (cryo insert
development and commercial optical RF synchronization system)
\item Final CC technology down-selection 
\item Preliminary Crab CM design – confirming dressed cavity
integration and compliance with beam-line specification
\item Final CM engineering design prior to production
\item Infrastructure for CC development and test in each region
\end{itemize}
\item Further ML\&SRF tasks associated with the Pre-Lab program (1 CM)
\begin{itemize}
\item Cavity (incl He tank) production (incl couplers and tuner),
magnetic shield for CM, high-pressure gas regulation, EP/HT/Clean
work, including VT 
\item Input coupler production including preparation/RF processing readiness (excluding klystron, baking furnace, clean room)
\item Prototype CM production including High-pressure gas, vacuum
vessel, cold-mass, and assembly (cavity-string, coupler/tuner, SCM and
tooling, \etc)
\item Prototype CM test including harmonized operation with two
cavities
\item Prototype CC-CM transport cage and shock damper design and
manufacture
\item Prototype CC-CM transport tests
\item Infrastructure for CM development and testing in each region
\end{itemize}
\item {\bf WP-4 (Sources)}:  Electron Source
\begin{itemize}
\item Drive laser system
\item HV Photogun
\item GaAs/GaAsP Photocathodes
\end{itemize}
\item {\bf WP-5 (Sources)}: Undulator Positron Source
\begin{itemize}
\item Simulation (field errors, masks, alignment)
\end{itemize}
\item {\bf WP-6 (Sources)}: Undulator Positron Source rotating target
\begin{itemize}
\item Design finalization, partial laboratory test, mock-up design
\item Magnetic bearings: performance, speciﬁcation, test
\item Full wheel validation, mock-up
\end{itemize}
\item {\bf WP-7 (Sources)}: Undulator Positron Source magnetic
focusing system 
\begin{itemize}
\item OMD design finalization with yield calculation
\item OMD with fully assembled wheel
\end{itemize}
\item {\bf WP-8 (Sources)}: Electron-Driven Positron Source rotating target
\begin{itemize}
\item Target stress calculation with FEM
\item Vacuum seal
\item Target module prototyping
\end{itemize}
\item {\bf WP-9 (Sources)}: Electron-Driven Positron Source rotating target
\begin{itemize}
\item Flux concentrator conductor
\item Transmission line
\item Flux concentrator system prototyping
\end{itemize}
\item {\bf WP-10 (Sources)}: Electron-Driven Positron Source capture system
\begin{itemize}
\item APS cavity for the capture linac
\item Capture linac beam loading compensation and tuning method 
\item Capture linac operation and commissioning
\item Power unit prototyping 
\item Solenoid prototyping
\item Capture linac unit prototyping
\end{itemize}
\item {\bf WP-11 (Sources)}: Positron Source target maintenance
\begin{itemize}
\item Target Maintenance
(a common issue for the undulator and electron-driven sources)
\end{itemize}
\item {\bf WP-12 (Damping Rings)}: System Design
\begin{itemize}
\item Optics optimization, simulation of the dynamic aperture with magnet model
\item Magnet design : Normal conducting magnet and SC wiggler
\item Magnet design : Permanent magnet
\item Prototyping of permanent magnet
\end{itemize}
\item {\bf WP-13 (Damping Rings)}: Evaluation of collective effects in the ILC damping ring
\begin{itemize}
\item Simulation : Electron cloud instability
\item Simulation : Ion-trapping instability 
\item Simulation : Fast ion instability (FII)
\item System design : Fast FB for FII
\item Beam test : Fast FB for FII
\end{itemize}
\item {\bf WP-14 (Damping Rings)}: System design of ILC DR injection/extraction kickers
\begin{itemize}
\item Fast kicker: System design of DR and LTR/RTL optics optimization
\item Fast kicker: Hardware preparation of drift fast step recovery diode pulser 
\item Fast kicker: System design and prototyping of induction kicker 
\item Fast kicker: Long-term stability test at ATF
\item E-driven kicker: System design,including induction kicker development
\end{itemize}
\item {\bf WP-15 (BDS)}: System design of ILC final focus beamline
\begin{itemize}
\item ILC-FFS system design: Hardware optimization
\item ILC-FFS system design: Realistic beam line driven / IP design
\item ILC-FFS beam tests: Long-Term stability
\item ILC-FFS beam tests: High-order aberrations
\item ILC-FFS beam tests: R\&D complementary studies
\end{itemize}
\item {\bf WP-16 (BDS)}: Final doublet design optimization
\begin{itemize}
\item Re-optimization of TDR FF design considering new coil winding technology and IR design advances
\item Assembly of QD0 prototype, connection to Service Cryostat and
measurement of warm/cold vibration stability with a sensitivity of a few nanometers
\end{itemize}
\item {\bf WP-17 (Beam Dump)}: System design of the main beam dump
\begin{itemize}
\item Engineering design of water flow system
\item Engineering design and prototyping of components; vortex flow in the dump vessel, heat exchanger, hydrogen recombiner  
\item Engineering design and prototyping of window sealing and remote exchange
\item Design of the countermeasure for failures / safety system
\end{itemize}
\item {\bf WP-18 (Beam Dump)}: System design of the photon dump for
the undulator positron source
\begin{itemize}
\item System design and component test of an open-window water dump
\item System design and component test of a graphite dump 
\end{itemize}
\end{itemize}

 The cost and required human resources required for the WPs are
 estimated in ~\cite{TPWP2021}.   The values given 
are initial estimates. The actual numbers will depend on the
laboratories  that will take the responsibility for the deliverables,
so these estimates will be 
re-evaluated later. Infrastructure associated with the series of
items mentioned above will need to be newly
 prepared and/or improved with each region taking responsibility for 
implementation and financial support.
The technical readiness scoped in each WP needs to be verified through
periodical reviews conducted by the ILC Pre-Lab. The ILC technical design
will need to be updated reflecting the progress on the WPs, and these
updates
 will be implemented/added to the engineering documents.
Stability and tuning issues in some WPs will also need to be
coordinated with the start-to-end accelerator design that will be done
as part of the "engineering design and documentation" activities of
the Pre-Lab. These linkages
will be carried out as a part of the ILC Pre-Lab responsibility. 

We expect the these activities can be completed within a  four-year
 preparation period. We divide the timeline into two categories: “Technical Preparation and Readiness” and 
“Engineering documentation”. Here is a plan showing how the 
 WP activities fit into the timeline, using the SRF and Positron Source
 work as examples:
 
\noindent \begin{tabular}{c|l|l}
Year &  Technical preparation & Engineering documentation \\ \hline
1   &   Continue cost-reduction R\&D for SRF cavities &  Start review
and update of TDR cost \\
 & Start pre-series production of SRF cavities &  \ \ estimates by an international
 team \\ 
 & \ \  in cooperation with industry & \\
 & Continue $e^+$ source development & \\  \hline
2 & Complete cost-reduction R\&D  &  Conduct a review on the progress
for \\ 
 & Determine production yield &   \ \ technical work and cost estimation  \\
 & Start assembling cavities into cryomodules & \ \ by an internal panel\\
 & Review $e^+$ source designs &\\ \hline
3 & Demonstrate overseas shipment of cryomodules & Complete cost
estimate and conduct  \\ 
 & \ \  taking all the safety and legal aspects &\ \ internal and
 external review \\ 
 & \ \ into account  & Complete risk
 analysis for the technical \\ 
  &Select $e^+$ source design and start prototyping  & \ \ and cost issues
 \\
  &  \ \ and cost issues of critical  items,  \eg, & Complete a draft of the
  Engineering\\ 
  & \ \  the  $e^+$ target
  & \ \ Design Report \\ \hline
4  & Evaluate cryomodules after shipment and  & Complete and publish
the Engineering  \\
 & \ \  demonstrate the quality assurance procedure  & \  \ Design Report  \\
& Establish regional organization for the ILC  & Start producing
specification documents
\\ 
& \ \ component production & \ \ and drawings of large items for \\
& Continue prototype work for critical components & \ \ tendering\\
& \ \ of the $e^+$ source,  \eg,  the $e^+$ target & \\
\end{tabular}

\noindent Progress in technical preparation activities will be monitored and evaluated through periodic reviews. The activities will be also synchronized with the engineering documentation.

\section{Opportunities for US contributions} 
\label{sec:acc-USopp}

US laboratories host world-class infrastructure and expertise in technology that is relevant for particle accelerators. This presents a number of opportunities for the US to make important contributions to the ILC accelerator that leverage existing capabilities. These contributions would help the project to go forward and position the US well for strong participation in ILC-based experiments.

\subsection{Superconducting linac}

%%%%%%%%%%%%%%%%%%%%%%%%%%%%%%%%%%%%%%%%%%%%%%%%%%%%%%%%%%%%%%%%%%%%%%%%%
\begin{figure}
\begin{center}
\includegraphics[width=0.60\hsize]{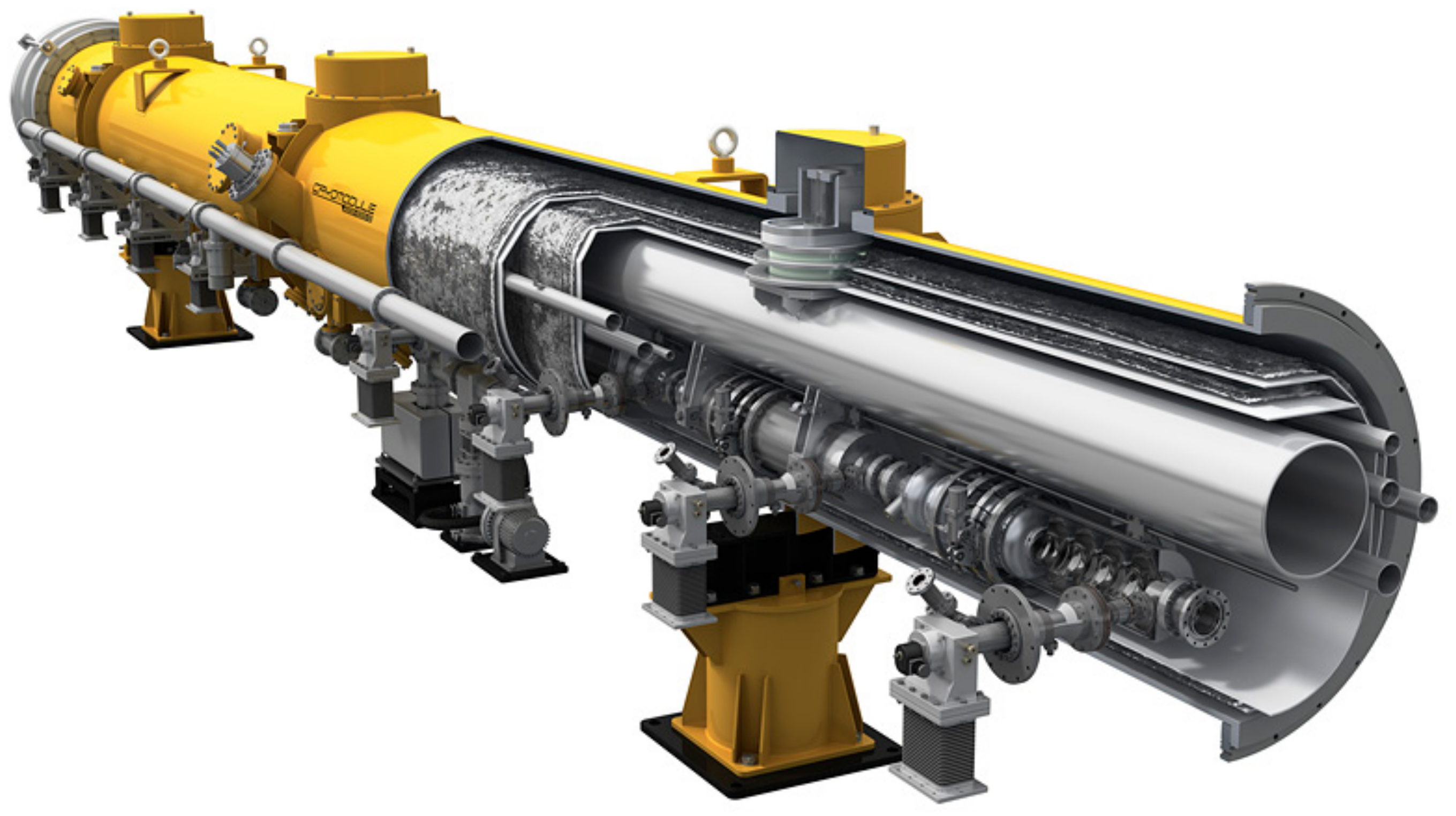}
\end{center}
\caption{Cutaway view of an ILC cryomodule.  (Image by Rey Hori \cite{HoriCryo}.)}
\label{fig:ILCcryo}
\end{figure}
%%%%%%%%%%%%%%%%%%%%%%%%%%%%%%%%%%%%%%%%%%%%%%%%%%%%%%%%%%%%%%%%%%%%%%%%%%%

\begin{figure}
\begin{center}
\includegraphics[width=0.95\hsize]{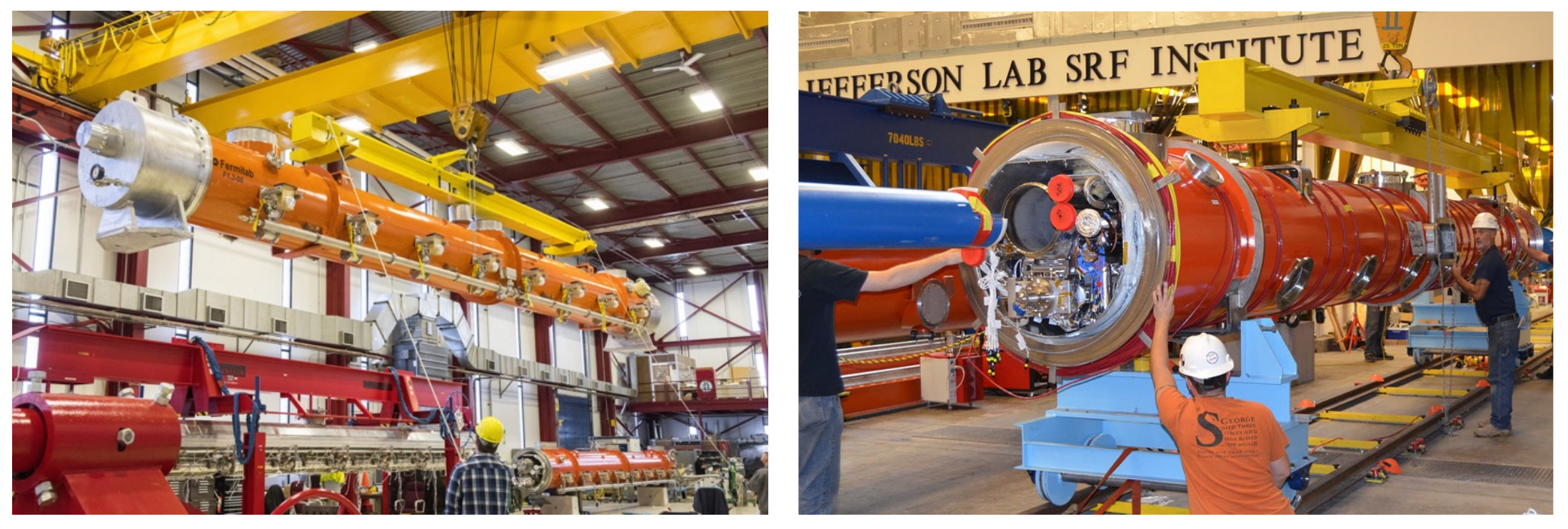}
\end{center}
\caption{View of some of the cryomodule assembly facilities at Fermilab (left) and Jefferson Lab (right).}
\label{fig:Cryoinstallation}
\end{figure}

The superconducting linear accelerator that drives the ILC requires ~1000 cryomodules to reach a center of mass energy of 250 GeV. Each cryomodule (see Fig.~\ref{fig:ILCcryo}) contains 8 or 9 superconducting radiofrequency (SRF) cavities, each about 1~meter long, which generate large amplitude electric fields to accelerate the beam. They also contain liquid-helium-based cryogenics to keep the cavities at 2~K, magnets, RF power couplers, frequency tuners, vacuum valves, and instrumentation. US labs have substantial experience with these sophisticated components from US-based accelerator projects including CEBAF, SNS, LCLS-II, and PIP-II. Large scale production facilities exist at Fermilab and at Jefferson Lab for assembling SRF cryomodules (see Fig.~\ref{fig:Cryoinstallation}). These facilities include large cleanrooms for making vacuum connections between cavities while minimizing the risk of generating particulates that can cause field emission, large fixtures for connecting cavity strings to cold masses and inserting cold masses into cryomodules, and equipment for welding, RF diagnostics, and coupler assembly. Fermilab and JLab also have existing cryomodule test facilities, which require 2~K refrigerators, dedicated radiation areas, and RF systems. These facilities have very recently been used for the mass production of cryomodules for LCLS-II, for which the cryomodule design was largely based on ILC. 
As such the production facilities have already been recently tested with a very relevant system, though ILC would require approximately 5 times as many modules to be produced as the entire production of LCLS-II and its high energy upgrade LCLS-II-HE combined. However, the Fermilab and JLab  teams would take on the larger production with enthusiasm and experience. The vast majority of the infrastructure is already in place, with some modifications required for the higher throughput required to meet the 1 cryomodule per week target for the Americas region at peak production. 

In addition to Fermilab and JLab, there are also SRF facilities at Argonne, Cornell, and FRIB, which are less specialized towards production of ILC-like cryomodules, but could be leveraged for example for cavity treatment. SLAC’s expertise in high power RF sources could be leveraged for driving the cavities as well as RF distribution. SLAC is also planning a relevant cryomodule test facility that could be used. LBNL’s expertise in low level RF could be leveraged for cavity control, particularly for resonance control at high accelerating gradients.

 %%%%%%%%%%%%%%%%%%%%%%%%%%%%%%%%%%%%%%%%%%%%%%%%%%%%%%%%%%%%%%%%%%%%%%%%%
%%%%%%%%%%%%%%%%%%%%%%%%%%%%%%%%%%%%%%%%%%%%%%%%%%%%%%%%%%%%%%%%%%%%%%%%%%%

US expertise can also contribute to advanced performance for ILC cryomodules. Since the 2012 TDR, significant progress has been made in SRF R\&D, including new procedures developed by researchers from US labs for reaching high gradients. Some of these developments could be implemented in ILC cryomodules to push performance by ~10\%, either resulting in fewer cryomodules required to reach the design center of mass energy, or else as a safety margin on top of the nominal energy and beginning towards first energy upgrades. The relevant new technologies include cold electropolishing~\cite{Furuta:2019epz} and the two step bake~\cite{Grassellino:2018tqg}.

Advances from US labs can also contribute to some of the auxiliary systems of the cryomodules. The tuner used in LCLS-II was an evolution of previous designs and is well suited to the short beamtubes of the ILC, while maintaining minimal backlash \cite{Pischalnikov:2015olg}. The quadrupole magnet used in LCLS-II is also an evolution of previous designs, with conduction cooling and a split design to allow it to be assembled outside of the cleanroom \cite{Kashikhin:2018jyy}. A system and procedure for plasma processing of SRF cavities was developed at ORNL \cite{Doleans2016}  and later adapted to 9-cell cavities by FNAL \cite{Berrutti:2019eto},   which may be useful for reducing effects such as field emission in some cases. 

US labs are expected to also play a leading role in developing technologies for energy upgrades to the ILC to reach the 380 GeV-1 TeV energy range beyond the baseline ILC and the multi-TeV energy range in the future. This includes SRF R\&D, such as development of advanced superconductors including Nb3Sn for cavities \cite{Posen:2020kei}, advanced geometries \cite{Kostin:2015tws}, and a plasma accelerator that leverage the SRF-based ILC baseline system. For more details on these upgrades, see Sec.~\ref{chap:farfuture}.

\subsection{Electron and positron sources}

Many US labs have capabilities in sources from their own facilities. The plan for ILC has a positron source, which can be accomplished in different ways. One of these employs superconducting undulators, the other targets, and both subjects can benefit from expertise at a number of US labs.

\subsection{Damping ring, beam delivery system, and beam dump}

US accelerator scientists have extensive experience also in the technologies needed for the damping ring, beam delivery system, and beam dump.

The damping ring is expected to be similar to multiple US facilities, such as the APS upgrade at Argonne, CESR at Cornell, and NSLS-II at Brookhaven. 

For beam dynamics and lattice development, researchers at nearly all US labs with accelerators have substantial relevant experience as well as specialized tools and codes such as ACE3P, ELEGANT, and BLAST. 

Expertise in superconducting magnets at labs such as FNAL, Berkeley, and BNL can be applied to the magnets needed for the final focus at the interaction point. A similar task is ongoing at US labs for production of magnets for the high luminosity upgrade of the LHC.

For research and development related to plasma-accelerator-based multi-TeV upgrades to ILC, US labs host multiple accelerator facilities that could be used for relevant R\&D including AWA at Argonne, FACET at SLAC, ATF at BNL, BELLA at LBL, and FAST at Fermilab.

\subsection{Summary}

The US National Laboratories are anticipating a wide range of contributions to the ILC accelerator. These contributions are synergistic, both from past programs---\ie, they leverage existing infrastructure and expertise in US labs---and for developments for the future---\ie, much of the needed R\&D for the US contribution to ILC has application to other accelerator projects that the laboratories are involved in.

By virtue of this, there is a broad interest among all of the US National Laboratories invested in accelerator physics in participating in ILC. In addition to synergies with US labs, there is also synergy with US industry. A substantial part of the US funds for ILC construction will be put towards procurements from US companies for high-tech components that will be used in cryomodules and other accelerator elements.

\chapter{General Aspects of the ILC Physics Environment} 
\label{chap:gen-phys}

This  chapter gives a general orientation to the physics of the ILC.  We will describe the major physics processes that the ILC will allow us to study, and the reactions that appear as backgrounds in the analyses discussed in Chapters 8--10.  

We will also call attention to the effects of beam polarization.   The reaction cross sections at an $\ee$ collider have direct and strong dependence on the beam polarizations.   In the SM, highly relativistic left- and right-handed polarized electrons are essentially different species, with different electroweak quantum numbers.  Thus, measurements with different beam polarization measure different reactions, and the comparison of these reaction rates can give direct insight into the physics.  Longitudinal polarization is maintained in linear acceleration, so that a highly polarized source of electrons or positrons produces a comparable effect of polarization in collisions.  Thus, at linear colliders, beam polarization works as a new tool for discovery, one not available at proton colliders or circular $\ee$ colliders.  It plays a large role in the complementarity of the various types of machines.

%Figure 1
%%%%%%%%%%%%%%%%
\begin{figure}
 \begin{center}
\includegraphics[width=0.72\hsize]{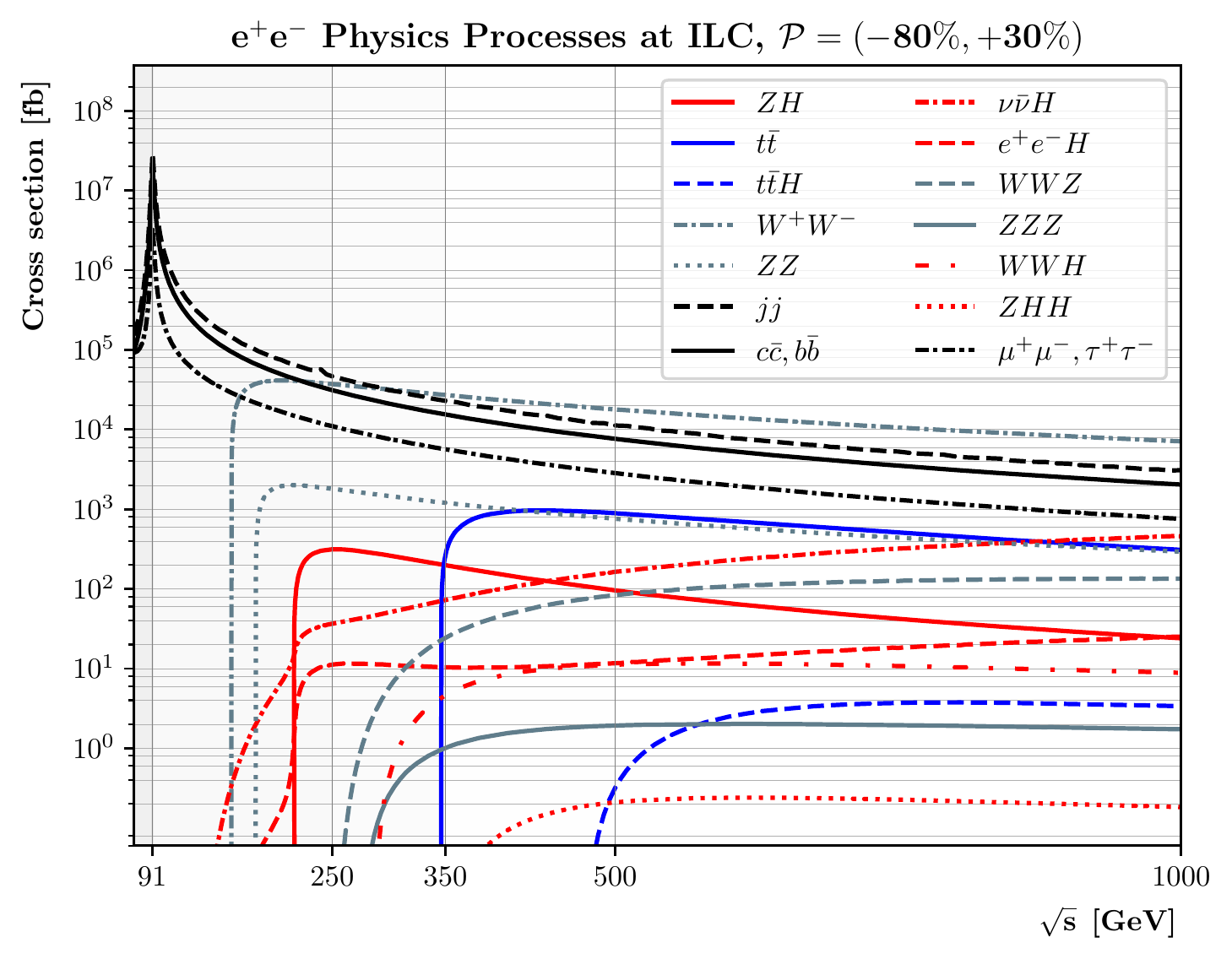} \\ 
\includegraphics[width=0.72\hsize]{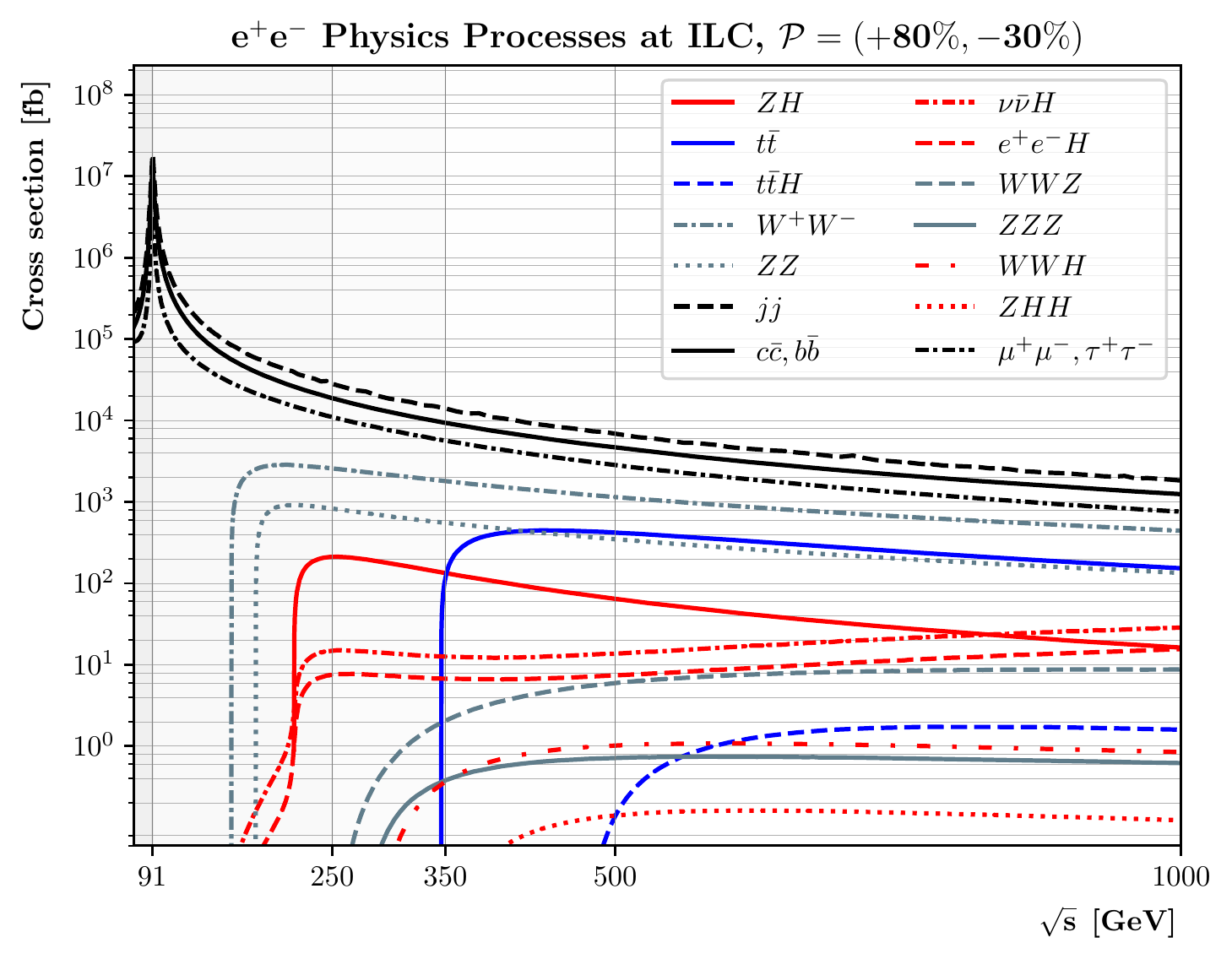} \\ 
\end{center}
\caption{Cross sections of the most important Standard Model processes in $\ee$ annihilation in the energy range of the ILC.  Initial state radiation is included, and cross section are plotted for reactions in which the annihilation retains $>90\%$ of the nominal CM energy.  The cross sections are shown for predominantly left-handed beam polarization ($-80\% /+30\%$ for $e^-/e^+$) (top) and for  predominantly right-handed beam polarization ($+80\% /-30\%$) (bottom).  It is instructive to compare the two plots, which have subtle and 
not-so-subtle differences.}
\label{fig:crosssections}
\end{figure}
%%%%%%%%%%%%%%%%%%%%%%%%%%%%%%%%%%%%%%%%%%%%%%%%

\section{Key Standard Model processes}
\label{sec:SMprocesses}

The major reactions at $\ee$ colliders in the center of mass (CM) energy range of 100~GeV to 1~TeV are shown in Fig.~\ref{fig:crosssections}.   The typical size of a cross section in $\ee$ annihilation is the point
cross section
\beq
           \frac{4\pi \alpha^2}{3E_{CM}^2} =  1.4 ~\mbox{pb} \cdot \biggl( \frac{250~\mbox{GeV}}{ E_{CM} } \biggr)^2 \ ,
\eeqn
 This corresponds to the middle region ($10^{3}$  fb) of the plots in Fig.~\ref{fig:crosssections}, corresponding, for example, to several million events in a data set of 2~ab$^{-1}$ at 250~\mbox{GeV}. The most important  2-body reactions are enhanced over this value by the strength of the weak interaction couplings $g^2/e^2$, color factors, and spin factors.  Still, this is a much smaller number of events than is typically collected by a hadron collider experiment.  However,
these events appear individually in $\ee$ bunch collisions, are essentially free of extra tracks and debris associated with the beams, and are reconstructable with high efficiency over an angular region that extends to within milliradians of the beam directions.  The simplicity of typical events allows the use of detector technologies with high degrees of discrimination and precision, as will be discussed in the next chapter.

Since the electron is an elementary particle, the basic parameters of the ILC beams are very well understood.   Though the beams contain a distribution of electron and positron energies due to 
initial state photon radiation and radiation of photons in the beam-beam interaction (``beamstrahlung''), these are minor and computable effects.   There is no analogue of the nonperturbative parton distributions needed for the interpretation of cross sections at hadron colliders.  Thus, the ILC can carry out measurements of absolutely normalized cross sections at the part-per-mil level, adding a dimension for precision tests of the SM. The beam polarizations can be measured both by dedicated detectors and through SM processes with large cross sections.  

Each of the reactions shown in the figure has its own individual role in the program of the ILC.  Each reaction gives access to its own set of precision tests of the Standard Model and searches for the effects of new physics.   It is important to understand the hierarchy of reactions to understand the important sources of background that enter the various analyses.  At an $\ee$ collider, backgrounds from simple QED and QCD processes are readily eliminated.   The major backgrounds to processes with the production of heavy particles---within the SM, $W$, $Z$, Higgs, and top---are other reactions with heavy SM particles in the final state.

Each reaction also has its own characteristic dependence on beam polarization, as is shown in the figure.  This will be an important theme of the discussion in this Chapter.

The simplest reaction in $\ee$ annihilation is that of $\ee\to f\bar f$, where $f$ can be a quark or a lepton.   Even for the hadronic reactions, the final state is typically two narrow jets and is easily discriminated from reactions of electroweak bosons.   At the tree level in the SM, the differential cross section is very simple.  For example, for 100\% left-handed polarized beams and $s \gg m_Z^2$, the differential cross sections are  
\beq
    \frac{d\sigma}{d\cos\theta} =  {\frac{\pi \alpha^2}{2s}} \bigg[ (\half I^3_{fL}  + \half Y_{fL})^2) (1 + \cos\theta)^2 + (\half Y_{fR})^2 (1 - \cos\theta)^2 \biggr] \ ,
\eeq{FBQnos}
where $(I^3_{fL}, Y_{fL})$ are the electroweak quantum numbers of $f_L$ and $(0, Y_{fR})$ are the 
electroweak quantum numbers of $f_R$.   Note that  the production of the two helicity states of $f$ separates into the two hemispheres. Thus, with two  different values for the beam polarization and separate measurement of the forward and backward cross sections, it is possible to probe all four 
individual helicity amplitudes contributing to this reaction.  This provides a powerful and specific probe for new physics, as we will discuss in Sec.~\ref{sec:pairs500}.  Bhabha scattering ($\ee\to \ee$) has a more complex differential cross section, but this reaction is extremely well understood within the SM, leading to its own set of new physics tests.

The reaction $\ee\to$ hadrons is also an exceptionally clean setting for studies of jets and the measurement of fragmentation functions.  The potential ILC contributions to QCD, including new observables sensitive to jet substructure, are described in Sec.~\ref{sec:QCD}.

The annihilation reaction with the largest cross section in the 250--500~GeV CM energy region is $\ee\to W^+W^-$.   This reaction is forward-peaked, due to the contribution from a diagram with $t$-channel neutrino exchange.   The reaction can be reconstructed in all $W$ decay modes, with the most complex final states having 4 jets.  This reaction obtains contributions from diagrams with the triple gauge couplings $WW\gamma$ and $WWZ$.  Because of a cancellation among the SM diagrams required by the unitarity of that theory, the angular distributions and polarization effects in this reaction are exceptionally sensitive to  new physics contributions to the triple gauge couplings.   These effects are most pronounced in the central and backward $W^+W^-$ production. We will discuss the measurement of 
these effects in Secs.~\ref{sec:Wboson250} and \ref{sec:Wboson500}.   In contrast, the forward production is essentially  model-independent. Because the neutrino exchange diagram requires left-handed electrons and right-handed positrons, the forward production has a large polarization asymmetry and so provides a very useful {\it in situ} measurement of beam polarization.

The other vector boson pair production reactions, $\ee\to \gamma\gamma, Z\gamma, ZZ$, do not involve triple gauge couplings in the Standard Model.   It can be shown that the new physics corrections to these reactions are also suppressed in the description of new physics by Effective Field Theory. Thus, these reactions can provide fundamental test of the Effective Field Theory framework, and, in some cases, tests of general  positivity theorems of Quantum Field Theory.  We will discuss these 
issues in Sec.~\ref{sec:BSMscale}.

The reaction $\ee\to \gamma Z$ with the photon almost collinear to the beam direction provides a large source of $Z$ bosons that can be used to probe the $Z$ properties even at CM energies well above the $Z$ resonance. In the ILC run at 250~GeV, we expect to study about 90 million $Z$ bosons in this ``radiative return'' reaction, leading to an improvement of a factor of 10  in the precision of $\sstw$ even without running at the $Z$ resonance.   The study of this reaction will be discussed in 
Sec.~\ref{sec:radreturn}.

At 250~GeV, the dominant reaction for production of the Higgs boson is $\ee\to ZH$.    This process is expected to produce about half a million Higgs bosons in the 250~GeV run of the ILC, with each Higgs boson tagged by a recoiling $Z$ boson.   This will give an excellent setting for the  measurement of SM and non-Standard Higgs boson decays.  That study will be described in Secs.~\ref{sec:Higgs250} and 
\ref{sec:HiggsExotic}.

The ILC also expects a number of reactions with photons in the initial state.   The photons arise as virtual photons from initial-state radiation and as real 
beamstrahlung photons emitted in the beam-beam 
interaction.   For the ILC accelerator parameter sets, these two sources contribute roughly equally to the spectrum of initial photons.   Important reactions due to initial-state photons are single $W$ production ($e  \gamma \to W \nu$) and single $Z$ production ($e \gamma \to Z e$).
Reactions with two photons in the initial state include photon annihilation to lepton pairs, quark pairs, and $W^+W^-$.   The single boson production reactions have a role in the precision determination of 
the $W$ and $Z$ masses, as will be described in Sec.~\ref{sec:WZmasses}.  All of these processes appear as the major backgrounds to new particle searches involving missing energy, as discussed 
particularly in Secs.~\ref{sec:newparticles} and \ref{sec:dark500}.

The cross sections for $\gamma\gamma$ production at large angle decrease as $1/s(\gamma\gamma)$.   The converse of this statement is that there is a large cross section for 
$\gamma\gamma$ annihilation to quarks and leptons at the lowest possible CM energies.   This leads to an ``underlying event'' giving  a few tracks in each $\ee$ bunch crossing.  We  find that this 
background has a negligible effect on our analyses.

At energies  above 250~GeV, the initial electrons and positrons can radiate $W$ and $Z$ bosons and these can interact to produce SM and, possibly, new particles by vector boson fusion.   The cross sections for these processes   rise as  $\log(s/m_W^2)$ and so above 500~GeV they become the dominant modes of heavy particle 
production.   The coupling of the  electron to the $Z$ is rather small, so the ratio of the $ZZ$ to $WW$ luminosities is 
\beq
       \biggl[ \frac{ (\half - \sstw)^2 + (\sstw)^2}{\cstw} / {\half} \biggr]^2  =  1\%
\eeqn
for unpolarized beams, and even smaller for polarized beams enhanced in the  $e^-_Le^+_R$ initial state.  Thus, $WW$ fusion plays the dominant role.   The most important processes here are $WW$ fusion to a single $Z$ ($\ee\to \nu\bar\nu Z$) and to a single Higgs boson ($\ee\to \nu\bar\nu H$).

The process $\ee\to \nu\bar\nu H$ begins to dominate the $\ee\to ZH$ process at about 400~GeV.
Above this energy, the $WW$ fusion process provides a second, independent data set for the study of Higgs boson couplings. In the $WW$ fusion events, the Higgs boson appears as a heavy, centrally-produced particle with no other visible activity in the event.  The fact that the Higgs boson can be produced in two distinct ways at $\ee$ colliders allows cross-checks of any anomalies with the same experimental program.   This is another of the special benefits of studying the Higgs boson through $\ee$ annihilation.   The study of the Higgs boson in $WW$ fusion will be discussed in detail in Sec.~\ref{sec:HiggsWW}.

The threshold for top quark pair production $\ee\to t\bar t$, occurs in the region around of CM energy of 340--345~GeV. Because the top quark threshold is a very narrow feature, the measurement of the threshold shape can give a very direct and accurate measurement of the top quark mass.  At and above the top quark threshold, the ILC can study the couplings of the top quark with  high precision.  Of special interest are the electroweak couplings of the top quark, which have secondary importance at hadron colliders but provide the primary pair production mechanism at $\ee$ colliders. These couplings can be especially sensitive to new physics corrections, especially in models in which the Higgs boson is composite. There is a significant advantage in measuring these couplings well above threshold, because the axial vector current terms in the top quark vertices are very small near threshold, and because the matching of predictions for the $t\bar t$ continuum to the rather different theory of the threshold region introduces extra theory uncertainties.
We will discuss all of these issues in Sec.~\ref{sec:top}.

At the highest ILC energies, it is also possible to access multi-Higgs boson production processes.  The most important of these are the reactions $\ee\to Z H H $ and $\ee\to \nu\bar\nu HH$, which depend
directly on the Higgs boson self-coupling, and $\ee\to t\bar t H$, which directly measures the Higgs boson coupling to the top quark.   We will discuss these analyses in Sec.~\ref{sec:Higgs500}.

Thus, each separate ILC reaction has a role to play in challenging the predictions of the SM.  Even 
further, it is best not to interpret the individual processes in isolation from one another.  By 
representing the SM and its possible corrections using Effective Field Theory, the contributions from the 
different reactions can be brought together and applied in a unified way.    The whole set of collider
measurements is then more powerful than the simple sum of its parts.   We will discuss this 
strategy of interpretating the ILC measurements in some detail in Chapter 12.

%%%%%%%%%%%%%%%%
\begin{figure}
\begin{center}
\includegraphics[width=0.90\hsize]{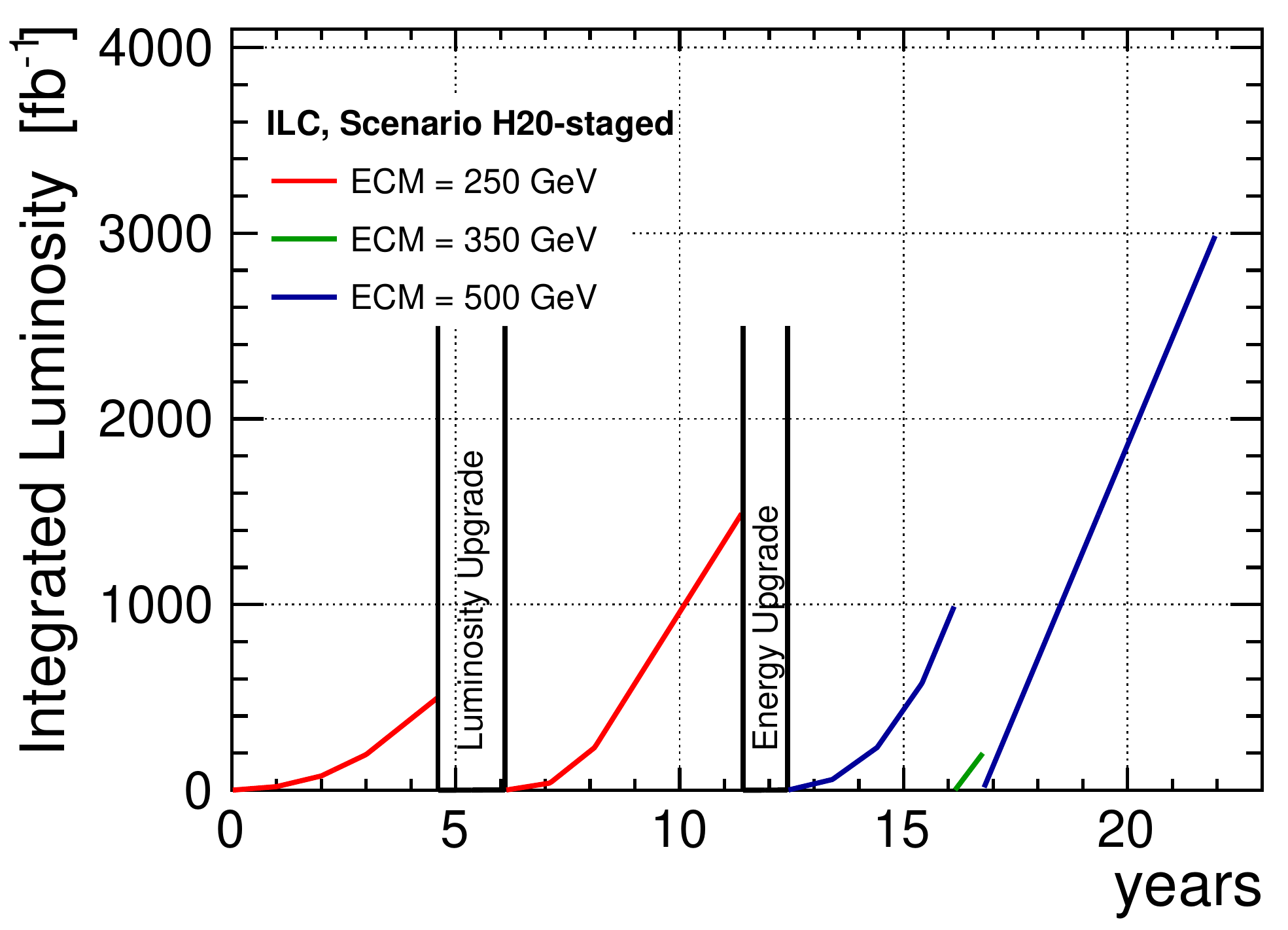}
\end{center}
\caption{The plan for the operation of the ILC through its various stages from 250~GeV to 500~GeV that is used in this report for projections of the physics results expected from the ILC. 
The details of this program most relevant for physics studies are shown in Table~\ref{tab:ILC-staging}.   The detailed accelerator parameters for each stage are given in 
Table~\ref{tab:ilc-params}.   The total length of the program is 22~years.   Additional stages at the $Z$ boson resonance and at 1~TeV could be added to this plan.   Parameters for these programs are also presented in Table~\ref{tab:ILC-staging}.}
\label{fig:ILC-staging}
\end{figure}
%%%%%%%%%%%%%%%%%%%%%%%%%%%%%%%%%%%%%%%%%%%%%%%%

\section{Energy and luminosity}
\label{sec:EvsL}

As we have discussed already in Sec.~\ref{sec:acc-staging}, the ILC is designed to be upgraded, in stages, in energy and luminosity.  Our current plan for the energy and luminosity evolution of the ILC is shown in Fig.~\ref{fig:ILC-staging}.    The parameters of the successive stages that are most important for understanding the physics studies are shown in Table~\ref{tab:ILC-staging}.  

%%%%%%%%%%%%%%%%%%%%%%%%%%%%%%%%%%%%%%%%%%%%%%%%%%%%%%
\begin{table}
\begin{center}
\begin{tabular}{lccccc}
                                     &  91 GeV &     250 GeV  &   350 GeV   &   500 GeV &   1000 GeV \\ \hline
$\int  \L$ (ab$^{-1}$)     &      0.1      &    2           &        0.2       &     4         &      8    \\     
duration (yr)                  &        1.5         &       11          &    0.75                &     9           &      10      \\ 
beam polarization ($e^-/e^+$; \%)   &   80/30  &  80/30    &  80/30  &   80/30  &  80/20  \\
(--, -+, +-, ++)  (\%)          &  (10,40,40,10) & (5,45,45,5) &  (5,68,22,5) &  (10,40,40,10) 
               & (10,40,40,10) \\
$\delta_{ISR}$ (\%)       &    10.8      &    11.7              &         12.0               &      12.4             &                13.0                \\ 
$\delta_{BS}$  (\%)       &   0.16    &   2.6            &       1.9                 &       4.5            &      10.5          
\end{tabular}
\end{center}
\caption{Parameters of the ILC stages most relevant for physics studies. The values given here are those 
actually used for the results to be quoted in this report.   The fourth line 
gives the fraction of the total running time spent in each of the four possible beam polarization 
orientations.  The fifth and sixth lines give the average energy loss in the electron or positron energy spectrum due to initial state radiation
and beamstrahlung, respectively. }
\label{tab:ILC-staging}
\end{table}
%%%%%%%%%%%%%%%%%%%%%%%%%%%%%%%%%%%%%%%%%%%%%%%%%%%%%%%%%%%%% 

The last lines of the table show the average fractional energy loss to initial state radiation and 
beamstrahlung.   These values reflect the long tails to low energy; for most accepted events, the loss is a few percent.

%  ref in the text to the figure 
% below  (?)

%%%%%%%%%%%%%%%%%%%%%%%%%%%%%%%%%%%%%%%%%%%%%%%%%%%%%%%
%\begin{figure}
%\begin{center}
% \includegraphics[width=0.90\hsize]{chapters/gen-phys/figures/eglumis.pdf}
% \end{center}
%\caption{Luminosity spectra for $e^+e^-$, $e\gamma$, and $\gamma\gamma$ annihilation 
%reactions for each of the designs in Table~\ref{tab:ILC-staging}.    The $x$ axis in each graph is the effective 
%CM energy as a fraction of the nominal CM energy.}
%\label{fig:eglumis}
%\end{figure}
%%%%%%%%%%%%%%%%%%%%%%%%%%%%%%%%%%%%%%%%%%%%%%%%%%%%%%%%%%%%% 

The ILC will begin with collisions at 250~GeV and a 
modest design luminosity of  $1.35\times 10^{34}$~cm$^{-2}$sec$^{-1}$.    This luminosity would then be 
doubled by doubling the number of accelerated bunches per RF pulse, an upgrade that only requires the addition of RF power.  
In about 11 years, the ILC will have accumulated a total integrated luminosity of  2~ab$^{-1}$.
This will be followed by an energy upgrade, which entails lengthening the linear accelerators to double their energy.   We
assume the current ILC accelerator parameters for this upgrade, but this 
will be less costly if higher-gradient superconducting RF cavities are available at that time.   The 
500~GeV stage will accumulate 4~ab$^{-1}$ of integrated luminosity at 500~GeV, with also 200~fb$^{-1}$ of luminosity near 350~GeV to measure the top quark mass to the level of
the theoretical systematic errors.

Two additional runs could be added to this plan.  The first is a run at the $Z$ boson resonance, accumulating about $5\times 10^9$ $Z$ bosons.   The accelerator parameters 
for this ``Giga-Z'' program have been discussed in Sec.~\ref{subsec:Zpole_accelerator}.
If this run is done after the installation of RF for the 250~GeV luminosity upgrade, the $Z$ program would take about 1.5 years.    The second is a run at 1~TeV, requiring a second lengthening of the 
linear accelerators.  Since the luminosity of a linear collider naturally increases roughly linearly with 
the center  of mass energy, we expect that the 1~TeV operator will accumulate 8~ab$^{-1}$ of
integrated luminosity in a 10-year program.

All of this data-taking will benefit from the expected high degree of polarization of the electron beam and the planned polarization of the positron beam.    At  1~TeV, where the main object of study will be the $WW$ fusion reactions, this benefit can be enhanced by chosing to operate predominantly with 
$e^-_Le^+_R$-polarized beams.

This plan for the evolution of the ILC is based on a detailed optimization study based on the physics 
goals, carried out in 2015~\cite{Barklow:2015tja}.   Of course, the plan can be revised according to new developments in particle physics and superconducting RF technology.

\section{Beam polarization}
\label{sec:polarization}

The ILC design includes beam polarization for both electron and positrons.  Beam polarization can be used in experiments in a number of different ways.   In this section, we will summarize these and provide the polarization measures appropriate to each case.   More details can be found
in the individual analysis described below.

The ILC design calls for electron polarization of 80\% and positron polarization of 30\%.  Both signs of the polarization will be available in each run.  It is important that the polarization be flipped as often as possible to cancel systematic errors from drifts of detector parameters.   For both beams, the polarization can be flipped pseudo-randomly bunch train by bunch train.  For the electron beam, this is done at the source by controlling the polarization of the laser used to stimulate electron emission from the cathode. This is similar to the scheme used in the SLD experiment~\cite{Woods:1996ph} and now applied with very high rate polarization flipping in the JLab program~\cite{MOLLER:2014iki,CREX:2021dix}.  For the positron beam, the 
polarization is flipped by spin rotators 
%appropriately 
placed  downstream of the helical undulator~\cite{Malysheva:2016jdr}.

The polarization of a beam containing $N_L$ left- and $N_R$ right-handed particles is given by 
\beq
                 P =    \frac{N_R - N_L}{N_L + N_R}
\eeqn
Then a beam of polarization $P$ contains the fractions of particles of each helicity
\beq
             f_L = \frac{1 - P}{2}           \qquad     f_R = \frac{1 +  P}{2} \ .
\eeqn
For beams that contain dominantly  $e_L^-$ over $e_R^-$,  $P$ is negative and therefore $f_L$ is larger than $f_R$.
The ILC will have four different possible polarization configurations.  We will refer to the one with $-80\%$ electron polarization and $+30\%$ positron polarization as $-80/{+30}$ and the other configurations similarly as $-80/{-30}$, $+80/{+30}$, and $+80/{-30}$.   For the $-80/{+30}$ beam configuration, the content in terms of the electron and positron helicity states is
\beq
         f_L(e^-) = 90\%  \quad   f_R(e^-) = 10\%  \quad  ; \quad
         f_R(e^+) =  65\%  \quad     f_L(e^+) = 35\%  \   ,
\eeqn
so the collisions are dominantly from the $e^-_Le^+_R$ initial state.
Since the $e^-_R$ and $e^-_L$ have different $SU(2)\times U(1)$ quantum numbers, each of the four polarization settings is effectively a different 
scattering experiment.   The results of the four experiments can be combined in various ways for different purposes.   We describe four of these 
here.

{\bf Cross section asymmetries:}
Because of helicity conservation in vector boson couplings, $\ee$ annihilation reactions proceed only from the $e^-_Le^+_R$ and $e^-_Re^+_L$ initial helicity 
combinations.   Typically, in annihilation to fermions, the first cross section
is larger than the second by about a factor of 2.  (Specifically for $\ee\to ZH$, the $e^-_Le^+_R$ cross section is larger by a factor  1.4.).  
If we write the two cross sections for 100\% polarized initial states as
\beq
        \sigma = \sigma_0 (1 \pm   {\cal A} )
\eeqn
with $+$ for pure $e^-_Le^+_R$ and $-$ for pure $e^-_Re^+_L$,
then the cross section for electron and positron polarizations  $P_{e-}$ and $P_{e+}$ is 
\beqa
   \sigma(P_{e-}, P_{e+}) &= &   f_L(e^-) f_R(e^+) \ \sigma_0 (1 +{\cal A} ) +  f_R(e^-) f_L(e^+)\ \sigma_0 (1 -  {\cal A} )  \CR
   &=&   \biggl(\frac{ 1  - P_{e-} P_{e+} } {2}\biggr) \sigma_0  - \biggl(\frac{P_{e-} - P_{e+} }{2}
   \biggr)\sigma_0 {\cal A}  \ .
\eeqa{CSpol}
The asymmetry $A$ between the cross sections with $-+$ and $+-$ polarized beams is then
\beq
   A  =   - P_{eff}  {\cal A} \ , 
\eeqn 
with 
\beq
   P_{eff} =  \frac{ P_{e-}- P_{e+}}{1  - P_{e-} P_{e+}  }   =  \mp 89\% 
\eeq{CSasym}
for the $-80/{+30}$ and $+80/{-30}$  beam configurations, respectively, at the ILC.   For this measurement of the intrinsic polarization asymmetry, many sources of systematic uncertainty cancel out, including the absolute luminosity and the absolute detector acceptance.   It is necessary that the detector performance be the same for left- and right-handed
beams, which is insured if the polarization is flipped rapidly. 
The measurement of ${\cal A}$ does depend strongly on the absolute knowledge of the polarization. As will be discussed in more detail in Sec.~\ref{subsec:pol_prec}, the Compton polarimeters up- and downstream of the IP will monitor the time variations, while the absolute scale of the long-term average polarization values will be directly determined from $e^+e^-$ collision data. Thus, the lion share of the uncertainty will decrease with the size of the data set, down to a floor which is given by the remaining uncorrectable point-to-point fluctuations. These are estimated to be between one and two orders of magnitude smaller than the total polarimeter uncertainty, thus at the level of $10^{-4}$.

It is clear already from Eqn.~\leqn{FBQnos} that the polarization asymmetry ${\cal A}$ gives direct information on the quantum numbers of the particles participating in an $\ee$ annihilation reaction. As we will see in Sec.~\ref{sec:SMEFTexpectations}, the polarization asymmetry in the reaction $\ee\to ZH$ also plays an outsize role in the global analysis using Effective Field Theory that determines the Higgs boson couplings.  It is then remarkable that this quantity can be obtained so precisely using polarized beams.

{\bf Cross section enhancements:}
Another result of Eqn. \leqn{CSpol} is that, if the physics of a process very much favors the $e^-_Le^+_R$ helicity state, beam polarization gives an 
enhancement of the effective luminosity.   For $WW$ fusion reactions, which appear only from the  $e^-_Le^+_R$ initial state, the effective luminosity
for $-80/+30$  polarized beams is enhanced from that for unpolarized beams by the factor
\beq  
    {\cal L}_{eff}/{\cal L}   =  (1 + P_e)(1-P_p)  =   2.3\  .
\eeq{lumienhance}

In practice, one should not try to achieve the full promised luminosity enhancement.  Each physics process has its own dependence on polarization, and it is also important to reserve some of the luminosity for data on the modes with smaller production cross sections.  This is reflected in our choice of the 
division of polarization modes in line 4 of Table~\ref{tab:ILC-staging}.   The run at the top quark threshold has a quite specific goal, and running with mainly $-80/+30$ polarized beams is optimal for this.  At other energies, where the physics program is more general, the fraction of polarizations used should be optimized taking into account also the uses that we will describe next.

{\bf Background reduction:}
Especially for reactions that include neutrinos or other sources of missing energy, the process $\ee \to W^+W^-$ is a the dominant source of background.   It is therefore important that the cross section for $\ee\to W^+W^-$ is dramatically reduced in the $+80/-30$ initial state. Backgrounds from photon-induced processes such as $\gamma\gamma\to \ell^+\ell^-$  and single $W$ production are still present in the $-80/-30$ and 
$+80/+30$ samples, while the annihilation reactions are highly suppressed.   This ability to reduce some relevant backgrounds and to directly measure others can be crucial in measuring the rates of these processes precisely or, in the case of particle searches, establishing strong limits.

\begin{figure}
\begin{center}
\includegraphics[width=0.48\hsize]{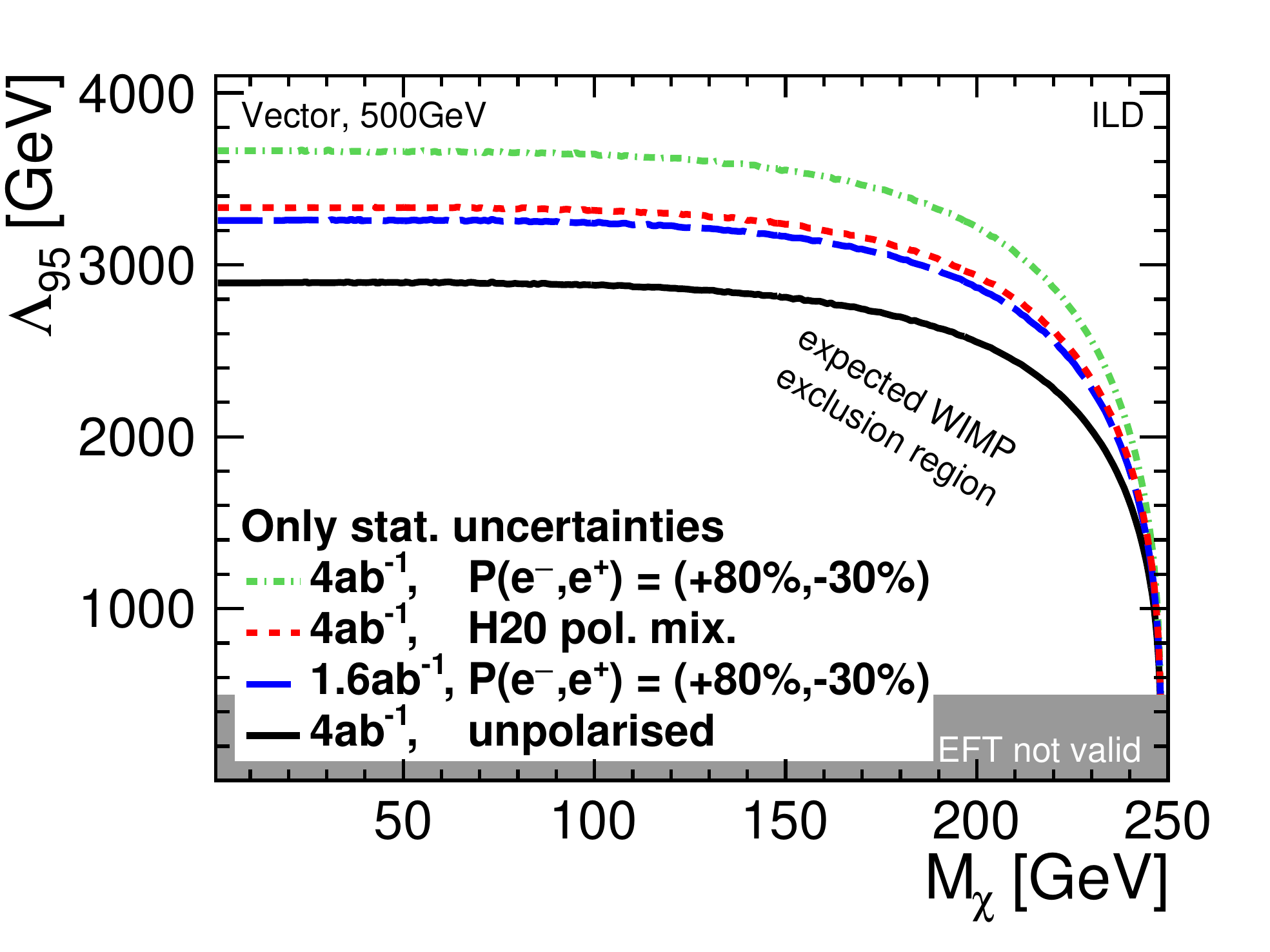}\ \ 
\includegraphics[width=0.48\hsize]{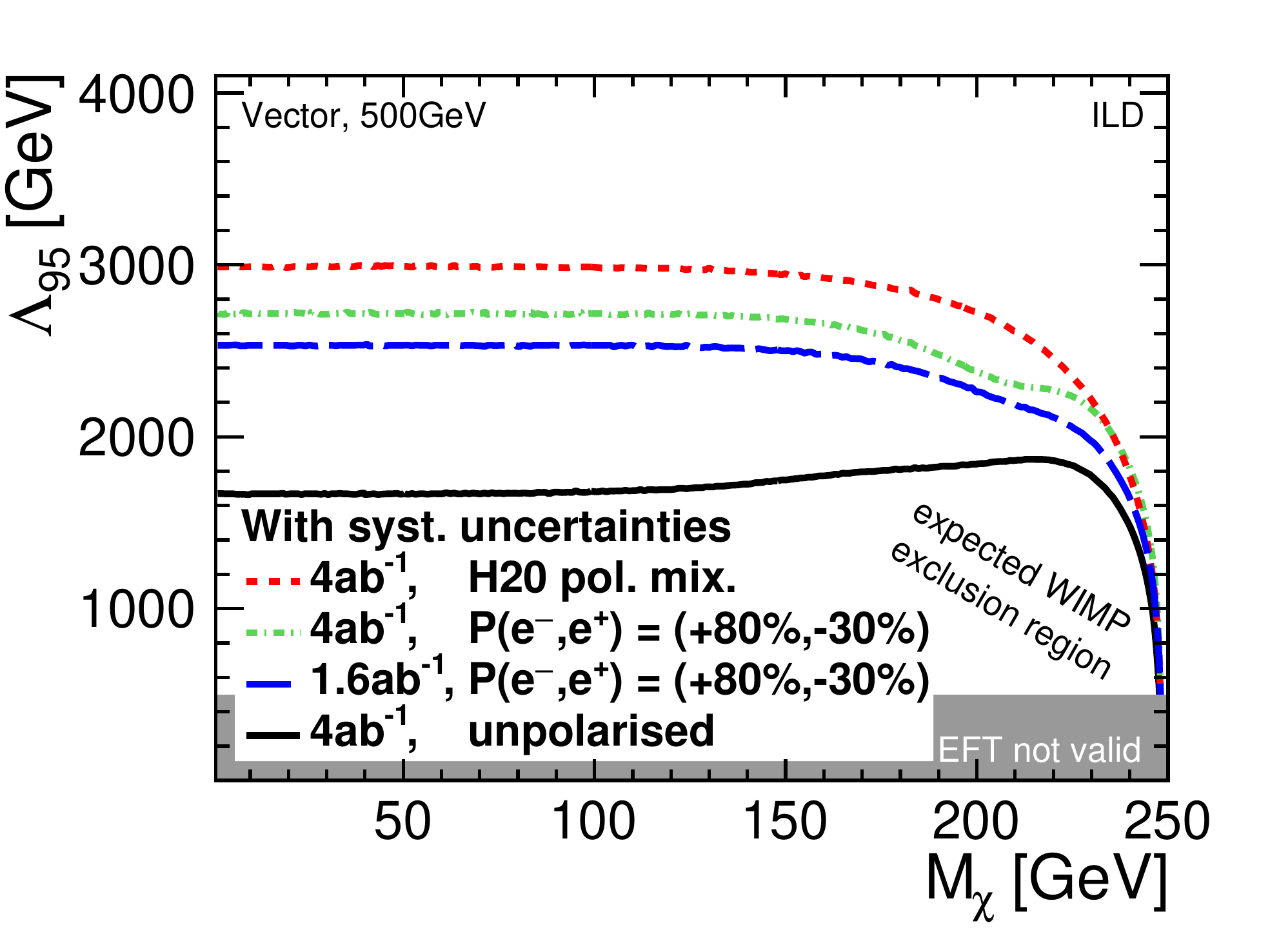}
\end{center}
\caption{Limits on an Effective Field Theory mass scale associated with dark matter particle pair production in $\ee$ annihilation from a search simulation at 500~GeV, showing the dependence of the results on beam polarization, from \cite{Habermehl:2018yul}.  Higher limits are better.   In both figures, the black curve is an analysis for unpolarized beams, the red (short-dash) curve corresponds to the mixture of  polarization states  in Table~\ref{tab:ILC-staging}.    Left: analysis with 
statistical errors only; Right:  analysis including both statistical and systematic errors.}
\label{fig:DMpol}
\end{figure}

This is illustrated for a search for dark matter pair production that will be described in Sec.~\ref{sec:newparticles}.   Fig.~\ref{fig:DMpol} shows the results of a simulated search for dark matter pair production $\ee\to \chi\chi$  at
 500~GeV~\cite{Habermehl:2018yul}.  The analysis assumes no signal and puts a lower limit on an Effective Field Theory mass scale $\Lambda$.  What concerns us now
 is the left-hand plot,  which  includes statistical errors only.   The red (short-dash) curve shows the limit from the 
mixture of polarization states in Table~\ref{tab:ILC-staging}.  The figure shows 
 that almost all of the exclusion comes from the 40\% of the run that is collected with the $+80/-30$ beam configuration.
However, there is a second half to this story, which is explained below.

%%%%%%%%%%%%%%%%
%%%%%%%%%%%%%%%%%%%%%%%%%%%%%%%%%%%%%%%%%%%%%%%%

{\bf Control of systematic uncertainties:}
With its four configurations for the polarization of the electron and positron beams, the ILC will be  carrying out four  different experiments simultaneously.  These four data samples have very different mixes of physics processes, with $\ee$ annihilation reactions essentially missing from the $-80/-30$ and $+80/+30$ samples while non-annihilation processes remain.  However, with rapid polarization flipping, the experiments will be done in the same detector.  This allows nuisance parameters associated with detector acceptance and energy response to measured by comparison of the different samples.  The potential 
systematic uncertainties associated with these parameters can thus be greatly reduced.

As an example of an application of this strategy, look now at the right-hand plot in Fig.~\ref{fig:DMpol}.   Nominally, uncertainties from knowledge of the detector would weaken the observed limits, and this effect is visible in the black curve 
giving the result for unpolarized beams and in the curves for individual polarization states.  However, the use of a  mix of 
polarization states, including the nominally unproductive helicity-violating configurations, can be used to evaluate these 
uncertain detector  parameters and retain most of the power of the analysis that included only statistical uncertainties.

In processes subject to smaller systematic uncertainties, an advantage from the enhancement of cross
sections is often compensated by the loss of productive luminosity in measuring the samples with
helicity-violating polarization configurations.  However, the use of positron polarization leads to double the number of individual data sets, adding redundancy and cross-checks.  The goal of the ILC is to 
demonstrate that the physics of $\ee$ annihilation differs from the predictions of the Standard Model.  These checks could prove 
essential in making that case.

\section{Control of luminosity, beam energy and polarization}
\label{sec:lep_precision}

In this section, we will discuss the control of the top-level accelerator parameters---luminosity, beam energy and polarization. 

\subsection{Luminosity measurement}
\label{subsec:lumi_prec}
The luminosity will be measured via low-angle Bhabha scattering. The forward regions of the ILC detectors will be  equipped with dedicated luminosity calorimeters, described in Sections~\ref{sec:ILD-forward} and~\ref{subsub:det:forward}. Recent test beam results obtained with prototype detectors~\cite{Abramowicz:2018vwb} underpin the assumptions on the hardware made in previous detailed simulation studies, which show that the luminosity can be measured to  $2.6\times 10^{-3}$ and    $1.6\times 10^{-3}$ at center-of-mass energies of  $500$ and 1000 GeV, respectively~\cite{BozovicJelisavcic:2013lni}. The energy calibration of the calorimeter contributes to the total uncertainty at the level of $1\times 10^{-3}$.  The largest contribution, however, originates from the residual physics backgrounds from $e^+e^- \to e^+e^-e^+e^-$ and $e^+e^- \to e^+e^-q\bar{q}$, which in these studies are not corrected for but taken as full-scale contribution to the error budget.
Other substantial contributions to the total uncertainty arise from beamstrahlung and ISR, as well as from the electromagnetic deflections of the Bhabhas while they traverse the colliding bunches. It is assumed that these effects can be corrected from simulations. The above numbers include the uncertainty corresponding to a knowledge of the bunch parameters at the 20\% level. As will be discussed below, this is a rather conservative assumption. Without the these simulation-dependent corrections, the numbers above would increase by about $50\%$. At $\sqrt{s}=250$\,GeV, the (uncorrected) effect of the electromagnetic deflection is about twice as large as at $500$\,GeV, but shrinks to half a per-mil with the simulation-based correction~\cite{Bozovic-Jelisavcic:2014aza}. From the results of this paper and~\cite{BozovicJelisavcic:2013lni}, we estimate the luminosity uncertainty at $250$\,GeV to be less than $4\times 10^{-3}$.

The in-situ determination of the properties of the colliding bunches will be performed by the so-called beam calorimeters (BeamCals); see Secs.~\ref{sec:ILD-forward} and~\ref{subsub:det:forward}. Beyond their role as veto taggers for high-energy electrons, the BeamCals record the energy depositions of the enormous numbers $e^+e^-$ pairs created by beamstrahlung. From the pattern of these energy depositions, the relevant parameters of the colliding bunches can be reconstructed, e.g.\ in terms of number of particles, horizontal and vertical emittances, beam positions and bunch sizes at the IP~\cite{Grah:2008zz}. For most parameters, precisions of 10\% are reached for each bunch crossing in single-parameter analyses. A full multi-parameter analysis would profit from additional information, e.g.\ from an optional GamCal further down-stream or external constraints on the emittances or bunch charges. The precisions achieved in~\cite{Grah:2008zz} have been propagated to the luminosity spectrum via GuineaPig~\cite{Schulte:1999tx} in the context of a mono-photon WIMP search~\cite{Habermehl:2020njb}. This results in precisions of better than 10\% per bin from 200 bunch crossings, showing a clear potential to monitor time-dependences within a bunch train.  
The long-term average luminosity spectrum can be determined also from Bhabha scattering. This approach has been pioneered for CLIC in~\cite{Poss:2013oea}, reaching a few percent per bin even in the more challenging beam conditions at CLIC. Combining these complementary online and offline methods for monitoring the luminosity spectrum together with accelerator instrumentation data, e.g.\ from the downstream energy spectrometer, will be an interesting study to pursue in the future.

\subsection{Beam energy measurement}
\label{subsec:ecm_prec}
The beam energy will be monitored upstream and downstream of the $e^+e^-$ interaction point by complementary techniques~\cite{Boogert:2009ir}: the upstream energy spectrometer measures the deflection of the beam in a magnetic chicane with high-resolution beam position monitors, while the downstream spectrometer detects the synchrotron radiation emitted in a chicane in the extraction beamline. Both systems are designed to reach relative precisions of $10^{-4}$ ($100$\,ppm). The downstream spectrometer can provide information on the beam energy spectrum, complementing the method for beam energy determination described in the previous subsection. Both systems are part of the beam delivery system and the machine-detector interface described in Sec.~\ref{subsubsec:bds_mdi}.

The long-term average center-of-mass energy can be controlled to at least one order of magnitude better 
from $e^+e^- \to \mu^+\mu^- (\gamma)$ events. As detailed in~\cite{Wilson:2020arh,Madison:2022spc}, the excellent momentum resolution of the detectors proposed for the ILC combined with a calibration to the $J/\psi$ mass will allow a calibration of the center-of-mass energy to a few ppm for the $Z$ pole of the ILC, and to $10$\,ppm at higher $\sqrt{s}$.

\subsection{Polarization measurement}
\label{subsec:pol_prec}

The beam delivery system and the extraction line of the ILC are equipped with laser-Compton polarimeters, providing instantaneous measurements of the beam polarizations at about $1800$\,m upstream and $150$\,m downstream of the main $e^+e^-$ interaction point. The expected polarimeter precision of $\delta P /P$ = 0.25\% was presented in Sec.~\ref{subsubsec:bds_mdi}. The decisive polarization values for physics analyses are, however, the luminosity-weighted long-term average values at the interaction point. 
These will be obtained by combining three main ingredients: the time-resolved polarimeter measurements~\cite{Vormwald:2015hla}, simulations of the spin transport along the beam delivery system and the extraction line and the beam-beam interaction~\cite{Beckmann:2014mka}, and a long-term reference scale obtained from the $e^+e^-$ data themselves, predominantly employing physics processes with large left-right asymmetries like $W$ pair production~\cite{Marchesini:2011aka, Karl:2019hes, List:2020wns}.  However, we will argue that for ultimate precision, the polarization values should be included as free (or somewhat pre-constrained) extra parameters in the actual extraction of physics parameters, which will then directly determine the \emph{residual impact} on the physics parameters as well as any possible correlations.

\begin{figure}[htb]
\begin{center}
\begin{subfigure}{0.99\textwidth}
\centering
 \includegraphics[width=0.99\textwidth]{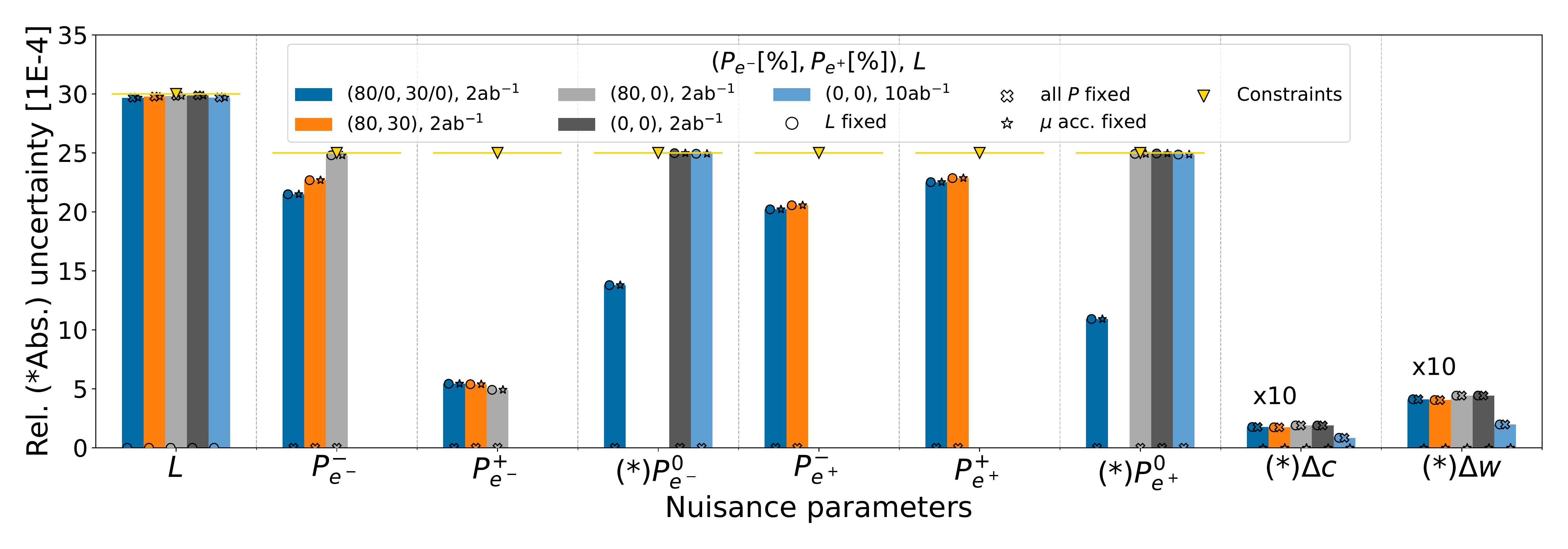}
\end{subfigure}
\begin{subfigure}{0.99\textwidth}
\centering
 \includegraphics[width=0.8\textwidth]{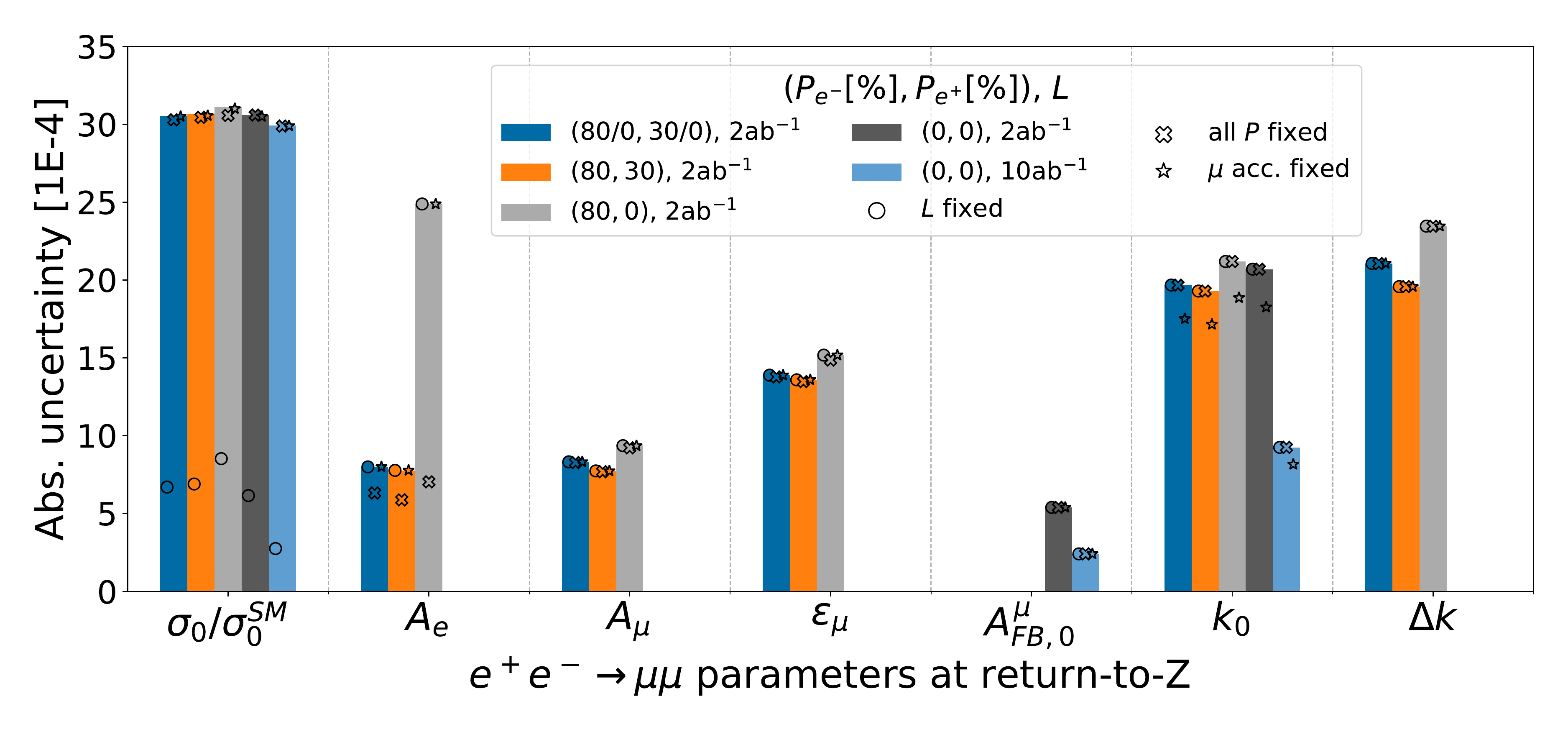}
\end{subfigure}
\end{center}
\caption{Precision to which (a) nuisance parameters, including the beam polarizations, and (b) 2-fermion physics parameters are determined from the combined $e^+e^-\to \mu^+ \mu^-$ and $e^+e^-\to \mu^{\pm} \nu_{\mu} jj$ fit under different assumptions on the integrated luminosity and the beam polarizations. The yellow triangles with horizontal lines indicate the level of external constraints from the luminosity measurement and the polarimeters.}
 \label{fig:nuis}
\end{figure}

Such an approach has been pioneered in a recent study~\cite{Beyer:2022xyz,Beyer:2022nyr}, which employs observables from 2-fermion and 4-fermion processes at $\sqrt{s}=250$\,GeV to constrain the relevant physics parameters (cross-sections, asymmetries, anomalous triple gauge couplings etc) as well as the beam polarizations and parameters modeling the detector acceptance. For the 2-fermion distributions, three additional free parameters allow us to describe deviations from the tree-level helicity amplitudes due to ISR/beamstrahlung ($k_0$ and $\Delta k$ in Fig.~\ref{fig:nuis}) and $\gamma$-exchange ($\epsilon_{\mu}$). The study considers various running scenarios of a generic $e^+e^-$ Higgs factory. The results are shown in Fig.~\ref{fig:nuis}. The orange bars correspond to the standard ILC configuration with both beams polarized and an integrated luminosity of $2$\,ab$^{-1}$. It is still a prototype analysis, and as such includes so far only the channels $e^+e^-\to \mu^+ \mu^-$ and $e^+e^-\to \mu^{\pm} \nu_{\mu} jj$, thus the absolute precisions on physics observables should not be taken from Fig.~\ref{fig:nuis}, but rather from the relevant sections in Chapters~\ref{ILC250}, \ref{chap:PEW}, and~\ref{chap:ILC500}. Extrapolating these results to include all relevant final states indicates that the control of all four polarisation values (for electron/positron beams with positive/negative signs) at the level of a few $10^{-4}$ should be feasible. 

The precision of such extractions of the average polarization values from collision data will increase with the size of the data set, thus in particular at the $Z$ pole, the precision will improve further, even without $WW$ processes being available~\cite{Beyer:2022xyz}. The final systematic limitation will be the residual point-to-point uncertainties of the polarimeter measurements, which limit the ability to correct for time variations of the polarisations based on the polarimeters. As discussed in Sec.~\ref{subsubsec:bds_mdi}, these uncorrectable point-to-point uncertainties are expected to be between one and two orders of magnitude smaller than the total polarimeter uncertainty, thus at the level of $10^{-4}$. 

It should be noted that, for ultimate precision physics and a reliable discrimination of genuine deviations from the SM expectations from instrumental effects, any residual polarization in a nominally unpolarized beam, denoted by $P^0_{e^{\pm}}$ in the figure, must also be included in such global interpretations. This has been realised more than 20 years ago~\cite{Rowson:2000qn} based on SLD experience, and has been revisited more recently in~\cite{Karl:2019hes} for the ILC and in~\cite{Wilkinson:2022epol} for the FCCee. As Fig.~\ref{fig:nuis}(a) shows, these residual polarizations are basically impossible to constrain from collision data beyond the information provided from polarimeters (yellow markers/lines), unless at least part of the data is taken with non-zero beam polarizations (dark blue bars). The corresponding effect on the 2-fermion physics observables is seen in Fig.~\ref{fig:nuis}(b), most strikingly in the column showing the initial state asymmetry $A_e$. If the data are taken with both beams polarized, there is only a tiny residual impact of the finite knowledge on the beam polarizations (the difference between the open cross symbols and the full bars for the orange/dark blue cases), while it is huge in case of electron polarization only (light grey bar). In case of no beam polarisation at all, the set of independent physics observables is reduced, and in particular the initial and final state asymmetries as well as the $\epsilon$ parameter which accounts for $Z-\gamma$-interference effects collapse into one parameter ($A_{FB}^{\mu}$). Thus, the ability to polarize both beams adds extra physics observables and reduces the impact of systematic uncertainties, as discussed in the previous section.

\chapter{ILC Detectors} 
\label{chap:detectors}

\section{Detector requirements for the physics program} 
\label{sec:requirements}

The ILC accelerator design allows for one interaction region, equipped
for two experiments. The two experiments are swapped into the
Interaction Point within the so-called ``push-pull" scheme. The
experiments have been designed to allow fast move-in and move-out from
the interaction region, on a timescale of a few hours to a day. In
2008 a call for letters of intent was issued to the
community. Following a detailed review by an international detector
advisory group, two experiments were selected in 2009 and invited to
prepare more detailed proposals.  These  are the SiD detector and the ILD
detector described in this section. Both prepared detailed and costed proposals which were
scrutinised by the international advisory group and included in the
2013 ILC
Technical Design Report~\cite{Behnke:2013lya}.  These specific detector designs have been critical input to the design of the ILC itself. A future process is expected in which detector designs will be reconsidered, with optimisations of these two designs and alternative designs which are proposed. In this chapter the two TDR detector proposals are
described.

The ILC detectors are designed to make precision measurements on the Higgs boson,  $W$- and $Z$-boson, the top quark and other particles. The detector performance requirements are more ambitious than in the LHC experiments, as the experimental conditions are naturally very much more benign and because the detector collaborations have developed technologies specifically to take advantage of these more forgiving conditions. In particular, an $\ee$ collider provides much lower collision rates and events of much lower complexity than a hadron collider, 
and detectors can be adapted to take advantage of this. The radiation levels at the ILC are equivalent to approximately $10^{11} n/cm^2/year$ of NIEL (Non Ionising Energy Loss) dose, very modest compared with the LHC, where NIEL doses of up $10^{16} n/cm^2$ are accumulated over the lifetime of the innermost tracking elements. One exception is the special forward calorimeter system very close to the
beamline, where radiation exposure will be an issue.

The stringent requirement on the momentum resolution for charged particle tracks is driven by the need to  precisely reconstruct the $Z$-boson mass in the Higgs recoil analysis. This requirement translates into an asymptotic momentum resolution for high-momentum tracks that is nearly an order of magnitude better than achieved in the LHC experiments. It moreover requires that the detector material be kept to a minimum, to maintain excellent momentum resolution also for lower-momentum tracks.

The identification of jets that originate from the fragmentation of bottom and charm quarks, known as flavour tagging, plays an important role in the scientific program of the ILC. An excellent separation of bottom and charm jets from each other and from gluon jets is crucial for the measurement of the Higgs couplings. This requires a vertex detector with a much improved performance in comparison to the pixel detectors installed at the previous generation of electron-positron colliders and the LHC. The relatively low radiation levels are particularly relevant for the design of the innermost vertex detector elements that can be located very close to the beam.

At the same time, although they are studying electroweak particle
production, it is essential that the ILC detectors have excellent
performance for jets.   At an $\ee$ collider, $W$ and $Z$ bosons are
readily observed in their hadronic decay modes, and the study of these
modes plays a major role in most analyses.    To meet the requirements
of precision measurements, the ILC detectors are optimized from the
beginning to enable jet reconstruction and measurement using 
the particle-flow algorithm (PFA). This drives the goal of $3\%$
 jet mass resolution at energies above 100 GeV,  a resolution
about twice as good as has been achieved in  the LHC
 experiments.

Finally, while the LHC detectors depend crucially on multi-level
triggers that filter out only a small fraction of events for analysis,
the  rate of interactions at the ILC is sufficiently low to allow
running without a trigger.     The ILC accelerator design is based on
trains of electron and positron bunches, with a repetition rate of
5~Hz, and with 1312 bunches (and bunch
collisions) per train. 
The 199 ms interval between bunch trains provides ample time for a full
readout of data
 from the  previous train.  While there are background processes arising
 from  beam-beam interactions, the detector occupancies arising from these 
have been shown to be manageable.

The combination of extremely precise tracking, excellent jet mass
resolution, and triggerless running gives the ILC, at 250 GeV and at
higher energies, a superb potential for discovery.

Quantitatively the requirements on the detectors may be summarised by the following points: 
\begin{itemize}
    \item {\bf Impact parameter resolution:}  An impact parameter resolution of 
%    $ 5~\mu \mathrm{m} \oplus 10~\mu \mathrm{m} / [p~[{\mathrm{GeV}/c}]\sin^{3/2}\theta$] 
   $ 5~\mu \mathrm{m} \oplus \frac{10~\mu \mathrm{m} \; \mathrm{GeV}/c}{ p\;\sin^{3/2}\theta}$  
   has been defined as a goal, where $\theta$ is the angle between the particle and the beamline. 
    \item {\bf Momentum resolution:} An inverse momentum resolution of $\Delta (1 / p) = 2 \times 10^{-5}~\mathrm{(GeV/c)}^{-1}$
    % no need for [] in my opinion, we do not put it with \mu m above
    asymptotically at high momenta should be reached. Maintaining excellent tracking efficiency and very good momentum resolution at lower momenta will be achieved by an aggressive design to minimise the detector's material budget.
    \item {\bf Jet energy resolution:} Using the paradigm of particle flow, a jet energy resolution $\Delta E/ E = 3-4\%$ for light flavour jets should be reached. The resolution is defined in reference to light-quark jets, as the R.M.S. of the inner $90\%$ of the energy distribution. 
    \item {\bf Readout:} The detector readout will not use any trigger, ensuring full efficiency for all possible event topologies.  The readout should provide precision signal measurements with high channel granularity and dynamic range.
    \item {\bf Powering} To allow a continuous readout, while also minimizing the amount of inactive material in the detector, the power of major systems will be cycled between bunch trains. 
\end{itemize}

To meet these goals an ambitious R\&D program has
 been pursued for more than a decade to develop and
 demonstrate the needed technologies. The results of this program are
 described in some detail in Ref.~\cite{janstrube_maximtitov_2021}. 
 The two experiments proposed for
 the ILC, SiD and ILD, utilise and 
rely on the results from these R\&D efforts.

Since the goals of SiD and ILD in terms of material budget, tracking
performance, heavy-flavor tagging, and jet energy resolution are very 
demanding, it is important to provide information about the level
of detailed input that enters our performance estimates.   These are
best
discussed together with the event reconstruction and analysis
framework that we will present in Chapter~\ref{chap:sim}.   In that
section, we will present estimates of detector performance as
illustrations at the successive stages of event analysis. 

\section{The ILD Detector} 
\label{ILD}

The International Large Detector, ILD, is a proposal for a multi-purpose detector at the ILC. The design of ILD is the result of more than a decade of work by an international group of scientists and engineers. Throughout this time ILD has profited from and at the same time driven extensive technological developments which make the advanced ILD design possible. 

\begin{figure}[tb]
 \begin{center}
 \begin{tabular}{lr}
 \includegraphics[width=0.48\hsize,clip]{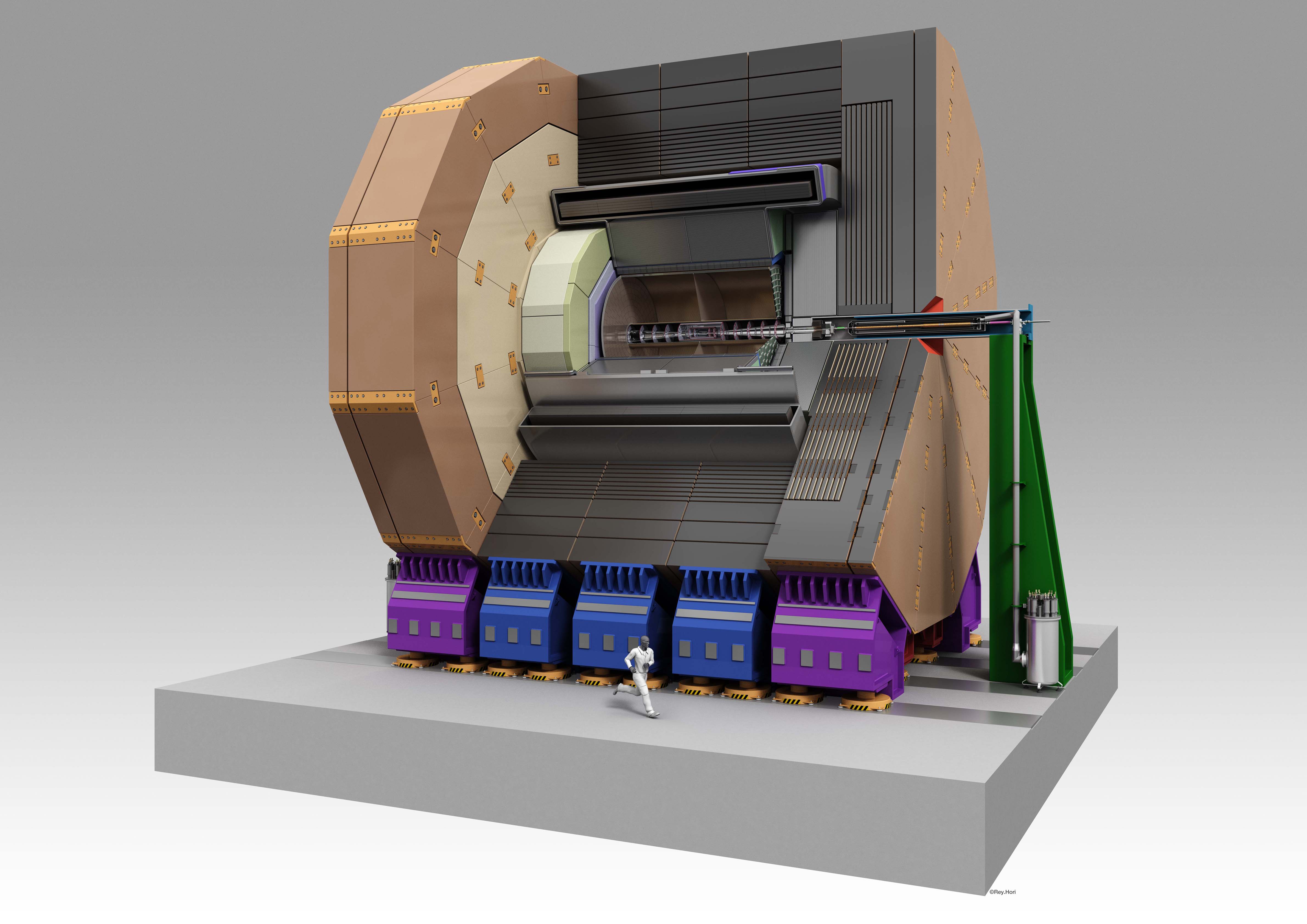} & 
 \includegraphics[width=0.35\hsize]{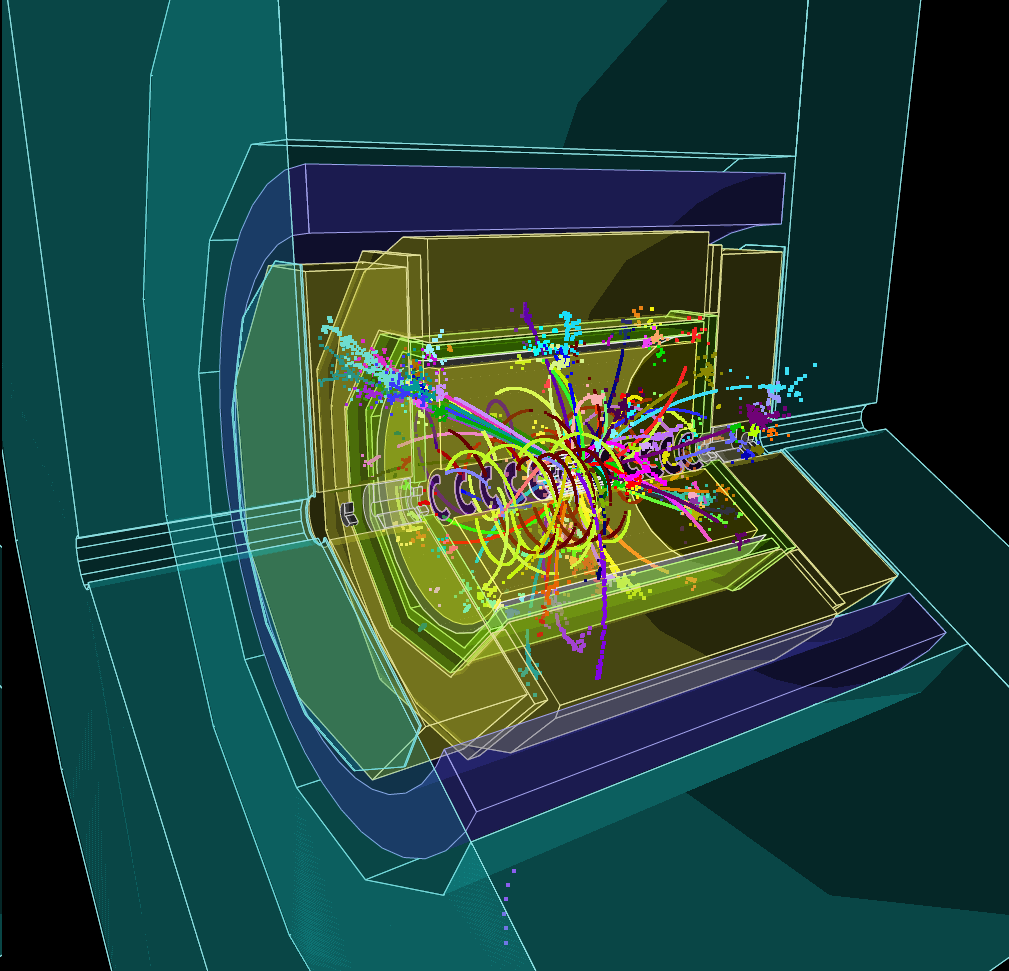}
 \\
 \end{tabular}
\caption{Left: Three-dimensional rendering of the ILD detector. Right: Event display of a simulated hadronic decay of a $t \bar t$ event in ILD. The colors of the tracks show the results of the reconstruction, each color corresponding to a reconstructed particle.
\label{ild-fig-ILD}}
 \end{center}
 \end{figure}

The particle flow concept~\cite{ild:bib:PandoraPFA} plays a central role in the ILD design,
% The ILD concept has been
described in a number of documents. 
The basic concept and its validation were extensivly discussed in the ILD Detector Baseline Document (DBD) in 2013~\cite{Behnke:2013lya}. ILD has recently, in 2020 published an update to the DBD, the Interim Design Report, IDR~\cite{ILDConceptGroup:2020sfq}.
A three-dimensional image of the detector is shown in Figure~\ref{ild-fig-ILD}, together with an event display of a simulated top--anti-top event within it. 
Detailed full-simulation studies~\cite{Behnke:2013lya,ILDConceptGroup:2020sfq} show that the ILD detector concept can reconstruct complex events with unprecedented precision, meeting all the requirements listed in section \ref{sec:requirements} above.

%A quadrant view of the large detector model is shown in Figure \ref{fig:ILD}~(left), together with an event display in this detector of a top event Figure~\ref{fig:ILD}~(right).

% \subsection{Requirements for the ILD detector}
\subsection{Concept of the ILD Detector}
The science which will be done at the ILC has been summarised earlier in this document. It is strongly dominated by the quest for ultimate precision in measurements of the properties of key particles like the Higgs boson, the weak gauge bosons, and, once the center-of-mass energy is beyond its production threshold, the top quark (see for example~\cite{Fujii:2017vwa} or \cite{ILCESU1} for recent summaries). 

%A more detailed description of the ILD detector and its design philosophy is available in \cite{Behnke:2013lya}.

The anticipated precision physics program drives the requirements for the detector. The reasoning resulted in the conceptual design of a particle flow detector have been discussed above. ILD thus has the rather starndard layout of a tracker and a claorimeter all inside a magnetic field, instrumentation down to rather low solid angles, and a powerful muon system surrounding the detector outside of the coil.

ILD is different from in the specific choice which has been made for the central tracker. Here ILD chose a large volume hybrid tracking system, with a silicon tracking system with excellent position resolution, combined with a large gaseous tracker which promises excellent efficiency combined with low material, together with a highly granular calorimeter in both the electromagnetic and hadronic sections. To ease linking between the tracker and the calorimeter, the calorimeter is placed within the solenoid magnet which provides a 3.5~T field. This choice is driven by the need to provide extremely high efficiency tracking over a large momentum range. The low material budget in a gaseous tracker combined with a large number of three-dimensional space points give an excellent performance for a wide range of topologies and energies.

A number of highly relevant physics processes require the precise reconstruction of exclusive final states containing heavy flavour quarks. This translates into the need for very precise reconstruction of the decay vertices of decaying particles, and thus implies a high resolution vertexing system close to the interaction region. 

The ultimate performance of the detector system depends critically on the amount of material in the inner part of the ILD detector. The total material budget in front of the calorimeter should be below 10\% of a radiation length, for the barrel part of the detector acceptance. As a consequence, this requires that the coil be located outside of the calorimeter system. The main parameters of the ILD detector concept are summarised in Table~\ref{tab-ILD-size}.

\begin{table}[t]

    \centering
    \begin{tabular}{llrrrr}
    \hline
    Barrel & Technology & $r_{in}$/mm & $r_{out}$/mm & $z_{max}$/mm &\\
    \hline
    VTX & Silicon pixel & 16& 60 & 125 &\\
    SIT & Silicon pixel & 153& 303& 644&\\
    TPC & Gas & 329 & 1770 & 2350  &\\
    SET & Silicon strip& 1773& 1776 & 2300 & \\
    \hline
    ECAL & Silicon pads& 1805& 2028& 2350 &\\
    HCAL & scintillator or RPC & 2058 & 3345 & 2350 &\\
    Coil & 4 Tesla Solenoid & 3425 & 4175 & 2350 &\\
    Muon & Scintillator& 4450 & 7755 & 4047  &\\
    \hline\hline
    Endcap & Technology & $z_{min}$/mm & $z_{max}$/mm & $r_{in}$/mm & $r_{out}$/mm\\
    \hline
    FTD 1 & Silicon pixel & 220 & 37 & - & 153\\
    FTD 1 & Silicon strip& 645 & 2212 & - & 200\\
    \hline
    ECAL & Silicon pads & 2411 & 2635& 250 & 2096\\
    HCAL & scintillator or RPC & 2650 & 3937 & 350 & 3226\\
    Muon & Scintillator & 4072 & 6712 & 350& 7716 \\
    \hline
    BeamCal & GaAs pads & 3115 & 3315 & 18 & 140 \\
    LumiCal& Silicon pads & 2412 & 2541 & 84& 194\\
    LHCAL & Silicon pads & 2680 & 3160 & 130 & 315\\
    \hline
    \end{tabular}
    \caption{Main parameters of the ILD detector for the barrel and the endcap part.}
    \label{tab-ILD-size}

\end{table}

The whole detector should be operated without a hardware trigger to maximise the sensitivity to new physics signals. This in turn places stringent requirements on the readout electronics, in terms of both speed and power consumption. The integration of ILD is faced with the additional complexity to allow for a rapid movement of the detector in and out of the interaction region, the so-called push-pull scheme~\cite{Parker:2008zza}. 

The ambitious requirements on the performance of the ILC detectors has sparked a broad R\&D program, as described above. ILD has traditionally maintained very close and collaborative relations with these R\&D collaborations. 

The ILD concept from its inception has been open to new technologies. 
No final decision on subdetector technologies has been taken, and in many cases several options are currently under consideration. ILD is actively inviting new groups to join the effort and propose new ideas or improvements to the current concept~\cite{Fujii:2020pxe}. 

In the following paragraphs, the different components of the ILD concept are introduced and discussed. 

\subsection{ILD vertexing system}
The system closest to the interaction region is a pixel detector designed to reconstruct decay vertices of short lived particles with great precision. ILD has chosen a system consisting of three double layers of back-thinned pixel detectors. The innermost layer is only half as long as the others to reduce the exposure to background hits. Each layer will provide a spatial resolution around  {3}~$\mu\mathrm{m}$ at a pitch of about {17}~$\mu\mathrm{m}$, and a timing resolution per layer of around 2--4~$\mu\mathrm{s}${, possibly lower}. Current technological developments will most likely make it possible to resolve single bunch crossing. R\&D is ongoing to explore the option of a significantly better timing resolution. Since the layers are specifically designed with a very low material budget, of close to $0.15\%$ of a radiation length per layer, the vertex detector also serves as an efficient tracker for low momentum tracks, which due to the magnetic field do not reach the inner tracking system.

ILD is exploring several technological options for the vertex detector, and has not yet decided on a baseline. 
Some of the considered technologies are listed below.

Over the last 10 years the CMOS pixel technology has matured close to a point where all the requirements (material budget, readout speed, granularity) needed for an ILC detector can be met. The technology has seen a first large scale use in the STAR vertex detector~\cite{ild:bib:VTXcps3}, and more recently in the upgrade of the ALICE {Inner Tracking System (ITS-2)}~\cite{ALICE:2013nwm}. 

Other technologies under consideration for ILD are DEPFET, which is also currently being deployed in the Belle II vertex detector ~\cite{Luetticke:2017zpx}, fine pitch CCDs ~\cite{fineCCD}, and also less far developed technologies such as SOI (Silicon-on-insulator) and Chronopix~\cite{RDliaision}.

Very light weight support structures have been developed, which bring the goal of 0.15\% of a radiation length per layer within reach.  Structures {that reach 0.21\% X$_0$ in most of the fiducial volume} are now used in the Belle II vertex detector~\cite{PLUME:2011rwc}. 

\begin{figure}
    \centering
    \includegraphics[width=0.3\hsize]{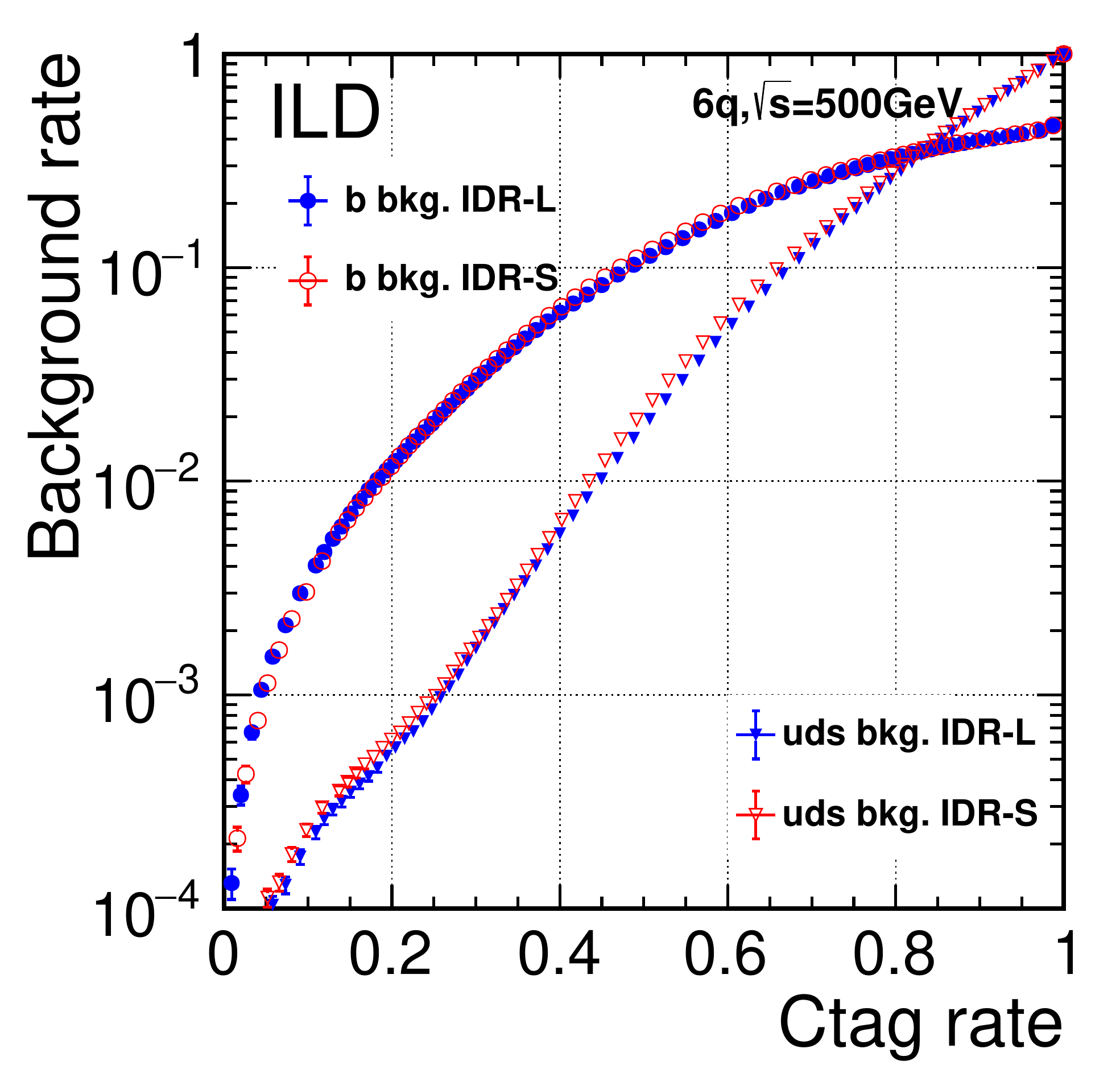}
    \includegraphics[width=0.3\hsize]{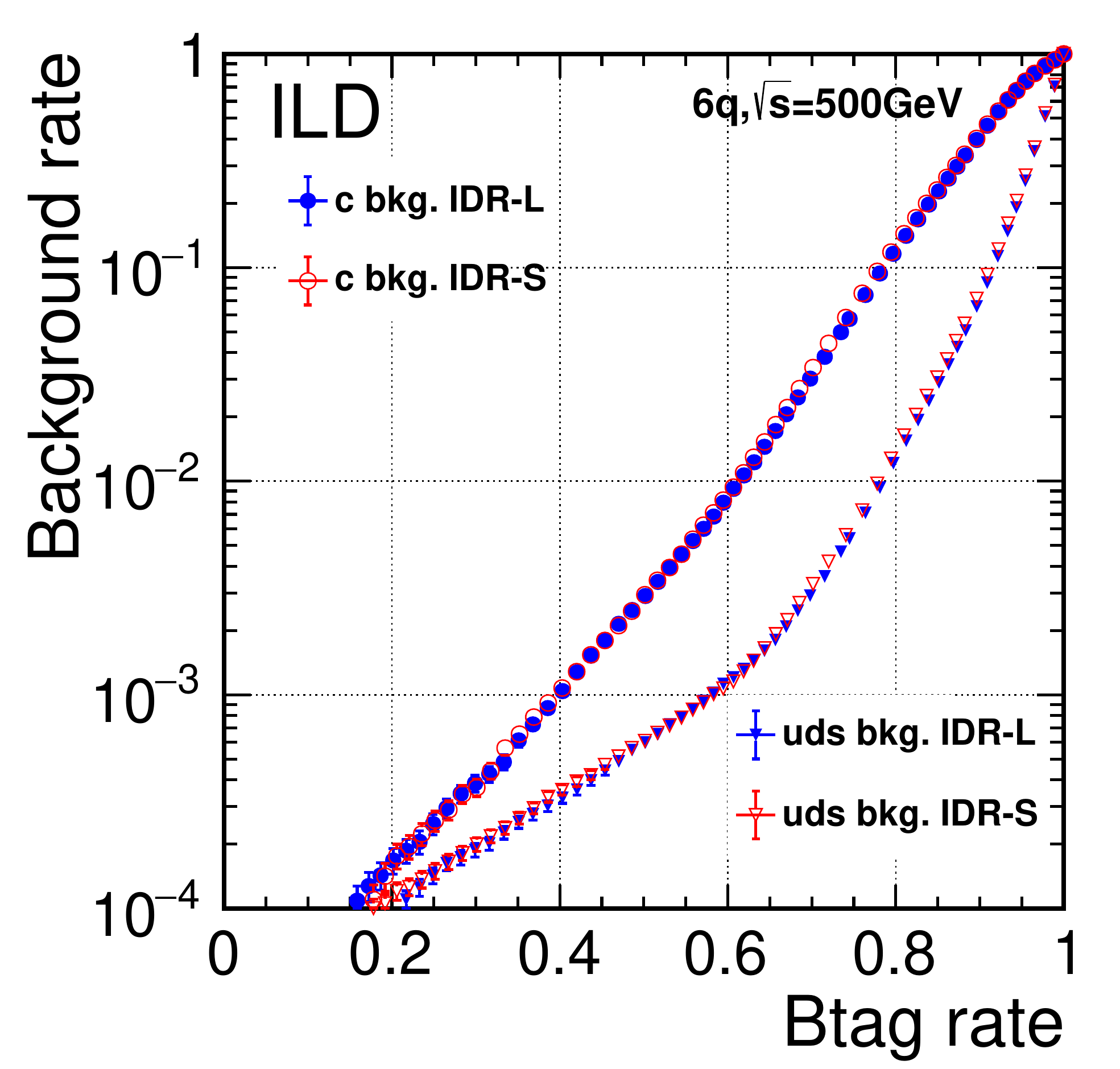}
    \includegraphics[width=0.3\hsize]{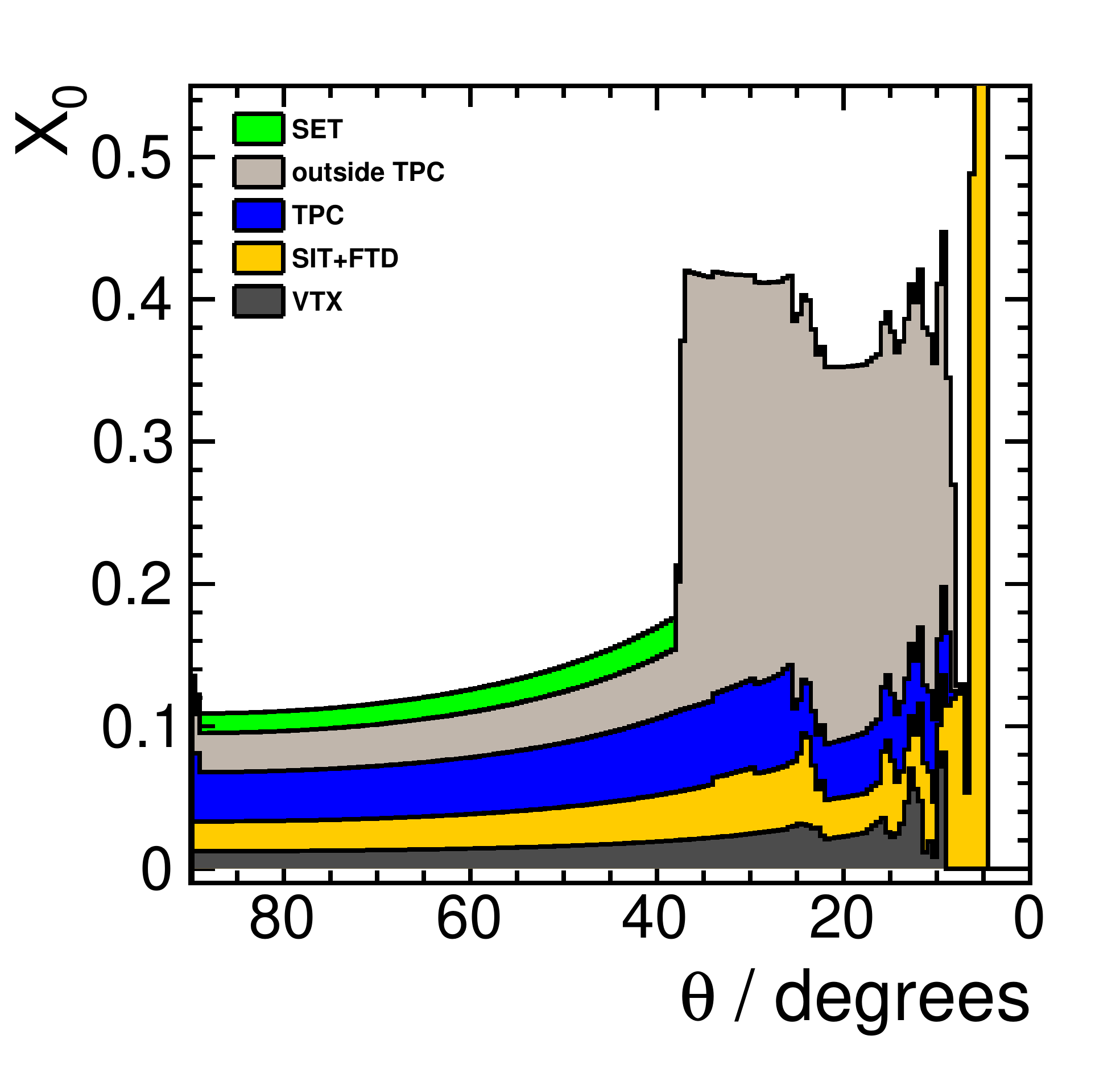}
    \caption{Flavour tag performance for the large and small ILD detector models.
    Background rate as a function of the c-tagging (left) and b-tagging (middle) efficiency
  for heavy quark and light flavour quark jets.
  %Purity of the flavour tag as a function of the efficiency, for different flavours tagged.
  Right: Material budget in ILD up to the calorimeter, in fraction of a radiation length .The different contributions are summed up to represent the cumulative radiation length at a given polar angle.(Figures from \cite{ILDConceptGroup:2020sfq})}
    \label{fig-btag}
\end{figure}

In Figure~\ref{fig-btag} the purity of the flavour identification in ILD is shown as a function of its efficiency.
The performance for b-jet identification is excellent, and charm-jet identification is also good, providing a purity of about 70\% at an efficiency of 60\%.
 The system also allows the accurate determination of the charge of displaced vertices, and contributes strongly to the low-momentum tracking capabilities of the overall system, down to a few 10's of\,MeV. An important aspect of the system leading to superb flavour tagging is the small amount of material in the tracker. This is shown in Figure~\ref{fig-btag} (right).

\subsection{ILD tracking system}

\begin{figure}[t!]
\centering
\includegraphics[width=0.5\hsize]{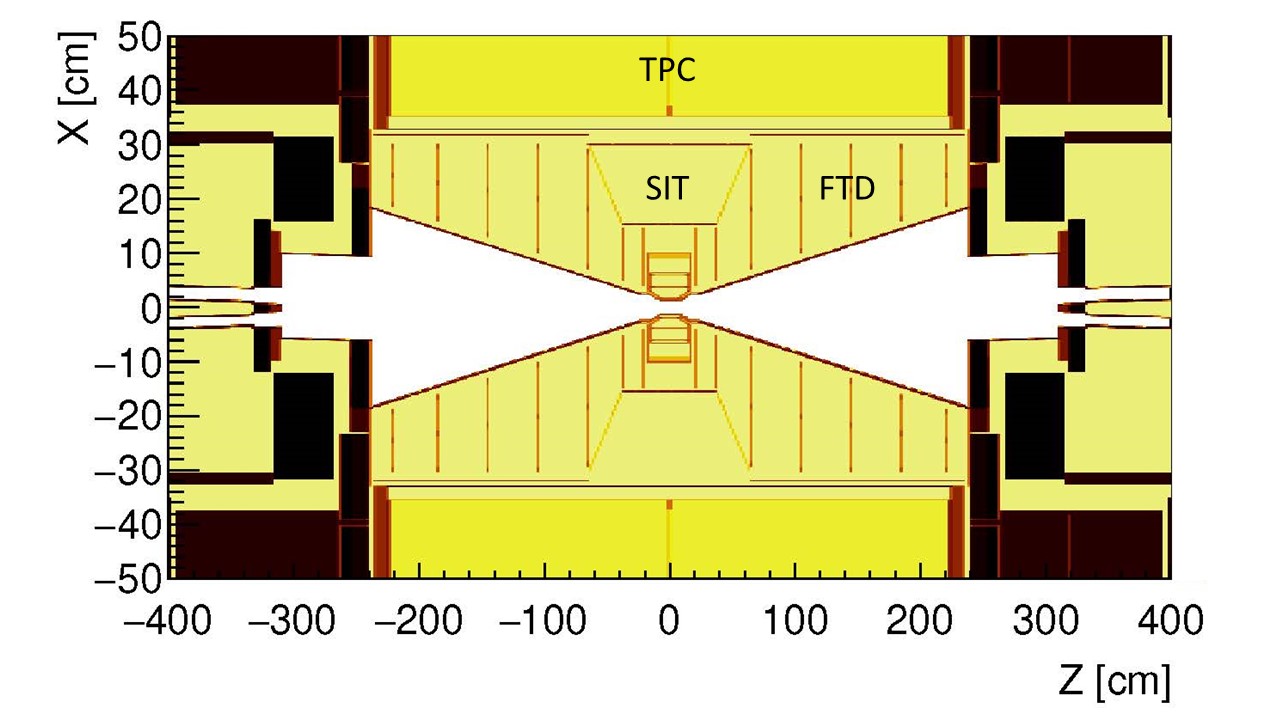}
\caption{Layout of the inner Silicon (SIT) and forward Silicon (FTD) trackers surrounding the vertex detector.}
\label{fig:det:silicon}
\end{figure}

 ILD has decided to approach the problem of charged particle tracking with a hybrid solution, which combines a high resolution time projection chamber (TPC) with a few layers of strategically placed Silicon strip or pixel detectors before and after the TPC. The technologies of the different tracking layers have not been decided yet. The baseline design calls for strips for the three intermediate tracking layers in front of the TPC and the external silicon tracker between the TPC and the ECAL. Recent advances in Silicon pixel technologies make it likely that the Silicon tracker part will be realised as a full pixel tracker. The layout of the inner tracking section is shown in Figure~\ref{fig:det:silicon}.

The time-projection chamber will fill a large volume about 4.6\,m in length, spanning radii from 33 to 180\,cm. In this volume the TPC provides up to 220 three dimensional points for continuous tracking with a single-hit resolution of better than 100~$\mu\mathrm{m}$ in $r \phi$, and about 1\,mm in $z$. This high number of points allows a reconstruction of the charged particle component of the event with high accuracy, including the reconstruction of secondaries, long lived particles, kinks, etc.. For momenta above 100\,MeV, and within the acceptance of the TPC, greater than 99.9\% tracking efficiency has been found in events simulated realistically with full backgrounds. At the same time the complete TPC system will introduce only about 10\% of a radiation length into the detector~\cite{Diener:2012mc}. 

\begin{figure}
    \centering
%    \begin{tabular}{c{0.45\hsize}c{0.55\hsize}}
    \begin{tabular}{cc}
    \includegraphics[width=.45\hsize]{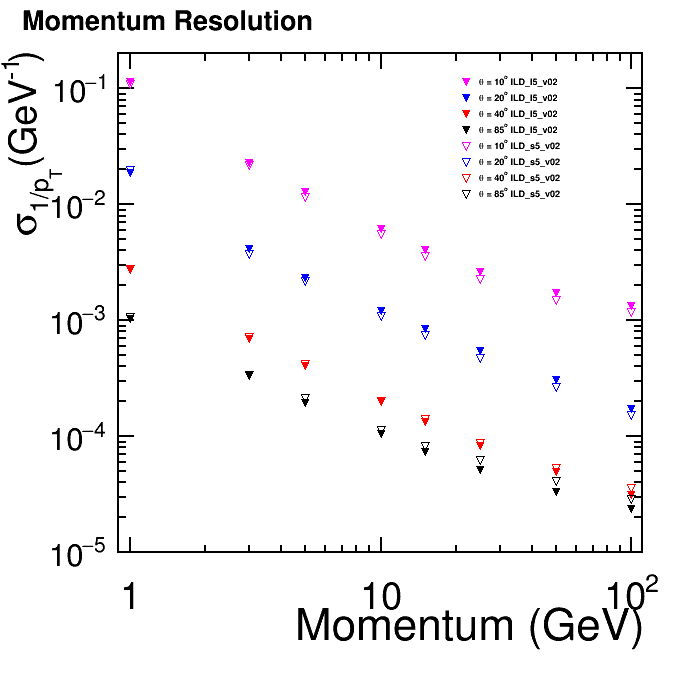} &
        \includegraphics[width=.49\hsize]{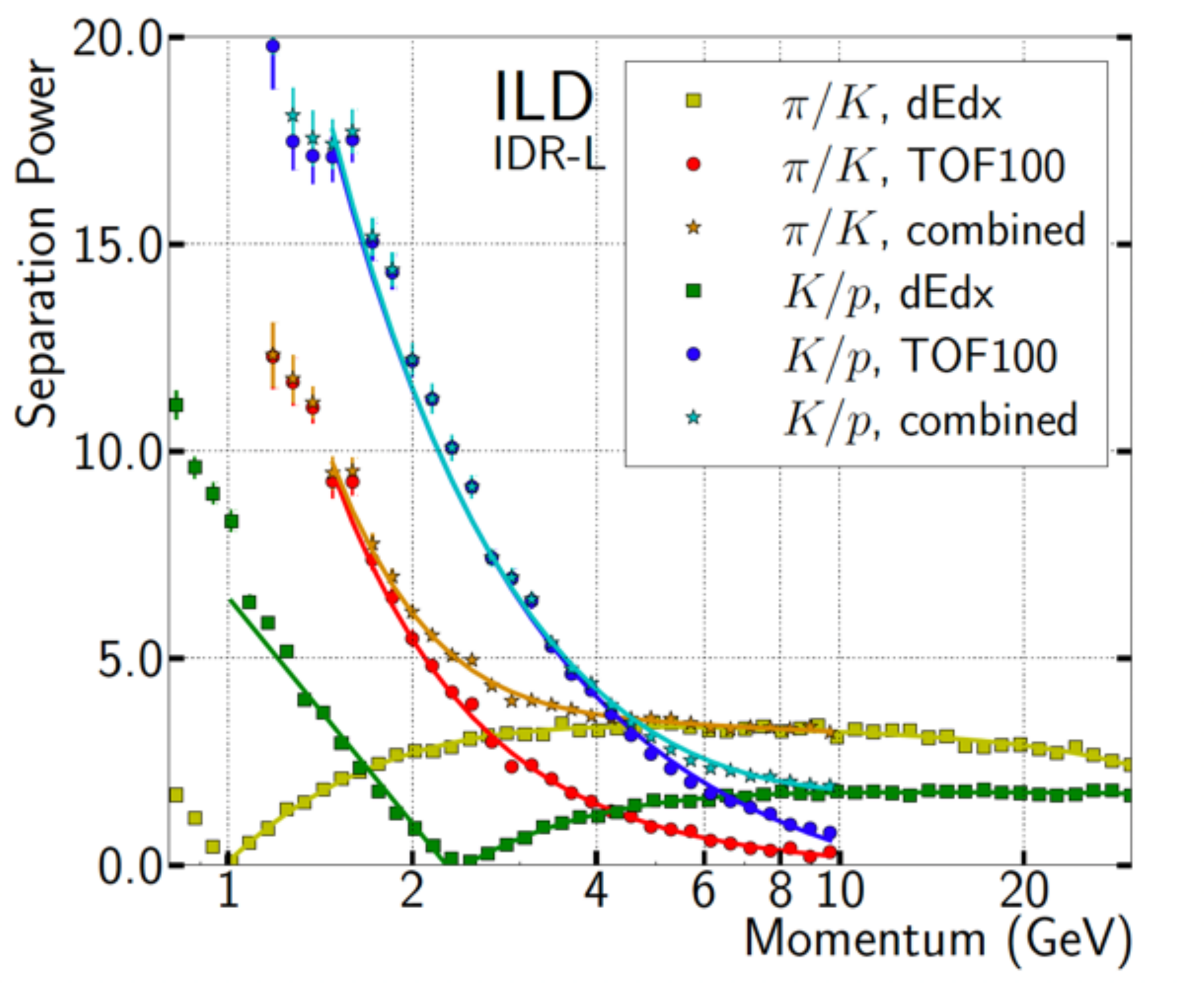}
    \end{tabular}
    \caption{ Left: Simulated resolution in $1/p_t$ as a function of the momentum for single muons. The different curves correspond to different polar angles. Right: Simulated separation power 
%    (probability for a {particle} to be reconstructed as {another one})
    between pions and kaons {and between kaons and protons}, from $dE/dx$ and from timing, assuming a 100~ps timing resolution of the first ECAL layers (Figures from \cite{ILDConceptGroup:2020sfq}).}
    \label{fig:momentumvsp}
\end{figure}

Inside and outside of the TPC volume a few layers of Silicon detectors provide high resolution points, at a point resolution of $10\mu \mathrm{m}$. In the forward direction, extending the coverage down to the beam-pipe, a system of two pixel disks (point resolution $5 \mu$m) and five strip disks (resolution $10 \mu$m, provide tracking coverage down to the beam-pipe. Combined with the TPC track, this will result in an asymptotic momentum resolution of $\delta p_t / p_t^2 = 2 \times 10^{-5}$ (GeV/c)$^{-1}$ for the complete system. Since the material in the system is very low, a significantly better resolution at low momenta can be achieved than is possible with a current Silicon-only tracker. The achievable resolution is illustrated in Figure~\ref{fig:momentumvsp}, where the $1/p_t$-resolution is shown as a function of the momentum of the charged particle. 

The time projection chamber enables the identification of the particle type of the crossing particle through the measurement of the specific energy loss, $dE/dx$, for tracks at intermediate momenta~\cite{Hauschild:2000eg}. The achievable performance is shown in Figure~\ref{fig:momentumvsp} (right). ILD wants to achieve a goal of $5\%$ relative dE/dx resolution in the TPC.  Time of flight measurements can provide additional information, which is particularly effective in the low-momentum regime which is problematic for $dE/dx$. Figure~\ref{fig:momentumvsp} (right) {shows in addition the effect of including time information (resolution 100~ps) from the first ECAL layers.} 

The design and performance of the TPC has been the subject of intense {{R\&D}} over the last 15 years~\cite{Fusayasu:2011sia, RDliaision}. A TPC based on the readout with micro-pattern gas detectors has been developed, and tested in several technological prototypes. The fundamental performance has been demonstrated, and solutions to construct a TPC with the required low mass have been developed. Most recently the performance of the specific energy loss, $dE/dx$, has been validated in test beam data. Based on these results, the TPC technology is sufficiently mature for use in the ILD detector, and can deliver the required performance (see e.g. \cite{Attie:2016yeu,Bouchez:2007pe}).

\subsection{ILD calorimeter- and muon  system}
\label{sec:ILD-forward}
A very powerful calorimeter system is essential to the performance of a detector designed for particle flow reconstruction. Particle flow stresses the ability to separate the individual particles in a jet, both charged and neutral. This puts the imaging capabilities of the system at a premium, and pushes the calorimeter development in the direction of a system with very high granularity in all parts of the system, both transverse to and along the shower development direction. A highly granular sampling calorimeter is the chosen solution to this challenge~\cite{Sefkow:2015hna}. The conceptual and technological development of the particle flow calorimeter have been largely done by the CALICE collaboration (for a review of recent CALICE results see {e.g.} \cite{Grenier:2017ewg,Ootani:2021qna}). 

ILD has chosen a sampling calorimeter {equipped} with silicon diodes as one option for the electromagnetic calorimeter. Diodes with pads of about $(5 \times 5)$ mm$^2$ are used to sample a shower up to 30 times in the electromagnetic section.  {A self-supporting carbon-fiber-reinforced-polymer (CFRP) incorporating tungsten plates supports the detector elements while minimizing non-instrumented spaces}. In 2018 {beam tests of detection elements in stacks and chained together into long cassettes made important steps towards the demonstration of the large scale feasibility of this technology.
Extended tests in 2021 and 2022, including a compact DAQ compatible with the ILD design, are expected to assess the performance with high energy particles.} A very similar system has been adopted by the CMS experiment for the upgrade of the endcap calorimeter, and will deliver invaluable information on the scalability and engineering details of such a system.
 The implementation of precise timing probably mostly in the first calorimeter layers, and the expected performances for single particles are currently under study. Adding timing capabilities of around 100~ps resolution or better to the first layers of the calorimeter would contribute to the capabilities of the ILD detector to identify particle types, in particular at low energies (see Figure~\ref{fig:momentumvsp}(right)).

As an alternative to the silicon based system, sensitive layers made from thin scintillator strips are also investigated. Orienting the strips perpendicular to each other has the potential to realize an effective cell size of $5\times 5$mm$^2$, with the number of read-out channels reduced by an order of magnitude compared to the all silicon case.  A fully integrated technological prototype with 32 layers has been constructed by a joined effort of the R\&D groups for the ILD Sc-ECAL and the CEPC-ECAL. It is currently under commissioning and will be tested in particle beam soon.

For the hadronic part of the calorimeter of the ILD detector, two technologies are studied, based on either silicon photo multiplier (SiPM) on scintillator tile technology~\cite{Simon:2010mi} or resistive plate chambers~\cite{Baulieu:2015pfa}. The SiPM-on-tile option has a  moderate granularity, with $3 \times 3$ cm$^2$ tiles, and provides an analogue readout of the signal in each tile (AHCAL). The RPC technology has a better granularity, of $1 \times 1$ cm$^2$, but provides only 2-bit amplitude information (SDHCAL). For both technologies, significant prototypes have been built and operated. Both follow the engineering design anticipated for the final detector, and demonstrate thus not only the performance, but also the scalability of the technology to a large detector. As for the ECAL the SiPM-on-tile technology has been selected as baseline for part of the upgrade of the CMS hadronic end-cap calorimeter, and will thus see a major application in the near future.

\begin{figure}[th]
    \centering
    \begin{tabular}{lcr}
    \includegraphics[width=0.51\hsize]{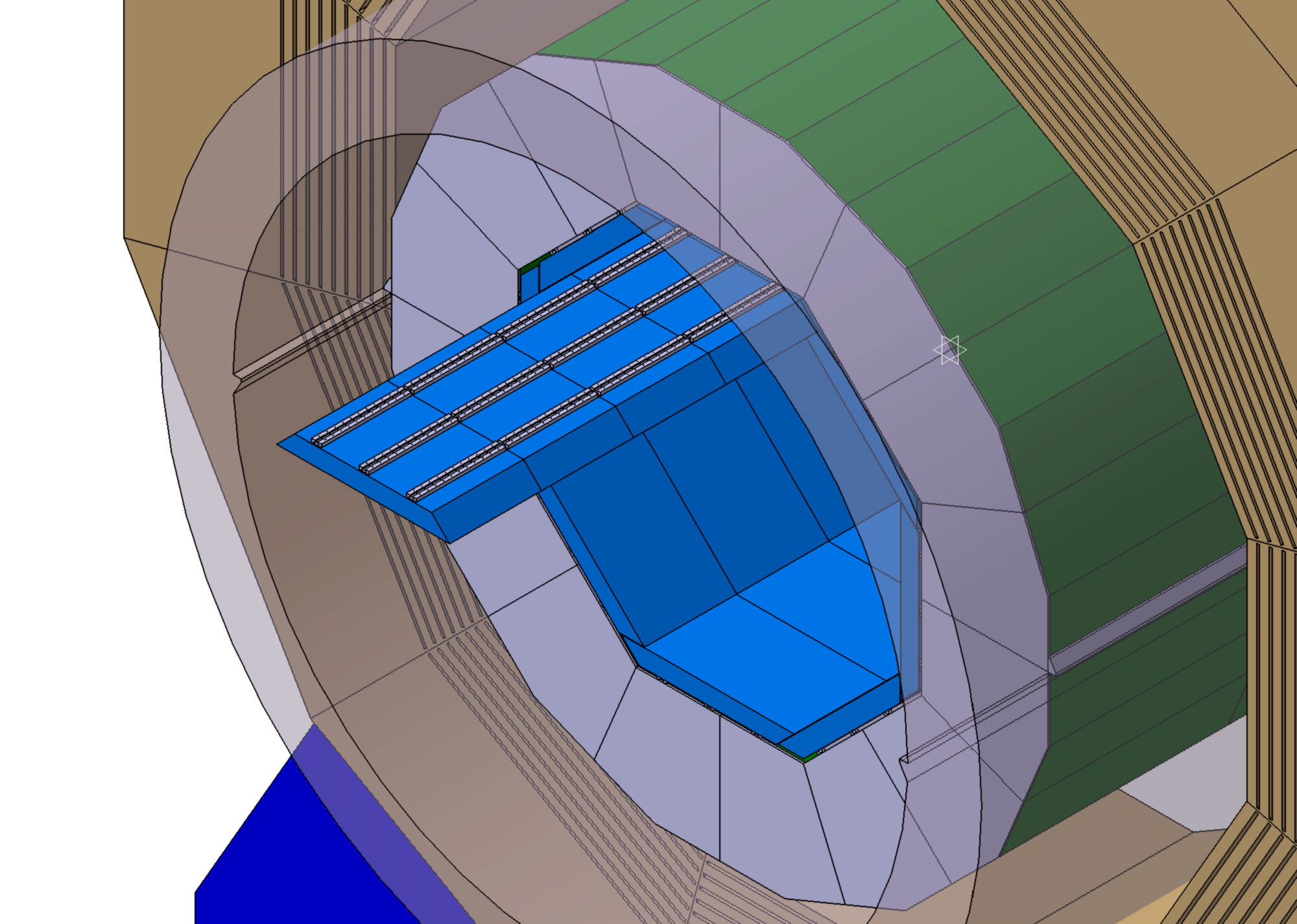} & ~~~~ &
    \includegraphics[width=0.33\hsize]{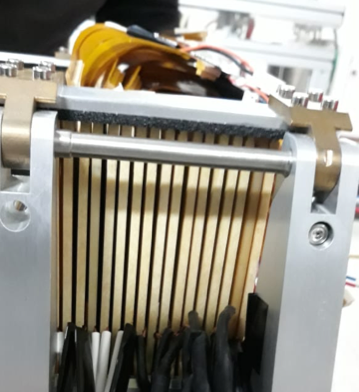}\\
    \end{tabular}
    \caption{Left: Three-dimensional rendering of the barrel calorimeter system, with one ECAL stave partially extracted. Right: Prototype module of the lumical calorimeter.}
    \label{ild-fig-CALO}
\end{figure}

%The simulated particle flow performance is shown in Figure~\ref{fig:pflow}~(right).
A rendering of ILD's barrel calorimeter is shown in Figure~\ref{ild-fig-CALO} (left). 

The iron return yoke of the detector, located outside of the coil, is instrumented to act as a tail catcher and as a muon identification system. Both RPC chambers and scintillator strips read out with SiPMs have been investigated as possible technologies for the system. Up to 14 active layers, located mostly in the inner half of the iron yoke 
%(see Table~\ref{ild:tab:barrelpara} and 
(see Figure~\ref{ild-fig-CALO} for more details) will be instrumented.  To minimize the number of readout channels a new readout scheme~\cite{Patent:WowenStrips}~\cite{ILDTalk:WowenStrips} has been  developed within the RPC readout option. In this scheme, pads and pixels are interconnected in a special way which allows a precise position measurement based on at least 3 strips under different direction, achieving a very good granularity with limited number of electronic channels.

Three rather specific calorimeter systems are foreseen for the very forward region of the ILD detector~\cite{Abramowicz:2010bg}. LumiCal is a high precision fine sampling silicon tungsten calorimeter primarily designed to measure electrons from Bhabha scattering, and to precisely determine the integrated luminosity as discussed in Sec.~\ref{subsec:lumi_prec}. The LHCAL (Luminosity Hadronic CALorimeter) just outside the LumiCal extends the reach of the endcap calorimeter system down to smaller angles relative to the beam, and closes the gap between the inner edge of the ECAL endcap and the LumiCal.  Below the LumiCal acceptance, where background from beamstrahlung rises sharply, BeamCal, placed further downstream from the interaction point, provides added coverage and is used to provide fast feedback on the beam position at the interaction region. As the systems move close to the beam-pipe, the requirements on radiation hardness and on speed become more and more challenging. Indeed this very forward region in ILD is the only region where radiation hardness of the systems is a key requirement. A picture of a prototype of the Lumical calorimeter is shown in Figure~\ref{ild-fig-CALO}(right).

\subsection{ILD detector integration and costing}
From the beginning, one of the major goals of the ILD concept group was  to move the detector concept from a collection of technological ideas to a real detector that can actually be built, commissioned, and operated within given engineering and site-dependent constraints. 

The main mechanical structure of the ILD detector is the iron yoke that consists of three barrel rings and two endcaps. The yoke provides the required shielding for radiation and magnetic fields and supports the cryostat for the detector solenoid and the barrel detectors, calorimeters and tracking system. 

A common concept for the detector services such as cables, cooling, gases and cryogenics has been developed. The requirements are in many cases based on engineering prototypes of the ILD sub-systems. 

The main detector solenoid is based on CMS experience and can deliver magnetic fields up to 4~T. A correction system for the compensation of the crossing angle of the ILC beam, the Detector Integrated Dipole, has been designed and can be integrated into the main magnet cryostat.

The cost of the ILD detector has been estimated at the time of the ILD interim design report, IDR~\cite{ILDConceptGroup:2020sfq}. The total detector cost is about $379$~Million EUR in 2018 costs. The cost of the detector is dominated by the cost of the calorimeter system and the yoke, which together account for about $60\%$ of the total cost. A slightly smaller version of ILD, where the outer radius of ILD has been reduced by about 10\%, but the length of ILD remains unchanged, results in a reduction of the cost by about 50~Million EUR. 

\subsection{Future developments of the ILD detector}
The ILD detector group is actively investigating where new technologies might deliver significantly improved performance, expand the capabilities of the detector, or deliver equal performance at lower cost. 

The fundamental design criteria of ILD - particle flow as a basis for a complete event reconstruction, excellent pattern recognition capabilities with high efficiency and coverage of the largest possible solid angle - are not at question and remain the basis of any design decisions. The studies on optimizing ILD summarised in the ILD IDR~\cite{ILDConceptGroup:2020sfq} have pointed out a number of areas of high potential where next-generation technologies might have a large impact. 

Timing in a number of different sub-systems is one key development direction. Timing at the level of a few 10 ns is already part of the concept. Pushing this to below 100~ps will contribute significant additional capabilities in particle identification and in background reduction. Technologically this is a significant challenge. This option is explored in the tracking system, and in the calorimeter system.

Timing capabilities in the silicon detectors might go hand in hand with in increased integration of functionality into the sensor. Moving to silicon systems with smaller feature size might allow the implementation of complex clustering or even tracking algorithms on individual pixels, which could change significantly the way these detectors are operated. 

The current layout of the inner tracking system in ILD was optimized for acceptance, robustness towards background and low material budget. With new pixel technologies an all Pixel forward tracker with an optimized layout is an attractive option, which would also ease the transition from the current vertex detector to the forward tracker. 

The further reduction of the material budget in the tracker remains a central goal of ILD. Experience from ongoing detector construction projects as the LHC upgrade detectors will provide valuable input, however, new approaches to support structures etc will be needed to really improve the situation further. 

The current choice of ILD to implement a gaseous time projection chamber as central tracker remains a very attractive solution, where clear advantages have been demonstrated. The rapid development of silicon technologies on the other hand might open the way to find non-gaseous solutions which offer similar benefits. The combination of a gaseous detector with a highly granular silicon readout over large areas could point into a direction which will combine the best of both worlds. 

The calorimeter continues to be an area of very active research, and many improvements to the current technologies are expected. The application of these technologies to the LHC detectors will provide very important input.

A few rather concrete examples of R\&D which could shape the development of ILD is summarised in the following section. 

\subsubsection{New technologies in ILD}
The CMOS detector technology is seeing rapid developments. Based on ever smaller feature size very small pixels might be realised, anticipated to provide a spatial resolution of $<$3$\mu$m. They also open the perspective of achieving large multi-reticular sensors, which may be exploited to suppress considerably the material budget of the detector layers, which may become nearly unsupported. The evolution of the CMOS technology also prepares for breakthroughs in terms of time resolution, with projections going well bellow 1ns. 

Another relevant R\&D area is the possibility of including precise timing information of the individual signals in the calorimeter readout, turning the calorimeters into a 5D device. This can improve the shower reconstruction to identify the type of particles and also to reduce the noise. Like silicon and scintillator, RPC and, more precisely MRPC, are excellent fast timing detectors which can be exploited by equipping their readout electronics with fast timing capability. This R\&D is starting now, studies using silicon systems with integrated amplification and explorative studies of detectors based on 65~nm feature size are being setup.

The general move to extremely large granularity comes at the prize of vast increase of the number of channels and the data volume to be handled and the power consumption of the system. Innovative ways to reduce the number of channels in areas of relatively low occupancies without sacrificing the individual precision will be an important challenge.

\subsection{Science with ILD}
ILD has been designed to operate with electron-positron collisions between 90 GeV and 1 TeV. The science goals of the ILC have been described in detail in \cite{ILCESU1}, 
% and will not be repeated here. 
{and results of numerous studies are reported in the following chapters of this document.}
It should be pointed out that the analyses which have been performed within the ILD concept group are based on fully simulated events, using a realistic detector model and advanced reconstruction software, and in many cases include estimates of key systematic effects. This is particularly important when estimating the reach the ILC and ILD will have for specific measurements. Determining, for example, the branching ratios of the Higgs at the percent level depends critically on the detector performance, and thus on the quality of the event simulation and reconstruction. 

In many cases 
%  the performance used in the physics analyses has been tested against prototype experiments.
  {the performance assumed in the detector simulation has been cross checked with prototype test results.}
The key performance numbers for the vertexing, tracking and calorimeter systems are all based on results from test beam experiments. The particle flow performance, a key aspect of the ILD physics reach, could  not be fully verified in the absence of   {a large scale detector prototype}, but key aspects have been shown in experiments. This includes the single particle resolution for neutral and charged particles, the particle separation in jets, the linking power between tracking and calorimetry, and
% key aspects of detailed shower analyses 
  {detailed shower reconstruction} 
important for particle flow. 

While the physics case studies are based on the version of the ILD detector presented in the detector volume of the ILC DBD~\cite{ild-dbd}, ILD  initiated a systematic benchmarking effort to study the performance of the ILD concept, and to determine in particular the correlations between science objectives and detector performance. The list of benchmark   {processes} which   {have been studied} is given in Table \ref{tab-benchmark}. Even if the ILC will start operation at a center-of-mass energy of 250\,GeV, the ILD detector is being designed to meet the more challenging requirements of higher center-of-mass energies, since major parts of the detector, e.g.\ the coil, the yoke and the main calorimeters will not be replaced when upgrading the accelerator. Therefore, most of the detector benchmark analyses   {were} performed at a center-of-mass energy of 500\,GeV, and one benchmark even at 1\,TeV. The assumed integrated luminosities and beam polarization settings   {followed} the canonical running scenario~\cite{Barklow:2015tja}. 
In addition to the well-established performance aspects of the ILD detector, the potential of new features not yet incorporated in the existing detector prototypes, e.g.\ time-of-flight information,   {have also been} evaluated. 

The results of these studies   {were} published in the ILD Interim Design Report~\cite{ILDConceptGroup:2020sfq}. They form the basis for the definition of a new ILD baseline detector model, which   {has been} used for a new physics-oriented Monte-Carlo production for 250\,GeV.   {Sample production with} the most recent beam parameters of the accelerator~\cite{Evans:2017rvt} and significantly improved reconstruction algorithms is expected to lead to further improvements of the expected results of the precision physics program of the ILC~\cite{ILCESU1}.

  {Further ILD performance and physics potential studies are ongoing. Special attention is paid to understanding of systematic effects.
Significant reduction of systematic uncertainties is possible in combined analysis of different channels, in particular when combining data taken with different beam polarization settings. 
}

\begin{table}[thb!]
    \centering
    \begin{tabular}{|p{4cm}|p{5cm}|p {5cm}|}
\hline
{\bf    Measurement}     & {\bf Main physics question} & {\bf main issue addressed} \\
\hline
Higgs mass in $H\rightarrow b {\bar b}$         &  Precision Higgs mass determination &Flavour tag, jet energy resolution, lepton momentum resolution  \\
\hline
Branching ratio $H \rightarrow \mu^+\mu^-$ & Rare decay, Higgs Yukawa coupling to muons & High-momentum $p_t$ resolution, $\mu$ identification \\
\hline
Limit on $H \rightarrow$ invisible & Hidden sector / Higgs portal & Jet energy resolution, $Z$ or recoil mass resolution, hermeticity\\
\hline
Coupling between $Z$ and left-handed $\tau$ & Contact interactions, new physics related to 3rd generation & Highly boosted topologies, $\tau$ reconstruction, $\pi^0$ reconstruction \\
\hline
$WW$ production, $W$ mass & Anomalous triple gauge couplings, $W$ mass&  Jet energy resolution, leptons in forward direction \\
\hline
Cross section of $e^+e^- \rightarrow \nu \nu qqqq$ & Vector Bosons Scattering, test validity of SM at high energies&  $W/Z$ separation, jet energy resolution, hermeticity\\
\hline
Left-Right asymmetry in $e^+e^- \rightarrow \gamma Z$ & Full six-dimensional EFT interpretation of Higgs measurements &  Jet energy scale calibration, lepton and photon reconstruction \\
\hline
Hadronic branching ratios for $H\rightarrow b \bar b $ and $c \bar c$ & New physics modifying the Higgs couplings &  Flavour tag, jet energy resolution\\

\hline
$A_{FB}, A_{LR}$ from $e^+e^- \to b\bar{b}$ and $t \bar t \rightarrow b\bar{b} qqqq / b \bar{b} qql\nu$ & Form factors, electroweak coupling &  Flavour tag, PID, (multi-)jet final states with jet and vertex charge\\
\hline

Discovery range for low $\Delta M$ Higgsinos & Testing SUSY in an area inaccessible for the LHC& Tracks with very low $p_t$, ISR photon identification, finding multiple vertices\\
\hline
Discovery range for WIMPs in mono-photon channel & Invisible particles, Dark sector & Photon detection at all angles, tagging power in the very forward calorimeters\\
\hline
Discovery range for extra Higgs bosons in $e^+e^- \rightarrow Zh$ & Additional scalars with reduced couplings to the $Z$ & Isolated muon finding, ISR photon identification.\\
\hline

%\hline
%\multicolumn{3}{|l|}{Running above the top threshold:}\\

    \end{tabular}
    \caption{table of benchmark reactions which are used by ILD to optimize the detector performance. The analyses are mostly conducted at 500\,GeV center-of-mass energy, to optimally study the detector sensitivty. The channel, the physics motivation, and the main detector performance parameters are given.}
    \label{tab-benchmark}
\end{table}

\subsection{Integration of ILD into the experimental environment}
ILD is designed to be able to work in a push-pull arrangement with another detector at a common ILC interaction region. In this scheme ILD sits on a movable platform in the underground experimental hall. This platform allows for a roll-in of ILD from the parking position into the beam and vice versa within a few hours. The detector can be fully opened and maintained in the parking position.

The current mechanical design of ILD assumes an initial assembly of the detector on the surface, similar to the construction of CMS at the LHC. A vertical shaft from the surface into the underground experimental cavern allows ILD to be lowered in five large segments, corresponding to the five yoke rings.

ILD is self-shielding with respect to radiation and magnetic fields to enable the operation and maintenance of equipment surrounding the detector, {e.g.} cryogenics. Of paramount importance is the possibility to operate and maintain the second ILC push-pull detector in the underground cavern during ILC operation.

\subsection{The ILD Concept Group}
The ILD collaboration initially started out as a fairly loosely organised group of scientists interested to explore the design of a detector for a linear collider like the ILC. With the delivery of the DBD in 2013, the group re-organised itself more along the lines of a traditional collaboration. The group imposed upon itself a set of by-laws which govern the functions of the group, and define rules for the membership of ILD. 
% Daniel removed the following sentence, which is no longer correct in my understanding
%Groups who want to become members of ILD must sign a memorandum of participation, a first step towards an eventual memorandum of understanding to construct ILD, as soon as the ILC has been approved. 

\begin{figure}
    \centering
    \includegraphics[width=0.9\hsize]{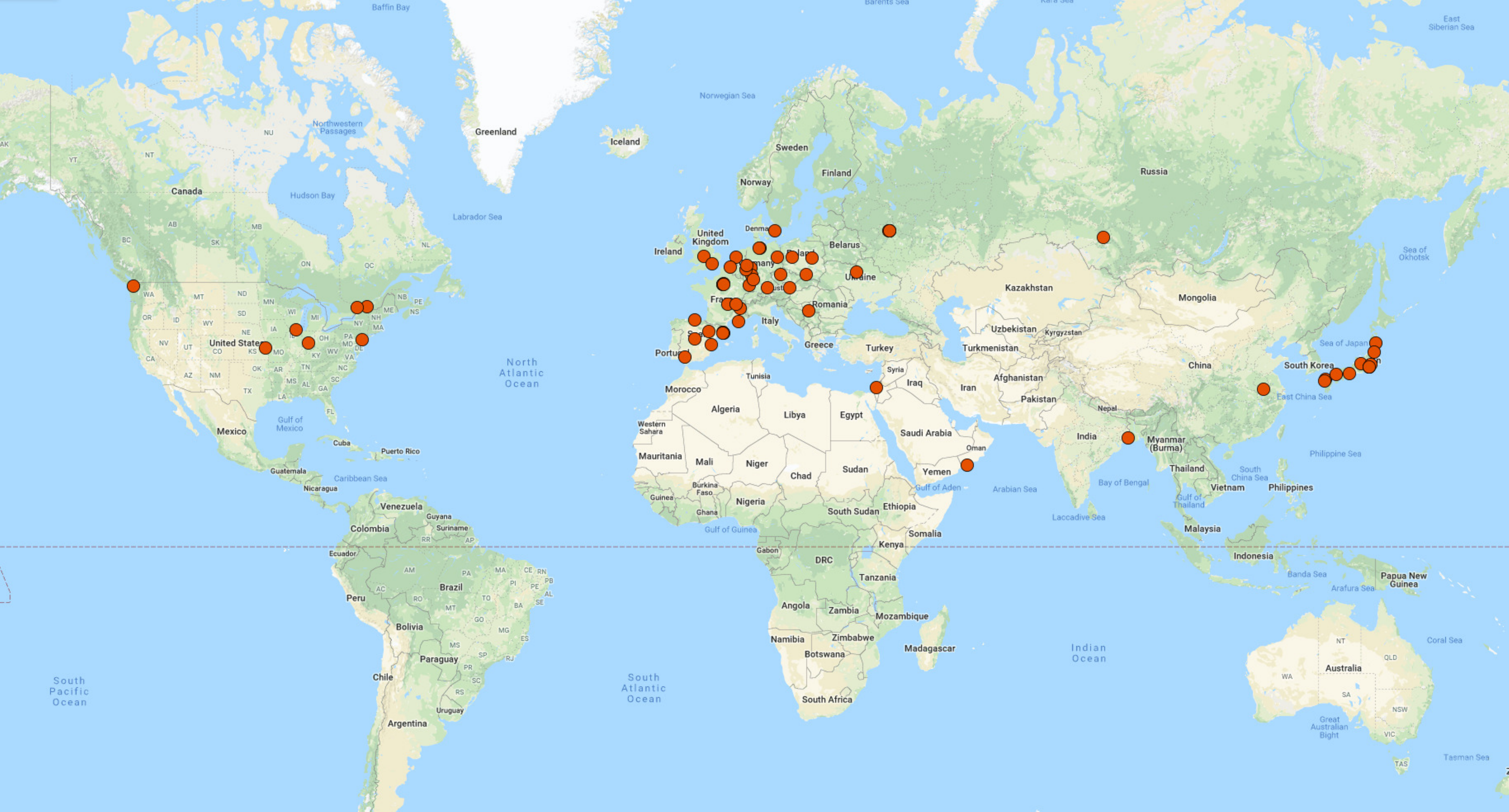}
    \caption{Map with the location of the ILD member institutes indicated.}
    \label{ild-fig-membermap}
\end{figure}

In total 65 groups from 30 countries signed the letter of participation in 2015. At present (2021), 68 institutions are members, and a number of individuals have joined as guest members of ILD. A map indicating the location of the ILD member institutes is shown in Figure~\ref{ild-fig-membermap}.

\subsection{Conclusion and outlook}
The ILD detector concept is a well developed integrated detector optimised for use at the electron-positron collider ILC. It is based on advanced detector technology, and driven by the science requirements at the ILC. Most of its major components have been fully demonstrated through prototyping and test beam experiments. The physics performance of ILD has been validated using detailed simulation systems. A community interested in building and operating ILD has formed over the last few years. It is already sizeable, encompassing 68 institutes from around the world. The community is ready to move forward once the ILC project receives approval. 

%\section{References}

%% MCruz Fixing bibliography problem ====
%\bibliography{chapters/detectors/ILD.bib}
%\printbibliography %Prints bibliography 
%% ========

\section{The SiD Detector} 
\label{SiD}

\subsection{Detector description and capabilities}
\label{SiD-gen}

\begin{figure}[tb]
  \begin{center}
 \includegraphics[width=0.8\hsize]{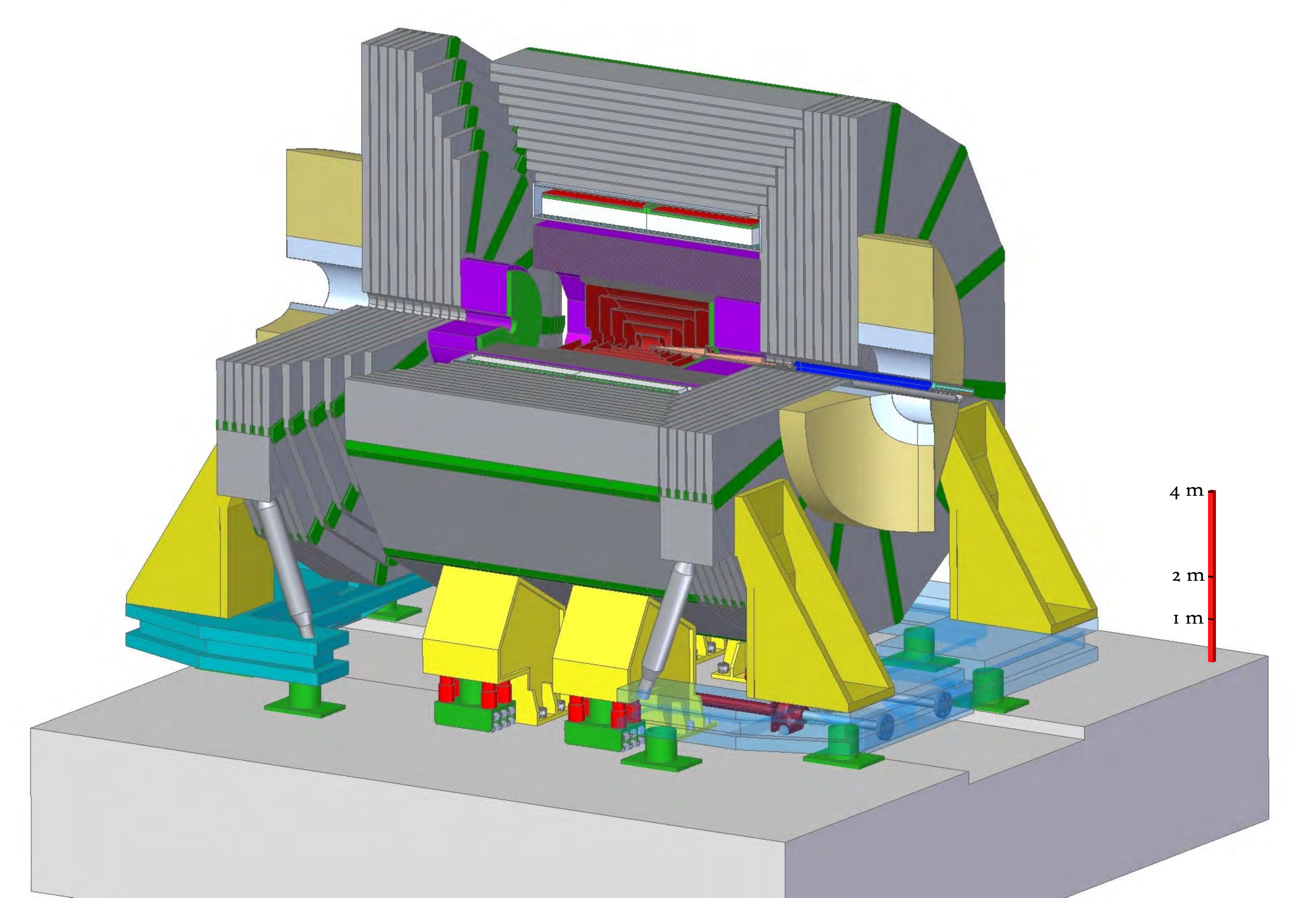}
\caption{The SiD detector concept.
\label{fig:fig_sid}}
 \end{center}
 \end{figure}

The SiD detector concept is a general-purpose experiment designed to perform
 precision measurements
at the ILC. It satisfies the challenging detector requirements resulting from the full range of 
ILC physics processes. SiD is based on the paradigm of particle flow, an algorithm by which
the reconstruction of both charged and neutral particles is accomplished by an optimised
combination of tracking and calorimetry. The net result is a significantly more precise jet
energy measurement than that achieved via conventional methods and which results in a di-jet mass resolution good enough to distinguish
between $W$s and $Z$s.
The SiD detector (Fig.~\ref{fig:fig_sid})  is a compact detector based on a powerful silicon
pixel vertex detector, silicon tracking, silicon-tungsten electromagnetic calorimetry, and
highly segmented hadronic calorimetry. 
SiD also incorporates a high-field solenoid, iron
flux return, and a muon identification system. The use of silicon 
sensors in the vertex, tracking,
and calorimetry enables a unique integrated tracking system ideally suited to particle
flow.

The choice of silicon detectors for tracking and vertexing ensures that SiD is robust
with respect to beam backgrounds or beam loss, provides superior charged particle momentum
resolution, and eliminates out-of-time tracks and backgrounds. The main tracking
detector and calorimeters are “live” only during a single bunch crossing, so beam-related
backgrounds and low-pT backgrounds from $\gamma\gamma$ processes will be reduced to the minimum
possible levels. The SiD calorimetry is optimised for excellent jet energy measurement
using the particle flow technique.
 The complete tracking and calorimeter systems are contained
within a superconducting solenoid, which has a 5 T field strength, enabling the overall
compact design. The coil is located within a layered iron structure
that returns the magnetic flux and is instrumented to allow the
identification of muons. 
All aspects of SiD are the result of intensive and leading-edge research aimed at achieving
performance at unprecedented levels. At the same time, the design represents a balance between cost
and physics performance. Nevertheless, given advances in technologies it is now appropriate to consider updates to the SiD design as discussed below. First, we describe the baseline SiD design for which the key parameters are
listed in  
Table~\ref{tab:Ovw_sidparams}.
The design is expected to meet all the requirements listed in section \ref{sec:requirements} above.

%%%%%%%%%%%%%%%%%%%%%%%%%%%%%%%%%%%%%%%%
%%%%%%%%%%%%%%%%%%%%%%%%%%%%%%%%%%%%%%%%%%%%%%

%%%%%%%%%%%%%%%%%%%%%%%%%%%%%%%%%%%%%%%%%%%%%%%%
\begin{table}
\begin{center}
\begin{tabular}{l l r r r}
 \hline
    \sid Barrel& Technology& In rad& Out rad& z extent \\
    \hline
    Vtx detector& Silicon pixels& 1.4& 6.0& $\pm \quad 6.25$ \\
    Tracker& Silicon strips& 21.7& 122.1& $\pm \quad 152.2$ \\
    ECAL& Silicon pixels-W& 126.5& 140.9& $\pm \quad 176.5$ \\
    HCAL& Scint-steel& 141.7& 249.3& $\pm \quad 301.8$ \\
    Solenoid& 5 Tesla SC & 259.1& 339.2& $\pm \quad 298.3$ \\
    Flux return& Scint-steel& 340.2 & 604.2& $\pm \quad 303.3$ \\
   \hline
 \sid Endcap& Technology& In z& Out z& Out rad \\
    \hline
Vtx detector& Silicon pixels& 7.3& 83.4& 16.6 \\
Tracker& Silicon strips& 77.0& 164.3& 125.5 \\
ECAL& Silicon pixel-W& 165.7& 180.0& 125.0 \\
HCAL& Scint-steel& 180.5& 302.8& 140.2 \\
Flux return& Scint/steel& 303.3& 567.3& 604.2 \\
LumiCal& Silicon-W& 155.7& 170.0& 20.0 \\
BeamCal& Semicond-W& 277.5& 300.7& 13.5 \\
    \hline
\end{tabular}
   \end{center}
    \caption{Key parameters of the baseline SiD design. (All dimension
are given in cm).}
\label{tab:Ovw_sidparams}
\end{table}
%%%%%%%%%%%%%%%%%%%%%%%%%%%%%%%%%%%%%%%%%%%%

\subsection{Silicon-based tracking}
\begin{figure}[tb]
 \begin{center}
 \includegraphics[width=0.7\hsize]{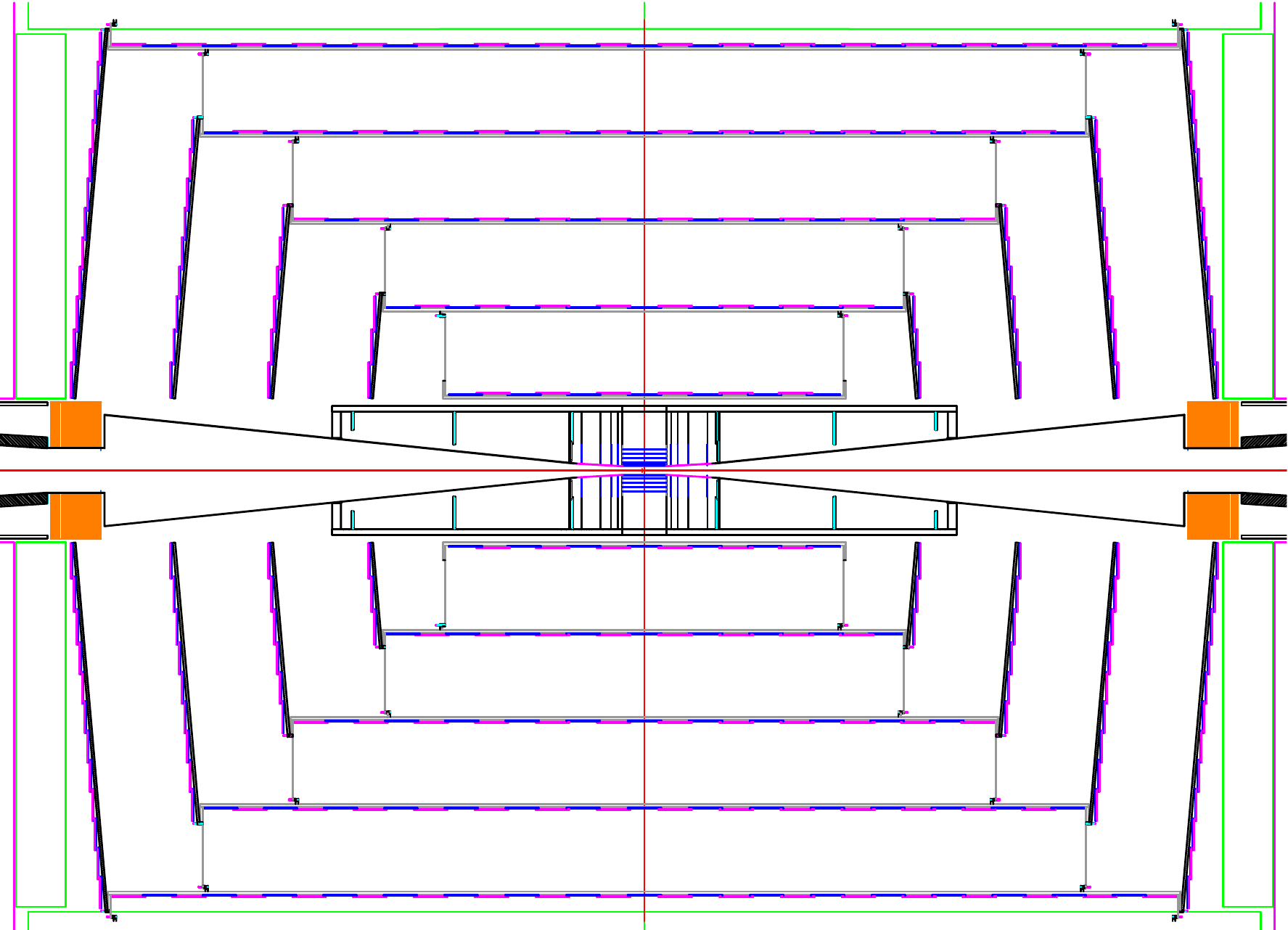}
 \end{center}
\caption{r-z view of vertex detector and outer tracker.
\label{fig:fig_vxdtrk}}
  \end{figure}

The tracking system (Fig.~\ref{fig:fig_vxdtrk}) is a key element of the SiD detector concept. The
particle flow algorithm requires excellent tracking with superb efficiency and
two-particle separation. The requirements for precision measurements, in
particular in the Higgs sector, place high demands on the momentum resolution at
the level of $\delta (1/\pT)  \sim 2-5 \times 10^{-5}/$GeV/$c$.

Highly efficient charged particle tracking is achieved using the pixel detector
and main tracker to recognise and measure prompt tracks, in conjunction with the ECAL, which can
identify short track stubs in its first few layers 
to catch tracks arising from secondary decays of long-lived particles. With
the choice of a 5~T solenoidal magnetic field, in part chosen to control the $\ee$-pair
background, the design allows for a compact tracker design. 

\subsection{Vertex detector}

To unravel the underlying physics mechanisms of new observed processes, the
identification of heavy flavours will play a critical role. One of the main
tools for heavy flavour identification is the vertex detector. The physics goals
dictate an unprecedented spatial three-dimensional point resolution and a very
low material budget to minimise multiple Coulomb scattering. The running 
conditions at the ILC impose the readout speed and radiation tolerance. 
These requirements are normally in tension. High
granularity and fast readout compete with each other and tend to increase the
power dissipation. Increased power dissipation in turn leads to an increased
material budget. The challenges on the vertex detector are considerable and
significant R\&D is being carried out on both the development of the sensors and
the mechanical support.
The SiD vertex detector uses a barrel and disk layout. The barrel section
consists of five silicon pixel layers with a pixel size of
$20~\times~20~\micron^2$. The forward and backward regions each have four
silicon pixel disks. In addition, there are three silicon pixel disks at a
larger distance from the interaction point to provide uniform coverage for the
transition region between the vertex detector and the outer tracker. This
configuration provides for very good hermeticity with uniform coverage and
guarantees excellent charged-track pattern recognition capability
 and impact parameter resolution 
over the full solid angle. 
This enhances the capability of the integrated tracking system and, 
in conjunction with the high magnetic field, makes for a very compact
system, thereby minimising the size and costs of the calorimetry.

To provide for a very robust track-finding performance the baseline 
choice for the vertex detector has a sensor technology that provides
time-stamping of each hit with sufficient precision to assign it to
a particular bunch crossing. This significantly suppresses
backgrounds. 

%Several technologies are being developed. One of them is a CMOS-based
%monolithic pixel sensor called Chronopixel. The main goal for the design is a
%pixel size of about $10~\times~10~\micron^2$ with 99\% charged-particle
% efficiency. Prototype devices have demonstrated that the concept works; 
%what should be a fully functional chip is presently under test. More 
%challenging is the 3D vertical integrated silicon technology, for which a full 
%demonstration is also close.

Several vertex detector sensor technologies are being developed.  One of these is a 
monolithic CMOS pixel detector with time-stamping capability (Chronopixel~\cite{Sinev:2015iwr}),
being developed in collaboration with SRI International. 
The pixel size is about  $10~\times~10~\micron^2$ with a design goal of 99\% charged-particle
 efficiency.
The time-stamping feature of the design means each hit is accompanied by a time tag with sufficient precision to assign it to a particular bunch crossing of
the ILC -- henc the name Chronopixel. This reduces the occupancy to negligible levels, even in the
innermost vertex detector layer, yielding a robust vertex detector which operates at background
levels significantly in excess of those currently foreseen for the ILC. Chronopixel differs from the
similar detectors developed by other groups by its capability to record time stamps for two hits in
each pixel while using standard CMOS processing for manufacturing. 
Following a series of prototypes, the Chronopixel has been proven to be
a feasible concept for the ILC. The three prototype versions
were fabricated in 2008, in 2012, and in 2014.
The main goal of the third prototype was to test possible solutions for a high capacitance problem
discovered in prototype 2. The problem was traced to the TSMC 90 nm technology design rules,
which led to an unacceptably large value of the sensor diode capacitance. Six different layouts
for the prototype 3 sensor diode were tested, and the tests demonstrated that the high capacitance
problem was solved.

With prototype 3 proving that a Chronopixel sensor can be successful with all known problems solved, optimal sensor design would be the focus of future tests.
The charge collection efficiency for different sensor diode options needs to be measured to determine
the option with the best signal-to-noise ratio. Also, sensor efficiency for charged particles with sufficient energy to penetrate the sensor thickness and ceramic package, along with a trigger telescope measurement, needs to be determined. Beyond these fundamental measurements, a prototype of a few cm$^2$ with a final readout scheme would
test the longer trace readout resistance, capacitance, and crosstalk.

A more challenging approach is the 3D vertical integrated silicon technology, for which a full 
demonstration is also close.

Minimizing the support material is critical to the development of a high-performance 
vertex detector. An array of 
low-mass materials such as reticulated foams and silicon-carbide
materials are under consideration. An alternative approach that is being pursued very actively is the
embedding of thinned, active sensors in ultra low-mass media. This line of R\&D
explores thinning active silicon devices to such a thickness that the silicon
becomes flexible. The devices can then be embedded in, for example, Kapton
structures, providing extreme versatility in designing and constructing a vertex
detector.

Power delivery must be accomplished without exceeding the material budget and
overheating the detector.  The vertex detector 
design relies on power pulsing during bunch trains to minimise heating 
and uses forced air for cooling. 

\subsection{Main tracker}
\begin{figure}
\begin{center}
\includegraphics[width=0.60\hsize]{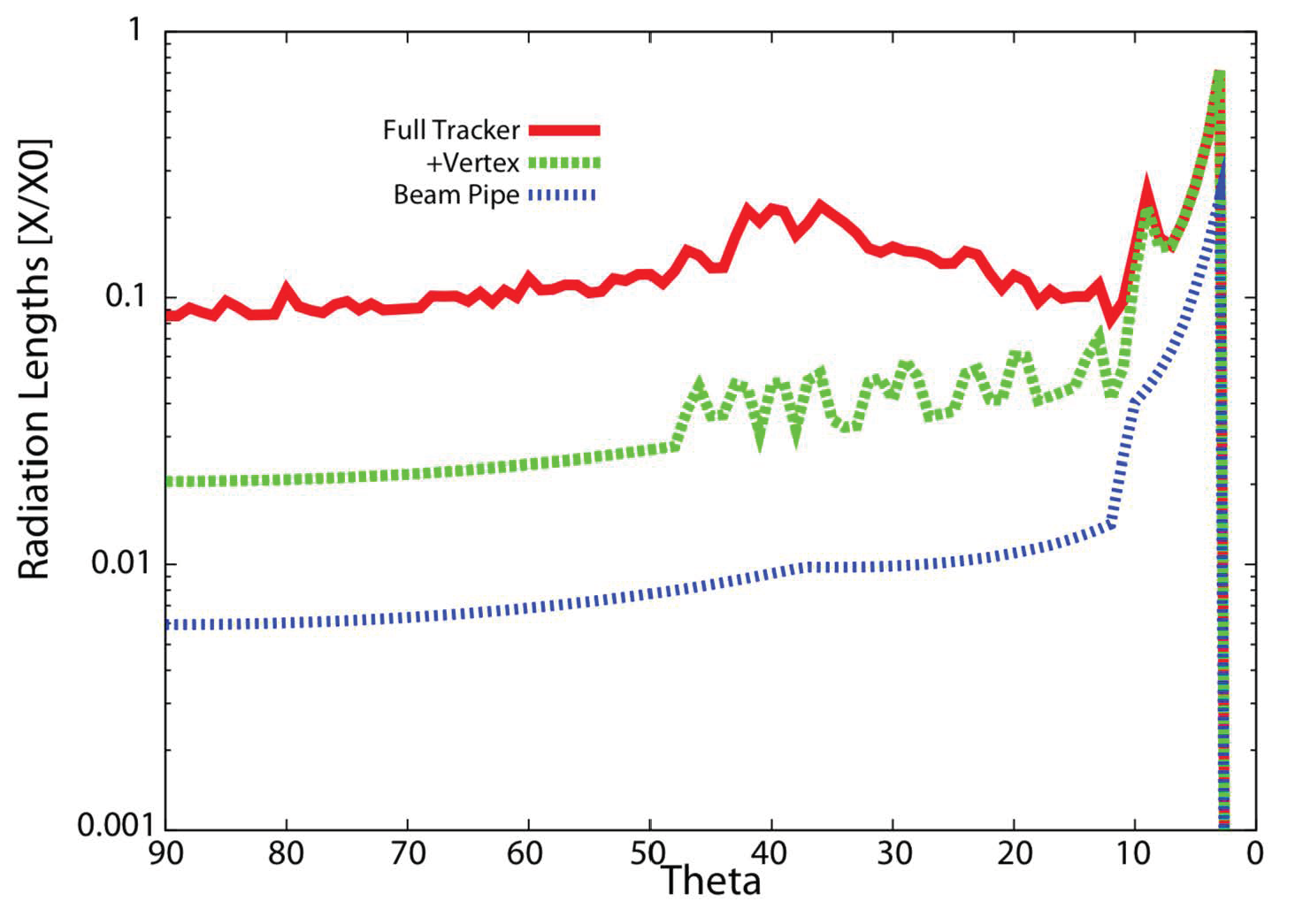}
\end{center}
\caption{Material in the SiD detector, in terms of fractions of a radiation length, as a function of the polar angle.}
\label{fig:sid_mat_budget}
\end{figure}

The main tracker technology of
choice is silicon strip sensors arrayed in five nested cylinders in the central
region and four disks following a conical surface with an angle of 5 degrees
with respect to the normal to the beamline in each of the end regions. The geometry of the endcaps
minimises the material budget to enhance forward tracking. The detectors are
single-sided silicon sensors, approximately 10 $\times$ 10 cm$^2$ with a readout
pitch of 50~$\micron$. The endcaps utilise two sensors bonded back-to-back for
small angle stereo measurements. With an outer cylinder radius of 1.25~m
and a 5~T field, the charged track momentum resolution will be better than
$\delta (1/\pT) = 5 \times 10^{-5} $/(GeV/$c$) for high momentum tracks with coverage down to polar angles of 10 degrees.
A plot of the material budget as a function of polar angle is shown in Fig.~\ref{fig:sid_mat_budget}.

%%%%%%%%%%%%%%%%
%%%%%%%%%%%%%%%%%

The all-silicon tracking approach has been extensively tested using full Monte-Carlo
simulations including full beam backgrounds. Besides having an excellent momentum resolution
it provides robust pattern recognition even in the presence of backgrounds and has a
real safety margin, if the machine backgrounds will be worse than expected.

\subsection{Main calorimeters}

The SiD  baseline design incorporates the elements needed to
successfully implement the PFA approach. This imposes a number of
basic requirements on the calorimetry. The central calorimeter
system must be contained within the solenoid in order to reliably associate
tracks to energy deposits. The electromagnetic and hadronic sections
must have imaging capabilities that allow both efficient
track-following and correct assignment of energy clusters to tracks. These
requirements imply that the calorimeters must be finely segmented both
longitudinally and transversely. In order to ensure that no significant amount
of energy can escape detection, the calorimetry must extend down to small
angles with respect to the beampipe and must be sufficiently deep to prevent
significant energy leakage. Since the average penetration depth of a hadronic
shower grows with its energy, the calorimeter system must be designed for the
highest-energy collisions envisaged.

In order to ease detector construction the calorimeter mechanical design consists of a series of modules of
manageable size and weight. The boundaries between
modules are kept as small as possible to prevent significant non-instrumented
regions. The detectors are designed to have excellent long-term stability and reliability,
since access during the data-taking period will be extremely limited, if not
impossible.

The combined ECAL and HCAL systems consist of a
central barrel part and two endcaps, nested inside the barrel. The entire barrel system is contained
within the volume of the cylindrical superconducting solenoid. 

%The
%electromagnetic calorimeter has silicon active layers between tungsten absorber
%layers. The active layers use 5$\times$5~mm$^2$ silicon pixels, which provide excellent spatial resolution.
%The structure has 30 layers in total, the first 20 layers having a
%thinner absorber than the last ten layers. This configuration is a 
%compromise between cost, electromagnetic shower radius, sampling frequency, and
%shower containment. The total depth of the electromagnetic calorimeter is 26
%radiation lengths (\xo) and one nuclear interaction length. 

SiD's reliance on particle flow calorimetry to obtain a jet energy resolution of  $\sim$3\% demands a highly segmented (longitudinally and laterally) electromagnetic calorimeter. It also calls for a minimized lateral electromagnetic shower size, by minimizing the Moliere radius to efficiently separate photons, electrons and charged hadrons.

The SiD ECal design employs thirty longitudinal layers, the first twenty each with 2.50 mm tungsten alloy thickness and 1.25 mm readout gaps, and the last ten with 5.00 mm tungsten alloy.  The total depth is 26 radiation lengths, providing good containment of electromagnetic showers.

Simulations have shown the energy resolution for electrons or photons
to be well described by 0.17 / $\sqrt{E}$ $\oplus$ 0.009, degrading a
bit  at higher energies due to changes in sampling fraction and a small leakage.

The baseline design employs tiled, large, commercially produced silicon sensors (currently assuming 15 cm wafers). The sensors are segmented into pixels that are individually read out over the full range of charge depositions. The complete electronics for the pixels is contained in a single chip, the KPiX ASIC~\cite{Brau:2013yb}, which is bump bonded to the wafer. The low beam-crossing duty cycle ($10^{-3}$) allows reducing the heat load using power pulsing, thus allowing passive thermal management within the ECal modules.

Bench tests of the KPiX bonded sensor  with a cosmic ray telescope trigger yielded 
a Landau distribution with a peak of the signal at about 4 fC is consistent with our expectation for minimum-ionizing particles (MIP) passing through the fully-depleted 320 $\mu$m thick sensors. Crosstalk between channels has been managed and the 
 noise distribution shows an RMS of 0.2 fC, well below the 4 fC MIP signal, and exceeding the ECal requirement.

The overall mechanical structure of the ECal barrel has been designed for minimal uninstrumented gaps. Input power and signals are delivered with Kapton flex cables.
The KPiX chip has an average power less than 20 mW, resulting in a
total heat load  that is managed with a cold plate and water pipes routed 
into the calorimeter.

A first SiD ECal prototype stack of nine (of thirty) layers has been constructed and was exposed to a 12.1 GeV electron beam at the SLAC End Station Test Beam Facility. 
This data collection demonstrated good measurements of multiple particle overlap and reconstruction of overlapping showers~\cite{Steinhebel:2017qze}.  Comparison of the deposited energy distribution in each of the nine layers also agrees well with simulations.
An algorithm developed to count the number of incident electrons in each event was used to assess the ability of the calorimeter to separate two showers as a function of the separation of the showers, achieving 100\% for separations of $>$10 mm.

The hadronic
calorimeter has a depth of 4.5 nuclear interaction lengths, consisting of
alternating steel plates and active layers. The baseline choice for the active
layers is scintillator tiles read out via silicon photomultipliers. For this approach SiD is closely following the analog hadron calorimeter developments within the CALICE collaboration. In this context, the simulated HCAL energy resolution has been shown to reproduce well the results from the CALICE AHCAL prototype module exposed to pion beams.

\subsection{Forward calorimeters}
\label{subsub:det:forward}
Two special calorimeters are foreseen in the very forward region: LumiCal for a precise luminosity measurement as discussed in Sec.~\ref{subsec:lumi_prec}, and BeamCal for the fast estimation of the collision parameters and tagging of forward-scattered beam particles. LumiCal and BeamCal are both compact cylindrical electromagnetic calorimeters centered on the outgoing beam, making use of semiconductor-tungsten technology. BeamCal is placed just in front of the final focus quadrupole and LumiCal is aligned with the electromagnetic calorimeter endcap. 

LumiCal makes use of conventional silicon diode sensor readout. It is a precision device with challenging requirements on the mechanics and position control, and must achieve a small Moliere radius to reach its precision targets. Substantial work has been done to thin the silicon sensor readout planes within the silicon-tungsten assembly. Dedicated electronics with an appropriately large dynamic range is under development.

BeamCal is exposed to a large flux of low-energy electron-positron pairs originating from beamstrahlung. These depositions, useful for a bunch-by-bunch luminosity estimate and the determination of beam parameters, require radiation hard sensors. The BeamCal has to cope with 100\% occupancies, requiring dedicated front-end electronics. A challenge for BeamCal is to identify sensors that will tolerate over one MGy of ionizing radiation per year. Sensor technologies under consideration include polycrystalline chemical vapor deposition (CVD) diamond (too expensive to be used for the full coverage), GaAs, SiC, Sapphire, and conventional silicon diode sensors. The radiation tolerance of all of these sensor technologies has been studied in a high-intensity electron beam. 

For SiD, the main activities are the study of these radiation-hard sensors, development of the first version of the so-called Bean readout chip, and the simulation of BeamCal tagging for physics studies. SiD coordinates these activities through its participation in the FCAL R\&D Collaboration.

\subsection{Magnet coil}

The SiD superconducting solenoid is based on the CMS solenoid
design philosophy and construction techniques, using a slightly modified CMS
conductor as its baseline design. Superconducting strand count in the coextruded
Rutherford cable was increased from 32 to 40 to accommodate the higher 5~T
central field. 

Many iron flux return configurations have been simulated in two
dimensions so as to reduce the fringe field. An Opera 3D calculation with the Detector
Integrated Dipole (DID) coil has been completed.
Calculations of magnetic field with a 3D ANSYS program
are in progress. These will have the capability to calculate forces and stress
on the DID as well as run transient cases to check the viability of using the
DID as a quench propagator for the solenoid. Field and force calculations with
an iron endcap HCAL were studied. The field homogeneity improvement was found
to be insufficient to pursue this option. 

Conceptual DID construction and
assembly methods have been studied. The solenoid electrical power system,
including a water-cooled dump resistor and grounding, was established.
Significant work has been expended on examining different conductor stabiliser
options and conductor fabrication methods. This work is pursued as a cost- and
time-saving effort for solenoid construction.

\subsection{Muon system}
The flux-return yoke is instrumented with position sensitive detectors to
serve as both a muon filter and a tail catcher. The total area to be
instrumented is very significant -- several thousand square meters. Technologies
that lend themselves to low-cost large-area detectors are therefore under
investigation. Particles arriving at the muon system have seen large amounts of
material in the calorimeters and encounter significant multiple scattering
inside the iron. Spatial resolution of a few centimetres is therefore
sufficient. Occupancies are low, so strip detectors are possible. The SiD 
baseline design uses scintillator technology, with RPCs as an alternative. 
The scintillator technology uses extruded scintillator readout with wavelength 
shifting fibre and SiPMs, and has been successfully demonstrated. 
Simulation studies have shown that nine or more layers of sensitive detectors 
yield adequate energy measurements and good muon detection efficiency and purity.
The flux-return yoke itself has been optimised with respect to the
uniformity of the central solenoidal field, the external fringe field,
and ease of the iron assembly. 
This was achieved by separating the  barrel and end sections of the
yoke along a 30 degree line.

\subsection{The machine-detector interface}
A time-efficient implementation of the push-pull model of
operation sets specific requirements and challenges for many detector and
machine systems, in particular the interaction region (IR) magnets, the
cryogenics, the alignment system, the beamline shielding, the detector design
and the overall integration. The minimal functional requirements and interface
specifications for the push-pull IR have been successfully developed and
published~\cite{Parker:2009zz,Buesser:2012et}.  All further IR design
work on both the detectors and machine sides are constrained by these 
specifications.

\subsection{R\&D issues for the SiD design}
\label{SiD-RandD}

\subsubsection{Monolithic Active Pixel Sensors}
MAPS technology is being actively studied for the \sid~ tracking and electromagnetic calorimeter systems, with initial prototyping underway.
For larger-scale objects like a full tracker or an ECAL sensor, larger structures than the usual full-reticle size (roughly 2.5 $\times$ {2.5}{cm$^2$}) units are required.
Reticles would be stitched together with balcony circuitry on one or two (opposing) edges. 

In terms of general MAPS R\&D required for \sid, mastering of the stitching technology is required, 
as for such large areas - O($\sim${100}{m$^2$}) for the 
tracker and  O($\sim${1000}{m$^2$}) for the ECAL - yield becomes an issue. The distribution of power and data over such a large area sensor will be a challenge 
as well and dedicated R\&D is needed.

Given the timescales involved for the construction of an ILC detector like \sid , with the mainstay of construction happening at the end of decade, 
investment into new processes are needed, as the presently available processes will most likely not be available anymore. The most probable technology 
for a next-generation MAPS process are the $\sim${65}{nm} CMOS processes that are just becoming available to the community. As CMOS processes use larger wafers 
(ten or twelve inch wafers) as well as taking advantage of a fully industrial process, the move to MAPS also has clear advantages in terms of a cost reduction for both the 
tracker and the ECAL.

Simulation studies of electromagnetic showers have demonstrated that the ILC TDR level resolutions, and even better, can be achieved with a digital hit/no-hit threshold MAPS ECAL.~\cite{Brau:2021} The pixel structure of \SI{25}{\micro\meter} x \SI{100}{\micro\meter} is chosen to optimize tracking and ECAL applications.

\subsubsection{Hadron Calorimeter}
Extensions to and optimization of the hadron calorimeter design will also address the following:
 \begin{itemize}
\item inclusion of timing layers to assist the particle flow algorithm in separating the delayed shower components from slow neutrons from the prompt components.
 
\item potential cost saving by making some of the outer layers thicker if there is no significant degradation in energy resolution.

\item optimization of the boundary region between the ECAL and the HCAL and optimization of the first layers of the HCAL to best assist with the measurement of electromagnetic shower leakage into the HCAL.

\item reconsideration of the effects of projective cracks between modules. There is some indication from earlier studies that projective cracks have no negative effect on energy resolution, but this needs further verification.

\item exploration of alternative layouts for HCAL sectors in the end-caps.

\item optimization of the boundary between the HCAL barrel and end-caps.

\end{itemize}

\subsubsection{Muon system}
\begin{itemize}
\item Optimization of number of instrumented layers, barrel and end-caps.
\item Optimization of strip lengths, mainly for barrel system.
\item Design for muon endcaps - twelve-fold geometry.
\item Occupancies at inner radius of muon end-caps versus strip widths.
\item Role of muon system as tail-catcher for HCAL. Consideration and implications of CALICE ECAL + HCAL + Tail-catcher test beam results.
\item Potential for use of muon system in search for long-lived particles; timing and pointing capabilities.
\end{itemize}
 
\subsubsection{Forward Calorimeters}
Tasks remaining for the forward calorimeters, with participation in the FCAL R\&D 
Collaboration, include:

\begin{itemize}
\item LumiCal: complete development of large dynamic range readout electronics.

\item LumiCal: develop and demonstrate the ability to position and maintain the position of the calorimeter, particularly at the inner 
radius, in view of the steep dependence of the rate of Bhabha events on polar angle.

\item BeamCal: continue the search for and testing of suiTable sensor technology(s) capable of sustained performance in the very high radiation environment.

\item BeamCal: continue the study of recognizing single electron shower patterns for tagging for physics studies in the face of high radiation background.
\end{itemize}

\section{New Technologies for ILC Detectors} 
\label{sec:detectortech}

\subsection{Introduction}
\label{sec:det-intro}

The global particle-physics community continues to develop new ideas for improved sensors and detector systems. In this section, several promising new developments are briefly discussed. Some of these are new technologies that can be integrated in the existing detector concepts, others represent alternatives to the baseline choices made by ILD and SiD.

Since funding for detector R\&D is scarce, it is important that the global program covers the essential R\&D for the ILC. In Europe, CERN~\cite{Aglieri:2764386} and the ECFA detector R\&D panel~\cite{Detector:2784893} have published road maps for the effort in instrumentation. A large EU Horizon 2020 project, AIDA Innova~\cite{AIDAINNOVA}, unites the effort of seven European national laboratories, 30 universities and institutes and eight industrial partners. In the US, important directions for detector R\&D are outlined in the report of the Office of Science Workshop on Basic Research Needs for HEP Detector Research and Development~\cite{osti_1659761}. The "instrumentation frontier group" in the Snowmass process will draft a road map for detector R\&D in the US. 

Especially important is the synergy with detector construction projects on the intermediate time scale. These projects can validate promising new ideas, with sufficient resources for complete engineering designs and extensive prototyping. The construction phase provides valuable lessons about their practicality in large-scale production. We envisage that projects such as the upgrades of the LHC experiments, and the construction of specialized experiments such as Mu3e and experiments at FAIR and the EIC can act as ``stepping stones" in the development of the optimal solutions for the ILC experiments. Smaller experiments, such as for example the LUXE experiment proposed at DESY, might provide an interesting platform to test specific technologies~\cite{Abramowicz:2021zja}.

\subsection{Low-mass support structures for Silicon trackers}
\label{sec:det-thinsilicon}

The very strict performance requirements of the silicon tracking systems and vertex detectors has pushed the field to develop active and monolithic silicon sensors that can be thinned to 50 $\mu\mathrm{m}$ or less. To build a superb transparent tracking system this innovation in silicon sensors must be accompanied by important advances in the support structures and cooling systems that make a very important contribution to the material of today's state-of-the-art detector systems. Integrated support and cooling solutions are required to meet the very challenging material budget of the ILC experiments.

An important step towards the integration of support structures was made by the DEPFET collaboration~\cite{DEPFET:2012apm}, with the development of the all-silicon ladder concept~\cite{Fischer:2007zzm}. In this ladder design, all on-detector electronics and power and signal lines are integrated on the silicon sensor itself. A robust and stiff mechanical structure is obtained by selective etching of the handle wafer, such that an integrated ``support frame" surrounds the thin sensor. The all-silicon ladder concept was proven in the Belle 2 vertex detector~\cite{Belle-II:2010dht}. Similar self-supporting all-silicon structures can be produced for CMOS active pixel sensors by stitching multiple reticles. The development of the CMOS multi-chip ladder is part of the R\&D for the upgrade of vertex detector envisaged in 2027 or 2028. 

The PLUME collaboration has developed a double-sided CMOS ladder concept. The ladder
design follows a classical approach: six sensors are connected to a low-mass flex-cable to
form a module, then two modules are glued on both sides of a mechanical support to form
the double-sided ladder. Most of the stiffness of this sandwich-type layer stems from the
two modules rather than from the support, which serves essentially as a spacer and is made of a low-density open Silicon Carbide foam~\cite{Baudot:2012mg}. 

A more aggressive approach is followed by the Mu3e experiment~\cite{Mu3e:2020gyw} that envisages a Kapton support structure for their thinned CMOS sensors. The ALICE upgrade of the Inner Tracking System~\cite{ALICE:2013nwm} envisage CMOS sensors thinned to approximately 50~$\mu\mathrm{m}$. Innovative solutions to the support structures are being pursued, including a study of large, stitched sensors that are thinned and bent to form cylindrical structures around the beam pipe. The experience gained in these construction projects can have important implications in the ILC vertex detector and tracker design.

\subsection{Integrated micro-channel cooling}
\label{sec:det-micro-channels}

The cooling of these ultra-low-mass detector systems represents an important challenge.
Cooling by a loosely guided gas flow has been demonstrated by the heavy flavour tagger in the STAR experiment~\cite{STAR:2002eio}. Gas-based cooling is also used to complement a traditional bi-phase cooling system in the Belle 2 pixel detector. The heat generated by the pixel sensors is effectively removed by a gas flow at several meters per second. Tests of the mechanical stability of prototypes in gas flows have been performed at CERN by ALICE and CLIC and at DESY by Belle 2. A facility is available for users at the University of Oxford under AIDA innova funding. The magnitude of vibrations induced by the gas flow in realistic prototypes can be kept at the level of a few $\mu\mathrm{m}$.

Micro-channel cooling promises to bring down the material involved in traditional liquid or bi-phase cooling systems. The use of active silicon cooling plates has been pioneered by the NA62 experiment~\cite{NA62:2017rwk} that has operated the GigaTracker successfully for several years. Micro-channel cooling with evaporative $CO_2$ at pressures up to 60 bar is part of the Vertex Locator upgrade of the LHCb experiment. The production of VELO modules based on hybrid pixel detectors glued onto silicon micro-channel cooling plates produced at CEA-LETI was successfully completed in 2021~\cite{Francisco:2021tda}. Installation in the LHCb experiment was still ongoing at the time of writing. Integration of micro-channels directly in the active sensor wafer~\cite{Andricek:2016rsq, Mapelli:2712079} offers the best possible cooling contact, with a thermal Figure-of-merit close to 1~K/W. 

Micro-channel cooling is being considered for FCC-ee~\cite{Barchetta:2021ibt}, where a vertex detector with fast read-out could be positioned close to the beam. In combination with a relatively high-temperature cooling system based on super-critical $CO_{2}$, it might offer a relatively low-mass solution, that brings better control of the temperature than can be achieved with a forced gas flow. The engineering and implications on the material budget need to be studied further.

\subsection{Dual read-out calorimetry}
\label{sec:det-dual-readout-calorimetry}

The 20-year-long $R\&D$ program on Dual-Readout Calorimetry (DR, DRC) of the DREAM/RD52 collaboration~\cite{DRC_Wigmans,RD52_emshow,RD52_emperf,RD52_hadperf,RD52_MC,RD52_PID,RD52_SiPM,IDEA_SiPM} has shown that  the effects of the fluctuations in the EM fraction of hadronic showers can be canceled by the independent readout of scintillation (S) and \v{C}erenkov (C) light. The DR fibre-sampling approach achieves a high sampling frequency leading to a competitive EM energy resolution $\sim 10\%/\sqrt{E}$. Application of the DR  procedure gives a stochastic term of the hadronic resolution close to or even below $30\%/\sqrt{E}$ with a small constant term. Test beam results also show excellent particle-ID performance.

The advancements in solid-state light sensors such as SiPMs have opened the way for highly granular fibre-sampling detectors with the capability to resolve the shower angular position at the mrad level or even better.
In the present design 1-mm diameter fibres are placed at a distance  of 1.5-2 mm (center to center) in a metal absorber. Brass, copper, iron and lead are currently under study. The lateral segmentation could then reach the mm level, largely enhancing the resolving power for close-by showers with a significant impact on $\pi^0$ and $\tau$ reconstruction quality. In addition the high Photon Detection Efficiency of SiPMs provide high light yields, thus reducing the effect of photon statistics.
Readout ASICs providing time information with $\sim$~100 ps resolution may allow the reconstruction of the shower position with $\sim$~5 cm of longitudinal resolution.

The large number and density of channels call for an innovative readout architecture for efficient information extraction. Both charge-integrating and waveform-sampling ASICs are available on the market and candidates for tests have been identified: the Weeroc Citiroc 1A charge integrator and the Nalu Scientific system-on-chip digitisers.  A first implementation of a scalable readout system has been tested with a calorimeter prototype on particle beams. Looking further ahead, digital SiPMs (dSiPMs) could provide significant simplification of the readout architecture, but the technology is still in an early $R\&D$ phase.

The performance of a 4$\pi$  DR calorimeter for an FCC-ee experiment has been studied with full GEANT4 simulation with good results on key physics processes.  This is now the baseline choice for the IDEA~\cite{IDEA_tb1} detector concept. Significant performance improvements have also been shown using deep-learning algorithms. Studies of the potential addition of a dual-readout crystal calorimeter in front find superb EM resolution, while maintaining the hadronic performance and even improving it by applying simple particle flow algorithms~\cite{Lucchini_2020}. A more detailed description is found in Ref.~\cite{Aleksa:2021ztd}.

\subsection{Crystal electromagnetic calorimetry}

As noted above, the CALICE and RD52/DREAM collaborations have demonstrated that both designs can achieve a jet energy resolution of 3-4\% for jets expected from $W/Z\to qq$ decays~\cite{Sefkow:2015hna,Antonello:2021tsz}. However, the EM energy resolution is expected to be $\sim 15\%/\sqrt{E}$ for Particle Flow  and $\sim 10\%/\sqrt{E}$ for DREAM, largely because of the small sampling fractions. These resolutions are significantly worse than those of crystal ECALs~\cite{L3BGO:1993tta,CMS:2013lxn}. Thus, it is interesting to study the combination of the DREAM fiber HCAL with a crystal ECAL.   This can potentially  maintain or even improve the jet energy resolution while attaining the EM resolution of $<3\%/\sqrt{E}$~\cite{Lucchini:2020bac}. A consortium of US teams is leading this R\&D.

\subsection{Liquid Argon calorimetry}

Noble-liquid calorimeters have been successfully used in many high-energy collider experiments, such as ATLAS, D0 or H1. They feature high energy resolution, excellent linearity, uniformity, stability, and radiation hardness. These properties make a noble-liquid calorimeter an appealing candidate for an experiment at the next-generation Higgs factories, especially in the case of a program of high precision physics at the $Z$ pole where an excellent control of the systematic uncertainties is required to match statistical precisions as low as $10^{-5}$.

A design of a highly granular sampling noble liquid calorimeter was first proposed in the context of a FCC-hh experiment~\cite{Aleksa:2019pvl}, and is now being revisited and optimised for a Higgs factory machine. In the central region, it consists of a cylindrical stack of 1536 lead absorbers (2mm thick), readout electrodes (1.2mm thick) and liquid argon active gaps, arranged radially but azimuthally inclined by $\sim 50^o$ with respect to the radial direction. This design allows for reading out the signals without creating any gaps in the acceptance, high sampling frequency, uniformity in $\phi$, and can be build with very good mechanical precision to minimise the constant term of the energy resolution. The use of liquid krypton as active material and of tungsten absorbers is being studied as it could result in a more compact design with better shower separation.

The use of readout electrodes allows to optimise the granularity of each of the 11 longitudinal compartments for the needs of particle-flow reconstruction and particle-ID. A total number of a few million cells can be achieved by using seven-layer PCBs, where the outermost layers provide the high voltage field in the noble-liquid gap, and next layers are signal pads, connected to the central layer where traces bring the signals to the outer edges of the electrodes. The trade-offs between granularity, noise and cross-talk in the design of the PCBs are being studied.

The expected noise levels assuming readout electronics outside the cryostat should allow
the tracking of single particles and yield a total noise of about 50\,MeV for an
electromagnetic cluster. The alternative of using cold readout electronics placed inside
the cryostat is also studied as it would achieve a much lower noise, and could simplify
the design of the feedthroughs. R\&D on high-density feedthroughs is indeed ongoing to
allow the analogue readout of millions of channels without any performance degradation.
A reduction of the amount of dead material in front of the calorimeter can be achieved
thanks to the progress on 'transparent' cryostats using carbon or sandwiches of
materials.

Better estimates of the expected performance (using the calorimeter alone and with particle-flow reconstruction), and answers on the feasibility of the designs of the PCBs, the readout electronics and the feedthroughs, will be available in the next months and years.

\subsection{Digital pixel calorimetry}

Initial proof-of-concept demonstrations of the use of digital electromagnetic calorimetry (DECAL) were made in the framework of ILC detector development \cite{Ballin:2008db,Ballin:2009yv,Dauncey:2010zz}.  The first proof-of-principle of a DECAL was made in the context of the ALICE experiment forward calorimeter proposal (FoCal) \cite{ALICE:2020mso}, with the design and fabrication of a multiple-layer prototype and corresponding measurements, proving the viability of the concept \cite{deHaas:2017fkf}.

The fundamental principle underlying a DECAL is to measure energy by counting the number of charged shower particles using very high transverse granularity sampling layers.  To avoid saturation effects and ensure competitive resolution and linearity, binary-readout CMOS pixels are used; these must be sufficiently small that the double-hit probability is negligible even in the core of high-energy electromagnetic showers.  The small pixel size also has clear benefits in dense particle environments for pattern recognition algorithms such as particle flow, e.g.\ \cite{Brient:2001fow}.

Digital calorimeters use a sandwich structure of silicon and tungsten layers, with Monolithic Active Pixel Sensors (MAPS) being a natural choice in terms of granularity and cost.
%% , e.g.\ the use of the TPAC sensors \cite{Dauncey3} for . 
The proof-of-principle prototype \cite{deHaas:2017fkf}, also called EPICAL-1, required a total sensor area of almost 400~cm$^2$, and therefore the  PHASE2/MIMOSA23 chip from IPHC \cite{Winter:2010zz} with a pixel size of $30 \times 30 \, \mu \mathrm{m}^2$ was used for this R\&D.
%, despite being too slow for use in an experiment, 

Current R\&D is performed using a second generation DECAL prototype, the EPICAL-2, which has an active area of approx.\ $3\times 3$~cm$^{2}$ per sensitive layer. These comprise state-of-the-art ALPIDE sensors, developed for the ALICE ITS and MFT \cite{AglieriRinella:2017lym}, which have a similar pixel size to the MIMOSA.  Measurements with this prototype have been performed with cosmic muons in the lab, and with test beams at both DESY in 2020 and the CERN SPS in 2021. This prototype has performed extremely well, surpassing the performance of EPICAL-1. The substantial experience with two prototypes using sensors from different developers and foundries has reliably demonstrated that the technology is a very good candidate for future calorimeters.

Full analysis of data from EPICAL-1 and preliminary results from EPICAL-2 show:
\begin{enumerate}
\item CMOS MAPS sensors work reliably in the high particle density environment of high-energy electromagnetic showers;
\item there are no substantial saturation effects due to shower particle overlap in the pixels up to energies of at least a few hundred GeV;
\item the energy resolution from test beam results at DESY is very similar to that from state-of-the-art in analogue silicon-tungsten calorimeters;
\item  the single-shower position resolution is of the order of the pixel size or better.  
\end{enumerate}
The data obtained with EPICAL-2 will improve the understanding of the detection process and allow the development of improved reconstruction algorithms, including the correction of possible residual saturation effects.

While the work discussed above concentrates on a small scale prototype to further develop the new technology, parallel R\&D activities are ongoing to solve the challenges related to scaling this up to a reasonable size. A very similar detector concept using the same technology is the basis for a development of a proton CT scanner for medical applications \cite{Alme2020}---this addresses the development of large area pixel sensor layers for use in a calorimeter, including full connectivity and services.

\subsection{Low gain avalanche detectors}
\label{sec:det-timing}

Low-Gain Avalanche Diode (LGAD) sensors~\cite{Pellegrini:2014lki} are a promising technology that, due to their intrinsic signal amplification, could significantly reduce the sensor substrate thickness and hence the material budget of the Silicon tracking systems of the ILC experiments. The large signal-to-noise ratio and short rise time of the LGAD signal make them also suitable for the precise time stamping of charged particles~\cite{Sadrozinski:2016xxe,Cartiglia:2016voy}. A tracker system based on these technologies could provide high-precision tracking and a timing resolution of the order of tens of picoseconds that would significantly enhance particle-identification capabilities, in particular for low-momentum charged-particle tracks.

LGADs are the current reference technology for timing detectors for charged particles in preparation for the high-luminosity upgrades of the LHC experiments. Large-area timing detectors based on this technology are envisaged for the CMS~\cite{Butler:2019rpu} and ATLAS~\cite{Allaire:2018bof} HL-LHC upgrades.

The adaptation of LGAD technology to the requirements of a high-precision tracking system for the ILC detectors involves two main challenges with respect to the current state of the art: high fill factor and large area detector fabrication. Dedicated R \& D activities are being carried out to address these challenges on the basis of specialized developments of the LGAD concept: inverted LGAD (iLGAD~\cite{Pellegrini:2014lki,Curras:2019aky}, trench-isolated LGAD (TI-LGAD~\cite{paternoster}), and AC-coupled LGADs (AC-LGAD or RSD~\cite{Mandurrino:2019csy}).

%[1] G.Pellegrini and et al., Recent technological developments on LGAD and iLGAD detectors for tracking and timing applications, Nucl. Inst. Meth. A 831 (2016) 24{28.
%[2] G. Paternoster and et al., Trench-Isolated Low Gain Avalanche Diodes (TI-LGADs), IEEE Electr. Device L. 41 (2020) no. 6, 884.

%[3] M. Mandurrino and et al., Demonstration of 200-, 100-, and 50- um Pitch Resistive AC-Coupled Silicon Detectors (RSD) With 100% Fill-Factor for 4D Particle Tracking, IEEE Electr. Device L. 40 (2019) no. 11, 1780.

\subsection{New sensor technologies for highly compact electromagnetic calorimeters }

%The preferred technology for luminometers for future electron-positron Colliders are highly compact and finely grained sandwich calorimeters, used to measure precisely the number of low-angle Bhabha scattering events. A small Moliere radius is essential to match the requirements on performance and keep the fiducial volume small. Tungsten will serve as absorber material. Essential is to keep the gaps for sensor planes small. Ultra-thin GaAs detector planes have been developed, with pad sizes of 5$\times$5 mm$^2$ and read-out traces integrated on the sensor.  

The luminosity is a key parameter of any collider. For electron-positron colliders Bhabha scattering at small angles is the gauge process to approach a precision
of 10${-3}$ or better. To count Bhabha events compact electromagnetic calorimeters are the preferred technology. A small Moliere radius is of advantage, in particular in the presence of background. In addition, it keeps the size of the calorimeters small, and allows to define precisely the fiducial volume, important for the precision of the measurement. Tungsten is an absorber material with a very small Moliere radius. To instrument the calorimeter with sensors, gaps between tungsten plates have to be foreseen. In order to keep the Moliere radius near that of tungsten, these gaps must be very small, requiring thin assembled sensor planes. For this purpose GaAs sensors with aluminum traces integrated on the sensors are developed. These traces connect the sensor pads with bonding pads on the edge of the sensor. A flexible Kapton PCB with copper traces is bonded to the sensor and feeds the signals to the FE ASICs.     

\begin{figure}[htb]
\begin{center}
    \includegraphics[width=0.45\columnwidth]{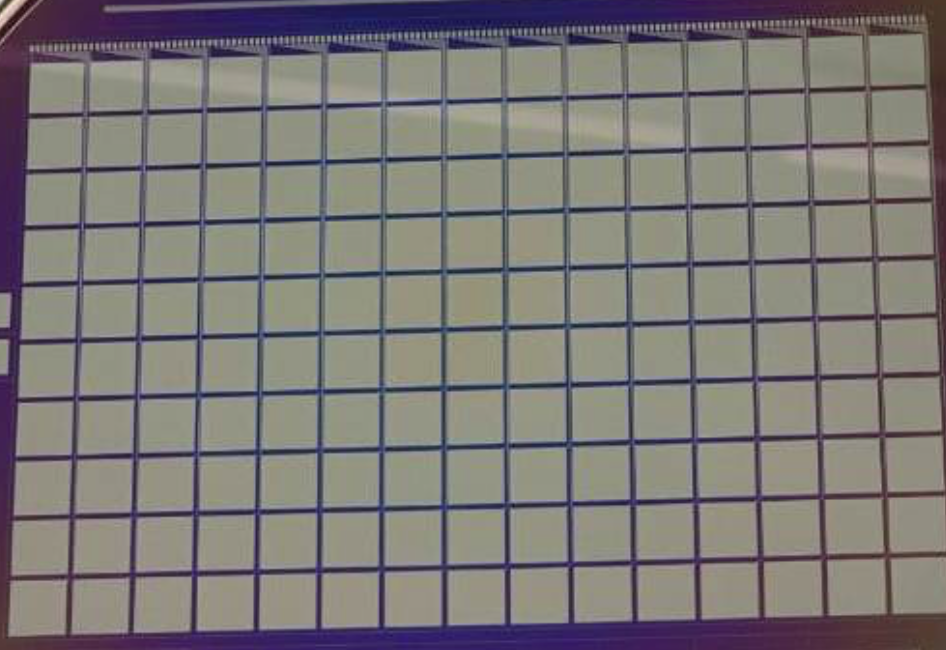}
    \hfill
    \includegraphics[width=0.45\columnwidth]{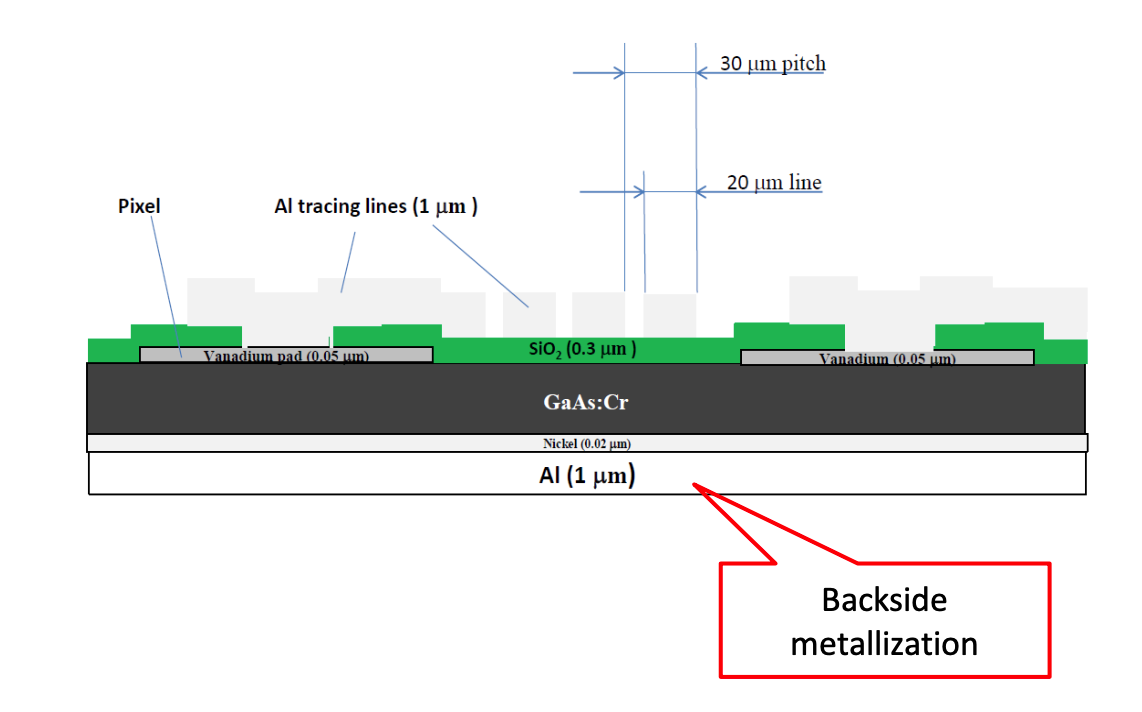}
    \caption{Left: Picture of a GaAs sensor. The bond pads are visible on top of the sensor, Right: Cross-profile of a GaAs sensor. The aluminum traces are positioned between the pads, on the top of the passivation layer.}
    \label{ECAL_GAAS_pic}
  \hspace{0.025\textwidth}
  \end{center}
\end{figure}

GaAs sensors are made of single crystals. High resistivity of $10^9 \Omega {\rm m}$ is reached by compensation with chromium. The pads are $4.7 \times 4.7$~mm$^2$, with $0.3$~mm gap between pads. Pads consist of a $0.05 \micron$ vanadium layer, covered with a $1 \micron$ aluminum, made with electron beam evaporation and magnetron sputtering. The back-plane is made of nickel and aluminum of $0.02$ and $1 \micron$ thickness, respectively. 
The sensors are $550 \micron$ thick with overall sizes of $51.9 \times 75.6$~mm$^2$.  The active area is $74.7 \times 49.7$~mm$^2$ leading to $15\times 10$ pads without guard rings. 
The signals from the pads are routed to bond pads on the top edge of the sensor      
by aluminum traces implemented on the sensor itself, thus avoiding the presence of a flexible PCB fanout.
The traces are made of $1 \micron$ thick aluminum film deposited on the silicon dioxide passivation layer by means of magnetron sputtering. 
A prototype sensor is shown in Fig.~\ref{ECAL_GAAS_pic} (left).

\begin{figure}[ht!]
\begin{center}
    \includegraphics[width=0.35\columnwidth]{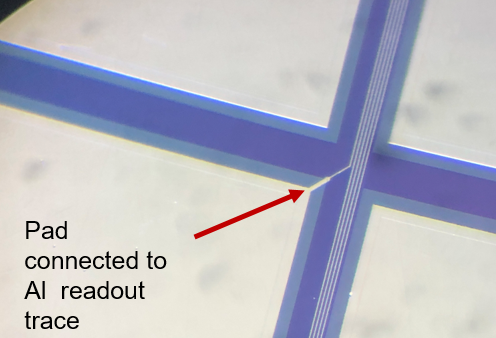}
    \hfill
    \includegraphics[width=0.55\columnwidth]{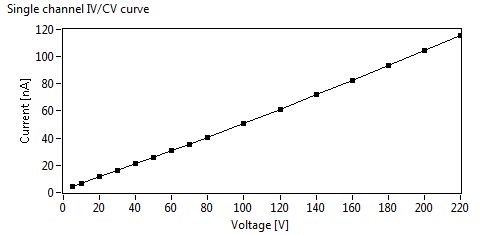}
    \caption{Left: Picture of the surface of a GaAs sensor. The aluminum traces are positioned between the pads, on the top of the passivation layer, Right: The leakage current of a pad as a function of the bias voltage, measured at $20^\circ$C.}
    \label{ECAL_traces_detail}
\end{center}
\end{figure}

Details on the sensor structure can be seen in the cross-profile shown in Fig.~\ref{ECAL_GAAS_pic} (right).
More details on the aluminum traces are illustrated in Fig.~\ref{ECAL_traces_detail}(left).

Using several prototype sensors, the leakage current of all pads was measured as a function of the bias voltage. A typical example is shown in Fig.~\ref{ECAL_traces_detail}(right). At a bias voltage of 100~V the leakage current amounts to about 50~nA.
\begin{figure}[htb]
\begin{center}
    \includegraphics[width=0.6\columnwidth]{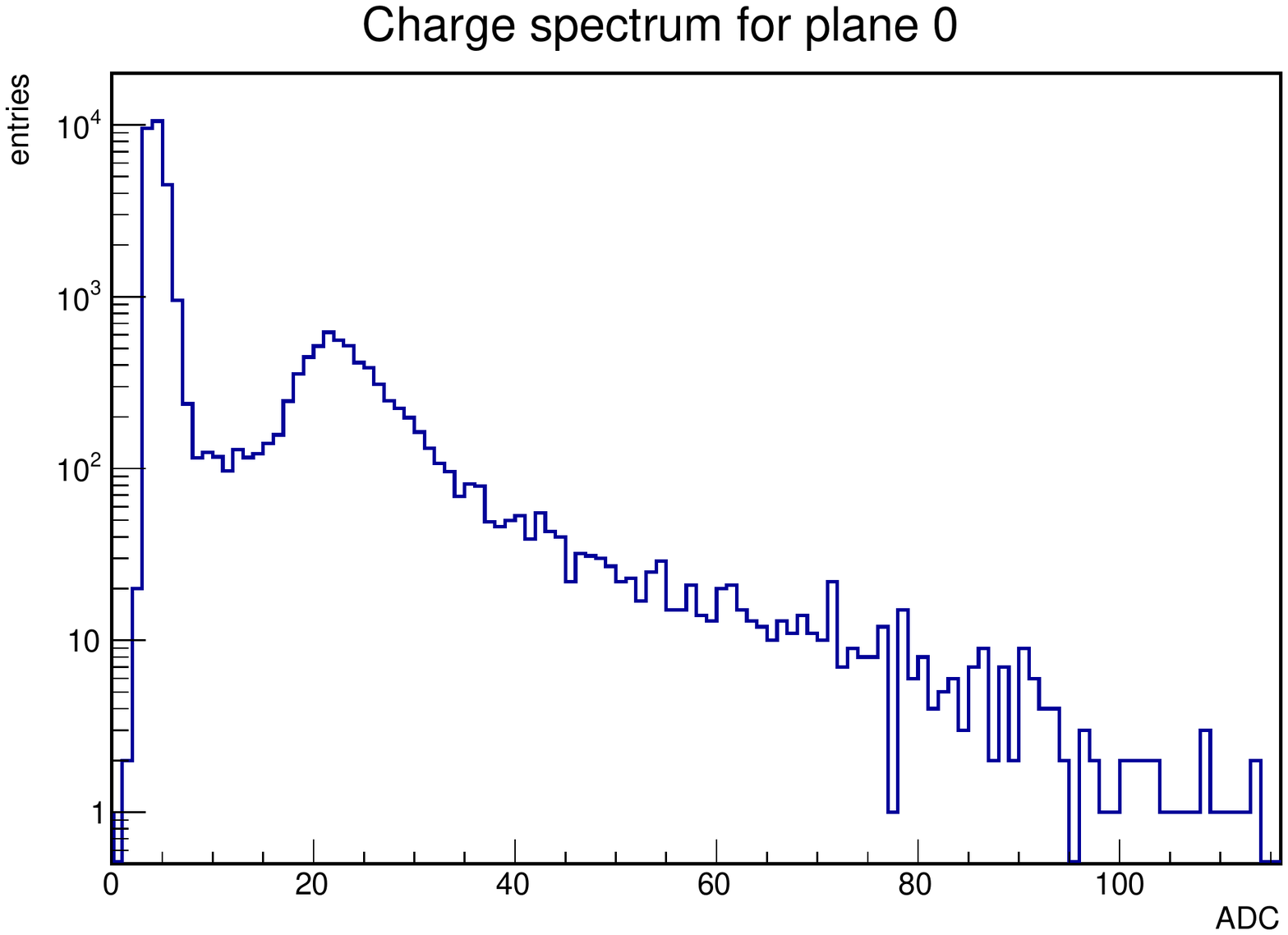}
 \caption{Distribution of signals measured with the GaAs pad sensors in an electron beam of 5~GeV.}
   \label{ECAL_signal}
\end{center}
\end{figure}

The implementation of the aluminum traces is a new technology. The response to relativistic electrons was measured in a test-beam of 5~GeV at DESY. A clear signal was observed, as can be seen in Fig.~\ref{ECAL_signal}

The first results from test-beam measurements are very promising. detailed studies on homogeneity of the response, and cross talk are still ongoing. Currently these sensors are the baseline option for the electromagnetic calorimeter of the LUXE experiment.

\subsection{Single crystal sapphire sensors for charged particle detection }

For the operation in a harsh radiation environment, typical for near-beam detectors at LHC or free electron lasers like FLASH and XFEL, extremely radiation hard sensors are needed. In the past often CVD grown diamond sensors are used in such environment.
%\cite[1, 2]. 
Regardless of the excellent radiation hardness and low leakage current at room temperature, the application of diamond sensors is limited due to high cost, relatively small size and low manufacturing rate. As an alternative we suggest using sapphire sensors. Optical grade single crystal sapphire is industrially grown in practically unlimited amounts and the wafers are of large size and low cost. Sapphire sensors have been used so far in cases where the signal is generated by simultaneous hits of many particles, i.e., in the beam halo measurement at at FLASH, XFEL and the CMS experiment at the LHC. It was found that the time characteristics of signals from sapphire sensors are similar to the ones from CVD diamond sensors
%\cite[2]. 
The radiation hardness of sapphire sensors was studied in a low
energy electron beam up to an absorbed dose of 12~MGy. 
%\cite[3]. 

A key parameter of the measurements is the Charge collection efficiency, CCE, defined as the ratio of the measured to the expected signal charge\footnote{The CCE corresponds to the effective drift path of charge carriers released by an ionising particle in the electric field in the sensor volume.} The expected signal is determined from the energy loss on 5~GeV electrons in sapphire, and the energy needed to create an electron-hole pair.
The  detector is composed of metallized sapphire plates of 10×10 mm$^2$ area and 525~$\micron$ thickness. The total thickness of this detector amounts to 14\% of a radiation length. Since the response is depending on the direction with respect to the plane axis of the particles crossing it, interesting fields of applications are beam-halo rate or low angle scattering measurements. Basic characteristics, like the dependence of the CCE on the applied voltage and position resolved sensor response. More details can be found in Ref.~\cite{Karacheban:2015jga}.

Sapphire is a crystal of aluminum oxide, Al$_2$O$_3$3. Wafers were obtained from the CRYSTAL company. 
%\cite[4]. 
Single crystal ingots were produced using the Czochralski method and
cut into wafers of 525~$\micron$ thickness. Contamination of other elements are on the level of a few ppm. The wafer was cut into quadratic sensors. Each sensor has dimensions $10 \time 10 \time 0.525$~mm$^3$, metallized on both sides with consecutive layers of Al, Pt and Au of 50~nm, 50~nm and 200~nm thickness, respectively.

\begin{figure}[ht!]
\begin{center}
    \includegraphics[width=0.45\columnwidth]{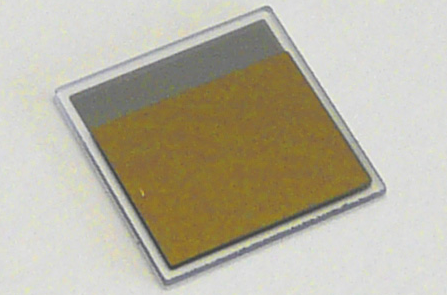}
    \includegraphics[width=0.45\columnwidth]{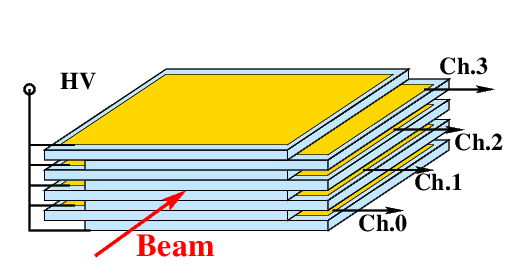}
    \caption{Left: Picture of a metallised sapphire sensor, Right: Schematic view of the sapphire sensor stack, consisting of 8 sensors.The direction of the beam electrons is indicated by the arrow and represents the z-coordinate. The y coordinate is perpendicular to the sensor plane.}
    \label{sapphire_stack}
  \hspace{0.025\textwidth}
  \end{center}
\end{figure}

To enhance the signal size, the orientation of the sapphire plates
in the test beam measurements was chosen to be parallel to the beam direction. In addition, this orientation leads to a direction sensitivity. Only particles crossing fully the sensor parallel to the surface create the maximum signal. A stack of eight plates were assembled together, as shown in Fig.~\ref{sapphire_stack} (left).
The leakage current of the sensors was measured as a function of the bias voltage. It amounts to less that 10~pA at 1000~V. 

In a first measurement sensors were exposed to a high-intensity electron
beam at the linear accelerator DALINAC at TU Darmstadt. The beam energy was
8.5~MeV. The response of the sensors was measured
as the signal current. The relative drop of the signal current, interpreted as the relative drop
of the charge collection efficiency, CCE, is about 30\% of the initial CCE after a dose of 12~MGy.

\begin{figure}[t]
\begin{center}
    \includegraphics[width=0.45\columnwidth]{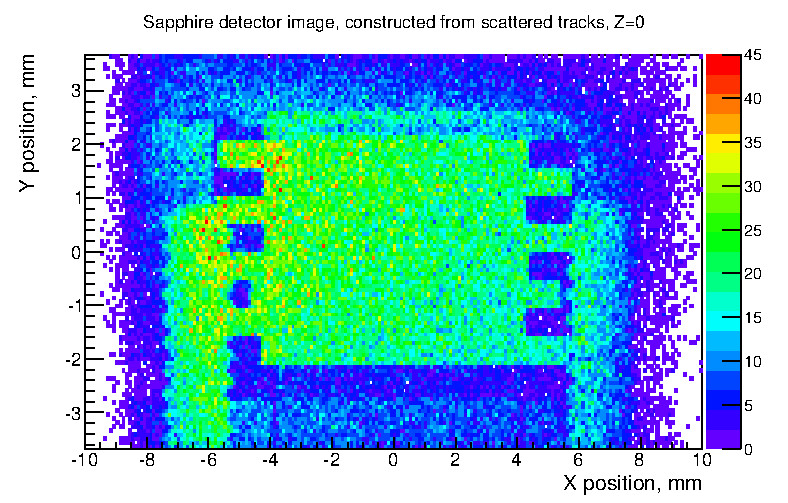}
    \caption{Image of the stack using electron tomography}
    \label{sapphire_stack_image}
  \hspace{0.025\textwidth}
  \end{center}
\end{figure}

The stack, as shown in Fig.~\ref{sapphire_stack} (left), was studied in a 5~GeV electron beam at DESY. The trajectory of each beam electron was precisely measured in a pixel telescope before and after the stack. The impact point on the stack was predicted with a precision of better that 10$\mu$m, and the scattering angle with a precision better that 50 $\mu$rad.
Several millions of triggers were recorded at several bias voltages.
Firstly, an electron tomographic picture of the stack was taken, as shown in Fig.~\ref{sapphire_stack_image}.
In this Figure the density of impact points is shown only for beam electrons deflected by an angle larger than 0.5~mrad. The position of the 5 sensor plates is clearly visible.  

\begin{figure}[t]
\begin{center}
    \includegraphics[width=0.45\columnwidth]{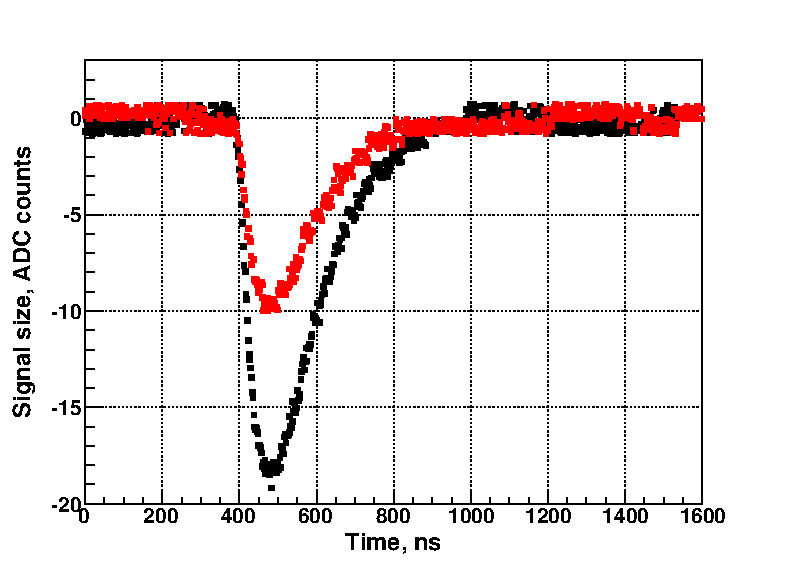}
    \hfill
    \includegraphics[width=0.45\columnwidth]{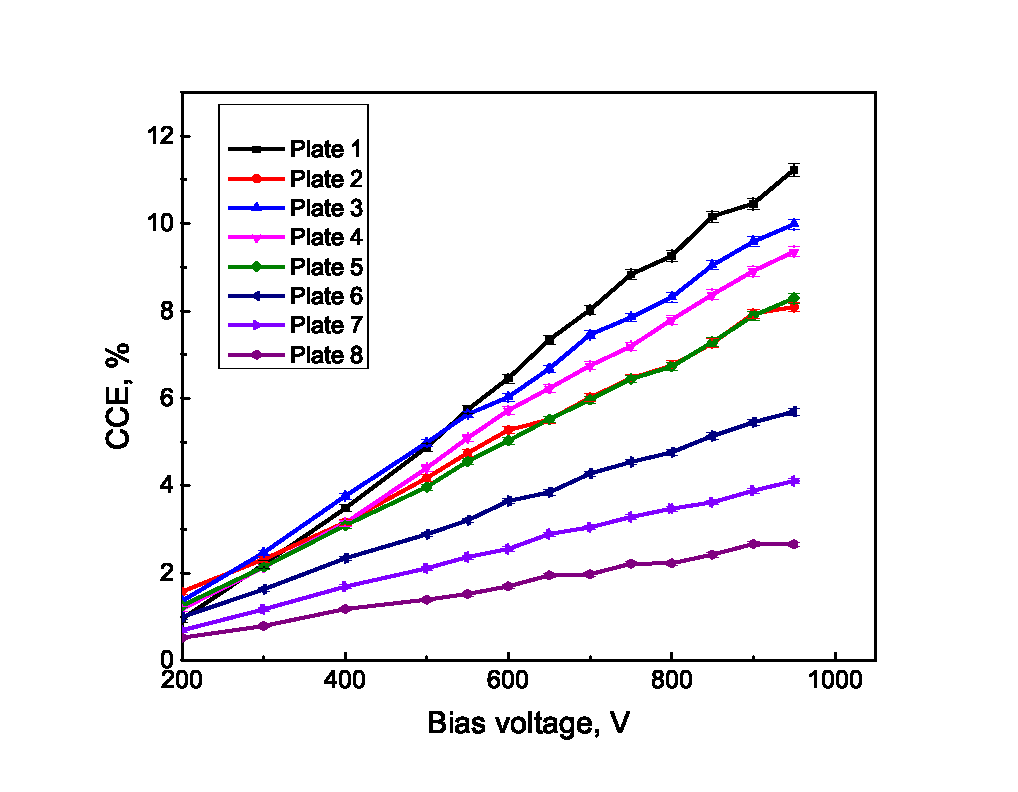}
    \caption{Left: Digitised analog signals for bias voltages of 550 (red) and 950~V (black) as a function of the time, Right: The CCD measured for all sensor plates a function of the bias voltage.}
    \label{sapphire_analog_signal}
  \hspace{0.025\textwidth}
  \end{center}
\end{figure}

The signals from the sapphire sensors are amplified and digitised. An example of signals averaged over several triggers is shown in Fig.~\ref{sapphire_analog_signal} (left) for bias voltages of 550 V and 950 V.

In almost cases a linear rise of the CCE is observed, reaching at 950~V e.g. for plane 1 a value of 10.5\%. The measured CCE varies from sample-to-sample reflecting variation of the substrate quality. As can be seen, 5 out of the 8 sensor plates have a relatively high and similar CCE of about 7-10\%, while three other plates have lower and different CCE values. 
The CCE was also measured as a function of the local y coordinate and described by a linear model of electron and hole drift taking into account recombination, trapping and space charges leading to a polarization field. As a result, the  drift length of electrons is more than 10 times larger then the one of holes at the same field strength. About 50\% of the produced electron-hole pairs recombine immediately.

\subsection{Other novel sensor technologies}

The Snowmass contributed paper ~\cite{Hoeferkamp:2022qwg} describes additional
novel sensor technologies the might have important advantages for future 
$\ee$ experiments.
Drivers of the technologies include radiation hardness, excellent position, vertex, and timing resolution, simplified integration, and optimized power, cost, and material.  We describe these briefly in this section.  These technologies are at different R \& D stages, from early research to final operating scale; please see the individual references for more
details.

{\bf Silicon sensors with 3D technology:}
Silicon sensors with 3D technology~\cite{Parker2019} have electrodes oriented perpendicular to their wafer surfaces.  Due to the short drift lengths, 
these are very promising for compensation of lost signal in high radiation environments and for separation of pileup events by precision timing. New 3D geometries involving p-type trench electrodes spanning the entire length of the detector, separated by lines of segmented n-type electrodes for readout, promise improved uniformity, timing resolution, and radiation resistance relative to established devices operating effectively at the LHC. Present research aims for operation with adequate signal-to-noise ratio at fluences approaching  $10^{18}n_{\rm eq}/{\rm cm}^2$ with timing resolution on the order of 10 ps.

{\bf 3D diamond detectors:}
The 3D technology is also being realized in diamond substrates~\cite{Tsung:2012gz}, where column-like electrodes are placed inside the detector material by use of a 130 fs laser with wavelength 800 nm. When focussed to a 2 micron spot, the laser has energy density sufficient to convert diamond into an electrically resistive mixture of different carbon phases. The drift distance an electron-hole pair must travel to reach an electrode can be reduced below the mean free path without reducing the number of pairs created. Initial tests have shown that after $3.5 \times 10^{15}$ n/cm$^2$, a 3D diamond sensor with $50~\mu{\rm m} \times 50~\mu{\rm m}$ cells collects more charge than would be collected by a planar device and shows less damage due to the shorter drift distance. 

{\bf Beyond CMOS: submicron pixels for vertexing:}
A pixel architecture named DoTPiX~\cite{Fourches2017} has been proposed on the principle of a single n-channel MOS transistor, in which a buried quantum well gate performs two functions---as a hole-collecting electrode and as a channel current modulation gate. The quantum well gate is made with a germanium layer deposited on a silicon substrate. The active layers are of the order of 5 microns below the surface, permitting detection of minimum ionizing particles. This technology is intended to achieve extremely small pitch size to enable trigger-free operation without multiple hits in a future linear collider, as well as simplified reconstruction of tracks with low transverse momentum near the interaction point. The necessary simulations have been made to assess the functionality of the proposed device. The next step is to find out what is the best process to obtain the functionality and to reach some required specifications.

{\bf Thin film detectors:}
Thin film detectors~\cite{Metcalfe:2014nma} have the potential to be fully integrated, while achieving large area coverage and low power consumption with low dead material and low cost. Thin film transistor technology uses crystalline growth techniques to layer materials, such that monolithic detectors may be fabricated by combining layers of thin film detection material with layers of amplification electronics using vertical integration.

{\bf Scintillating quantum dots in GaAs for charged particle detection:}   Lastly, a technology is under development in which a novel ultra-fast scintillating material employs a semiconductor stopping medium with embedded quantum dots~\cite{Oktyabrsky:2016ard}. The candidate material, demonstrating very high light yield and fast emission, is a GaAs matrix with InAs quantum dots. The first prototype detectors have been produced, and pending research goals include demonstration of detection performance with minimum ionizing particles, corresponding to signals of about 4000 electron-hole pairs in a detector of 20 micron thickness. A compatible electronics solution must also be developed. While the radiation tolerance of the device is not yet known, generally quantum dot media are among the most radiation hard semiconductor materials.

These sensor technologies and others still to be developed offer the promise of still
higher performance in the ILC detectors.  We encourage further development, with 
new collaborators, in all of these directions.

\begin{figure}
\centering
\includegraphics[width=0.8\textwidth]{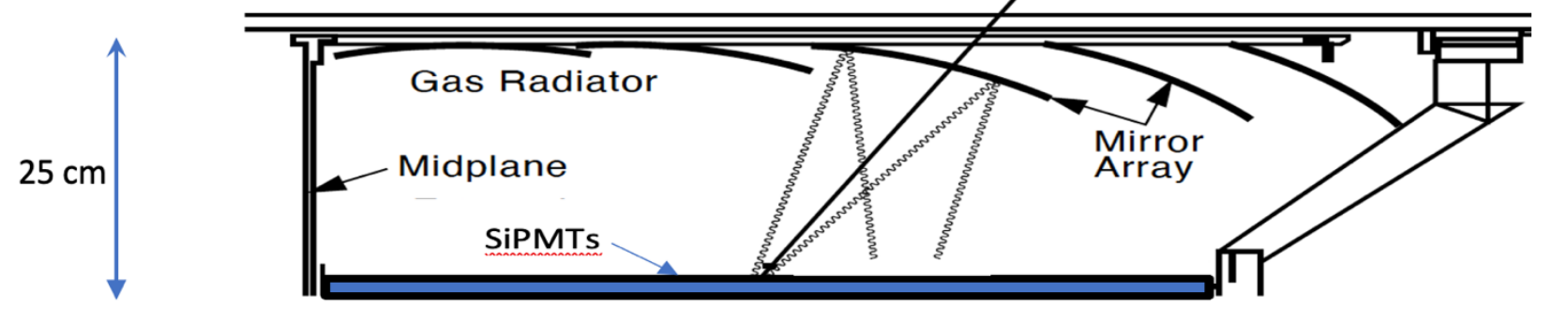}
\caption{Proposed gaseous RICH detector addition to SiD/ILD~\cite{CairoVavra}.}
\label{fig:RICH} 
\end{figure}

\subsection{Gaseous RICH detector for particle ID at ILC}
\label{sec:RICH4ILC}

Particle ID can be important for some ILC analysis, in particular, the measurement of $\Gamma(H\to s\overline{s}$ described in Sec.~\ref{sec:Hhadronic}.  Here we describe a  possible RICH detector for $\pi/K$ separation  up to 25 GeV/c~\cite{CairoVavra}. It is well known that a gaseous RICH detector is the only way to reach $\pi/K$ separation up to 30-40 GeV/c. The detector concept is shown in Fig.~\ref{fig:RICH}. An initial choice for the RICH detector thickness is 25 cm active length to minimize magnetic field smearing effects. The RICH detector uses spherical mirrors and SiPMT  photon detectors. The design in the figure resembles the SLD CRID gaseous RICH detector; however, introducing SiPMT-based design improves the performance substantially. Although we have selected a specific type of SiPMT to make our estimates, we believe that the photon detector technology will improve over the next 15 years, both in terms of noise performance, timing capability, pixel size and detection efficiency. The overall aim is to make this RICH detector with as low mass as possible in order not to degrade the calorimeter
performance. This requires for mirrors made of beryllium and a structure made of low mass carbon-composite material. Another important aspect is to make the RICH detector depth as thin as possible to reduce the cost of the calorimeter.

\chapter{ILC Detector Simulation} 
\label{chap:sim}

%(This section will describe the ILC fast simulation frameworks used for SiD, ILC, and Snowmass projects;  see~\cite{Fujii:2020pxe}.)

%\newcommand{\delphes}{\textsc{Delphes}}
\newcommand{\CPP}{C\nolinebreak\hspace{-.05em}\raisebox{.4ex}{\tiny\bf +}\nolinebreak\hspace{-.10em}\raisebox{.4ex}{\tiny\bf +}}
\newcommand{\delphes}{{\normalfont\textsc{Delphes }}}
\newcommand{\delphesnsp}{{\normalfont\textsc{Delphes}}}

\section{ILC Fast Simulation Frameworks}
\label{sec:fastsim}
As a first step to get started with ILC physics one can use fast simulation tools 
that can be used to quickly generate substantial samples of simulated and reconstructed events.
Situations where this is desirable include detector optimisation
and new physics searches. In these cases,
similar processes need to be simulated and reconstructed at
a, potentially very large, number of different conditions.
In the first case, one needs to  modify various aspects of the detector in steps, in the latter,
one needs to  explore the entire allowed parameter space
of a theory for new physics.
In addition to these cases,
fast simulation is also an asset for simulating high cross section
SM processes, such as $\gamma\gamma$ processes, where the investment 
in processor power and intermediate storage might be
prohibitively large to attain the goal that simulation statistics
should be a negligible source of systematic uncertainty.
The ILC community uses two tools for fast simulation that are described in the following:

\subsection{\delphes for ILC}

\begin{figure}[htbp]
\begin{subfigure}{0.49\hsize}
 \includegraphics[width=\hsize]{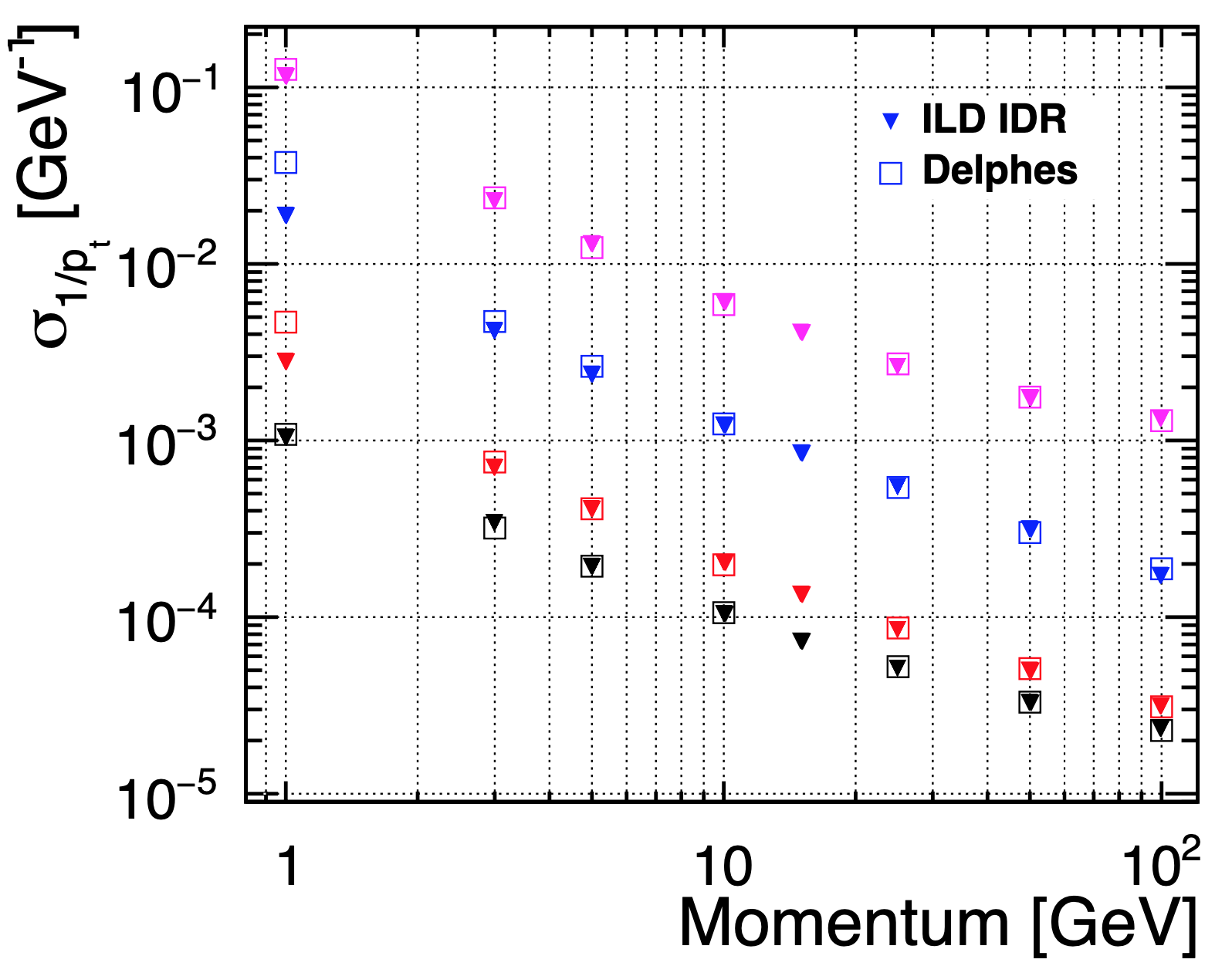}
 \caption{ \label{fig:sim:delphes_trk}}
 \end{subfigure}
\begin{subfigure}{0.49\hsize}
 \includegraphics[width=\hsize]{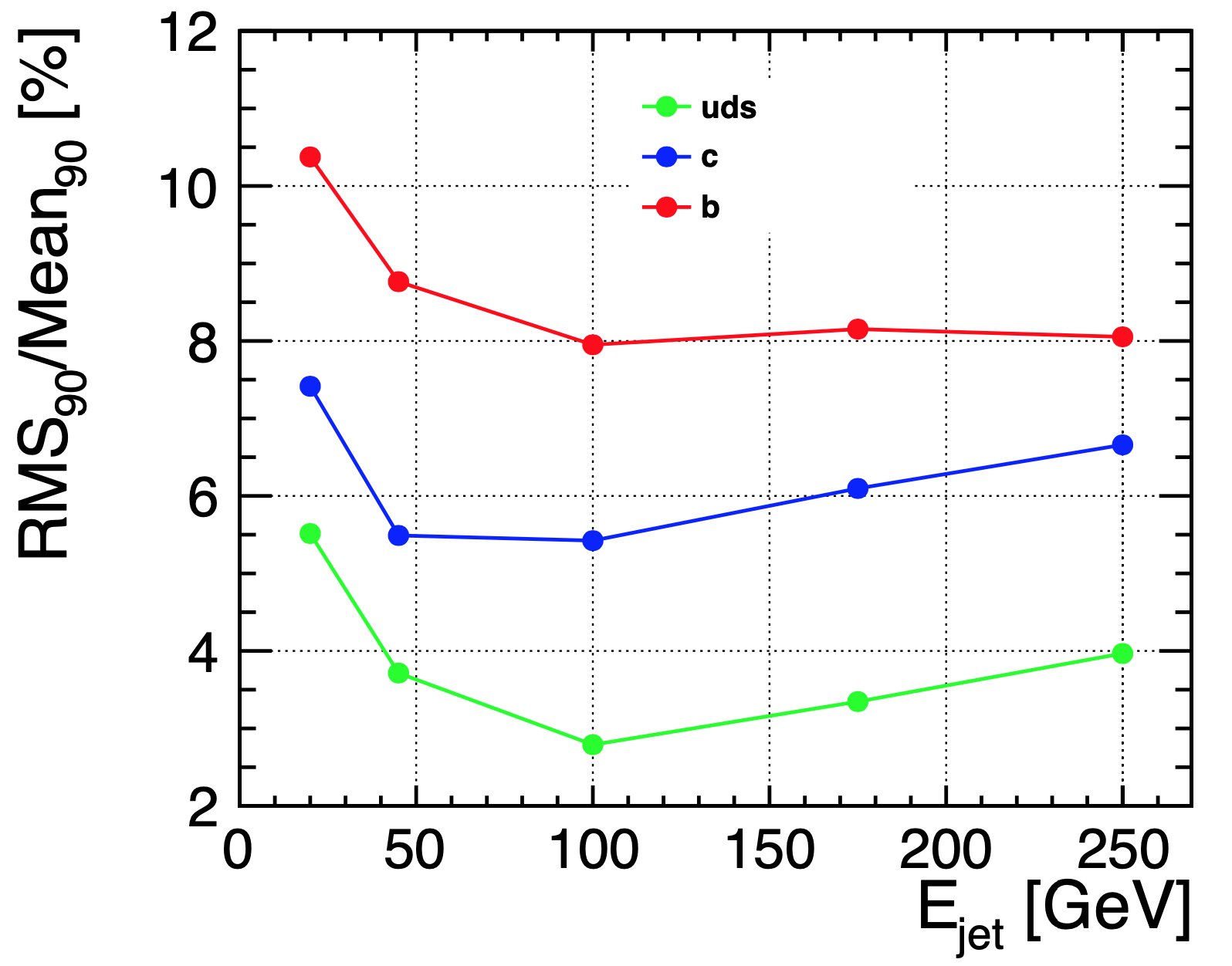}
 \caption{  \label{fig:sim:delphes_pfa}}
 \end{subfigure}
\caption{(a): transverse momentum resolution for different polar angles for the ILD full simulation and for the
  ILCgen \delphes parameterization. (b): jet energy resolution for the ILCgen simulation }
\end{figure}

\delphes\cite{deFavereau:2013fsa} is a fast, parameterized simulation framework for generic collider detectors,
developed originally for phenomenological studies at hadron colliders like the LHC. In its recent incarnation
the \delphes framework has been modularized and an attempt has been made to roughly emulate a particle-flow
reconstruction philosophy~\cite{Ovyn:2009tx} a feature that is crucial for its applicability to the ILC.
\delphes also integrates the FastJet~\cite{Cacciari:2006sm} package allowing to directly run the most common
jet clustering algorithms in use for the ILC. A specific collider detector is mimicked in \delphes via the
specification of efficiencies and resolutions for the long lived final state particles,
based on their charge, momentum, polar angle\footnote{\delphes uses pseudo-rapidity $\eta$ instead of polar angle}
and type (charged/neutral hadron, photon, electron or muon).
A dedicated \delphes card: \emph{delphes\_card\_ILCgen.tcl} with parameterizations for a generic ILC detector has been
created~\cite{ILC-DELPHES,bib:ILCgen} and is shipped with the \delphes source code~\cite{bib:delphes:github}.
The parameterization of the detector and reconstruction performance is based on the latest results of the
ILD-IDR~\cite{ILD:2020qve}, where due to the nature of the rather coarse simulation accuracy of the \delphes approach any
potential differences to the SiD detector performance can be neglected for studies carried out with \delphesnsp.
Fig.~\ref{fig:sim:delphes_trk} shows a comparison of the transverse momentum resolution for charged particles at different angles as simulated
with \delphes compared to a full simulation and reconstruction for the ILD detector as well as the jet energy resolution for di-jet events of different quark flavors (Fig.~\ref{fig:sim:delphes_pfa}).
\delphes can read many standard generator formats such as \emph{stdhep} and produces ROOT~\cite{Brun:1997pa} output files by default. With the
\emph{delphes2lcio}~\cite{bib:delphes2lcio} tool standard LCIO miniDST can be created (see section~\ref{sec:miniDST}).

\subsection{SGV}

The SGV program\cite{Berggren:2012ar} used at ILC has a more sophisticated way
of simulating the response to charged particles than the Delphes program described above.
The time to simulate and reconstruct an event is similar to
the time it takes to generate it ($\sim 1-10$~ms).
The response of the detector is as far as possible calculated
from the detector design (so there is no need to parameterize
pre-existing full simulation results).  SGV
has been shown to compare well  both with full simulation
and with  real data~\cite{Abdallah:2003xe}.

The program uses a simplified ``{\it cylinders-and-discs}'' description
of the detector,
which is used to calculate the Kalman-filtered track-helix covariance matrix
of each generated charged particle.
By Cholesky~\cite{bib:cholesky} decomposition of the covariance matrix,
the track-parameters are simulated in a way such that all correlations
are respected.
The calorimetric response is calculated from the expected single-particle
performance of the different components of the calorimetric system,
for each particle impinging on it. Optionally,
the effects of shower-confusion can be included.
To reduce the needed storage for a Giga-event size sample,
event filtering can be applied at different steps of the processing,
directly after generation, after the detector response is known,
or after higher-level event analysis is done.
Events passing all filters are output in LCIO DST-format,
and can seamlessly be further analyzed within the Marlin framework.

\section{ILCSoft framework}
\label{sec:ILCSoft}

Accurate and detailed modeling of the physics interactions as well as the detector
response are crucial for making realistic predictions about the expected physics and detector
performance. The ILC software for detector simulation, reconstruction and analysis is entirely
based on the common linear collider software ecosystem called \emph{iLCSoft}~\cite{ilcsoft}.
The main core software tools in iLCSoft are the common event data model and persistency tool LCIO~\cite{Gaede:2003ip},
the \CPP\ application framework Marlin~\cite{Gaede:2006pj} and the generic detector description toolkit
DD4hep~\cite{Frank:2014zya,Frank:2015ivo}. DD4hep provides a single source of information for describing the detector geometry, its
materials and the readout properties of individual sub detectors. Various components of DD4hep provide different functionalities.
Here we use DDG4, the interface to full simulations with Geant4~\cite{Agostinelli:2002hh} and DDRec the specialized view into the
geometry needed for reconstruction. 
In the following we briefly describe the main features of the full simulation and reconstruction tools in use for ILC and SiD, more details can
be found in the corresponding chapters of~\cite{ILD:2020qve} and~\cite{Bambade:2019fyw}.

\subsection{Simulation models}

Both ILC detector concept groups have developed detailed and realistic simulation models with realistic geometrical dimensions, material budgets,
imperfections and cables and services. Wherever possible, realistic simulations and parameterizations for the individual sub detectors have been
implemented based on available test beam results for the proposed technology.
Great care has been taken to include realistic material estimates, established by the detector R\&D groups,
in particular in the tracking region where the material budget has a direct impact on the detector performance.
\begin{figure}[b!]
  \begin{subfigure}{0.4\hsize}
    \includegraphics[width=\textwidth]{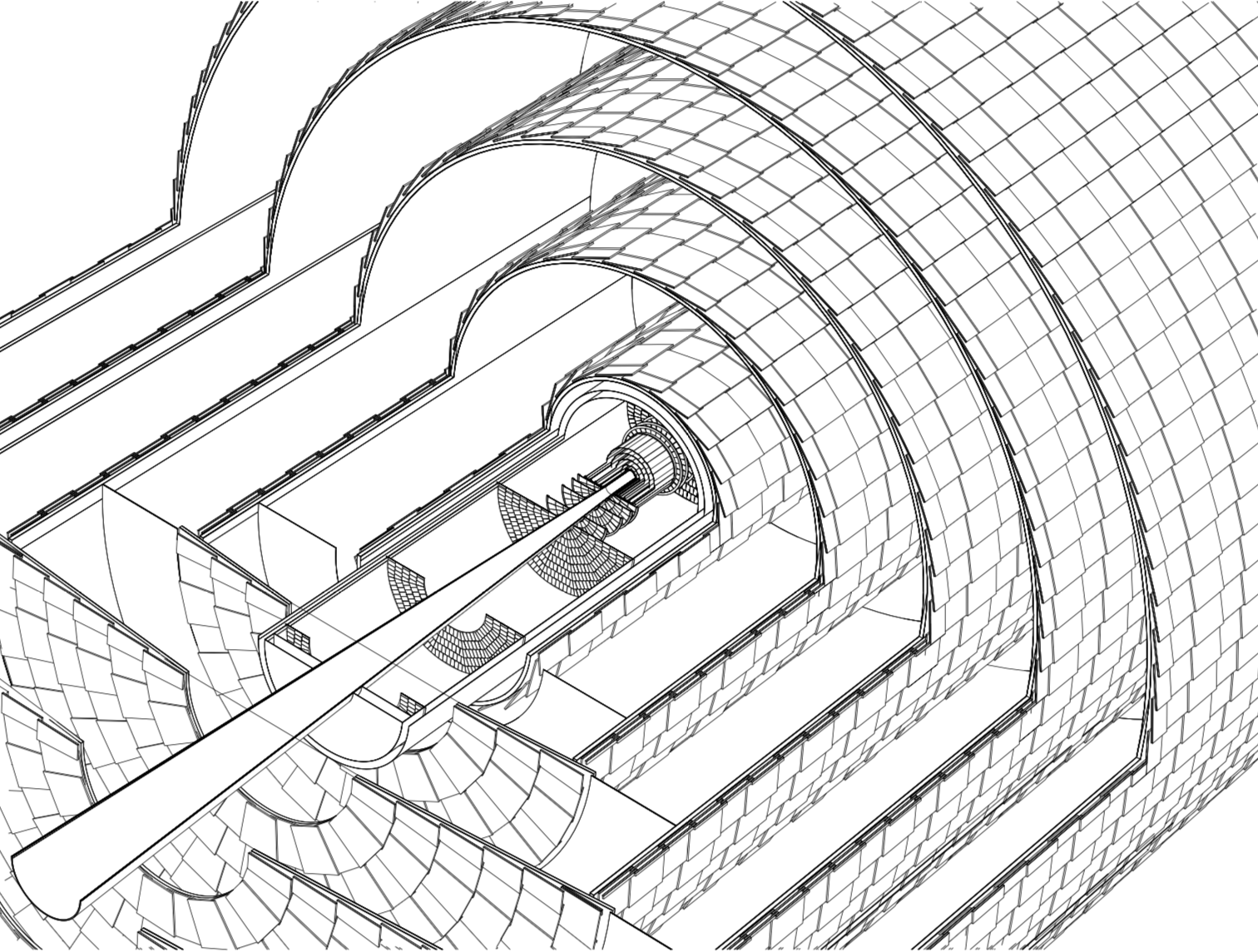}
    \caption{ \label{fig:sim_sid_tracker}}
  \end{subfigure}
  \begin{subfigure}{0.59\hsize}
    \includegraphics[width=\textwidth]{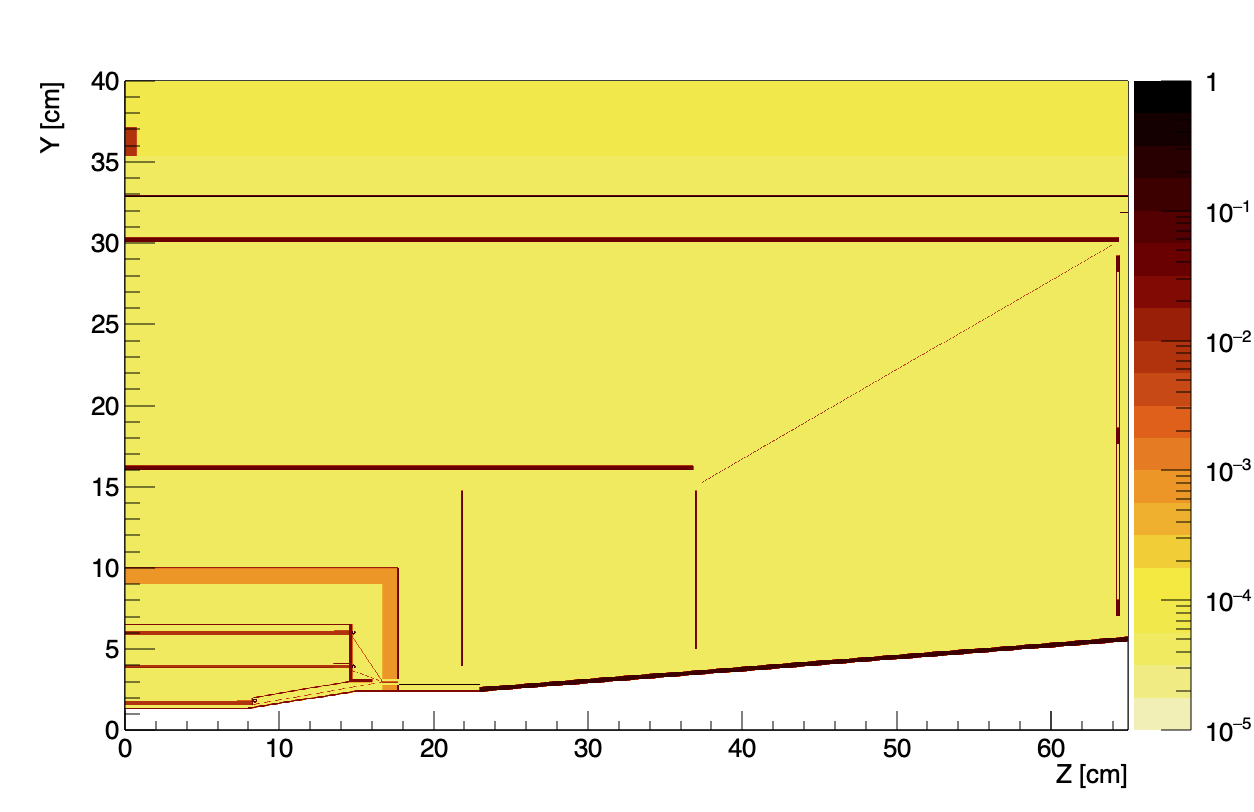}
    \caption{ \label{fig:sim_materialscan_vxd}}
  \end{subfigure}
  \caption{(a): cut-away view of the tracking system as implemented in the \emph{SIDLOI3} simulation model (from~\cite{Behnke:2013lya}).
      (b): Material scan in inner tracking region of the ILD simulation model showing detector components of the VTX, SIT and FTD as well
    as dead material from the beam pipe, support structures, cables and services. Plotted is the local material budget per bin in units of X0
    with an arbitrary scaling factor applied.)\label{fig:inner_trackers} }
\end{figure}
Examples of the inner tracking regions as implemented in the realistic simulation models for SiD and ILD are shown in Fig~\ref{fig:inner_trackers}.

\subsection{Event reconstruction}

The reconstruction of simulated events in the ILC detectors is done with a number of dedicated algorithms implemented in Marlin.
For the digitization of tracking detectors and calorimeters dedicated modules exist that provide a parameterization of the expected resolutions as
established by the R\&D collaborations taking into account effects like cross talk, electronic noise and signal collecting efficiencies.
The reconstruction of charged particle tracks is performed with a variety of pattern recognition algorithms implemented in the
MarlinTrk~\cite{Gaede:2014aza} package. This is followed by sophisticated clustering and particle flow algorithms from PandoraSDK~\cite{Marshall:2015rfa}
that delivers a complete collection of reconstructed particles or so called \emph{particle flow objects}. Additional high level reconstruction algorithms,
like jet clustering with FastJet~\cite{Cacciari:2006sm}, vertexing and flavor tagging with LCFIPlus~~\cite{Suehara:2015ura}, particle identification using $dE/dx$-information and time-of-flight measurements complete the event reconstruction for final physics analyses.

\section{ILC SM Background Samples}
\label{sec:SMbacgrounds}

\subsection{Event generation}

ILC physics sample generation is typically done with the Whizard~\cite{Kilian:2007gr} event generator providing crucial features
like correct treatment of ISR and FSR via creation of photons as individual final state particles.
Whizard uses tree-level matrix elements and loop corrections to generate events with the final state partons and leptons
based on a realistic beam energy spectrum, the so called \emph{hard sub-process}. The hadronization into the visible final state
is performed with Pythia~\cite{Sjostrand:2006za} tuned to describe the LEP data.
The correct beam energy input spectrum for a given collision energy and set of accelerator parameters is created with
Guinea-Pig~\cite{Schulte:1998au}, a dedicated simulation program for computing
beam-beam interactions at linear colliders.

\subsection{Beam induced background}

\begin{figure}[h!]
  \begin{subfigure}{0.45\hsize}
    \includegraphics[width=\textwidth]{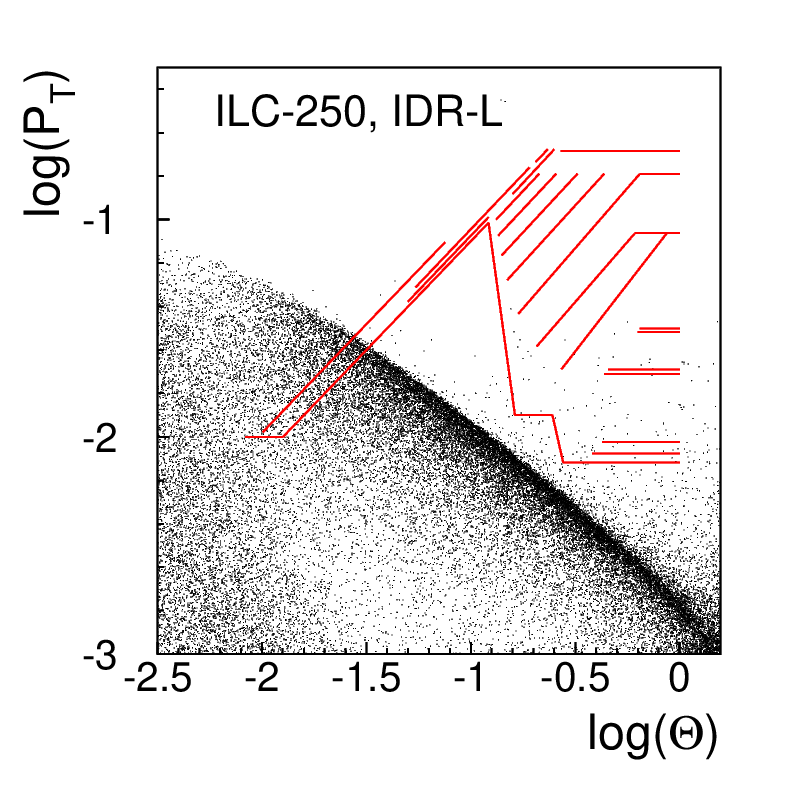}
    \caption{ \label{fig:pair_bg_cone_ild}}
  \end{subfigure}
  \begin{subfigure}{0.55\hsize}
    \includegraphics[width=\textwidth]{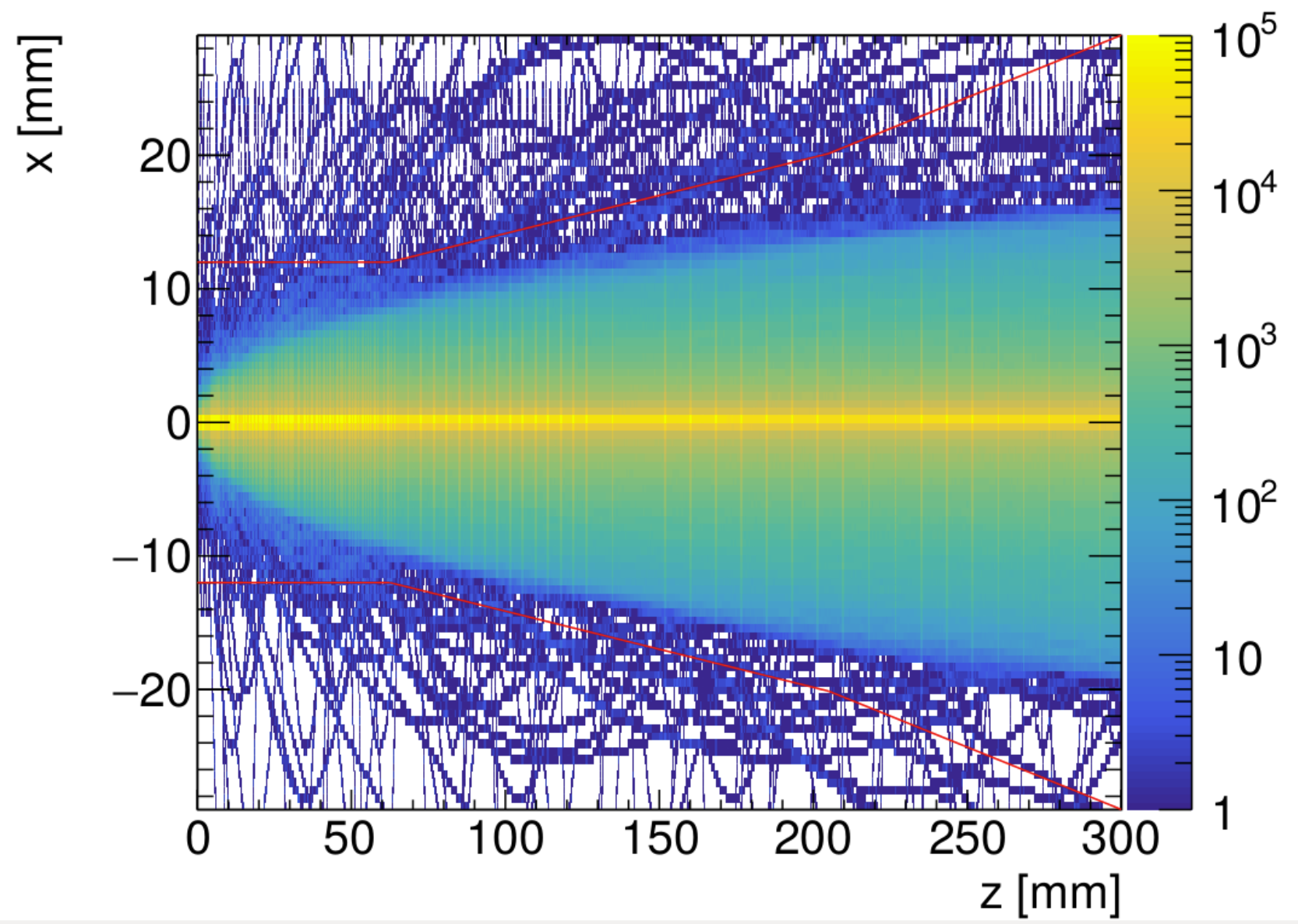}
    \caption{ \label{fig:pair_bg_cone_sid}}
  \end{subfigure}
  \caption{\label{fig:pair_bg_cone} (a): Cones of incoherent  $e^+e^-$-pairs in the ILD detector for $E_{cms}=250~GeV$ as created with GuineaPig.
    Shown is $\log{p_t}$ of the particles (radius of the helical trajectory) as a function of $\log{\theta}$.
    Also shown are the inner detector elements of the ILD detector (horizontal lines represent
    barrel elements and diagonal lines represent end-cap elements).
    %All detector layers are well outside of the background cone,
    %except for the face of the BeamCal and the beam pipe endcap in front of it (the two leftmost diagonal lines in the plot).
    (b): Cone of background from incoherent $e^+e^-$-pairs, generated with Guinea-Pig and simulated in the 5 T B-field of the SiD
  detector (from~\cite{Schutz:2017ihd}).
  }
\end{figure}

The strong beam-beam interactions lead to two distinct sources of backgrounds:
\begin{itemize}
\item the creation of incoherent $e^+e^-$-pairs that are the source of the dominating background at the ILC.
These electrons and positrons are predominantly created in a forward cone as shown in Fig~\ref{fig:pair_bg_cone} for the ILD detector.
It is this cone that restricts the minimal allowed radius of the innermost layer of the vertex detector of any linear collider detector
as can be seen in Fig.~\ref{fig:pair_bg_cone}.

\item creation of $\gamma\gamma \to hadrons$ events, due to the interaction of beamstrahlung photons.
This type of events is generated for $\gamma\gamma$ cms-energies from 300~MeV to 2~GeV with a dedicated generator based
on ~\cite{Chen:1993dba}, whereas for higher energies Pythia is used.

\end{itemize}
For realistic physics analyses and detector studies for the ILC it is important to take these backgrounds into account. This is typically done
through event overlay techniques in the iLCSoft based full simulation and reconstruction chains of ILD and SiD.
%
%%%%%%%%%%%%%%%%
%%%%%%%%%%%%%%%%%

\subsection{Event Samples and data formats}
\label{sec:miniDST}
Large sets of SM samples for the ILC have been generated for $E_{cms}=250, 350, 500, 1000~GeV$ are available at~\cite{bib:ILCsnowmass}.
Data sets with miniDSTs created with \delphes and SGV of these generated samples are also available at this web site.
Access to more realistic fully simulated and reconstructed event samples from ILD or SiD is possible via a lightweight guest membership.
The web-site~\cite{bib:ILCsnowmass} provides additional information on ILC simulation resources and tools.

\begin{figure}[h!]
  \begin{subfigure}{0.45\hsize}
    \includegraphics[width=0.85\textwidth]{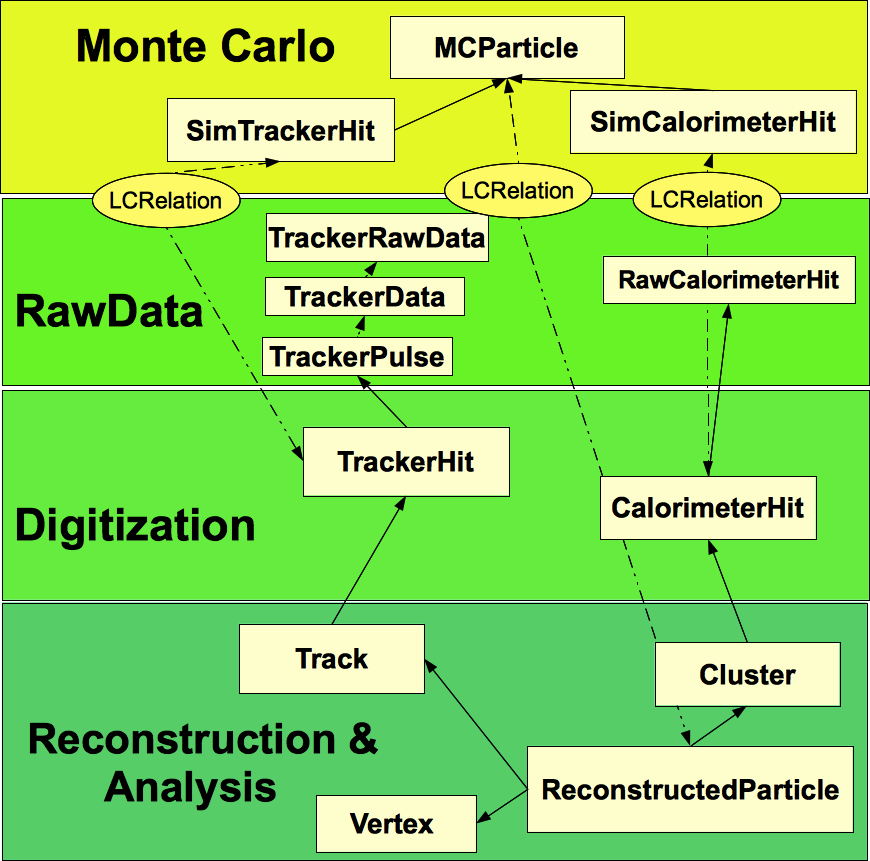}
    \caption{ \label{fig:lcio_schema}}
  \end{subfigure}
  \begin{subfigure}{0.55\hsize}
    \includegraphics[width=0.9\textwidth]{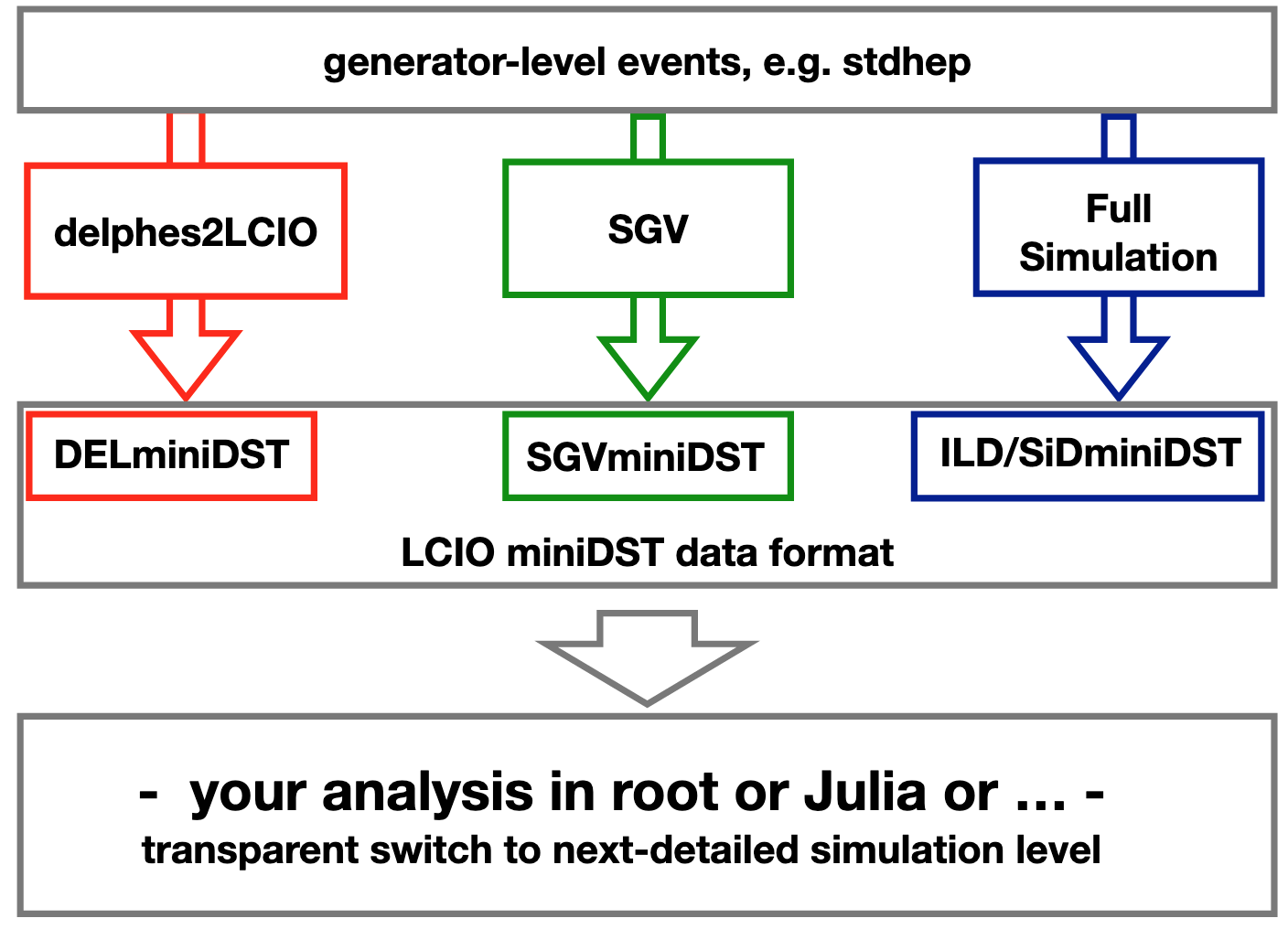}
    \caption{ \label{fig:minidst_workflow}}
  \end{subfigure}
  \caption{(a): Schematic view of the hierarchical event data model of LCIO.
    (b): Using the miniDST format a common analysis code can be developed that works with all simulation and reconstruction tools presented above.
  }
\end{figure}

All ILC fast and full simulation and reconstruction tools can provide the common data format LCIO as output. The LCIO event data model
(see Fig.~\ref{fig:lcio_schema})
is the de facto standard for ILC physics and detector studies.
Recently a particularly lightweight set of output collections for ILC event data has been defined, the \emph{miniDST}-format. By starting out with
developing an analysis with \delphes or SGV based on \emph{miniDST}s one can later easily move to a more realistic analysis based on full simulation
using the same format with only minor modifications as shown in Fig.~\ref{fig:minidst_workflow}.
%%%%%%%%%%%%%%%%
%%%%%%%%%%%%%%%%%

\chapter{ILC Physics Measurements at 250 GeV} 
\label{ILC250}

The first stage of the ILC will be at a CM energy of 250 GeV.   In this chapter, we will describe aspects of the ILC experimental program that are specific to 250~GeV, in particular, the study of the Higgs boson in the process $\ee\to ZH$.  We will also discuss precision SM tests that are available this energy---in particular, the measurement of the triple gauge couplings through $\ee\to W^+W^-$ and tests of QCD in $\ee\to$~jets.  Aspects of the ILC program that benefit from higher energy---in particular, searches for new particles in pair-production and fermion-fermion scattering--will be discussed over the whole ILC program in Chapter 10. 

\section{Higgs -- conventional decays}
\label{sec:Higgs250}

The precise measurement of ``conventional'' Higgs decay branching ratios is key to
probing virtual effects of new physics in the Higgs sector.   The
value of the Higgs boson mass is now known from the LHC to part per mil
precision, and this precision will be improved at the ILC.  By
combining this value with other precisely known SM inputs, it will be 
possible to predict the absolute strengths of Higgs boson couplings to
the 0.1\% level.
Many models of new physics lead to
variations in the Higgs couplings, typically leading to few-\% variations of
Standard Model Higgs couplings for new physics at the TeV scale.
Thus, the measurement of these couplings to the \%-level precision or
better is one of the major goals of
the Higgs program at high energy electron--positron colliders such as the ILC.

\begin{figure}[!ht]
\begin{center}
\includegraphics[width=0.4\hsize]{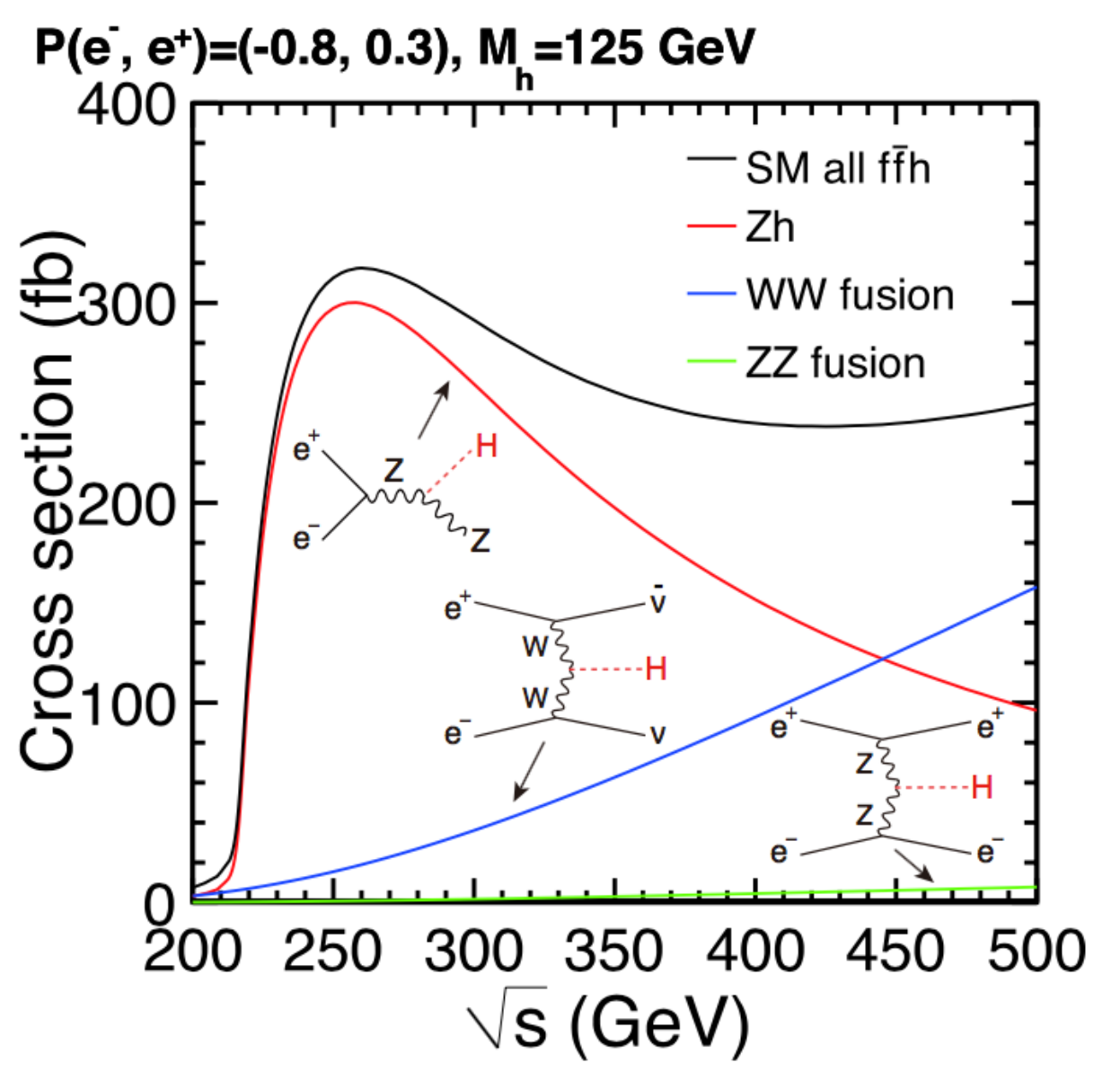}
\includegraphics[width=0.5\hsize]{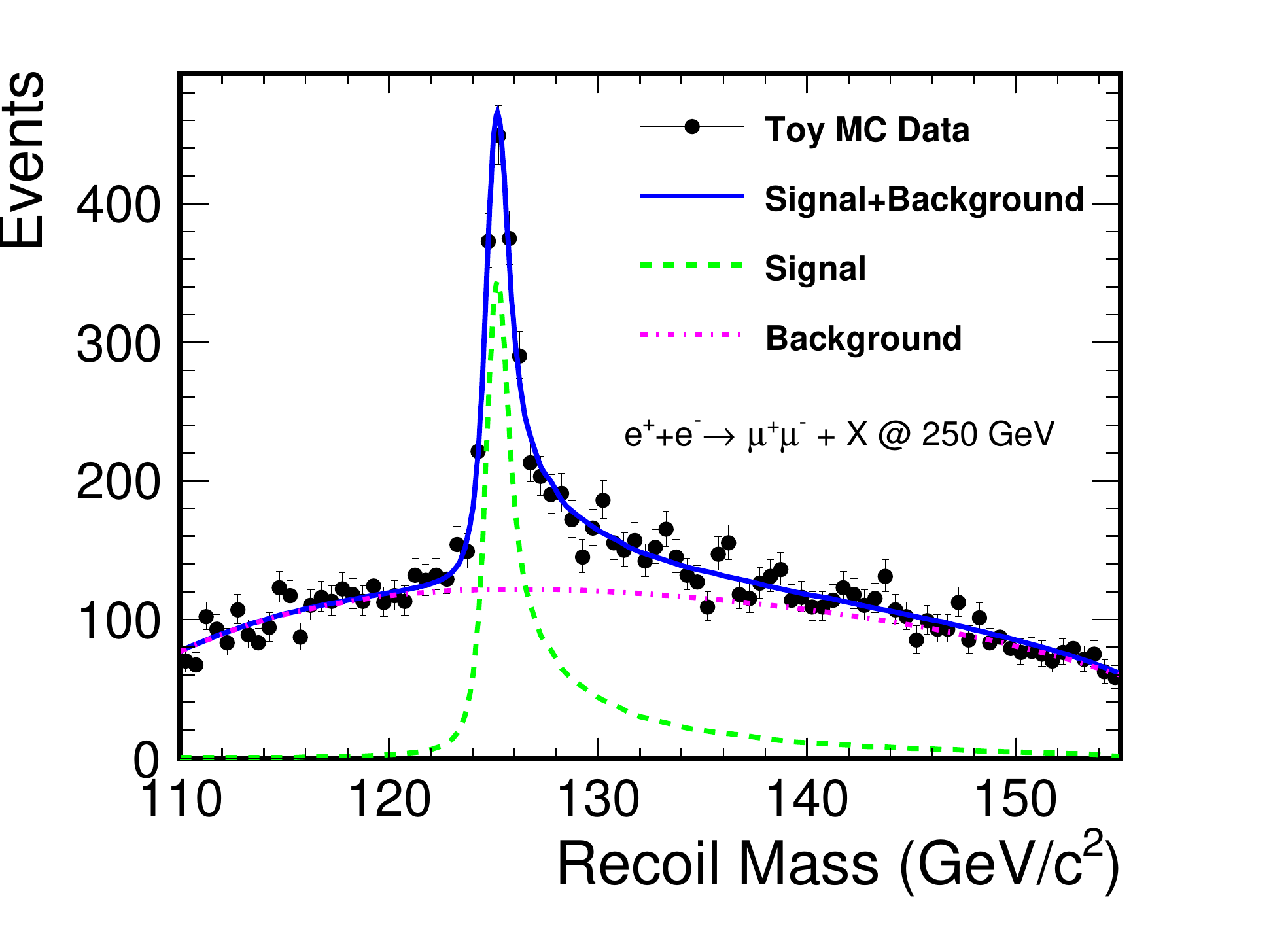}
\end{center}
  \caption{
Left: Cross sections for the three major Higgs production processes
  as a function of center of mass energy~\cite{Baer:2013cma}.
The Zh ``Higgs-strahlung'' process dominates at 250~GeV.
Right: Recoil mass spectrum against $Z\to\mu^+\mu^-$ for signal $e^+e^-\to Zh$ and SM background
  at 250 GeV~\cite{Yan:2016xyx}.}
  \label{fig:Higgsstrahlung}
\end{figure}

Higgs production in electon--positron collisions at 250~GeV is dominated by the
associated production of Higgs and Z bosons (``Higgs-strahlung''), as
shown 
in Fig.~\ref{fig:Higgsstrahlung} \cite{IoffreKhoze:1978}.
Because electron-positron collisions provide an initial state with well-defined four-momentum,
this process allows the identification of Higgs bosons by considering the mass
recoiling against an identified Z boson, without any reference to the decay products of the Higgs.
A typical reconstructed recoil mass distribution is shown in Fig.~\ref{fig:Higgsstrahlung}.

Higgs-strahlung events at ILC250 in which the Z decays to hadrons or charged leptons will provide the experimenter
a sample of about half a million Higgs bosons that is almost completely unbiased with respect to the Higgs decay mode.
Such a sample is very useful for making precise and unbiased measurements of the Higgs boson's properties,
for example the partial cross-sections to different Higgs decay modes
$\sigma_{ZH} \times BR(H \to X)$.   At the same time, this sample can
provide a precise value of Higgs boson mass from the position of the
recoil mass peak.  An order of magnitude improvement in precision over
the current measurement is needed because
the Higgs branching ratios to $WW^*$ and $ZZ^*$ depend strongly on the
Higgs boson mass, and the recoil technique can meet this goal.

Because the identification of the Higgs boson does not depend on the
decay mode, it is also possible at an $\ee$ collider to measure  the
total Higgs-strahlung production cross-section
in the different ILC beam polarization setups.  Combining these
results with other ILC measurements, it is possible to extract
absolutely normalized values for the couplings of the Higgs boson to
its various final states, and for the Higgs boson width.   We will
present the precision on these quantities expected from a global fit
to ILC data in Chapter~\ref{chap:SMEFT}. 

ILC also presents an opportunity to probe the Higgs boson's CP properties,
a key to understanding the potential for baryogenesis at the electroweak scale, in its interaction
both with $\tau$ leptons and with massive vector bosons.

Projections for the experimental precisions attainable at the ILC are based on full simulation studies
which take into account experimental conditions such as beam energy spread and beam background processes,
as well as detailed simulation of the experimental apparatus and realistic data analysis techniques.

\subsection{Zh cross-section and Higgs mass}

The recoil mass distribution shown in Fig.~\ref{fig:Higgsstrahlung} can be used to extract the
total Zh production cross-section and the Higgs boson mass, by consideration
respectively of the area and position of the signal peak~\cite{Yan:2016xyx}.
The cross-section will be measured in all ILC beam polarization combinations,
switching between dominantly left- and right-handed electrons and positrons.
The cross-section $\sigma_{Zh}$ in the two major polarization combinations will be measured to a precision of 1\%.
The asymmetry between these measurements in different polarizations offers an important additional
input to the global understanding of Higgs couplings.  At ILC-250, 
the precision on the Higgs mass is expected to reach 14~MeV using the recoil mass method~\cite{Yan:2016xyx}.

The Higgs mass can also be directly reconstructed from its decay products, providing
complementary measurements.
A demonstration in the case of the dominant Higgs decay to $b \overline{b}$ can be found in
\cite{ILD-PHYS-PUB-2019-001}, while rare Higgs decays to
final states which can be very precisely measured, such as two or four muons and/or electrons,
can also provide very competitive precision despite the
limited numbers of events~\cite{gwwilson-hmass}.

\subsection{Hadronic decays}
\label{sec:Hhadronic}
The majority of Higgs bosons will decay into hadronic final states; within the
SM we expect dominant contributions from b-quarks, c-quarks, and gluons.
The experimental separation of these hadronic contributions relies on jet flavor tagging.
The keys to distinguishing jet flavor are the identification of displaced vertices
produced in the decay of meta-stable particles,
of leptons within hadronic jets originating from massive hadron decays,
and particle identification, in particular the ability to identify kaons.

\begin{figure}[!ht]
    \centering
    \includegraphics[width=0.45\textwidth]{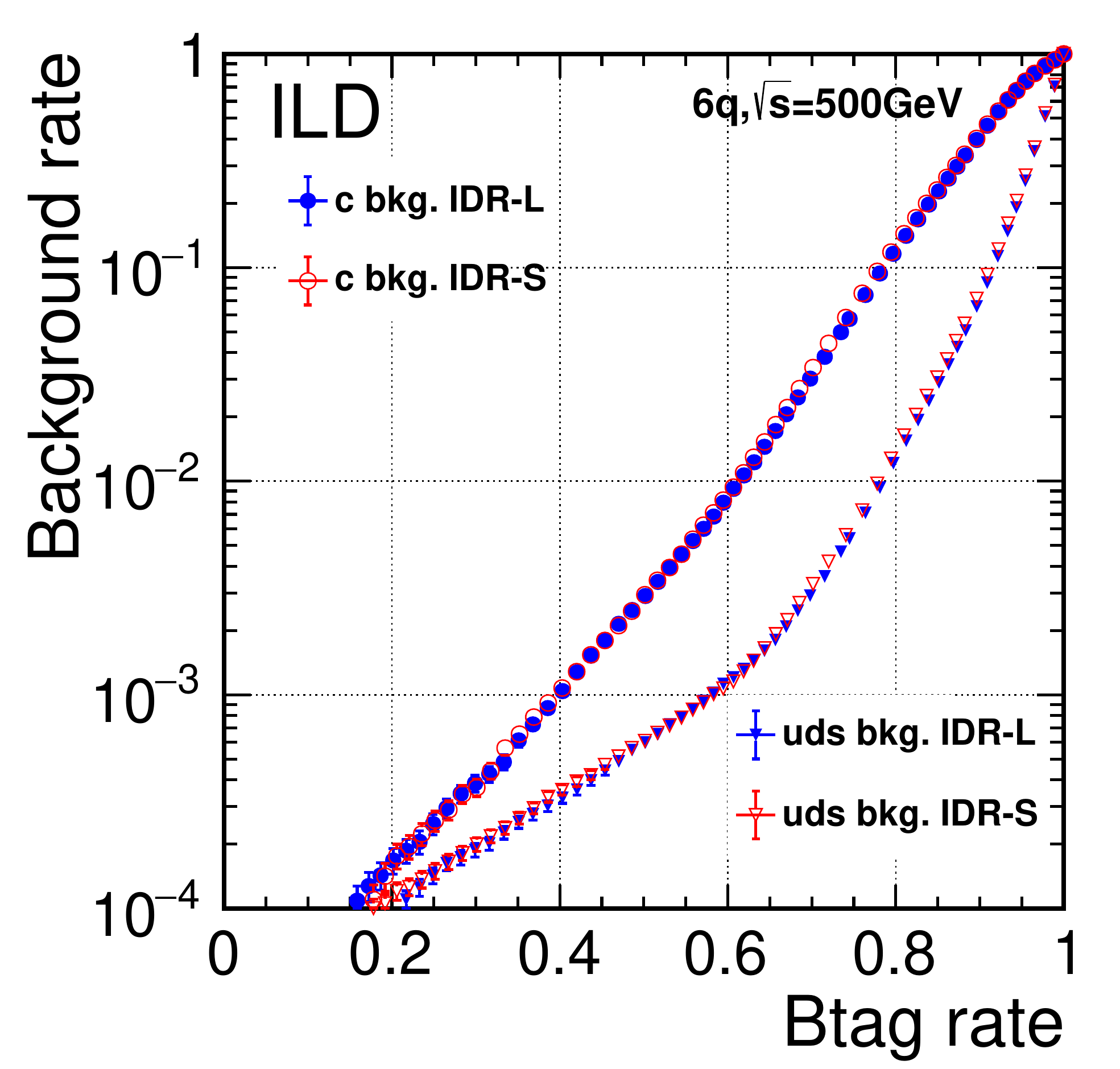}
    \includegraphics[width=0.45\textwidth]{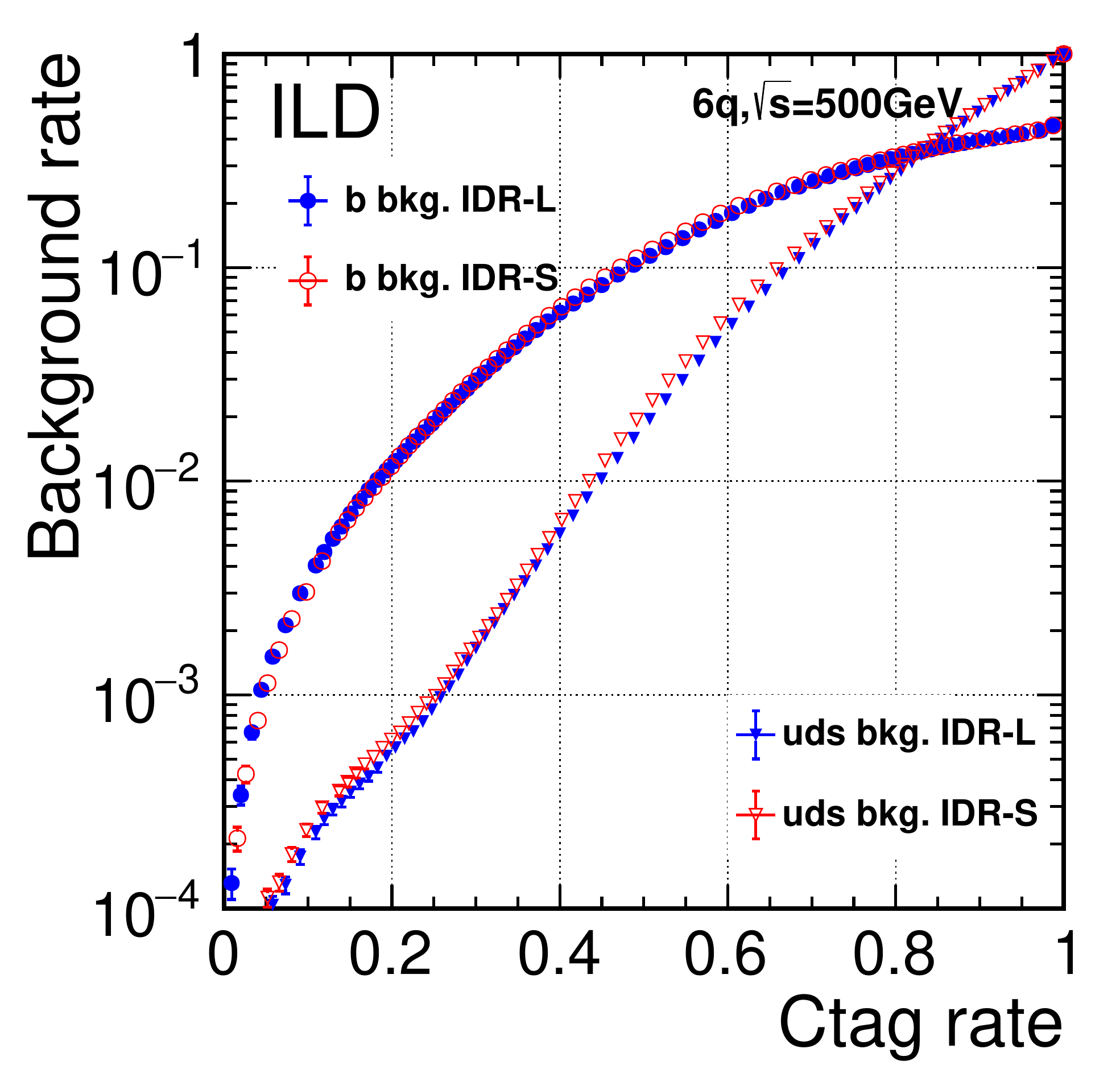}
    \caption{B-tag (left) and c-tag (right) performance in full-simulation studies of two variants
    of the ILD concept, IDR-L and IDR-S (figure from \cite{ILDConceptGroup:2020sfq}).}
    \label{fig:btag}
\end{figure}

The reconstruction of displaced vertices is aided by the tiny ILC interaction region
and the vertex detector, with its few-micron hit position resolution and first layer placed only $\sim 15$~mm from the IP.
Figure~\ref{fig:btag} shows the excellent
b- and c-tagging performance achieved by the LCFIplus algorithm in full simulation studies
of the ILD concept at ILC.

\begin{figure}[!ht]
    \centering
    \includegraphics[width=0.6\textwidth]{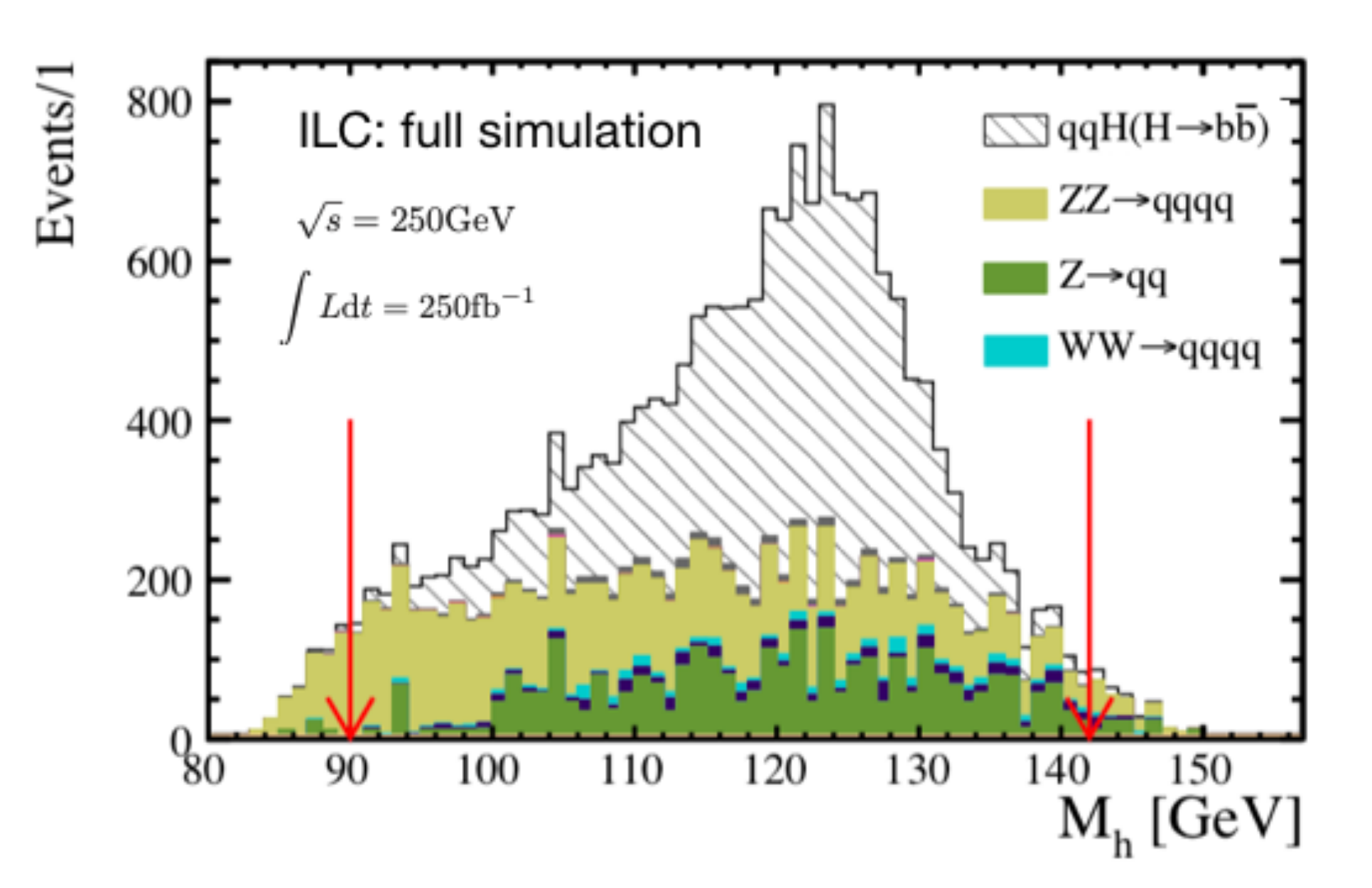}
    \caption{Comparison of signal and backgrounds from ILD full
      simulation for the measurement
      of the $\sigma\times BR$ for $H\to b\overline{b}$, for
      250~fb$^{-1}$ of ILC data at 250~GeV, from~\cite{Ogawa:2018ssv}.}
    \label{fig:bb250}
\end{figure}

Applying the LCFIplus algorithm to hadronically decaying Higgs bosons produced at ILC250,
assuming the nominal $2~ab^{-1}$ total integrated luminosity,
the partial cross--section $\sigma_{Zh} \times BR(H \to bb)$ can be measured to 0.7\%,
and $\sigma_{Zh} \times BR(H \to cc, gg)$ to around 4\%
precision~\cite{Ono:2012oyw} in each of the major polarization combinations. 
The expected signal and major
backgrounds, from a full simulation study of $H \to bb$, are shown in Fig.~\ref{fig:bb250}.
For $cc$, especially, the direct nature of the
measurement as well as the high level of precision contrasts markedly with the
situation at the LHC.

The identification of $H \to s\overline{s}$ decays presents a significant experimental challenge due both to its subtle signature and its small expected branching ratio.
Huge multi-jet backgrounds  make it nearly impossible to probe the strange Yukawa coupling via direct $H \to s\overline{s}$ searches at the LHC.
At the ILC, studies in progress indicate that the observation of this process, though challenging, might be possible.
Potentially useful experimental techniques and detector capabilities include the reconstruction of decaying $V^0$  mesons as well as $K/\pi$ particle discrimination through measurements of the energy loss $dE/dx$ of charged particles, and the use of time-of-flight or modern and compact Cherenkov detectors (see Sec.~\ref{sec:RICH4ILC}). A limit on the $s\overline{s}$ branching ratio at a factor of about 5 above the SM value would already be a significant constraint on flavor models (see Sec.~\ref{sec:ILCflavor}), and the studies in~\cite{Albert:2022mpk} confirm that this is achievable. Better understanding and thus rejection of the backgrounds from $Z$ bosons and other Higgs decays could further improve this measurement.

\subsection{Leptonic decays}

The measurements of Higgs decays to $\tau$ leptons and muons are
feasible at ILC if these  branching fractions are roughly at the
levels predicted in the SM. 
The decay branching ratio to $\tau$ leptons is relatively large in the
SM.  In addition, Higgs decays to $\tau^+\tau^-$ are straightforward
to recognize in the ILC experimental environment, allowing
identification of these decays with high precision.  This should lead
to  a precision at ILC-250 of better than 2\% in the measurement of the
partial cross--section $\sigma_{Zh} \times BR(H \to \tau\tau)$~\cite{Kawada:2015wea} in each major polarization combination.

The small branching ratio to muons limits the statistics available at ILC.
The predicted precision on $\sigma_{Zh} \times BR(H \to \mu\mu$) at ILC-250 is 38\%
in each polarization set~\cite{Kawada:2020lxr}. It should be noted that an
LHC measurement of the ratio of branching ratios $BR(H\to
\mu\mu)/BR(H\to ZZ^*)$ can be combined with ILC data to produce an
absolutely normalized value of the Higgs boson coupling to muons.

Direct observation of the Higgs coupling to electrons is essentially impossible at ILC
if the branching fraction is that predicted by the SM. The final state can in principle
be well reconstructed, so if this channel is very significantly enhanced with respect to the SM,
for example to a level similar to the branching ratio to muons, it can be observed at ILC.

\subsection{Electroweak boson decays}

The measurements of the Higgs branching ratios to $WW^*$ and $ZZ^*$ play an important
role in the global probing of the Higgs sector, since these same couplings are involved in
Higgs production via $WW$-fusion and Higgs-strahlung, respectively. These therefore
allow direct extraction of the total Higgs decay width $\Gamma_h = \Gamma_{WW[ZZ]} / BR_{WW[ZZ]}$.

The large number of different final states make for a complex analysis. A recent
study of Higgs decay to ZZ* estimated that $\sigma_{Zh} \times BR(H \to ZZ)$ can be measured
to a precision of 8\% in each of the major polarization
stages of ILC-250 by making use of a variety of Z and h decays modes~\cite{Antonov:2021dwd}.
A precision of 2.4\% is expected on the corresponding measurements of $\sigma_{Zh} \times BR(H \to WW)$.

Rare loop-induced Higgs decays to $\gamma \gamma$ and $\gamma Z$ can also be
sought at ILC-250, although the small SM branching ratios will
severely restrict the statistical precision of these measurements.
In the case of $\gamma \gamma$, a precision of 18\% on the
partial cross-section is expected at ILC-250 in each of the two major polarization samples.
As for muons, though, an LHC measurement of a ratio of branching ratios can be combined with
ILC data to obtain a normalized coupling value.

The $h \gamma Z$ coupling can also be probed via the $e^+ e^- \to \gamma h$
process, whose cross-section is also maximal around 250~GeV.
The cross-sections in the SM are rather small, for example 0.20~fb for
the beam polarization $P(e^-, e^+) = (-0.8, +0.3)$.
Upper limits at 95\% on the production cross-sections
in the different polarization scenarios can be set at 1.8~fb
for the same beam polarization~\cite{Aoki:2021khh}.   The polarization-dependence of this limit and the implications for limits on SMEFT parameters are discussed in~\cite{Aoki:2022dxg}.

\subsection{CP properties}

CP properties of the Higgs boson can be probed in its decays to $\tau$ leptons~\cite{Jeans:2018anq}, or
in its coupling to the EW bosons W and Z~\cite{Ogawa:2017bmg}.

In the $\tau$ decay channel, the $\tau$ decay products act as polarimeters, providing an estimate of the spin orientation.
The correlation between the two $\tau$s' polarimeter components perpendicular to the $\tau$ momentum direction
is sensitive to their CP state. The clean experimental environment at ILC and the high precision
detectors being developed are conducive to accurate reconstruction of $\tau$ lepton decays, allowing
good reconstruction of $\tau$ polarimeter information.
Mixing between odd and even CP components of the $\tau$ pair can be probed with a
precision of 75~mrad at ILC-250~\cite{Jeans:2018anq,Bozovic-Jelisavucic:2022ivd}.

The couplings of the Higgs to $WW$ and $ZZ$, both in decay and production, also provide sensitive probes
of CP violation effects.
Anomalous CP-violating couplings can affect angular correlations between vector boson decay planes.
Limits of around 7\% on CP violating terms in the $HVV$ coupling can be
achieved at ILC-250, and further improved at higher ILC energies~\cite{Ogawa:2017bmg, Ogawa:2018ssv}.

\section{Higgs -- exotic decays}
\label{sec:HiggsExotic}

Higgs exotic decays provide unique opportunities to probe a broad
class of new physics models~\cite{Curtin:2013fra}. Studying the Higgs
exotic decay precision would help reveal new physics, especially
hidden sector dynamics through this generic Higgs portal. The physics
we can learn from the Higgs exotic decay program is also complementary
to the Higgs coupling precision measurements.

The fact that the Higgs boson is produced at 250~GeV mainly in association with a $Z$ boson implies that events with a single  isolated
$Z$ boson provide a tag for Higgs decay to completely invisible final states.  There is a SM decay $H\to 4\nu$, but this has a predicted branching fraction of 0.1\%, giving a significant window for the discovery of invisible decays due to new physics.  At the ILC, it is expected that, if this mode is not observed, it will be possible to place a 95\% confidence upper limit on this branching ratio of 0.16\%~\cite{Kato:2020pyl,Potter:2022shg}. 

\begin{figure}[!ht]
\centering
\includegraphics[scale=0.45,clip]{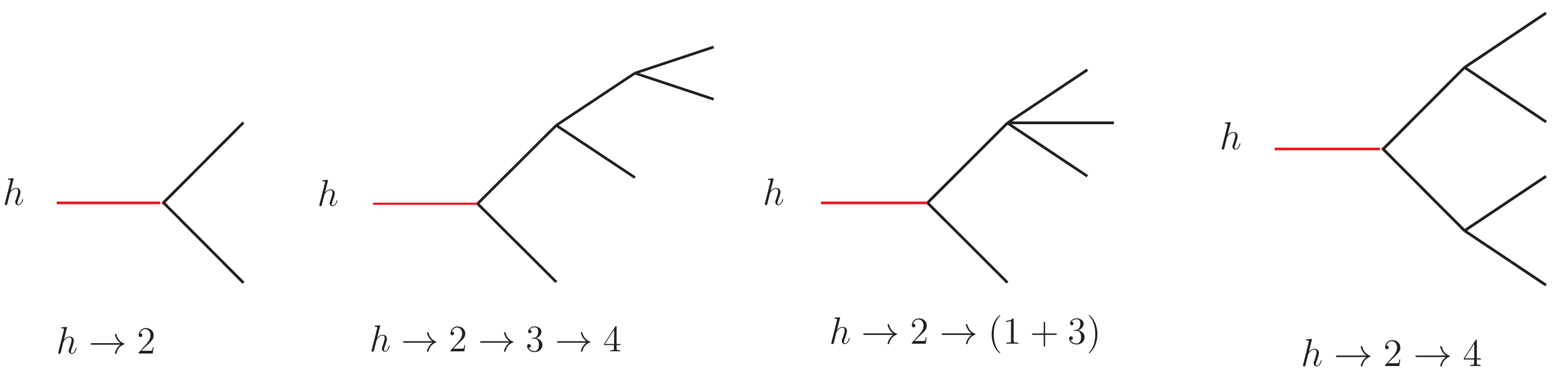}
\caption{\label{fig:topo}Representative topologies of the Higgs exotic decays. }
\end{figure}

In addition to the completely invisible decay, the Higgs boson has many possible modes of exotic decay that are forbidden in the SM, including partially invisible decays and flavor-changing decays.   Many of these decay modes, especially hadronic decays, are very challenging to observe at hadron colliders.
An initial  survey of the these possibilities
has been carried 
out in~\cite{Liu:2016zki}, and this study has shown  promising sensitivities at lepton colliders across the range of these modes.
The study focuses on two-body Higgs decays into BSM particles, dubbed as $X_i$, $h\to X_1 X_2$, which are
allowed to decay further, to up to four-body final states.
The cascade decay modes are classified into four cases, schematically
shown in Fig. \ref{fig:topo}.
Decays with these topologies are 
 motivated by a large class of BSM physics, such as singlet
 extensions, 
two-Higgs-doublet-models, SUSY models, Higgs portals,
gauge extensions of the SM~\cite{Curtin:2013fra,Liu:2016zki,deFlorian:2016spz,Cepeda:2021rql}.

For ILC running at the center of mass energy $250$~GeV, the essential Higgs production mechanism is $Z$-Higgs associated production $\ee\to Zh$. The $Z$ boson
with visible decays
enables Higgs tagging using the ``recoil mass'' technique.
A cut around the peak of the recoil mass spectrum would remove the
majority of the SM background. To demonstrate a typical Higgs exotic
search at ILC, we discuss here the analysis of  benchmark
processes  $H \to jj+\missET$.
In the last part of this section, we present the summary for 
Higgs exotic decay physics potential at ILC for an integrated
 luminosity of 2~ab$^{-1}$ and provide an outlook for future studies and improvements.

For event simulation,  we generate both the signal and the background events for a 250 GeV
electron-positron collider with {\tt MadGraph5} at parton level \cite{Alwall:2014hca}
Our parameter choices for the detector effects and our pre-selection
cuts are chosen to be  universal for the analyses for all Higgs exotic decay modes.
All of the visible particles in the final state are required to have
$|\cos\theta|<0.98$. The final state particles are 
required to be well-separated with
\beq
y_{ij}\equiv {2\min\left(E_i^2,E_j^2\right)\left(1-\cos\theta_{ij}\right)}/{E_{vis}^2}\geq~0.001 \ .
\eeqn
We only study the cases where the $Z$ boson decays into $\ell^+\ell^-$
where $\ell^\pm=e^\pm, \mu^\pm$.
The signal events are required to contain at least one pair of
opposite-sign, same-flavor charged leptons with an opening angle greater than
$80^\circ$ and satisfying $E_{\ell}>~5~\mbox{GeV}$ and
$|m_{\ell\ell}-m_Z|<10~{rm GeV}$, 
where $m_{\ell\ell}$ is the invariant mass of the di-lepton system.
The recoil mass is defined as
$m_{\rm recoil}^2 \equiv s-2\sqrt{s}E_{\ell\ell}+m_{\ell\ell}^2$
where $E_{\ell \ell } = E_{\ell^+} + E_{\ell^-}$.
The recoil mass is required to satisfy $\left|m_{\rm recoil}-m_h\right|< 5$~GeV. To suppress the ISR contribution to the backgrounds\footnote{Corrections from beamstrahlung effect~\cite{Xiu:2015tha} and ISR effect~\cite{Greco:2016izi} need to be carefully taken into account for certain processes relying on a precise reconstruction of the recoil mass.}, for Higgs exotic decay modes without missing energy, we require the events to have the total visible energy $E_{vis}>~225~{\rm GeV}$.
We mimic the detector resolution effect by adding Gaussian smearing
effects on the four-momentum of the particles, 
as detailed in Ref.~\cite{Liu:2016zki}.

For the  $H \to jj+\missET$ analysis, we assume that 
the SM-like Higgs boson decays into $X_2X_1$ with $X_1$ invisible and
$X_2$ having the decay $X_2\to X_1 j j$ through an off-shell intermediate state.
Beyond the pre-selection cut and the recoil mass cut,
we require that there are two additional jets that satisfy
$E_j>~10~{\rm GeV}~{\rm and}~|\cos\theta_j|<0.98$.
The dominant background after the recoil mass cut will be the Higgsstrahlung process with
$h\to ZZ^* \to q \overbar q \nu \overbar{\nu}$.

\begin{figure}[!ht]
\centering
\includegraphics[width=8cm]{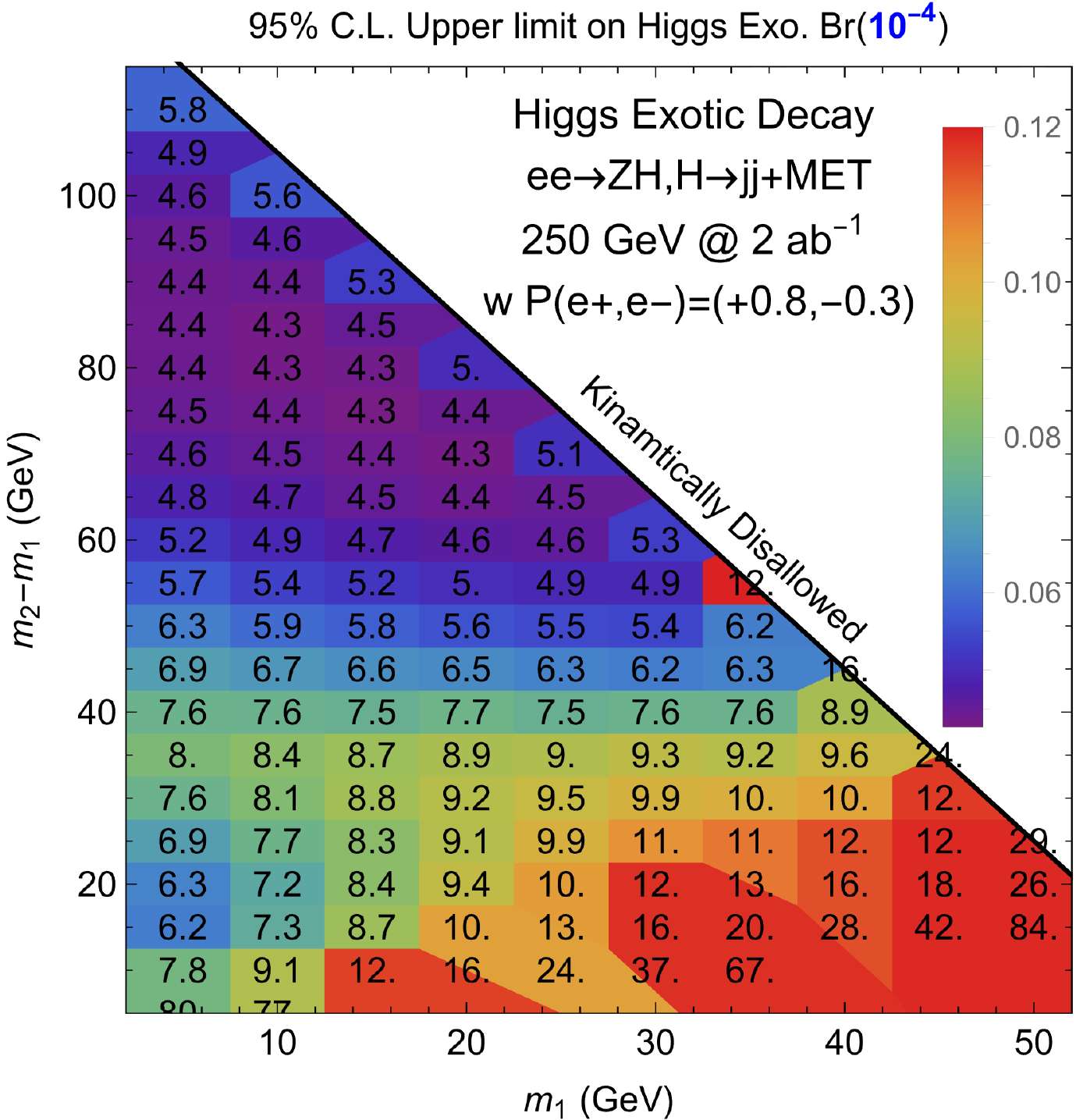}
\caption{\label{fig:zvvjj2}The 95\% C.L. upper limit on the Higgs exotic decay branching
fractions into $jj+\missET$ arising from a decay $H\to X_2X_1$, as
described in the text, with $X_1$ a stable particle of mass $m_1$ and
$X_2$ higher in mass by  $m_2-m_1$.}
\end{figure}

We use the likelihood function of the
$m_{jj}$-$\missET$ distribution to derive the
exclusive limit. The results are
shown in Fig.~\ref{fig:zvvjj2} in the plane of $X_1$, mass $m_1$, and the mass splitting
between $X_2$ and $X_1$, $m_2-m_1$ for $h\to jj+\missET$. The exclusion limits on the branching fraction
in the bulk region of the parameter space reach $3\times10^{-4}\sim 8\times
10^{-4}$ for $h\to jj+\missET$.
We can see that when the mass splitting
$m_2-m_1$ is around 80~GeV, the future lepton colliders have the strongest sensitivities
on these Higgs exotic channels, reaching around $4.3\times 10^{-4}$ for $h\to jj +\missET$.
When $X_1$ is light and $m_2-m_1$ is large, the energy is shared by the two jets and the $X_1$.
Consequently, when the mass splitting $m_2-m_1$ is around 80~GeV, the dijet invariant mass
will be around 40--60~GeV, falling in the ``valley'' of low SM background.
For heavier $X_1$, the MET will be lower due to less momentum available for the LSP. 

\begin{figure}[!ht]
\begin{center}
\includegraphics[width=0.90\hsize]{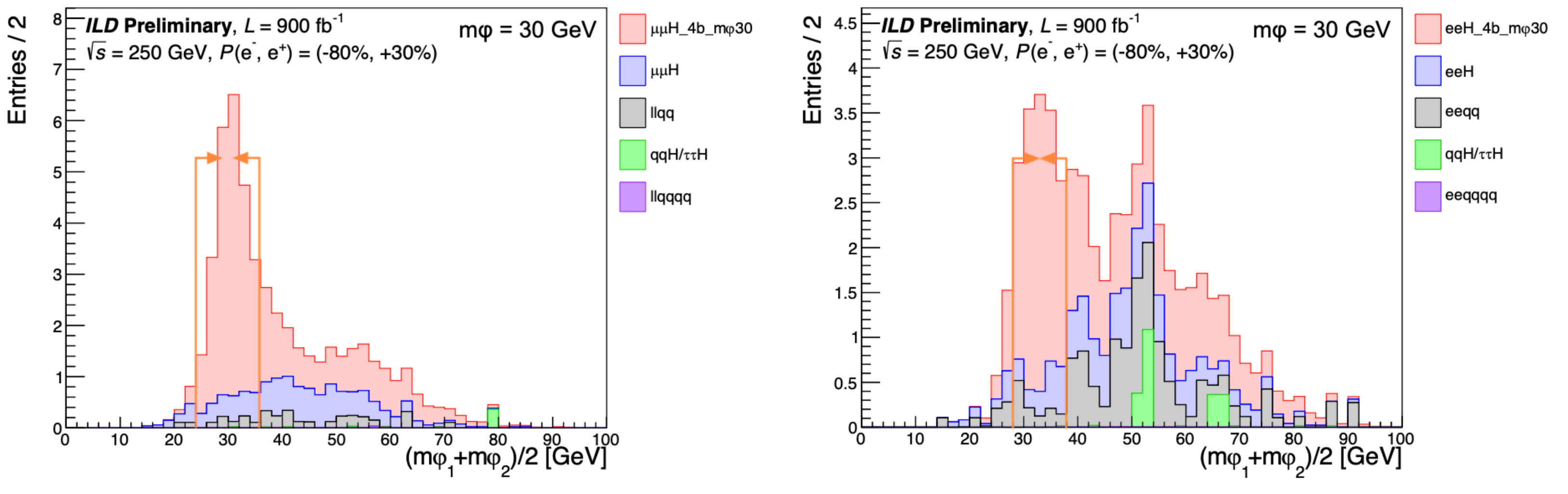}
\end{center}
\caption{Stacked distribution of reconstructed average scalar mass 
for the signal $e^+e^-\to l^+l^- H$, $H\to \phi\phi\to (b\bar{b})(b\bar{b})$ and 
background events based on full detector simulation, 
at $\sqrt{s}=250$ GeV with an integrated luminosity of 900 fb$^{-1}$. 
Left figure is for $P(e^-,e^+)=(-0.8,+0.3)$ and right is for
$P(e^-,e^+)=(+0.8,-0.3)$.}
\label{fig:exotic_H4b}
\end{figure}

To further demonstrate the search of Higgs exotic decay, a more 
realistic analysis based on full detector simulation 
has been carried out at the ILC~\cite{Kato:2022},
focusing on one of the representative channel where Higgs decays into 
a pair of light new scalars ($\phi$) both of which decay into a pair of $b$-quark.
A set of masses for the scalars from 15 GeV to 60 GeV are studied as benchmarks.
Higgs-strahlung process $Z\to e^+e^-/\mu^+\mu^-$ is employed as the Higgs production channel. The signal final state consists of two isolated leptons and four 
$b$-jets. This is a very clean channel thanks to the excellent 
$b$-tagging performance and very narrow leptonic recoil mass spectrum. 
It turns out that the dominant background after 
all the selection cuts are from SM Higgs decay.
The distributions of the reconstructed average of two scalar masses 
are shown in Fig.~\ref{fig:exotic_H4b} for the remained signal and background 
events assuming a branching ration of 1\% for $H\to 4b$ and the scalar mass of
30 GeV, where the signal resonance peak from exotic decay can be clearly seen. 
By combining two leptonic channels, the 95\% confidence level 
uppper limit of $BR(H\to 4b)$ is expected to be around 0.1\%
for all the four benchmark scalar masses (15/30/45/60 GeV) with an integrated
luminosity of 900 fb$^{-1}$ for each left-handed and right-handed polarization
at the ILC250.

%%%%%%%%%%%%%
%%%%%%%%%%%%%

\begin{figure}[!ht]
\centering
\includegraphics[width=16cm]{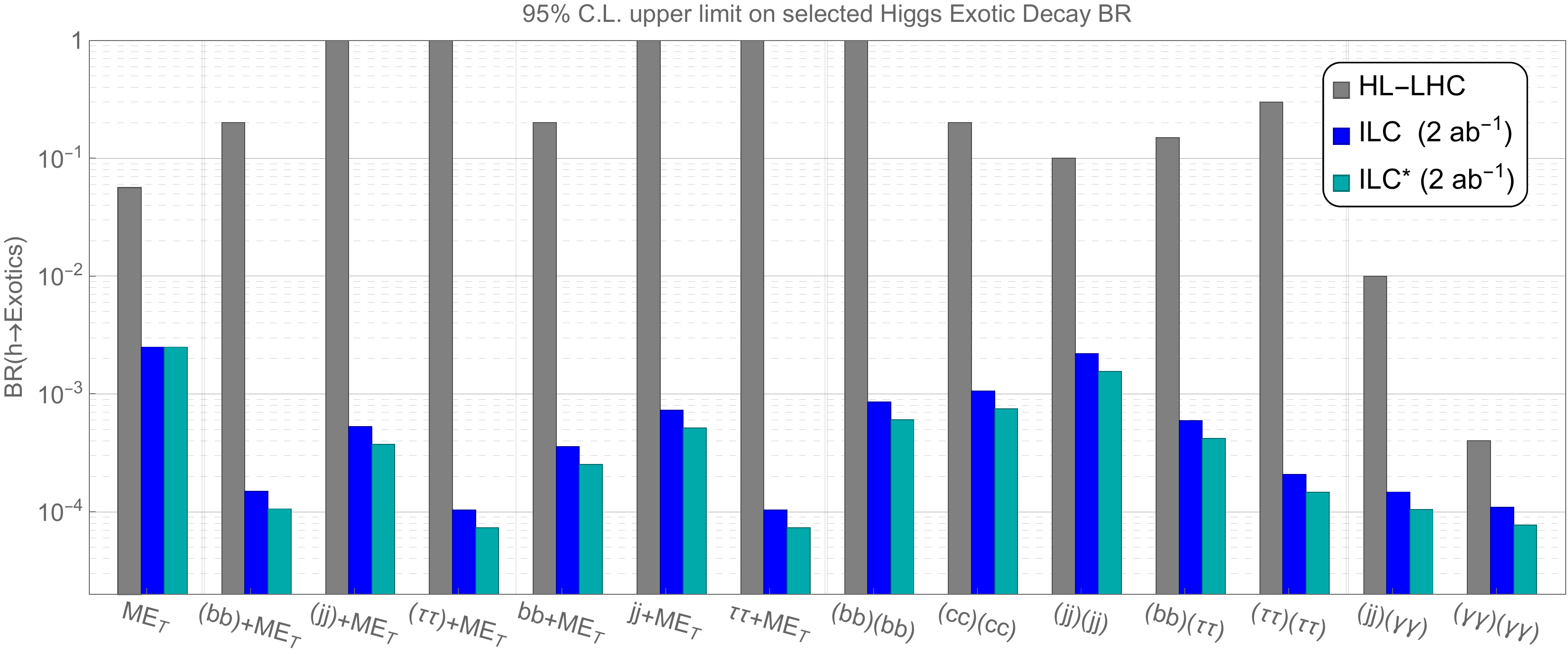}
\caption{\label{fig:ExoticHiggssummary} The 95\% C.L. upper limit on selected Higgs exotic decay
branching fractions at HL-LHC and ILC, based on
Ref~\cite{Liu:2016zki}. The ILC curves are derived using results from
Ref~\cite{Liu:2016zki} with leptonic decaying Z boson in the
$e^-e^+\rightarrow ZH$ process. The ILC$^*$ scenario further utilizes
the hadronically decaying Z boson and includes an estimated
(indicative) improvement of 40\%. Each set of three bars describes a
different topology of exotic Higgs decay. For a recent review on current LHC constraints, see Ref.~\cite{Cepeda:2021rql}.}
\end{figure}

A large number of similar analyses are described in \cite{Liu:2016zki}.
We summarize the results  in Fig.~\ref{fig:ExoticHiggssummary}, giving the
expected limits  for the ILC with 2~ab$^{-1}$ integrated luminosity. We
also include the projected LHC sensitivities  in gray bars.  We use
the up-to-date projected sensitivities for the LHC constraints, but
many do not exist or are very conservative. More recent studies, e.g., Ref.~\cite{Shelton:2021xwo} on $h\to 4\tau$, and Ref.C~\cite{Kato:2022} on $h\to 4b$, show consistent projection on sensitivities as well. 

 The LHC will provide strong constraints on many 
many channels that can be
characterized with muons, electrons, and photons.
In the summary Fig.~\ref{fig:ExoticHiggssummary},
the exotic Higgs decay channels are some of those that are difficult
to constrain at 
the LHC.   For these channels, which rely on signals from jets, heavy
quarks, and $\tau$s, the  improvements over the LHC expectations
vary from one to four orders of magnitude.
This great advantage benefits from the 
low QCD background and Higgs tagging from the recoil mass technique at future lepton colliders.
For the Higgs exotic decays without missing energy, the improvement varies between two to
three orders of magnitude, except for the one order of magnitude improvement for the
$(\gamma\gamma)(\gamma\gamma)$ channel.   Here the possibility at the
LHC of
reconstruction of the Higgs mass from
the final state particles provides additional signal-background discrimination
power.   Channels with  electrons, muons, and photons, which 
are relatively clean objects at the LHC,  can  take advantage of the
higher statistics available from the HL-LHC.

Many new and interesting channels deserve  further study.
Higgs exotic decays of $H\to XX \to 4f$ where the intermediate
resonant particle $X$ mass is below 10~GeV is one of these 
channels. This scenario is particularly motivated by the recent discussion
of the connection between  Higgs exotic decay and  strongly first order
electroweak phase
transitions~\cite{Carena:2019une,Kozaczuk:2019pet}. In this
region, the particle $X$ can be long-lived, so the study should be
extended  
 into long-lived particle regime~\cite{Alipour-fard:2018mre}.

Another example is the Higgs decay into a dark shower, that is, a
shower of dark-sector particles.\footnote{These can be bosons or be fermions, for
example, composite neutrinos~\cite{Chacko:2020zze}.}  These   can either
decay promptly or be long-lived, and their decay back to visible SM
particles 
can be either hadronic or leptonic. The process is motivated by
generic considerations of hidden sector strong dynamics.  It also
appears in  the discussion of neutral
naturalness~\cite{Craig:2015pha}.   Current
studies have been focusing on the Higgs decays into a pair of twin
glueballs~\cite{Curtin:2015fna,Liu:2018wte,Alipour-fard:2018mre,Liu:2020vur},
but this  is only a subclass of the generic Higgs decays into these final
states. This dark shower channel is also motivated by
the class of models with large number of light scalars~\cite{Jung:2021tym}, e.g.,
NNaturalness~\cite{Arkani-Hamed:2016rle}, EW scale as a
trigger~\cite{Arkani-Hamed:2020yna}, and delayed or non-restored electroweak 
symmetry~\cite{Meade:2018saz,Baldes:2018nel,Glioti:2018roy,Matsedonskyi:2020mlz}.
 The phenomenological study of this general
class of models is complex, and this is especially true  at the LHC due to the challenges
of triggers and backgrounds. The triggerless operation of the ILC
detectors, where any novel Higgs decay can be recorded and  identified,  provides  a strong advantage.

\section{Triple gauge couplings}
\label{sec:Wboson250}

A second major goal of the 250~GeV ILC program is to carry out
precision measurements on the $W$ boson.   Improvements in quantities
relevant to precision electroweak observables---the $W$ mass, width,
and decay branching ratios---will be discussed in
Chapter~\ref{chap:PEW}.   Here we discuss the improvement in the
direct measurement of the $W$ boson interactions.

The electroweak interactions of a pair of $W$ bosons of lowest
dimensionality are 
described by the Lagrangian
\beqa 
\Delta L &=&   i g_V \biggl\{ V^\mu (W^-_{\mu\nu} W^{+\nu} - W^+_{\mu\nu}
W^{-\nu} ) + \kappa_V  (W^+_\mu W^-_\nu V^{\mu\nu} ) \CR
      & &  \hskip 0.5in + \frac{\lambda_V}{m_W^2}\, W^-_\mu{}^\nu
      W^+_\nu{}^\rho V^{\mu\rho} \biggr\}    \ , 
\eeqan
where $V = \gamma, Z$,  and $W^\pm_{\mu\nu}$ and $V_{\mu\nu}$ are the
gauge boson field strengths. Always,  $g_\gamma = e$, reflecting
the electric charge of the $W$ boson.  
In the SM,  $g_Z = -\frac{c_w}{s_w}$,  $\kappa_\gamma = \kappa_Z = 1$,
and  $\lambda_\gamma = \lambda_Z = 0$, with $s_w, c_w$ the sine and
cosine of $\theta_w$.   In the most general setting,
the 5 free 
coefficients are all independent of one another.  However,   there are
two $SU(2)\times U(1)$ relations, 
\beq
(\kappa_Z - 1) =  - \frac{s_w^2}{c_w^2} (\kappa_\gamma - 1 ) \qquad
\lambda_\gamma = \lambda_Z  \ .
\eeqn
This leaves three free parameters to be determined.  

  New interactions at the TeV scale, especially mixing of the $W$ bosons
with new elementary or composite vector bosons, can generate 
small corrections to these couplings, suppressed by a factor
$m_W^2/M^2$, where $M$ is the heavy mass scale.  Electroweak loop
diagrams give computable  corrections to the triple gauge
coupling (TGC) parameters at the $10^{-3}$  level.  Searches for
these effects require high precision in the reconstruction of the full
reaction producing  the $W$ bosons.

The TGC parameters also play an important role in the Higgs boson
program.   As we will discuss in Chapter~\ref{chap:SMEFT}, our
determination of the Higgs boson width and the normalization of Higgs
couplings is based on a fit using SM Effective Field Theory.   This
fit relies not only on measurements of Higgs processes but also on
other ILC electroweak measurements.  The TGC parameters play an
important role.   For this reason also, it is essential to have
precise determinations of the TGC parameters at 250~GeV.

In this section, we will discuss the measurements of the TGC
parameters at the ILC at 250~GeV, based on $W^+W^-$ production and on
single $W$ production.   The impact of higher center-of-mass energy
measurements will be described in
 Section~\ref{sec:Wboson500}. 

At LEP,  the three couplings $g^Z$, $\kappa_{\gamma}$ and
$\lambda_{\gamma}$ have  been constrained a the level of a few
$10^{-2}$. Limits have been derived in fits of individual parameters,
fixing the other two to their SM values~\cite{ALEPH:2013dgf}, as well
as in two- and three-parameter fits, which allowed two or all three
couplings to 
vary simultaneously~\cite{ALEPH:2004klc, DELPHI:2001kqh, L3:2004ulv, OPAL:2003xqq}.
The same three parameters are currently being studied at the LHC, for
example, in~\cite{CMS:2019ppl},  reaching precisions between $6$ and
$8\times 10^{-3}$ in single-parameter fits and between $7$ and
$12\times 10^{-3}$ in two-parameter fits. In these analyses,  $g_Z$ and
$\kappa_{\gamma}$ show a strong, almost $100\%$ correlation. For the
HL-LHC, generator-level projections of three-parameter fits have been
performed based on NLO cross-sections and assumptions on efficiencies
derived from the corresponding $8$\,TeV ATLAS and CMS
analyses~\cite{Azzi:2019yne}. This study projects precisions between
$2$ and $5\times 10^{-3}$, with the same strong correlation between
$g^Z_1$ and $\kappa_{\gamma}$. It also evaluated the effect of non-SM
Z-fermion couplings (in particular the $q$-$\bar{q}$-$Z$ couplings) by
letting them float in the fit within $2\sigma$ bounds from fits to LEP
data. This has a huge impact on the to ability extract $g^Z$ and
$\kappa_{\gamma}$: their constraints weaken to the level of $1$-$2\times
10^{-2}$. This highlights an important area of ILC-LHC interplay: the
couplings of the $Z$ boson to fermions will be measured to
unprecedented precision both at the $Z$ pole and -- more relevant here
-- at higher energies, as 
discussed in sections~\ref{sec:radreturn},~\ref{sec:Zpole} and~\ref{sec:pairs500} of this report.

Most studies of the capability of future $e^+e^-$ linear colliders to
constrain triple gauge vertices have been performed at a
center-of-mass energy of $500$\,GeV. These range from studies based on
full, Geant4-based simulations of the ILD detector concept focusing on
the $WW \to \mu \nu qq$ and $WW \to e \nu qq$ channels and the
determination of the three LEP couplings~\cite{Marchesini:2011aka} to
theory-level studies showing that with polarized beams, all 28 real
parameters of the most general possible Lagrangian for triple gauge
interactions can be determined~\cite{Diehl:2003qz, Diehl:2002nj,Diehl:1997ft}. The results of the full simulation studies, which
included only a subset of channels and observables, as will be
discussed in section~\ref{sec:Wboson500}, have been extrapolated to
$\sqrt{s}=250$\,GeV in Sec.~2.3.3.2 of~\cite{Karl:2019hes}, with
rather conservative assumptions on the change in the impact of
detector effects with center-of-mass energy. This extrapolation yields
precisions between $8$ and $10\times 10^{-4}$. Notably it also shows
that the correlation between $g^Z_1$ and $\kappa_{Z}$ in $e^+e^-$
collisions depends on the center-of-mass energy and the beam
 polarizations, and can even change sign. 
Thus, runs with different energies and polarizations can eliminate any blind direction.  

Finally, the expected impact of including all channels and using an
unbinned log-likelihood fit to all observables
 (instead of binned fit to a reduced set of observables) improves the
 projections for ILC250 to the level of 
$4-6\times 10^{-4}$ --- nearly a full order of magnitude better than
the previously 
discussed HL-LHC expectations, even when fixing Z-fermion couplings for the HL-LHC.
A comparison with the higher energy stages of the ILC is shown in Fig~\ref{fig:TGC_ILC_allECM}. 

\begin{figure}
\begin{center}
\begin{subfigure}{0.525\textwidth}
\includegraphics[width=0.95\hsize]{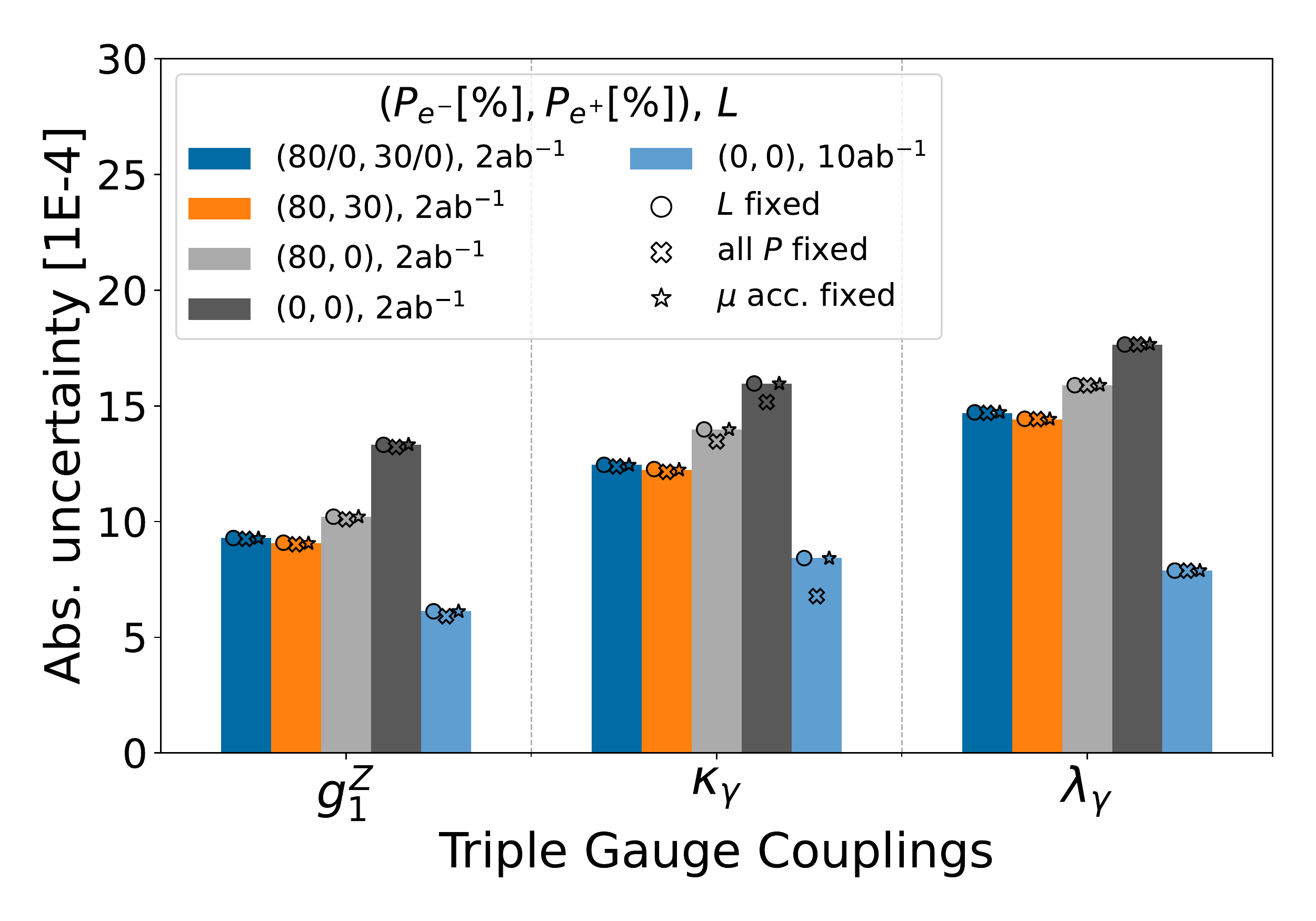}
\caption{\label{fig:tgcs_fitJakob:a}}
\end{subfigure}
\begin{subfigure}{0.465\textwidth}
\includegraphics[width=0.95\hsize]{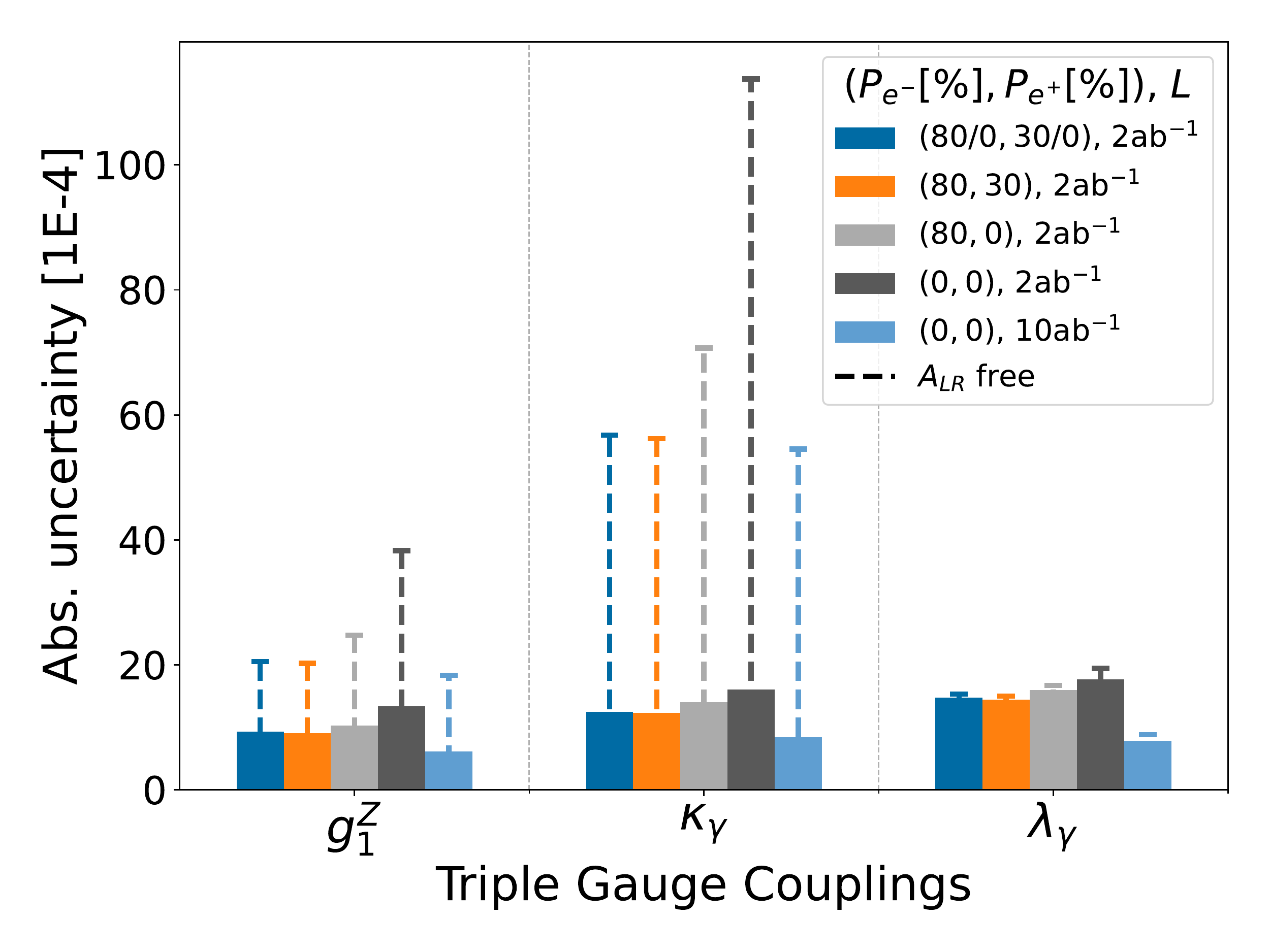}
\caption{\label{fig:tgcs_fitJakob:b}}
\end{subfigure}
\end{center}
\caption{(a) Expected precisions at $\sqrt{s}=250$\, GeV of charged TGCs for different assumptions on the beam polarizations and the integrated luminosity. Note that these appear more pessimistic than in Fig.~\ref{fig:TGC_ILC_allECM} since here only the $\mu \nu qq$ channel is used in a binned three-angle analysis (b) Effect of floating the left-right asymmetry of the the $Zee$ coupling in the fit.  This underlines, first, the importance of independently measuring this this quantity --- not only on the $Z$ pole, but at $250$\,GeV --- and, second, the increased robustness of the fit due to beam polarization, which reduces the dependence on the left-right asymmetry considerably. Both from~\cite{Beyer:2022xyz, List:2022xyz}.}
\label{fig:tgcs_fitJakob}
\end{figure}

At the ILC, triple gauge vertices can also be studied in  single-$W$
production, $e^+e^-\to e^{\pm}W^{\mp}\nu_e (\bar{\nu}_e)$.   This adds
another independent data set to the determinations.   It also brings
up another issue.  At the ILC, the 
single and pair production of $W$ bosons are used not only to measure
the TGCs but also to serve as standard candles to
gauge the luminosity-weighted 
and long-term averaged beam polarization values.   This  raises the question of
whether effects of anomalous couplings and beam polarization can be
reliably disentangled, and whether beam polarizations introduce an
additional uncertainty. Furthermore, there is the question of possible
impact from the other involved vertices, namely the $e-\nu-W$ vertex
in all $t$-channel contributions and 
the $e-e-Z$ vertex in $s$-channel $WW$ and $t$-channel single-$W$ production. 
In order to address these questions, a fit to a variety of binned
generator-level $e^+e^- \to f \bar{f}$, $e^+e^- \to W^+W^-$ and
single-$W$ distributions has been pioneered in~\cite{Karl:2019hes} and
further developed in~\cite{Beyer:2020eas, Beyer:2022xyz,
  List:2022xyz}. The results of a fit to differential distributions
from $e^+e^- \to \mu^+ \mu^- $ and $e^+e^- \to \mu \nu qq$ 
at 250~GeV, which
treats not only the three triple gauge couplings, but also
(unpolarized) total cross-sections, left-right asymmetries, the
angular acceptance and the beam polarizations as free parameters, is
displayed in Fig~\ref{fig:tgcs_fitJakob} for various assumptions on
the integrated luminosity and beam polarizations. For the ILC-like configuration
(orange bars), the triple gauge couplings are determined at the level
of $10$ to $15\times 10^{-4}$ in this much more general fit from the
muon final state only.   With the final state with electrons also
included, 
this would  correspond to precisions of $7$ to $11\times 10^{-4}$. 
 Within the uncertainty of the
extrapolation and the different number of observables and free
parameters, this compares very well with the $8$ to $10\times 10^{-4}$
from the extrapolation of the 
full simulation analysis and shows that the ILC measurements will be 
extremely robust against consideration of additional free parameters.

We emphasize that the use of beam polarization plays an important role.
Since the effect of anomalous TGCs on the differential cross-sections
differs between $e^-_Le^+_R$ and $e^-_Re^+_L$, the ability to take
data in different polarization configurations adds qualitatively new
information. This leads to improvements of 40\%, 30\% and
20\%, respectively, for $g_Z$, $\kappa_{\gamma}$ and $\lambda_{\gamma}$,
compared to the case of unpolarized beams. Thus, the
additional information provided by the polarized beams is equivalent
to a factor 1.5 to 2 more luminosity. Figure~\ref{fig:tgcs_fitJakob} 
also shows another
important aspect, concerning the robustness against finite knowledge
of other SM parameters.   When the left-right
asymmetry  in the $Zee$ coupling 
is set free in the fit, the uncertainties on $g_Z$ and
$\kappa_{\gamma}$ for the unpolarized case are a factor of two larger
than in the polarized case.  This  shows that the additional
information from the polarization reduces the dependency on residual
parametric and theoretical uncertainties that enter the analysis of
$WW$ production.

\section{Precision QCD} 
\label{sec:QCD}

%This section will describe not only the precision measurement of $\alpha_s$ from jets at ILC, but also new observables sensitive to jet substructure and other detailed features of QCD.  See, for example, \cite{Chen:2019bpb,Chen:2020vvp}.

The theory of Quantum Chromodynamics is one of the central elements of the Standard Model, and plays a dominant role in understanding a wide range of collider experiments. Due to their QCD neutral initial state, $e^+e^-$ colliders are the simplest setting in which to study the dynamics of energy flow in QCD, enabling precision measurements well beyond what is possible in hadron colliders. While $e^+e^-$ colliders such as the ILC allow the precision measurement of QCD parameters, such as $\alpha_s$, their legacy is much larger due to the notions of factorization and universality, which allow detailed measurements of QCD final states to be applied in the search for new physics in hadronic colliders. 

To emphasize the immense impact that the ILC would have on studies of QCD, it is worth recalling the impact that LEP has had, as well as emphasizing some of its shortcomings that the ILC would be able to fill. While a wide variety of event shapes were measured at LEP, since LEP ran at the $Z$-pole, these were primarily dijet event shapes measured  on quark jets. These data have had a profound impact on the study of jets at the LHC in the fact that quark jets are extremely well modelled in parton shower Monte Carlo programs, since their non-perturbative effects can be tuned against this rich dataset. On the other hand, gluon jets, which were not produced that often at LEP, but are copiously produced at the LHC, are poorly modeled. The precision measurement of event shapes have also enabled precision extractions of the strong coupling constant, $\alpha_s$ \cite{Abbate:2010xh,Hoang:2015hka}.

In the time since LEP there has been massive theoretical progress, driven by a renewed interest in studying the dynamics of jets in jet substructure at the LHC. The high energies and remarkable angular resolution of the LHC have enabled the detailed structure of energy within jets to be measured, requiring new calculational techniques to be developed. This was originally driven by the fact that the energy pattern within jets can be used to distinguish jets initiated by a light quark or gluon from jets initiated by an electroweak scale boson. The techniques developed with this initial motivation in mind have enabled a variety of new ways of understanding the dynamics of QCD with increasing sophistication. This includes both qualitative advances in the design of observables for probing specific features of QCD, as well as advances in theoretical techniques for event shape calculations. For a review, see \cite{Larkoski:2017jix,Marzani:2019hun} It is therefore worth asking what one could do with a fresh slate if one had a new $e^+e^-$ machine for understanding QCD.  This section provides a brief overview of some such possibilities, as well as more detailed references for the interested reader,  emphasizing that the higher energies and better resolution calorimeters of the ILC would be transformative for QCD. 

%%%%%%%%%%%%%%%%%%%%%%
\subsubsection*{Energy Flow Observables in QCD}
%%%%%%%%%%%%%%%%%%%%%%

Measurements of the flow of radiation in collider events provide one of the most interesting tests of our understanding of QCD. High energy collisions are particularly interesting, since they provide a probe of the dynamics of QCD from asymptotically free quarks and gluons, through the confining phase transition to free hadrons at asymptotic infinity. Energy flow in colliders can be studied either using event shapes, which can be thought of as resolution variables about an underlying $S$-matrix element of quarks and gluons, or using correlation functions, which measure statistical properties of the radiation. Both approaches have seen significant progress driven by jet substructure at the LHC, giving rise to many interesting new observables that could be measured at the ILC, providing a significantly extended understanding of energy flow in quantum field theory.

%%%%%%%%%%%%%%%%%%%%%%
\subsubsection*{New Event Shape Observables}
%%%%%%%%%%%%%%%%%%%%%%

Event shape observables, which constrain radiation about a particular $S$-matrix element are particularly useful at the LHC for identifying boosted electroweak scale bosons decaying into jets. There has therefore been significant progress in their understanding, and many new observables have been proposed. In particular, one of the most important outputs of the jet substructure program is the ability to design event shape observables with specific properties. Such observables were simply not available in the LEP era, and would therefore be extremely interesting to measure at the ILC. 

While there are endless examples of such observables, here we content ourself with describing one particular class of observables, namely ``groomed" observables \cite{Dasgupta:2013ihk,Larkoski:2014wba}. One of the insights of the jet substructure program has been the introduction of grooming algorithms that systematically remove low energy soft radiation. Such low energy soft radiation generically contributes the leading hadronization corrections, and therefore grooming can significantly reduce non-perturbative effects. For a generic infrared and collinear safe observable, one can then measure its ``groomed" counterpart, which will also be IRC safe. Although these observables are theoretically cumbersome, due to the fact that they reduce non-perturbative corrections they can be practically useful, for example for measurements of $\alpha_s$. 

These ``groomed" observables have received significant theoretical attention due to their use in jet substructure. However, since they were introduced post-LEP, they have not been measured in an $e^+e^-$ environment. An example of a theoretical prediction for a groomed observable is shown in Fig. \ref{fig:new_obs}, taken from \cite{Kardos:2020gty}. Measurements of these observables are useful for fundamental studies of QCD, and also would provide insights into their behavior at the LHC, but in a simpler context.

%%%%%%%%%%%%%%%%%%%%%%
\subsubsection*{Characterizing QCD with Correlation Functions}
%%%%%%%%%%%%%%%%%%%%%%

Since the LEP era there has been a significant improvement in our understanding of energy flow in collider experiments, driven quite interestingly, by purely formal developments. While the observables in the previous section were so called ``jet shape" observables, if the goal is to understand the structure of the underlying theory, one may be curious why one does not proceed in the standard manner taken for other physical systems, namely measuring correlation functions. Unlike for condensed matter systems where one typically characterizes systems by correlation functions of local operators, building up from low point correlators, in collider experiments one cannot measure correlation functions of local operators. However, instead, one can measure certain non-local lightray operators called energy flow operators, defined as integrals of the stress tensor along null infinity in a direction characterized by a unit vector $\vec n$ \cite{Hofman:2008ar}
\begin{align}\label{eq:ANEC_op}
\mathcal{E}(\vec n) = \lim_{r\to \infty}  \int\limits_0^\infty dt~ r^2 n^i T_{0i}(t,r \vec n)\,.
\end{align}
One can then characterize the system by measuring correlation functions of $\langle \mathcal{E}(\vec n_1) \mathcal{E}(\vec n_2) \cdots \mathcal{E}(\vec n_k) \rangle$ of these operators. These objects are particularly simple theoretically, since they exhibit symmetry properties similar to standard correlation functions of local operators, and are also governed by an operator product expansion. This enables one to discuss jet phenomenon in the language of correlation functions, and there has been a significant program to make this a phenomenological reality \cite{Chen:2022jhb,Chen:2021gdk,Chen:2020adz,Chen:2019bpb,Chen:2020vvp,Holguin:2022epo}. Furthermore, they exhibit simple structures in perturbation theory. One can show that jet shape observables are infinite sums over these correlation functions, and hence they lose many of these desirable theoretical properties.

\begin{figure}
\begin{center}
\includegraphics[width=0.95\hsize]{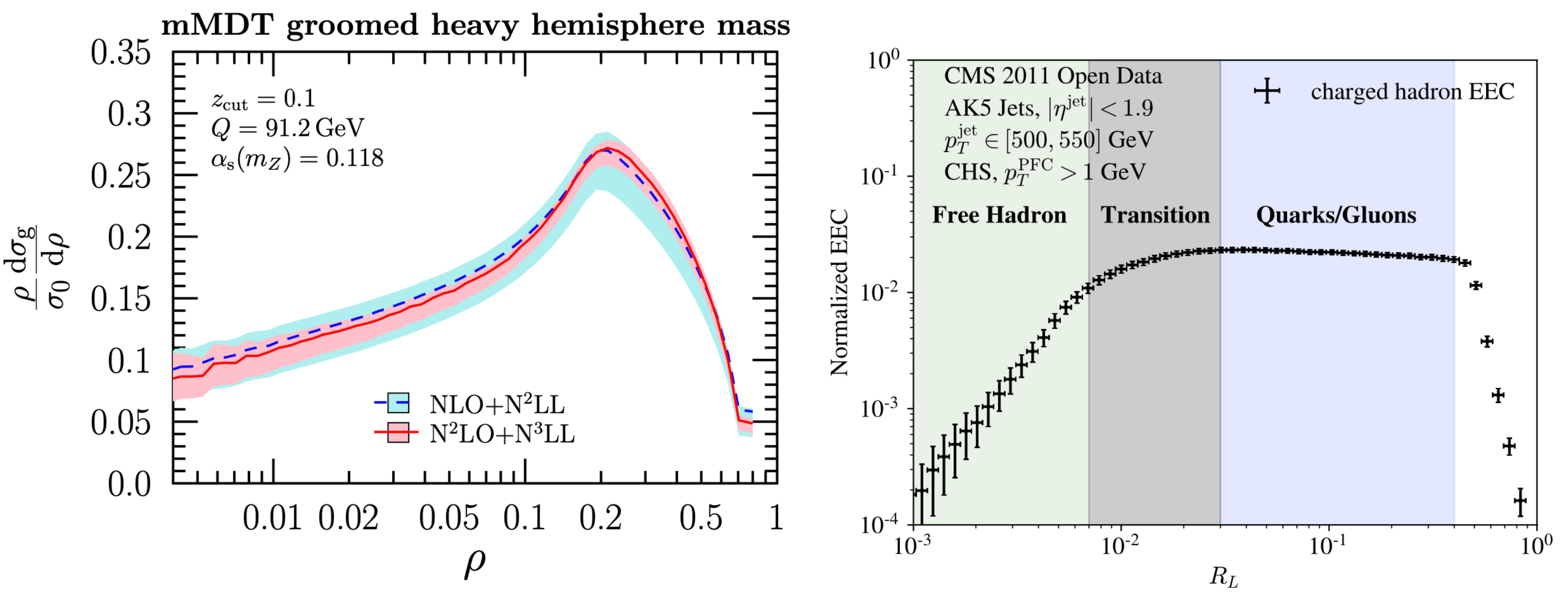} 
\end{center}
  \caption{(a) The groomed mass observable in $e^+e^-$ \cite{Kardos:2020gty}.  (b) A precision measurement of the two-point correlator in the collinear limit at the LHC \cite{Komiske:2022enw}. Both of these new observables provide interesting probes of $\alpha_s$}
  \label{fig:new_obs}
\end{figure}

Although the two-point correlator was measured at LEP, it was not studied in detail in the OPE limit to look for scaling behavior, and higher point correlators, which probe more interesting features of the theory, such as spin correlations, were never measured. A measurement of the two-point correlator using Open Data from the CMS experiment is shown in Fig. \ref{fig:new_obs}, illustrating beautiful scaling behavior of weakly coupled quarks and gluons, and a transition to the scaling of free hadrons \cite{Komiske:2022enw}. Measurements of this quality in the ILC environment would provide remarkable insights into the dynamics of QCD jets, and the hadronization transition.

The ILC would provide a beautiful environment where one can rethink how jets are studied and measure in detail the structure of multi-point correlators in QCD. These are of significant interest for understanding QCD, but also provide insight into the behavior of perturbative nearly conformal theories more general, and have been the focus of much recent interest of the theoretical community (see e.g. \cite{Kologlu:2019mfz}). Precision measurements of these correlators would build a bridge between the QCD phenomenology and formal theory communities which would result in significant progress.

%%%%%%%%%%%%%%%%%%%%%%
\subsubsection*{Precision Extractions of $\alpha_s$ with Old and New Observables}
%%%%%%%%%%%%%%%%%%%%%%

One of the key parameters of interest in QCD is the value of the strong coupling constant, $\alpha_s$. Since this is an arbitrary parameter of the theory, it can only be obtained by comparison precision theoretical predictions with experimental measurements. While there are many different possible observables that can be used to constrain the value of $\alpha_s$, measurements of the distribution of energy in $e^+e^-$ have proven to be competitive. However, there is currently an unresolved tension between extractions of $\alpha_s$ from event shape extractions at LEP as compared with lattice based extractions. Resolving this tension is important to illustrate a consistent understanding of QCD at the percent level.

The most precise extractions of $\alpha_s$ from event shapes are based on thrust and the $C$-parameter \cite{Abbate:2010xh,Hoang:2015hka}, which are closely related double logarithmic observables. To understand any possible issues in these extractions, it is crucial to have measurements based on other observables. Two observables that can be computed to high accuracy that exhibit significant differences from thrust/ $C$-parameter are the groomed thrust event shape, and the energy-energy correlators.

One of the complexities in extractions of $\alpha_s$ from event shapes is that one has to incorporate non-perturbative power corrections. These power corrections cannot be computed from first principles, and therefore must be simultaneously fit for along with the value of $\alpha_s$. One approach to reducing this potential uncertainty is to use grooming algorithms, inspired by the study of jet substructure at the LHC, to reduce non-perturbative corrections from low energy soft radiation. This makes the groomed thrust a potentially appealing observable for extractions of $\alpha_s$. Much like the thrust observable, its resummation is governed by the cusp and collinear anomalous dimensions, but grooming reduces it to a single logarithmic observable, and reduces the non-perturbative corrections. Due to this differing theoretical structure, an extraction of $\alpha_s$ from the groomed thrust would provide a relatively independent measurement of the value of $\alpha_s$. The groomed thrust can be computed to high perturbative accuracy, using a factorization formula. This is shown in Fig. \ref{fig:new_obs}. Furthermore, non-perturbative corrections to the groomed thrust distribution have been studied in \cite{Hoang:2019ceu}.

While the groomed thrust provides many complementary features to the standard thrust based extraction of $\alpha_s$, it is ultimately based on the same event shape paradigm, and therefore similar assumptions enter in the treatment of non-perturbative effects. Another interesting complementary measurement would be to perform a measurement of the two-point energy correlator in the collinear limit \cite{Dixon:2019uzg}. The collinear limit is described by completely different physics (fixed spin DGLAP) than the Sudakov region, and furthermore, since the energy correlators are not event shape observables, they have a different structure for their non-perturbative effects. However, despite being an old observable that was measured at LEP,  extractions of $\alpha_s$ from the collinear limit were never performed at LEP. We believe that this is partially due to the angular resolution of the calorimeters. Comparing the measurement of the two-point correlator at LEP vs. using the modern calorimetry of the LHC shows a completely different understanding of the collinear limit. Achieving a similarly precise measurement in the clean $e^+e^-$ environment of the ILC would be extremely valuable for precision measurements of $\alpha_s$, and would hopefully resolve the longstanding tensions in its extracted values.

In addition to the precision determination of $\alpha_s$, the ILC will also add to our knowledge of another basic QCD parameter, the bottom quark mass.  Methods for improved measurements of $m_b$ and, in particular, the evolution of this parameter under QCD running, are described in~\cite{Aparisi:2022yfn}.

%%%%%%%%%%%%%%%%%%%%%%
\subsubsection*{Gluons from the Higgs}
%%%%%%%%%%%%%%%%%%%%%%

\begin{figure}
\begin{center}
\includegraphics[width=0.45\hsize]{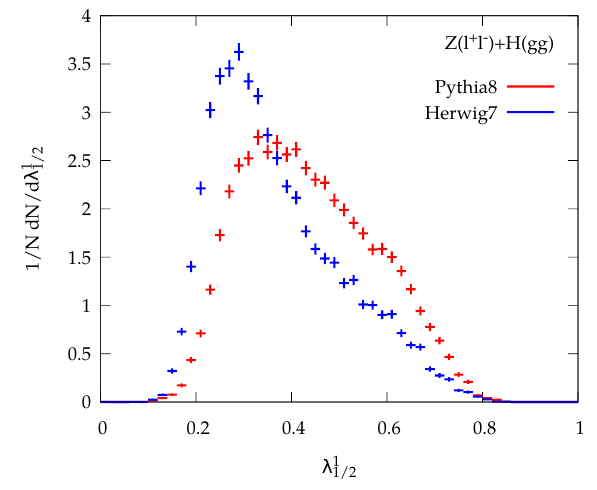} \label{fig:gluon_a}
\includegraphics[width=0.45\hsize]{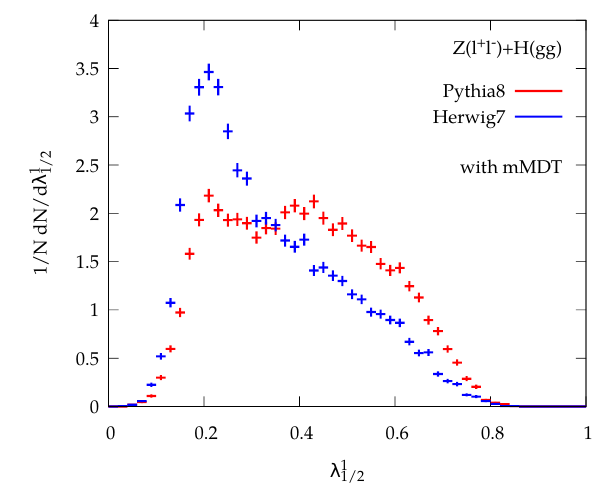} \label{fig:gluon_b}
\end{center}
  \caption{Generalized angularities (a), and groomed angularities (b) measured on gluon jets with different event generators. Large discrepancies are observed due to a lack of tuning data. Figures courtesy of Gregory Soyez.}
  \label{fig:gluons}
\end{figure}

As compared with LEP, a particular advantage of the ILC for the study of QCD is the availability of pure samples of gluon jets through the process $\ee\to HZ$, with $Z$ decay to leptons and Higgs boson decay to $gg$~\cite{Gao:2019mlt}. This would be a unique feature of the ILC. At the LHC, one of the primary issues in understanding precision jet substructure is the difficulty in disentangling quark and gluon jets. As such the study of properties of gluon jets in QCD is extremely poor; this is reflected in large discrepancies in different parton shower simulations shown in Fig.~\ref{fig:gluons}. This lack of understanding of gluon jets, and in particular their non-perturbative properties, is a major issue and a leading uncertainty in many new physics searches at the LHC. One of the promises of jet substructure is that it offers the potential of discriminating quark vs. gluon jets to identify new physics signals. However, this requires a detailed understanding of both quark and gluon jets. Currently, quark vs. gluon tagging has not fulfilled its promise due to large uncertainties in the modelling of gluon jets. Having pure samples of gluon jets in QCD would significantly change this situation and have a major impact on the LHC physics program.

\begin{figure}
\begin{center}
\includegraphics[width=0.45\hsize]{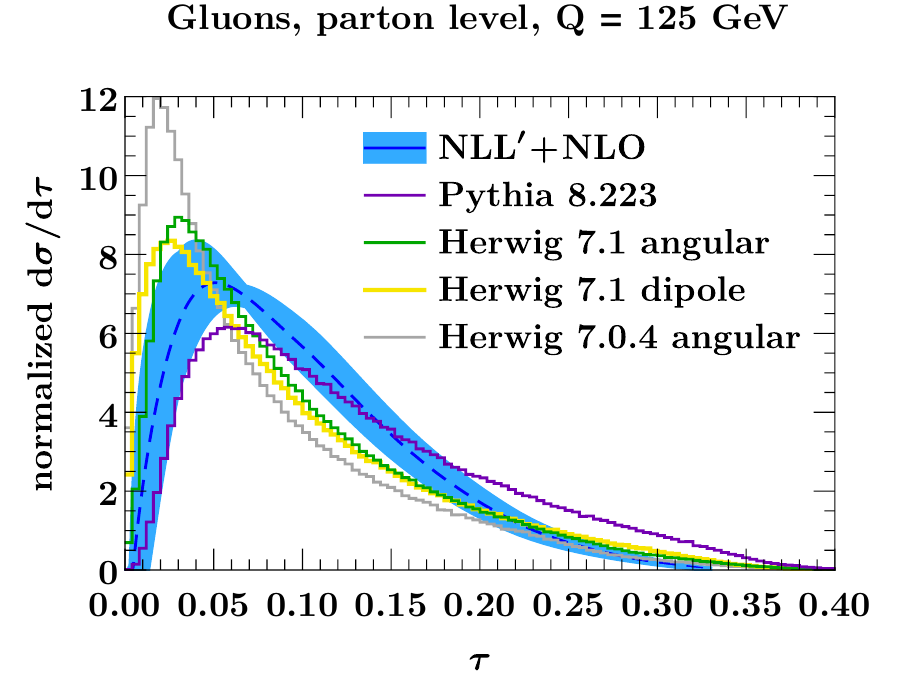} \label{fig:gluon_calc_a}
\includegraphics[width=0.45\hsize]{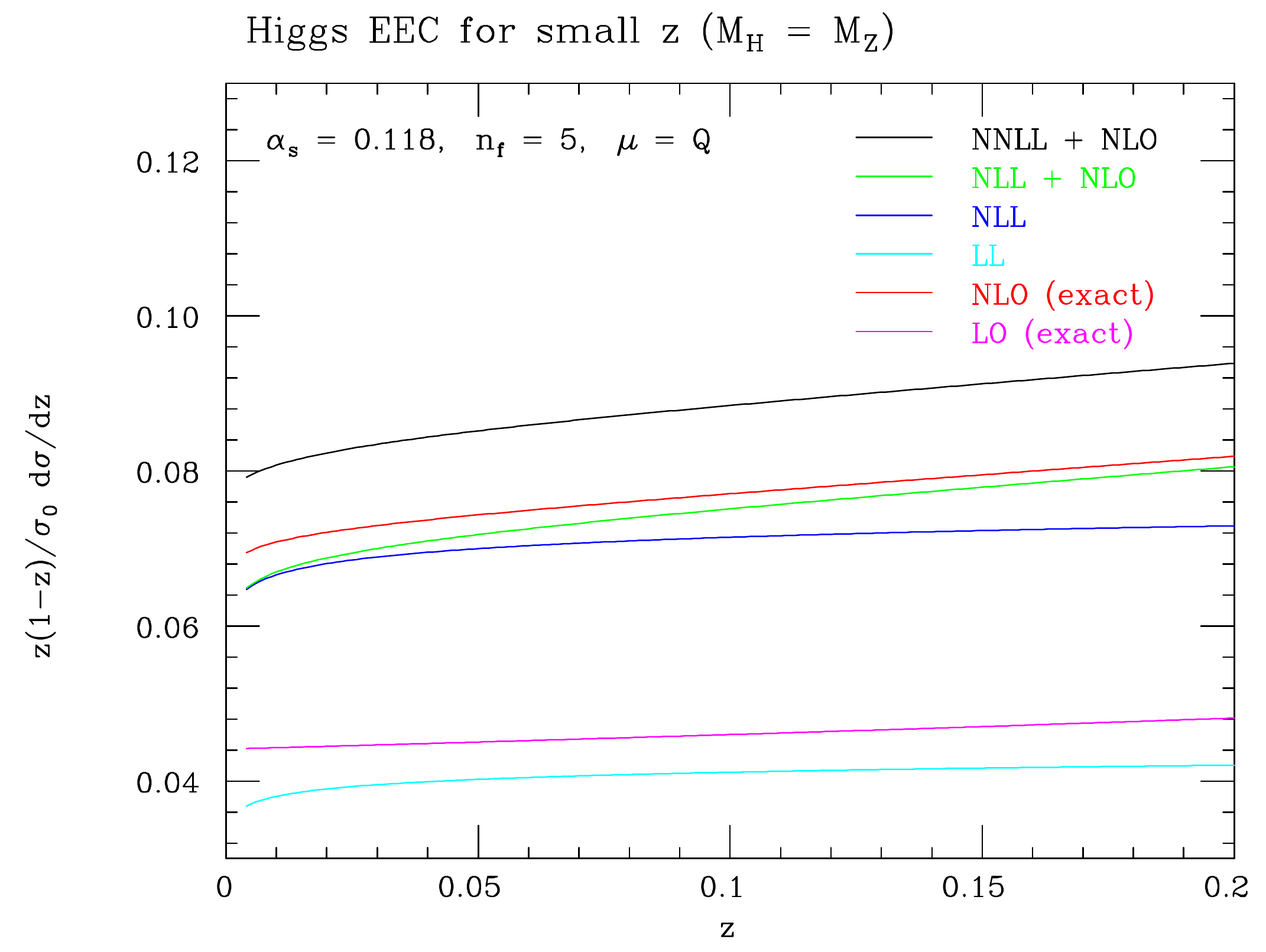} \label{fig:gluon_calc_b}
\end{center}
  \caption{Precision perturbative calculations in $H\to gg$. In (a) we show a double logarithmic Sudakov event shape observable \cite{Mo:2017gzp}, and in (b) and single logarithmic collinear observable \cite{Dixon:2019uzg}.}
  \label{fig:gluon_calc}
\end{figure}

Although the understanding of gluon jets is quite poor, there in fact exist a wide range of precision perturbative calculations of event shapes on $H\to gg$, which have never been compared to data. Two examples, the thrust event shape and the energy-energy correlator, are shown in Fig. \ref{fig:gluon_calc}. These predictions have never been compared with data. Since the perturbative features of gluon jets are well understood, and already available to high accuracies, comparison with data would enable detailed studies of the non-perturbative structure of gluon jets.

%%%%%%%%%%%%%%%%%%%%%%
\subsubsection*{QCD for the Higgs}
%%%%%%%%%%%%%%%%%%%%%%

Although the primary focus of this section is on the use of ILC for learning about QCD, due to its close relation to the other topics in this section, it is also interesting to briefly mention how newly developed jet substructure tools, in particular quark vs. gluon tagging just discussed, can be used to provide new handles on the Higgs boson at the ILC. One of the interesting questions about the Higgs boson that is difficult to study directly at the LHC are its couplings to light (u,d,s) quarks. While these couplings can be probed at the LHC by precise measurements of the $p_T$ spectrum of the Higgs, potentially enabling measurements at the level of $y_s \leq 0.5 y_b$, this requires a precise understanding of the gluon/quark luminosities. 

\begin{figure}
\begin{center}
\includegraphics[width=0.45\hsize]{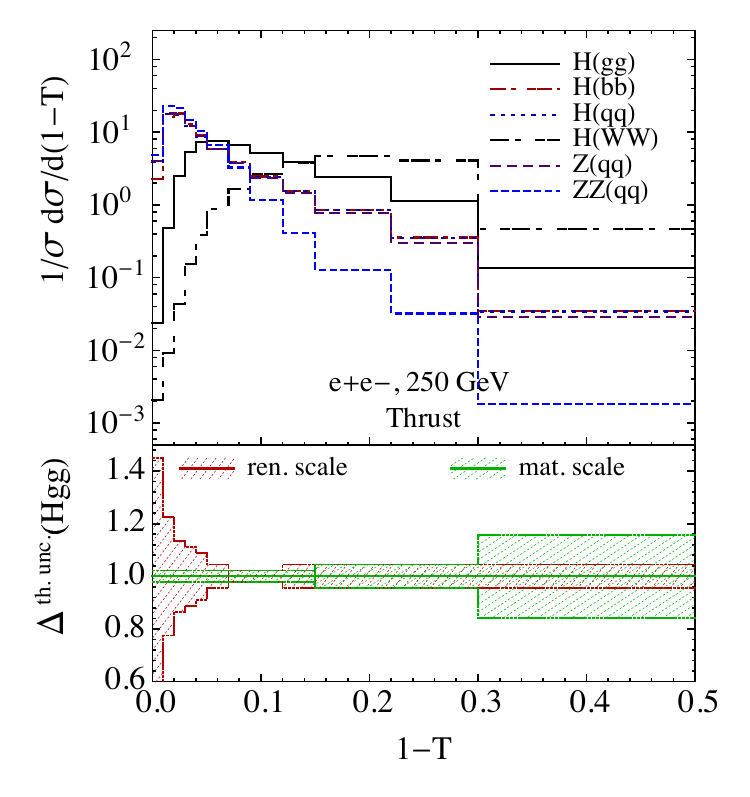} \label{fig:gao_a}
\includegraphics[width=0.45\hsize]{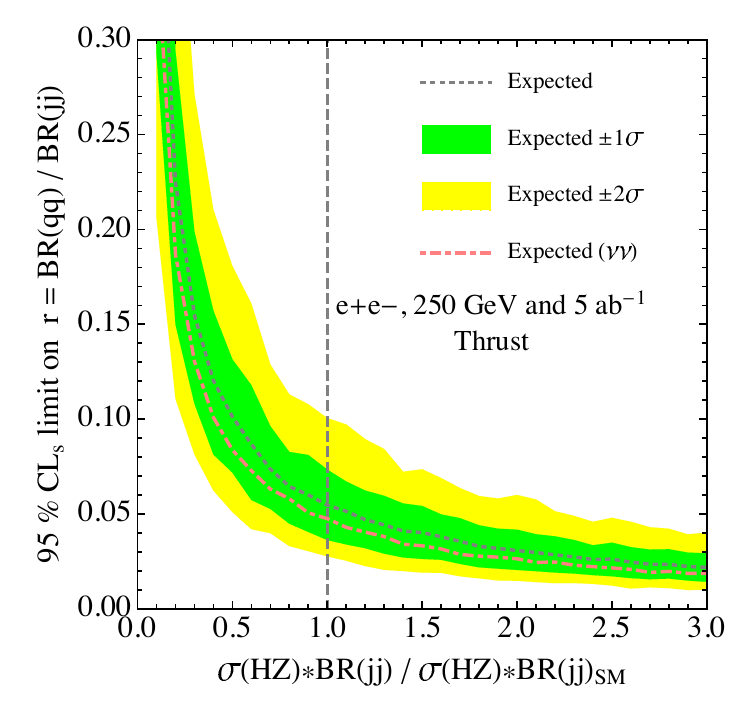} \label{fig:gao_b}
\end{center}
  \caption{Precision measurements of event shapes on Higgs decays used to bound the light quark Yukawas \cite{Gao:2016jcm}.}
  \label{fig:gao}
\end{figure}

At the ILC, precision measurements of event shapes on Higgs decays can provide much more precise handles on the light quark Yukawa couplings due to the differing radiation patterns of quark and gluons. In particular, \cite{Gao:2016jcm} was able to achieve $y_{u,d,s}\leq 0.091 \, y_b$ and $95\%$ confidence. This was using a fairly conservative approach of a single event shape, thrust. This is shown in Fig. \ref{fig:gao}. Almost certainly with more sophisticated event shapes, this bound could be significantly improved, and would provide an interesting example of the interplay between precision QCD measurements and the Higgs program at the ILC.

%%%%%%%%%%%%%%%%%%%%%%
\subsubsection*{New Non-Perturbative Inputs}
%%%%%%%%%%%%%%%%%%%%%%

Another important part of the legacy of LEP is the measurement of universal non-perturbative functions of QCD. Although there are currently no methods for first principles calculations of non-perturbative Lorentzian observables in QCD, much of the predictivity of QCD comes from factorization theorems which express cross sections in terms of universal non-perturbative functions. Famous examples measured at LEP include fragmentation functions. These can then be used in other colliders, such as at the LHC, and have had a large impact on the collider physics program. While it is clear that measurements at the ILC would improve our understanding of fragmentation functions, since these functions have been discussed extensively in the literature, here we focus on universal non-perturbative inputs that were not defined at the time of LEP, which could be measured precisely at ILC, and would have a significant impact on the LHC physics program.

Measurements at the LHC rely strongly on the use of tracking information. This is both due to the fact that tracking is less sensitive to  pile-up, which will become increasingly important in the high-luminosity era, and because of the  remarkable angular precision of the tracker , which is particularly important for many jet substructure measurements. However, only observables that are completely inclusive over the spectrum of final states can be computed purely from perturbation theory, and therefore the calculation of track based observables (which distinguish final state particles based on their charge) requires some non-perturbative input. Still, the situation is not hopeless, since this non-perturbative information is universal.  It can be parametrized by  so-called ``track functions" \cite{Chang:2013rca,Chang:2013iba}, which describe the fraction of energy carried by charged particles from a fragmenting quark or gluon. These can be viewed as generalizations of fragmentation functions, since they encode correlations among the final state hadrons due to the fact that a single fragmenting quark or gluon can decay into multiple charged hadrons. The track functions must be measured experimentally using well-defined quark and gluon jet sources.

Recently it has been shown how to compute jet substructure observables at high precision incorporating track functions \cite{Jaarsma:2022kdd,Li:2021zcf}, which gives promise for precision jet substructure measurements at the LHC. However, despite this new found theoretical understanding, track functions have never been experimentally measured. Furthermore, such measurements in archived LEP data are difficult due to the lack of gluon jets.

The ILC would provide very clean sources of quark and gluon jets on which one could precisely measure the quark and gluon track functions. This would be synergistic with the high luminosity LHC physics program, and would be a particularly valuable contribution with many applications.

%%%%%%%%%%%%%%%%%%%%%%
\subsubsection*{QCD in $\gamma\gamma$ Interactions}
%%%%%%%%%%%%%%%%%%%%%%

Beyond $\ee$ annihilation, there are additional reactions at an $\ee$
collider that are fascinating from the point of view of QCD.   These
include, especially, the regimes 
where a fixed-order perturbative approach needs to be supplemented by one or more all-order resummations.
In the high-energy or \emph{Regge} limit, large center-of-mass energy
logarithms enter the perturbative series
 with a power that increases with the order, thus compensating the
 smallness of the strong coupling. Therefore, the convergence of the
 perturbative series is spoiled and an all-order resummation of these
 large logarithms must be carried out. 
The most powerful tool to perform this resummation is the Balitsky--Fadin--Kuraev--Lipatov
(BFKL)
approach~\cite{Fadin:1975cb,Kuraev:1976ge,Kuraev:1977fs,Balitsky:1978ic}
(see Ref.~\cite{Celiberto:2020wpk} for recent applications), where
amplitudes for high-energy reactions are elegantly factorized via a
convolution between two impact factors, 
describing the production of an identified final-state object from the
corresponding intial-state particle, 
and a process-independent Green's function. This factorization has been proven up to the next-to leading logarithmic accuracy (NLA).

The best opportunities to test this resummation come in the scattering
of small transverse-size objects, in particular, objects with no
hadronic activity in the initial state. A particularly interesting
setting is the case of the total hadronic cross section hadronic  two virtual photons.
In this  $\gamma^* \gamma^*$ reaction, the BFKL theory predicts a
growth of the total cross section with
energy~\cite{Brodsky:1998kn,Brodsky:2002ka,Caporale:2008is,Zheng:2013uja,Chirilli:2014dcb,Ivanov:2014hpa}. Unfortunately,
the only data available to test this prediction is that from LEP~2.  A
sample of events with higher collision energy and better detector
performance for forward reactions would be very advantageous.
Another reaction that probes this physics is  $(\gamma^* \gamma^* \to
VV$ reaction, with two vector mesons ($V$) detected in the final
state. NLA predictions for the diffractive electroproduction of $\rho$
mesons were provided in
Refs.~\cite{Ivanov:2005gn,Ivanov:2006gt,Enberg:2005eq}, and
comparisons with LEP~2 data for the double diffractive $J/\Psi$ photoproduction were done in Ref.~\cite{Kwiecinski:1998sa}.
The impact factor for the production of forward heavy-quark pair was
recently calculated~\cite{Celiberto:2017nyx,Bolognino:2019pba}, and
first predictions of cross sections and azimuthal-angle correlations
for the double heavy-quark pair photoproduction were presented for
LEP2 energies as well as for the energies of future linear $e^+ e^-$
colliders~\cite{Celiberto:2017nyx,DafneBolognino:2019ccd}.
The BFKL resummation is an important aspect of QCD that is very
difficult to test in hadronic collisions, but these reactions give the
opportunity for tests in which the dependence on the photon virtuality
gives additional powerful information.

%\section{Dark Sector}
%\label{sec:darksector}

%[corresponding editor: Maxim Perelstein (m.perelstein@cornell.edu)]

%(This section will describe dark sector searches in $\ee\to \gamma + X$ and related processes at 250~GeV.  More dark sector material appears in Sec.~\ref{sec:HiggsExotic} above and in Chap. 10 and 11.)

%\bibliography{ILC250}

\chapter{ILC Precision Electroweak Measurements} 
\label{chap:PEW}

\newcommand{\qq}{\mathrm{q}\overline{\mathrm{q}}}
\newcommand{\uubar}{\mathrm{u}\overline{\mathrm{u}}}

% Move each sub-section into separate files

\section{Introduction}
\label{sec:PEWintro}

The ILC, which will provide a  thousand-fold increase in the  accumulated data-sets 
compared to prior $\ee$ experiments, polarized beams, and the potential 
to take data at a wide range of center-of-mass energies, will offer the 
opportunity to advance knowledge of the precision electroweak (PEW) 
sector. The measurement precision will far exceed the precision achieved in 
the legacy measurements from LEP/SLC near the $Z$ pole~\cite{ALEPH:2005ab} and 
from LEP at higher center-of-mass-energies up to 208~GeV~\cite{ALEPH:2013dgf}.
The underlying assets are much higher statistics, precise modern 
detectors with much improved reconstruction of particle and jet momenta, 
polarized beams, and improved theoretical modeling.
 
Adequate control of systematic uncertainties is mandatory. 
This includes detector calibration and alignment, control of 
reconstruction efficiencies and geometrical acceptance, determination 
of the center-of-mass energy, differential
luminosity, 
integrated luminosity, and the polarization of the beams. These 
need to be maintained over years of accelerator operation 
and are a critical prerequisite for extracting the maximum physics out 
of dedicated physics runs at special center-of-mass energies such as 
the $WW$ threshold and especially at the $Z$ pole. Such running is already 
feasible with the initial 250~GeV ILC accelerator described in Chapter~4 
as discussed in~\cite{Yokoya:2019rhx}.
% Chapter 4 should be beefed up with description of Z running parameters. - Graham

ILC operation must be crafted so as to exploit opportunities for 
special runs to address individual electroweak quantities such 
as the $W$ and $Z$ masses, while maintaning sufficient luminosity and time to 
deliver on the Higgs, top, and Higgs self-coupling programs. 
The various ILC data-taking configurations will impact 
the knowledge gained for each PEW observable in different ways. 
Having several measurement methods for the same observable with complementary 
systematic uncertainties should lead to improved knowledge. 
%An important 
%goal for the next years is to further explore these possibilities 
%to better understand their relative merit and inform accelerator 
%and detector design work.

Examples of the configurations under consideration are:
\begin{itemize}
\item Running synergistic with the core physics program. A good example is a 
LEP2-style measurement of the $W$ mass that is well suited to exploiting data 
collected at the center-of-mass energy of 250~GeV.
\item Complementary methods enabled by high energy running, such as 
measurement of $Z$ properties using radiative return events.
\item A dedicated physics run using a polarized scan near the $Z$ pole 
accumulating a data sample of 
$100\,{\mathrm{fb}}^{-1}$ 
%100~$fb^{-1}$ 
and up to 4 B $Z$s.
\item Short few day pilot runs near the $Z$ pole, each accumulating at least 
10M hadronic $Z$s,  for detector calibration and alignment, and 
for physics.  Each such sample would be roughly comparable to the 
whole LEP-1 program and would permit calibration of the tracker momentum-scale 
to a statistical uncertainty of 2.5~ppm.
\item A dedicated physics run with a polarized scan near the $WW$ threshold.
\end{itemize}

After describing some of the measurement techniques and prospects, we 
will revisit these issues more quantitatively. For now let us summarize 
our current thinking:
\begin{enumerate}
\item An accelerator built for running above $ZH$ threshold 
should be exploited as much as possible using 
data that can be collected along with Higgs production. 
So a clear case needs to be made for the added 
benefit of dedicated extensive runs at lower energies.
\item The opportunities to make large improvements in the $Z$ observables with 
a dedicated scan are obvious and warrant dedicated exploitation 
once the accelerator has been upgraded in luminosity by bunch doubling. 
In order to evaluate better the 
eventual reach and required running time, the $Z$ pilot runs should 
be used early in the ILC program to gain valuable experience. They will also 
serve as a rich physics sample, a valuable resource for calibration and 
alignment for the higher energy running, and a high statistics benchmark 
for the tuning of hadronization models.
\item  
The $W$ mass can already be measured rather
well with 
the standard ILC program, likely obviating the need for 
substantial time investment in a dedicated run near threshold. 
Nevertheless, the potential for such a dedicated run 
with as high as possible beam polarizations should be retained 
given the perceived uniqueness of the threshold-based observables.
\end{enumerate}

Our expectations for the precisions with which the ILC will measure electroweak observables are summarized in Table~\ref{tab:PEWresults}.

\begin{table}[p]
\begin{center}
  \begin{tabular}{cc|c|cc|cc}
    \hline
    Quantity & Value &  current &  \multicolumn{2}{c}{Z pole}     & 
  \multicolumn{2}{c}{ILC250}    \\ 
 &   &    $\delta[10^{-4}]$&
                                                 $\delta_{stat}[10^{-4}]$&
               $\delta_{sys}[10^{-4}]$ & $\delta_{stat}[10^{-4}]$ & $\delta_{sys}[10^{-4}]$ \\ 
   \hline
   boson properties \\ \hline
   $m_W$       &     80.379   &  1.5  &  - & -  &   &  0.3  \\
   $m_Z$       &     91.1876  & 0.23  &    &  0.022 &  0.08   &  -   \\
  $\Gamma_Z$   &     2.4952   &  9.4  &  0.5   &  -  &   6  &  - \\
  $\Gamma_Z (had)$  &  1.7444      &  11.5     &      & 4.  & -
                                                                  &   - \\ \hline
    $Z$-e couplings \\  \hline
    $1/R_e$     &   0.0482   & 24.    & 2. &  5  &  5.5   & 10 \\
    $A_e$       &   0.1513   & 139.    &    1.5    &  1.2  & 12.    &  9. \\    
    $g^e_L$     & -0.632     &  16.    &    1.0    &  3.2  &  2.8   &  7.6 \\ 
    $g^e_R$     & 0.551      &  18.    &    1.0    &  3.2  &  2.9   &  7.6 \\ 
   \hline    
     $Z$-$\ell$ couplings \\  \hline 
   $1/R_{\mu}$ & 0.0482 & 16.    &  2.  & 2.  &     5.5 &  10  \\
    $1/R_{\tau}$ & 0.0482 &   22.    & 2. &  2. &    5.7 &  10  \\
    $A_{\mu}$ & 0.1515 &   991. & 2. &  5   &
                                          54.&  3. \\
   $A_{\tau}$ & 0.1515  &   271.  &   2.  &   5.   &
                                         57. & 3 \\
    $g^\mu_L$ & -0.632 &  66.   &     1.0      &     2.3   & 4.5 & 7.6\\ 
    $g^\mu_R$ & 0.551&     89.    &     1.0     &   2.3    &5.5 &  7.6  \\ 
    $g^\tau_L$ & -0.632 &    22.   &      1.0   &    2.8    & 4.7 & 7.6 \\ 
    $g^\tau_R$ & 0.551 &   27.     &      1.0      &    3.2   & 5.8
                                                          &  7.6 \\  \hline
   $Z$-$b$ couplings  \\ \hline
   $R_b$ & 0.2163 & 31.   &  0.4 &   7.   & 3.5  &   10  \\ 
    $A_b$ & 0.935 &   214.      &  1.   &  5.   &  5.7  &   3 \\ 
    $g^b_L$ & -0.999 & 54.&0.32 &4.2  & 2.2& 7.6\\ 
    $g^b_R$ & 0.184 &1540 &7.2 &  36.& 41.& 23.\\  \hline
    \hline
  $Z$-$c$ couplings \\ \hline  
  $R_c$ & 0.1721 & 174.  & 2.  &  30   &   5.8  & 50   \\
    $A_c$ & 0.668 & 404.  & 3.   &  5  &  21. &  3  \\
    $g^c_L$ & 0.816 & 119. &1.2 & 15. &5.1  &  26. \\ 
    $g^c_R$ & -0.367 & 416. &3.1  &  17. &21. & 26. \\ 
    \hline
  \end{tabular}
\end{center}
\caption{Projected precision of precision electroweak quantities
  expected from the ILC.  Precisions are given as {\it relative}
  errors ($\delta A = \Delta A/A$) in units of $10^{-4}$. The column labelled ``Z pole'' refers to 
  the dedicated $Z$ pole run described in Sec.~\ref{subsec:Zpole_accelerator}; the column labelled ``ILC250'' refers to values that 
  can be obtained from radiative return events at 250~GeV. 
  %It should be noted that the dominant systematic uncertainty in 
  %all asymmetry ($A_f$)  measurements is the knowledge of the magnitude of the polarization. 
  % I commented this out as this problem disappears when one combines helicity configurations in methods like the so-called modified Blondel scheme.
  % Graham
  REMARK: The table is taken from Ref.~\cite{LCCPhysicsWorkingGroup:2019fvj} but also  reflects some updates and corrections.}
   \label{tab:PEWresults}
 \end{table}

\section{Radiative Return to the $Z$}
\label{sec:radreturn}

The ILC  running at 250~GeV will already produce a data set that will allow
substantial improvements of our knowledge of precision electroweak
observables.
One of the high-cross-section reactions at 250~GeV is the radiative
return to the $Z$,  $\ee\to Z\gamma$.  In this reaction, the $Z$ is
produced in the forward direction but still accessible to the ILC
detectors.  We will explain in a moment that the photon, which is
produced in 
the opposite forward
direction, does not need to be observed to provide a very clean event sample.
  The ILC program, with
2~ab$^{-1}$ of data, will produce roughly  77 million hadronic $Z$s
and 12 million leptonic $Z$s, a substantial increase over the event
sample of LEP.   Further, these events are produced with polarized
beams, so that, for polarization observables, the event
sample to compare with is that of SLC.   The full power of the ILC
detectors can be used for flavor identification. 

We tag the signal events for the radiative return analysis 
 based on the polar angles 
of the two fermions from $Z\to f\bar{f}$. 
To describe the method simply, we will use the approximations that the fermions
 are massless and the photon is 
collinear to the beam directions. This is already quite close to
realistic, and the approximations can be relaxed with small
corrections.
Then let $E_i$ and $\theta_i$, $i=1,2$, 
 denote the energy and polar angle, respectively, of each final lepton
 or jet.   Transverse momentum conservation implies that 
$E_1\sin\theta_1=E_2\sin\theta_2$.
The fermion pair is boosted only in the beam direction.  
The boost
factor can be determined as
\beq
|\beta|=\frac{|E_1\cos\theta_1+E_2\cos\theta_2|}{E_1+E_2}
=\frac{|\sin(\theta_1+\theta_2)|}{\sin\theta_1+\sin\theta_2}.
\eeqn 
It is interesting that the $E_i$ cancel out, so 
 $\beta$ only depends on $\theta_1$ and $\theta_2$. 
The invariant mass of the fermion pair, $m_{12}$, can then be reconstructed as 
\beq
m^2 _{12}= \frac{1-|\beta|}{1+|\beta|} \cdot  s \ ,
\eeqn
where $\sqrt{s}$ is the center-of-mass energy. For the signal events
 we expect that $m_{12}$ peaks at $m_Z$ 
and, for $\sqrt{s} = 250$~GeV,  $|\beta|$ peaks at 0.76. The angles 
$\theta_1$ and $\theta_2$ can be measured very
precisely at the ILC detectors, so that the signal events can be
tagged 
without the need to observe the ISR photon.

This method was actually used at LEP2 ~\cite{ALEPH:1998ac}, 
though mainly for calibrating the beam energy due to the limited statistics. 
But at ILC250, we will expect $90$ million of such radiative events, a
factor of 5
 (100) more than the total number of $Z$ produced at LEP (SLC). 
 
\begin{figure}
\begin{center}
\includegraphics[width=0.70\hsize]{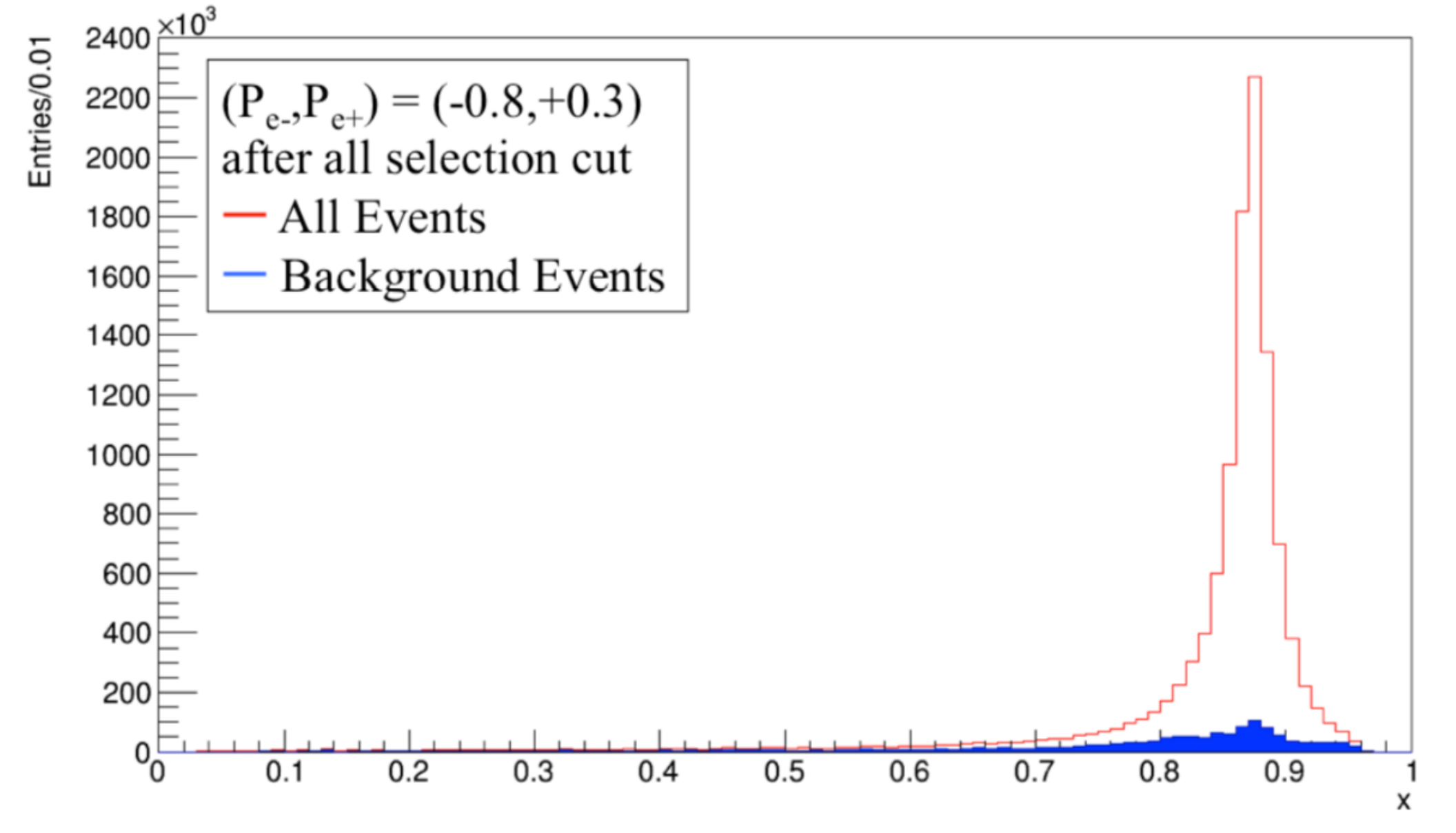}
\end{center}
\caption{Reconstructed distribution of $x\equiv\frac{2|\beta|}{1+|\beta|}$ 
for the signal $e^+e^-\to\gamma Z$, $Z\to q\bar{q}$ and 
background events, 
at $\sqrt{s}=250$ GeV with an integrated luminosity of 250 fb$^{-1}$.}
\label{fig:beta_az}
\end{figure}

\begin{figure}
\begin{center}
\includegraphics[width=0.90\hsize]{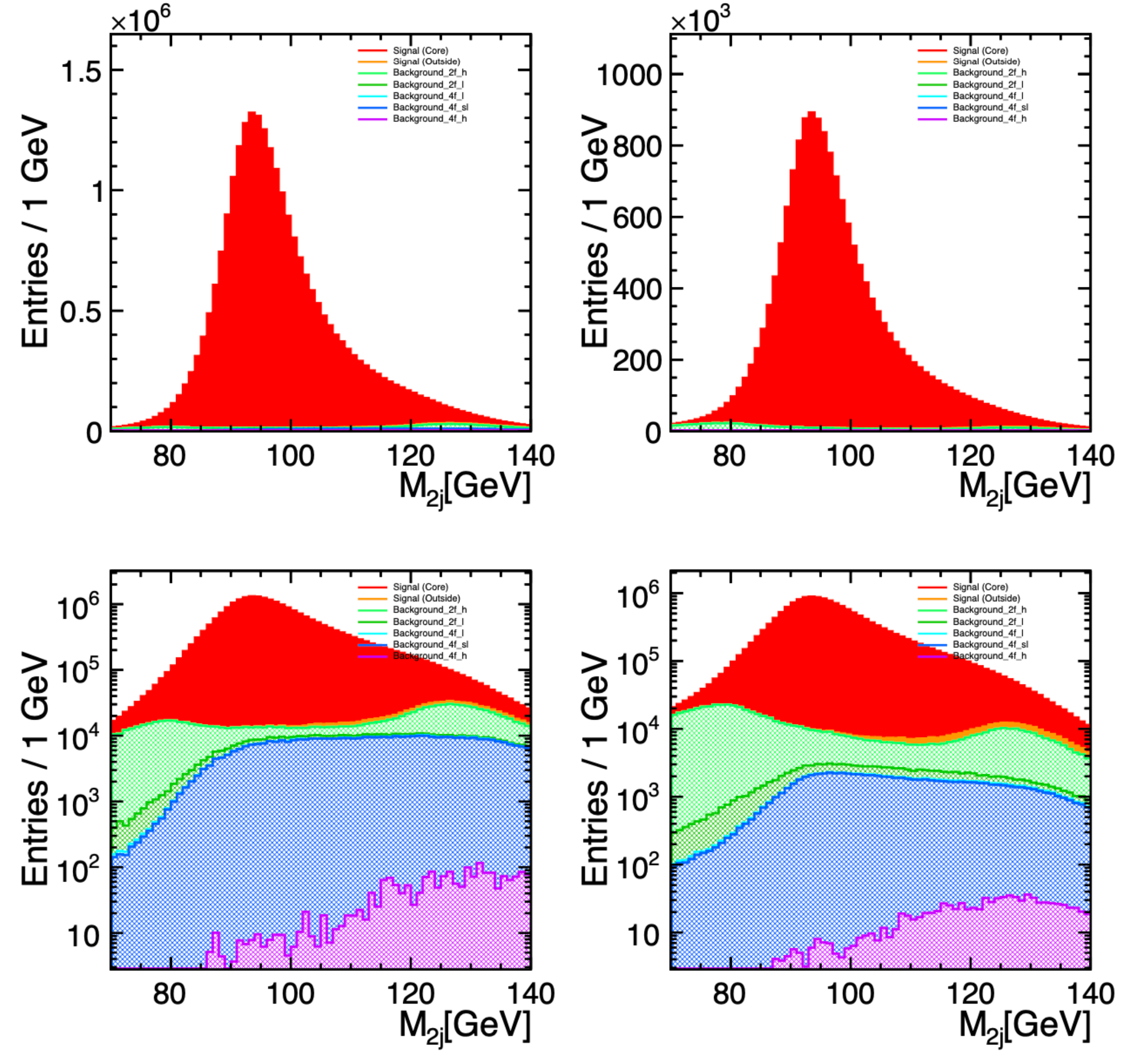}
\end{center}
\caption{Reconstructed distribution of 2-jet invariant mass 
for the signal $e^+e^-\to\gamma Z$, $Z\to q\bar{q}$ and 
background events based on full detector simulation, 
at $\sqrt{s}=250$ GeV with an integrated luminosity of 900 fb$^{-1}$. 
Top two figures are in linear scale and bottom two are in logarithmic scale.
The left two are for $P(e^-,e^+)=(-0.8,+0.3)$ and the right two are for
$P(e^-,e^+)=(+0.8,-0.3)$.}
\label{fig:m2j_az}
\end{figure}

A fast simulation study has been performed for the $A_e$ measurement
using the 
$e^+e^-\to\gamma Z$, $Z\to q\bar{q}$ channels and the full SM 
background~\cite{Ueno:2019}. This has now been followed by a full-simulation analysis~\cite{MizunoThesis:2022, Mizuno:2022xuk}.
After all of the selection cuts, the signal efficiency is 53\% and the
remaining
background events,  due to systems with
approximately the $Z$ mass from other processes,  are almost negligible,
as shown in Fig.~\ref{fig:beta_az}.  For the results shown, 
 realistic effects from finite fermion mass
and beam crossing angles have already been taken into account. 
From the measured cross sections for the left- and right-handed beam
polarizations,
%$A_e$ can be determined from Eq.~\leqn{ALRmeas}.
the statistical error on $A_e$ for 2 ab$^{-1}$ data in the ILC250
scenario is 
estimated to be 0.00018, including also an
estimate of the contribution of leptonic 
$Z$ decays.
This is a relative error of $\delta A_e= 11.9\times
10^{-4}$.
This is a factor of 9 improvement over the current uncertainty  on $A_e$.
By taking into account all realistic SM
$2f$ and $4f$ background events, the signal over background ratio can be
controlled to better than 20/1. After all selection cuts 
the distribution of the reconstructed 
invariant mass of two jets for remained signal and background events
are shown in Fig.~\ref{fig:m2j_az} for $P(e^-,e^+)=(-0.8,+0.3)$ (left)
and $P(e^-,e^+)=(+0.8,-0.3)$ (right), and linear scale (top) and logarithmic 
scale (bottom). 
Most of the systematic errors in the cross section measurement for each beam polarization, such as those from uncertainties on 
event selection efficiency and integrated luminosity, 
are correlated.  Thus they will essentially be cancelled out in
the measurement of the cross section asymmetry for $A_e$. 
It is pointed out in~\cite{MizunoThesis:2022} that in order to match the
expected statistical error of $A_e$, the uncorrelated relative uncertainty 
on efficiency and luminosity should be controlled to 0.016\%. 
The systematic uncertainty due to finite knowledge of the beam polarizations has been discussed in Sec.~\ref{subsec:pol_prec}, we have
explained that all four polarizations values as well as further nuisance parameters can be determined simultaneously with $A_e$ from the collision data themselves. From the comparison of the $A_e$ precision obtained when all polarisation values are free parameters to a fit where they are fixed to their true values~\cite{Beyer:2022xyz}, the residual impact of the polarization uncertainty on $A_e$ has been estimated to be $1.2\times 10^{-4}$. Note that, as also discussed in Sec~\ref{subsec:pol_prec}, this uncertainty reduces with increasing statistics, down to an even smaller residual from uncorrectable point-to-point uncertainties.  

%%%%%%%%%%%%%
%%%%%%%%%%%%%
%%%%%%%%%%%%%
%%%%%%%%%%%%%

In principle, the value of $A_e$  also depends on 
 the CM energy in the $\ee\to Z\gamma$ reaction.  The polarization
asymmetry actually measured in this reaction has the form~\cite{Rowson:2001cd}
\beq
A_{obs} =A_e+\Delta A,
\eeqn
where $\Delta A$ is a correction due to interference between the
contributions  to the  $\ee\to
f\bar f \gamma$ from the resonant diagram with an intermediate  $Z$ and the
nonresonant diagram with an intermediate  $\gamma$. At the $Z$ pole,
the interference term has significant energy-dependence, requiring
excellent knowledge of the CM energy.   This will be an issue in
Sec.~\ref{sec:Zpole}.    However, for the radiative return
process at 250~GeV, the dependence $\Delta A_e/\Delta E_{CM}$ is negligible, 3
orders of magnitude smaller, because the signal events are populated over
the full $Z$ mass peak which is far away from the nominal center-of-mass energy.

For $A_f$ measurements other than $A_e$, we need to measure the
left-right forward-backward asymmetry $A_{LRFB}$. A
dedicated simulation study for $A_f$ ($f=b/c/\mu/\tau$) has not yet been performed. Nevertheless we can estimate the signal
efficiency in two steps based on existing simulation analyses.
 The first step is to tag the signal events as from radiative return,
just as in the $A_e$ measurement. The second step is to identify
the
 flavor and charge of the fermion. 
For example,  the efficiency for  the $A_b$ measurement can be estimated
to 
be $73\%\times 40\%$ in which the 73\%, for 
tagging the hadronic radiative return event, is from fast simulation
analysis described above~\cite{Ueno:2019}, 
and the 40\%, for $b$-tagging and $b$ charge identification, is from a
full simulation
 analysis described in ~\cite{Bilokin:2017lco}. 
The statistical error of $A_b$ is then estimated to be $\Delta
A_b=0.00053$, a relative uncertainty of $\delta A_b=5.7\times 10^{-4}$.
Similarly, the efficiencies for $A_c$, $A_\tau$ and $A_\mu$ can be
derived from full simulation 
results in~\cite{Roman:2019b, Jeans:2019brt, Deguchi:2019tvp}.  These
are estimated to be 
 $73\%\times 10\%$, 80\%,  and 88\%,  respectively. 
Their statistical errors are summarized in Table~\ref{tab:PEWresults}.
The impact of the polarization uncertainties on the final-state asymmetries is negligible, as can be seen for the example of $A_{\mu}$ in Fig.~\ref{fig:nuis} by comparing the full bars (free polarization parameters) to the open crosses (fixed polarizations).

The measurements of 
\beq
R_q \equiv \Gamma(Z\to q\bar{q})/\Gamma(Z\to\ \mbox{hadrons})\ \ \mbox{and}
\ \ 1/R_\ell \equiv \Gamma(Z\to \ell^+\ell^-)/\Gamma(Z\to\ \mbox{hadrons})\ , 
\eeqn
for $q=b/c$ and 
$\ell =e/\mu/\tau$,
are  simpler to describe, since we only need to measure the total rate
for each flavor 
without the
need to identify the charge. The signal efficiencies can be 
estimated based on the same analyses cited above
\cite{Ueno:2019, Bilokin:2017lco, Roman:2019b, Jeans:2019brt, Deguchi:2019tvp}. 
The expected statistical errors are summarized in
Table~\ref{tab:PEWresults}. The 
dominant systematic errors would
come from the uncertainties in the 
 flavor-tagging efficiencies,  estimated in~\cite{Asner:2013psa} to be 0.1\% for $f=e/\mu/\tau/b$ and 
0.5\% for $f=c$.  

Noting that  $R_e$ is expected to be improved by a factor of 2 over
the current uncertainty~\cite{ParticleDataGroup:2020ssz}, it is interesting
to convert this to a value of the quantity
$\Gamma_e\equiv\Gamma(Z\to e^+e^-)$, which is a useful input to the
 SMEFT global fit for Higgs boson couplings that will be
discussed in Sec.~\ref{sec:SMEFTfit}.  $\Gamma_e$ can be derived from 
the measurements of the cross section of $Z$ to hadrons
$\sigma_{\mathrm{had}}$,  the $Z$ total width $\Gamma_Z$,  and $R_e$, with
the uncertainty estimated as 
\beq
\delta\Gamma_e=\frac{1}{2}\delta\sigma_\mathrm{had}
\oplus\frac{1}{2}\delta\Gamma_Z\oplus\frac{1}{2}\delta R_e \ ,
\eeqn
where $\delta$ denotes a relative uncertainty: $\delta A = \Delta
  A/A$. 
With the current uncertainties for $\sigma_\mathrm{had}$ and
$\Gamma_Z$ 
from ~\cite{ParticleDataGroup:2020ssz}, 
and expected uncertainty for $R_e$ at ILC250, we expect the precision
for $\Gamma_e$ to decrease to 
$\delta\Gamma_e=0.86\times 10^{-3}$.  This improves over the current
relative 
uncertainty  on $\Gamma(Z\to \ell^+\ell^-)$ of $1.02\times 10^{-3}$ and
also allows us to relax the assumption of lepton universality in this
input to the SMEFT fit.

\section{Di-fermion production}
\label{sec:Zpole}

Electroweak di-fermion production is the process $\ee \to \ffbar$, with $f \neq e$.  At general CM energies, this process is dominated by $s$-channel photon and $Z$ exchange. Thus, the cross section depends on the couplings $Q_{e,f}^\gamma$ of the initial and final state particles to the photons and the couplings $Q_{fL}^{Z}$, $Q_{fR}^{Z}$,of fermions with left-handed  and right-handed helicity to the $Z$ boson.  There may also be contributions from a potential $Z’$ boson of new physics.  In this section, we will discuss the general case and then specialize to the study of fermion  couplings at the $Z$ pole.

\subsection{Di-fermion production at general CM energies}
\label{sec:gendifermions}

At a general $E_{CM}$, the tree-level production amplitude for $\ee\to f \overline{f}$ is given by ${\cal M} =   e^2 {\cal Q}_{e_i f_j}/s$, where 
\begin{small}
\begin{equation}
{\cal Q}_{e_i f_j} = Q_e^\gamma Q_f^\gamma + 
\frac{Q_{e_i}^Z Q_{f_j}^Z}{\sin^2\theta_W \cos^2\theta_W}\frac{s}{s-M^2_Z +i\Gamma_Z M_Z} +   
\sum \frac{Q_{e_i}^{Z'} Q_{f_j}^{Z'}}{\sin^2\theta_W \cos^2\theta_W}\frac{s}{s-M^2_{Z'} +i\Gamma_{Z'} M_{Z'}} 
\label{pew:elwcouplings}
\end{equation}
\end{small}
with $i,j=L,R$, $Q_{e,f}^\gamma$ the electromagnetic charges, $Q_{e,f_i}^Z$ the helicity-dependent charges for the $Z$ boson couplings, and $\theta_W$  the weak mixing angle at Born level. The first two terms come from $s$-channel photon and $Z$ diagrams. This second term may be affected by $Z-Z'$ mixing as for example suggested in~\cite{Djouadi:2006rk}. The third term takes into account couplings to new vector bosons  $Z'$,  as for example heavy Kaluza-Klein recurrences included in Randall-Sundrum models with warped extra dimensions. The relative importance of the contributions is determined by the Breit-Wigner functions.  We will discuss the ILC sensitivity to $Z'$ resonances in Sec.~\ref{sec:pairs500}.

%The first part describes the couplings to the photon, the second part that to the $Z$ and the third part that of the $Z’$. 
%The parity violating interactions with the Z-boson are of type vector-axial vector (V-A). At tree level the couplings gL, gr are related to the vector and axial-vector couplings via F1V=0.5(gL+gR) and F1ZA=0.5(gL-gR). 
The differential cross section of the process $\ee \to \ffbar$ for relativistic polarized electron, with polarization $\pmi$ and positron beams, with polarization $\ppl$, can be expressed as~\cite{ALEPH:2005ab, MoortgatPick:2005cw, Schmidt:1995mr}.
\begin{equation}
\begin{split}
 \frac{d\sigma}{d \cos \theta} = & \frac{3}{4}(1+|\pmi||\ppl|)(1-\peff\alr)\left(\frac{1}{2}\sigma_{0,HC}(1+\cos^2 \theta)+(\sigma_{0,HV}/\gamma_f)\sin^2 \theta \right) \\
 & +\left[\sigma_0(1+|\pmi||\ppl|)(\afbnull - \peff \alrfb)  \right] \cos \theta
\end{split}
\label{pew:diffcrossff}
\end{equation}
where $\theta$ is the polar angle of the final state fermion, $\peff = (\pmi-\ppl)/(1+|\pmi||\ppl|)$ is the effective polarization, as in \leqn{CSasym}, and $\gamma_f$ is the boost of the final state fermions. 

The differential cross section contains four linearly independent quantities:
\begin{itemize}
    \item The total unpolarized cross section $\sigma_0$ split into a helicity-conserving, $\sigma_{0,HC}$, and a helicity-violating part $\sigma_{0,HV}$. In the Standard Model the helicity violating part vanishes at relativistic energies of the final state fermion. In practice the actual cross-section for a given fermion is often normalised to the total hadronic cross section $\sigma_{had}$ yielding $R_q =\sigma_q/\sigma_{had}$ and $1/R_\ell =\sigma_\ell/\sigma_{had}$ in case of final state quarks and leptons, respectively; 
    \item The unpolarized forward backward asymmetry $\afbnull$; For given beam polarization `-+' it is defined as
    \begin{equation}
       \afbnull =  \frac{(\sigma_F-\sigma_B)_{-+} + (\sigma_F-\sigma_B)_{+-}}
       {2\sigma_0 (1+|\pmi||\ppl|)}         
       \label{eq:afbnull}
    \end{equation}
    with $\sigma_{F,B}$ being the cross sections in the forward and backward hemisperes with respect to the electron beam direction. 
    \item The left-right asymmetry $\alr$ defined as 
    \begin{equation}
     \alr = \frac{1}{|\peff|}\frac{\sigma_{-+}- \sigma_{+-}}{2\sigma_0(1+|\pmi||\ppl|)} = \frac{1}{|\peff|} \cdot \alr^{obs.}   
     \label{eq:afblr}
    \end{equation} 
    The superscript `obs' indicates the measured quantity. 
    \item The left-right-forward-backward asymmetry $\alrfb$ defined as  
    \begin{equation}
     \alrfb =  \frac{1}{|\peff|} \frac{(\sigma_F-\sigma_B)_{-+} - (\sigma_F-\sigma_B)_{+-}}
       {2\sigma_0 (1+|\pmi||\ppl|)} = \frac{1}{|\peff|} \cdot \alrfb^{obs.}
       \label{eq:afblrfb}    
    \end{equation}
\end{itemize}
These quantities depend on the combinations $Q_{e_i f_j}$ defined in \leqn{pew:elwcouplings}. These or similar quantities derived from \leqn{pew:diffcrossff} can be used to determine independently four different individual couplings or  four combinations. In all observables, the couplings to the $Z$ enter linearly through the $\gamma/Z$ interference for CM energies away from the $Z$ pole.  This  allows us to determine the actual sign of the couplings.  
The two asymmetries $\alr$ and $\alrfb$ are only available with polarized beams. In the Standard Model these are of the form $(Q_{e_L}^Z - Q_{e_R}^Z) \times A(Q_e^\gamma,\,Q_f^\gamma, Q^Z_{e_{L,R}},\,Q_{f_{L,R}}^Z)$
in case of $\alr$ and $\beta_f (Q_{f_L}^Z - Q_{f_R}^Z) \times A'(Q_e^\gamma,\, Q_f^\gamma,\, 
Q^Z_{f_{L,R}},\,Q_{e_{L,R}}^Z)$, with $\beta_f = \sqrt{(\gamma^2_f-1)/\gamma^2_f}$, in case of $\alrfb$. 
In addition $\afbnull$ is of the form $\beta_f (Q_{f_L}^Z - Q_{f_R}^Z)(Q_{e_L}^Z - Q_{e_R}^Z)\hat{A}(Q_e^\gamma,\, Q_f^\gamma,\,Q^Z_{f_{L,R}},\,Q_{e_{L,R}}^Z)$.
All asymmetries thus vanish close to the production threshold of the fermions yielding reduced sensitivity to the weak part of the interaction.

On the $Z$ pole, these expressions simplify considerably.  $\alr$ depends only on the couplings of the electrons while $\alrfb$ depends only of the final state fermion couplings to the $Z$. More precisely,
\begin{equation}
\afbnull = \frac{3}{4}{\cal A}_e{\cal A}_f,\,\alr={\cal A}_e,\,\alrfb=\frac{3}{4}{\cal A}_f
\label{eq:afpol}
\end{equation}
with the fermion asymmetries ${\cal A}_{e,f}$ given as 
% changed Q_f to |Q_f|    (Graham)
\begin{equation}
{\cal A}_{e,f} = \frac{\left(Q_{(e,f)_L}^Z\right)^2 - \left(Q_{(e,f)_R}^Z\right)^2}{\left(Q_{(e,f)_L}^Z\right)^2 + \left(Q_{(e,f)_R}^Z\right)^2} 
\approx 8\, (1/4- |Q_f| \sin^2\theta^{{\rm (e,f)}}_{{\rm eff.}}).    
\label{eq:afgf}
\end{equation}
Here, $\theta^{{\rm (e,f)}}_{{\rm eff.}}$ is the effective weak mixing angle.
This is close to the underlying parameter  $\theta_W$ but differs by  radiative corrections that are in general different for each fermion species. Usually, all leptons are assigned a single value $\theta^{{\rm \ell}}_{{\rm eff.}}$. Inspecting  \leqn{eq:afpol} and  \leqn{eq:afbnull} through~\leqn{eq:afblr} and remembering that ${\cal A}_e \approx 0.15$ in the Standard Model, we see that  stronger beam polarization compensates smaller luminosity. That is why SLD delivered as precise or even more precise results on asymmetries than LEP 1 despite having about  30 times fewer produced $Z$s.

The fractional production for each fermion flavor is given by 
\begin{equation}
R_q,\,1/R_\ell \propto \left(Q_{f_L}^Z\right)^2 + \left(Q_{f_R}^Z\right)^2  
\label{eq:rqlcoupl}
\end{equation}
On the $Z$ pole, the clean separation between initial and final state couplings allows for the determination of the initial and final state couplings without the assumption of flavor universality. 

The argumentation above relies heavily on the availability of initial state beam polarization. In principle the electroweak couplings can also be extracted by analysing the final state polarization. This is readily possible for $\tau$ leptons (and $t$ quarks at higher energies), for which the polarization can be derived from the decay particles. At the ILC, the analysis of the final state polarization could be useful as an independent cross-check.    

\subsection{Di-fermion production at the $Z$ pole}
\label{sec:difermionatZ}

\begin{figure}
\centering
\includegraphics[width=0.7\textwidth]{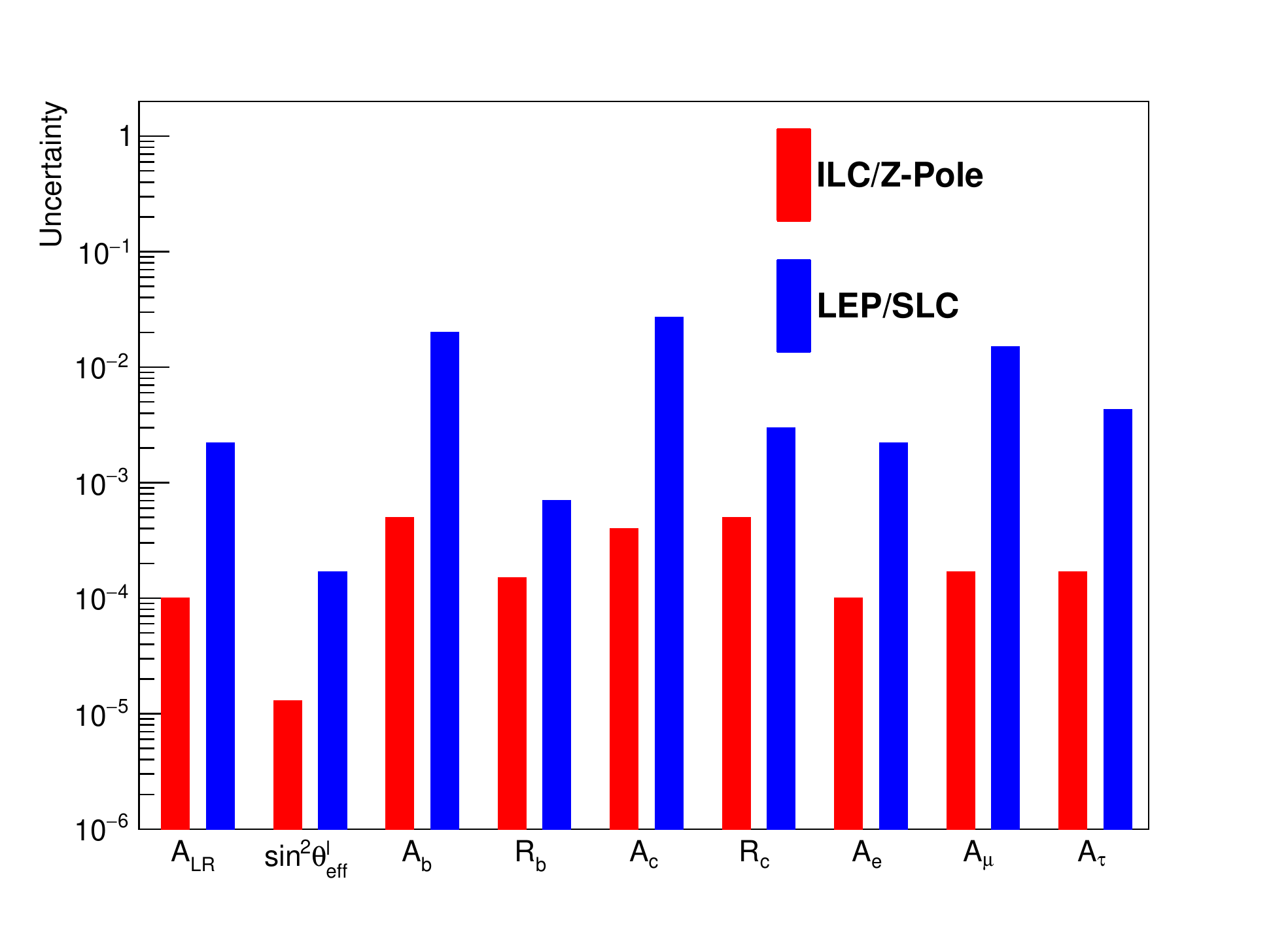}
\caption{\label{pew:alr-gigaz} \sl Summary of the precision achievable for ILC $Z$-Pole running compared  with LEP/SLC results~\cite{ALEPH:2005ab} 
%and FCCee projections~\cite{FCC:2018evy}
for observables and derived quantities that are described in the text.}
\end{figure}

We now concentrate on the results expected from the program of dedicated running at 
the $Z$ pole. 
% (GigaZ).  I prefer not to use this nomenclature given that ILC now can provide more than 10 Giga Z ! (Graham)
Figure~\ref{pew:alr-gigaz} shows the precision that can be expected for key quantities from ILC running at the $Z$-pole. 

%%%%%%%%%%%%%%%%%%%%%%%%%%%%%%%%%%%%%%%%%%%%%%%%%%%%%%%
%%%%%%%%%%%%%%%%%%%%%%%%%%%%%%%%%%%%%%%%5

An important benchmark quantity is the measurement of $\alr$ that 
can be used to determine $\sin^2\theta^{{\rm \ell}}_{{\rm eff.}}$ 
as in \leqn{eq:afgf}. The precisions are related through 
%expected at the ILC 
%from Z-pole running is 
$\Delta \sin^2\theta^{{\rm \ell}}_{{\rm eff.}} \approx \Delta \alr / 8$. 
Projected uncertainties on $\alr$ are shown in Table~\ref{tab:ALR} for 
two scenarios of assigned integrated luminosity and for two scenarios of available beam polarizations.

\begin{table}[!htbp]
\begin{center}
\begin{tabular}{lccccc} L (fb$^{-1}$)  &  $N_{Z}^\mathrm{had}$  & $|P(e^{-})|$ (\%) &  $|P(e^{+})|$ (\%) &  $\Delta \alr$ (stat.)  & $\Delta \alr$ (syst). \\ \hline
%100    &   $3.3  \times 10^{9}$ & 80  & 30 & $4.3 \times 10^{-5}$  &  $ 1.3 \times 10^{-5}$ \\
%100    &   $4.2  \times 10^{9}$ & 80  & 60 & $2.4 \times 10^{-5}$  &  $ 1.3 \times 10^{-5}$ \\
%250    &   $8.4  \times 10^{9}$ & 80  & 30 & $2.7 \times 10^{-5}$  &  $ 1.3 \times 10^{-5}$ \\
%250    &   $1.1 \times 10^{10}$ & 80  & 60 & $1.5 \times 10^{-5}$  &  $ 1.3 \times 10^{-5}$ \\ 
100    &   $3.3  \times 10^{9}$ & 80  & 30 & $2.3 \times 10^{-5}$  &  $ 1.9 \times 10^{-5}$ \\
100    &   $4.2  \times 10^{9}$ & 80  & 60 & $2.0 \times 10^{-5}$  &  $ 1.7 \times 10^{-5}$ \\
% Jenny are you sure this next line is correct?
250    &   $8.4  \times 10^{9}$ & 80  & 30 & $1.4 \times 10^{-5}$  &  $ 1.3 \times 10^{-5}$ \\
250    &   $1.1 \times 10^{10}$ & 80  & 60 & $1.3 \times 10^{-5}$  &  $ 1.3 \times 10^{-5}$ \\  \hline
\end{tabular}
\end{center}
\caption{Estimated uncertainties on $\alr$ for four different scenarios of Z-pole running 
with data-taking fractions at $\sqrt{s}=91.2$~GeV in each of the four helicity 
configurations $(-+), (+-), (--), (++)$ chosen to minimize the statistical 
uncertainty on the asymmetry. The quoted statistical uncertainty includes Bhabha statistics 
for relative luminosity. The systematic uncertainty includes 5 ppm uncertainty on the absolute center-of-mass energy and a 1\% understanding of beamstrahlung effects.
}
\label{tab:ALR}  
\end{table}

The $\alr$ uncertainty is dominated by how well one can determine the effective beam polarization using the cross-sections measured from 
each of the four helicity configurations available with polarized electron and positron beams. 
The basic method, described in~\cite{Hawkings:1999ac}, relies on the availability of positron polarization, and 
benefits from higher positron polarizations.
Systematic uncertainties are expected to be controlled sufficiently for the data-sets envisaged at ILC.
In the most conservative scenario (100 fb$^{-1}$ and standard beam polarizations), 
the combined statistical and systematic uncertainty on $\alr$ is $3.2 \times 10^{-5}$ corresponding approximately to an uncertainty 
of $4.0 \times 10^{-6}$ on $\sin^2\theta^{{\rm \ell}}_{{\rm eff.}}$.
The former precision is about a factor of 50 better than SLD and the latter is 
about a factor of 30  better than that from LEP1/SLD %see Ref.~\cite{ParticleDataGroup:2020ssz} 
and a factor six better than the current best Standard Model prediction for $\sin^2\theta^{{\rm \ell}}_{{\rm eff.}}$~\cite{ParticleDataGroup:2020ssz}.
%Further details can be found in~\cite{LCCPhysicsWorkingGroup:2019fvj}. 
 
Figure~\ref{pew:alr-gigaz} illustrates also that the expected precision on the asymmetries are similar for the three leptons 
in contrast to LEP/SLD where they differed by up to a factor of seven. The results for ${\cal A}_\mu$ and, especially,   ${\cal A}_\tau$ are supported by full simulation studies at higher energies~\cite{Jeans:2019brt,Deguchi:2019tvp}. The study in Ref.~\cite{Jeans:2019brt} shows that the $\tau$ polarization can be measured with a statistical precision of 0.5\%--1\% even at 500\,GeV. At this energy the $\tau$ decay products are extremely collimated. The reconstruction will therefore be easier at smaller centre-of-mass energies.     
This uncertainty translates into precision on the couplings at the $Z$ pole given in Table~\ref{tab:PEWresults}.  Figure~\ref{pew:alr-gigaz} and Table~\ref{tab:PEWresults} also include results on the heavy quarks $c,b$ that will be discussed next.

\begin{figure}[!pt]
\begin{center}
  \begin{tabular}{c}
    \includegraphics[width=0.42\textwidth]{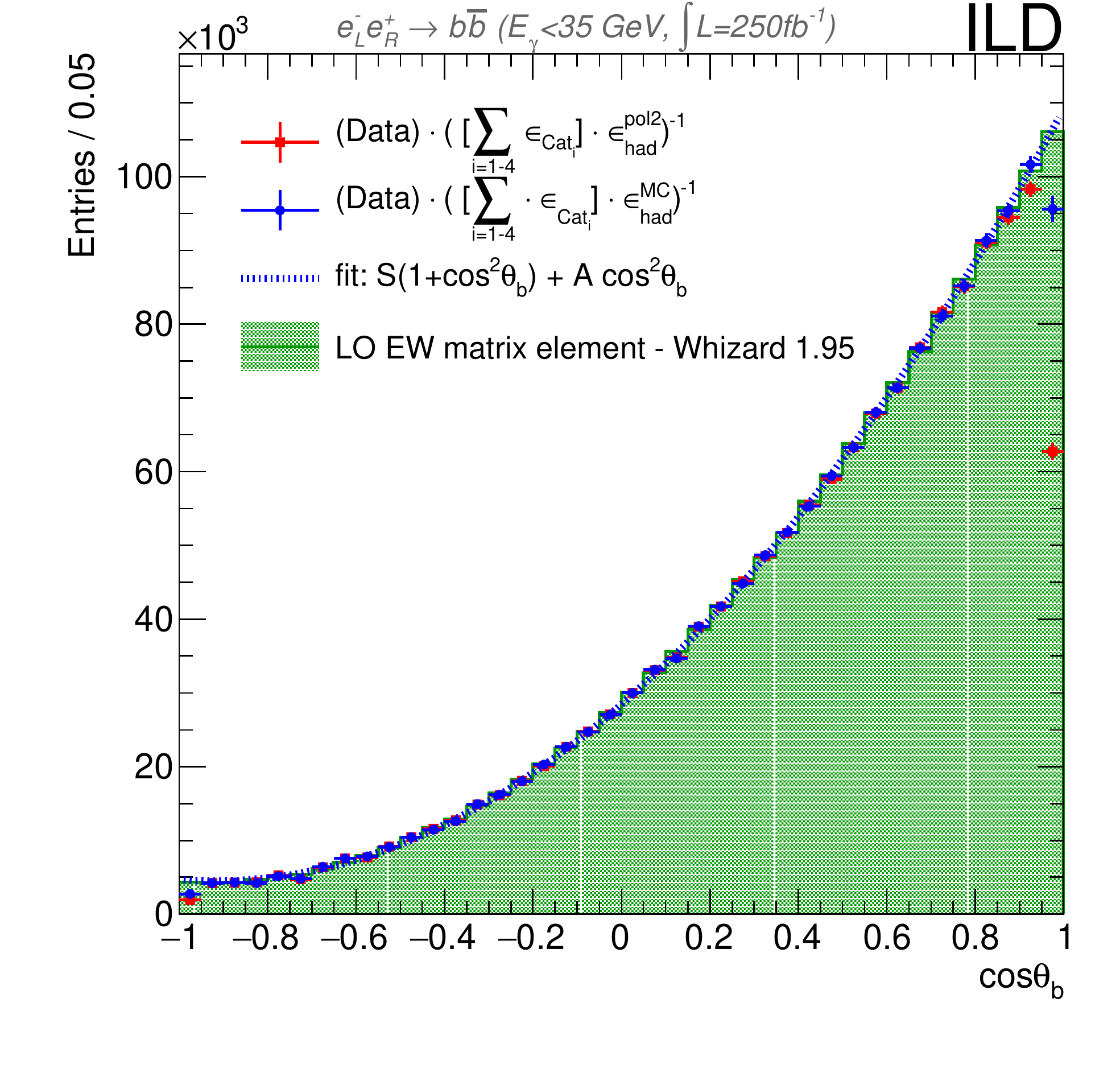} 
    \includegraphics[width=0.42\textwidth]{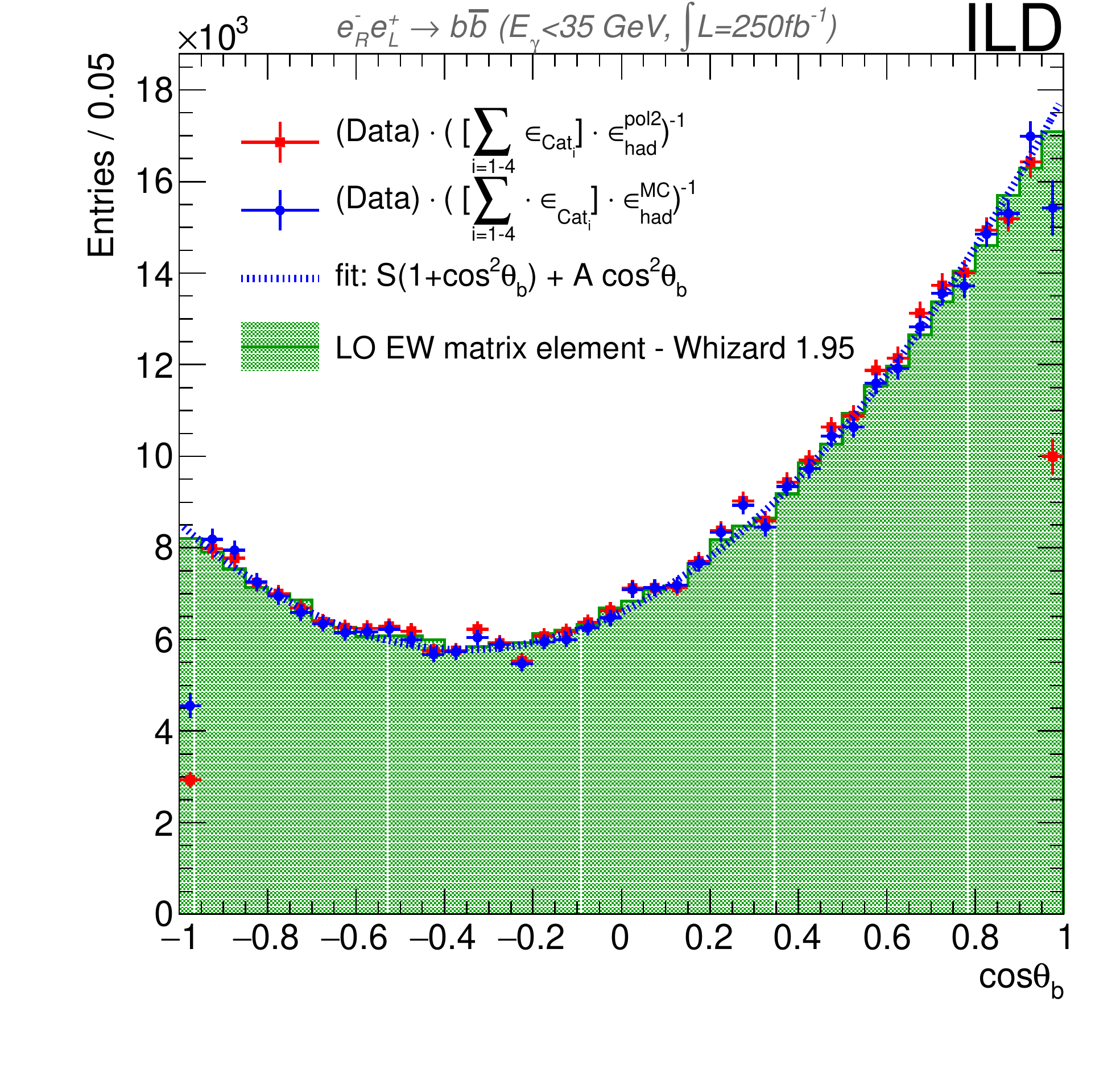} 
  \end{tabular}
\caption{Distributions of $\cos\theta_b$  obtained for $\ELER$ (left) and $\EREL$ (right). The generator distribution is the green histogram and the red and blue dots show the reconstructed distributions after correction charge for charge migration and two different methods for the correction for efficiency and acceptance.  }
%  showing very good agreement between the corrected reconstructed data (blue) and the 
%  generated distribution (green).
%  With the points we show the results using a parametrization of the $\epsilon_{had}$ b\todored{between $|\costhetab|>0.85$. With the blue points we show the results 
%  using a 100\% knowledge of the $\epsilon_{had}$, bin by bin. The difference 
%  between the blue dots and the green curves in the forward region are associated to the correlation factors assumed negligible during the correction procedure but 
%  that in fact make a sizable contribution in the forward regions.}
%  One assumes full beam polarization for an integrated 
%  luminosity of 250 \fb  for each case and no background.
  \label{pew:fit}
\end{center}
\end{figure}

In recent years the community has carried out detailed studies of the processes $\ee \to \bbbar$ and $\ee \to \ccbar$ at $\sqrt{s}=250$\,GeV. The 
expected polar angle distribution for $\ee \to \bbbar$ with the two ILC polarization settings is shown in Fig.~\ref{pew:fit}~\cite{bib:irles-susy21}. It illustrates very clearly that the two combinations of beam polarization yield different sensitivities for the underlying electroweak couplings. 

The result makes use of the correct determination of the charge of the final state quark and is therefore an important benchmark for detector optimisation in terms of measuring secondary vertices and particle ID. A careful analysis of systematic uncertainties has been carried out that includes the uncertainties on the hadronic 2-jet cross section from the normalisation in $R_q$, the beam polarization, and the influence of initial state radiation. The last of these requires  the detection of the ISR photon among the two jet final state, which calls for the availability of highly granular calorimeters that allow for efficient particle  separation.
Overall, the statistical and systematic uncertainties on the observables $\afb$ and $R_q$ are of the order of  1--3 parts per mil and is dominated by uncertainties on selection efficiencies due to ISR. On the $Z$ pole, these uncertainties can be neglected. The impact of the polarization uncertainty on the forward-backward (or final-state) asymmetries is negligible, as can be seen for the example of $A_{\mu}$ in Fig.~\ref{fig:nuis} by comparing the full bars (free polarization parameters) to the open crosses (fixed polarizations).

%\subsection{Outlook}
%The previous paragraphs focused on results obtained for the production of bottom of charm quark pairs. These studies are currently complemented by a study of the process $\ee \to \ssbar$. The set of results for bottom, charm and strange-quark pair production may then allow for estimating also the precision on the measurement of electroweak for the lightest quarks u, d. For a complete picture of di-fermion production the precision on electroweak couplings for final state leptons will have to evaluated. For this, studies presented in~\cite{Deguchi:2019tvp} will have to be carried out in more detail or existing results would have to be extended~\cite{Jeans:2019brt}.
%The precision expected for the $Z$-Pole running are based on extrapolations from full simulation studies at a centre of mass energy of 250\,GeV. Given the high precision it will be very important to carry out the studies for $Z$ pole running with full simulation.

\section{$W$ and $Z$ boson masses}
\label{sec:WZmasses}
Improved measurement of the masses and widths 
of the $W$ and $Z$ bosons are primary goals for 
precision tests of the SM in the electroweak sector.
The ILC program with polarized beams and all standard 
stages of the machine is well suited to measurements in the $W$ sector, 
and especially at $\sqrt{s}=250$~GeV where data can be collected 
synergistically with Higgs boson related studies.
Ultimate precision on $Z$ observables benefits from dedicated 
running at the $Z$.

\subsection{Measurement of $\mw$}

The $W$ mass is a prime target for the ILC.  Improvements of the measurement are
understood to be very tractable based on extrapolations of measurements 
from LEP2 both well above the $W$-pair threshold, and at the $W$-pair threshold. 
Prior prospects for such measurements are summarized in Tables 1-9 
and 1-10 in~\cite{Baak:2013fwa} and discussed 
in~\cite{LCCPhysicsWorkingGroup:2019fvj}.

Measurements of $m_W$ from LEP2, the Tevatron, ATLAS, and LHCb  have 
led to today's PDG precision of 12~MeV, with the best single 
experiment measurements being those from D0, 
with a precision of 23~MeV~\cite{D0:2012kms}, and from 
ATLAS, with a precision of 19~MeV~\cite{ATLAS:2017rzl}.
Recently, a new measurement from CDF has given a value with a
precision of 9~MeV~\cite{CDF:2022hxs}.
However, it is in  significant tension  with the earlier results. All of
these measurements are dominated by systematic uncertainties.
Further improvements from long-existing hadron 
collider data sets at the Tevatron and LHC are possible; it was suggested in 
\cite{Baak:2013fwa} that the LHC could eventually 
improve the uncertainty on the $W$ mass to 5~MeV. 
But, given the predominant systematic uncertainties, and limited 
progress over many years, this aspirational goal remains very challenging. 

It is then interesting to study the challenges to a high-precision 
measurement of $\mw$ at lepton colliders.  Data sets 
at LEP2 totalled 0.7 $\rm{fb}^{-1}$ per experiment, leading to 
statistically limited measurements. The ILC  is
expected to produce much larger data sets of 2 ab$^{-1}$ 
at $\sqrt{s}=250$~GeV and 4 ab$^{-1}$ 
at $\sqrt{s}=500$~GeV both with polarized beams.
These samples should lead to over 140~M produced $W$ bosons.

There are a number of  promising approaches to measure
the $W$ mass at an $\ee$ collider such as ILC.
The statistical power of the anticipated 
data sets is illustrated in Table~\ref{tab:Wmass}. 
Shown are the statistical 
uncertainties on $\mw$ and $\Gamma_W$ from fitting the 
measured $W$ boson invariant mass distribution based on 
100M reconstructed $W$ bosons for a range of assumed 
experimental mass resolutions. We see that, for a 
typical mass resolution of 2.5~GeV, the statistical uncertainty 
on $\mw$ would be 0.35~MeV from such a sample.
Additionally, by fitting the mass, width, and Gaussian experimental 
mass resolution, the total width could similarly be determined from the 
lineshape with a statistical uncertainty of 1.0~MeV.
With non-negligible systematic uncertainties associated with 
measurement of the center-of-mass energy scale, the luminosity spectrum, 
the lepton and jet energy scales, overlay, hadronization, the integrated luminosity, 
radiative corrections, and backgrounds, it will be 
challenging at any $\ee$ collider to reach the 1 MeV overall uncertainty scale on $\mw$.

% Update to 100M rather than 10M.
% Should double-check the rounding that was done when scaling up to 100M. 
\begin{table}[!htbp]
\begin{center}
\begin{tabular}{lccc}
   $\sigma_{M}$ (GeV)     &  $\Delta \mw$ (MeV)    &  $\Delta \Gamma^{a}_{W}$ (MeV)  &  $\Delta \Gamma^{b}_{W}
$ (MeV) \\ \hline
1.0 & 0.21 & 0.41 & 0.63 \\
% 2.0 & 0.98 & 1.7 & 2.7 \\
2.5 & 0.35  & 0.63 & 1.0 \\
% 3.0 & 1.3  & 2.3 & 3.7 \\
4.0 & 0.50  & 0.89 & 1.6 \\ \hline
\end{tabular}
\end{center}
\caption{Statistical uncertainties for 
$\mw$ and $\Gamma_{W}$ for 100M W bosons. These are estimated 
from a simple parametric fit of the Breit-Wigner 
lineshape convolved with a range of constant Gaussian 
experimental mass resolutions, $\sigma_M$, ranging from 1 to 4 GeV. 
The $\mw$ uncertainty is evaluated with a one parameter fit with 
the mass resolution and width fixed. The corresponding uncertainties on 
the $W$ width are evaluated either with the mass resolution fixed and 
known perfectly from a 
two parameter fit ($\Gamma^{a}_{W}$), or more realistically, 
from a three parameter fit ($\Gamma^{b}_{W}$) that also fits for 
the mass resolution.
}
\label{tab:Wmass}  
\end{table}

Given this expectation that measurements will be systematics limited, 
it is appropriate to also consider $W$ mass measurement methods that have promising  
systematic behavior in this high statistics regime.
The various methods for $\mw$ measurement are as follows:

\begin{enumerate}
\item\label{item:kf} {\bf Constrained reconstruction.}
Kinematically-constrained reconstruction of $W^{+}W^{-}$ 
using constraints from {\it four-momentum conservation} and optionally 
mass-equality, as was done at LEP2.

\item\label{item:mhad} {\bf Hadronic mass.} 
Direct measurement of the {\it hadronic mass}. This can be applied 
particularly to single-$W$ events decaying hadronically 
or to the hadronic system in 
semi-leptonic $W^{+}W^{-}$ events. 
This method does not rely directly on knowledge of 
the beam energy or its distribution. A full simulation 
study~\cite{Anguiano:2020qpk} indicates a statistical sensitivity of 2.4~MeV on $\mw$ 
using the hadronic mass in semi-leptonic $W^{+}W^{-}$ events 
at $\sqrt{s}$=500~GeV using the favorably polarized data-set.

\item\label{item:endpoint} {\bf Lepton endpoints.}
The two-body decay of each $W$ leads to endpoints 
in the lepton energy spectrum at 
\beq
E_{\ell } = E_{\rm{b}} (1 \pm \beta)/2 \ , 
\eeqn
where  $\beta$ is the $W$ velocity. These can be used to infer $\mw$. 
The endpoints correspond to leptons parallel and anti-parallel 
to the $W$ flight direction.
This technique can be used for both semi-leptonic and fully-leptonic 
$WW$ events with at least one prompt electron or muon.

\item\label{item:pm} {\bf Di-lepton pseudo-mass.}
In $WW$ to dilepton events, with electrons or muons, 
one has six unknown quantities, namely, the three-momenta of each neutrino.
Assuming four-momentum conservation and equality of the two $W$ masses, 
one has five constraints. By assuming that both neutrinos are in the 
same plane as the charged leptons, the kinematics can be solved to 
yield two ``pseudo-mass'' solutions that are sensitive to the 
true $W$ mass. This technique was discussed in Appendix B
of~\cite{Hagiwara:1986vm}  and 
used along with the lepton endpoints by the OPAL experiment at
LEP2~\cite{OPAL:2002hhr}. 

\item\label{item:threshold} {\bf Polarized Threshold Scan.}  
Measurement of the $W^{+}W^{-}$cross-section near 
threshold with longitudinally 
polarized beams is discussed in~\cite{Wilson:2016hne} and 
references therein. 
The ability to ``turn-on'' and ``turn-off'' the signal with polarized 
beams, a capability unique to ILC,  allows a precise in-situ 
measurement of the background.
\end{enumerate}

Methods~\ref{item:kf}, \ref{item:mhad}, \ref{item:endpoint}, \ref{item:pm} 
can all exploit the standard ILC program at 250~GeV and above. 
Method~\ref{item:threshold} needs 
dedicated running near $\sqrt{s}=161$~GeV.
Methods for measuring the $W$ mass at $\rm{e}^+\rm{e}^-$ colliders were 
explored extensively in the LEP2 era; 
see~\cite{Kunszt:1996km,Stirling:1995xp} and references therein. 

For ILC-sized data sets, the constrained reconstruction 
approach (method~\ref{item:kf}) will likely be restricted 
to semi-leptonic events so as to  
avoid the final-state interaction issues 
that beset the fully hadronic channel. 
With the 
large data-sets of $WW$ events expected above threshold, the expectation 
is that this measurement will be systematics limited. 
With much improved detectors compared to LEP2 and with much better 
lepton and jet energy resolution, it is expected 
that uncertainties at the few MeV level can be targeted. Table 1-9 
in~\cite{Baak:2013fwa} estimates an uncertainty of 2.8~MeV 
at $\sqrt{s}=250$~GeV based on extrapolating LEP2 methods using only 
the semi-leptonic channels with electrons or muons.

Method~\ref{item:mhad} is based purely on the hadronic mass 
and was not used explicitly at LEP2. 
With the increased 
cross-section for singly-resonant events 
($\ee \rightarrow W e \nu$) at higher $\sqrt{s}$, 
the excellent resolution for particles in jets expected from 
particle-flow detectors, and the availability of control channels 
with hadronic decays of the $Z$, an opportunity exists to make a 
competitive measurement also using this method. 
However the demands on the effective jet energy scale calibration are 
very challenging. It was estimated 
(Table 1-10 in ~\cite{Baak:2013fwa}) that a $\mw$ 
uncertainty of 3.7~MeV could be reached.  This is likely to be dominated 
by the hadronic energy scale systematic uncertainty.

The endpoints method~\ref{item:endpoint} was only used for 
fully leptonic events at LEP2. It has the inherent advantage that 
the systematic uncertainties are dominated simply by 
the uncertainties on the lepton energy scale 
and the beam energy, given that one can express $\mw$ in terms of the 
endpoints as follows:
\beq
\mw^2 = 4 E_l ( E_{\rm{b}} - E_l )  \ .
\eeqn
This can also be used as a simple complementary method 
for semi-leptonic events, but of course it will be strongly correlated with 
the constrained reconstruction method.

The pseudo-mass method and the endpoints method were applied to the 
fully leptonic channel in~\cite{OPAL:2002hhr}. Little 
correlation (+11\%) was found between the two methods, indicating 
that the two  methods can be independently effective and can be 
combined. 
%The OPAL result achieved a statistical 
%uncertainty of 390 MeV on $\mw$ using 0.7 fb$^{-1}$ of data.
%The lepton energy resolution for ILC detectors is about 0.15\% 
%based on momentum measurements; this is much better than 
%the 3\% energy (for electrons) and 8\% momentum (for muons) 
%resolutions for OPAL. 
%Assuming a factor of two improvement for ILC detectors,  
%(note that resolutions much less than $\Gamma_W$ are not necessary), and the 
%statistics of the 2 ab$^{-1}$ data set at ILC250, we project 
%a statistical uncertainty on $\mw$ of around 3.6~MeV.
An updated study including beamstrahlung effects 
based on~\cite{Wilson:2019} using these leptonic observables with 
just the 2 ab$^{-1}$ of $\sqrt{s}=250$~GeV data projects a statistical 
uncertainty on $\mw$ of 4.4~MeV (including the semi-leptonic channel endpoints). 
Experimental systematic uncertainties are expected to be under good control.
%straightforward. 
%With the standard 10 ppm uncertainty 
%on center-of-mass energy and detector momentum scale, this approach 
%promises to be very fruitful with the full ILC program.

\begin{figure}[!pt]
\begin{center}
  \begin{tabular}{c}
    \includegraphics[width=0.333\textwidth]{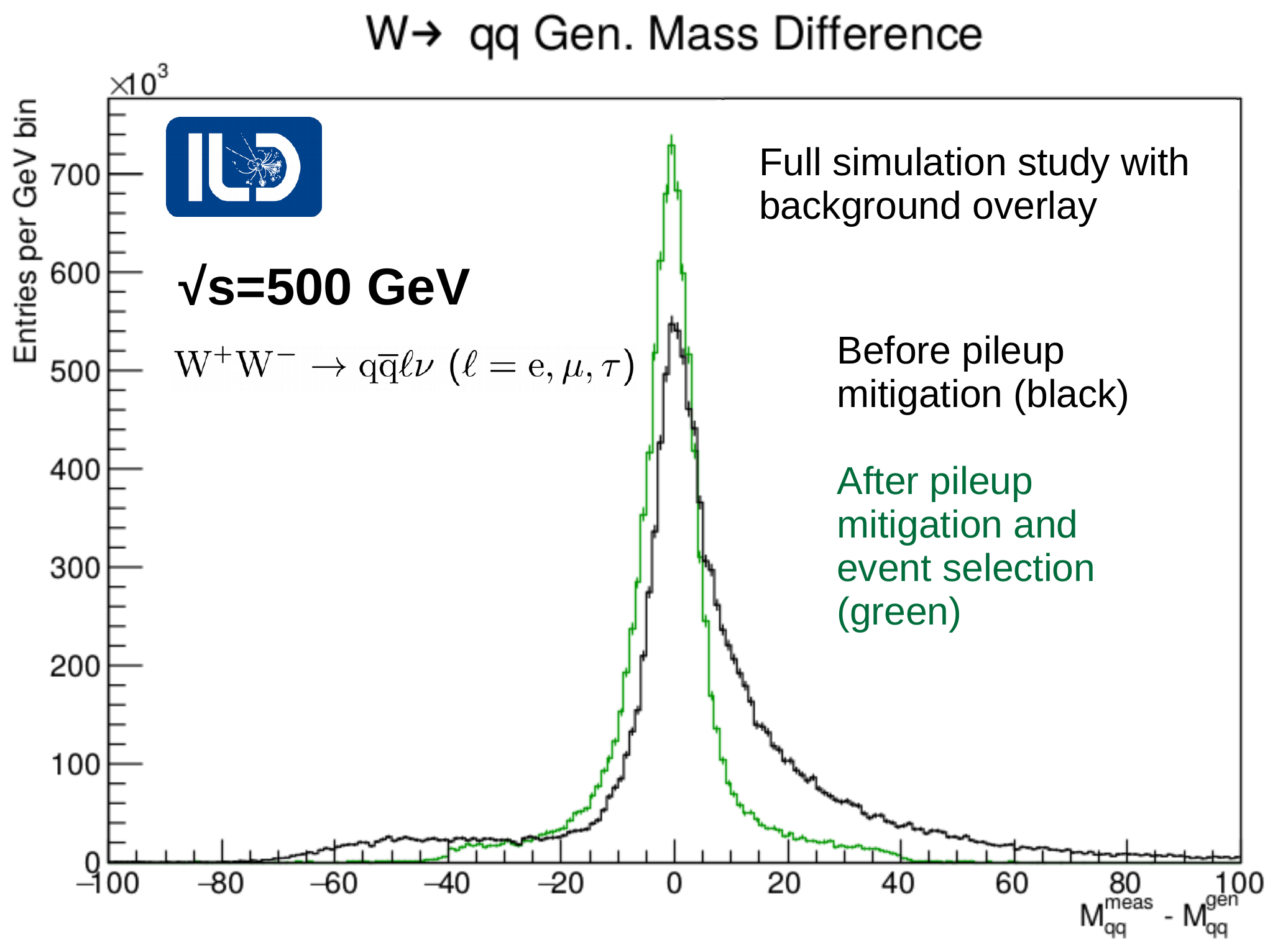} 
    \includegraphics[width=0.333\textwidth]{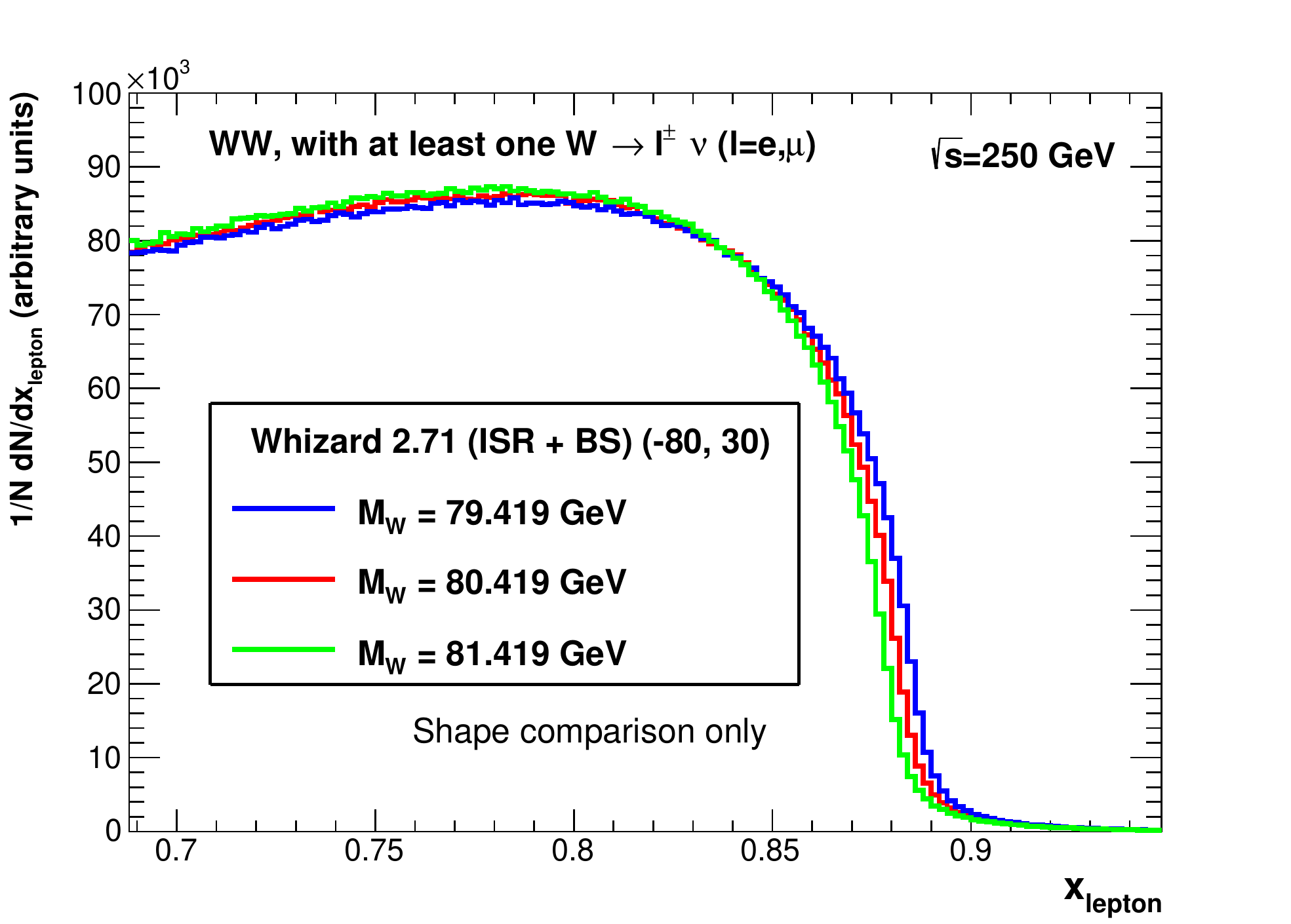}
    \includegraphics[width=0.333\textwidth]{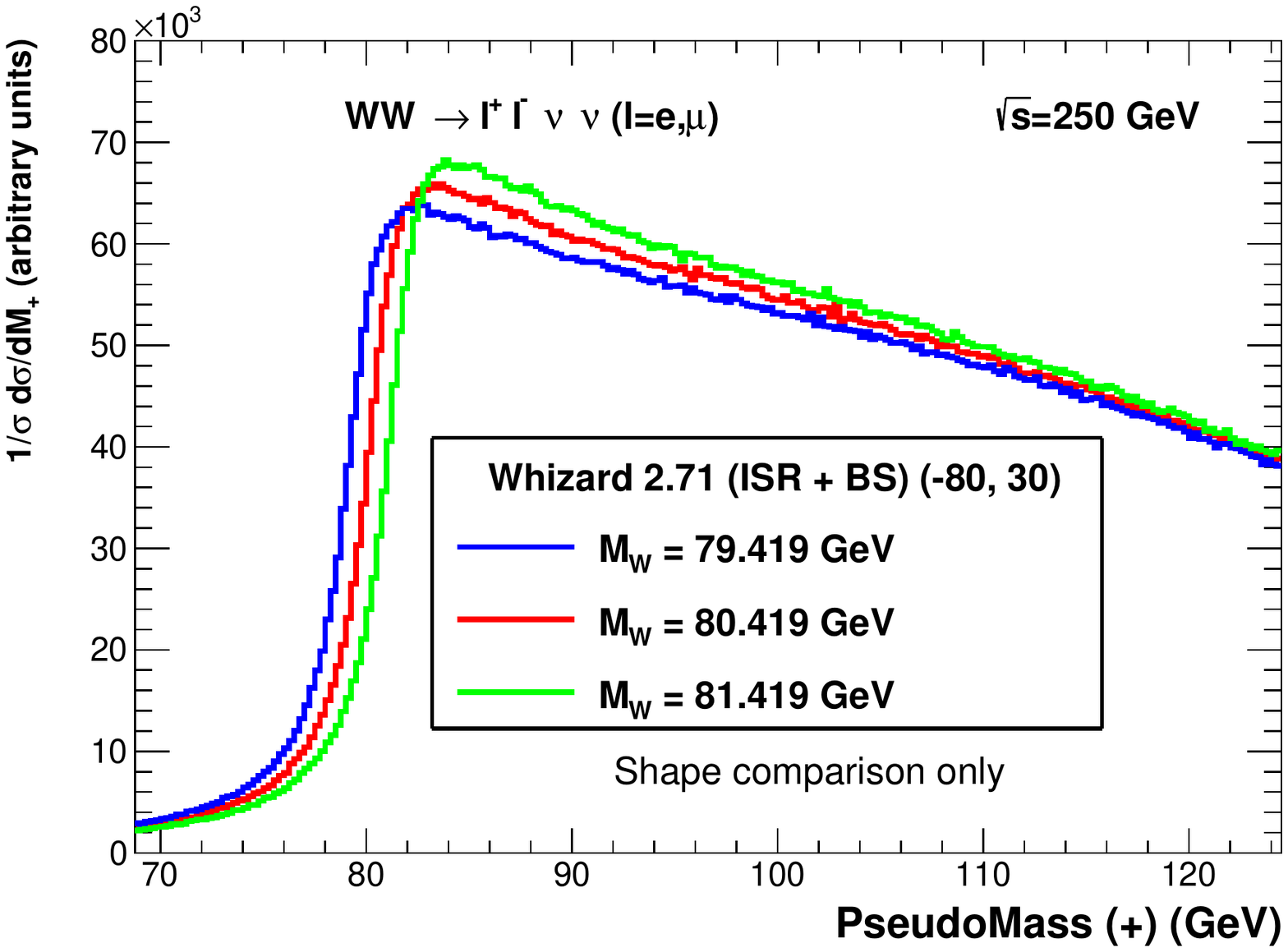}     
  \end{tabular}
\caption{Illustrations of sensitivity to $\mw$. Hadronic mass 
resolution (left) in semi-leptonic events from full simulation study at $\sqrt{s}=500$~GeV~\cite{Anguiano:2020qpk}. 
Endpoints and pseudomass methods at $\sqrt{s}=250$~GeV (right).}
  \label{pew:wmass}
\end{center}
\end{figure}

Method~\ref{item:threshold} needs dedicated running near 
$\sqrt{s}=161$~GeV. This is now feasible for 
the ILC machine.  The expected integrated luminosity is
about 125 fb$^{-1}$/year if the run is done after the bunch doubling luminosity
upgrade. 
The use of a threshold scan with polarized electron and positron beams 
to yield a precision measurement of $\mw$ at ILC was studied in ~\cite{Wilson:2016hne}.
One of the potentially dominant systematic uncertainties,
the background determination, is under very good experimental control 
because of the polarized beams. This is difficult to achieve with 
an unpolarized collider. Errors at the few MeV level can be envisaged. 
With 100~fb$^{-1}$, 
and polarization values of (90\%, 60\%), 
the estimated uncertainty is 
\beq
 \Delta \mw (\rm{MeV}) = 2.4 \: \rm{(stat)} \oplus 3.1 \: (syst) 
\oplus 0.8 \: (\sqrt{s}) \oplus \rm{theory}  \ , 
\eeqn
with these values added in quadrature,
amounting to an experimental uncertainty of 3.9~MeV. With 
standard ILC polarization values of 80\% and 30\% the 
estimated precision is 6.1 MeV.
Eventual experimental precision approaching 2 MeV from 
this approach can be considered at ILC if one is able to 
dedicate 500 fb$^{-1}$ to such a 
measurement, and the physics perspective of the day demands it. 
There are excellent prospects for very competitive ILC 
measurements of the $W$ mass from the four other methods using data 
collected above the production threshold for Higgs bosons, and so 
it would seem premature to make exclusive 
running at $W$-pair threshold a requirement for the ILC run plan. 
Nevertheless, given the complementary nature of a threshold-based 
measurement it is prudent to retain accelerator 
compatibility with such a scenario.

Given that the leading experimental 
systematic uncertainties for the different methods are 
reasonably complementary, it is expected that, 
with the combination of these five different methods, 
ILC will be able to measure $\mw$ to at least 2.5~MeV. 
This uncertainty can potentially already be reached with 
data-taking at $\sqrt{s}=250$~GeV.
    
\subsection{Measurement of $\mz$ and $\gz$}
Measurements sensitive to $\mz$ and $\gz$ are feasible with modest improvements over 
LEP at center-of-mass energies such as $\sqrt{s}=250$~GeV using the reconstructed 
di-lepton invariant mass in radiative return events. 
A study using $\ee \to \mu^+\mu^-$ at $\sqrt{s}=250$~GeV indicates a statistical 
sensitivity to $\mz$ of 1.0~MeV and 2.2~MeV on $\gz$~\cite{Wilson:2021}.
Including similar sensitivity in the $\ee$ decay channel, leads 
to corresponding estimated uncertainties of 0.7~MeV and 1.5~MeV respectively from 
the two lepton channels combined.

Ultimate precision on both quantities would benefit from 
dedicated running at the Z pole. Significant improvement 
on $\mz$ requires excellent control of the absolute scale of the 
collision center-of-mass energy. There are prospects for statistical control 
of the center-of-mass energy scale at the 1 ppm level or below based 
on measuring dimuon events, and relying on exquisite knowledge 
of the tracker momentum scale.
Presuming that one can reach a 2~ppm target uncertainty, a polarized 
lineshape scan with 100~$\mathrm{fb}^{-1}$ would allow 
measurement of $\mz$ to 0.2~MeV with the uncertainty dominated by this scale uncertainty. 
The measurement of $\gz$ depends on point-to-point systematic uncertainties on 
the center-of-mass energy scale; it 
is expected that these should be under good control leading to a $\gz$ uncertainty 
of 0.12~MeV driven mostly by statistics.
Related studies are discussed in more detail in~\cite{Wilson:2022}.

\section{$W$ boson branching fractions}
\label{sec:Wbfs}

With the large data sets envisaged at ILC, one can 
also target much improved measurements of 
the $WW$ production cross section and the individual $W$ decay branching 
fractions. This would use the ten different final state cross sections
available from $WW$ production:  
the six $WW$ final states associated with fully leptonic 
final states with two charged leptons (dielectrons, dimuons, 
ditaus, electron-muon, electron-tau and muon-tau), the three 
semileptonic $WW$ 
final states, one for 
each lepton flavor, and the fully hadronic $WW$ final state.  This 
follows the methodology used at  
LEP2~\cite{ALEPH:2004dmh, DELPHI:2003ftu, L3:2004lwm, OPAL:2007ytu}.

The ten measured event yields can be fitted for 
an overall $WW$ cross section, $\sigma_{WW}$, at each center-of-mass energy, 
and the three individual leptonic branching fractions, $B_{e}$, $B_{\mu}$, and $B_{\tau}$ 
with the overall constraint that
\beq
B_{\rm{had}} = 1 - B_{e} - B_{\mu}  - B_{\tau} \ , 
\eeqn
taking into account background contributions in each channel.
With ten channels and four fit parameters, there is some 
redundancy in the fitting procedure.  This means that the parameters can 
be determined well even if the more challenging channels, 
namely the fully hadronic, the semileptonic with a tau,  
and the di-tau channel would end up being systematically limited.
At LEP2, the signal process was modelled simply through the three
dominant, doubly resonant Feynman diagrams (so called CC03 process), 
while other diagrams and their interferences resulting in the 
same four fermion final state, such as those for 
$ZZ$ or $We\nu$,  were treated as background. 
These ``4f-CC03'' corrections were typically about 10\% depending on 
final state.  A complete calculation of $\ee\to 4f$
final states would need to be used in the high statistics regime.

We have looked into the feasibility of this method for ILC250, 
building on LEP2 studies 
at $\sqrt{s}\approx 200$~GeV, by putting together 
a fit ansatz that  assumes 
identical efficiencies and mis-classification 
probabilities for all ten $WW$ channels~\cite{OPAL:2002hhr}. 
For the purpose of making an estimate for
this report, we concentrate on the impact of a single subsample of the
data.  The actual analysis at the ILC will be based 
on a global fit to the 
results from 
all polarization modes at all center-of-mass energies.
% along the lines described in Sec.~\ref{sec:pol}. 

Of the total 2 ab$^{-1}$ to be collected at
ILC250,  
0.9 ab$^{-1}$  is to be collected 
with $\ELER$ enhanced collisions.  These benefit from a $WW$ cross section 
enhancement over unpolarized beams of a factor of 2.32 for $-80\%, +30\% $
beam polarizations. 
The estimated background per selection channel 
depends on the polarization asymmetry of the different backgrounds
and is estimated to be about +29\% for the important 
two-fermion background from hadronic events. 
Taking this effect that leads to an increased background, 
and the decreased background estimated from $1/s$ scaling, 
we find that the unchanged 
OPAL background estimate is a good first estimate, and believe 
that this is a reasonably conservative estimate.   We have based our 
estimates of statistical uncertainties on the size of this subsample. 
We assume  that the other 55\% of the data set with 
the less favorable beam polarization configurations is used to measure 
and test the background modeling and have neglected it for now in 
estimating statistical sensitivity.

We also include the 6\% reduction in unpolarized cross section 
at $\sqrt{s}=250$~GeV. Given that ILC detectors will have 
much improved forward hermeticity, jet and lepton 
energy resolutions, vertexing, and 
electron, muon, and tau identification, it is very 
reasonable to expect that the efficiency and background 
performance would be much better.
One effect that is more difficult at higher $\sqrt{s}$ 
is a more forward polar angle distribution of the W decay products. 
We find that 94.7\% of leptons in semi-leptonic events have a polar angle 
satisfying, $|\cos{\theta}| < 0.975$, whereas at $\sqrt{s}=200$~GeV, 
the corresponding fraction is 96.7\%.

It is straightforward to estimate statistical uncertainties 
and we have done so for a number of scenarios. For systematic 
uncertainties, there are five that come to mind:
\begin{itemize}
\item absolute integrated luminosity: The precision is likely limited
  to about 0.1\%; however, to a great extent, this value cancels out
  of the determination of branching ratios.

\item lepton efficiencies: This can be measured with high precision using 
control samples of di-leptons as was done for precise $Z$ lineshape 
measurements preferably using a tag-and-probe method. The key element 
is efficiency within the geometrical acceptance. With control samples 
totalling $10^{7}$ leptons, statistical uncertainties
of $3\times 10^{-5}$ can be targeted assuming highly efficient lepton 
identification.

\item hadronic system modeling: Uncertainties of order 0.03\% seem 
feasible based on LEP1 hadronic $Z$ studies targeted at estimating the 
hadronic efficiency/acceptance.

\item fake $\tau$ candidates from hadronic events:
One needs to be able to model the rate of isolated tracks from 
hadronic systems that can fake tau candidates. 
This should be easier to reduce than at LEP2 given the excellent 
vertexing performance envisaged.

\item background estimation:  This will be  controlled with the less 
signal-favorable beam polarization configurations.
\end{itemize}

% Graham 2022. The last digits of these relative numbers are updated based on calculating these directly
% changes are of order 1 in the second significant figure.
% Also add Rhad = (Bhad/(Be+Bmu+Btau))
\begin{table}[t]
\begin{center}
\begin{tabular}{lcccccc}
  Event selections       &   $B_e$    &  $B_{\mu}$  &  $B_{\tau}$  & $R_{\mu}$ & $R_{\tau}$ & $R_{\mathrm{had}}$ \\ \hline
  All 10          &   4.2   &    4.2   &    5.1             &  6.1  &  7.5   & 3.0 \\
  9 (not fully-hadronic)      &    5.9    &     5.9    &   6.4  &  6.1  &  7.5 & 6.7  \\
  9 (not tau-semileptonic)     &    4.6   &     4.6    &   7.9 &  6.1  & 10.8  & 3.2 \\
  8 (not f-h and not $\tau$-semileptonic) & 8.3   &   8.3  &   7.9 &  6.1  & 12.8 & 7.6  \\
  7 (not f-h and not $\tau$-sl and not di-$\tau$) & 9.1  &  9.1  & 10.6 &  6.1  & 16.7 & 7.6  \\ \hline
\end{tabular}
\end{center}
\caption{Statistical uncertainties, expressed as relative uncertainties in units of $10^{-4}$ for 
the leptonic branching fractions  of the $W$ boson ($B_e$, $B_{\mu}$
and $B_{\tau}$)
and the ratios of branching fractions $R_\mu = B_\mu/B_e$, $R_\tau =
B_\tau/B_e$, and $R_{\mathrm{had}} = B_{\mathrm{had}}/(B_e + B_\mu + B_\tau)$. 
The lines of the table refer to different choices of 
the included event selections. The values  assume ILC measurements 
at $\sqrt{s}=250$~GeV using the 45\% of the 2 ab$^{-1}$ integrated 
luminosity with enhanced $\ELER$ collisions, with the same 
efficiencies and the same background 
cross sections as in the OPAL measurement~\cite{OPAL:2007ytu}.
The uncertainties given for $R_\mu$, $R_\tau$ are from 
a separate fit using the ($B_e$, $R_{\mu}$ and $R_{\tau}$) parametrization. 
Similarly the uncertainty for $R_{\mathrm{had}}$ is from a fit 
using a ($B_e$, $B_{\mu}$ and $R_{\mathrm{had}}$) parametrization.}
\label{tab:WWBRs}  
\end{table}

In Table~\ref{tab:WWBRs} we show expected absolute 
statistical uncertainties for three different parameterizations, one based on 
the three leptonic branching fractions, ($B_e$, $B_{\mu}$ and $B_{\tau}$), 
one based on $B_{e}$ and the ratios $B_{\mu}/B_e$ and 
$B_{\tau}/B_e$ and one based on $B_{e}$, $B_{\mu}$, 
$R_{\mathrm{had}} = B_{\mathrm{had}}/(B_e + B_\mu + B_\tau)$ that 
is appropriate for sensitivity to $\alpha_{S}$.
Five different configurations of included event 
selections are considered, indicating a reasonable degree of robustness.
The fits also fit for the cross section but the absolute 
value is likely to be systematics limited.
It can be seen that fractional statistical uncertainties
on $B_e$ below 0.1\% and as low as 0.03\% can be envisaged. 
The fits do not assume lepton universality. 
The data set considered consists of 29.7 million 
$WW$ candidates. The efficiency systematics seem not to be limiting. 
The main systematic issue is likely to be the background 
estimation that should be facilitated with the various polarized 
data sets. The event selection purity 
will likely need to be tightened to reduce systematics from backgrounds, 
but the current statistical estimates should be a reasonable 
starting point. Furthermore the data-sets at $\sqrt{s}=500$~GeV 
would double the useful statistics for such measurements.

% End of file

\chapter{ILC Physics Measurements at 350, 500, and 1000 GeV}
\label{chap:ILC500}

At the  higher-energy stages of the ILC at 350~GeV, 500~GeV, and 1~TeV, new processes appear in $\ee$ annihilation, giving new opportunities for consequential discoveries.   At 350~GeV and above, the ILC can study the top quark in pair-production, bringing the precision capabilities of $\ee$ colliders to bear on this heaviest SM particle.   In Higgs physics, the new reaction of Higgs boson production by $WW$ fusion opens up, providing a second large data set to explore the Higgs boson properties.  Also, the processes of Higgs pair production, sensitive to the Higgs boson self-coupling, and the Higgs boson associate production with top quarks, become accessible.   In this chapter, we will review the prospects for the study of these reactions.  We will also revisit the study of triple gauge couplings (TGCs) and discuss in some detail the study of $\ee$ annihilation to quark and lepton pairs.  Finally, we will review the prospects for the ILC to discover new particles hidden from searches at the LHC.

\section{Top quark}
\label{sec:top}

Pair production of the top quark can be studied at the ILC in two distinct regimes, first, at the threshold, and, second, at energies where the top quarks have  relativistic velocities.    These programs complement one another in addressing the principal  open issues for the top quark---determining its mass with ultimate precision, exploring its connection to the electroweak interactions, and exploring its role in models of flavor.  A fourth crucial issue, the  measurement of the top quark coupling to the  Higgs sector 
will be discussed in Sec.~\ref{sec:Higgstop}.

\subsection{Top quark mass}
\label{sec:topmass}

The top quark mass is one of the fundamental parameters of the Standard Model that must be determined experimentally. Direct measurements at hadron colliders based on Monte Carlo template fits to the reconstructed top quark decay products reach a precision down to 600~MeV at the LHC~\cite{ATLAS:2018fwq,ATLAS:2014wva} and the Tevatron~\cite{CDF:2016vzt}. Combinations can further improve ~\cite{Zyla:2020zbs,ATLAS:2014wva}. Extractions of the top quark pole mass from measured cross sections using first-principle, fixed-order calculations have reached GeV precision~\cite{Aad:2019mkw}. 

Top quark mass measurements at the HL-LHC are expected to reach an experimental precision of a few hundred MeV~\cite{Azzi:2019yne}, while work is ongoing to improve Monte Carlo generators~\cite{Jezo:2016ujg,FerrarioRavasio:2018whr} and to provide a robust interpretation of the Monte Carlo mass parameter in a field-theoretical mass scheme~\cite{ATL-PHYS-PUB-2021-034,Butenschoen:2016lpz,Kieseler:2015jzh}. A complete and recent review can be found in Ref.~\cite{Hoang:2020iah}.   However, this review also discusses the 
limitations of interpreting on-shell top quark mass values in terms of more the fundamental short-distance top quark mass parameter.   These come both from QCD uncertainties in the relation~\cite{Marquard:2015qpa} and from possible non-pertubative contributions.

An electron-positron collider with sufficient energy to produce top quark pairs has excellent potential to measure the top quark mass with even better precision and to make a tighter connection to the underlying short-distance value. It was realized even before the discovery of the top quark that a scan of the center-of-mass energy of the collider through the top quark pair production threshold yields a very precise top quark mass 
measurement~\cite{Gusken:1985nf, Fadin:1987wz,Fadin:1988fn,Strassler:1990nw}, with a rigorous interpretation. Since then, the theory predictions for the threshold scan have reached NNNLO precision~\cite{Beneke:2015kwa} and an NNLL resummation~\cite{Hoang:2001mm} has been performed. The threshold mass that is most naturally extracted from a comparison to the theory can be converted to the $\bar{MS}$ scheme (or any other scheme) at four-loop accuracy~\cite{Marquard:2015qpa}, with an intrinsic uncertainty due to missing higher orders of $\cal{O}$(10 MeV) and a parametric uncertainty of $\cal{O}$(50~MeV) with the currrent $\alpha_s$ world average~\cite{Zyla:2020zbs}.

\begin{figure}
\begin{center}
\includegraphics[width=0.50\hsize]{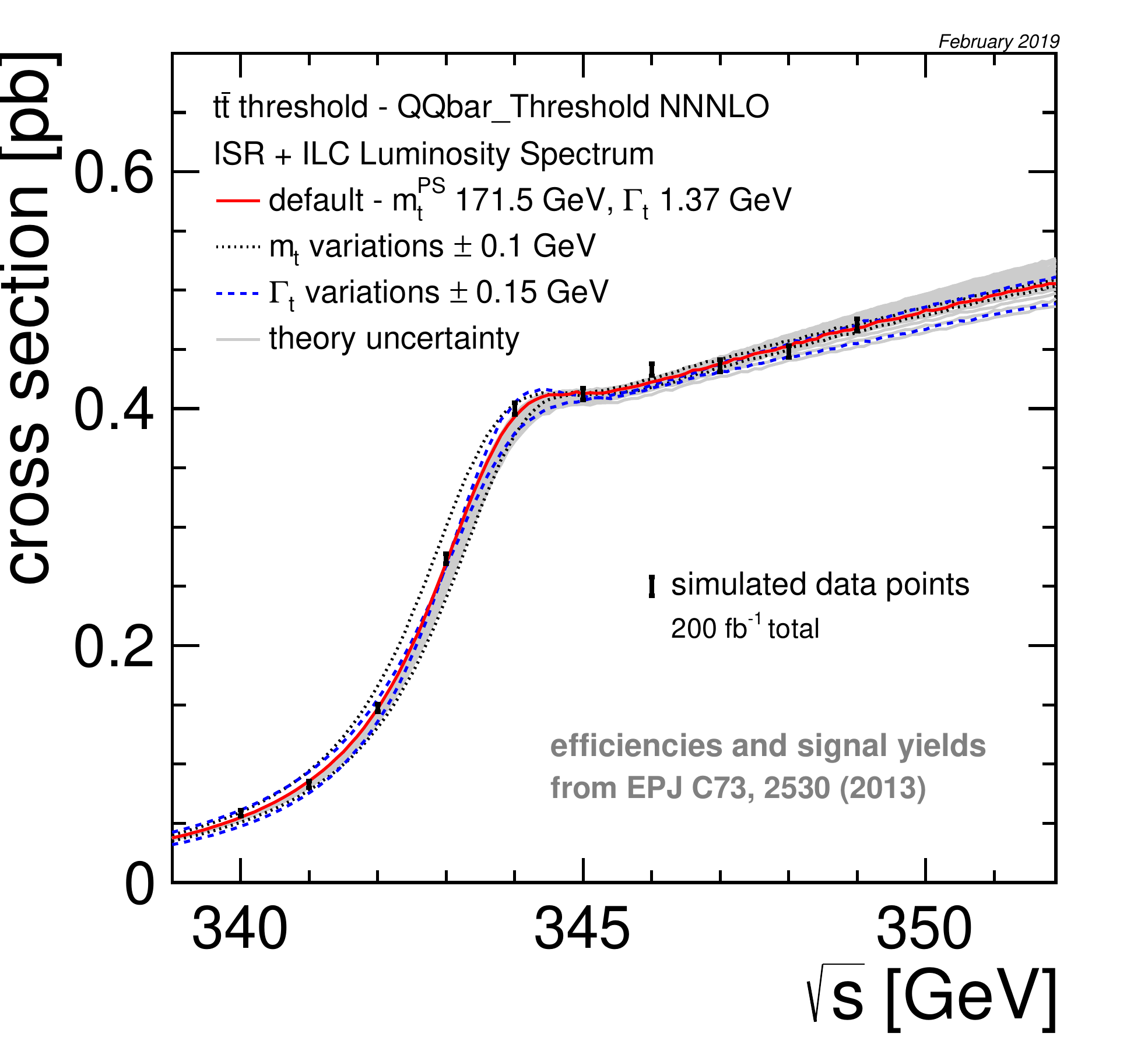}
\end{center}
  \caption{A simulated top quark threshold scan with a total integrated luminosity of 200 fb$^{-1}$.  The bands around the central cross section curve show the dependence of the cross section on the top quark mass and width, illustrating the sensitivity of the scan.  The error bars show  the statistical uncertainties, taking into account signal efficiencies and backgrounds.  From \cite{CLICdp:2018esa}.}
  \label{fig:top_scan}
\end{figure}

Phenomenological studies of the threshold scan in realistic conditions have been performed by several groups~\cite{CLICdp:2018esa,Seidel:2013sqa,Horiguchi:2013wra,Martinez:2002st}. Fits are performed on pseudo-experiments with an integrated luminosity of 100--200~fb$^{-1}$ divided over up to 10 center-of-mass energies. Apart from the top quark mass, the threshold scan is sensitive to the strong coupling, the top quark Yukawa coupling and the top quark width. Typically, several parameters are floated simultaneously in the fits.\footnote{These studies are valid within the SM, releasing only the relation between the width and the mass. The interplay between the top-quark mass extraction and electroweak coupling uncertainties (parameterized in an effective field theory) has not been studied yet.}  Importantly, recent studies take into account the theory uncertainty~\cite{Simon:2016htt,Vos:2016til}, that is expected to be the dominant source of uncertainty for a top quark mass measurement in a threshold scan at an $e^+e^-$ collider. The statistical uncertainty can be reduced to approximately 20~MeV, depending on the number of free parameters and the number and range of the energy points~\cite{Simon:2019axh,Nowak:2021xmp}. The systematic uncertainty from missing higher orders in the prediction and the parametric uncertainty due to the strong coupling constant add up to approximately 50~MeV, with the current state-of-the-art calculations and world average for $\alpha_s$. A simulated scan of the top quark threshold, from~\cite{CLICdp:2018esa}, is shown in Fig.~\ref{fig:top_scan}.

While the threshold is generally considered to be the ``golden'' top quark mass determination, alternative methods have been studied by several groups. A direct mass measurement can be performed at any center-of-mass energy above the top quark pair production threshold.  The comparison of this measurment to the threshold value will 
 provide important information on the interpretation of the MC mass parameter. 
The CLIC-DP collaboration has estimated that a statistical uncertainty of 30~MeV (40~MeV) is expected in the $l+$jets (all-hadronic) channel after collecting  500~fb$^{-1}$ at $\sqrt{s}=$ 380~GeV~\cite{CLICdp:2018esa}. 

A measurement of the differential cross section of radiative $e^+e^-  \rightarrow t\bar{t}\gamma$ events, where the top quark pair is produced in association with a hard photon from Initial State Radiation (ISR) can yield a top quark mass determination~\cite{Boronat:2019cgt}. The measurement of the photon energy gives an event-by-event determination of the effective center-of-mass energy and allows to map out the $t\bar{t} $ threshold with data collected at any center-of-mass energy below $\sim$ 1~TeV. The expected precision is approximately 110~MeV for CLIC380 (1~ab$^{-1}$ at $\sqrt{s}=$ 380~GeV and approximately 150~MeV for ILC500 (4~ab$^{-1}$ at $\sqrt{s}=$ 500~GeV), including theoretical and experimental  systematic uncertainties. This approach is competitive with the  HL-LHC expectation, and the method maintains flexibility in, and control over, the field-theoretical mass scheme.  Moreover,  a combination with the mass obtained from the threshold scan moreover enables a study of the scale dependence (``running'') of the top quark mass, testing the evolution predicted by renormalization group evolution.

Operation of the ILC at the top mass threshold and beyond can thus provide a top quark mass measurement with a precision well beyond what is achievable at hadron colliders and also clarify the various top quark mass definitions in terms of a well-understood  field-theoretical framework.

\begin{figure}
\begin{center}
\includegraphics[width=0.45\hsize]{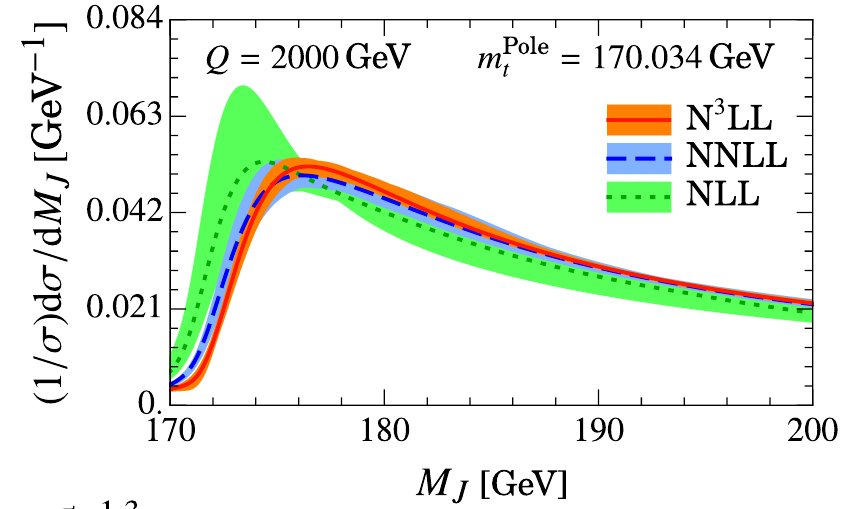} \label{fig:adi_a}
\includegraphics[width=0.45\hsize]{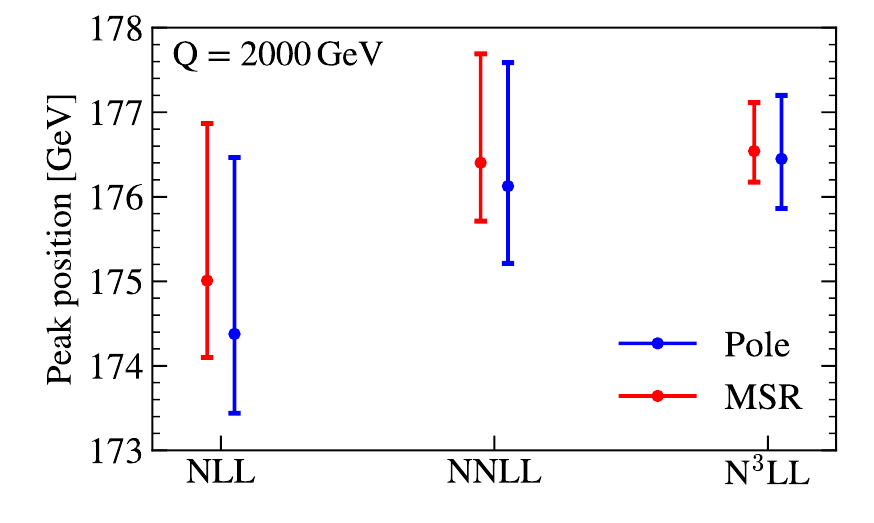} \label{fig:adi_b}
\end{center}
  \caption{(a) The jet mass distribution on boosted top quark jets. (b) The peak position extracted using Pole mass or MSR mass schemes. \cite{Bachu:2020nqn}}
  \label{fig:top_adi}
\end{figure}

\begin{figure}
\begin{center}
\includegraphics[width=0.30\hsize]{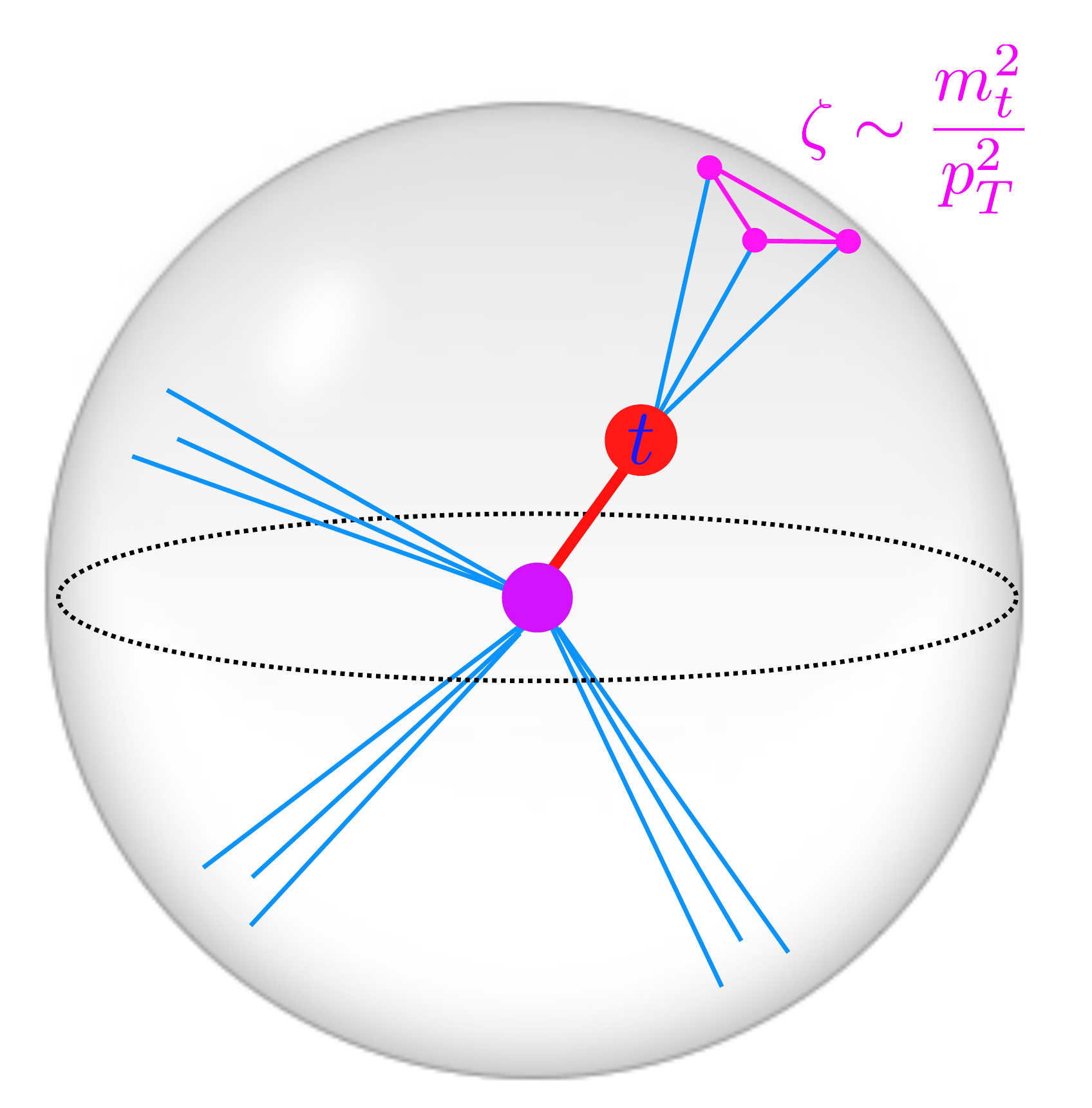} \label{fig:EEECtop_a}
\includegraphics[width=0.45\hsize]{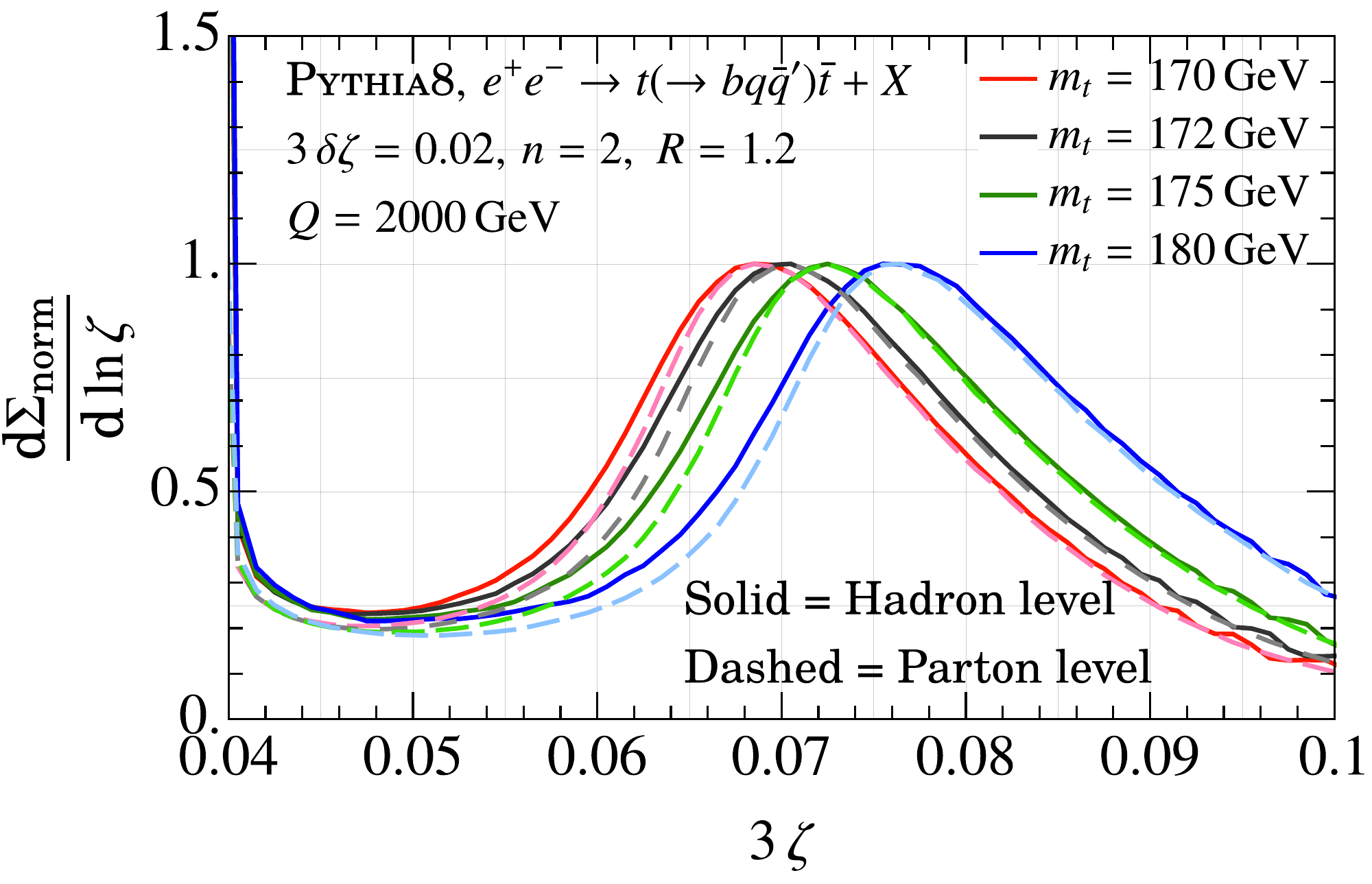} \label{fig:EEECtop_b}
\end{center}
  \caption{(a) The angular scale of the top quark as probed by a three point correlation function. (b) The angular distribution for different top masses. Hadronization has an extremely minor impact on the peak position, leading to a clean probe of the top quark mass with perturbative physics \cite{Holguin:2022epo}.}
  \label{fig:EEECtop}
\end{figure}

\subsection{Boosted top quarks}
\label{sec:boostedtop}

Additional tools for the top quark mass measurement become available in the highest energy running of the ILC at 1~TeV.
At these energies, we enter the ``boosted top" regime in which the top quark decay products are boosted in to a single fat jet.  This brings in interesting QCD issues and also gives two new,  complementary methods for the top quark mass measurement. The first is a measurement using the two-jettiness observable, and the second is a measurement using the three-point energy correlator.

The most well established program to measure the top quark mass using event shapes in $e^+e^-$ is to use the two-jettiness (thrust) observable \cite{Bachu:2020nqn}. In the limit of boosted top quarks, the two-jettiness observable effectively measures the sum of the masses of the two jets arising from the decay products of the boosted top quarks. It thus exhibits strong sensitivity to the value of the top quark mass. Since it is an inclusive event shape observable, one can derive a rigorous factorization theorem for the observed distribution using a combination of SCET and HQET, which factorizes the dynamics at different scales, allowing in particular for a rigorous field theoretic treatment of the top quark mass in a short distance scheme. This distribution has been computed at next-to-next-to-next-to leading logarithmic accuracy, and suggests that perturbative uncertainties at the order of $100$ MeV can be achieved for the top quark mass. A plot of the distribution is shown in Fig. \ref{fig:top_adi}.

More recently, an alternative approach to measuring the top quark mass was put forward \cite{Holguin:2022epo}, particularly motivated by developing a clean understanding in the complex LHC environment. One of the issues with measurements based on the jet mass is that the mass is sensitive to soft contamination and non-perturbative effects. Instead of measuring the mass directly, one can flip the measurement of the mass into a measurement of the angular scale of the top decay products as measured by a three-point correlation function. Unlike the behavior in a conformal theory, these correlators will exhibit a sharp peaked structure at the angular scale $\zeta \sim m_t^2/Q^2$. The location of the peak therefore provides direct sensitivity to the top quark mass. However, unlike the jet mass, the location of the peak is unaffected by soft contamination and hadronization. This is seen in Fig. \ref{fig:EEECtop}. Furthermore, this measurement is quite interesting from the point of view of precision QCD, since it probes the structure of multipoint energy correlations on top decays.

\subsection{Top quark electroweak couplings}
\label{sec:topEW}

%\subsubsection{Relevance of EW couplings}
In many extensions of the Standard Model, the top quark plays a special role. Composite Higgs models, for instance, generally predict sizeable deviations for the top quark electro-weak couplings~\cite{Richard:2014upa}. Precise measurements of these couplings can therefore constrain broad classes of composite Higgs scenarios~\cite{Berger:2005ht,Grojean:2013qca,Durieux:2018ekg}.

%\subsubsection{State of the art and HL-LHC prospects}
Since the top quark could not be studied at the previous generation of electron-positron colliders, its interactions with the neutral gauge bosons of the Standard Model are relatively poorly constrained. Studies of top quark pair production at hadron colliders have characterized the strong interaction of the top quark in detail, and single top quark production and top quark decay are a sensitive probe of the charged-current interaction. The interactions with the photon and $Z$-boson have only become accessible with the observation of rare associated production processes, such as $ pp \rightarrow t\bar{t}X$ and $pp \rightarrow tXq$, with $X= \gamma, Z$.  (The coupling of the top quark to the Higgs boson will be discussed in Sec.~\ref{sec:Higgstop}.)  Recent comparisons of cross section measurements to SM predictions have reached a precision of 10-15\%, with statistical, experimental and theoretical uncertainties contributing with roughly equal weight~\cite{Miralles:2021dyw}. Top quark EW operators can also be constrained through loop-level effects of off-shell degrees-of-freedom in the top quark pair production rate~\cite{Martini:2019lsi}, which could provide complementary bounds of competitive precision for some operator coefficients.

The potential of LHC run 3 and the HL-LHC stage to improve these measurements has been studied in Ref.~\cite{Azzi:2019yne} for $t\bar{t}V$ production and EW single top production. A complete set of estimates for the HL-LHC expectations can be found in Ref.~\cite{Miralles:2021whitepaper}, refining earlier results of \cite{Durieux:2019rbz}.
These studies adopt  the S2 scenario also used for the Higgs sector~\cite{Cepeda:2019klc}.  That is, they extrapolate LHC run 2 results by scaling the statistical and experimental sytematics uncertainties with the inverse square root of the luminosity, while assuming that the uncertainties in the theoretical SM predictions and uncertainties due to Monte Carlo modelling are reduced by a factor 2 with respect to today's state of the art.

\begin{figure}
    \centering
    \includegraphics[width=\linewidth]{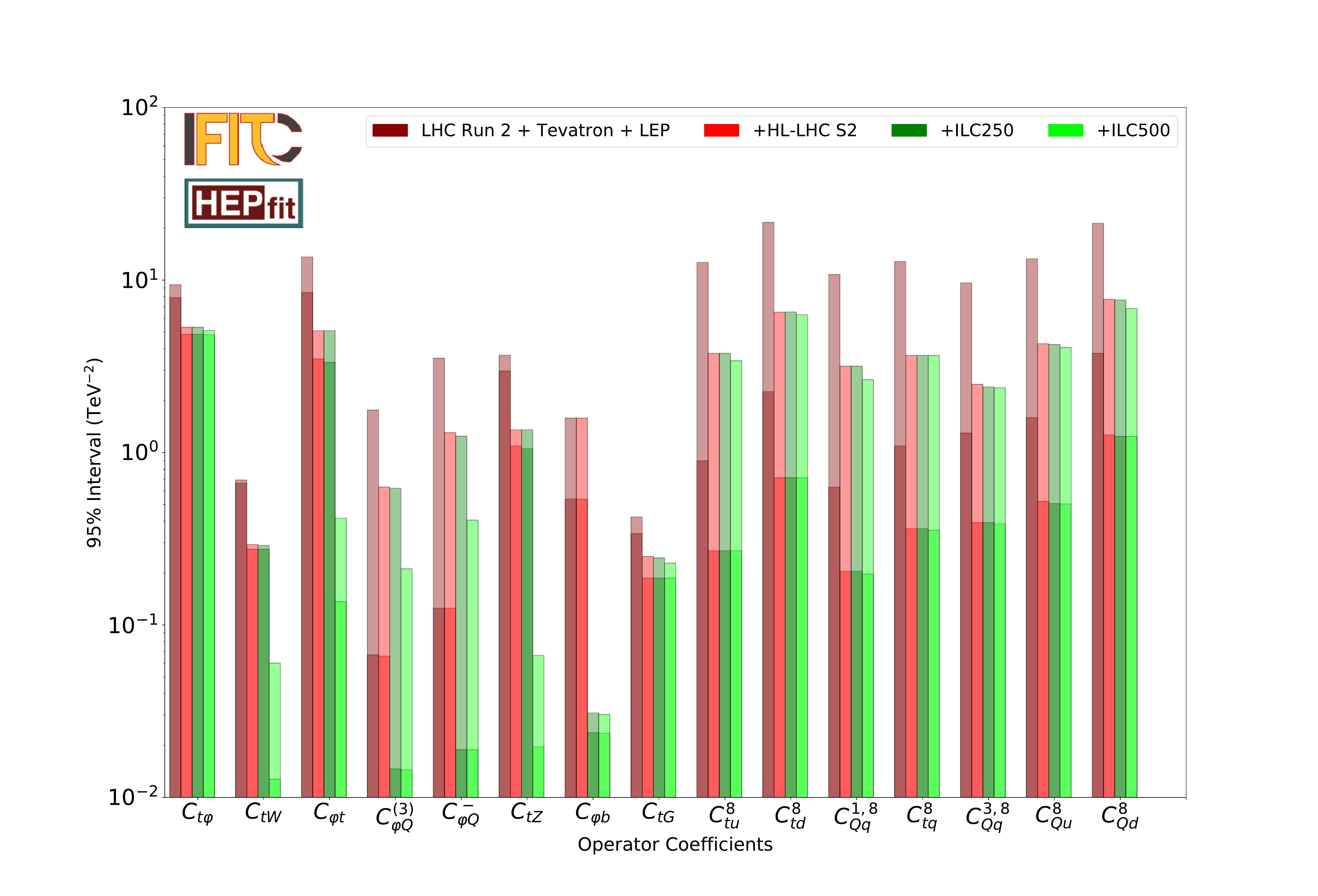}
    \caption{Comparison of current 95\% CL bounds on the coefficients of the two-fermion SMEFT operators  that affect the top and bottom quark EW couplings and the four-fermion operators $q\bar{q}t\bar{t}$. The LHC bounds correspond to the results of Ref.~\cite{Miralles:2021dyw}, the HL-LHC S2 projection follows Ref.~\cite{Durieux:2019rbz} and the HL-LHC Higgs report~\cite{Cepeda:2019klc}, while the ILC prospects are based on Ref.~\cite{Durieux:2018tev}.  Bars with dark shading show constraints on individual couplings considered separately,  while bars with light shading show constraints from a global fit to the full set of operators.}
    \label{fig:top_ew_coupling_prospects}
\end{figure}

The ILC offers a unique opportunity to go beyond these studies in measuring  the electro-weak couplings of the top quark~\cite{Amjad:2015mma,Amjad:2013tlv}. These measurements are among the prime targets of the ILC top physics programme. The pair-production process in $e^+e^-$ collisions probes the $t\bar{t}Z$ and $t\bar{t}\gamma$ vertices directly. The contributions from the photon and $Z$-boson are disentangled by using the two polarization configurations. Figure~\ref{fig:top_ew_coupling_prospects} shows the comparison of the ILC expectations to current results from LHC and the prospects for HL-LHC; the comparison is expressed in terms of bounds on the coefficients of 2-fermion dimension-6 operators in the Effective Field Theory description of the top quark couplings. The measurements of top quark production rates at the ILC improve the measurement of the EW couplings and the corresponding bounds on the relevant EFT operator coefficients by two orders of magnitude with respect to the current LHC results, and by well over an order of magnitude with respect to HL-LHC expectations. Data above the top quark pair production threshold are clearly required to provide tight bounds on the operator coefficients that affect the top quark couplings. 

Measurements at two center-of-mass energies above the $t\bar{t}$ threshold allow to disentangle contributions of the relevant two-fermion and four-fermion operators in the SMEFT~\cite{Durieux:2018tev}. The prospects for constraints on the $e^+e^-t\bar{t}$ four-fermion operators with the 1~TeV run envisaged at the ILC yield 68\% CL bounds of order $C/\Lambda^2 \sim 10^{-3} TeV^{-2}$~\cite{Durieux:2018tev} and form a powerful test for scenarios with composite (right-handed) top quarks~\cite{Matsedonskyi:2020mlz} for compositeness scales well beyond the center-of-mass energy.

Dedicated CP-odd observables yield powerful constraints on CP violation in the top sector~\cite{Bernreuther:2017cyi}. Other processes, such as single top quark production and vector-boson-fusion production at high energy provide complementary information~\cite{CLICdp:2018esa}.

%\subsubsection{ILC prospects for top EW couplings}

There is a subtle interplay between the Higgs and top physics programmes, since the  top quark couplings affect the loop diagrams for $gg\rightarrow H$ production at the LHC and $H\rightarrow \gamma \gamma$ and $H\rightarrow Z\gamma$ decays at the LHC and ILC~\cite{Durieux:2018ggn}. Precision measurements of tree-level processes, such as $e^+e^- \rightarrow ZH$ production, gain a sensitivity to top EW couplings through loop corrections. Precise measurements of top quark couplings are required to fully constrain all of the degrees of freedom of the Higgs EFT~\cite{Jung:2020uzh}.
% A further discussion is presented in Sec.~\ref{sec:heavyquarks}.

\subsection{Searches for FCNC interactions of the top quark}

Processes with flavour-changing neutral currents (FCNC) are forbidden at tree level in the SM and are strongly suppressed at higher orders by the Glashow-Iliopoulos-Maiani (GIM) mechanism. The branching fractions for top quark FCNC decays $t\rightarrow qX$,  where $q=u,c$ and $X=\gamma, g, Z, H$, are of the order of $10^{-12}$--$10^{-16}$. Some extensions of the SM predict a strong enhancement of the FCNC top quark decay rates, increasing the branching fraction up to $10^{-4}$.

The search for FCNC interactions of the top quark at the LHC has reached excellent sensitivity for the $tqX$ vertex. The current 95\% CL bounds based on searches for top decays and single top production with the partial run 2 data are equivalent to branching fractions of $10^{-3}-10^{-5}$ and are expected to improve significantly with the HL-LHC data~\cite{Azzi:2019yne,ATL-PHYS-PUB-2019-001, ATL-PHYS-PUB-2016-019}. 

An $e^+e^-$ collider has a very specific role in the search programme for FCNC couplings. The LEP bounds from searches for $e^+e^- \rightarrow t\bar{q}, \bar{t}q$ remain competitive for $tqZ$ and $tq\gamma$ and in particular the $tqll$ operators~\cite{Durieux:2014xla}. The 250 GeV phase of a Higgs factory is expected to improve the LEP bounds by one to two orders of magnitude~\cite{Shi:2019epw}, yielding competitive results in comparison with the full HL-LHC prospects. The higher-energy stages of the ILC are particularly relevant for the bounds on four-fermion operators $e^+e^-tq$. The sensitivity to these operators increases very strongly with the higher-energy operation~\cite{deBlas:2018mhx}. 

\begin{figure}
    \centering
    \includegraphics{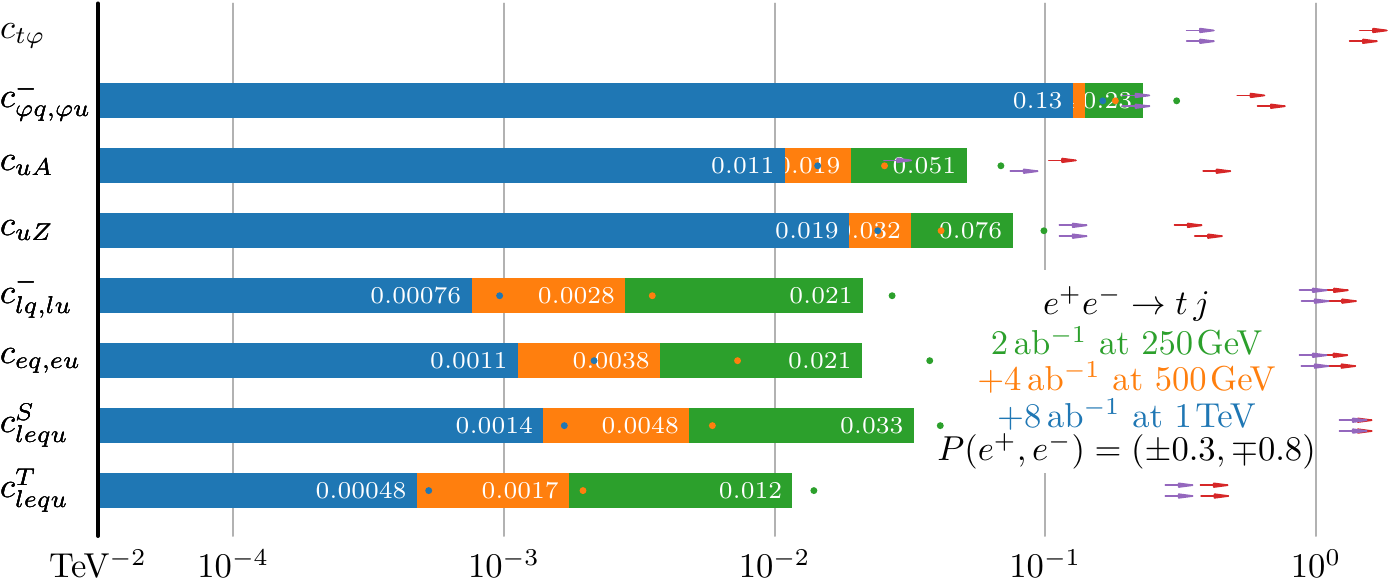}
    \caption{The projected 95\% C.L. bounds on the EFT operator coefficients that give rise to the FCNC $e^+e^- \rightarrow tq$ production process. The bounds are given in units of TeV$^{-2}$ for the LHC run 2 (dark red arrows), for the HL-LHC (purple arrows) and for the three nominal ILC stages: 250~GeV (green bars), 500~GeV (orange bars) and 1~TeV (blue bars). The round markers of the same color represent the expected bounds without beam polarization. }
    \label{fig:top_fcnc_projections}
\end{figure}

 The current 95\% CL bounds on the EFT operator coefficients are compared to the prospects of the HL-LHC (3~ab$^{-1}$ at 14~TeV), and three energy stages of the ILC (2~ab$^{-1}$ at 250~GeV, 4~ab$^{-1}$ at 500~GeV and 8~ab$^{-1}$ at 1~TeV) in Fig.~\ref{fig:top_fcnc_projections}. The current LHC bounds and HL-LHC projections from~\cite{} are indicated as dark red and purple arrows, respectively, where the upper arrow corresponds to up quarks and the lower one to charm quarks. The expected bounds for the several ILC energy stages, shown as solid bars, are extrapolated from the study of Ref.~\cite{Hesari:2014eua,Aguilar-Saavedra:2001ajk}. More details of the procedure are given in Ref.~\cite{deBlas:2018mhx}. The increase in sensitivity is particularly pronounced for the $e^+e^-tq$ operators, that are found to scale roughly as $s^{-3/2}$. 

\section{Higgs} 
\label{sec:Higgs500}

In this section, we discuss additional Higgs boson reactions that become 
accessible to the ILC at 500~GeV.

\subsection{$WW$ fusion}
\label{sec:HiggsWW}

As well as providing additional Higgs-strahlung events, ILC collisions at 500~GeV will
provide a large sample of Higgs bosons produced via the $WW$ fusion process.
While the resulting set of Higgs bosons cannot be identified using the unbiased
recoil mass method applicable to Higgs-strahlung, they none the less provide
an important sample to further probe the Higgs sector.
The number of Higgs bosons produced at ILC-500 will be similar to the number at ILC-250,
providing comparable statistical power as at ILC-250 for all the measurements at 250~GeV discussed in Sec.~\ref{sec:Higgs250}.
The experimental techniques and background composition are different at the different energies,
production methods and beam polarizations,
providing a range of systematic checks by comparing measurements of
related observables made under different conditions, before
combining the measurements to achieve optimal sensitivity while also
testing the internal consistency of the measurements when interpreted
within the Standard Model.

The comparison of Higgs production in the $Zh$ and $WW$-fusion processes,
enabled respectively by the $hZZ$ and $hWW$ couplings,
with the measured decay branching ratio to $WW^*$ and $ZZ^*$ will allow
independent checks of the Higgs couplings to $V (=W/Z)$. 
The experimental sensitivity to anomalous $HVV$ couplings,
whose effects typically grow with energy, will be significantly
enhanced at ILC-500.

The impact of ILC-500 data on the understanding of the Higgs 
sector is clearly demonstrated later in this report, for example in Fig.~\ref{fig:Higgsresults}.

\subsection{Higgs self-coupling}
\label{sec:HiggsSelf}

\begin{figure}
\begin{center}
\includegraphics[width=0.70\hsize]{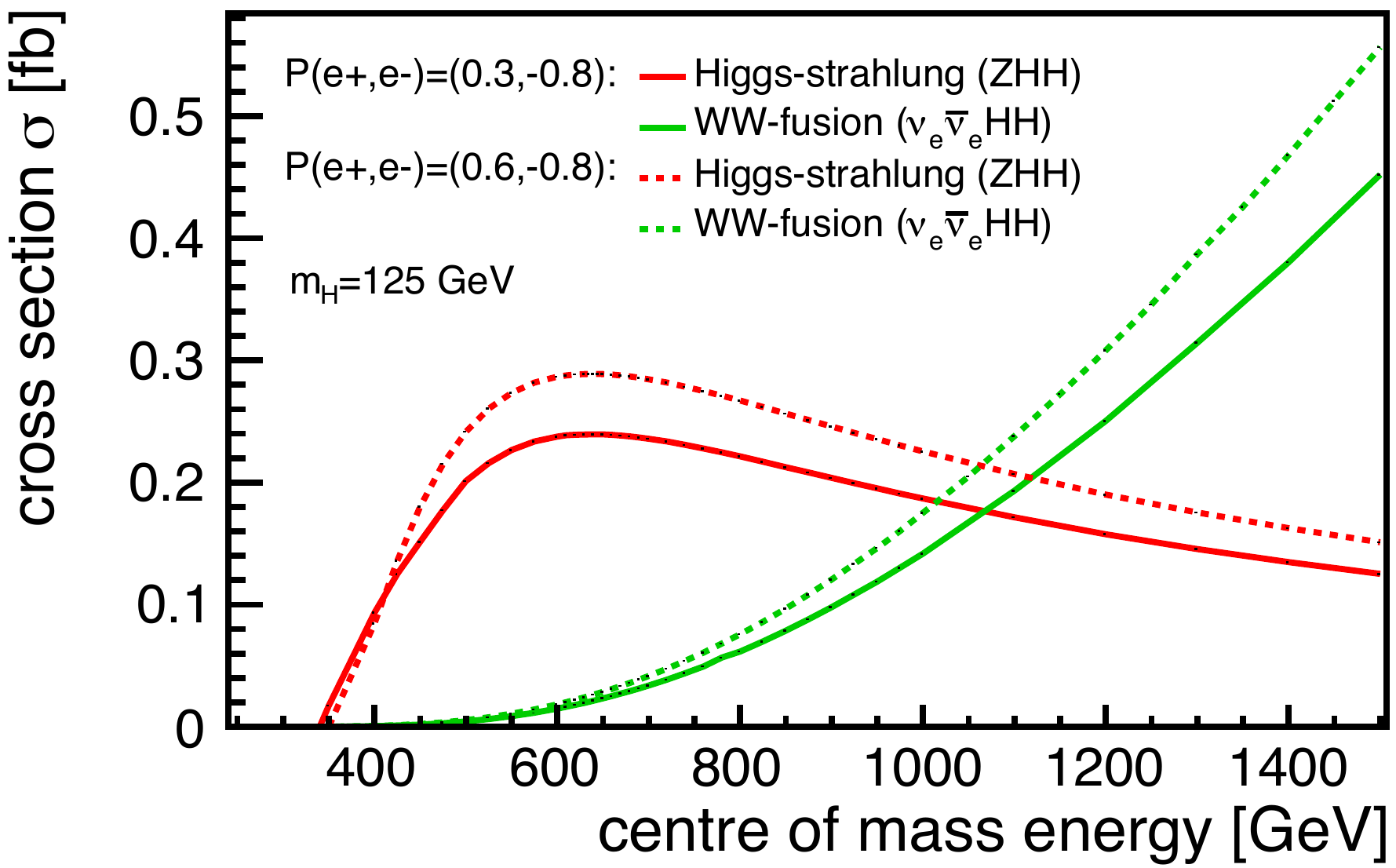}
\end{center}
\caption{Cross-sections for double Higgs production in the SM via Higgs-strahlung and $WW$ fusion as a function of the center-of-mass energy~\cite{Durig:2016jrs}.}
\label{fig:xsecHHvsECM}
\end{figure}

At center-of-mass energies of at least $500$\,GeV, the self-interaction of the Higgs boson, in particular the triple-Higgs coupling $\lambda$, can be probed directly by studying the production of Higgs boson pairs.  There are two relevant di-Higgs production processes,  double Higgs-strahlung, $\ee \to ZHH$, and di-Higgs production in $WW$ fusion, $\ee \to \nu\bar{\nu}HH$.  The cross sections for these reactions  as a function of the center-of-mass energy are shown  in Fig.~\ref{fig:xsecHHvsECM}. While the $WW$ fusion becomes important at and above $1$\,TeV, the cross-section for double Higgs-strahlung reaches a maximum around $500-600$\,GeV.   We will argue that it is important to  provide running both near 500~GeV and at 1~TeV so that both of these reactions can be studied.

%%%%%%%%%%%%%%%%%%%%%%%%%%%
%%%%%%%%%%%%%%%%%%%%%%%%%%%%%%%%%%

The prospects for measuring double Higgs production through these two reactions was studied at the time of the ILC TDR.
The reactions were simulated using 
 full, Geant4-based simulation of the ILD detector, using the state-of-the-art reconstruction tools at the time~\cite{Behnke:2013lya}, both at $\sqrt{s}=500$\,GeV~\cite{Durig:2016jrs,tian_LC_REP:2013} and $1$\,TeV~\cite{tian_LC_REP:2013}. These studies found that, if the self-coupling takes its SM value, then  double Higgs-strahlung can be observed at $E_{CM} = 500$~GeV  with a significance of $8\,\sigma$,  combining the $HH\to b\bar{b} b \bar{b}$ and $HH\to b\bar{b} WW^*$ channels.  This would translate into a measurement precision on $\lambda$ of $27\%$.  The analysis assumed that all other couplings of the Higgs bosons take their 
SM values, but subsequently it was shown that the  ILC single-Higgs measurements  strongly control possible variations, enough to add only a negligible uncertainty~\cite{Barklow:2017awn}. We will discuss this point further  in Sec.~\ref{sec:expectself}.  When the ILC runs at $E_{CM} = 1$~TeV, it will be possible to add the measurement with the double Higgs production from $WW$ fusion. This will improve the determination of (the SM value of) $\lambda$ to a relative precision of $10\%$.

Over the past few years, there have been many improvements in the planned ILD detector that are relevant for these measurements.
The $b$-tagging efficiency in ILD has been improved by $5\%$ at the same level of purity~\cite{ILDConceptGroup:2020sfq}. This improvement and the inclusion of $HH\to \tau^+ \tau^- b \bar{b}$ have been estimated to improve the ILC500 precision on $\lambda$ from the $27\%$ mentioned above to $21-22\%$~\cite{Durig:2016jrs}. Another limiting factor for the double Higgs-strahlung analysis is the invariant di-jet mass reconstruction, important for separating $ZHH$ from $ZZH$ and $ZZZ$ backgrounds. New developments in correcting for missing energy from neutrinos in semi-leptonic heavy quark decays and kinematic fitting show striking improvements on the di-jet mass reconstruction~\cite{Radkhorrami:2021cuy}. Further improvements on the jet clustering and on the flavor tag are being expected from deep learning approaches~\cite{Goto:2021wmw} as well as from a full exploitation of the charged hadron identification capabilities of ILD~\cite{Einhaus:2021vcb}.
Propagation of all of these improvements of the high-level reconstruction to the full double Higgs-strahlung analysis carries the potential to bring the ILC500 sensitivity to better than $20\%$.

%%%%%%%%%%%%%%%%%%%%%%%%%%%%%%%%%%%%%%%%%%%%%%%%%%%%%
\begin{figure}
\begin{center}
\begin{subfigure}{0.495\textwidth}
\includegraphics[width=0.90\hsize]{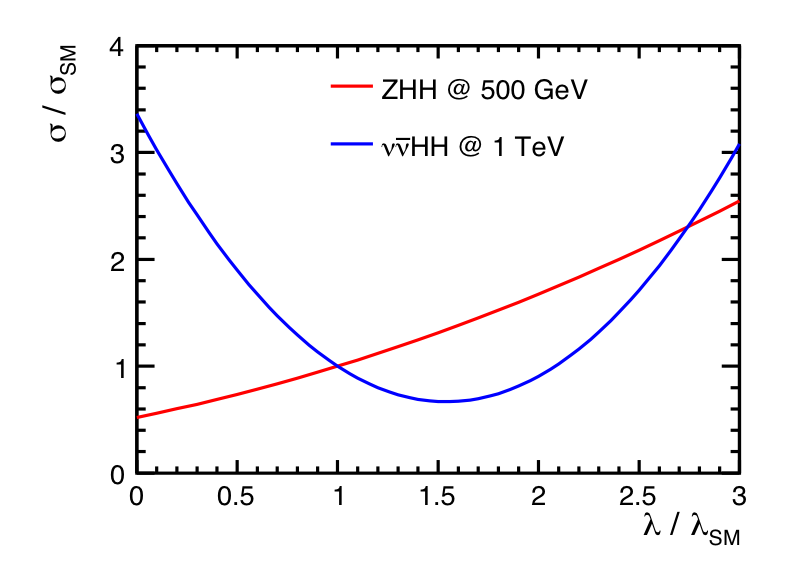}
\caption{}
\end{subfigure}
\begin{subfigure}{0.495\textwidth}
\includegraphics[width=0.90\hsize]{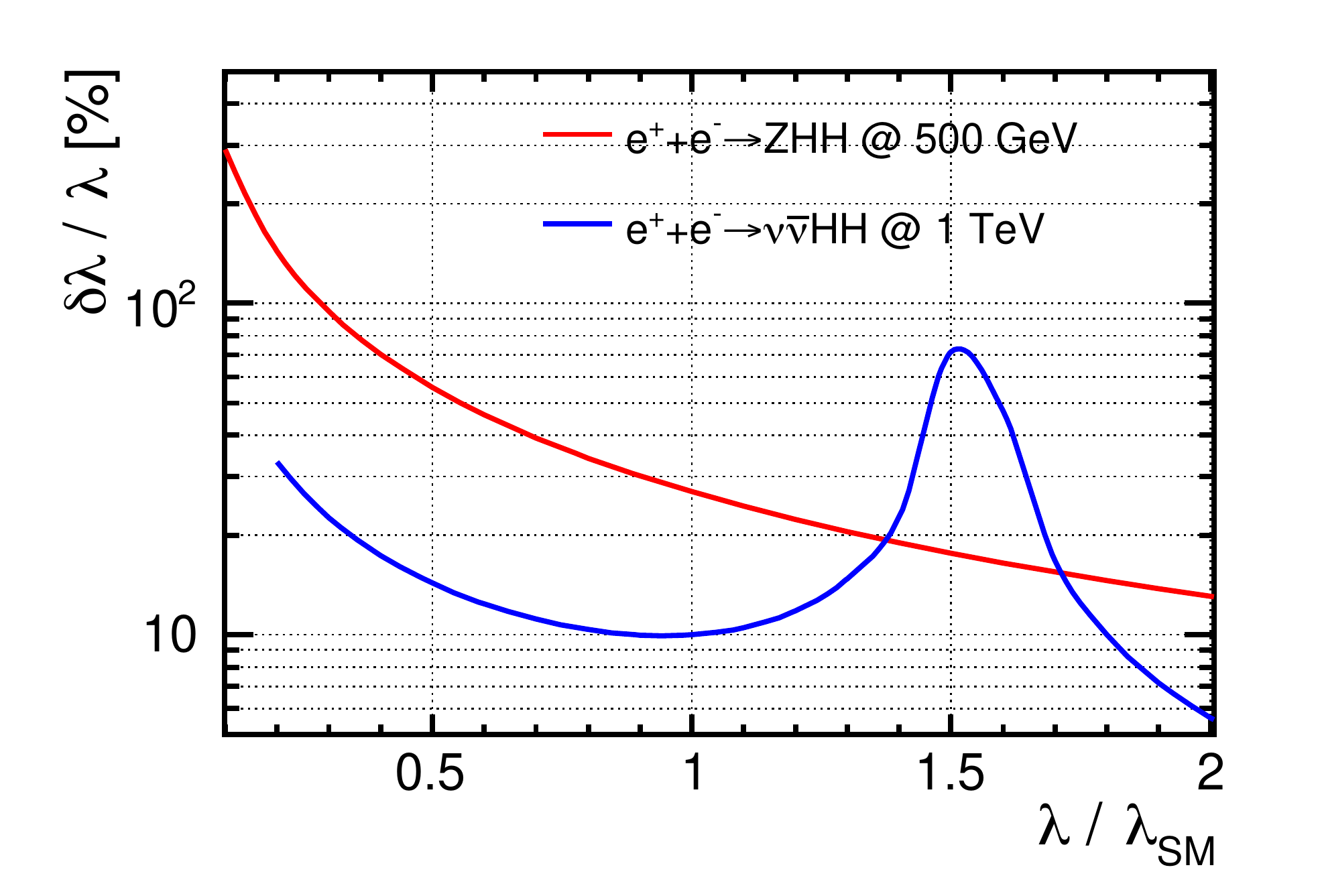}
\caption{}
\end{subfigure}
\end{center}
\caption{(a) Cross-sections for double Higgs production via Higgs-strahlung (at $\sqrt{s}=500$\,GeV) and $WW$ fusion (at $1$\,TeV) as a function of the triple-Higgs coupling (normalised to its SM value)~\cite{Durig:2016jrs}. (b) Expected precision from each of the two ILC reactions as a function of the actual value of $\lambda$ relative to the SM value.}
\label{fig:xsecHHvsLambda}
\end{figure}
%%%%%%%%%%%%%%%%%%%%%%%%%%%%%%%%%%%%%%%%%%%%%%%%%%

The availability at the ILC  of two different $HH$ production reactions becomes more important when one realizes that the real goal of this measurement is to demonstrate a Higgs self-coupling different from that of the SM.  In models with extended Higgs sectors,  the self-coupling can deviate significantly from its SM value, even if other Higgs couplings are rather SM-like. 
For instance in models with two Higgs doublets (2HDM) where all fermions couple only to one Higgs doublet (``Yukawa type~I'') values of $-0.5 \lsim \lambda/\lambda_{\mathrm{SM}} \lsim 1.5$ are possible even after taking into account theoretical and experimental constraints~\cite{Arco:2020ucn,Arco:2021bvf} (where higher-order corrections can enhance these values~\cite{Bahl:2022jnx}).    In models with singlet scalars that mix with the Higgs boson, even larger enhancements are possible.  Models of electroweak baryogenesis typically require large enhancements in $\lambda$, by a factor 1.5--2.5~\cite{Morrissey:2012db}.   We discuss this point  further in Sec.~\ref{sec:antimatter}.   On the other hand, there are models in which $\lambda$ decreases with respect to the SM value.

%%%%%%%%%%%%%%%%%%%%%%%%%%%%%%%%%%%
\begin{figure}
\begin{center}
\includegraphics[width=0.60\hsize]{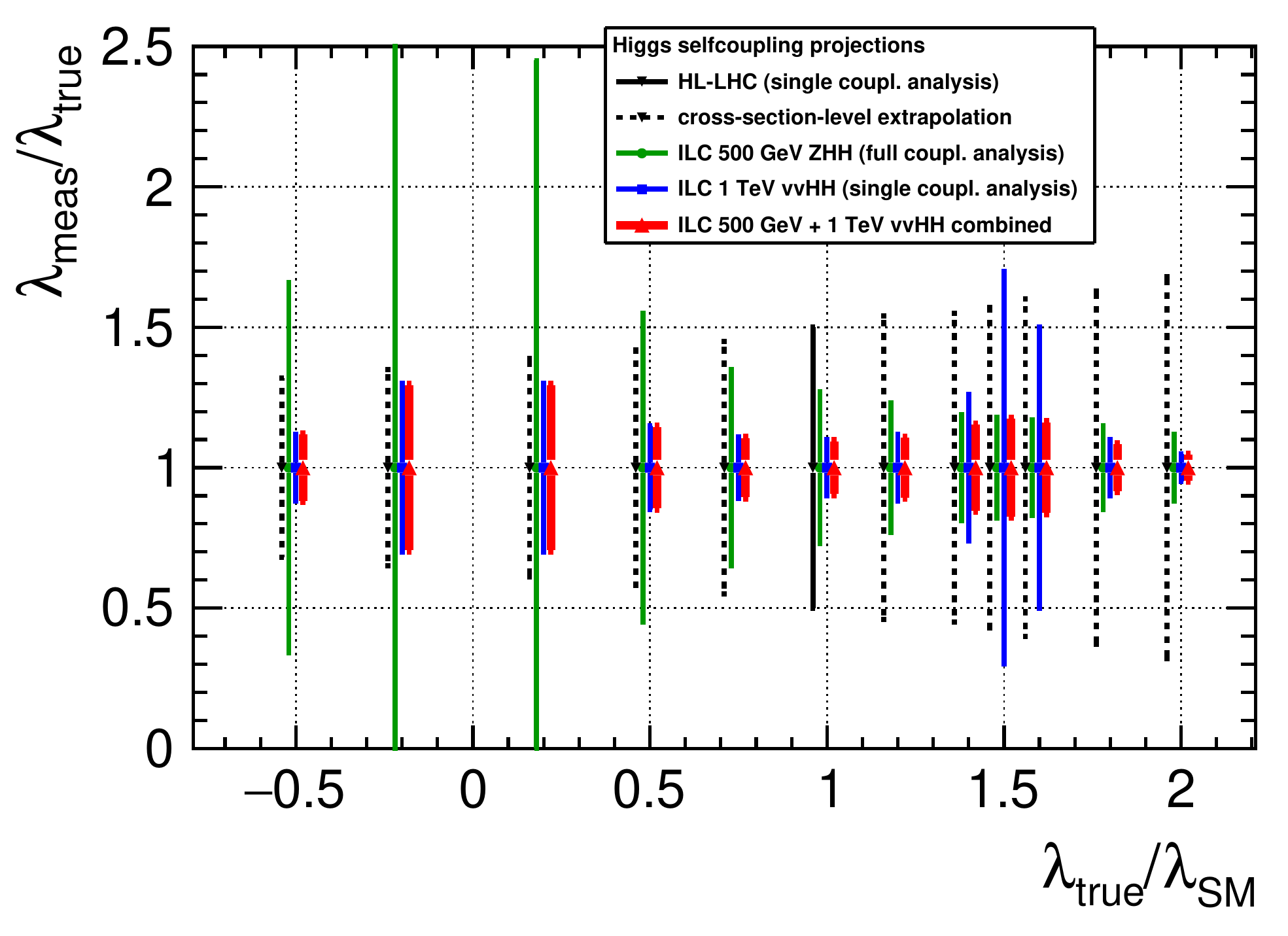}
\end{center}
\caption{Expected uncertainties in the determination of the Higgs self coupling at the HL-LHC and the ILC as a function of 
$\lambda/\lambda_{SM}$.  The HL-LHC value for the SM value of $\lambda$ is that projected by the ATLAS collaboration in ~\cite{ATLAS:2018rvj} that is then extrapolated to other values of the coupling. The ILC measurements at 500~GeV and 1~TeV are shown separately and, in red, in combination~\cite{Durig:2016jrs}.}
\label{fig:PrecisionvsLambda}
\end{figure}
%%%%%%%%%%%%%%%%%%%%%%%%%%%%%%%%%

Ideally, we would like to be maximally sensitive to modifications in $\lambda$ no matter what the sign turns out to be.   It is thus
remarkable that the two ILC reactions complement each other neatly in this respect.   In both double-Higgs reactions, the Higgs self-coupling gives only one contribution to the full amplitude and thus appears in interference with other SM effects.  For the $WW$ fusion reaction, this interference is destructive, so a small increase in $\lambda$ leads to a decrease in the cross section.  For the double-Higgs-strahlung process, the interference is constructive, leading to an increase in the cross section as $\lambda$ increases.  This effect is shown in Fig.~\ref{fig:xsecHHvsLambda}.   An increased cross section leads to greater sensitivity to the value of 
$\lambda$.  At the ILC,  whether $\lambda$ increases or decreases, one of the possible reactions will increase in cross section and reflect this improved sensitivity.   The situation is quite different at proton colliders such as the LHC.    The dominant process of double Higgs production, $gg\to HH$, is a fusion process with destructive interference.   The double-Higgs-strahlung process, which is higher order in the electroweak coupling, has a cross section smaller by many orders of magnitude.  At the ILC, though, there are 
two reactions whose results can be combined to  guarantee a measurement of the self-coupling at the level of at least $30\%$ \emph{for any value of the self-coupling nature might have chosen}.  Figure~\ref{fig:PrecisionvsLambda} shows the effect of this 
synergy in comparison with an extrapolation of the uncertainty projections from the ATLAS collaboration~\cite{ATLAS:2018rvj} to non-SM values of $\lambda$.

\subsection{Top quark Yukawa coupling}
\label{sec:Higgstop}

The top quark is the SM particle with the strongest coupling to the Higgs boson. The top quark Yukawa coupling has a value close to 1 in the SM.   It is therefore important to understand this value and its relation to the top quark mass. The Higgs boson discovery channels at the LHC are sensitive to this coupling indirectly, through Higgs production and decay channels such as $gg \rightarrow H$ and $H\rightarrow \gamma\gamma$; in the SM, these proceed primarily through top quark loops. Under certain assumptions, the Higgs production and decay rates can yield a precise bound on the top quark Yukawa coupling. A more direct, and more robust, measurement is possible in the associated $pp \rightarrow t\bar{t}H$ production process, observed in 2018~\cite{ATLAS:2018mme,CMS:2018uxb}. The projection for the HL-LHC envisages an uncertainty of approximately 3\% on the signal multiplier $\kappa_t$ dominated by theory uncertainties~\cite{Cepeda:2019klc}. Several groups have studied the interplay between 

At the ILC, indirect probes are also available: the $H\rightarrow \gamma\gamma$, $H\rightarrow gg$ and $H\rightarrow Z\gamma$ channels provide sensitivity to the top Yukawa coupling already in 250~GeV data. These measurements can determine the top Yukawa coupling with $\sim$1\% precision, under the assumption that no new particles enter in the loops. These measurements may therefore provide an early indication of new physics, but a deviation of the SM cannot be unambiguously pinpointed. In more general EFT fits, the constraint on the coefficient $C_t\phi$ of the operator that shifts the top Yukawa coupling obtained from these indirect probes is not robust, as its effect is degenerate with poorly bounded degrees of freedom~\cite{Jung:2020uzh}.

The $t\bar{t}$ threshold scan offers an indirect determination that is more specific for the top quark Yukawa coupling. The production rate close to threshold is sensitive to Higgs-exchange effects and can yield a competitive precision of 4\% on the top-quark Yukawa coupling~\cite{Yonamine:2011jg}. However, the current uncertainty in state-of-the-art calculation would add a 20\% theory uncertainty~\cite{Vos:2016til} and there is no clear perspective to reduce or circumvent this uncertainty.

The direct measurement in $e^+e^- \rightarrow t\bar{t}H$ production requires a center-of-mass energy of at least 500~GeV. The cross section rises sharply around that energy; raising the center-of-mass energy to 550~GeV enhances the production rate by a factor or approximately four and the measurement of the $t\bar{t}H$ coupling by a factor two. Several groups have performed detailed full-simulation studies at center-of-mass energies ranging from 500~GeV to 1.4~TeV~\cite{Abramowicz:2016zbo, Price:2014oca, Yonamine:2011jg}. With 4~ab$^{-1}$ at 550~GeV, a precision of 2.8\% is expected on the top Yukawa coupling, which could improve to 1\% with 8~ab$^{-1}$ at 1~TeV. Measurements at multiple center-of-mass energies and with different beam polarizations can further characterize the $t\bar{t}H$ coupling~\cite{Han:1999xd}. 

Another important target requiring center-of-mass energies between $600$\,GeV and $1$\,TeV are the $CP$ properties of the $t\bar{t}H$ coupling. Achievable constraints have been studied at the cross-section level~\cite{Godbole:2011hw}, showing a significant improvement due to polarized beams. A detailed detector-level study of the relevant observables remains an interesting task for future studies.

\section{Triple gauge couplings}
\label{sec:Wboson500}

\begin{figure}[htb]
\begin{center}
\includegraphics[width=0.75\hsize]{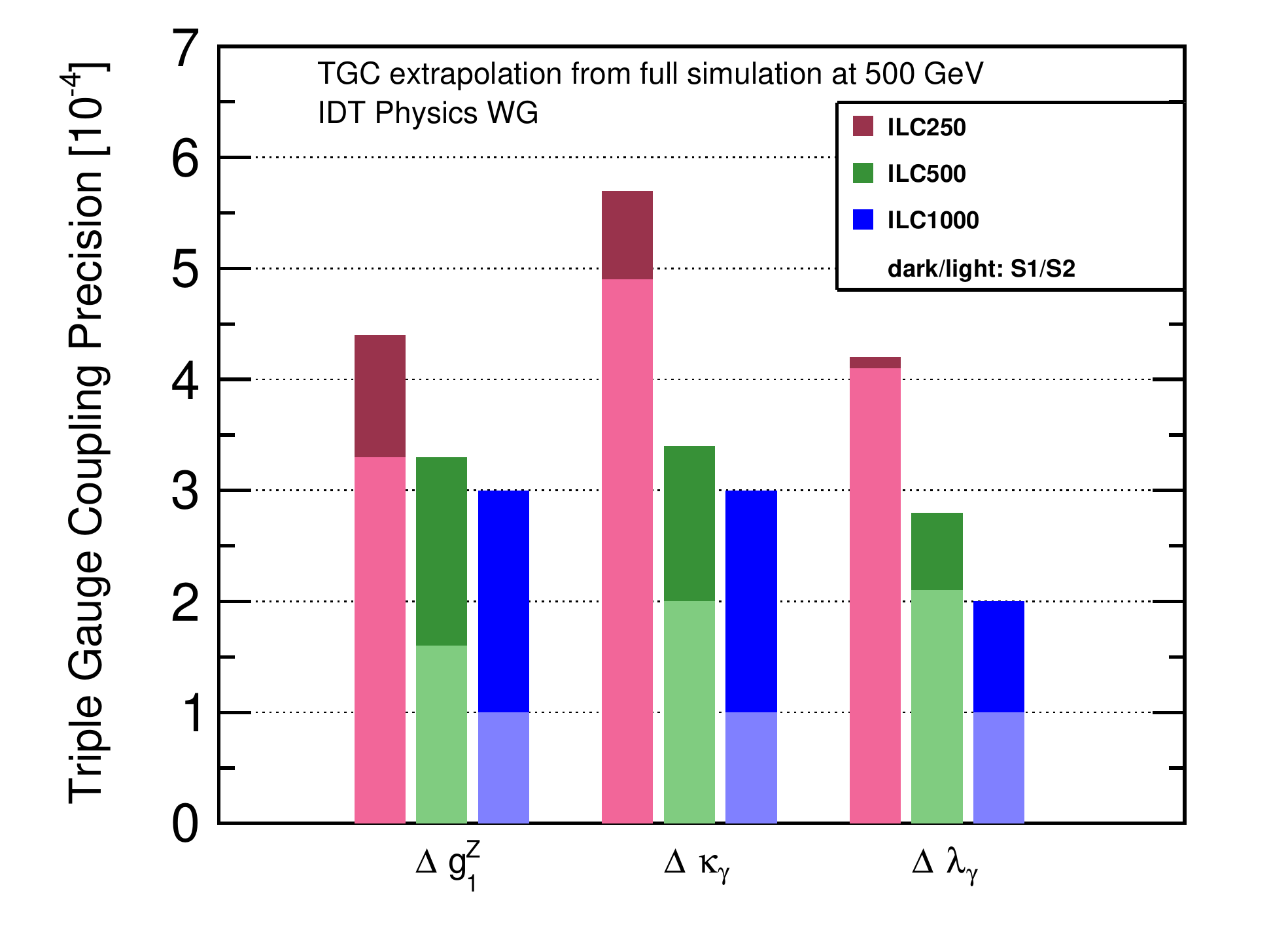}
\end{center}
\caption{Expected precisions on the three triple gauge coupling parameters at the three energy stages of ILC. The results at $500$\,GeV and at $1$\,TeV are based on the ILD full simulation analyses of semi-leptonic $W$ pair production, extrapolated to include improvements from the fully hadronic channel and single-$W$ production as well as for upgrading from a binned analysis of three angles to an optimal observable technique~\cite{Karl:2019hes}. The S1 scenario assumes the systematic uncertainties from~\cite{Karl:2019hes}, the S2 illustrates the hypothetical reduction by a further factor 2-3 to the level of $1 \times 10^{-4}$.
\label{fig:TGC_ILC_allECM}}
\end{figure}

The ILC prospects for triple gauge coupling measurements at $\sqrt{s}=250$\,GeV have been introduced in Section~\ref{sec:Wboson250}.
These have actually been extrapolated from previous studies at $\sqrt{s}=500$\, GeV and $1$\,TeV based on full simulation of the ILD detector concept~\cite{Marchesini:2011aka, Rosca:2016hcq}.  At higher energies, the relative effect on the differential cross section of the three TGC parameters $g_Z$, $\kappa_\gamma$, and  $\lambda_\gamma$ increases proportional to $s/m_W^2$. There are  a  number of experimental effects that become more challenging at higher energies --- more forward-boosted event topologies, higher pile-up from beamstrahlung pairs and photoproduction of low-$p_t$ hadrons ---  the fundamental gain in sensitivity with $s$ dominates by far. Figure~\ref{fig:TGC_ILC_allECM} summarizes the current state of the expected precisions, as discussed in more detail in the following paragraphs.

The full simulation study at 500\,GeV~\cite{Marchesini:2011aka} was limited to a binned analysis of three (out of five) angles in the $WW \to \mu\nu qq$ and $WW \to e\nu qq$ channels. For an integrated luminosity of 500\,fb$^{-1}$, this study found statistical uncertainties of $(6.1, 6.4, 7.2) \times 10^{-4}$ for $g^Z_1$, $\kappa_{\gamma}$ and $\lambda_{\gamma}$, respectively. 
An unbinned likelihood or optimal observable analysis of all five angles, including also fully hadronic $WW$ events as well as single-$W$ events has been estimated~\cite{Barklow:2018priv} to improve these numbers by a factor of 2.4 for $g^Z_1$ and by a factor of 1.9 for $\kappa_{\gamma}$ and $\lambda_{\gamma}$. Assuming the full integrated luminosity of ILC500 instead of only 500\,fb$^{-1}$ gives another factor of 2 improvement to $(1.3, 1.7, 1.9) \times 10^{-4}$. At this level of precision, systematic uncertainties need to be considered. As shown in~\cite{Beyer:2022xyz}, the effects of a finite knowledge of the luminosity and the beam polarizations are negligible when including them as nuisance parameters in a global fit. The effect of different per mil-level uncertainties on the selection efficiency and percent-level uncertainties on the residual background has been evaluated in~\cite{Marchesini:2011aka} by propagation through the whole analysis chain, thereby treating them as fully uncorrelated between data sets and observables, obviously a very pessimistic assumption. Based on considerations of correlated uncertainties and nuisance parameters in global fits, more recent studies expect that systematic uncertainties of $(3, 3, 2) \times 10^{-4}$ can be reached~\cite{Karl:2019hes}. In total, the expected precisions on the three couplings thus reach $(3.3, 3.4, 2.8) \times 10^{-4}$ for ILC500.

The full simulation study at 1\,TeV~\cite{Rosca:2016hcq} found statistical precisions of $(1.9, 1.7, 2.7) \times 10^{-4}$ for a luminosity of 1\,ab$^{-1}$ with the same analysis technique as at 500\,GeV (semileptonic $W$ pairs, binned analysis using three angles). A simple scaling to the full luminosity of 8\,ab$^{-1}$ renders the statistical uncertainty negligible with respect to the systematic uncertainties as given above. Thus, an adequate estimate of the $1$\,TeV prospects requires a thorough re-analysis of the systematic effects. It has already been shown that any global scaling as well as the variation of a simple angular cut-off can be determined from the data without any loss of precision on the TGCs~\cite{Beyer:2022xyz}.  But a  complete treatment of the remaining backgrounds in a multivariate fit still remains to be done.  In Fig.~\ref{fig:TGC_ILC_allECM} we present the expected precisions on the TGC paramters assuming the currently understood level of systematic uncertainties and the result of possible improvements by a factor 2-3 to the level of $1\times 10^{-4}$.

\section{Quark and lepton pair-production} 
\label{sec:pairs500}

The pair productions of leptons and quarks at ILC are also an important probe for new physics via precise measurements of total and differential cross sections. We have reviewed the formulae governing these processes in Sec.~\ref{sec:gendifermions}.

\subsection{Full simulation studies}

Lepton pair production, $e^+e^- \to \ell^+\ell^- (\ell = e, \mu, \tau)$, gives distinguishable signatures with pure-leptonic final states. Most of the SM background can be efficiently removed by selection of visible energies ($\sim 250$ GeV except for $\tau\tau$ final states) and angular selection (back-to-back topology). Since the high-energy phenomena like $Z^\prime$ and loop contribution of WIMPs are more sensitive to high $q^2$, cutting low $q^2$ events do not significantly degrade the physics reach.

A full simulation study has been done for $e^+e^-$ and $\mu^+\mu^-$ final states with $\sqrt{s} = 250$ GeV \cite{Deguchi:2019tvp}. The event selections of track parameters for particle ID, visible energy and rejecting forward tracks give almost pure signal samples without significantly cut effective signals. The signal efficiency is more than 98\% for $\mu$ pair final states and more than 97\% for $e$ pair final states at $|\cos\theta|< 0.95$ without significant dependence on the polar angle in the range. The remaining background is negligibly small.   Full-simulation studies for $c$ and $b$ pair production have been reviewed earlier in Sec.~\ref{sec:difermionatZ}.

Tau-pair production has been separately studied with $\sqrt{s} = 500$ GeV again with ILD detector simulation \cite{Jeans:2019brt,Yumino:2022vqt}. It shows that the selection efficiency of tau-pair events with at least one tau decaying hadronically can be as high as 70\% while remaining background is $< 10$\% of the signal contribution. Thanks to the highly-granular calorimeters, the tau decay channel can be identified by separating charged tracks and photons in the narrow jet of tau decay products. Expected performance of tau polarization is around 1\%, which adds another observable to explore BSM models. A study on further improvement on the polarization measurement using impact parameters of tau products is ongoing \cite{Yumino:2022vqt}. 

%[quark study - already discussed in section 9.3]

\subsection{$Z^\prime$ limits}

Given the ability of the ILC to measure these pair-production cross sections precisely, we can explore the possibility of searching for contributions from $s$-channel exchange of  $Z'$ bosons,  additional neutral vector gauge bosons coupled to SM fermions.
The coupling constants differ depending on the model.  Some standard models used as benchmarks are the Sequential Standard Model (SSM), the Alternative Left-Right symmetric model (ALR), and models in which the $Z'$ is a
 vector boson from the grand unification group $E_6$. The SSM assumes the same coupling constants as SM $Z$.   The ALR is  based on an model in which the electroweak symmetry is extended to 
 $SU(2)_L\times SU(2)_R\times U(1)$.
The $E_6$ model is a string-motivated model which naturally introduces $Z'$ as a linear combination of the two extra $U(1)$ gauge bosons $Z_{\psi}$ and $Z_{\chi}$ : $Z' =Z_{\chi}\cos\beta + Z_{\psi}\sin\beta$. IT is common to choose three values of the $\beta$ parameter: $\beta = 0$ ($\chi$ model), $\beta = \pi /2 $ ($\psi$ model) and $ \beta =\pi - \mathrm{arctan} \sqrt{5/3}$ ($\eta$ model).

For each of these models, we can use the ILC expectations for 
$e^+e^- \rightarrow f\bar{f}$ processes ($f = b, c, e, \mu, \tau$)  to derive exclusion and discovery limits. Here we assume fixed efficiency for each final state: $c_b = 0.29, c_c = 0.07, c_e = 0.97, c_\mu = 0.98, c_\tau = 0.65$,  based on full-simulation studies. The SM background is small enough compared with the signal events that it does not need to be  considered.
The expected signal events are estimated from tree-level differential cross section at the polar angle of $|\cos\theta| < 0.9$ with 19 points. Systematic uncertainty of 0.1 to 0.5\% depending on the final states are  included to calculate the mass limits. Table \ref{tab:2f-zprime} shows the obtained exclusion and discovery limit after combining all five final states \cite{Suehara:2022pwv}.

\begin{table}[t]
    \centering
    \begin{tabular}{l|rr|rr|rr}
         & \multicolumn{2}{c}{250 GeV, 2 ab$^{-1}$} & \multicolumn{2}{c}{500 GeV, 4 ab$^{-1}$} & \multicolumn{2}{c}{1 TeV, 8 ab$^{-1}$}\\
         Model & excl. & disc. & excl. & disc. & excl. & disc. \\ \hline
         SSM & 7.7 & 4.9 & 13 & 8.3 & 22 & 14 \\
         ALR & 9.4 & 5.9 & 16 & 10 & 25 & 18 \\
         $\chi$ & 7.0 & 4.4 & 12 & 7.7 & 21 & 13 \\
         $\psi$ & 3.7 & 2.3 & 6.3 & 4.0 & 11 & 6.7 \\
         $\eta$ & 4.1 & 2.6 & 7.2 & 4.6 & 12 & 7.8 
    \end{tabular}
    \caption{Projected limits on $Z^\prime$ bosons in standard models, from the study of $e^+e^- \to ff$. The values presented, given in TeV, are the 95\% exclusion limits and the 5$\sigma$ discovery limits for the successive stages of the ILC program up to 1 TeV.}
    \label{tab:2f-zprime}
\end{table}

\begin{figure}
\centering
\includegraphics[width=0.7\textwidth]{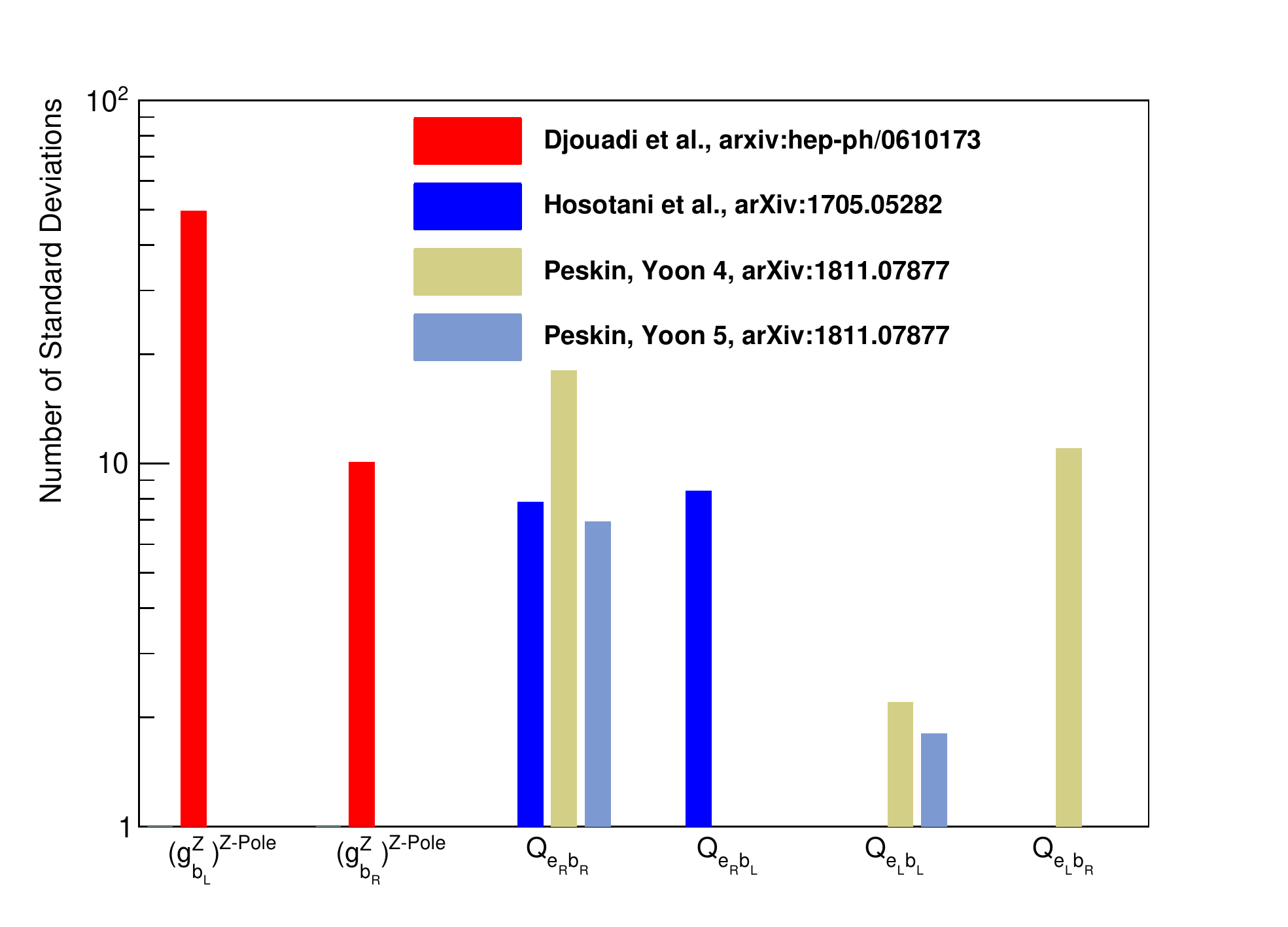}
\caption{Visibility of deviations from the SM predictions in $g_{b_j}^Z$, $g_{b_j}^Z$ (to do: still missing in this plot) and the helicity amplitudes $Q_{e_i b_j}$, in  standard deviations,  from combined ILC250/Z-Pole running, expected from new physics models with Randall-Sundrum extra dimensions~\cite{Djouadi:2006rk,Funatsu:2017nfm,Yoon:2018xud}.}
\label{pew:gigaz-rs} 
\end{figure}

In a similar way, we have investigated the sensitivity to $Z'$ bosons arising from specific models based on warped 5-dimension backgrounds, as in Randall-Sundrum theory~\cite{Randall:1999ee}. The results for models presented in~\cite{Djouadi:2006rk,Funatsu:2020haj,Yoon:2018xud} are shown in  Fig.~\ref{pew:gigaz-rs}. 
As is pointed out in~\cite{Funatsu:2020haj}, it is important to measure di-fermion production for all fermions and for different CM energies, since the effect of new $Z’$ bosons can dependent strongly on flavor  and beam polarization and will  increase with increasing beam energy.

\subsection{Indirect WIMP search}

One can also use $e^+e^- \to ff$ final states to carry out a generic search for a WIMP dark matter particle $\chi$.   If $\chi$ has nontrivial electroweak quantum numbers, it will appear
 in a $Z\rightarrow\chi\chi\rightarrow Z$ loop diagram and 
give a correction to the $Z$ coupling constants. The correction  depends only on the quantum numbers,  
spin, and mass of $\chi$ and is independent of model details \cite{Matsumoto:2017vfu,Harigaya:2015yaa}. We investigated three well-motivated types of WIMPs:
wino ($SU(2)_L$ triplet and $U(1)_Y$ hypercharge of 0), Higgsino ($SU(2)_L$ doublet and $U(1)_Y$ hypercharge of $\pm$1/2) and Minimal Dark Matter ($SU(2)_L$ pentet and $U(1)_Y$ hypercharge of 0).

Table \ref{tab:2f-wimp} shows the exclusion limits from this study,
based on  $e^+e^- \to ee, \mu\mu$ final states at $\sqrt{s} = 250$ GeV. We use 20 angular bins of $\cos\theta = -0.95$ to $0.95$  with systematic error assumed to be 0.1\% on each bin.  The limit on the $\chi$ mass from  direct production is $< \sqrt{s}/2$, so indirect search gives a larger discovery potential for these WIMPs. With enough statistics, it is possible to separate the effects of  WIMP models and $Z^\prime$ models using the  angular distribution of the deviation of the cross section from the SM prediction.

\begin{table}[h]
    \centering
    \begin{tabular}{l|r}
         Model & 2$\sigma$ exclsion  \\ \hline
         wino & 240 GeV \\
         Higgsino & 180 GeV \\
         Minimal Dark Matter & 500 GeV
    \end{tabular}
    \caption{Projected limits on WIMP indirect search with $e^+e^- \to ee, \mu\mu$ with $\sqrt{s} = 250$ GeV, 4 ab$^{-1}$ integrated luminosity. }
    \label{tab:2f-wimp}
\end{table}
\section{New Particle Searches -- TeV Scale} 
\label{sec:newparticles}

%\subsection*{\it The ILC strong points for searches}

In this section, we will discuss the prospects at the ILC for the
direct discovery of new particles.
Our discussion will of course be
given in the context in which the LHC experiments have carried out a
large number of new particle searches, some reaching deeply into the
mass region above 1~TeV.   Still, we will explain, experiments at
$\ee$ colliders can bring a complementary  approach to new particle searches and
open new and  very interesting windows for discovery~\cite{Fujii:2017ekh,Berggren:2020tle}. 

In general, the new particle searches done at the LHC have focused on 
scenarios within each theory of new physics that give the 
{\it best} possible experimental prospects to observe new physics.
This gives a chance to find such signs far out in  a
hitherto uncharted land.
However, there is no guarantee that new physics would be discovered
even if it is within the kinematic reach of the experiment.
The actual parameters of the theory might be far from the
ones giving the optimal signature sought in these analyses.

It is a rather different perspective to concentrate on the 
{\it worst} possible
points in the theoretical parameter space.
This clearly cannot reach as far out as in the previous case,
but now  a negative result
would make it possible to claim that the new physics theory is
ruled out at {\it all} possible parameter values below the
kinematic reach of the experiment.
It would also make discovery of the new physics {\it guaranteed}
if it is indeed energetically reachable.

Lepton colliders have a lower
reach in energy,
but excel in fully exploiting
all possible manifestations of new physics within
reach. 
As the $e^+e^-$ initial state implies electroweak production, the background rates will be quite low.
This has consequences for the detector design and optimization: The detectors can feature close to $ 4\pi$ coverage,
and they do not need to be  radiation hard, so that the tracking system in front of calorimeters
can have a thickness as low as a few percent of a radiation-length.
In addition, the low rates means that the detectors needn't be triggered, so that {\it all}
produced events will be available to analysis.
Furthermore, at an   $e^+e^-$ machine, point-like objects are brought into  collision,
meaning that the initial state is fully  known,
and that the full beam energy is carried by the interacting objects.
The beam-spot is sub-microscopic in size, allowing experimenters to find 
displaced vertices at much smaller distances, even in channels 
(like \stau~pair production),
where there is no reconstructable primary vertex.

\begin{figure}[t]%  \begin{wrapfigure}{r}{0.5\columnwidth}
   \begin{center}
     \includegraphics [align=t,scale=0.40]{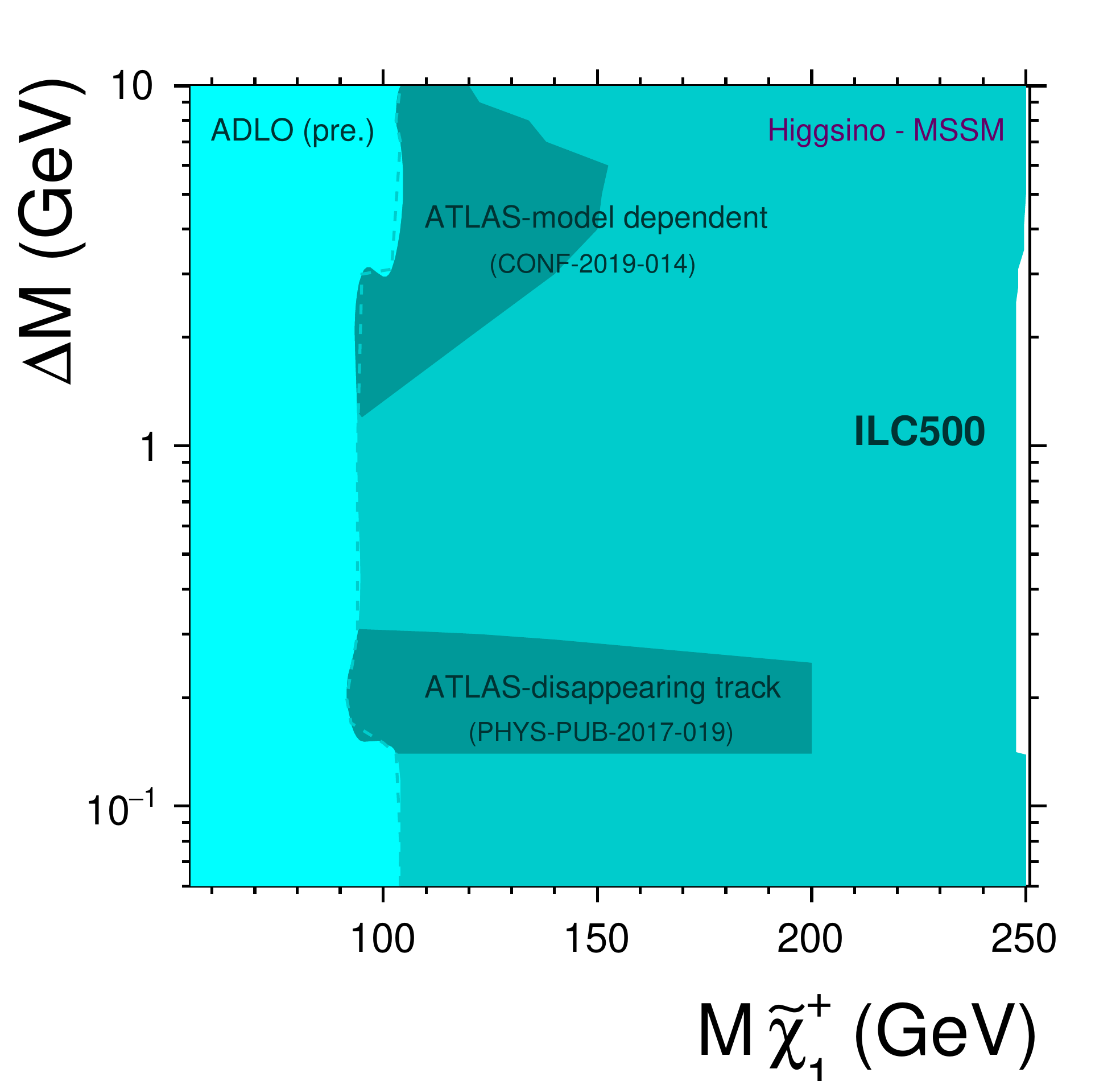}
     \includegraphics [align=t,scale=0.37]{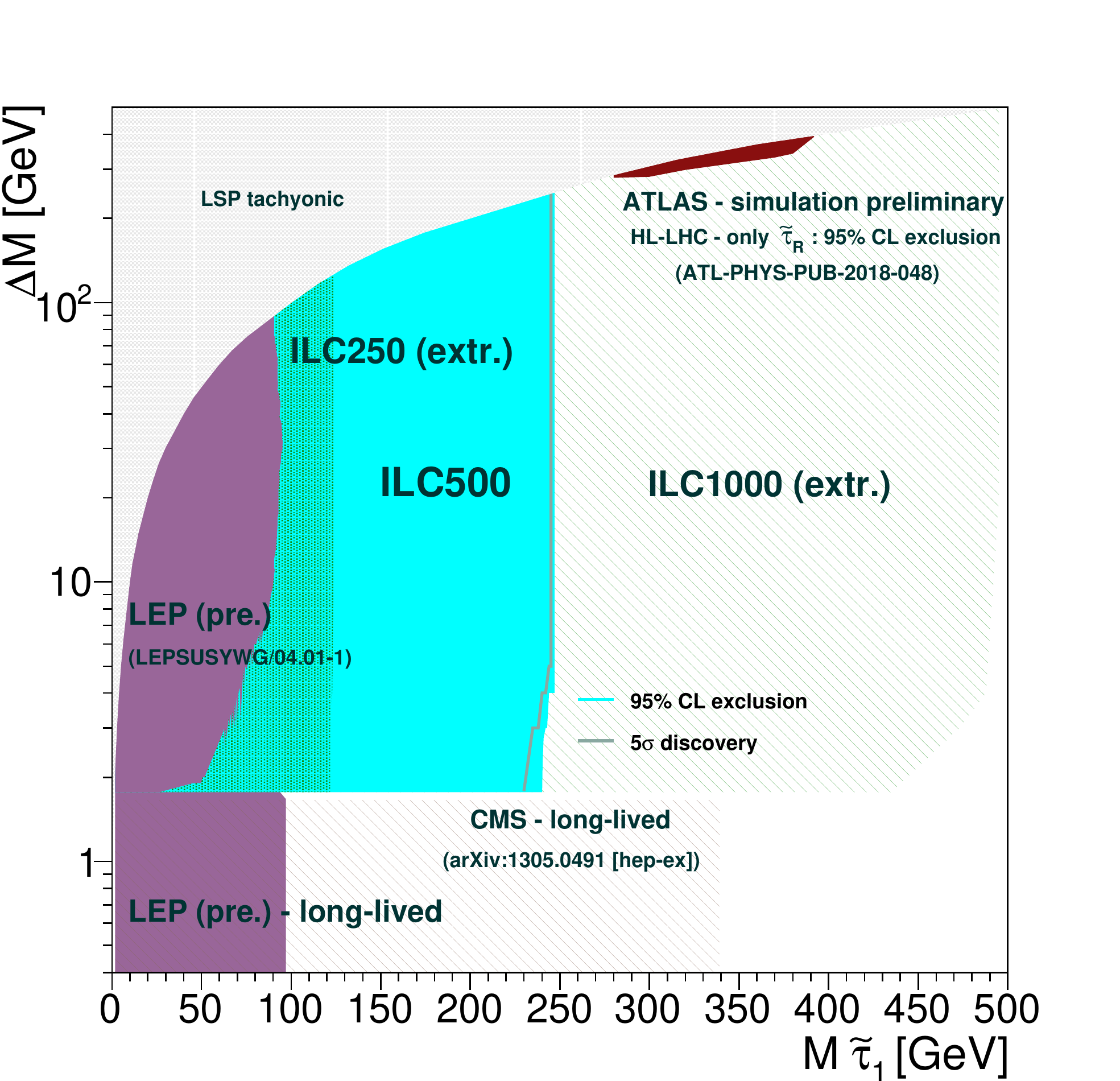}
\end{center}
\caption{ Exclusion and discovery reaches for a $\XPM{1}$ (left), or a $\stone$ (right).
  In both cases, the horizontal axis is the mass of the charged SUSY particle and the
  vertical axis is the mass splitting between this state and the (neutral) lightest SUSY
  particle.}
  \label{fig:C1stauexcl}
\end{figure} %\end{wrapfigure}

%% Many of 
These features also are relevant in exploiting the LHC's blind-spots, 
in particular, any signal stemming from processes without QCD interactions
or with only soft final states.
Here, trigger-less operation of almost fully hermetic detectors
is a great advantage.
Often, in reactions of this type,  only kinematic reconstruction of the full event
can reveal BSM physics.  These reactions 
can be studied powerfully at a lepton collider.

This section will mainly review studies of specific models of new physics.  However,  it is critical
 that a model-focused search program of future colliders be complemented by  model-agnostic strategies.  Machine learning tools are becoming increasingly powerful in searching for event classes that differ from those generated by the SM, efficiently exploring representations of the data in  high-dimensional feature spaces.   A variety of new methods have been
 proposed to  carry out anomaly detection beyond the standard classification approach (see, for example~\cite{Karagiorgi:2021ngt}).  The study~\cite{Gonski:2021jek} explores multiple ways to implement such searches at $\ee$ colliders, providing sensitivity to generic new hadronic resonances via training with imperfect or missing labels. With the triggerless operation of the ILC, we must be alert for surprises, so it will be important to adapt these methods to the $\ee$ environment.
 
Many studies of searches at ILC are done in full simulation using the 
full simulation tools presented in Chapter~\ref{chap:sim}.  
This is essential for searches with difficult signals or large backgrounds, 
in particular, studies of SUSY and dark sector models with very small
visible energy.
Also for channels with small expected signals, full simulation is
important, as it is expected that systematic errors would dominate
over statistical ones in such cases.

However, this chapter also includes studies done by members of the
broader community using fast simulation resources such as  \textsc{SGV} or \textsc{Delphes},
described in \ref{sec:fastsim}.
In each case, we will make clear what level of tool was used in the
analysis.

\subsection{SUSY}

%%%%%%%%%%%%%%%%%%%%%%%%%%%%%%
   \begin{figure}[t]
     \begin{center}
 \includegraphics [scale=0.38]{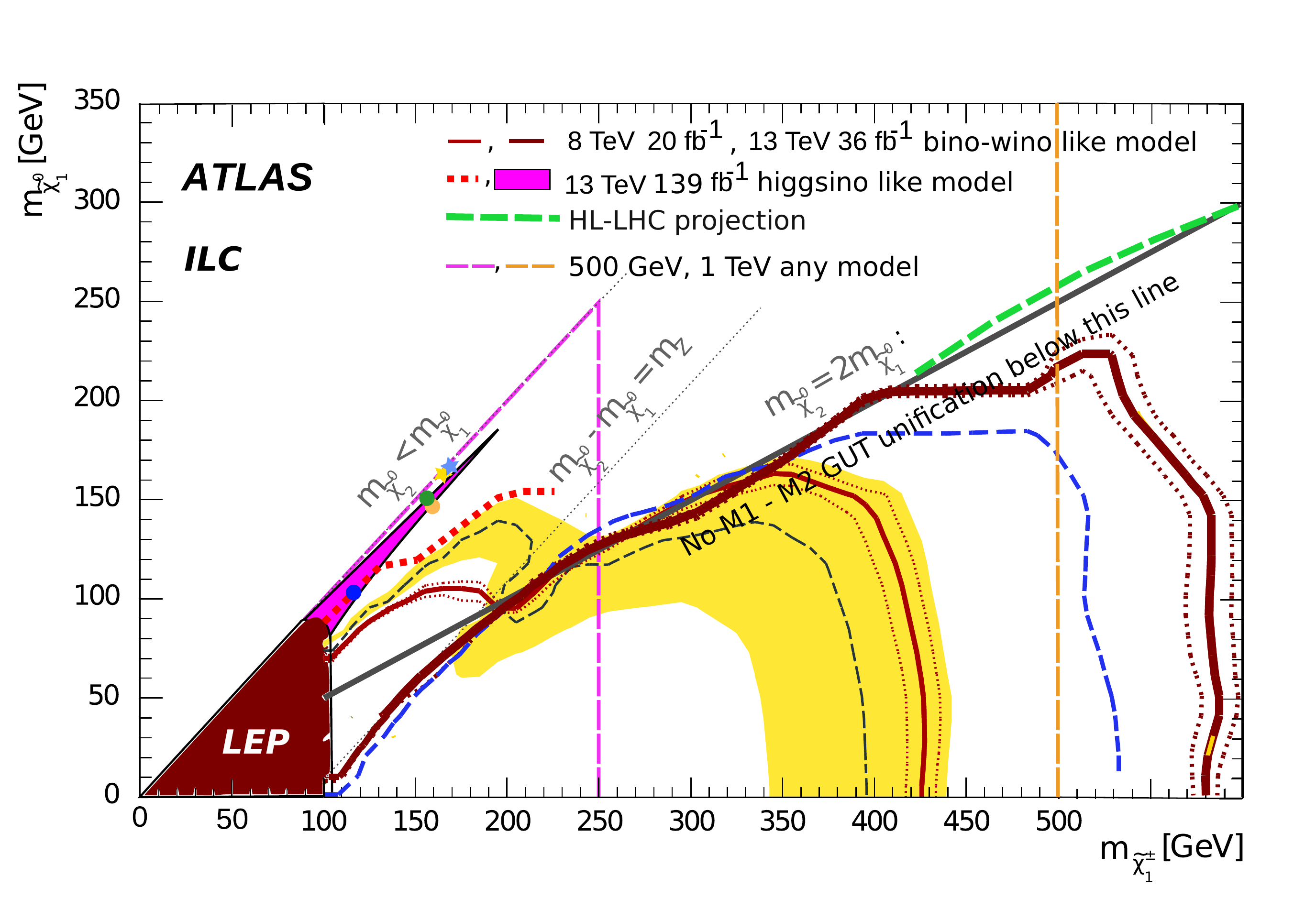}
 \end{center}
 \caption{ Observed or projected exclusion regions for a $\XPM{1}$ NLSP, for LEPII, LHC, HL-LHC.  The vertical lines indicate the model-independent reach of ILC-500 and ILC-1000.
   The symbols very close to the line $\MXN{1} = \MXC{1}$ indicate the location of the  Higgsino LSP models shown in Fig. \ref{fig:sleptC1N2}.} \label{fig:X1summary}
  \end{figure}
 %%%%%%%%%%%%%%%%%%%% 
  
  We begin our review with supersymmetry (SUSY)
%\tagged{biber}{\mcite{susy,*Wess:1974tw,*Nilles:1983ge,*Haber:1984rc,*Barbieri:1982eh},}
%\tagged{bibtex}{
\cite{Golfand:1971iw,Volkov:1973ix,Wess:1974tw,Nilles:1983ge,Haber:1984rc,Barbieri:1982eh},
%}%\tagged
  for several reasons.
   Not only is SUSY the most complete theory of BSM, it can also
   serves as a template for BSM in general, since almost any new topology can
   be obtained in some flavor of SUSY, in particular if also possible violation of
   R-parity and/or CP-symmetry,
   or non-minimal models
   are considered.
   In addition, it is the paradigm  that has been most studied
   with detailed detector simulation. In most cases, studies were done with
   full simulation with all SM backgrounds and all
      beam-induced backgrounds included.
    It is true that SUSY is under some stress by recent
    LHC results. However, ILC offers different angles to explore the properties
    of SUSY compared to LHC, with  loop-hole free searches and
    complete coverage of compressed spectra.  The missing corners of the SUSY model space
   have specific interest, and they can be covered by ILC.

  General principles such as naturalness, the hierarchy problem, and  the nature of dark matter (DM)
 still point to a light electroweak sector of SUSY.   In addition, experimental anomalies such as the 
   observed value of the magnetic moment of the muon~\cite{Muong-2:2021ojo} call for 
  a light electroweak sector of SUSY~\cite{Bagnaschi:2017tru,Chakraborti:2020vjp}.
  Except for the third generation squarks, the colored sector, 
   where pp machines excel,
  does not provide any insight into any of these issues.
\begin{figure}[p] %\begin{wrapfigure}{r}{\columnwidth}
   \begin{center}
     \includegraphics [scale=0.25]{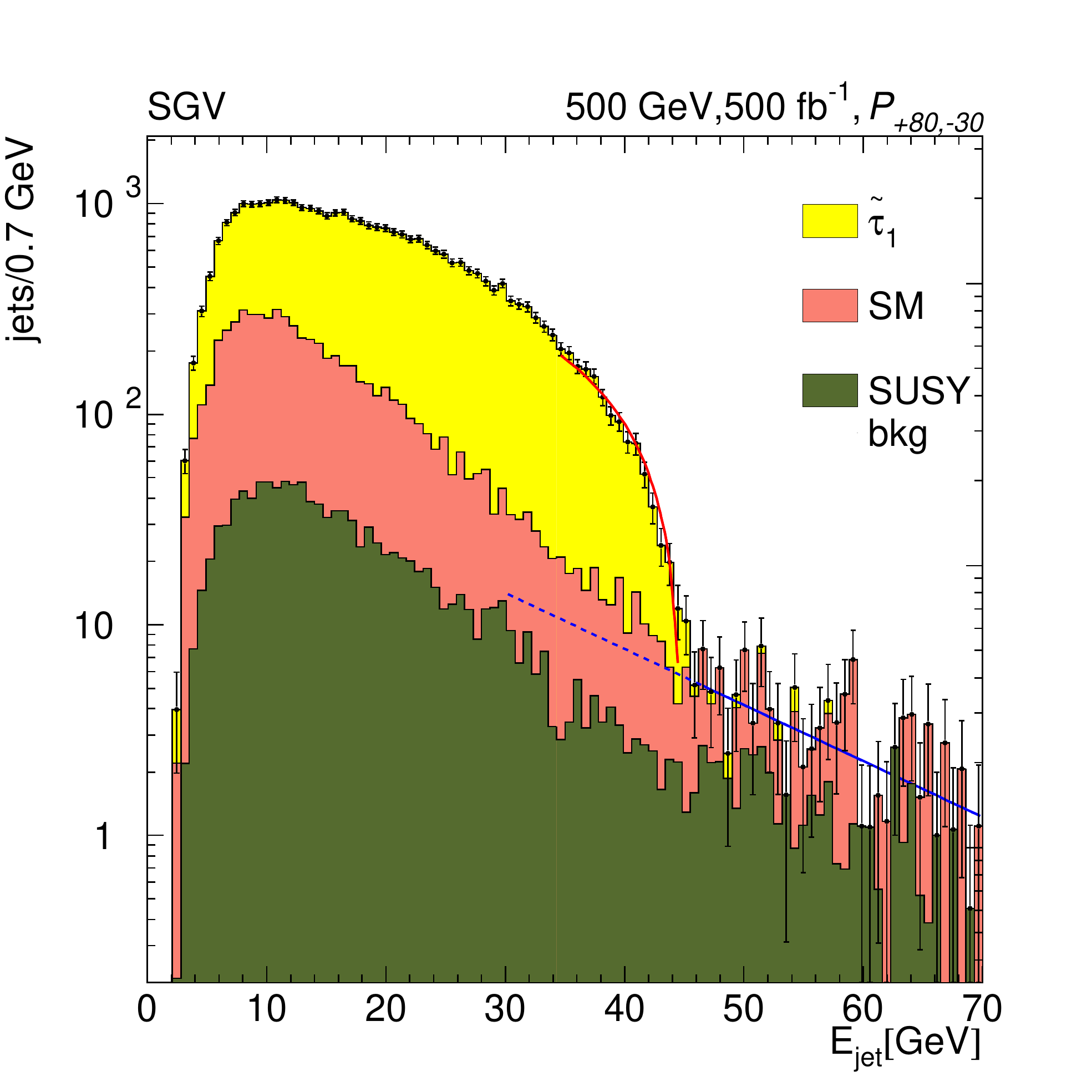}
     \includegraphics [scale=0.25]{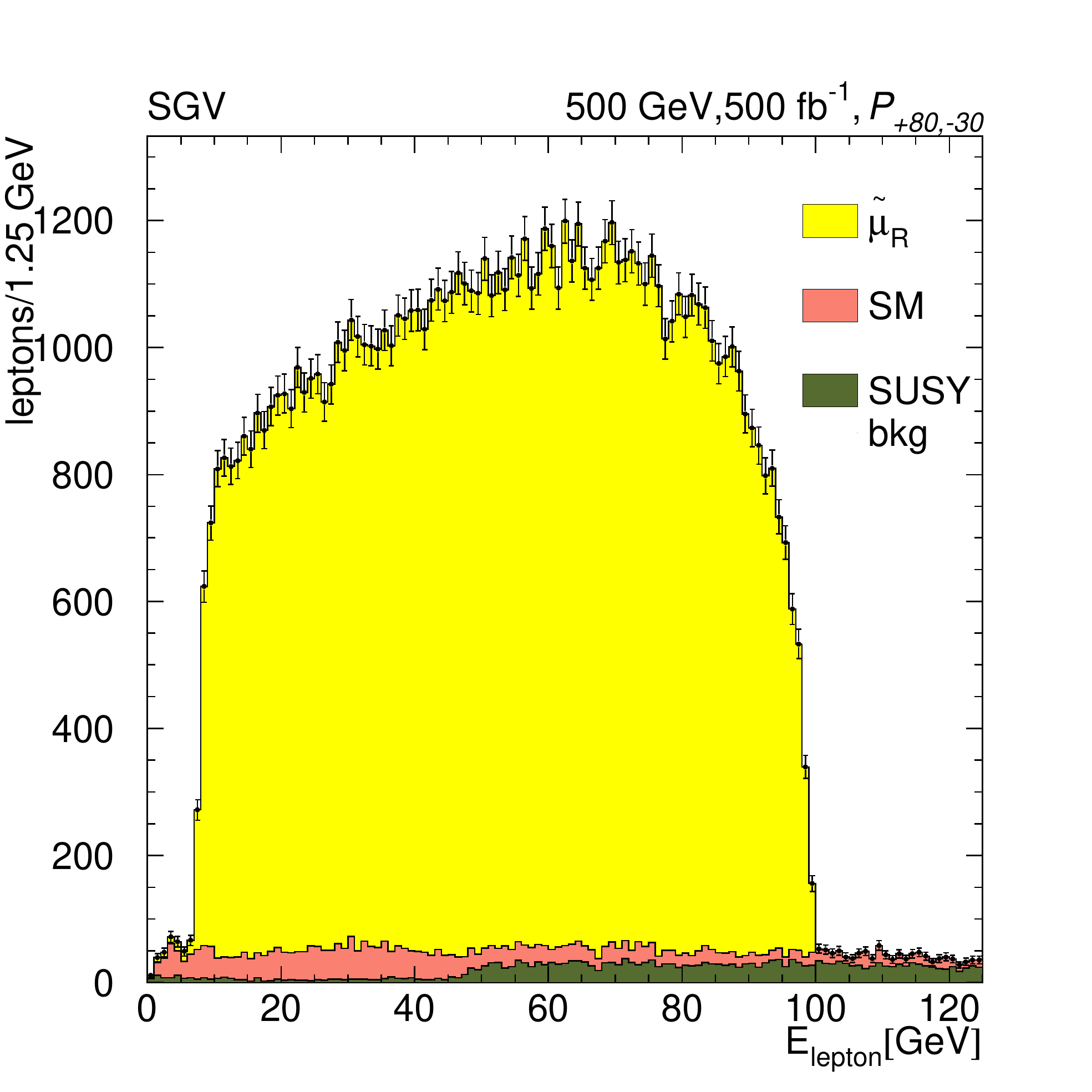}
     \includegraphics [scale=0.25]{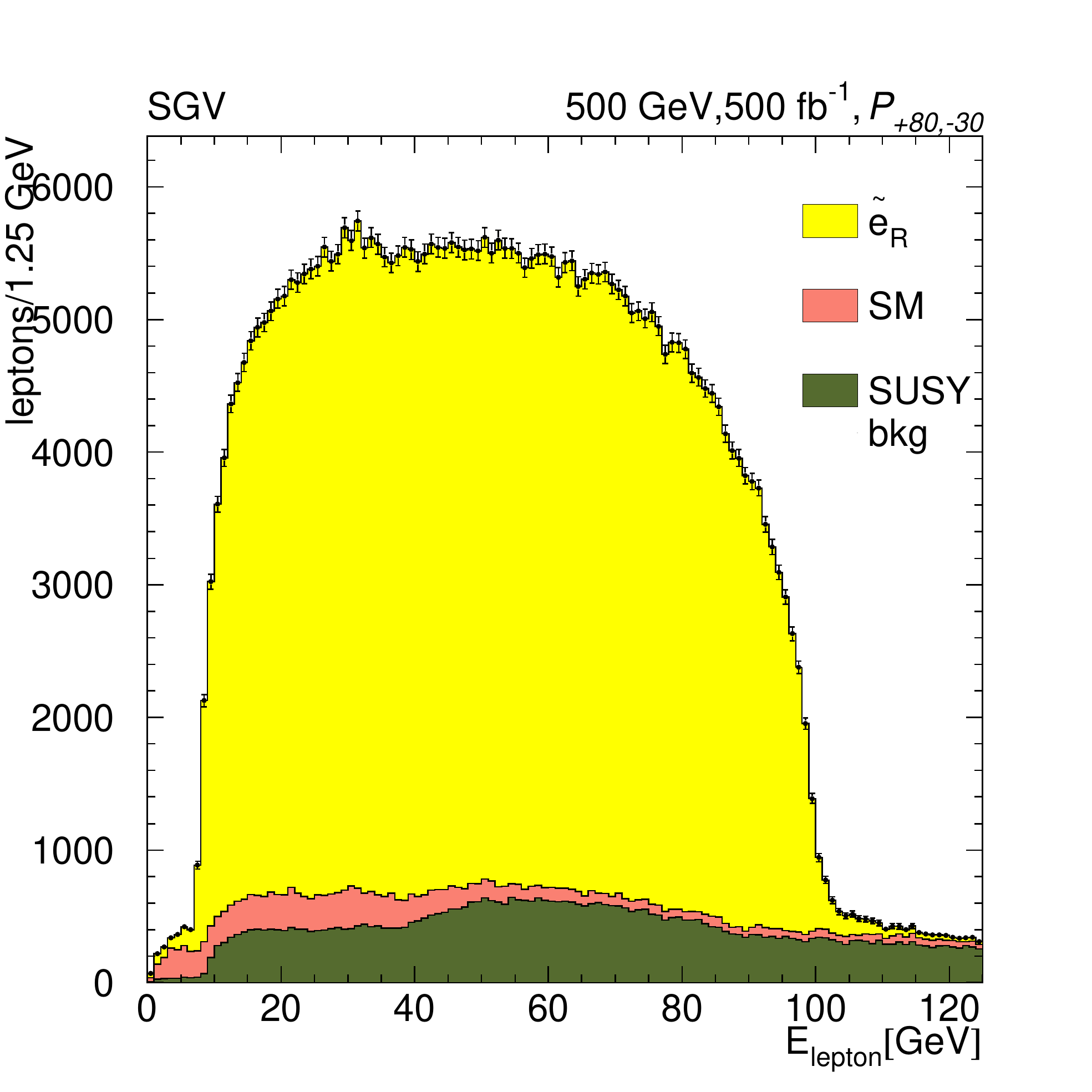}
     
      \includegraphics [align=c,scale=0.23]{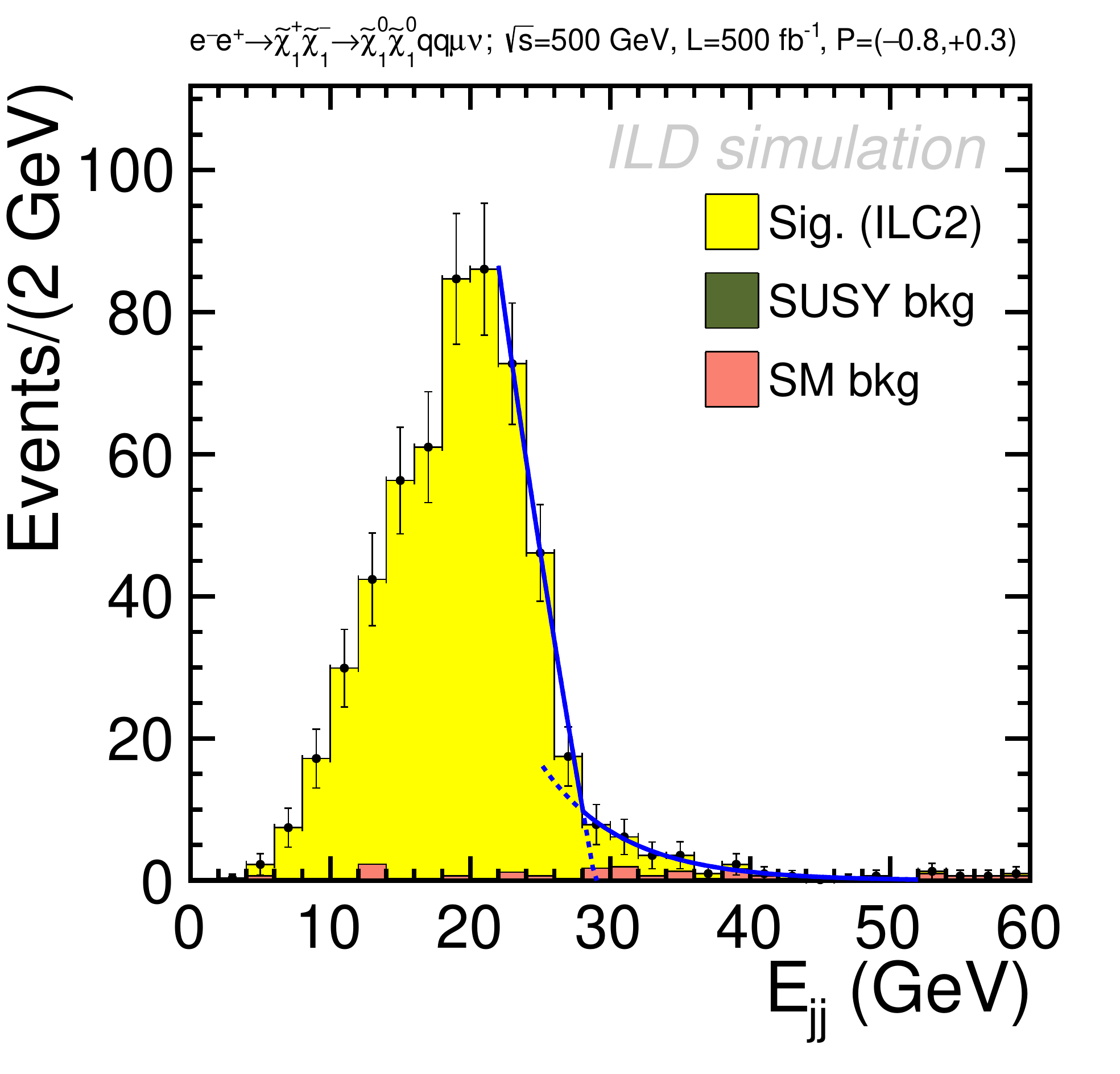}
      \includegraphics [align=c,scale=0.23]{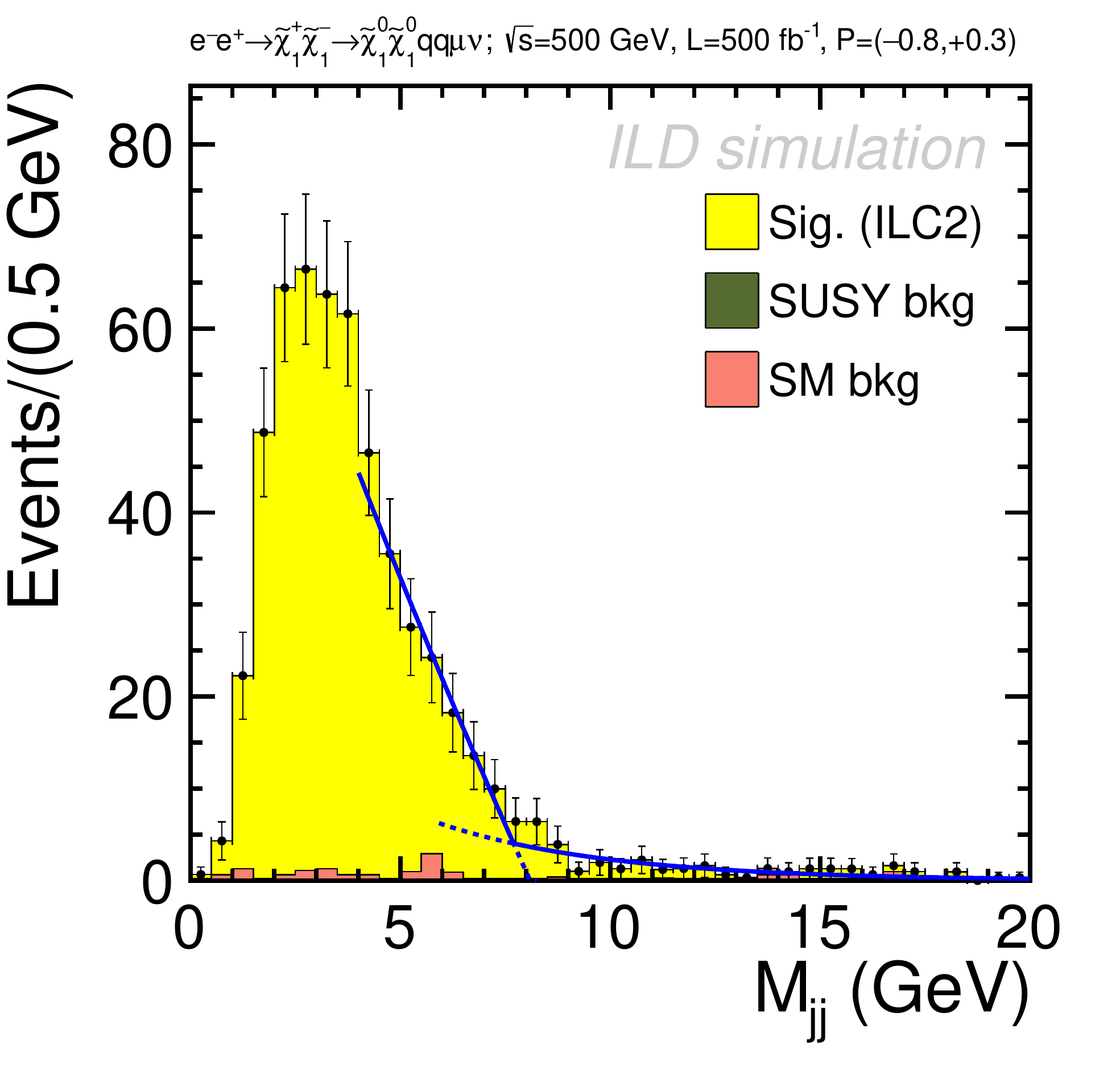}
      \includegraphics [align=c,scale=0.28]{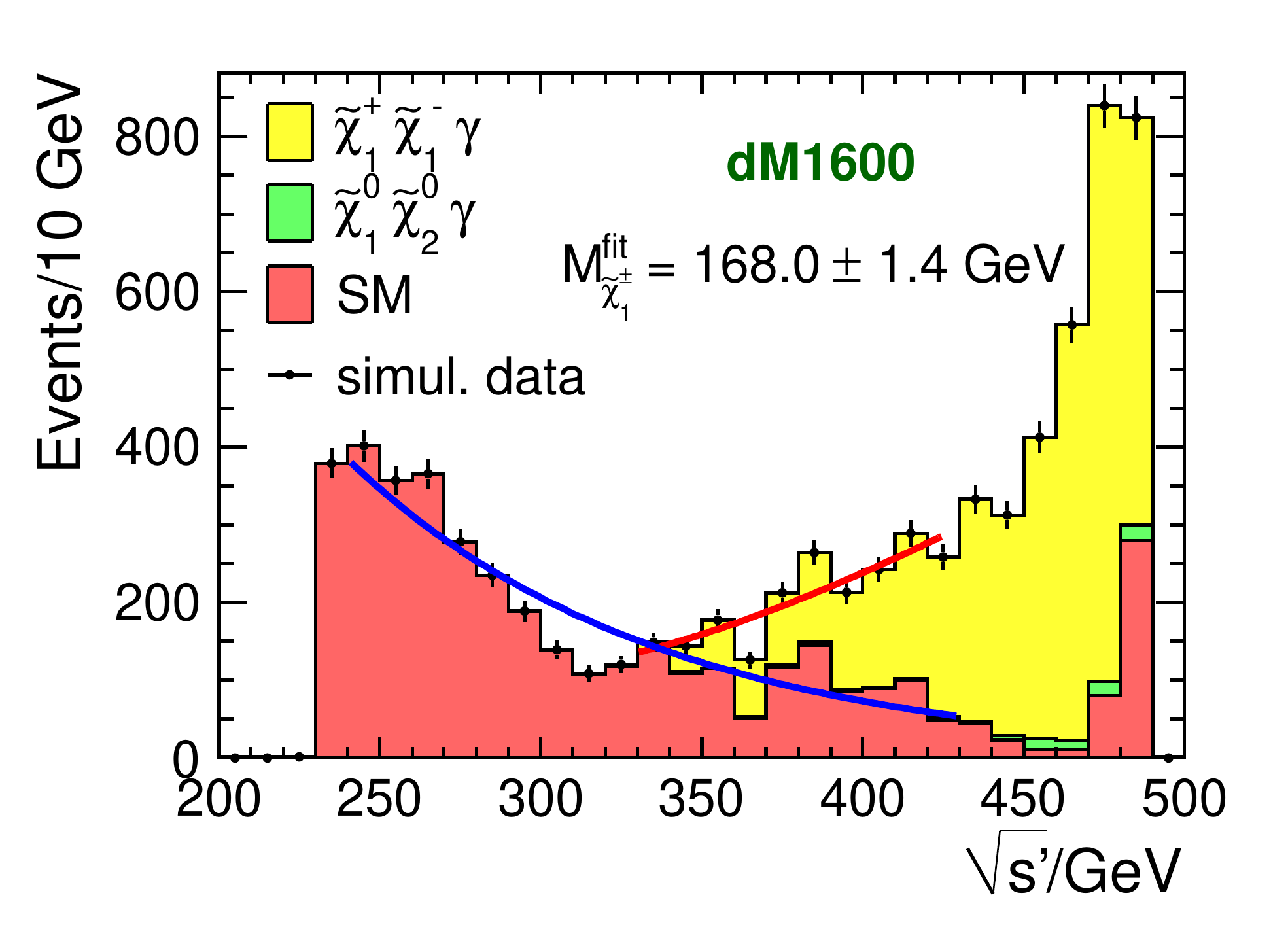}
      
      \includegraphics [align=c,scale=0.23]{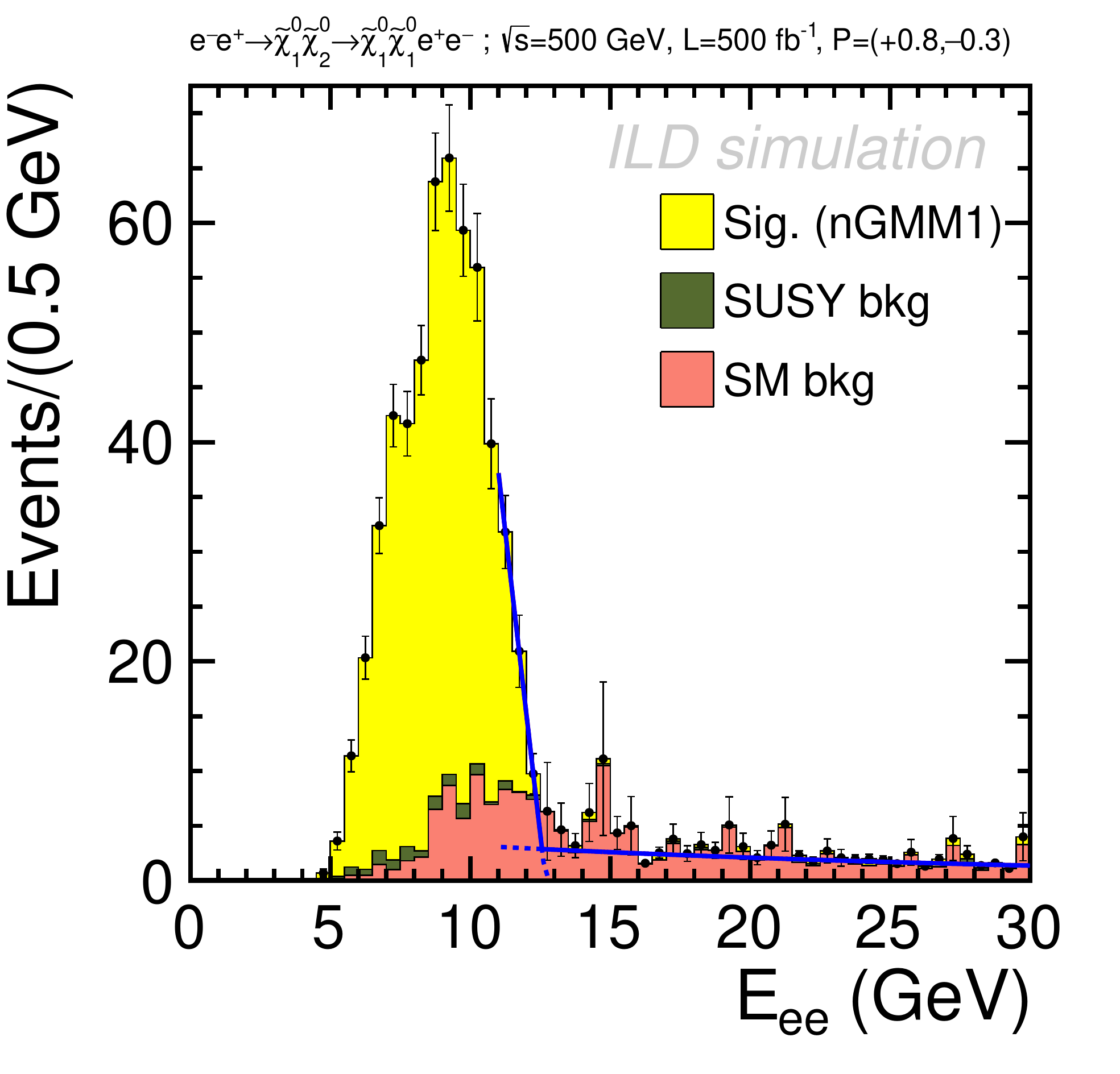}
      \includegraphics [align=c,scale=0.23]{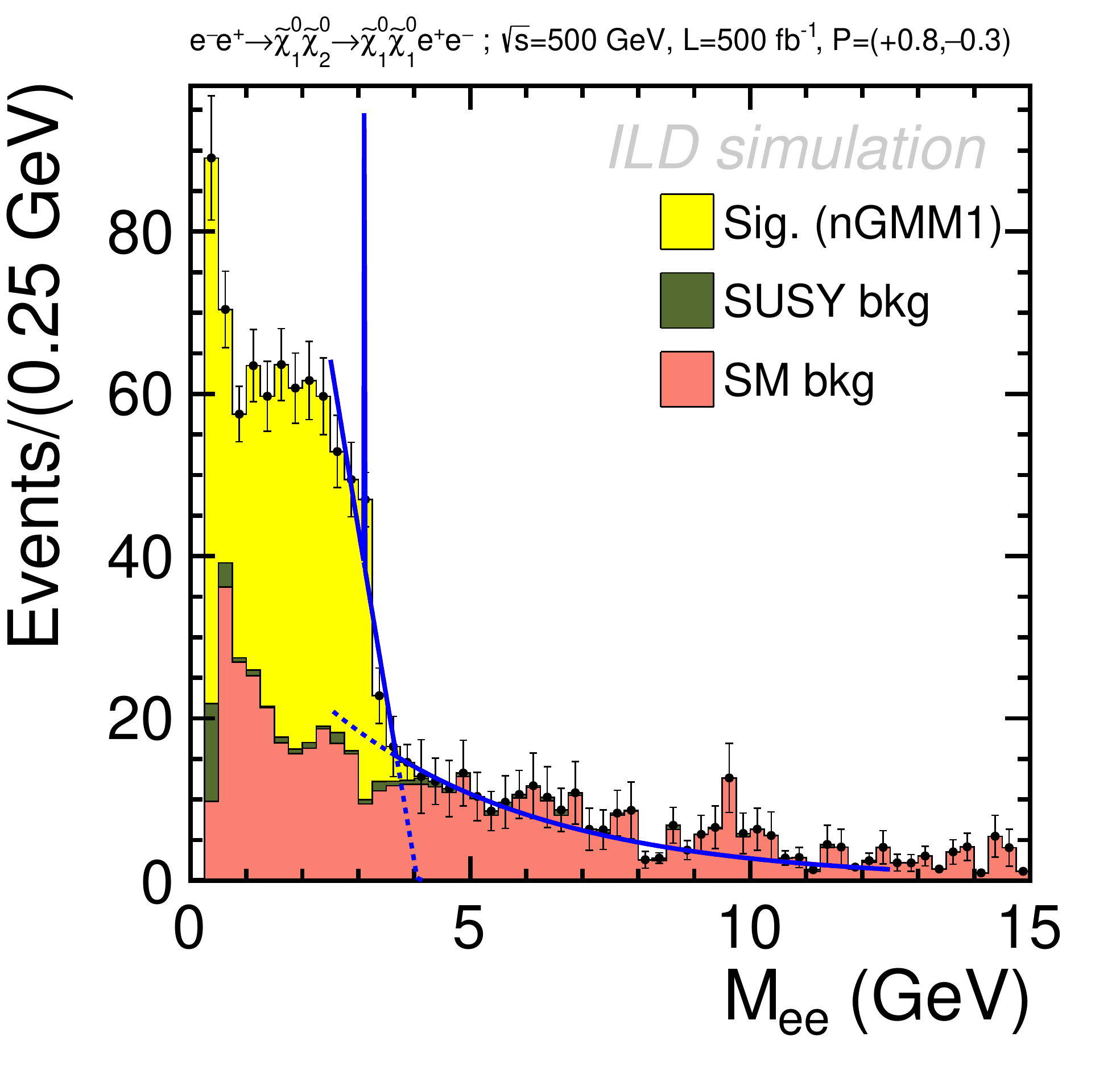}
      \includegraphics [align=c,scale=0.28]{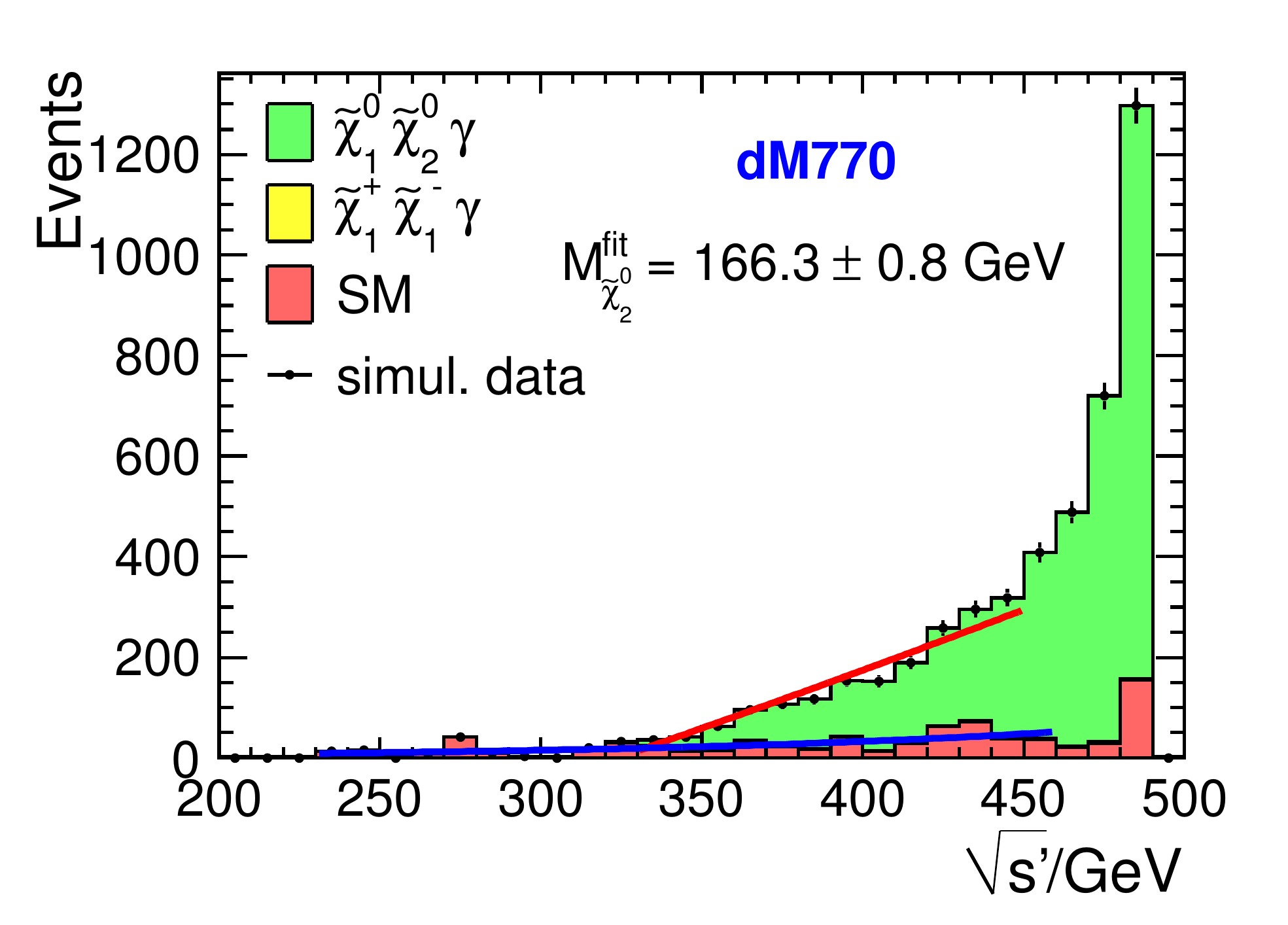}
    \end{center}
    \caption{Top row: $\stau$, $\smu$ and $\sel$ spectra. Middle and bottom rows: Observables
      for three different Higgsino-LSP models. The middle row shows the case of $\XPM{1}$ production, the bottom one 
      that of $\XN{2}$ production.  \label{fig:sleptC1N2}}
\end{figure}  %\end{wrapfigure}

 Particular attention, under the name ``Natural SUSY''~\cite{Baer:2006rs},
 has been given to the relation for the Higgsino parameters 
   \beq
    m_Z^2\;=\;2\frac{m_{H_u}^2
      \tan^2\beta-m_{H_d}^2}{1-\tan^2\beta}-2\,|\mu|^2 \ . 
\eeqn
   This implies that 
   that requiring low fine-tuning leads to
   the condition that the Higgsino mass-parameter 
   $\mu$ must be  $\mathcal{O}(m_Z)$, so that, whatever are the masses of 
 the colored SUSY particles, the Higgsino sector must be close to the weak scale.

   There are also theoretical arguments for small mass gaps in the electroweak SUSY 
spectrum. 
   If the lightest SUSY Particle (the LSP) is Higgsino or Wino, there must be other
   bosinos close in mass to the LSP, since the $\tilde{H}$ and $\tilde{W}$
   fields have several components with parametrically small splittings in mass.
   Only a Bino-LSP can have a large mass difference, $\Delta(M)$, between the LSP and the
   next to lightest
   SUSY Particle (the NLSP).
   However, in the case of a Bino LSP,
    an overabundance of DM  is expected~\cite{Roy:2007ay},
  and to avoid such a situation,
  a balance between early universe LSP production and
    annihilation is needed.   A method for enhancing the  annihilation of SUSY particles that is ready at 
hand  is 
 $\stau$~co-annihilation, and 
  for this process to operate,  the masses of $\stau$ and $\XN{1}$ should
  be within a few GeV of one another.

   In the case of such compressed, low $\Delta(M)$, spectra, most sparticle-decays are
   via cascades,
in which  the last decay in the cascade---that to SM particles and the LSP--features
     small $\Delta(M)$.
   For such decays, current LHC limits have many qualifications, 
   and only the limits from LEP 2
%\tagged{biber}{\mcite{lepsusywg,aleph,*Heister:2001nk,*Heister:2003zk,*Heister:2002mn,Abdallah:2003xe,Achard:2003ge,Abbiendi:2003ji}}
%\tagged{bibtex}{
\cite{LEPSUSYWG/01-03.1,Heister:2001nk,Heister:2003zk,Heister:2002mn,Abdallah:2003xe,Achard:2003ge,Abbiendi:2003ji}
%}%\tagged
   are model-independent.

 At ILC,  one can perform a loophole-free search for SUSY
because, in SUSY, the properties of the NLSP
production and decay are completely predicted
given the  LSP and NLSP masses, due to the 
  SUSY-principle that the couplings of particles and sparticles are related by symmetry.
Note that this does not depend on the (model dependent) SUSY breaking mechanism.
In R-parity conserving models, the final stages of a cascade decay are highly constrained.
By definition, there is only one NLSP, and this particle must have a 100\% BR
  to its (on- or off-shell) SM-partner and the (stable or unstable) LSP.
  Also, there is only a handful of possible candidates to be the NLSP.
  Hence by performing searches for every possible NLSP, we can obtain
  model independent exclusions  and  discovery reaches in the $M_{NLSP} - M_{LSP}$ plane,
separately for each NLSP candidate, or globally.   In models with R-parity violation, there are different decay patterns, but these can also be classified and, typically, lead to more stringent limits in a setting where 
the full event can be reconstructed.

\begin{figure}[t]
  \begin{center}
    \subcaptionbox{}{\includegraphics [align=c,scale=0.30]{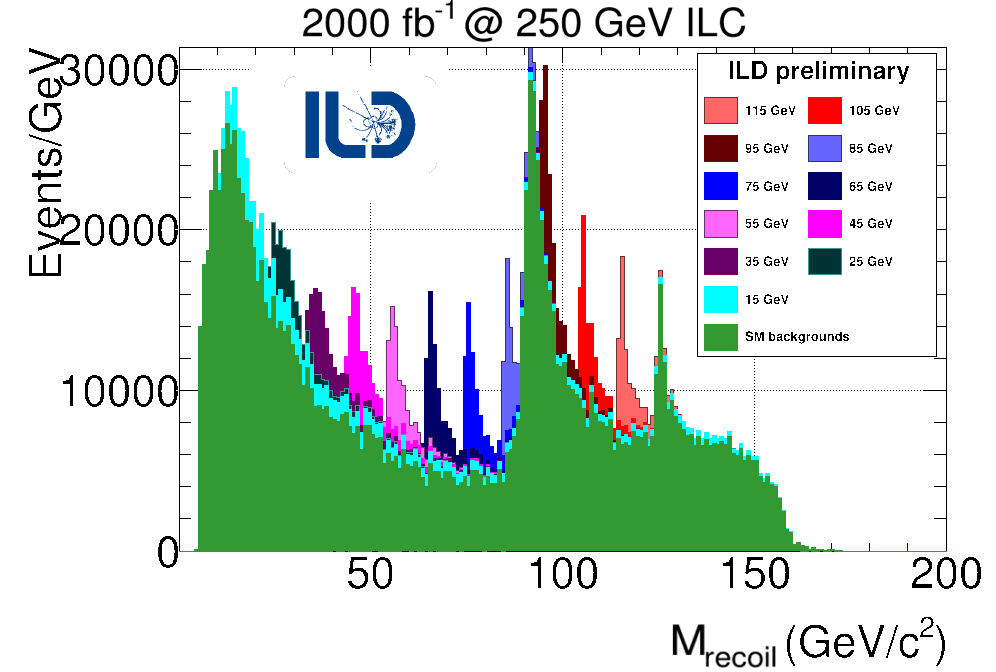}}
    \subcaptionbox{}{\includegraphics [align=c,scale=0.35]{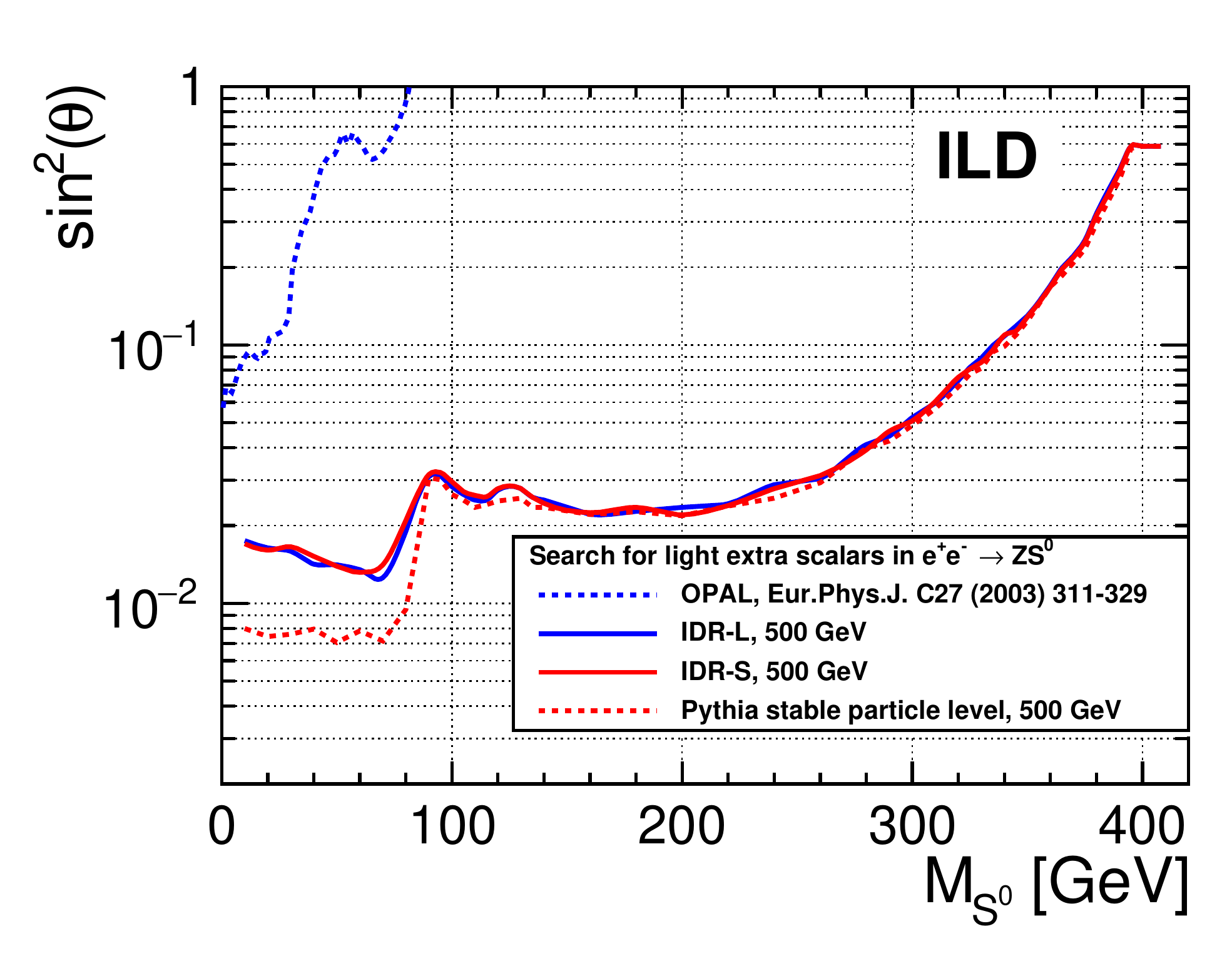}}
\end{center}
\caption{
  (a) Recoil mass distributions for several new scalars and the SM background.
  (b) Projected exclusion limit for
  new scalars, in terms of the coupling compared to the coupling an SM Higgs at the same mass would have.
\label{fig:otherbsm}}
 \end{figure}

Examples of this procedure are shown in Fig. \ref{fig:C1stauexcl} for the cases of a $\XPM{1}$~\cite{PardodeVera:2020zlr} or
a $\stone$~\cite{NunezPardodeVera:2021cdw,NunezPardodeVera:2022izz} NLSP.
The $\XPM{1}$ is a conservative extrapolation from the LEP results, while the
$\stone$ one is obtained with {\it full} detector simulation,
%of ILD, 
in which the $\stau$
and LSP properties were chosen such that the limit is the weakest possible one,
i.e. the experimentally ``worst possible'' case.
In the figure, it can be seen that the discovery and exclusion reaches are almost
the same, and the reach is quite close to the kinematic limit $2 M_{NLSP} = E_{CM}$.
It should also be noted that the HL-LHC projection from ATLAS is exclusion only,
and is for specific assumptions on the $\stau$ properties, assumptions that are
not the most pessimistic. In Fig.~\ref{fig:X1summary}, the various current or projected limits are
shown in a single plot
%\tagged{biber}{\mcite{PardodeVera:2020zlr,ATLAS:2018ojr,ATLAS:2019lng,ATLAS:2021moa,ATL-PHYS-PUB-2018-048,LEPSUSYWG/02-04.1}.}
%\tagged{bibtex}{
\cite{PardodeVera:2020zlr,ATLAS:2018ojr,ATLAS:2019lng,ATLAS:2021moa,ATL-PHYS-PUB-2018-048,LEPSUSYWG/02-04.1}.
%}%\tagged
It should be noted that below the heavy black line, GUT unification of the
Bino and Wino mass-parameters  $M_1$ and $M_2$
is not possible: The difference between $\MXN{1}$ and $\MXC{1}$ cannot be larger than what the line indicates,
if such a unification is realized.

In fact, at the ILC, SUSY discovery would happen at quite low levels of integrated
luminosity.  Either the
process is not in reach and there is no sign of it, or it will be discovered straightforwardly.
  This means that studies of SUSY at ILC would almost directly enter into the
realm of precision measurements. The plots in Fig. \ref{fig:sleptC1N2} show a number of examples of the
kind of signals that would be expected, as seen in
detector simulation studies.   The top row shows
slepton signals ($\stau$, $\smu$ and $\sel$)
 in a  $\stau$ co-annihilation model~\cite{Berggren:2015qua}.  The following rows show
       typical chargino and neutralino signals in different Higgsino LSP model.
       The left-hand two plots are models  with moderate (a few to some tens of GeV) 
$\Delta{M}$~\cite{Baer:2019gvu},
       while the right-hand plots are  for a  model with very low (sub-GeV)
 $\Delta{M}$~\cite{Berggren:2013vfa}.
      In all of the illustrated cases, it was found that the
 SUSY masses could be determined at the sub-percent level,
 the polarized production cross-sections to the level of a few percent.
 Many other properties could also be obtained from the same data, such as
 decay branching fractions, mixing angles, and sparticle spins.

\subsection{New scalars}
\label{sec:newscalars}

\begin{figure}[t]
\hspace{0.03\textwidth}
  \includegraphics[width=0.47\textwidth]{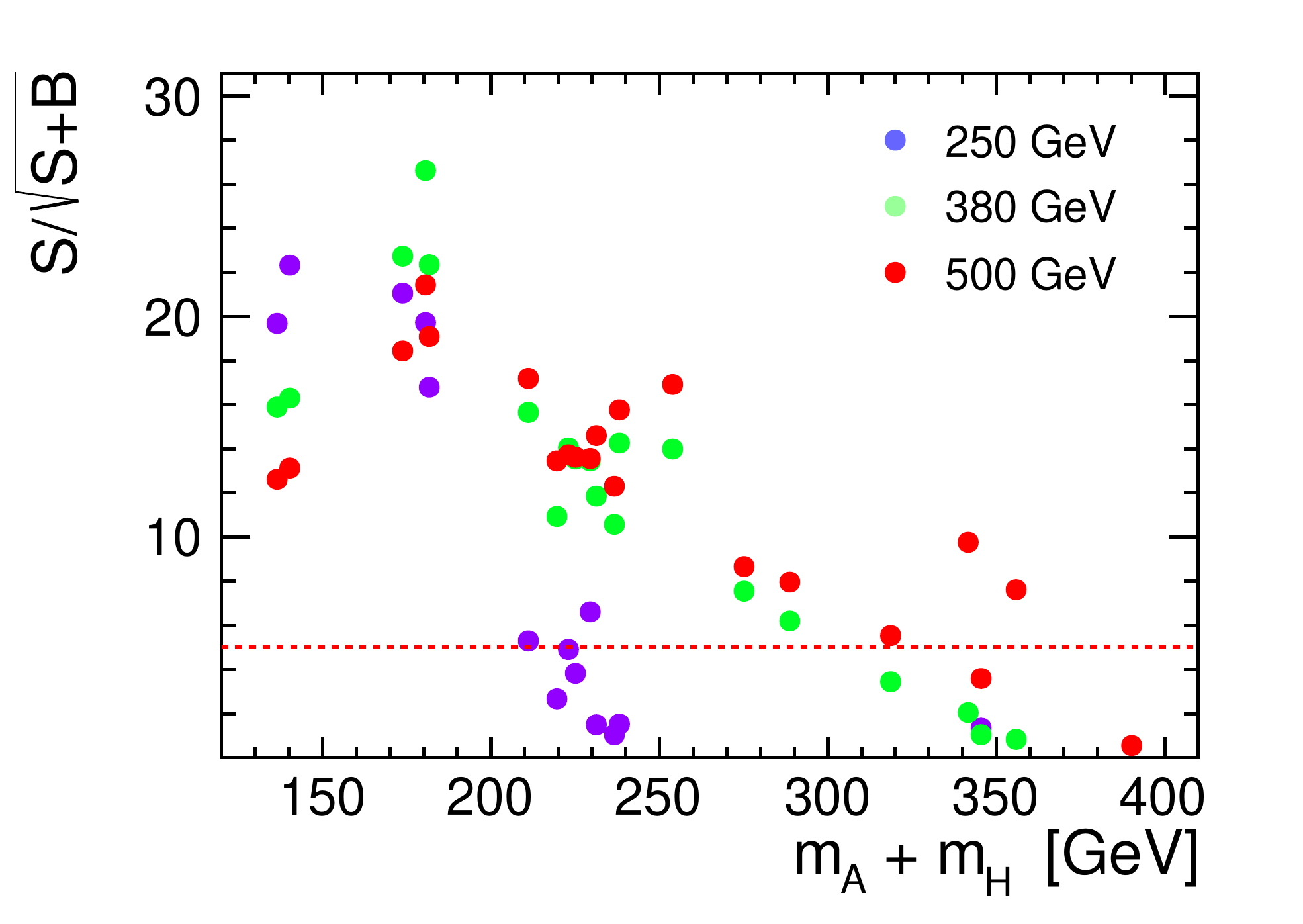}
  \includegraphics[width=0.47\textwidth]{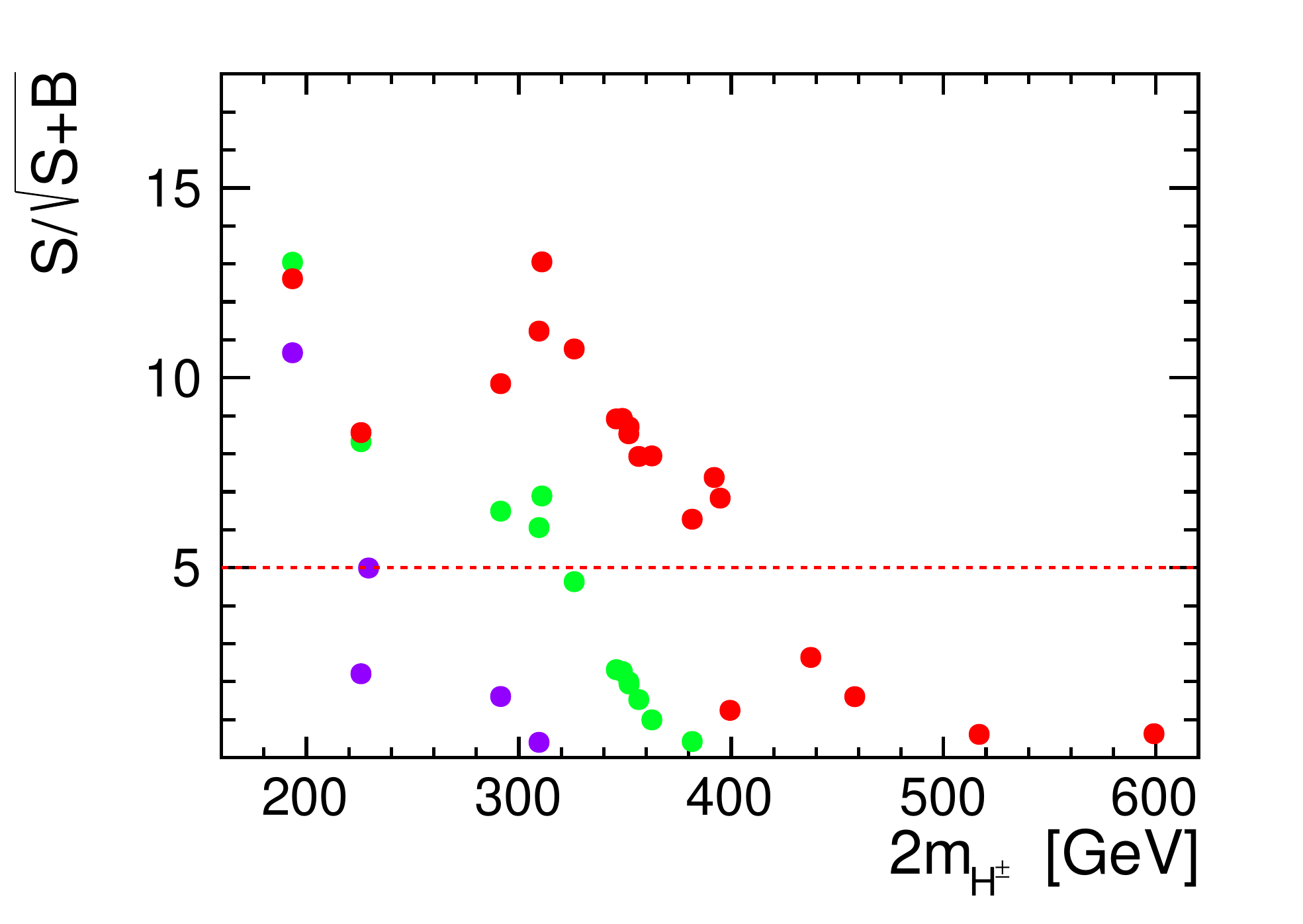}
\caption{Significance of the deviations from the Standard Model
  predictions for the extended Higgs bosons of the Inert Doublet Model, as expected for 1~ab$^{-1}$ of data collected at
  center-of-mass energy of 250\,GeV, 380\,GeV and   500\,GeV, from~\cite{Zarnecki:2019poj,Sokolowska:2019xhe}.  Left:  for 
events with two muons in the final state ($\mu^+\mu^-$), 
  as a function of the sum of neutral inert scalar masses; 
    Right: for events with an electron and a muon in the final state
  ($e^+\mu^-$ or $e^-\mu^+$) as a function of twice the charged scalar
  mass.  }
\label{fig:IDMleptonsig}
\end{figure}

Many BSM models predict the existence of a  new Higgs-like scalar ($S$), produced in $\eeto Z^* $ $\rightarrow Z S$, with the decay pattern of $S$ different from that of the Higgs boson.  
Such a state could have escaped detection at LEP if its production cross-section is much lower than that of a SM Higgs at the same mass.
Hence, a search for such a state should be done at all accessible masses, and without any assumption on the decay modes.
At an $e^+e^-$ collider this can be done using the recoil-mass, \ie,  the mass of the system recoiling against the measured $Z$.
%% In Fig. \ref{fig:otherbsm}(a), (from \cite{Wang:2020lkq})
%% we show how such a state would manifest itself, if its  coupling would be equal to the an SM-Higgs at the same mass.
%The analysis done  \cite{Wang:2020lkq} used full detector simulation of the ILD concept, and
In~\cite{Wang:2020lkq}, a full 
%ILD 
detector simulation study was performed, and it was found
that couplings down to a few percent of the  SM-Higgs equivalent can be excluded;
  see Fig.~\ref{fig:otherbsm}. %  (a).   % this is the only reference to this figure

In models with much smaller couplings of extended Higgs bosons to the $Z$ boson, the ILC can search for the pair production of the new bosons.   The primary reactions are  $\ee \to AH$, where $H$ is a heavy neutral CP-even boson and $A$ is a heavy neutral CP-odd boson, and $\ee\to H^+H^-$.   If the new Higgs bosons decay into final states that are completely visible in the ILC detectors, the discovery of these reactions is straightforward almost all of the way up to the production thresholds $E_{CM} = m_A + m_H$ or $E_{CM} = 2m_{H^+}$~\cite{Battaglia:2010qoc}.  However, there are more difficult cases, and these also have significant opportunity  for discovery at the ILC.   An example is the Inert Doublet Model (IDM)~\cite{Deshpande:1977rw,Cao:2007rm,Barbieri:2006dq,Ilnicka:2015jba}.
  This is a 2-Higgs doublet model with a $Z_2$ symmetry that prevents the scalars of the second doublet from coupling directly to the SM fermions.   This implies that the lightest particle in the second double will be stable and can be a candidate for the particle of dark matter.   Models with IDM scalar masses of the order of 100~GeV are still consistent with constraints from direct detection experiments, relic density of dark matter, as well as with all collider and low-energy limits. In this model, the final neutral boson (it is assumed to be $H$) is invisible. The visible signals are leptons emitted in the decay of the $A$ or $H^\pm$ to the $H$ through off-shell or on-shell $W$ or $Z$ exchange (depending on the scalar mass difference).  The phenomenology of this model has been studied extensively in \cite{Zarnecki:2019poj,Sokolowska:2019xhe,Kalinowski:2018kdn,Klamka:2022ukx,Kalinowski:2022fot}. Working with the benchmark scenarios proposed in \cite{Kalinowski:2018ylg,Kalinowski:2020rmb}, this study investigated the significance of  signatures with two muons or an electron and a muon in the final state.  The results are presented in Fig.~\ref{fig:IDMleptonsig}. For the integrated luminosity of 1~ab$^{-1}$, the expected discovery reach of 500~GeV ILC extends up to the sum of neutral scalar masses of 330~GeV and up to charged scalar masses of 200~GeV.   The ILC capabilities for models with light, weakly coupled scalar bosons are reviewed in~\cite{Robens:2022erq,Robens:2022zgk}.

 Another feature that can appear in models with extended scalar sectors that does not arise in SUSY is the possibility of scalars or fermions is higher representations of electroweak $SU(2)$, for example, 
 $I = \frac{3}{2}$ or 2.  In~\cite{Kumar:2021umc}, the phenomenology of a model of this type motivated by the muon $g-2$ anomaly is studied at $\ee$  colliders.

\subsection{WIMP dark matter}
\label{sec:WIMPDM}
\begin{figure}
%\setlength{\unitlength}{1.0cm}
%\begin{center}
\subcaptionbox{}{\includegraphics[align=c,width=0.54\linewidth]{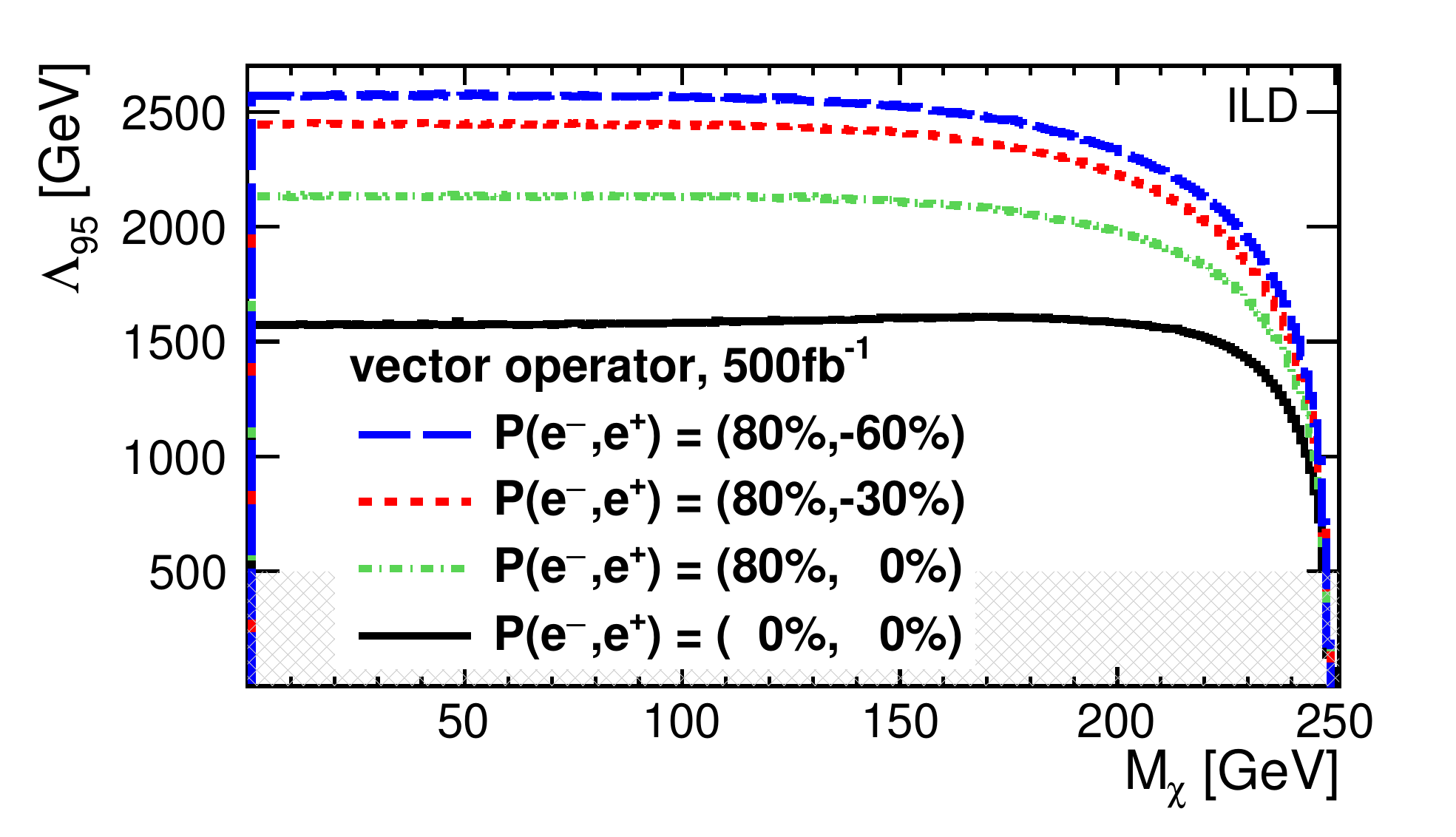}}
%\hspace{0.05cm}
%\subcaptionbox{}{\includegraphics[align=c,width=0.3\linewidth]{chapters/ILC500/figures/extrapolation_vector_pm}}
%%x \subcaptionbox{}{\includegraphics[align=c,width=0.36\linewidth]{chapters/ILC500/figures/dm_light_med_ilc_fig11a}}
\subcaptionbox{}{\includegraphics[align=c,width=0.36\linewidth]{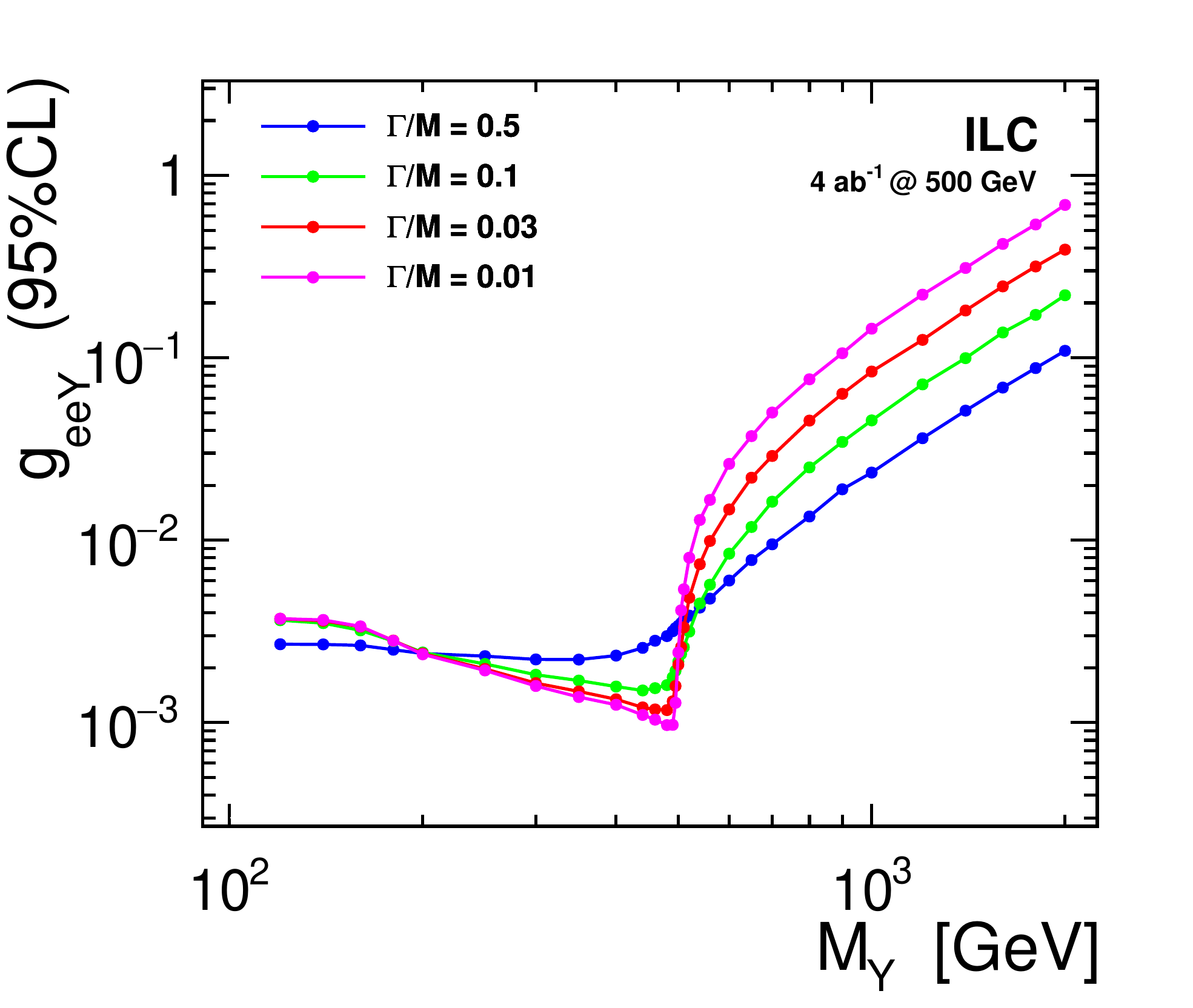}}
%\end{center}
\caption{\label{fig:searches_WIMPs} (a):
Observational reach (2$\sigma$) of the ILC for a fermionic WIMP with a WIMP-fermion vector coupling in terms of the WIMP mass for four different beam-polarization configurations~\cite{Habermehl:417605}.
%
%  Observational reach ($3\sigma$) of the ILC for a spin-1 
%  WIMP in terms of the  WIMP
%  mass and $\kappa_e$ for three different chiralities of the WIMP-fermion couplings~\cite{Habermehl:417605}.
 %  (b): Expected sensitivity for a vector operator in an EFT-based interpretation as a function of integrated
%   luminosity and centre-of-mass energy~\cite{Habermehl:417605}.
%%   x   (b): Expected excludable cross-section as a function of the mass of the mediator, for different mediator widths.
  (b): Expected limits on the vector mediator coupling to electrons for the
    ILC running at 500\,GeV and different fractional mediator widths, 
    as a function of the mediator mass \cite{Kalinowski:2021tyr}.
}
\end{figure}

The primary probe at the ILC for the direct production of {\it WIMP dark matter} are photons
emitted as initial-state radiation in association with the pair production of invisible dark matter particles.
Such a Mono-photon search is analogous  to Mono-$X$ searches at the LHC.
The main backgrounds to this search are the radiative neutrino production
and the radiative Bhabha scattering process, in which the outgoing electron and positron escape 
undetected in the beam pipe.   The neutrino production is irreducible,  but its dominant contribution, from 
$t$-channel $W$ exchange, is present only from the $e^-_Le^+_R$ initial state.
At LEP, searches for photon events with missing energy 
were performed~\cite{Abdallah:2003np,Abdallah:2008aa},
%}%\tagged
and were later re-analyzed within the  effective
operator framework.~\cite{Fox:2011fx}.\footnote{Note that under LEP or ILC conditions the 
effective field theory approximation is accurate, while this assumption is questionable
in similar analyses at hadron colliders.}

The prospects to detect WIMPs with such methods at the ILC and to determine their properties 
have been studied
for a center-of-mass energy of $500$\,GeV
in detailed detector simulation~\cite{Bartels:2012ex,Habermehl:417605}. 
Also at the ILC, the experimental sensitivity has been interpreted 
in the framework of effective
operators.
Figure~\ref{fig:searches_WIMPs}(a) shows the exclusion reach found at 500~GeV in the effective operator approach~\cite{Habermehl:417605}.
The importance of beam polarization in achieving these results has 
already been discussed in Sec.~\ref{sec:polarization}.
%, and 
%Figure~\ref{fig:searches_WIMPs}b shows the extrapolation of these
%results to a wide range of integrated luminosities and centre-of-mass energies~\cite{Habermehl:417605}.
For the full $500$\,GeV-program of the ILC, scales of new physics ($\Lambda$) 
of up to $3$\,TeV  can be probed,
while the $1$\,TeV-energy-upgrade of the ILC would extend this even 
to $4.5$\,TeV or more, 
depending on the integrated luminosity.
At 250 GeV, 
the full reach will be attained already at a modest integrated luminosity.

The EFT approach is only valid if the considered new physics mass scale (mediator mass) is much higher
than the collision energy. 
This is the case when we assume the mediator coupling to SM particles is large.
However, scenarios with light mediator exchange are still not excluded by the existing data for couplings
of the order of 0.01 and below.
Assuming that the total mediator width is dominated by the DM partial width, cross section limits
extracted from the analysis of mono-photon event spectra, calculated for fixed mediator mass and
width hardly depend on the DM particle type or coupling structure.
Limits on the vector mediator coupling to electrons, expected from the combined analysis of the data
taken with different beam polarizations at 500\,GeV ILC, with systematic uncertainties taken into account,
are presented in Fig.~\ref{fig:searches_WIMPs}(b) \cite{Kalinowski:2021tyr,Kalinowski:2022cnt}. 
For proper modeling of mono-photon events, dedicated simulation method was
proposed \cite{Kalinowski:2020lhp}.
The study was based on the Delphes fast simulation framework with dedicated ILC detector model~\cite{ILC-DELPHES,bib:ILCgen}, see section \ref{sec:fastsim}.

A broad parameter region of the light scalar mediator scenario is also not excluded,
with its thermal relic abundance being consistent with the observed dark matter density.
Since the mediator couples directly to the Higgs boson in a renormalizable way, another exciting
signal is expected at the ILC, the exotic Higgs decay into a pair of the mediators.
The ILC could efficiently search for the decay, especially when the mediator decays into a
pair of bottom quarks, as mentioned in section 8.2 \cite{Kato:2022}. On the other hand, the leptophilic
mediator is another interesting scenario. The mediator carries a lepton number and connects
the dark matter with SM leptons, making the dark matter leptophilic.
The scenario is also not excluded by the existing data,
with its thermal relic being consistent with the observed dark matter density.
The ILC will play a crucial role in searching for the scenario via the mediator pair
production \cite{Baum:2020gjj}, and the mono-photon (a pair production of the dark matters associated with a photon) process \cite{Horigome:2021qof} even at the 250 GeV running.

\subsection{Heavy neutrinos}

The ILC also has a role to play in testing models of neutrino mass.  It is possible to give a mass to 
neutrinos by adding to the SM a set of right-handed neutrinos with conventional Yukawa couplings.  However, 
this is  usually considered inadequate to explain the very small sizes of neutrino masses.  To solve this 
problem, new heavy particles are introduced such that,  when these are integrated out, a dimension-5 term 
called the Weinberg operator  is generated~\cite{Weinberg:1979sa},
\beq
             \Delta \L = 
      -  \alpha_{ij} { (\epsilon_{ab} H^\dagger_a \bar{L}_{bi})  (\epsilon_{a'b'}  H_{a'}^\dagger L^c_{b'j}) \over M } + h.c.
\eeq{Weinbergop}
When the Higgs field develops its vacuum expectation value, this leads to a Majorana-type mass matrix for the
light  neutrinos, with  $m_{ij} = \alpha_{ij} v^2/M$.  In most discussions, it is assumed that this operator arises
from integrating out heavy right-handed neutrinos.  With neutrino Yukawa couplings of order $10^{-3}$, the 
right-handed neutrinos would have masses of order $10^{10}$~GeV.  

However, in fact, there are three distinct
possibilities for the generation of the term \leqn{Weinbergop}~\cite{Foot:1988aq,Ma:1998dn}.  
 The case just discussed
 is the Type I  seesaw.  In the Type II seesaw, the heavy particle integrated out is a isospin triplet scalar; in the
Type III seesaw, the heavy particle is an isospin triplet Majorana fermion.    Like the vector bosons of 
extended gauge symmetries discussed in Sec.~\ref{sec:pairs500}, these particles may have their own
symmetry constraints that put their masses at the TeV scale and dimensionless couplings to neutrinos proportional 
to small mixing angles~\cite{Abada:2007ux,Maiezza:2015lza}.

%%%%%%%%%%%%%%%%%%%%%%%%%%%%%%%%%%%%%%%%%%%%%%%%%%%%%%%%%%
\begin{figure}[t]
 \begin{center}
\includegraphics[width=0.60\hsize]{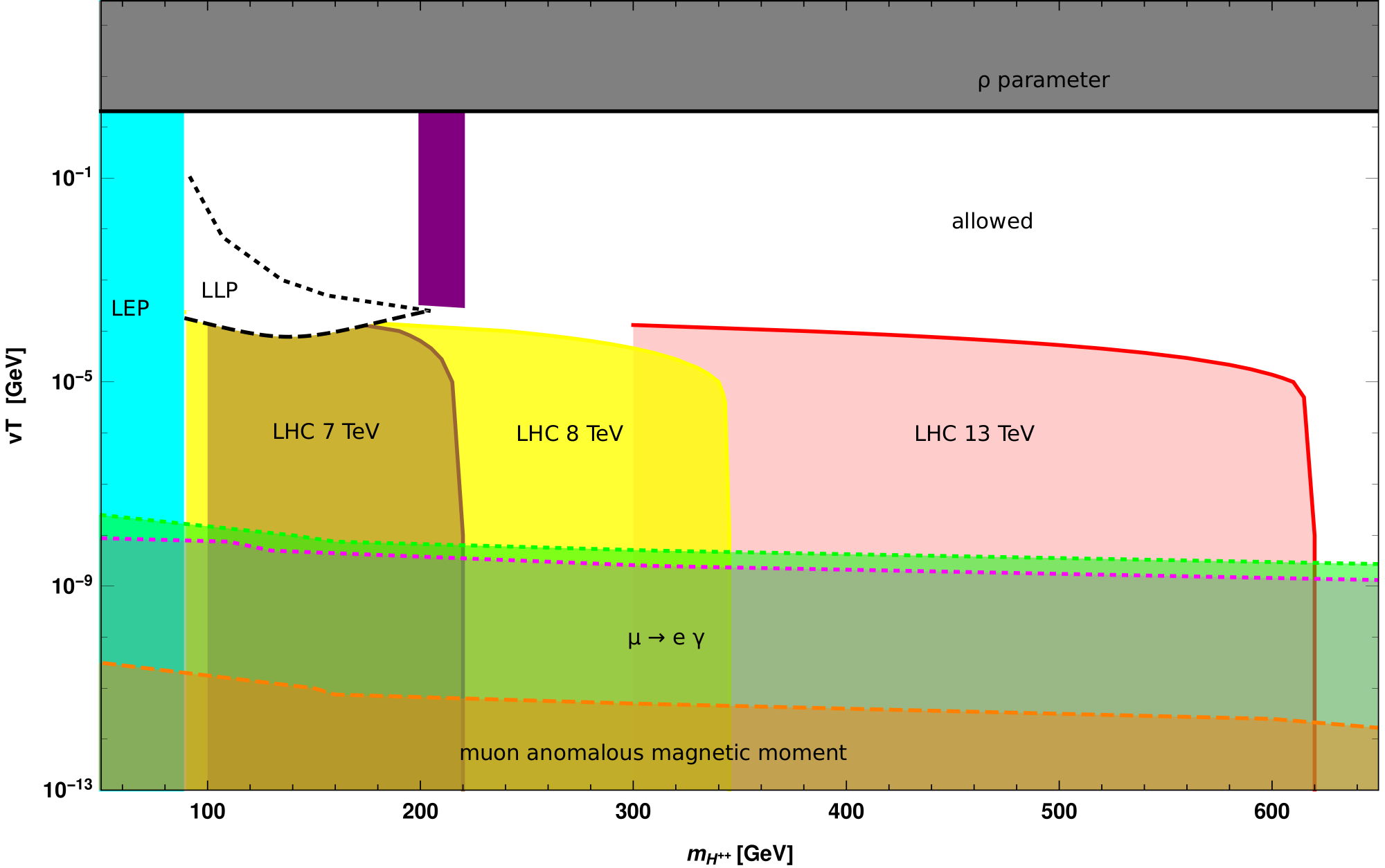}
\end{center}
 \caption{Parameter space of the type-II seesaw model in the plane of the mass of the doubly-charged scalar boson $H^{++}$ and the vacuum expectation value $v_T$ of the neutral component of the $I=1$ scalar 
field, from~\cite{Antusch:2018svb}.  The upper limit on $v_T$ comes from the value of the $\rho$ parameter.  The strongest lower limits come from direct searches for $H^{++}$ in its decay to like-sign dileptons. The region
labeled LLP can be explored at the LHC  in searches for displaced decays.  In the remaining allowed 
region, the dominant decays of the $H^{++}$ are 3-body decays to $W^+f\bar f$.}
\label{fig:Antusch}
\end{figure}
%%%%%%%%%%%%%%%%%%%%%%%%%%%%%%%%%%%%%%%%%%%%%%%%%%%%%%%%%%

Current constraints on the Type II models are shown in Fig.~\ref{fig:Antusch},
 from~\cite{Antusch:2018svb}. These models contain a 
doubly-charged scalar boson  $H^{++}$ that, for sufficiently low masses,  is directly pair-produced at 
colliders.  At the LHC, it is straightforward to search for this particle in its decay to like-sign dileptons, but 
this might not be the dominant mode.  The alternative decay mode, to the 3-body  $W^+ f\bar f$, will 
benefit from the ability at an $\ee$ collider to  completely reconstruct events in leptonic and multi-jet final 
states.

%%%%%%%%%%%%%%%%%%%%%%%%%%%%%%%%%%%%%%%%%%%%%%%%%%%%%%%%%%
\begin{figure}
\begin{center}
\includegraphics[width=0.46\hsize]{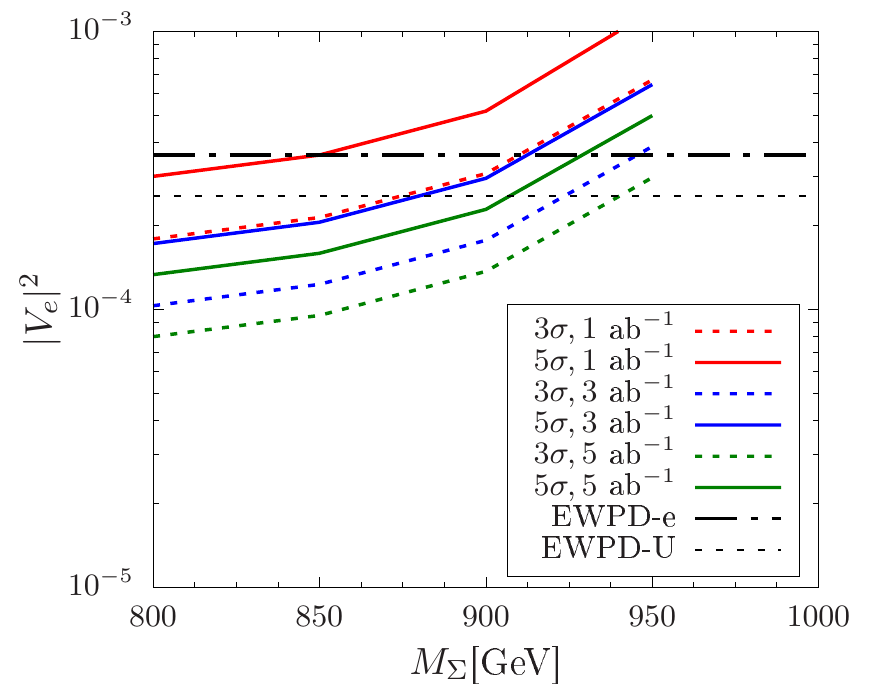} \ \ 
\includegraphics[width=0.46\hsize]{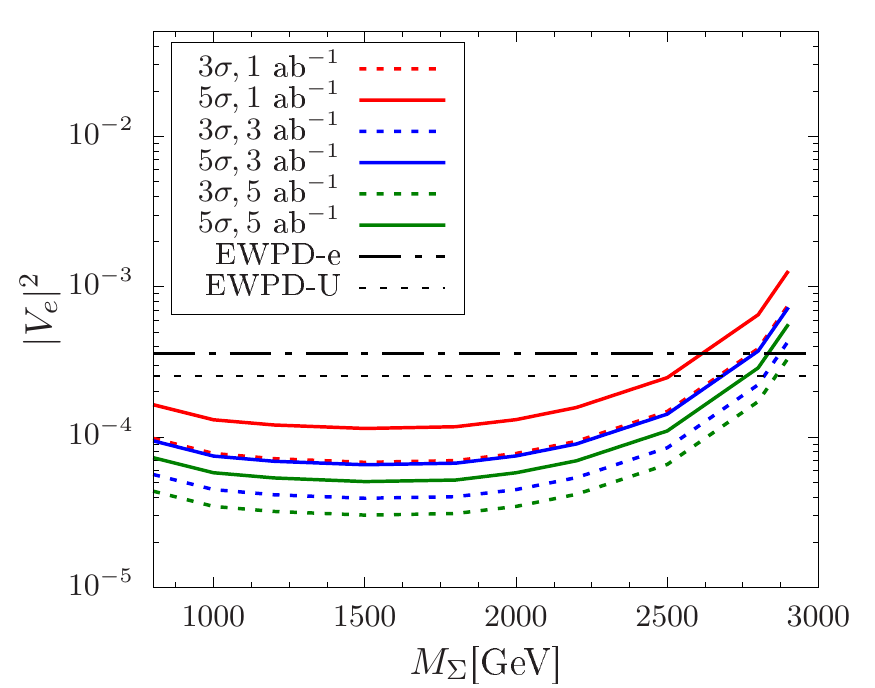}
\end{center}
 \caption{Sensitivity reach in the mixing angle for production of the heavy $I=1$ lepton in 
Type III seesaw models, using the final state $e + \missET+$ boosted jet, from \cite{Das:2020gnt}.
Left: ILC at 1~TeV with varying levels of integrated luminosity; Right: $\ee$ at 3~TeV with 
 varying levels of integrated luminosity.  The solid lines show upper bounds from  precision electroweak
observables, from~\cite{delAguila:2008pw}.}
\label{fig:DasILD}
\end{figure}
%%%%%%%%%%%%%%%%%%%%%%%%%%%%%%%%%%%%%%%%%%%%%%%%%%%%%%%%%%

In Type III models, the most stringent bounds come from searches for pair-production of the $I=1$
 fermions, for example, $q\bar q\to \Sigma^+\Sigma^0$.  The  phenomenology of these particles
has been studied at both $pp$ and $\ee$ colliders~\cite{Das:2020gnt,Das:2020uer}.  The $\Sigma$ particles decay to $W$ or $Z$ plus a lepton through the heavy/light lepton mixing.   At the LHC, one can search for a trilepton plus missing energy final states under the assumption that the $e$ and $\mu$ decay channels are dominant or at least democratically produced.  The lower limits on the $\Sigma$ masses are at roughly
 900~GeV~\cite{CMS:2019lwf,ATLAS:2020wop,ATLAS:2021xxb}.  At $\ee$ colliders,  it is also possible to search for the 
single production of the $\Sigma$ states together with a lepton using the $W$ hadronic decay mode, by 
searching for the final state $e + J + \missET$, where $J$ is a boosted jet with 2-jet substructure.   This 
search has been studied in parametric fast simulation at the 1 TeV ILC in \cite{Das:2020gnt}, leading to the
expected limits on the heavy/light mixing angle shown in Fig.~\ref{fig:DasILD}.   Results for 3~TeV are also shown.

%%%%%%%%%%%%%%%%%%%%%%%%%%%%%%%%%%%%%%%%%%%%%%%%%%%%%%%%%%
% Contribution by K.Mekala

\begin{figure}
    \centering
    \begin{subfigure}[b]{0.48\textwidth}
         \centering
         \includegraphics[width=\textwidth]{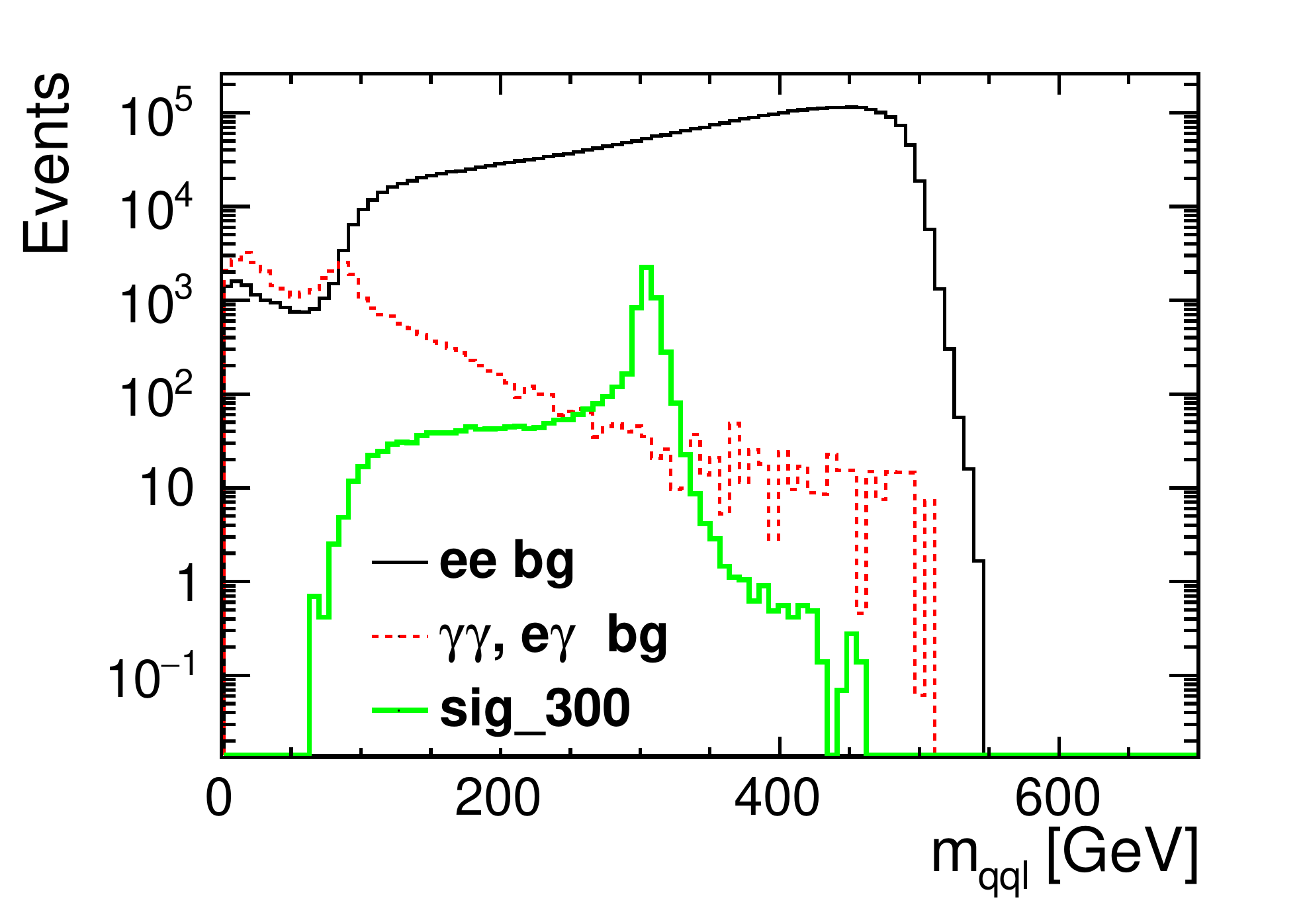}
         \caption{}
         \label{mekala_fig1}
     \end{subfigure}
     \begin{subfigure}[b]{0.48\textwidth}
         \centering
         \includegraphics[width=\textwidth]{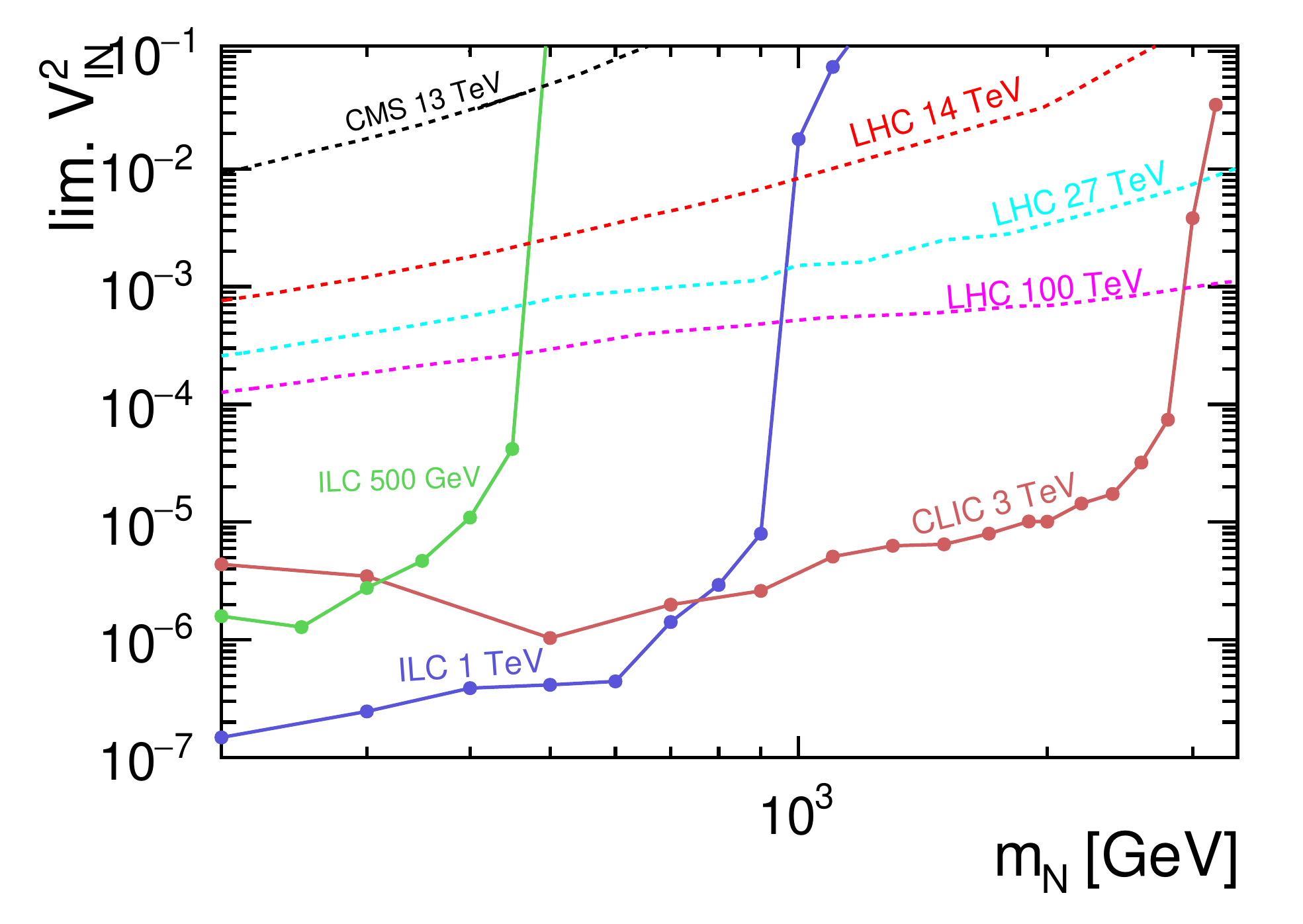}
         \caption{}
         \label{mekala_fig2}
     \end{subfigure}
    \caption{Left: \textit{qql} mass distribution for ILC500 for muons in the final state. Black solid line stands for the $e^+e^-$ background, red dashed line for the $\gamma$-induced background and thick green line for a reference signal scenario (Dirac neutrino with a mass of 300\,GeV). Right: limits on the coupling $V^2_{lN}$ for different colliders (solid lines: ILC500 -- green, ILC1000 -- violet, CLIC3000 -- dark red). Dashed lines indicate limits from current and future hadron colliders based on \cite{Sirunyan:2018mtv,Pascoli:2018heg}. Figure taken from \cite{Mekala:2022cmm}.}
\end{figure}

The possibility of searching for the heavy right-handed neutrino production at 500\,GeV and 1\,TeV ILC was also studied within the model assuming heavy neutrino mixing with all lepton flavors~\cite{Mekala:2022cmm}. 
The study was based on the Delphes fast simulation framework, using dedicated ILC detector model~\cite{ILC-DELPHES,bib:ILCgen}, see section \ref{sec:fastsim}.
For the light-heavy neutrino pair production, with the subsequent decay of the heavy neutrino into two quarks and a charged lepton, direct reconstruction of the neutrino mass is possible, see Figure \ref{mekala_fig1} for the channel involving muons in the final state. 
The limits on the heavy neutrino-lepton coupling (effectively the neutrino mixing angle) expected from the multivariate analysis of the collected data are presented in Figure \ref{mekala_fig2}. 
Results from ILC running at 500\,GeV and 1\,TeV are compared with the corresponding limits for 3\,TeV CLIC, current constraints from LHC and future limit estimates for hadron colliders.
Within the kinematic reach, the limits expected for lepton colliders are orders of magnitude better than those coming from current and future hadron colliders. 
The results were obtained for the Dirac neutrinos but it was verified that the limits for the Majorana particles would be of the same order. 

%%%%%%%%%%%%%%%%%%%%%%%%%%%%%%%%%%%%%%%%%%%%%%%%%%%%%%%%%%
\begin{figure}
\begin{center}
     \subcaptionbox{}{\includegraphics [align=c,scale=0.25]{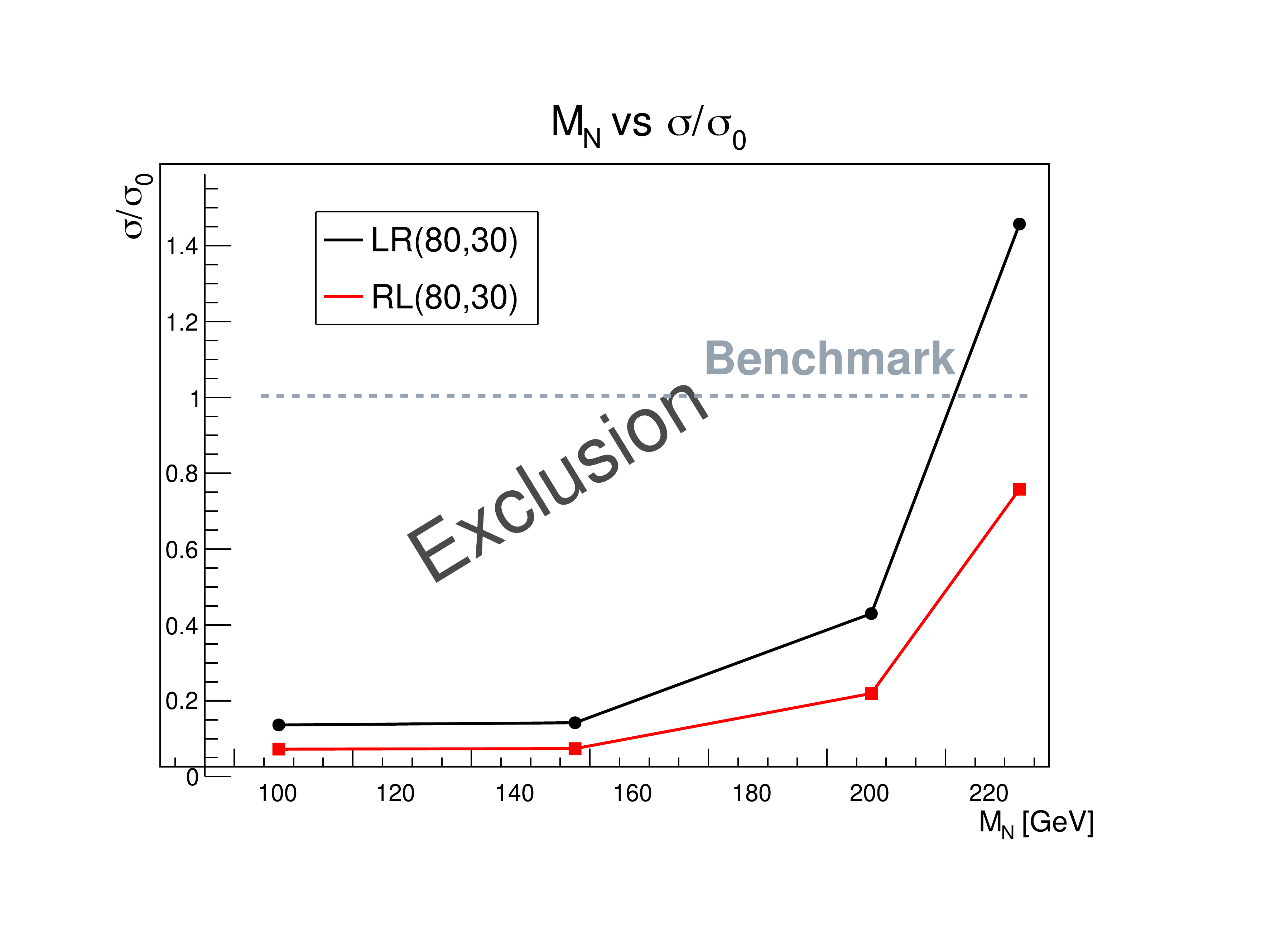}}
 \subcaptionbox{}{\includegraphics[align=c,scale=0.35]{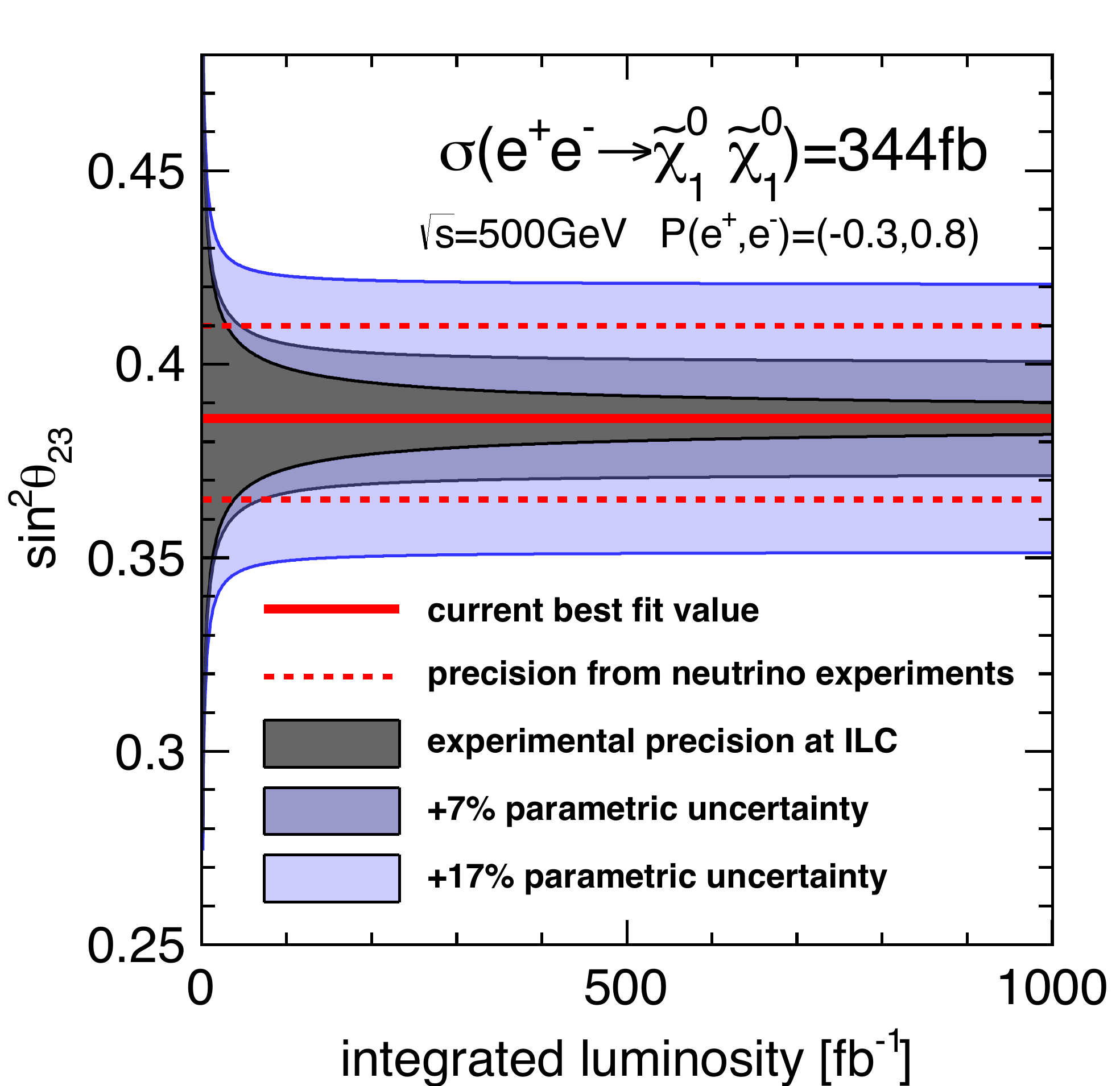}}
\end{center}
\caption{  (a) Exclusion reach for Right-handed Heavy Neutrinos.
  (b) Comparison of the value of the neutrino atmospheric mixing angle $\sin^2\theta_{23}$  to the
value obtained from 
the ILC measurements of the neutralino decay branching ratios in the R-parity  violating  model 
described in the text, from \cite{Vormwald:2013sfm}. 
}
 \label{fig:ListVormwald}
\end{figure}
%%%%%%%%%%%%%%%%%%%%%%%%%%%%%%%%%%%%%%%%%%%%%%%%%%%%%%%%%%

Another model that has been studied in {\it full} simulation is {\it heavy right-handed majorana neutrinos}~\cite{Das:2018tbd,Nakajima:2022pkd}.
In this model, the striking signal is the observation of same flavor, same charge (SFSC) leptons.
While SFSC as such are not rare in the SM, they will typically occur in association with neutrinos,
leading to missing energy.
In the studied model, due to the majorana nature of the heavy neutrino, they can decay completely
visibly, with no missing energy.
As missing energy is not observable at pp colliders (only missing {\it transverse} energy or momentum),
such signals are difficult to detect at LHC.
At ILC, on the other hand, the observation of missing energy is straight-forward, for reasons
explained above.
Figure ~\ref{fig:ListVormwald}(a) shows that the ILC running at $E_{CMS}$=500 GeV can exclude pair-produced
right-handed majorana neutrinos up to at least $M_N$= 225 GeV.

There is one more scenario for neutrino mass involving R-parity violating SUSY.   One possible R-parity 
violating term is the bi-linear coupling in the superpotential
\beq
   \Delta W = - \eps_{ab} \eps_i  L_{ia} H_{ub} 
\eeqn
Together with small induced sneutrino masses, this term leads to small neutrino masses of order 
$|\eps_i|^2$~\cite{Romao:1991ex}.
It is interesting that the same mixing angles appear in the neutrino mass matrix and the decay 
amplitudes for the lightest neutralino decay (which is now allowed) to $W\ell$~\cite{Hirsch:2003fe}.  With observation of 
neutralino pair production and decay, it is possible to test this theory by comparing the observed values of 
the mixing angles.   This analysis was also studied in {\it full} simulation in \cite{Vormwald:2013sfm}.   The 
comparison of the measurement that could be obtained at the ILC at 500~GeV with the observed value of the 
neutrino atmospheric mixing angle is shown in Fig.~\ref{fig:ListVormwald}(b).

\section{New particle searches -- Dark Sector} 
\label{sec:dark500}

Many extensions of the Standard Model contain fields that do not carry any SM gauge charges. Such fields are said to belong to the ``dark sector", and may include sterile neutrinos, additional gauge bosons, and particles responsible for dark matter. A brief review of the commonly used set of benchmark models for dark sectors and their interactions with the SM is contained in Section~\ref{sec:physfixedtarget}. The common feature of dark-sector particle candidates is their feeble couplings to the SM, typically orders of magnitude smaller than the SM gauge interactions. High luminosity and the clean environment of the ILC offer unique opportunities to search for such particles.  If signals of such particles are found,  the precisely characterized initial state and beam polarization may be crucial to determine their nature.  

\begin{figure}[t!]
\begin{center}
 \includegraphics[width=0.6\textwidth]{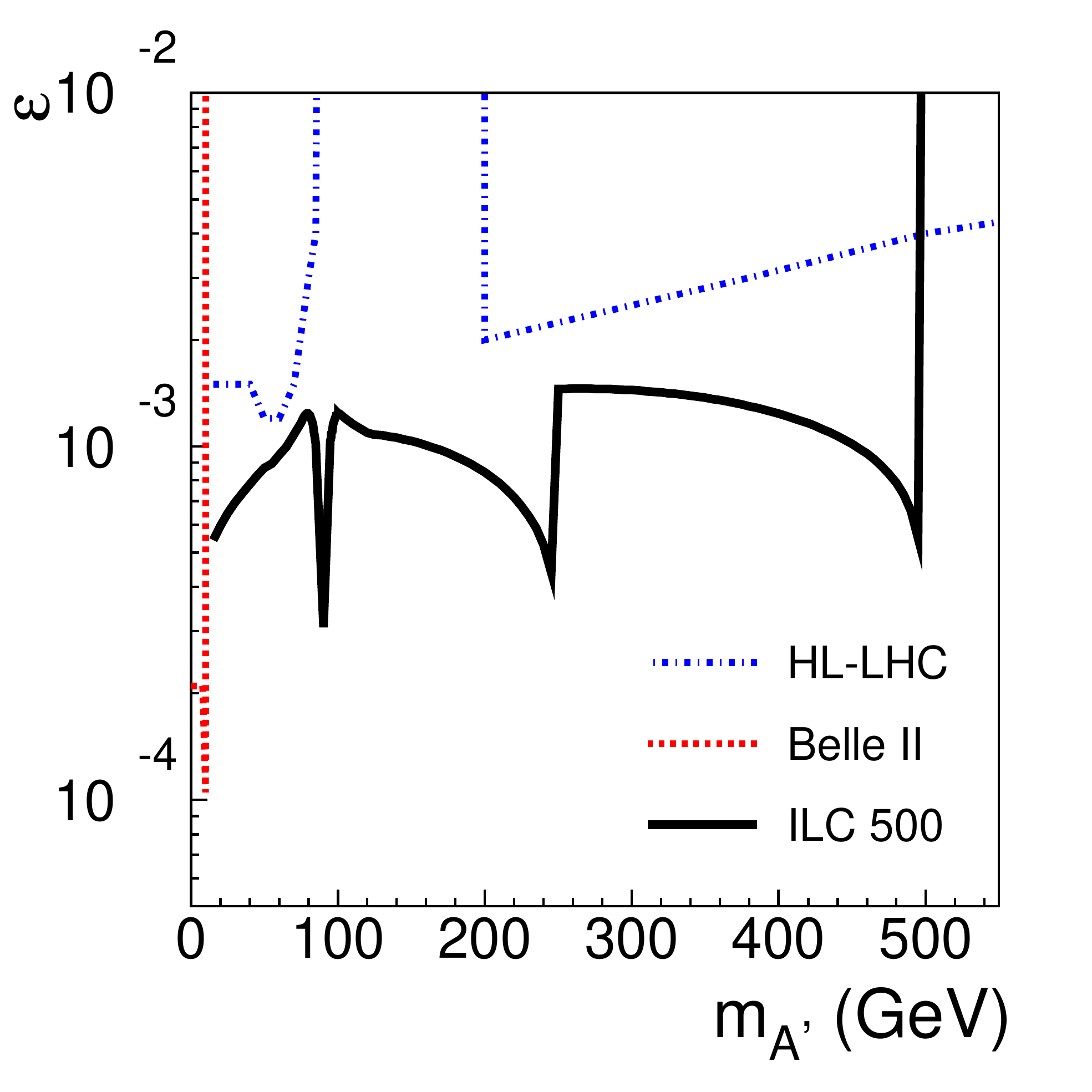}
\end{center}
\caption{Projected reach of HL-LHC, Belle-II and the ILC for the dark $Z$ gauge boson, in terms of its mass and the kinetic mixing parameter $\varepsilon$~\cite{EuropeanStrategyforParticlePhysicsPreparatoryGroup:2019qin,Berggren:2021lan}.
\label{fig:darkZreach}}
\end{figure}

\begin{figure}[t!]
\begin{center}
 \includegraphics[width=0.8\textwidth]{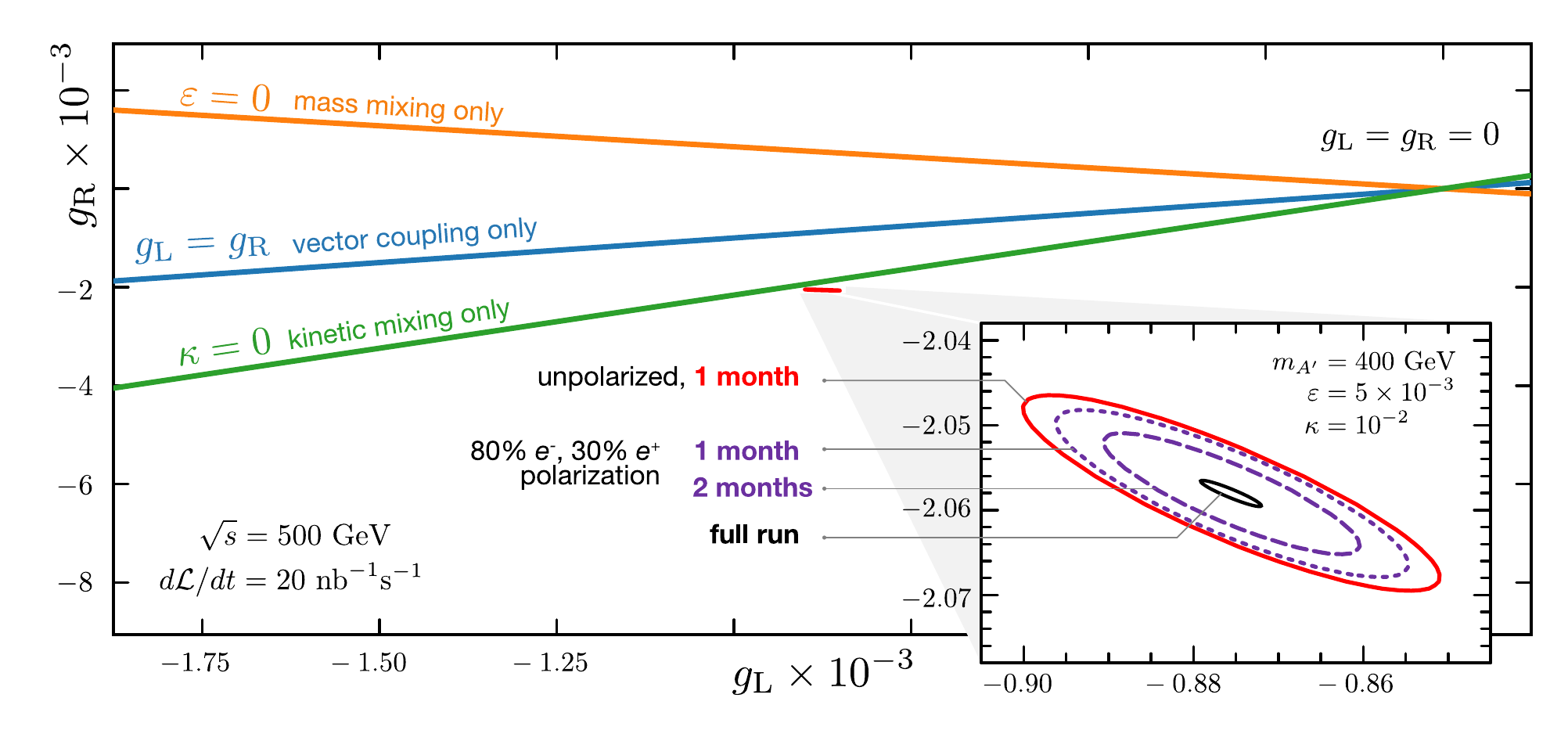}
\end{center}
\caption{Left- and right-handed couplings of dark $Z$ to leptons, measured by a short dedicated ILC run on the dark $Z$ resonance (assumed to be at 400 GeV in this illustrative example). The benchmark model generates parity-violating dark $Z$ couplings to matter through a combination of mass and kinetic mixing. Orange, blue and green lines correspond to alternative models that can be ruled out by this measurement. From Ref.~\cite{San:2022uud}.
\label{fig:darkZcouplings}}
\end{figure}

The ILC will offer two complementary avenues to search for the dark sector particles. First, additional detectors mounted at the ILC beam dumps will provide unmatched sensitivity to dark sector particles with masses below 10 GeV or so. This will be covered in detail in Section~\ref{sec:beamdump}. Second, the general-purpose detectors at the main IP have sensitivity to signals of the dark sector particles with masses up to the full ILC center-of-mass energy. Here, we focus on this latter case. As an example, consider a ``dark $Z$", the gauge boson of an additional $U(1)_D$ gauge group. The interactions of the dark $Z$ with the SM are induced by kinetic mixing of the $U(1)_D$ with the hypercharge, inducing mixing with the photon and $Z$.  It can also be produced through mass mixing if there exist additional Higgs field(s) charged under both $U(1)_D$ and the SM gauge groups. Unlike the $Z^\prime$ bosons that have been extensively considered in the literature, the dark $Z$ does not have direct gauge couplings to SM fermions, greatly weakening experimental constraints. The dark $Z$ can be produced at the ILC through $e^+e^-\to A^\prime + \gamma$. The projected reach of HL-LHC, Belle-II, and the ILC (combining 250 GeV and 500 GeV runs) are shown in Fig.~\ref{fig:darkZreach}.\footnote{This analysis assumes that the dark $Z$ decays back into SM particles.  An alternative scenario is that the dark $Z$ decays invisibly to other dark sector states. In this case, the $\gamma+$missing energy signature discussed in Sec.~\ref{sec:WIMPDM} can be used to search for the dark $Z$ at the ILC.} There exists a parameter range where the ILC would be the first experiment to discover this new physics. If the discovery is made, either at the HL-LHC or the ILC, the ILC will have a unique capability to precisely measure the dark $Z$ couplings and determine their chiral structure. The dark $Z$ will appear as a (very narrow) resonance in $\ee$ annihilation, so   short dedicated ILC run with $\sqrt{s}\approx m_{A^\prime}$ could provide this information, just as  LEP and SLD measuremed the SM $Z$ properties. This is illustrated in Fig.~\ref{fig:darkZcouplings}, which also demonstrates how this measurement can be used to discriminate among possible dark $Z$ models ({\it e.g.} parity-violating vs. parity-conserving, and kinetic vs. mass-mixing.) Dark $Z$ bosons could  also be 
observed through their coupling to the Higgs portal, leading to $H\to Z_D Z_D$. A study of Higgs decay to a dark $Z$ pair  followed by dijet and dilepton decays is presented in \cite{Snyder:2022uih}. 

 If the dark sector contains a scalar field $S$, couplings $S|H|^2$ or $S^2|H|^2$ are possible. If $m_S<m_h/2$, these couplings would induce exotic Higgs decays. The specific signatures depend on the details of the dark sector model, but we have seen in Sec.~\ref{sec:HiggsExotic} that the ILC can identify very many exotic Higgs decay modes.  In models where $S$ is stable or decays purely within the dark sector, such decays will appear as $h\to~$invisible. The ILC offers an exquisite sensitivity in this channel, extending the HL-LHC reach on the branching ratio by a factor of 20. On the other hand, in models where dark sector states can decay back to the SM, visible signatures may appear.  A well-motivated example is  the decay  $S\to   b\overline{b}$.  This is the dominant decay if $m_S>2m_b$ and the flavor texture of its couplings is aligned with the SM Yukawas. This results in a 4$b$ final state, which is notoriously difficult to discern at the LHC but will be accessible at the ILC. Another possibility is that $m_S>m_h/2$, so that no new Higgs decays are induced. This case is very challenging at hadron colliders, especially if the $S$ field carries quantum numbers that forbid its mixing with the Higgs (as happens, for example, in models where $S$ is the dark matter particle). The ILC will offer a unique window on this scenario through a very precise measurement of the $e^+e^-\to hZ$ cross section, sensitive to one-loop corrections induced by $S$ loops~\cite{Craig:2013xia}.

Another well-motivated experimental target is a pseudo-Goldstone boson of a spontaneously broken global symmetry in the dark sector, with coupling structure motivated by the familiar QCD axion. Such ``axion-like particle", or ALP, can be produced at the ILC in association with photons, $Z$, or Higgs, and detected through its decays to photons or $e^+e^-$ pairs. ILC searches will be sensitive to ALPs in the 1--500~GeV mass range, with couplings 1--2 orders of magnitude below the current limits~\cite{Bauer:2017ris,Bauer:2018uxu}. 

\begin{figure}[t!]
\begin{center}
 \includegraphics[width=0.48\textwidth]{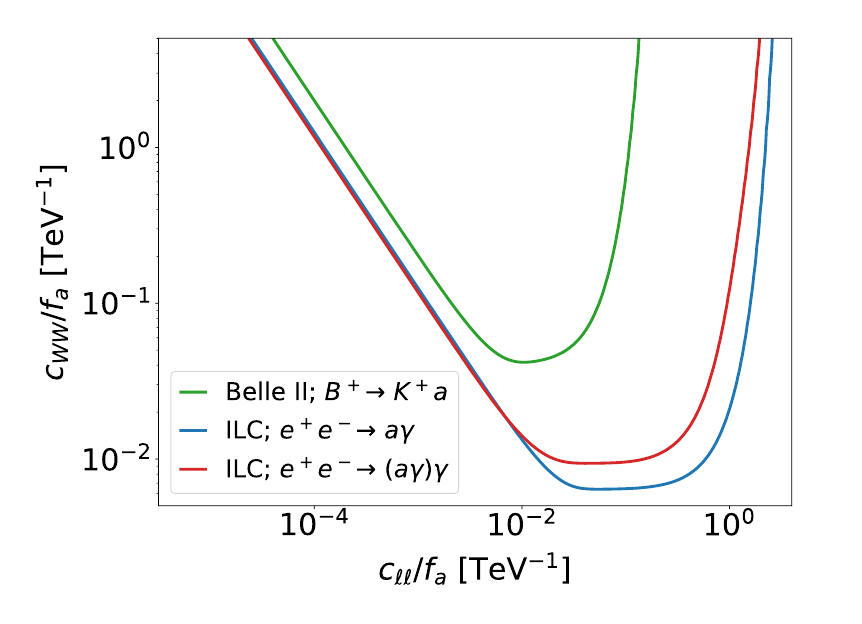}
 \includegraphics[width=0.48\textwidth]{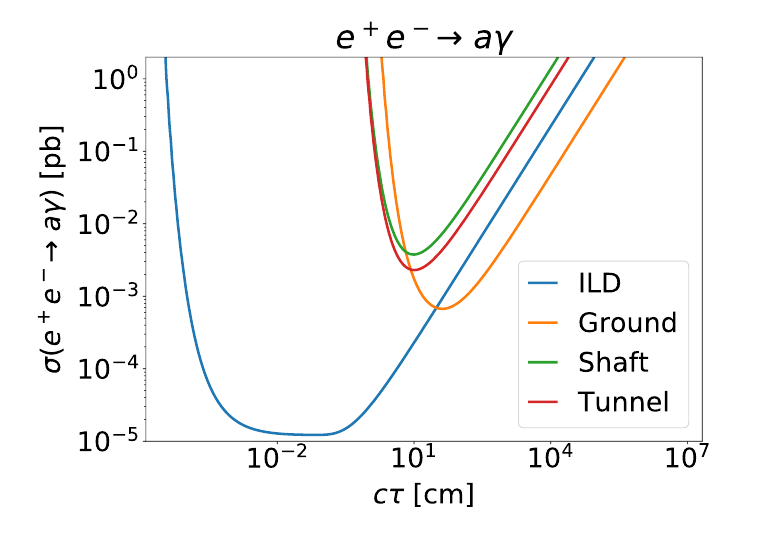}
\end{center}
\caption{Left panel: Sensitivity projections for long-lived ALPs with $m_a = 300$~MeV at the ILC and
at Belle II, as a function of the couplings $c_{WW}/f_a$ and $c_{ll}/f_a$. Right panel: Sensitivity projections for long-lived ALPs with $m_a = 300$~MeV at the ILC main detector (ILD), and three dedicated far-detector options (Ground, Shaft, and Tunnel). From Ref.~\cite{Schafer:2022shi}.  
\label{fig:LLPs}}
\end{figure}

Dark-sector models often contain long-lived particles (LLPs), which have macroscopic decay lengths due to their small couplings. The ILC offers excellent capabilities to search for the LLPs, thanks to clean environment and hermetic angular coverage of the detector. A recent study~\cite{Schafer:2022shi} found that for LLPs produced with a typical cross section of a few picobarns, the ILD detector could probe lifetimes up to 300 ns, or proper decay lengths up to 100 m. This capability will open up new opportunities to search for dark-sector models. For example, for long-lived ALPs with masses below 1 GeV, the sensitivity of this search will significantly exceed the reach of the search for displaced vertices in meson decays at Belle II; see Fig.~\ref{fig:LLPs}. Moreover, Ref.~\cite{Schafer:2022shi} compared the reach of the displaced-vertex search at the main general-purpose ILC detector to that of a potential dedicated ``far" detector, which could be placed in the planned underground cavities or on the ground specifically to search for LLP decays. It was found that a realistically sized far detector can provide at best a moderate improvement over the already impressive sensitivity of the main detector. These conclusions apply quite generally to searches for long-lived particles at the ILC.   

\begin{figure}[t!]
\centering
\includegraphics[width=0.4\textwidth,height=4.8cm]{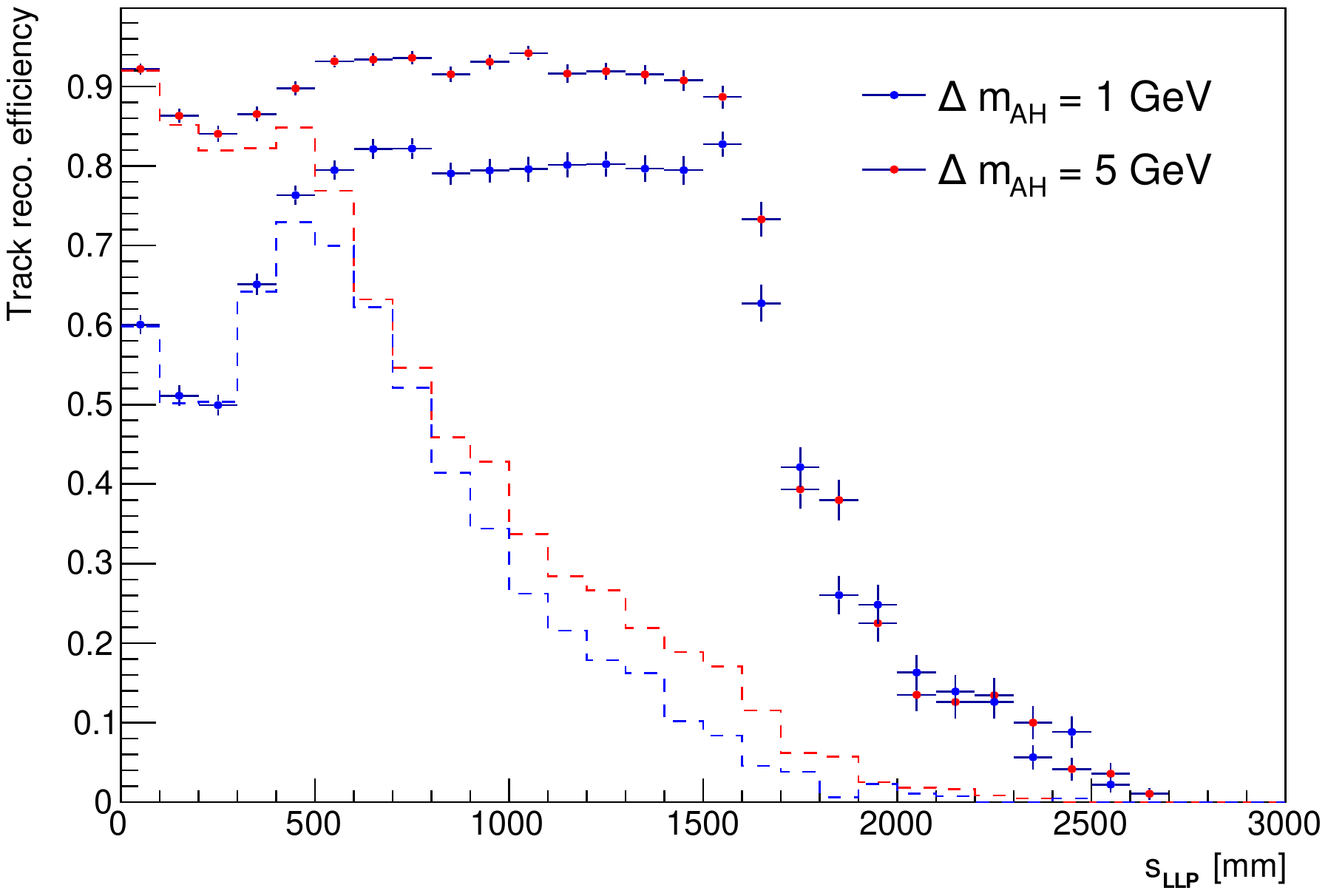}
\caption{Efficiency of track reconstruction as a function of the long-lived particle decay
  length s$_\textrm{LLP}$, for example scenarios with decays of heavy neutral scalar A to lighter heavy
  scalar H and a pair of muons, for two heavy scalar mass differences of 1\,GeV and 5\,GeV.
  \label{fig:llp_effi}
    }
\end{figure}

A high precision pixel vertex detector is crucial for reconstruction of LLP decays, both for SM states
(heavy flavour tagging) and possible BSM candidates, for the expected decay lengths in the range of micrometers to millimeters.
For larger decay lengths, of the order of centimeters and meters, reconstruction of LLP decays has to be based on central tracking.
Time Projection Chamber (TPC) of the ILD has strong advantages for this purpose, since  it has almost uniform response to
charged particles, independent on the particle production point and direction. This was confirmed with dedicated full simulation study,
preliminary results of which are shown in Fig.~\ref{fig:llp_effi}. The efficiency of track reconstruction is shown for two example
scenarios with decays of heavy neutral scalar A (LLP) to lighter heavy scalar H (DM candidate) and a pair of muons. High and
uniform reconstruction efficiency is expected for LLP decays inside the TPC volume even for very low mass difference between
the two exotic states (i.e. very soft visible final state). For global track reconstruction cut on the track impact parameter has to be released, otherwise the efficiency is
degraded significantly (dashed lines in Fig.~\ref{fig:llp_effi}).  A similar study of long-lived particle capabilities for the silicon tracker of the SiD detector is presented in \cite{Jeanty:2022cwr}.

\chapter{ILC Fixed-Target Program}  
\label{chap:fixedtarget}

In addition to its central collider, the 
ILC accelerator can host a number of additional detectors, including detectors for fixed-target experiments and beam dump experiments.
These can provide the setting for a multi-faceted program. The main purpose of these experiments will be to search for dark sector particles interacting only feebly with the Standard Model. %, as motivated by many dark matter models, as well as by models addressing the baryon-anti baryon asymmetry problem, the strong CP problem, and the hierarchy problem.   
The intense and high-energy electron and positron beams that the ILC makes available also have uses in nuclear and hadron physics and  in studies of strong-field QED.  They can also provide resources for developing advanced electron and positron accelerators.

%Dark sector particles could communicate with the SM in a number of ways, via a  dark photon $A^\prime_\mu$ interacting through the gauge portal $\epsilon A^{\prime}_{\mu\nu}F^{\mu\nu}$, via a dark scalar interacting through the Higgs portal $\kappa |S|^2|H|^2$, through a dark fermion interaction through the neutrino portal, $y N L H$, or through a pseudo-Goldstone boson interacting through the axion portal $({a}{\Lambda}) F_{\mu\nu}\tilde F^{\mu\nu}$.   We have already described ILC probes of these interactions in Chapter 8 and 10.  But all of the portals can also be explored powerfully in the fixed-target program, using a variety of different experimental approaches.

In this chapter, we will present the variety of fixed-target and remote experiments that could be mounted at the  ILC site and estimate their potential both for dark sector searches and for other physics questions.
 
%%%%%%%%%%%%%%%%%%%%%%%%%%%%%%%%%%%%%
\section{The physics of light Dark Sectors} 
\label{sec:physfixedtarget}

Many extensions of the Standard Model contain fields that do not carry SM gauge charges. Such fields are said to belong to the ``dark sector". From the observational point of view, such singlet fields are motivated by the existence of dark matter, as well as by the baryon-antibaryon asymmetry. From a more theoretical side, they appear frequently in models of gauge unification, string theory compactifications, {\it etc.} Dark sectors can also address some of the current SM anomalies such as the anomaly in $(g-2)_\mu$. Dark sector particles with masses at or below the GeV scale are particularly motivated since they can naturally lead to thermal dark matter scenarios.
Dark sector fields may still have non-gauge couplings to the SM, allowing them to be produced and detected in collider experiments. Particularly, as we describe below, dark sector particles can communicate with SM particles through the so-called ``portal interactions''. 

The field content of the dark sector and the structure of its interactions with the SM are not strongly constrained by theoretical considerations or by data, and a large variety of viable models are possible. Focusing on {\it renormalizable} couplings between dark sector and SM fields provides a useful set of benchmark models to explore this physics scenario.

\begin{itemize}
    \item {\bf Dark Photon Portal:} If the dark sector contains an abelian gauge group, $U(1)_D$, its gauge field can couple to the SM via the ``kinetic mixing" term ${\cal L}=\epsilon F^{\mu\nu}_D F_{Y\,\mu\nu}$, where $F_Y$ and $F_D$ are the $U(1)_D$ and the SM hypercharge field-strength tensors, respectively. The kinetic mixing induces a coupling of $A^\prime$, the gauge boson associated to $U(1)_D$, to the SM. If $m_{A^\prime}\ll M_Z$, $A^\prime$ simply couples to the electromagnetic current, while a heavier $A^\prime$ acquires $Z$-like couplings (the latter scenario is often described as a ``dark $Z$").

    A light dark photon ($m_{A^\prime}<10$~GeV) can be produced at the ILC beam dump through electron - positron pair-annihilation, and bremsstrahlung productions. Once produced, the dark photon can be long lived, propagate through the dump and then decay to SM particles like $e^+e^-$ in the decay volume. The dark photon can also decay invisibly (e.g. to DM). As we will discuss in Sec. \ref{sec:beamdump}, in both cases, detectors placed behind the dump will offer new sensitivity to the dark photon parameter space.
    
    \item {\bf Higgs Portal:} If the dark sector contains a dark scalar field $S$, the couplings $S|H|^2$ or $S^2|H|^2$ are possible, where $H$ is the SM Higgs doublet. If $m_S<m_h/2$, the $S^2|H|^2$ coupling induces exotic Higgs decays of the type $H\to SS$. (the ILC sensitivity to exotic Higgs decays is discussed in Sec.~\ref{sec:HiggsExotic}.)  Furthermore, relatively light scalars, $S$, could be produced in the dump through electron - positron pair-annihilation, the Primakoff process, and bremsstrahlung, thanks to the $S$ mixing with the SM Higgs possibly induced by the $S|H|^2$ and $S^2|H|^2$ operators. Because of this mixing, the dark scalar can decay back to SM particles. 
    
    \item {\bf Neutrino Portal:} A right-handed neutrino, $N$, is a SM singlet, and as such may be considered to belong to the dark sector, coupled to the SM through the neutrino portal interaction, $H L N$, where $L$ is the SM lepton doublet. This operator induces the mixing of the sterile neutrino with the SM active neutrinos, leading to the production of sterile neutrinos in the dump and to its subsequent decay into SM particles.
    
\end{itemize}

It is customary to add another benchmark to this list, which involves dimension-5 couplings but is very well theoretically motivated: the ``axion portal''. Finally, new dark gauge bosons arising from gauging anomaly-free approximate symmetries of the SM are also well studied in the literature:

\begin{itemize}

    \item {\bf Axion Portal:} A pseudo-scalar singlet, $a$, can couple to the SM via $a F \tilde{F}$, where $F$ is the EM (or other gauge) field strength tensor. This coupling is allowed if $a$ is a Nambu-Goldstone boson, such as the axion. While the original motivation comes from the ``QCD axion" solution to the strong CP problem, phenomenological studies also consider a more general possibility of ``Axion-Like Particle" (ALP), whose masses and couplings are not constrained by the QCD axion model. If sufficiently light, ALPs can be produced in the ILC dump through Primakoff production and then decay to photons thanks to the $a F \tilde{F}$ coupling. %in association with photons, $Z$, or Higgs, and detected through its decays to photons or $e^+e^-$ pairs. ILC searches will be sensitive to ALPs in the 1--500~GeV mass range, with couplings 1--2 orders of magnitude below the current limits. 
\item {\boldmath $U(1)_{e-\mu}$, $U(1)_{e-\tau}$, $U(1)_{\mu-\tau}$:} The corresponding gauge bosons, $Z^\prime$, couple to some of the leptons of the SM. Because of these couplings, they will be produced in the dump from electron (or positron) scattering with the dump nuclei. $Z^\prime$ will decay back to either the charged leptons or the neutrinos of the SM, giving rise to either a visible or invisible signature to be searched for in a detector placed after the dump.
\end{itemize}

%%%%%%%%%%%%%%%%%%%%%%%%%%%%%%%%%%%%
\section{ILC Facilities for fixed-target experiments} 
\label{sec:ILCfixedtarget}

%A tour of the ILC design, with possible locations of fixed-target and remote detectors, was given by 
%Kaoru Yokoya at the LCWS 2021  [https://indico.cern.ch/event/995633/].   A fully developed plan will be given in this section.

The ILC can provide very high energy, high intensity, low emittance electron and positron beams. 
The unique beams can also be used for purposes other than the collider 
experiments. The single-pass nature of ILC allows us to use the beams even destructively 
so long as the influence to the collider  experiments is not significant.

\begin{figure}[t!]
\includegraphics[width=\textwidth]{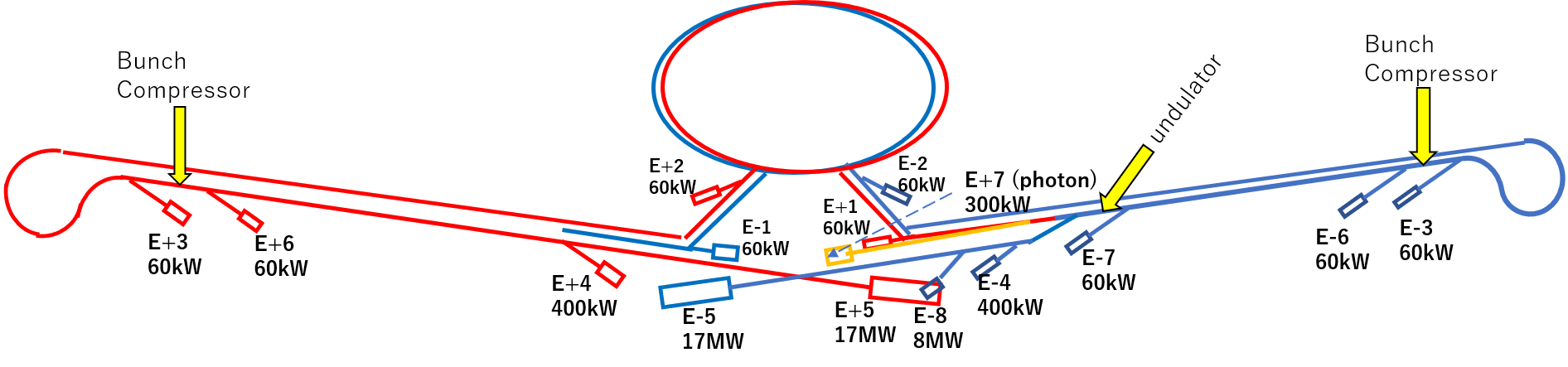}
\caption{Distribution of beam dumps over the ILC facility. The electron, positron and 
photon beamlines are colored blue, red and yellow, respectively.
\label{fig:DumpDistribution}}
\end{figure}

The most appropriate locations of using the beams are the beam dumps.  There, very high intensity 
electron and positron beams interact with thick targets, hopefully producing large numbers of highly penetrating particles.
There are 15 
beam dumps distributed over the entire facility. Their locations are schematically 
shown in Figure \ref{fig:DumpDistribution}. In this section, we will briefly describe only those which may be 
useful for some of the fixed target experiments.

\medskip
\noindent{\bf Main dumps (E$-$5,E+5)}

The main dumps (E$-$5 and E+5) are located about 300 m downstream of the interaction point (IP). 
Each of them accepts the full power beam (125 GeV, 2.5 MW) of the ILC250 beam.
The main body of the dump is a water tank of cylindrical shape, 1.8 m diameter, $\sim$10 m length, filled with high-pressure ($\sim$10 atm for ILC500) water. 
This is followed by a shield several tens of meter long, designed to absorb muons created in the dump. This muon shield can be split into many pieces so that appropriate locations can be chosen to insert the detectors for fixed target experiments. 
The accelerator carrying the opposite beam  to the IP is running nearby. The beam-center spacing is 0.014 (crossing angle) $\times$  300-400 m  = 4-6 m. This will limit the size of the region available for a fixed target experiment. 

There have already been several proposals to make use of the secondary particles from these dumps.
Experiments parasitic to the collider experiment are normally expected so that the beams come to the dumps after beam-beam interaction at the IP. 
It may also be possible to plan a dedicated machine time in principle but it is better to use the tune-up dumps (E$-$4, E+4) unless the full power beam is necessary.
Also it is almost impossible to make use of the beam between IP and the dump, by either placing a target or by extracting the beam, because  of the safety issue.  

\medskip
\noindent{\bf Tune-up dumps (E$-$4,E+4)}

Another location to make use of the full energy beam (but at lower power) is the tune-up dumps E$-$4 and E+4. These dumps are used for the commissioning of the main linacs. Here, the electron or positron beam can be extracted from the main beam line so that it does not go to the experimental hall. 
Up to 400 kW (including a 20\% margin) can be accepted. These dumps will also be used in case of an emergency. When a highly off-energy or other  erroneous beam is detected,  
 fast kickers are excited to eject the beam to these tune-up dumps. 
The field rise time is shorter than the bunch spacing (554 ns) and the duration is more than 
100 $\mu$s, corresponding to more than 200 bunches.

These dumps can be used in two different modes, dedicated and parasitic.
In the latter mode a part of the beam (some small number of bunches or pulses) is extracted during normal collider operation using fast kickers. However, for either mode of operation,  it must be recognized  the devices of the dumps 
(dump body, kickers, etc.) are not necessarily designed for routine operation at 5 Hz. Deliberate planning between the experimental and accelerator teams is mandatory. 

One possible proposal to make use of E$\pm$4 is the QED experiment (Sec.~11.4). In this case, the beam interacts with a high-power laser whose  repetition rate is limited,  so  it suffices to extract the last bunch in a 1312 bunch train during the regular collider operation. The required kicker is simple (500~ns rise time, no flat-top necessary, no constraint of fall time, 5 Hz) and can be installed in the main beamline. The emergency kicker need not be used for this purpose. A major challenge for this proposal is how to transport the laser beam deep underground if  the laser is housed on the surface. 

\medskip
\noindent{\bf Photon dump (E+7)}

The baseline design of ILC adopts a positron source using helical undulators. The 125 GeV electron beam 
emits photons which produce the positron beam. The energy of the photons is several MeV and 
the number of photons is $\sim 10^{17}$ per second. After producing positrons, these photons are dumped 
at $\sim 2$ km downstream. The total photon power is about 60 kW. (The design limit of the dump is 300 kW 
because of future upgrade.) This can be a unique source of gamma rays although the parameters are driven 
by the requirement of the collider operation.

\medskip
\noindent{\bf Beams with low bunch charge}

Colliders prefer high bunch charge because the luminosity is proportional to bunch charge square. However, some fixed target experiments may prefer a lower bunch charge with a shorter bunch spacing.
CW operation is impossible because the klystrons allow only pulsed operation (duty factor $\sim$1 \%). 
What may be done at most is to fill all buckets of 1.3~GHz with weak bunches (population up to $\sim 2\times 10^7 $) with the pulse length $\sim 0.7$ms by introducing a different electron gun. The damping ring is not compatible with this beam format, hence a beamline, a few hundred meter long, is needed to bypass the damping ring. There are several other issues expected (e.g., emergency issue) so that serious discussion with accelerator team is needed. A positron beam of such a format seems to be very difficult to produce due to the large emittance of the positron source.

  It is easy to reduce the bunch charge with the bunch spacing fixed. This is possible only in dedicated modes. The only issue is whether the beam is visible by the monitors for orbit control. It may be possible to add a `pilot bunch' with normal charge for orbit control. 
Another possibility, depending on the nature of the experiment, is to scrape the halo particles by a movable target during the normal collider experiment. This is appropriate in the tune-up dump line. The safety issue must be carefully considered.

%%%%%%%%%%%%%%%%%%%%%%%%%%%%%%%%%

\section{Dark Sector particle searches}
\label{sec:beamdump}

Dark sector particles could be produced from the interactions of either the $e^-$ or the $e^+$ beam with the corresponding beam dump. Since such particles are very weakly coupled to ordinary matter, they could propagate through the beam dump and the muon shield without interacting, and decay back to the SM after that. Such events could be probed by detectors located 50--100 m away from the beam dump and behind the muon shield, searching for visible decay products (e.g. signatures involving two or more leptons, or two or more photons)~\cite{Kanemura:2015cxa}. Alternatively, the dark sector particles could decay to other dark sector states, such as the stable DM particles. A detector could be mounted behind the muon shield to search for elastic scattering of DM particles on atomic electrons, similarly to what has been proposed for the BDX experiment at Jefferson Lab~\cite{BDX:2016akw}.

\begin{figure}[t!]
\includegraphics[width=0.45\textwidth]{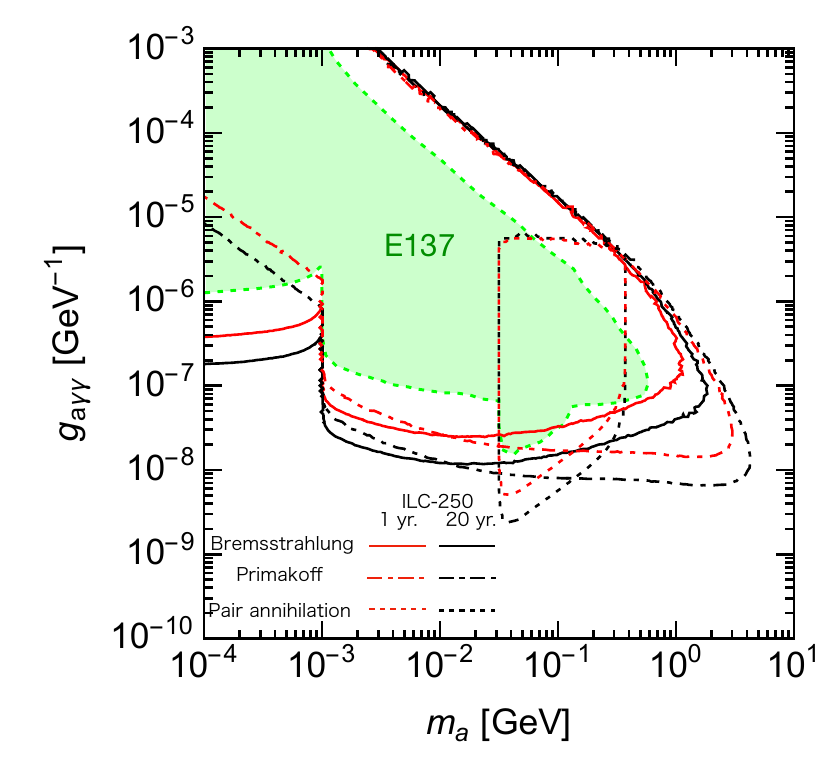}
\includegraphics[width=0.43\textwidth]{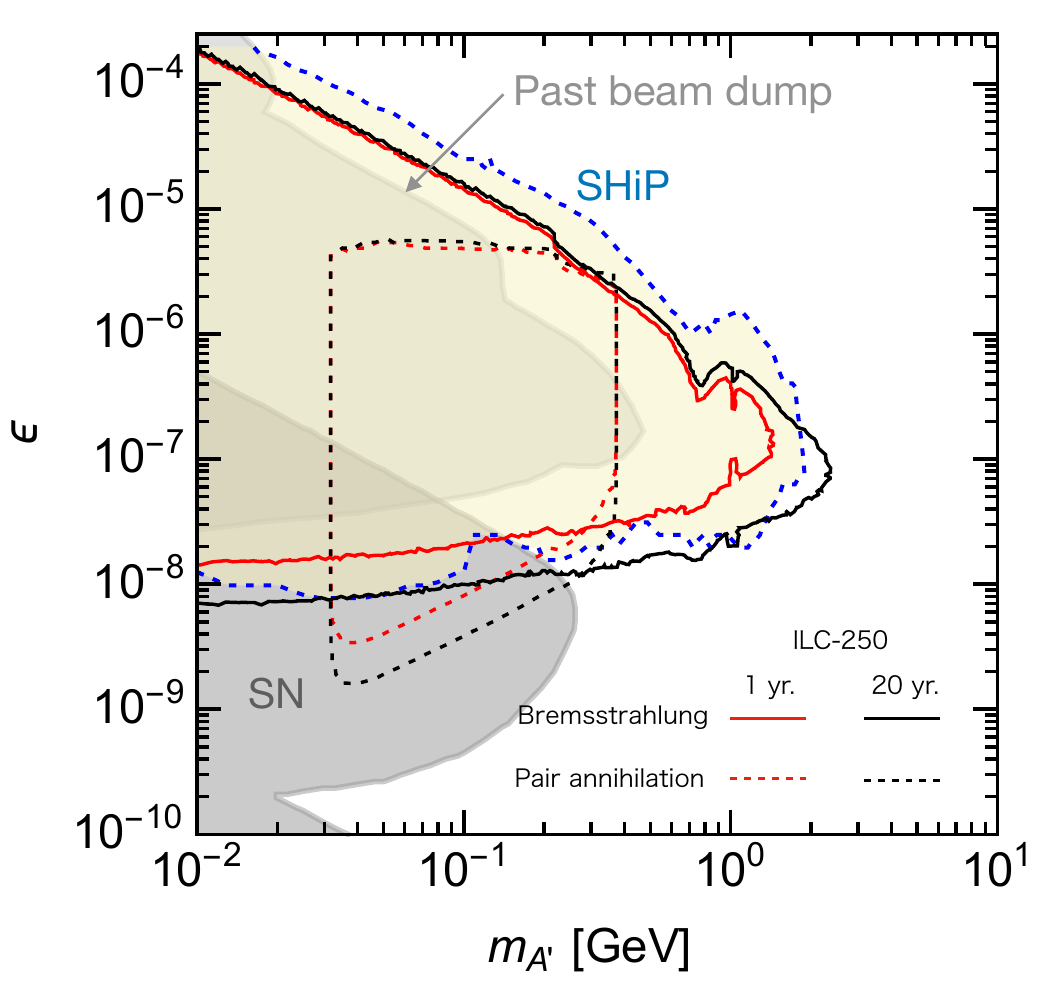}
\caption{The reach of a beam-dump experiment at the ILC-250 for axion-like particles (left) and dark photons (right). Taken from Ref.~\cite{Asai:2021ehn}. See also \cite{Sakaki:2020mqb} for another study of the reach on axion-like particles.} 
\label{fig:VisibleZPrimeALP}
\end{figure}

In this section, we discuss the discovery prospects of (both visible and invisible) dark particles at the ILC beam dump experiment. The ILC environment offers unique advantages for this type of physics: highest-energy lepton beams available at any existing or planned machine; very high integrated luminosity (about $4\times 10^{21}$ particles on target per year for the main beam dumps at ILC-250) and the availability of both electron and positron beams. These features will enable the ILC experiments to expand the reach of searches for dark particles to higher masses and smaller couplings.

As an example, the expected ILC reach for visibly-decaying ALPs and dark photons is shown in Fig.~\ref{fig:VisibleZPrimeALP}. In both cases, the ILC will greatly expand the reach of the currently available experiments, probing higher dark particle masses and smaller couplings. Similar improvements were demonstrated for leptophilic gauge bosons, such as $U(1)_{\mu-\tau}$~\cite{Asai:2021xtg}, or for dark scalars \cite{Sakaki:2020mqb}. If a dark-sector particle is discovered, the ILC can probe its nature and discriminate among theoretical models. Uniquely among the proposed experiments, the ILC can measure and compare the production rates at electron and positron beam dumps, as well as study the dependence on the rates on beam polarization. 

It is worth noting that the main ILC detector also has an impressive sensitivity to visibly decaying LLPs produced in hard processes at the collider IP; see section~\ref{sec:dark500}, in particular Fig.~\ref{fig:LLPs}. Dedicated ``far" detectors to search for LLPs produced at the main IP have also been explored, but were shown to not provide significant improvements in sensitivity for realistic parameters~\cite{Schafer:2022shi}.

\begin{figure}[t!]
\includegraphics[width=0.45\textwidth]{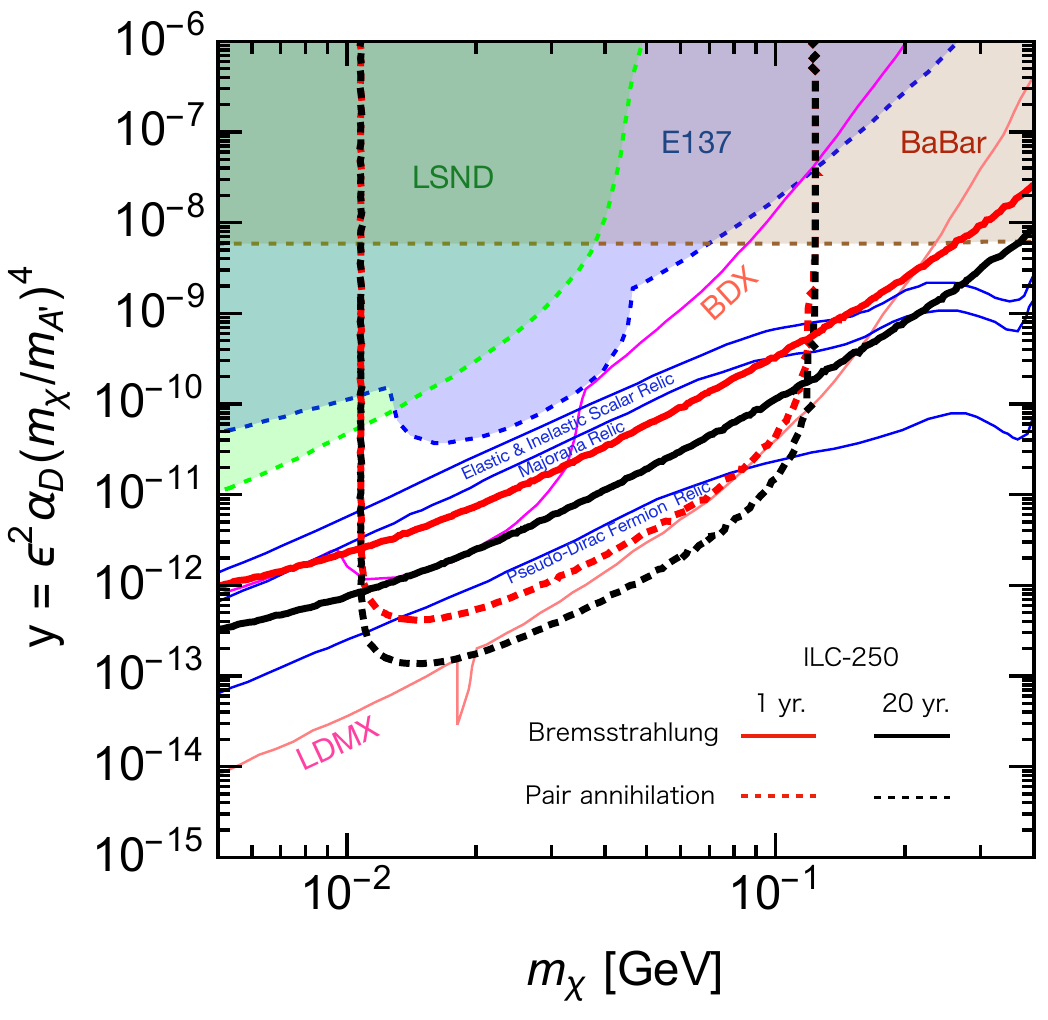}
\includegraphics[width=0.45\textwidth]{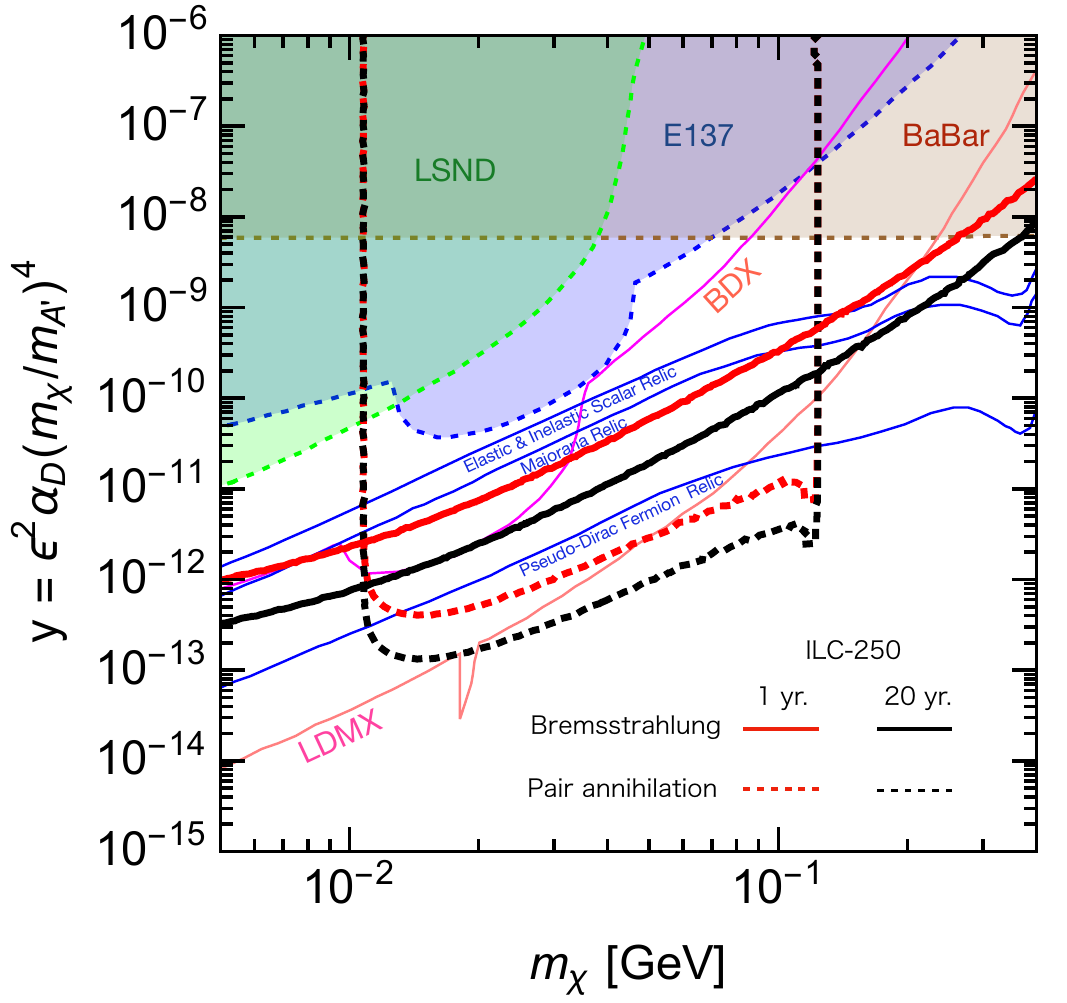}
\caption{Reach of the ILC search for dark photon decaying invisibly to a pair of stable dark matter particles. For comparison, the current constraints (shaded) and reach of proposed BDX and LDMX experiments are also shown. Blue lines indicate the parameters where the DM thermal relic density matches the observed value. 
\label{fig:InvisibleAPrime}}
\end{figure}

The reach of a search for a dark photon decaying invisibly to a pair of dark matter particles is shown in Fig.~\ref{fig:InvisibleAPrime}. This search relies on detection of elastic scattering of DM particles on an atomic electron in the detector placed 100 m downstream of the beam dump behind a muon shield, and is conceptually similar to the proposed BDX experiment~\cite{BDX:2016akw}. The ILC experiment will probe the parameter space of this model far beyond the current constraints. In particular, a broad range of parameters where the model can reproduce the observed DM relic density through thermal freeze-out can be probed (see the blue lines in the figure). Note that the experiment at the positron dump (right panel of Fig.~\ref{fig:InvisibleAPrime}) has a somewhat higher reach than the electron-dump counterpart, due mainly to the additional dark photon production channel $e^+e^-\to A^\prime$ (with $e^-$ being an atomic electron inside the dump). Once again, if a signal is discovered, the availability of $e^-$ and $e^+$ beams with closely matched parameters, as well as beam polarization, will give the ILC the unique ability to discriminate among possible theoretical interpretations.  

The ILC beam dumps also offer an excellent setting to search for heavy neutral leptons.  By following the production of neutral leptons from the primary ILC interaction point and from lepton decays and inelastic scattering in the beam dump, the studies\cite{Nojiri:2022xqn,Giffin:2022rei} show that the ILC beam dump experiments can be sensitive to mixing angles as small as $|U_\mu sim 10^{-{10}}$ and $|U_\mu sim 10^{-8}$. 

\begin{figure}[t!]
\begin{center}
 \includegraphics[width=0.6\textwidth]{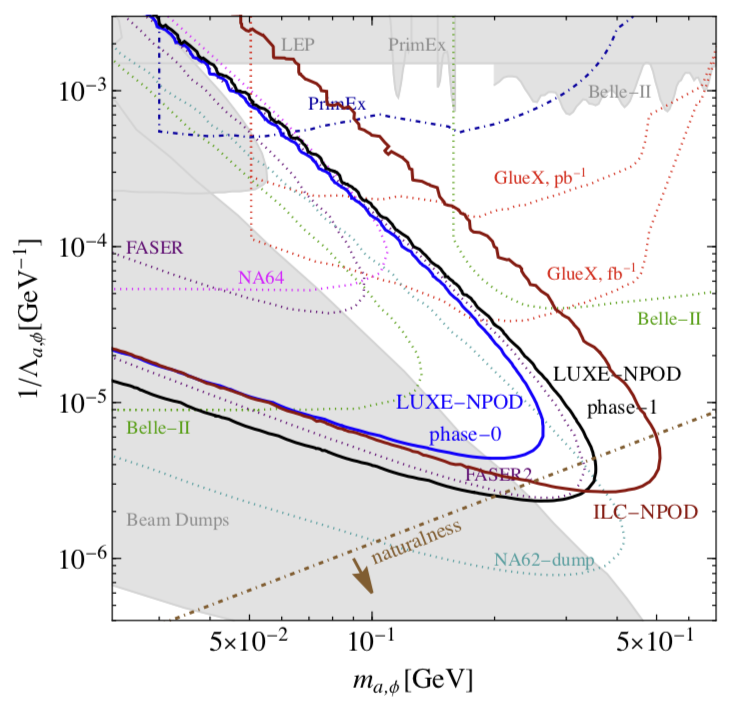}
\end{center}
\caption{Sensitivity of the ILC version of the LUXE-NPOD experiment~\cite{Bai:2021dgm} 
searching for scalar and axion-like particles that couples to photons.  The sensitivity is expressed in terms of parameters $\Lambda_a$ and $\Lambda_\phi$ defined in eq. \leqn{LforALPs}.}
\label{fig:LUXILC}
\end{figure}

In addition to using the main beam dumps, the ILC offers other interesting opportunities for novel dark sector particle searches. For example, a high-powered laser can be mounted at one of the tune-up beam dumps. This setup will enable the exploration of strong-field QED, as discussed below. Interactions of the electron beam with the laser field will also produce a high-luminosity photon beam, which can in turn interact with a target to produce dark sector particles such as ALPs.  An effective Lagrangian describing the coupling of an ALP $a$ or a scalar $\phi$ to the photon field can be written
\beq 
 {\cal L}= \frac{a}{4\Lambda_a} F_{\mu\nu}\tilde{F}^{\mu\nu} +\frac{\phi}{4\Lambda_\phi} F_{\mu\nu}{F}^{\mu\nu} \ .
\eeq{LforALPs}
and the sensitivity of such experiments can be described as limits on the parameters $\Lambda_a$, $\Lambda_\phi$.

This experimental concept was developed for the LUXE-NPOD experiment proposed at DESY~\cite{Bai:2021dgm}. The high energy and intensity of the ILC beams will greatly expand its sensitivity.
The limits expected from the ILC fixed target program are shown in  Fig.~\ref{fig:LUXILC}. Another interesting possibility is to use the photon beam produced by the positron source to search for ALPs based on the ``light-shining-through-the-wall" concept. This and other schemes are currently being investigated.

\section{Experiments on strong-field QED}
\label{sec:strongQED}

The  electron beam  of  120 GeV  available for fixed-target experiments will also provide another experimental program, one on QED in very strong fields.   At the {\it Schwinger critical field} of 
\beq
                     e E =   m_e^2 \ \mbox{or} \  E = 10^{18} \ \mbox{V/m} 
\eeqn
the QED vacuum becomes unstable with respect to spontaneous $\ee$ pair creation.  This suggests a new regime of QED that has not yet been studied in the laboratory.  The subject of QED in strong background fields has recently been reviewed in Ref.~\cite{Fedotov:2022ely}. 

A figure of merit is defined by 
\beq
                \chi =   eE_e / m_e^2
\eeqn
where $E_e$  is the external electric field measured in the electron rest frame. 
Currently, the highest $\chi$ achieved in the laboratory is $\chi \sim 0.3$ at the SLAC experiment on nonlinear QED E-144~\cite{E144:1996enr,Burke:1997ew,Bamber:1999zt}.   This experiment observed the nonlinear Compton and Breit-Wheeler processes
\beq
     e^- + n\,\gamma_l \to e^- \gamma  \qquad \mbox{and} \qquad  e^- + n\,\gamma_l \to e^- e^+ e^- \ ,
\eeqn
where $\gamma_l$ denotes a laser photon, up to $n = 4$. This experiment observed successively smaller rates for increasing $n$. However, as the laser field strength increases, it becomes necessary to resum contributions from all higher $n$ using dedicated nonperturbative analysis. The rates of these nonlinear QED phenomena become comparable to the single-photon rates at $\chi \sim 1$.

 Such large fields are not only of conceptual interest.  The corresponding magnetic fields of 
\beq
                        B  =  10^{14}\ \mbox{gauss} 
\eeqn
are observed in magnetars, pulsars with large magnetic fields that are responsible for Fast Radio Bursts and other extreme astrophysical phenomena, and such large fields are also likely to be present in active galactic nuclei.   These systems also host electron-positron plasmas that may have unique and surprising properties.  Such high fields also occur in the bunch-bunch collisions at TeV $\ee$ colliders.   In both cases, we need laboratory experiments to develop and calibrate the plasma evolution codes that are needed to model these systems.

To achieve fields above the Schwinger critical field in the laboratory, the best method is to interact a relativistic electron beam with an intense laser beam.    In a head-on collision with an electron of energy $\gamma_e m_e$, the intensity of the laser field is increased by $\gamma_e^2$ when viewed in the frame of the electrons, boosting the laser fields to very high intensity.  A 2~GeV electron beam on a focused pulse from a 10~PW laser can achieve $\chi\sim 1$ in the frame of the electrons, and we can imagine going higher both in the laser intensity and the in the electron beam energy.  The E-144 experiment collided a 50~GeV electron beam with pulses from a 1~TW laser.   Today, there are two new initiatives.   The experiment E-320, at SLAC’s FACET-II facility, now commissioning, will interact a 13~GeV electron beam with a 20~TW laser~\cite{Meuren:2019,Salgado:2021fgt}.   The LUXE experiment, planned at DESY, will interact a 16.5~GeV electron beam with
 a 40~TW laser~\cite{Abramowicz:2019gvx,Abramowicz:2021zja}. Both experiments should reach $\chi$ values above 1, with possible upgrades to reach $\chi \sim 5 - 10$.

In the mid-2030’s, we should have available 100  PW lasers at wavelengths of 1$\mu$.   Such high-power lasers are mainly limited in repetition rate, so one might imagine 100~PW pulses at 1~Hz or 10~PW pulses at 10~Hz.   We estimate the pulse sizes at 2~$\mu$  in diameter, with a pulse length of 40~fsec or 120~$\mu$.  For electron energies of 120~GeV, $\gamma_e = 2.4\times 10^5$, and ILC beam sizes, these conditions lead to 
\beq
                      \chi \sim 250
\eeqn
deep in the regime beyond the critical field.  In this strong field, the radiation length is about 
0.3~$\mu$. 

We envision three stages of strong QED experiments.   First, in normal incidence, high energy single electrons would pass through the laser bunch with an optical depth of a few radiation lengths.  With a tracker and calorimeter the interaction point to measure the final $e^+$, $e^-$ and $\gamma$ momenta and energies, this experiment would study the primary radiation processes at $\chi \sim 100 - 200$.   Second, in head-on collisions, single electrons would initiate QED showers leading to the coherent production of an $\ee$ plasma.  The features of this plasma have been simulated in \cite{Qu:2020gpy}.   It will be fascinating to observe the dynamics and modes of excitation of this plasma.   Third, an electron beam with bunches of $10^7$ particles or more would be collided head-on with the laser bunches.   This would produce a dense, incoherent $\ee$ plasma of astrophysical interest.   This three-stage program would enter and fully characterize this new regime of QED.

The requirements of the first stage of the program, for single- or few-electron collision and particle tracking and calorimetry, are very similar to the requirements for the LDMX-type dark matter experiment described in the previous subsection.  Thus, these experiments could be located in the same experimental hall, swapping targets but keeping much of the infrastructure in place. 

%\section{Dedicated Far Detectors}
\label{sec:secondary}

\chapter{Precision Tests of the Standard Model} 
\label{chap:SMEFT}

\section{Precision Standard Model theory for ILC} 
 \label{sec:precision}

To achieve the goals described in sections~\ref{ILC250} and \ref{chap:PEW}, precise predictions for the Standard Model (SM) expectations of the relevant observables are needed. A detailed discussion of the required theory work for studies at $\sqrt{s}\approx 91$~GeV, 160~GeV and 250~GeV can be found in Ref.~\cite{Freitas:2019bre} and references therein. The necessary improvement can be split into three categories:
%%%%%%%%%%%%%%%%%%%%%%%%%%%%%%%%%%%%%%%%%%%%5
\begin{table}[tb]
\begin{center}
\begin{tabular}{l|c|c}
\hline
& Current theory error &  Projected theory error \\
\hline
$M_W$ [MeV] & 4 & 1 \\
$\sin^2\theta^\ell_{\rm eff}$ [$10^{-5}$] & 4.5 & 1.5 \\
$\Gamma_Z$ [MeV] & 0.4 & 0.15 \\
$R_\ell$ [$10^{-3}$] & 6 & 1.5 \\
\hline
$\sigma(HZ)$ & 1\% & 0.3\% \\
$\Gamma[H\to b\bar{b}$] & $<0.4\%$ & 0.2\% \\
$\Gamma[H\to gg$] & 3\% & 1\% \\
$\Gamma[H\to WW^*$] & 0.5\% & $<0.3\%$ \\
\hline
\end{tabular}
\end{center}
\caption{Current and projected theory uncertainties from missing higher orders for the SM prediction of various key precision observables at the ILC (from Ref.~\cite{Freitas:2019bre}). The projected future scenario assumes the availability of N$^3$LO corrections and leading N$^4$LO corrections for Z-pole observables, and NNLO electroweak corrections and higher-order QCD corrections for Higgs observables.}
\label{tab:therr}  
\end{table}
%%%%%%%%%%%%%%%%%%%%%%%%%%%%%%%%%%%%%%%%%%%%%%%%%%
\begin{itemize}
    \item Fixed-order calculations: For the $Z$-pole program, electroweak N$^3$LO corrections as well as leading N$^4$LO corrections for the effective $Z$-fermion vertices are needed. Here ``leading'' refers to corrections enhanced  by powers of the top Yukawa coupling and/or QCD strong coupling. For the 250-GeV program and physics at the $WW$ threshold, NNLO electroweak corrections for $2\to2$ scattering processes are mandatory. In addition, calculations of Higgs decay amplitudes must be completed to NNLO order, in particular, for the Higgs decay $H\to 4f$.  Higher-order QCD corrections to $H\to gg$ and $H\to b\bar b$ are also needed.
    
    The estimated impact of these corrections on a few key quantities is illustrated in Tab.~\ref{tab:therr}. The values in this table should be taken with a grain of salt, since any theory error evaluations of currently unavailable calculations are somewhat speculative.
    
    \item To study effects of detector acceptance and background subtraction, Monte-Carlo tools need to be created with the precision of the expected measurements.  This requires an accurate treatment of multi-photon initial state radiation and awareness of beam polarization.  Furthermore, the Monte-Carlo programs must be matched to the fixed-order calculations discussed in the previous bullet point. Beyond the leading order QCD and electroweak corrections must be merged in an appropriate way.  For a more detailed discussion of QED effects, see Ref.~\cite{Jadach:2019bye}.
\item Theoretical predictions for the precision observables within the SM also require a range of SM parameters as inputs, most notably the top and bottom quark mass, $m_{t,b}$, the strong coupling $\alpha_s$, and the running electromagnetic coupling at the weak scale, $\alpha(M_Z$). $m_t$ can be measured with high precision at the ILC, but its extraction from the data requires resummed higher-order QCD corrections computed in an effective field theory framework (see section~\ref{sec:topmass} for more details). More precise determinations of $m_b$, $\alpha_s$ and $\alpha(M_Z)$ may be possible with improved lattice QCD calculations \cite{Aoki:2021kgd}.

The strong coupling $\alpha_s$ can also be extracted from measurements at ILC itself. One option is the analysis of jet rates and event shapes in $e^+e^- \to jets$ (see e.g.~Refs.~\cite{Verbytskyi:2019zhh,Marzani:2019evv} and references therein). These methods are subject to sizeable non-perturbative QCD effects that are not fully understood at this point, but further theory developments could make this an attractive option for a high-precision determination of $\alpha_s$. Another possibility is the determination of $\alpha_s$ from the branching ratio $\Gamma[Z\to\text{had.}]/\Gamma[Z\to\ell\ell]$, which is practically free of non-perturbative QCD effects.
However, new physics effects can also modify the Z-fermion couplings, so that this extraction method becomes model dependent. Both of these methods could complement a future lattice-QCD determination of $\alpha_s$ with improved precision.
\end{itemize}

Instead of running on the $Z$~pole, ILC can also produce high-precision measurements of $Z$~boson properties by using the radiative return method at $\sqrt{s}=250$~GeV, as described in section~\ref{sec:radreturn}. A detailed study of the theoretical needs for this program is still lacking. It will require the evaluation of multiple emissions of collinear initial state photons, see e.g.\ Ref.~\cite{Ablinger:2020qvo}, as well as full SM corrections to the process $e^+e^- \to \gamma Z$.

\section{Standard Model Effective Field Theory}
\label{sec:EFT}

To demonstrate that the SM is violated, it is only necessary to compare a precise theoretical calculation of an appropriate process to an experimental measurement.   Usually, though, we want more than this.   If a deviation from the SM is found, we would like to pinpoint its origin and express the deviation in such a way that it can be compared to theoretical models that extend the SM.  One way to do this is to follow the route that we took in our discussion of $WW$ pair production in 
Sec.~\ref{sec:Wboson250}, introducing new parameters into the expressions for the production amplitudes and then fitting these parameters to experiment.   For the Higgs boson couplings, there is a similar  approach, called the $\kappa$ parameterization~\cite{LHCHiggsCrossSectionWorkingGroup:2012nn}.   These approaches are frankly phenomenological.
In the two case just discussed, the parametrizations are not accurate beyond the leading order and so are inappropriate for precision studies.   In addition, these approaches are applied separately for each process under study and thus cannot 
take into account the synergies that result from  combining data from different reactions into a common fit.   To address these problems, we need a method that is better grounded in theory.

A powerful solution  to these problems is given by Effective Field Theory.   In this approach, we view the Standard Model as a part of a larger, more general, theory, that might contain many additional particles. This theory would be described by an underlying Lagrangian  $\L_0$.  The simplest way to proceed from this general starting point is to assume that all particles beyond those of the Standard Model itself are much heavier than the $W$ and $Z$; we will use $M$ to represent their mass scale.  It is then possible to integrate out all of the fields associated with the new heavy particles.   This produces an effective Lagrangian in which the corrections due to the heavy particles are represented by operators that depend on the light fields.   We can represent the result of this calculation as a sum over operators of successively higher dimensions,
\beq
     \L_{eff} =  \L_4 + \sum_i  {  b_i \over M} {\cal O}_i + \sum_j  { c_j \over M^2} {\cal O}_j   + \cdots \ . 
\eeq{Leffone}
In this expression, $\L_4$ contains all possible operators of dimension 4 and lower, the ${\cal O}_i$ are operators of dimension 5, the $ {\cal O}_j$ are operators of dimension 6, and so on.  The allowed operators are restricted by symmetry, since $\L_{eff}$ must have the symmetries of $\L_0$ after whatever symmetry breaking is generated at the scale $M$.   The factors of $M$ in \leqn{Leffone} reflect the dimensions of the operators and the requirement that $\L_{eff}$ has the units of 
(mass)$^4$.
 
There is a very attractive assumption that restricts this framework.   We can take the effective Lagrangian $\L_{eff}$ to have the gauge symmetry $SU(3)\times SU(2)\times U(1)$ and the field or particle content of the SM.   A stronger assumption is that the Higgs field is present in $\L_{eff}$ as an $SU(2)$ doublet field $\Phi(x)$ as in the SM.   Then the SM gauge symmetries are realized linearly on the fields in $\L_{eff}$.  Physically, this assumption treats the Higgs boson as light field at the scale $m_W$ while it treats all BSM effects as resulting from heavy fields at the scale $M$.   In the effective Langrangian, the Higgs field is the sole source of $SU(2)\times U(1)$ symmetry breaking.   Operators in $\L_0$ that obtain expectation values after electroweak symmetry breaking are represented in the effective theory as terms of the form
\beq
           {\cal O}_j \to    A  \Phi^\dagger \Phi  
\eeqn
where $A$ is the result of a loop calculation involving the heavy fields, either a c-number or a more general invariant function of the field $\Phi$.   This framework is called Standard Model Effective Field Theory  (SMEFT). 

These assumptions lead to some simplifications.  It is well-known that the SM is the most general renormalizable quantum field theory with the known particle content and the gauge symmetry $SU(3)\times SU(2)\times U(1)$.  Thus, $\L_4$ in \leqn{Leffone} is exactly the SM Lagrangian.   The only possible operators of dimension 5 are neutrino mass terms \leqn{Weinbergop}, and, similarly, operators with odd dimension parametrize fermion-number-violating interactions.   Thus, for the description of collider physics, we can restrict outselves to operators of even dimension.   We then rewrite \leqn{Leffone} as 
\beq
     \L_{eff} =  \L_{SM}  + \sum_j  {  c_j \over v^2} {\cal O}_j + \sum _k { d_k \over v^4} {\cal O}_k  + \cdots \ . 
\eeq{Leff}
To quote definite values for the coefficients $c_j$, $d_k$, \etc, we will use the electroweak symmetry breaking scale 
$v = 246$~GeV as the dimensionful quantity.    Then, if $M$ is at the TeV scale, the coefficients in \leqn{Leff} would naturally be of the order of magnitude:  $c_j \sim~10^{-2}$, $d_k \sim 10^{-4}$, $\ldots$, perhaps with additional suppression by small coupling constants.   In this naive method of  estimation, the 
Higgs boson couplings, TGCs, and 4-fermion couplings would be expected to receive corrections at the percent level from the  $c_j$ coefficients of dimension-6 operators while the effect of the $d_k$ coefficients would be ignorable.   Of course, the actual values of the SMEFT coefficients  would be determined in each particular theoretical model, which might provide 
large or small order-1 factors.    But we will take as a working assumption that dimension-6 coefficients give the only relevant corrections at the levels of precision that will be achieved by Higgs factories. This framework  has model-dependence, but that dependence is weak. In particular, it does not prejudice us toward any specific type of model but rather incorporates on an even footing models from weakly coupled scalar extensions, to supersymmetry, and to composite Higgs and extra dimensional models of new physics.

There might be several reasons to question these assumptions.   First, as we have emphasized in Sec.~\ref{sec:newparticles}, there might be new particles that exist in the mass region of a few hundred GeV but have not been discovered in LHC searches.  Such particles might give large corrections to $\L_{eff}$ that are not well described by dimension 6 operators alone.
The discovery of large corrections will of course be welcome; as for the interpretation, it should be noted that the effects of the top quark and the Higgs boson on precision electroweak observables are well-described by the $S$ and $T$ parameters, which are linear combinations of the dimension 6 coefficients, even though these particles do not have masses much greater than the $Z$ mass.   Second, there might be very light new particles, such as those described in Secs.~\ref{sec:dark500} and \ref{sec:physfixedtarget}.  We will assume that these particles can give additional contributions to the Higgs boson width but do not affect the precision electroweak observables.  This is plausible because of their very feeble couplings.

A more general issue is our assumption that the Higgs boson is the sole source of $SU(2)\times U(1)$ symmetry breaking.  It is possible that there is an additional $SU(2)\times U(1)$-breaking expectation value in what we have deemed the heavy sector. This might be reflected in a special treatment of the mass generation for the top quark with respect to that for the vector bosons.  Another possibility is that a new scalar obtains a large fraction of its mass from the Higgs boson.  In these cases, the appropriate effective Lagrangian would be one with only the subgroup $SU(3)\times U(1)$ realized linearly and the broken generators of $SU(2)\times U(1)$ realized nonlinearly.   This is called the Higgs Effective Field Theory (HEFT) or the Electroweak Chiral Lagrangian (EWChL)~\cite{Buchalla:2013rka,Buchalla:2015wfa}.    The HEFT allows considerably more parameter freedom than the SMEFT.  In particular, while in the SMEFT the deviations from the SM values of the Higgs boson couplings are naturally of the order of several percent, in the HEFT these deviations are unconstrained by dimensional analysis and can be of order 1.    The LHC experiments have  measured the Higgs boson couplings to be in good agreement with their SM values at the 10--20\% level.   This situation is expected in the SMEFT---and points to smaller corrections that can be measured at the next level of precision---but it requires special explanation in the HEFT.  For that reason, we consider the SMEFT to be preferred experimentally, and we will work in that context.   It is possible but rather difficult to distinguish the HEFT and SMEFT frameworks experimentally. This requires measurement of several  cross sections for multiple Higgs production.   Please see \cite{Alonso:2016oah,Helset:2020yio,Cohen:2021ucp} for a detailed discussion.

In this report, we will use the SMEFT as a practical tool for combining measurements from  a number of different ILC reactions in a coherent framework.   The next section will explain how we do this.  Other approaches to the global fitting of SMEFT parameters to data from $\ee$  Higgs factories are described in~\cite{DeBlas:2019qco,deBlas:2019rxi,deBlas:2022ofj}. Despite the differences in philosophy among these papers, the actual results are in good agreement, giving confidence in the validity of this approach.

\section{A practical SMEFT analysis for ILC}
\label{sec:SMEFTfit}

The ILC requires a model framework to make specific statements about the Higgs boson couplings and electroweak observables.   An important property of the Higgs boson is its total width $\Gamma_H$.   The total width of the Higgs boson must be known to interpret the data.  The most commonly measured observable  is the rate of a Higgs boson process, which is given by 
\beq
      \sigma \cdot BR (\ee\to A\bar A) =    \sigma(\ee\to H + X) \cdot {\Gamma(H \to A\bar A)
              \over \Gamma_H}
\eeqn
Theoretical predictions, both in the SM and in new physics models, are given for the absolutely normalized partial widths $\Gamma(H\to A\bar A)$.  To extract these, we need to know $\Gamma_H$.  On the other hand, $\Gamma_H$ has the SM value of 4.3~MeV for a 125~GeV Higgs boson mass.   This value is so small that it cannot be extracted with high precision directly from experiment, either at $\ee$ or at hadron colliders.  To determine $\Gamma_H$, we need a model. 

The model used to extract $\Gamma_H$ should on the one hand be general and model-independent, while on the other hand it should have few enough 
parameters that these can all be determined from data without degeneracies.   Such a model must be a compromise, but hopefully we can use theory insight to choose a model that satisfies both requirements as well as possible.

It is quite remarkable that the ILC provides a sufficiently large number of measurements of sufficient specificity that we can use SMEFT as a model to  reconstruct the Higgs width.   General SMEFT has of course an infinite number of parameters, and even truncating SMEFT to consider only dimension-6 baryon- and lepton-number conserving operators leads to 76 new coefficients for 1 generation and 2499 for three generations. However,  the set of coefficients involved in  ILC reactions at the tree level is much smaller.  We will argue in a moment that  18 operators suffice.  Choosing these as parameters of the model, we add 4 relevant SM parameters and 2 parameters representing the Higgs boson decay rates to invisible and unclassified exotic decay modes. Removing, for the moment, the Higgs self-coupling and a particular 4-fermion operator constrained by measurements in that sector, we arrive at a practical SMEFT fitting scheme with 22 parameters~\cite{Barklow:2017awn,Barklow:2017suo}.   These parameters can be fit to measurements of Higgs decays.  But also, since the  SMEFT Lagrangian is intended to be a complete low-energy representation of particle physics, we can add data from precision electroweak measurements, $\ee\to W^+W^-$, fermion pair production, and other reactions that can be studied at the ILC. With care, we can also make use of particular quantities measured at the LHC.  This gives a robust framework to use in translating the ILC data to absolutely normalized values of the Higgs boson  partial widths and the value of the total Higgs width $\Gamma_H$. 

The model prescriptions for this ``model-independent'' framework are:
\begin{enumerate}
\item We truncate the SMEFT to renormalizable and dimension-6 operators only.   The fit is done strictly at the linear level in SMEFT operator coefficients.
\item We calculate the new physics contributions to ILC processes at the tree level only, and drop all operators that do not contribute in the tree-level expressions.  It is consistent to drop all 4-fermion operators except for the operator that corrects $G_F$ and to drop all operators that contain quark and gluon fields except for the operators that correct the $W$ and $Z$ total widths.   In this framework, corrections to the Higgs boson self-coupling do not contribute to the set of observables that we consider.  We will discuss fits including the Higgs self-coupling in 
Sec.~\ref{sec:expectself}.
\item   Given the strong constraints that will result from measurements of  $\ee\to \mu^+\mu^-$, we  drop the 4-fermion contribution to $G_F$. 
\item We assume lepton universality.  That is, we assign the same coefficients to corresponding operators with $e$, $\mu$, and $\tau$. 
\item Results from the LHC and expected results from the HL-LHC can be added only if these do not expand the set of SMEFT operators included in the fit.  We expect that the ratios of Higgs boson branching ratios to $\gamma\gamma$, $ZZ^*$, $Z\gamma$, and $\mu^+\mu^-$, all measured in central Higgs boson production, meet this criterion, and we will include only these inputs.
\item We drop all CP-violating operators and all operators giving flavor-nonconserving Higgs boson decays.  This is justified because these  operators with coefficients $c_i$ contribute to the CP-conserving, flavor-conserving Higgs  observables only in order $c^2_i$, while we keep new physics contributions in linear order only. 
Of course, it is 
extremely important to search for these couplings, as we have emphasized in Sec.~\ref{sec:Higgs250}, but these searches
are outside the fit presented here.
\item We include invisible and unclassified exotic  decays of the Higgs boson with two parameters, the Higgs branching ratios to these modes.   We assume that the light states into which the Higgs boson could decay have no effect on precision electroweak observables.  Note that Higgs decays to invisible final states are directly measureable from  $\ee\to HZ$ by observing the $Z$ recoil against nothing.  Also,  very general classes of modes of Higgs decay to exotic final states are directly observable, as explained in Sec.~\ref{sec:HiggsExotic}.   Thus, leaving this branching ratio as a free parameter is a very conservative assumption.
\end{enumerate}

It can be shown that the assumptions 3 and 4 can be dropped from the analysis with almost no effect on the projections for Higgs couplings by including additional ILC measurements in the fit.  To analyze this point, we must carry out a more complicated fit with many additional parameters.  Such a fit would also include the separate precision electroweak results for $e$, $\mu$, and $\tau$ and estimates of the precision with which 4-fermion processes will be measured at the ILC. This fit is described in the presentation \cite{MEPatSnowmass}; it gives essentially the same results as those shown below.  It is noteworthy that our fits are so overconstrained that $G_F$ is not needed as an input. 

These assumptions do include the assumption of a clear separation in mass scale between the particles of the SM---including the Higgs boson---and particles mediating new interactions.  However, there is no assumption that the new physics model be of a specific type, for example,  weak or strong coupling, leptophilic or leptophobic, \etc \    The use of 
SMEFT has a clear advantage over other modelling schemes for the Higgs width in that it allows us to use constraints from the well-established gauge symmetry $SU(2)\times U(1)$ to reduce the number of parameters.

We then take the set of SMEFT operator coefficients used in the practical fit as a subset of the full set of dimension~6
operators in the Warsaw basis~\cite{Grzadkowski:2010es}.  Our effective Lagrangian is 
\beq
  \L = \L_{SM} + \L_{H} + \L_{W,B} + \L_{\Phi\ell}  + \L_{\Phi q} + \L_{\Phi f} + \L_g \ .
\eeq{refL}
Here $\L_{SM}$ is the Standard Model Lagrangian, $\Phi$ is the scalar Higgs doublet field.
For definitiveness, we take the defining mass 
scale of the dimension~6 operator coeffiicients to be $v = 246$~GeV.  Then 
 the dimension~6 terms are given by, for the terms depending only on the Higgs and vector boson
fields,
\beqa
   \L_{H} &=&   {c_{H}\over 2 v^2}   (\del_\mu \Phi^\dagger \Phi)^2  
             +{c_T\over 2 v^2} (\Phi^\dagger \Dlr^\mu \Phi)(\Phi^\dagger \Dlr_\mu \Phi)
                             - {\lambda c_6 \over v^2 }(\Phi^\dagger \Phi)^3 \CR
 \L_{W,B} &=& {g^2 c_{WW}\over v^2} (\Phi^\dagger \Phi)  W^a_{\mu\nu} W^{a \mu\nu} +    {2 g g' c_{WB}\over v^2} (\Phi^\dagger t^a \Phi)   W^a_{\mu\nu} B^{ \mu\nu} \CR
   & & \hskip 0.4in +   {g^{\prime 2} c_{BB}\over v^2} (\Phi^\dagger \Phi)  B_{\mu\nu} B^{ \mu\nu} + 
                     {c_{3W}\over 6 v^2} \eps_{abc} W^a_{\mu}{}^\nu W^b_\nu{}^\rho W^c_\rho{}^\mu  \ . 
\eeqa{cforbosons}  
The terms depending on the Higgs fields and the electron fields are 
\beqa
 \L_{\Phi \ell}  &=& {c_{\Phi L}\over v^2} (\Phi^\dagger\,  i \Dlr_\mu \Phi)   (\bar L^\dagger \gamma^\mu L) + 
     { 4 c_{\Phi L}^\prime\over v^2 } (\Phi^\dagger  t^a \,  i \Dlr_\mu \Phi)
       (\bar L^\dagger t^a  \gamma^\mu L)  \CR
                & & \hskip 1.0in    +  {c_{\Phi E}\over v^2} (\Phi^\dagger \, i\Dlr_\mu \Phi)   (\bar e^\dagger \gamma^\mu)   \ , 
\eeqa{cforHL}
where $L$ and $e$ are the left- and right-handed fields of the first lepton generation and $t^a = \sigma^a/2$ is the weak isospin generator.   The operators depending on the Higgs fields and other quark and lepton fields 
are defined similarly.  In these formulae, 
\beqa
  \Phi^\dagger\Dlr_\mu  \Phi &= &
   (\Phi^\dagger D_\mu  \Phi - (D_\mu \Phi^\dagger) \Phi ) \CR
    \Phi^\dagger\Dlr_\mu^a  \Phi &= &
   (\Phi^\dagger t^a D_\mu  \Phi - (D_\mu \Phi^\dagger t^a) \Phi ) 
 \eeqan
 
 The dimension-6 operator that shift the Higgs-$\tau$ Yukawa coupling is 
\beq
 \L_{\Phi\tau} =   { y_\tau c_\tau\over v^2}  (\Phi^\dagger \Phi)  (\bar L_\tau \cdot \Phi e_\tau)   \ , 
\eeq{cforHiggs}
and the other operators that contribute to scalar couplings are constructed in a similar way. There are dimension-6 
operators that couple to the Higgs boson through a magnetic moment interaction, but these do not contribute to 
Higgs couplings at the tree level.   The operator
\beq
 \L_{g} = {g_s^2 c_{gg}\over v^2} (\Phi^\dagger \Phi) 
 G^a_{\mu\nu} G^{a \mu\nu} 
\eeqn
shifts the Higgs boson partial width to gluons.   This partial width also receives corrections from loop diagrams involving the top quark and SMEFT operators associated with the top quark.  But we are concerned here with only one amplitude, the $Hgg$ coupling on the Higgs boson mass shell, so we will represent all of these effects by the single parameter $c_{gg}$.

Our practical SMEFT fit then contains 4 SM parameters, 6 parameters from \leqn{cforbosons}  (excluding $c_6$), 3 parameters
from \leqn{cforHL}, 2 additional combinations of coefficients that shift the $W$ and $Z$ widths, 5 parameters of the form \leqn{cforHiggs} for the Higgs couplings to $b$, $t$, $c$, $\mu$, and $g$, plus the two parameters for invisible and exotic Higgs decays mentioned in point 6 of our assumptions, for a total of 22 parameters.

\section{Expectations for the practical SMEFT fit}
\label{sec:SMEFTexpectations}

We are now ready to present the expected uncertainties on individual Higgs boson couplings that arise from the SMEFT fit described in the previous section.   The results here represent a minor update of the similar fit  presented
 in \cite{LCCPhysicsWorkingGroup:2019fvj}.   Similar fits with slightly different assumptions but very similar results have been carried out in \cite{deBlas:2019rxi}.

As we have described  in \cite{LCCPhysicsWorkingGroup:2019fvj}, the inputs to the fit are the defining SM observables $\alpha$, $m_Z$, $G_F$, and $m_h$, and the additional electroweak observables $m_W$, $A_\ell$, $\Gamma(Z\to \ell^+\ell^-)$, 
$\Gamma_Z$, and $\Gamma_W$.    For $\Gamma_W$, we expected that this can be improved to a determination at the level of $10^{-3}$ from the electroweak and the value of  $BR(W\to  \ell \nu)$ measured in $\ee\to W^+W^-$.   Our new, still preliminary, studies show that this is conservative.   We also include the measurements of the cross section for $\ee\to HZ$ and the various $\sigma\times BR$ values for Higgs decays in this mode, including the invisible mode, corresponding values of $\sigma\times BR$ in the $WW$ fusion reaction,  and the measurements of the TGC parameters. Our methods for obtaining estimates of uncertainties in these quantities were explained in Chapters 8 and 10.  In all cases, the integrated luminosities used are those in run plan shown in Fig.~\ref{fig:ILC-staging}.   Finally, to close the fit, we need some measurements from the LHC, in particular, the ratios of branching ratios of the Higgs boson to $ZZ$, $\gamma\gamma$, $\gamma Z$, and $\mu^+\mu^-$.    With these inputs, the 22-parameter fit has no unconstrained direction.  

\begin{figure*}[t]
\begin{center}
\includegraphics[width=0.8\hsize]{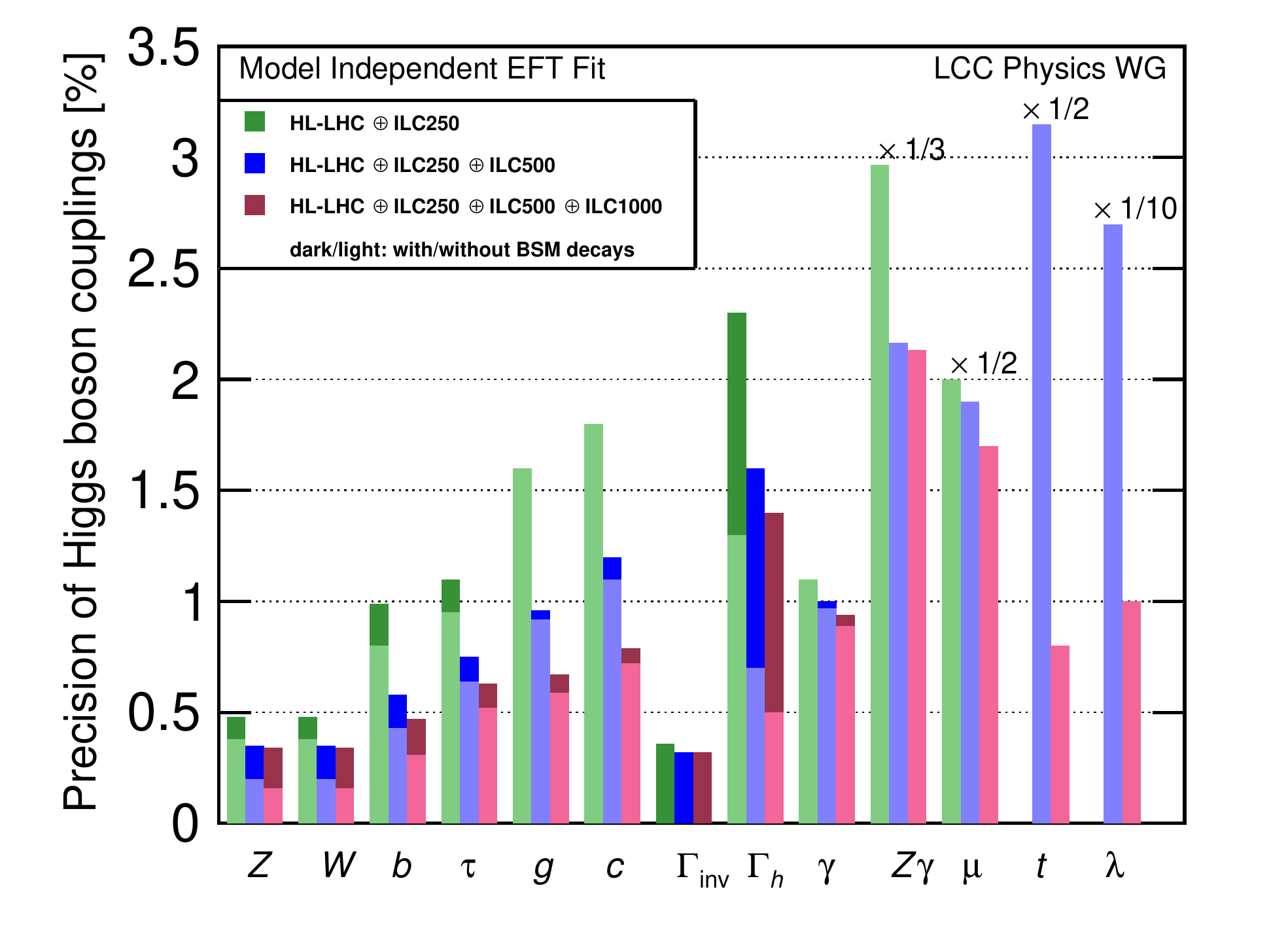}
\caption{Projected Higgs boson coupling uncertainties for ILC250,
  ILC500, and ILC1000, also  incorporating  results expected from the
  HL-LHC, based on the SMEFT analysis described in the text.   The
  darker bars show the results allowing invisible and exotic Higgs
  decay channels; the lighter bars assume that these BSM decays are
  not present.  The column $\lambda$ refers to the $HHH$ coupling.  In the last four columns, 
  all bars are rescaled by the indicated factor.   From \cite{LCCPhysicsWorkingGroup:2019fvj}.}  
\label{fig:Higgsresults}
\end{center}
\end{figure*}

The results from the fit are shown in Fig.~\ref{fig:Higgsresults} and in  Tables~\ref{tab:ILCHiggs} and \ref{tab:ILCHiggsEW}.   In the figure and in Table~\ref{tab:ILCHiggs}, we show the expectations from the 22-parameter fit, and, for comparison with SMEFT fits in \cite{deBlas:2019rxi}, and elsewhere, expectations from a 20-parameter fit that assumes that the Higgs boson has no exotic decays. We also include the improved estimates for and the Higgs self-coupling and the top quark Yukawa coupling that were presented in Secs.~\ref{sec:HiggsSelf} and \ref{sec:Higgstop}.

It is interesting to ask what level of precision in the precision electroweak observables is needed to achieve the values quoted in this table. Actually, the projected uncertainties in Table~\ref{tab:ILCHiggs} are obtained using only the level of precision that is achievable from running at 250~GeV and analyzing the radiative return reactions to improve the uncertainty in $A_\ell$.  It is difficult to see improvements beyond this point as long as the possibility of uncharacterized exotic decays is included in the fit.   So, in Table~\ref{tab:ILCHiggsEW},  we assume that there are no exotic decays and carry out the 20-parameter fit using the levels of precision expected from radiative return, from the dedicated $Z$ pole program discussed in Sec.~\ref{sec:Zpole}, and from the levels of precision expected from the TeraZ  program at the FCC-ee~\cite{FCC:2018byv}.

%%%%%%%%%%%%%%%%
%%%%%%%%%%%%%%%%%%%%%%%%%%%%%%%%%%%%%%%%%%%%%%%%%%%%%%%%%%%%%%
%

%%%%%%%%%%%%%%%%%%%%%%%%%%%%%%%%%%%%%%%%%%%%5
\begin{table}
\begin{center}
\begin{tabular}{l|cc|cc|cc}
      &  \multicolumn{2}{c}{ ILC250 }     &  
\multicolumn{2}{c}{ ILC500}  &
               \multicolumn{2}{c}{ ILC1000 }\\ 
coupling & full & no BSM & full & no BSM & full & no BSM \\ \hline 
$hZZ$            &             0.49 & 0.38 &            0.35 &  0.20    &
                 0.34  &  0.16   \\ 
$hWW$            &          0.48 & 0.38  &            0.35  &  0.20      &                                                
                              0.34 &  0.16 \\ 
 $hbb$            &     0.99  & 0.80  & 0.58 &0.43   &  0.47 & 0.31 \\ 
$h\tau\tau$    &          1.1 & 0.95 &   0.75 & 0.63 & 0.63 & 0.52 \\ 
$hgg$                  &  1.6 &  1.6    &   0.96  & 0.91   
 & 0.67 &  0.59 \\ 
$hcc$         &   1.8  &  1.7&  1.2   &   1.1 &    0.79 &   0.72  \\ 
$h\gamma\gamma$ &  1.1  & 1.0&   1.0 &  0.96 & 0.94  & 0.89 \\ 
$h\gamma Z$     &  8.9 &  8.9 &   6.5 &  6.5 &  6.4  & 6.4 \\
$h\mu\mu$ &  4.0  & 4.0 &  3.8  &  3.7 &     3.4  &  3.4     \\ 
$htt$  &   ---     &   --- &   6.3 & 6.3  &      1.0 & 1.0    \\ 
$hhh$  &  ---    &  ---& 20 &  20 &    10  & 10  \\ \hline 
$\Gamma_{tot}$ & 2.3 & 1.3 & 1.6 & 0.70 &  1.4  &  0.50\\  
$\Gamma_{inv}$ &   0.36 & ---   & 0.32 & --- &  0.32  & --- \\  \hline
\end{tabular}
\end{center}
\caption{Projected uncertainties in the Higgs
  boson couplings for the ILC250, ILC500, and ILC1000, with
  precision LHC input.  All values are  {\it relative}
  errors,
given in percent (\%).   The columns labelled ``full'' refer to a
  22-parameter fit including the possibility of invisible and exotic
  Higgs boson decays.   The columns labelled ``no BSM'' refer to a
  20-parameter fit including only decays modes present in the SM. }
\label{tab:ILCHiggs}  
\end{table}
%%%%%%%%%%%%%%%%%%%%%%%%%%%%%%%%%%%%%%%%%%%%%%%%%%

%%%%%%%%%%%%%%%%%%%%%%%%%%%%%%%%%%%%%%%%%%%%5
\begin{table}
\begin{center}
\begin{tabular}{l|ccc|ccc}
      &  \multicolumn{3}{c}{ ILC250 }     &  
\multicolumn{3}{c}{ ILC500}  \\ 
coupling & RadRtrn & GigaZ & TeraZ &  RadRtrn & GigaZ & TeraZ        \\ \hline 
$hZZ$            &   0.38   &   0.35       &  0.30      &   0.20    &    0.20       &  0.19  \\
 $hWW$            &  0.38  &   0.35     &   0.31     &   0.20      &    0.20        &  0.19   \\                                    
  $hbb$            &  0.80  &    0.78    &  0.77       &    0.43     &   0.43         &   0.43  \\     
$h\tau\tau$    &  0.95 &  0.94    &    0.92     &   0.63    &     0.63       &   0.63  \\   
$hgg$                 &   1.6  &  1.6    &   1.6      &  0.91    &     0.91       &  0.91   \\      
$hcc$                  &   1.7  &  1.7    &   1.7      &  1.1    &    1.1        &  1.1   \\   
$h\gamma\gamma$    &   1.0  &  1.0    &   1.0      &  0.96    &    0.96        &  0.96   \\   
$h\gamma Z$              &  8.9  &  8.5    &  7.9       & 6.5    &    6.4        &   5.8  \\    
$h\mu\mu$            &   4.0  &  3.9    &   3.9     & 3.7    &    3.7        &   3.7  \\    \hline
$\Gamma_{tot}$           &   1.29  &  1.26    &  1.21       & 0.70   &   0.70         & 0.69    \\    \hline
\end{tabular}
\end{center}
\caption{Projected uncertainties in the Higgs
  boson couplings for the ILC250 and  ILC500,  with
  precision LHC input, showing the dependence on precision electroweak measurenments.  All values are  {\it relative}
  errors,
given in percent (\%). The fit assumes that there are no exotic Higgs boson decays.   The columns labelled RadRtrn use the uncertainties expected from the radiative return events at 250 Gev.   The columns labelled GigaZ use the uncertainties 
expected from the ILC dedicated Z pole program discussed in Sec.~\ref{sec:Zpole}.  The columns labelled TeraZ use the 
uncertainties 
expected from the TeraZ program at the FCC-ee~\cite{FCC:2018byv}.}  
\label{tab:ILCHiggsEW}  
\end{table}
%%%%%%%%%%%%%%%%%%%%%%%%%%%%%%%%%%%%%%%%%%%%%%%%%%

%%%%%%%%%%%%%%%%%%%%%%%%%%%%%%%%%%%%%%%%%%%%5
\begin{table}
\begin{center}
\begin{tabular}{l|cc|cc}
      &  \multicolumn{2}{c}{ ILC500 }     &  
\multicolumn{2}{c}{ ILC500 w. $c_6$} \\ 
coupling & full & no BSM & full & no BSM  \\ \hline 
$hZZ$        &            0.35 &  0.20    &  0.37  &  0.21   \\ 
$hWW$            &       0.35  &  0.20      &     0.37 &  0.21\\ 
 $hbb$            &  0.58 &0.43   &  0.60 & 0.43 \\ 
$h\tau\tau$    &   0.75 & 0.63 & 0.78 & 0.64 \\ 
$hgg$                  &   0.96  & 0.91   & 0.97 &  0.92\\ 
$hcc$         &  1.2   &   1.1 &   1.2 &   1.1  \\ 
$h\gamma\gamma$ &    1.0 &  0.96 & 1.0  & 0.97 \\ 
$h\gamma Z$     &  6.5 &  6.5 &  7.0 & 6.4 \\
$h\mu\mu$ & 3.8  &  3.7 &     3.8  &  3.7     \\  \hline
$  c_6$      &     --  &   --   &    53.      &   52.        \\ \hline
$\Gamma_{tot}$ & 1.6 & 0.70 &  1.6  &  0.70\\  
$\Gamma_{inv}$ &   0.36 & ---   & 0.32 & ---  \\  \hline
\end{tabular}
\end{center}
\caption{Projected uncertainties in the Higgs
  boson couplings for the  ILC500, including loop effects proportional to  the Higgs self-coupling.  All values are  {\it relative}
  errors,
given in percent (\%).  The  rest of the notation is as in Table~\ref{tab:ILCHiggs}.}
\label{tab:ILCHiggswc6}  
\end{table}
%%%%%%%%%%%%%%%%%%%%%%%%%%%%%%%%%%%%%%%%%%%%%%%%%%

\section{Expectations for the Higgs self-coupling}
\label{sec:expectself}

Up to this point, we have not included in our fits the coefficient $c_6$ that modifies the Higgs self-coupling,
\beq
     c_6  =   \lambda_{eff}/\lambda  \ .
\eeqn
Still, the measurement of the Higgs self-coupling is an important goal for future colliders.  We have discussed already in Sec.~\ref{sec:HiggsSelf} how the ILC can determine the Higgs self-coupling through two separate processes for double Higgs production.   Here we will fill in some details of the interpretation of these processes in SMEFT.

The reaction of double Higgs-strahlung, $\ee\to Z HH$, contains both diagrams with double Higgs emission from the $Z$ as well as a diagram containing the Higgs self-coupling.   There are also several new vertices from the dimension-6 SMEFT Lagrangian that contribute to this process.   Thus, it is not correct to treat the effect of the self-coupling in isolation.  One must consider all possible effects on the cross section from dimension-6 SMEFT contributions, and it might not be that the variation of the self-coupling is the dominant one.   This question was addressed in \cite{Barklow:2017awn}. The intermediate result did not look promising.  For unpolarized beams at ILC500, the dependence of the cross section on SMEFT parameters includes
\beq
    \sigma/\sigma_{SM} = 1+  0.56 c_6 - 4.15 c_H + 15.1 c_{WW}  + 62.1 (c_{\Phi L} + c^\prime_{\Phi L}) - 53.5 c_{\Phi E} + \cdots\ , 
\eeq{sigforZHH}
In addition to $c_6$, many other dimension-6 coefficients affect this process, and some enter with very large numerical factors. 
However, all of the other  dimension-6 parameters in \leqn{sigforZHH} are strongly constrained by the SMEFT analysis of 
single-Higgs production, sufficiently so that the effect of those terms is completely negligible with respect to the statistical 
error on the value of the cross section.

%%%%%%%%%%%%%%%%
\begin{figure*}[t]
\begin{center}
\includegraphics[width=0.4\hsize]{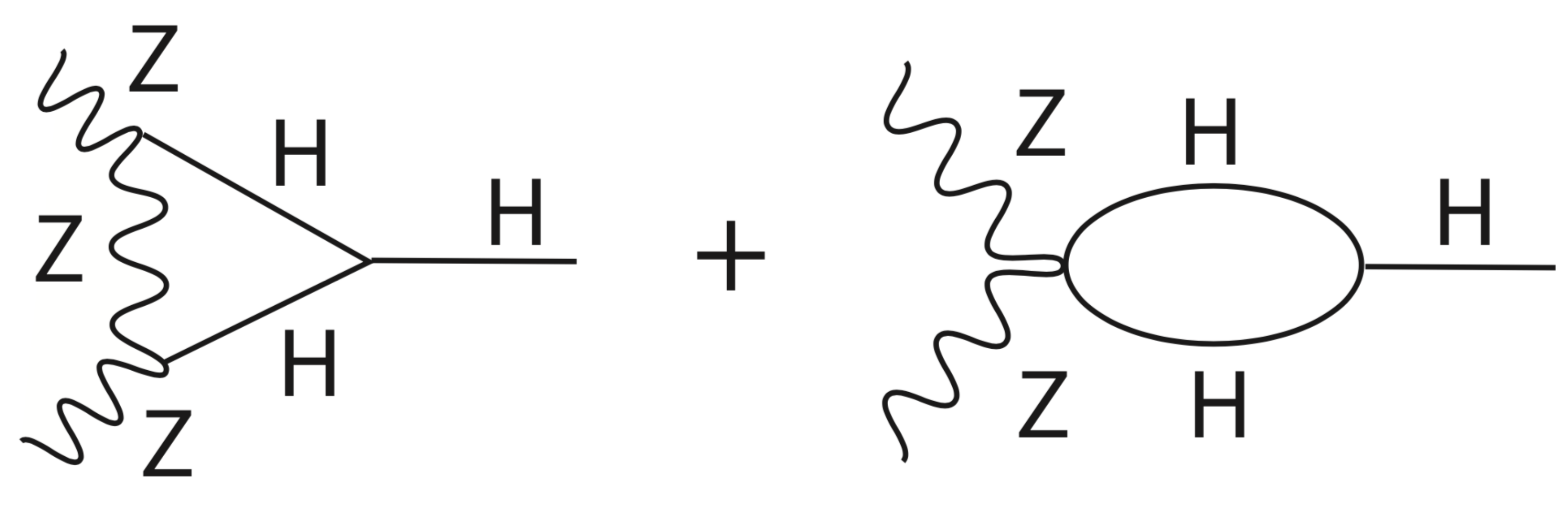}
\caption{Vertex correction giving a dependence of the $HZZ$ and $HWW$ vertices on the Higgs self-coupling.}
\label{fig:HiggsVertex}
\end{center}
\end{figure*}
%%%%%%%%%%%%%%%%%%%%%%%%%%%%%%%%%%%%%%%%%%%%%%%%%%%%%%%%%%%%%%
%

It is also possible to extract information on the Higgs self-coupling from the values of single-Higgs cross sections.   The 
Higgs self-coupling enters the cross section for $\ee\to HZ$ and the partial widths for $H\to ZZ^*$ and $H\to WW^*$  through a vertex correction shown in Fig.~\ref{fig:HiggsVertex}.   In \cite{McCullough:2013rea}, it was pointed out that the vertex loop correction has a sharp dependence on the momentum of the off-shell vector boson near the threshold for $V^*\to HV$ and, thus, the effect might be 
measurable at $\ee$ colliders.   At ILC250, the size of the enhancement to the $\ee\to HZ$ cross section is about 1.5\%.   Again, the effect 
can easily be obscured by variation of other SMEFT parameters.   The formula comparable to \leqn{sigforZHH} for  this cross section at 250~GeV is 
\beq
    \sigma/\sigma_{SM} = 1 + 0.015  c_6 - c_H + 4.7 c_{WW}  + 13.9 (c_{\Phi L} + c^\prime_{\Phi L}) - 12.1 c_{\Phi E} + \cdots\ , 
\eeq{sigforZH}
In particular, the effect at any single energy is highly degenerate with an enhancement of the $HZZ$ coupling, for example, through a change in $c_H$.   It is possible to gain some sensitivity by comparing the  $\ee\to HZ$ cross sections at two different energies, for example, 250 and 500~GeV.   Table~\ref{tab:ILCHiggswc6}  shows the effect of a SMEFT fit that adds the 1-loop contributions to the Higgs vertices to the calculation of tree diagrams.   This analysis gives an uncertainty of 53\% on $c_6$, which can slightly improve the  uncertainty from the more direct measurement of double Higgs production.   More relevant, though, is the fact that the determination of the other SMEFT parameters contributing to the Higgs boson couplings is very robust with respect to the addition of this parameter.

%  This section is not ready for prime time
% \section{Constraints on heavy-quark operators}
% \label{sec:heavyquarks}

% (The list of dimension-6 SMEFT operators expands greatly when operators that specifically involve heavy-quark fields are included. This section will discuss the relation more standard form-factor descriptions of BSM corrections to top quark properties to a full SMEFT analysis.  What experiments are required to resolve all of the ambiguities?)

\chapter{Big Physics Questions Addressed by ILC} 
\label{chap:bigquestions}

The discovery of the 125~GeV Higgs boson poses even more questions that it answers. Within the SM, the Higgs boson explains the the origin of all particle masses through the Higgs mechanism. The 125~GeV boson seems to fulfill this role, but still there remain many questions both about this boson and about the SM itself.
Is this boson solely responsible for the breaking of electroweak symmetry and the generation of mass?  Is it a singleton, or is it merely the first of several Higgs bosons?  What sets the mass parameter for this boson?  Can we  explain electroweak symmetry breaking in physical terms, with a theory in which that mass is computable?  If the SM is correct up to very high scales and the its parameters are equal to the current central values, the vacuum we see is unstable.  Is this the true situation, and, either way, what is the true behavior of the vacuum of the universe far in the future?   In addition, the discovery of the Higgs boson sharpens questions that have been asked since the SM was first formulated. What is the origin of flavor and the fermion generations? Why is there more matter than antimatter? What is the nature of dark matter?  What other types of new matter exist in nature?  

Through its comprehensive set of precision measurements of the couplings of the 125~GeV Higgs boson, and through its larger program of measurements of electroweak reactions at the weak-interaction scale, the ILC has the power to give insight into all of these questions.  In this chapter, we will outline these questions in more detail and describe their relation to ILC measurements.  In the next chapter, we will illustrate the insights from the ILC in a complementary way, through quantitative comparison of the ILC projected measurements with the predictions of models of physics beyond the SM. 

  \section{Can the Standard Model be exact to very high energies?}
  \label{sec:SMexact}
  
At TeV energies, the Higgs field quartic coupling increases with energy due to renormalization-group running.
However, it is a prediction of the SM that this coupling turns over and begins to decrease at very high energies.  For the current central values of SM parameters, the Higgs quartic coupling becomes negative at about $10^{11}$~GeV, leading to a vacuum instability, assuming that the SM is still exact at those energies.  Within the SM, the outcome depends sensitively on the values of the Higgs boson mass and the top quark mass.   We do not know today what the SM predicts for our universe.  

\begin{figure}
\begin{center}
   \includegraphics[width=0.5\hsize]{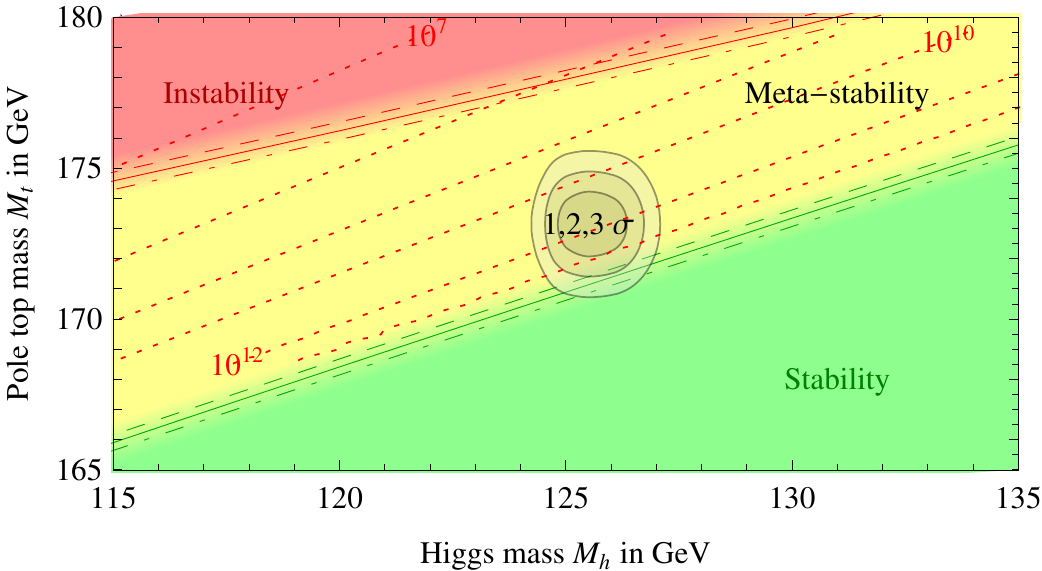} \\
  \includegraphics[width=0.3\hsize]{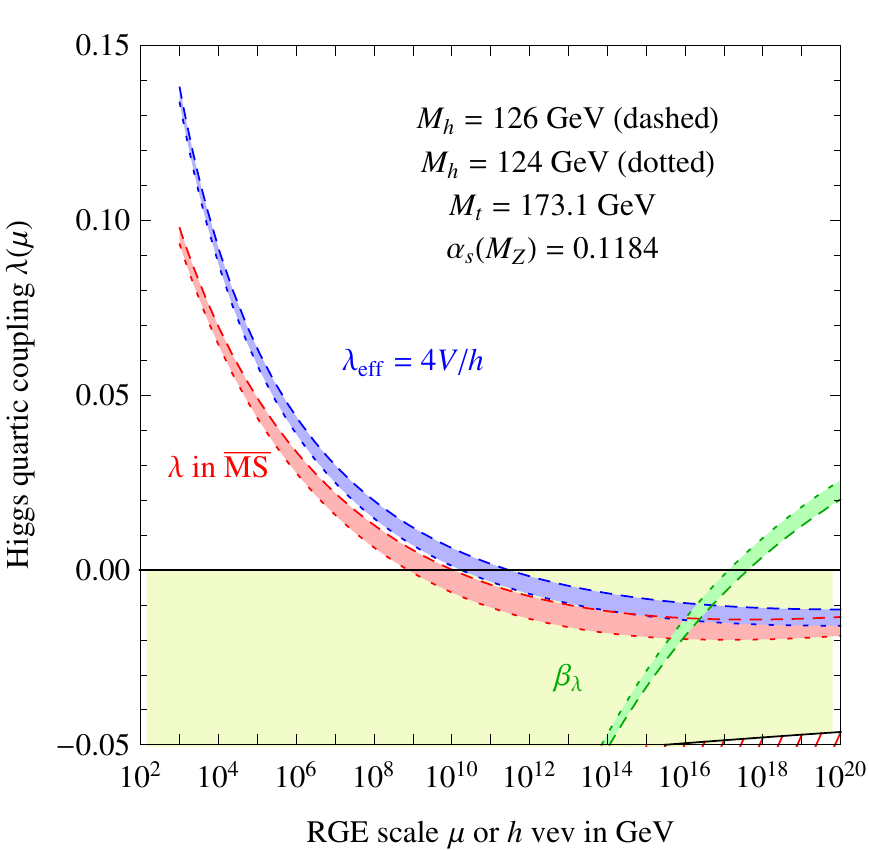} \quad 
 \includegraphics[width=0.3\hsize]{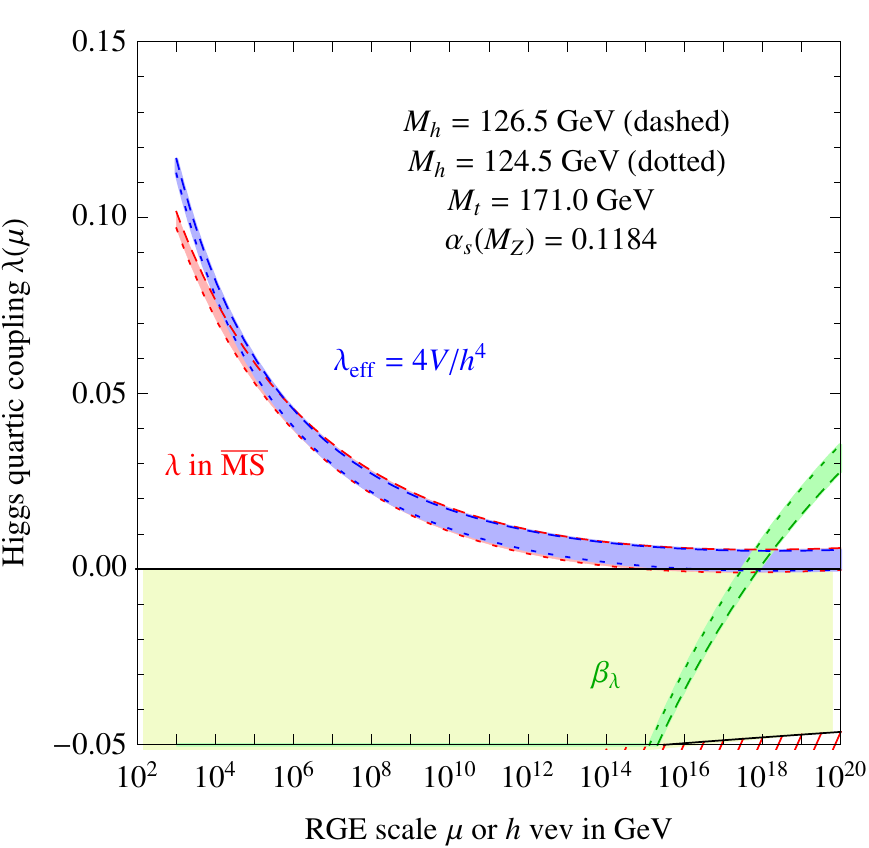} 
\end{center}
\caption{top: Regions of stability, metastability, and instability of the SM vacuum, shown as a function of $m_H$ and $m_t$,showing the current best values in the region of metastability. bottom: Renormalization-group evolution of the Higgs quartic coupling to large $Q$ assuming $m_t = 173.1$~GeV (left) and $m_t = 171.0$~GeV (right). From~\cite{Degrassi:2012ry}. }
\label{fig:deadoralive}
\end{figure}

Precision measurements of these two quantities to the accuracy projected for the ILC will resolve this.  The nature of the SM vacuum state as a function of the top quarks and Higgs boson masses has been studied in ~\cite{Degrassi:2012ry} and more recently in \cite{Chigusa:2018uuj,Chigusa:2017dux}.  The current situation is shown in  Fig.~\ref{fig:deadoralive}.   As we can see from  the bottom graphs in this figure, a change of 2~GeV in the central value of the top quark mass brings us from eventual instability to stability.  Since the calculation uses the short-distance value of the top quark mass, the uncertainty must include the error in converting the top quark mass as measured in experiment  (\eg, the pole mass) to a short-distance value (\eg, the $\MSB$ mass).   Thus, this calculation, within the SM, requires very precisely understood inputs at the energy of the electroweak scale.  We have seen above that the ILC will determine the Higgs boson mass to a precision of 15~MeV and the short-distance top quark mass to a precision of 40~MeV, well within the requirements for a definitive statement.

If the future experiment proves the SM vacuum metastable, two possibilities arise. On the one hand, the SM could be exactly correct up to the scale of the instability.  In that case, we will need to understand how the universe before the electroweak phase transition settled down to the metastable vacuum of today.    Alternatively, new physics may arise  below the energy scale of $10^{10}$ GeV, where the value of the four-point interaction of the Higgs boson becomes negative, 
and this could change the physics of the Higgs potential in such a way as to make the vacuum state stable.  Such new physics may exist above the scale of 1 TeV or so, which is directly or indirectly explored in current particle experiments, but it may also occur at lower scales, since the nature of the Higgs boson remains largely unexplored. 
 
Another intriguing possiblity is that the Higgs boson and top quark masses are such that the balance point toward instability is moved just to Planck scale, as indicated in the right-hand graph in Fig.~\ref{fig:deadoralive}. In this case, it is possible to arrange that the Higgs field is the inflaton which is responsible for generating cosmic structure~\cite{Bezrukov:2007ep,Bezrukov:2009db}.  

We do not know whether the Standard Model is correct up to high energy scales.  If we relax this assumption, there are 
relatively straightforward extensions of the Standard Model that can make the vacuum stable. For example, in a model where singlet scalar fields interact with the Higgs boson, the vacuum can be stable for some parameter regions of the model. It is even possible that such extension of the SM  can accommodate dark matter by requiring $Z_2$ symmetry. The Higgs boson couplings can be different from the standard model ones, and such deviations may be detected by the precision measurement of the Higgs bosons.  These models can contain additional first-order phase transitions. In this case,  significant gravitational waves may be produced by a phase transition in the early universe and observed as a background in low-frequence gravitational wave observations.

It is also possible that the Higgs sector is stabilized by high symmetry. Such a symmetry would require many new particles to completely change the Higgs boson interaction and its high-energy behavior.
An example of such a scenario is the appearance of  supersymmetry at high energies. In a supersymmetric model, all bosons have partner fermions and vice versa due to the symmetry of the theory.  The model also relates Higgs four-point couplings to the fourth power of gauge couplings so that the scalar potential is bounded from below. The supersymmetic models have at least two Higgs doublets,  namely, five Higgs bosons. In addition, the down-type quarks and leptons can have large Yukawa coupling. The Higgs boson decay can receive significant corrections detectable by the Higgs factories if the masses of the additional Higgs bosons are around 1  TeV.
In addition to that, the predicted partners can be directly searched for at a linear collider or though the measurement of oblique corrections. 

The other new physics possibility between the Planck scale to the weak scale is the change of space-time. In the warped extra-dimensional model, the Higgs boson can be the field in the IR brane. Yukawa coupling to the fermions is determined by the overlap of the fermion wave function in 5 dim to the Higgs boson on the brane. The effective field theory involving Higgs boson higher-order terms can express the physics picture, and the precision study of Higgs interaction can provide crucial information. 

It is quite generally true that the high-precision measurement of Higgs boson and top quark masses can give profound insight into all of these possibilities.  The measurement must be carried out with a high degree of confidence and control of experimental and theoretical systematic errors.  That is possible uniquely at an $\ee$ collider such as the ILC.

   \section{Why is there more matter than antimatter?}
\label{sec:antimatter}
      
The origin of matter is no less compelling a mystery than the origin of mass. Assuming inflationary cosmology, the universe began in a state with equal amounts of matter and antimatter.  From this starting point, the abundance of matter over anti-matter can be explained starting from symmetric initial conditions if some epoch in the early universe satisfies the Sakharov conditions---$B$ violation, $C$ and $CP$ violation, and loss of thermal equilibrium. These ingredients seem suggestively present in the quark sector of the SM itself, but, quantitatively, the asymmetry generated is too small by 10 orders of magnitude.  The problem is that the quarks that are sensitive to the CP-violating CKM angles are very light compared to the Higgs vacuum expectation value.  So  it is possible to  generate the observed baryon asymmetry in simple extensions of the Standard Model in which there are new particles and new sources of CP violation at or above the weak interaction scale.  These models must also include a mechanism for taking the universe out of thermal equilibrium, such as a first-order phase transition or late-decaying particles. Models in which the out-of-equilibrium events take place at or below the TeV scale can be directly tested at the ILC.  A prominent class of models is that in which the electroweak transition itself becomes first-order due to the coupling of the Higgs boson to other new particles.   Another interesting class of model involves dark sector particles or heavy neutrinos that would be revealed at the ILC.

In the SM, the electroweak phase transition (EWPT) is predicted to be a second-order, or nearly so.  A first-order phase transition, necessary for electroweak baryogenesis, requires a substantial modification of the SM Higgs potential at finite temperature. Generically, this is only possible if new particles with substantial couplings to the Higgs boson, and with masses below the TeV scale, are present. Such particles can be searched for directly at the LHC, and some possibilities (for example, top quark partners in supersymmetric models) are already strongly constrained. However, other options, such as new gauge-singlet scalar fields coupled to the Higgs, remain wide open. Precision Higgs measurements at the ILC will be sensitive to such scenarios. In particular, the $e^+e^-\to Zh$ cross section will be measured at the level sensitive to generic one-loop corrections to the Higgs propagator. This measurement will probe a wide range of first-order EWPT models, including those with a gauge-singlet scalar. Likewise, models with a first-order EWPT typically predict significant deviations in the Higgs cubic coupling, which can be discovered at the 500 GeV or 1 TeV ILC upgrade.   

\begin{figure}
\begin{center}
   \includegraphics[width=0.7\hsize]{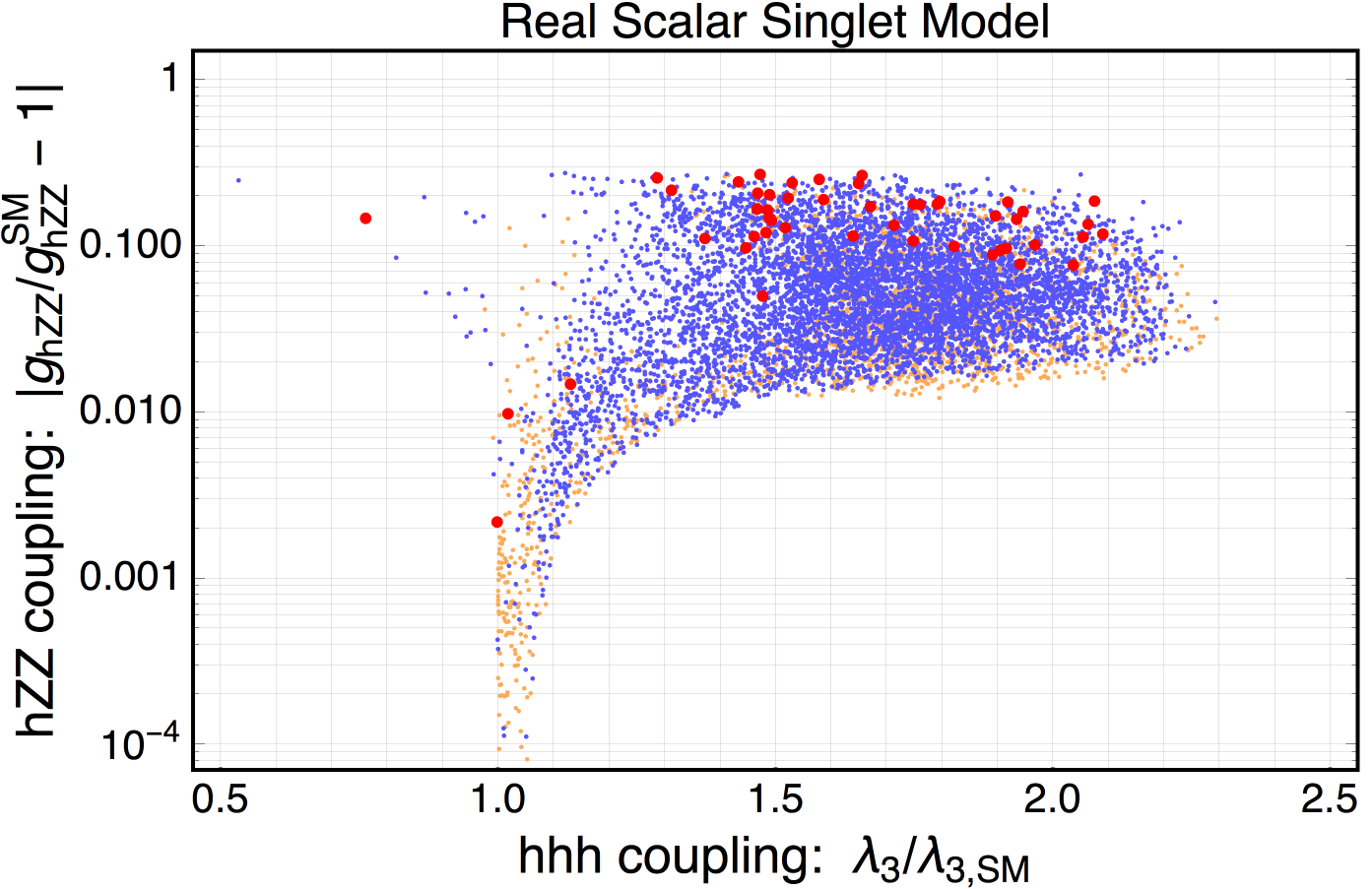} 
\end{center}
\caption{Scan of the parameter space for a model of baryogenesis at the electroweak scale with one new electroweak singlet Higgs field mixing with the SM Higgs doublet, from \cite{Huang:2016cjm}. Blue points represent models with a strong first-order electroweak phase transition, required for successful baryogenesis.}
\label{fig:Huang}
\end{figure}

An illustrative scan of the parameter space of a model with a single real scalar mixing with the SM Higgs boson is shown in Fig.~\ref{fig:Huang} \cite{Huang:2016cjm}.   The blue points represented models with a strongly first-order electroweak phase transition.  In this class of models, the Higgs self-coupling is enhanced almost by a factor of 2, and the Higgs couplings to $ZZ$ is has a relatively large correction (about 5\%) compared to the SM prediction.   With the precisions explained in previous sections, uncertainties of 23\% on the Higgs self-coupling and 0.4\% on the $HZZ$ coupling after the 500~GeV stage, the ILC will be able to discover these effects with high confidence.

The exploration of models of electroweak baryogenesis will also include tests for CP violation in Higgs boson and top quark decays.  There is an alternative class of baryogenesis models, called ``leptogenesis'',  in which the CP violation and the out-of-equilibrium dynamics occurs in the neutrino sector.  This can also be tested at the ILC if the relevant heavy neutrinos are at the weak scale.  We will discuss both these issues in the following chapter.

% \begin{itemize}
%     \item Electroweak baryogenesis and Higgs coupling deviations. Use Higgs+singlet model as an example of ILC capabilities beyond LHC. \todo{Make a plot: $hZZ$ and $h^3$ coupling deviations needed for first-order PT vs. ILC sensitivity.}   MN: should we focus on $\sqrt{s}= 250$ GeV or as large as 1TeV?  
%     \item CPV in Higgs couplings
%     MN:  $h\rightarrow \tau^+\tau-$  any update from D. Jeans LCWS 2016?   $hVV$  T Ogawa  EPS 2017.  Ge et al 2012.13922.Not much collider study but 2003.00099 Fuchs for b, t CP violation as well.    ask plot from them? 
%     \item WIMPy baryogenesis and direct searches?
%     \item MN: Leptogenesis? Collider implications of leptogenesis with 5-50~GeV heavy neutrinis is studied in 1710.03744 
% \end{itemize}

   \section{What is the dark matter of the universe?}
   \label{sec:darkmatter}

Perhaps the most compelling evidence for physics beyond the Standard Model comes from the sky, with a host of concordant observations indicating that baryons comprise only a fraction of the matter in the universe. Although viable dark matter candidates span many decades in mass, the near-coincidence of dark matter and baryon abundances suggests a non-gravitational mechanism to connect the two. This singles out dark matter candidates at or below the weak scale that interact with the Standard Model through one of several possible portals.  We have discussed in Chapters 8, 10, and 11 that these models often have special difficulties for the discovery of new particles at hadron colliders, difficulties that can be overcome at the ILC.

\begin{figure}
\begin{center}
  \includegraphics[width=0.45\hsize]{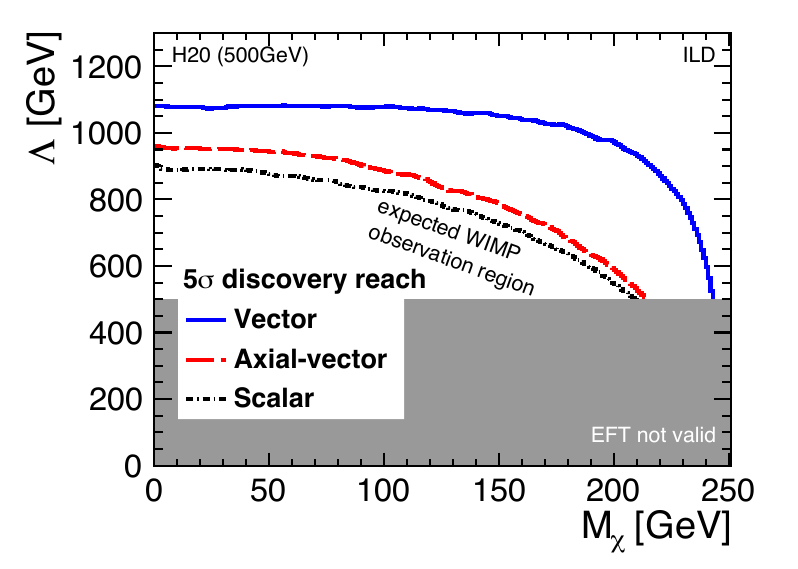} \quad 
 \includegraphics[width=0.35\hsize]{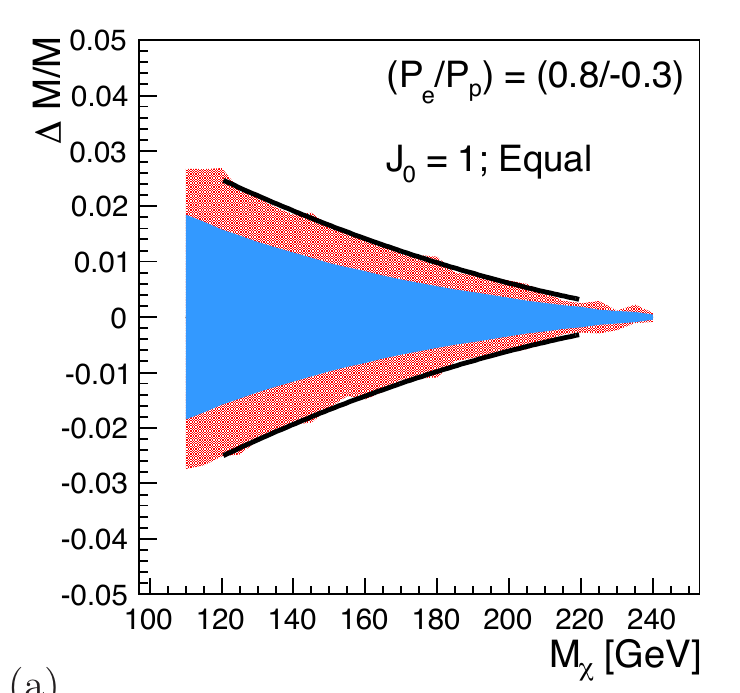} 
\end{center}
\caption{Left: ILC reach for dark matter particle coupled to electrons through an effective dim.-6 operator of various spin structures. Right: Fractional accuracy of WIMP mass determination at the ILC using the fit to a photon spectrum in the $\gamma+$inv. final state.}
\label{fig:DM_EFT}
\end{figure}

Famously, a particle with a mass in the GeV-TeV range, coupled to the SM via weak-scale interactions, naturally has the right relic density to explain the observed DM. Such Weakly-Interacting Massive Particles (WIMPs) can be pair-produced at colliders. Once produced, WIMPs escape the detector, leading to a missing energy signature. The reach of the ILC to WIMPs in the model-independent $\gamma+$MET channel is shown in Fig.~\ref{fig:DM_EFT}. As we have shown in Fig.~\ref{fig:DMpol} and the associated discussion, the ILC beam polarization can be used to analyze and control backgrounds, adding power to this search. The ILC is sensitive to the lepton (specifically, electron) coupling of the WIMP, making the ILC search complementary to those at hadron colliders and nuclear-recoil direct detection searches which are primarily sensitive to the WIMP coupling to quarks and gluons. 

The WIMP can also be produced in decays of other, heavier BSM particles. A well-studied example of this production mechanism occurs in supersymmetric models, where the lightest supersymmetric particle (LSP) can play the role of WIMP dark matter. In many models, the LSP is nearly degenerate in mass with other electroweak-ino states, while strongly-interacting superpartners are much heavier. Such models pose difficulties for searches at hadron colliders due to small cross sections and soft visible energy deposits. The democratic production and clean environment in the ILC collisions allow for  efficient searches for this physics. We have discussed the experimental aspects of this search in Sec.~\ref{sec:newparticles}. 

While WIMP paradigm is attractive, there are many alternative scenarios for microscopic origin of dark matter. The ILC will be able to shed light on many of these alternatives. For example, the DM may reside in a ``dark sector", a set of fields with no SM gauge interactions (but potentially rich structure of interactions among themselves). Such dark sectors are connected to the SM via a ``portal" interaction. A simple and natural portal to DM can be provided by a dark photon, a new $U(1)$ gauge boson which couples both to the SM (via kinetic mixing with the SM $U(1)$ gauge group) and to the dark sector. The ILC will be able to search for the dark photon in two ways. First, it can be produced at the main interaction point, and detected either through its decays back to the SM or the missing-mass peak in the spectrum of the associated SM photon.  Second, an additional detector placed 10-50 m downstream of the ILC beam dump can exploit the high current end energy of the ILC beams to extend the sensitivity to sub-GeV dark photons. We have discussed the experimental aspects of this search in Sec.~\ref{sec:beamdump}.

Another natural candidate for a portal to the dark sector is the Higgs boson. Higgs decays into dark-sector states can provide a window into the dark sector. Such decays may result in an invisible Higgs decay signature, which can be accessed at the ILC with sensitivity a factor of 20 better than that expected at HL-LHC. Alternatively, some of the produced dark-sector states can decay back to SM particles, lead to exotic multi-particle final states in Higgs decays. For example, in models of asymmetric dark matter consisting of bound states of a confining gauge group (similar to QCD) in the dark sector, Higgs decays may produce events known as ``dark showers", characterized by multiple displaced vertices. Thanks to the large sample of Higgs bosons that will be collected and clean environment with low track multiplicity, the ILC offers unparalleled opportunities to search for such phenomena. We have reviewed searches for such exotic Higgs decays in Sec.~\ref{sec:HiggsExotic}.

If a signature of the dark matter particle (or an associated mediator particle) is discovered, either at the ILC or in another experiment, the ILC can play a crucial role in determining the properties of this particle such as its mass and spin, as well as strength and structure of its couplings to the SM. For example, the WIMP mass can be determined with a 1-2\% accuracy by fitting the photon spectrum from the model-independent $\gamma+$invisible signature; see Fig.~\ref{fig:DM_EFT}. Such measurements are challenging at hadron colliders. Further, the polarized beams at the ILC may help to disentangle the chiral structure of the couplings. In some models, the ILC may even provide enough information to calculate the relic abundance of the discovered stable particle(s), and to test whether it is indeed responsible for the observed dark matter.  

We will bring together all of the ILC approaches to the search for dark sector particles and summarize their sensitivity in the next chapter.

% \begin{itemize}
%     \item 
%    Minimal (electroweak) dark matter. \todo{Use plots from Jenny and Co for the ILC. Update direct detection comparisons, include new DM-electron scattering searches such as SENSEI which are sensitive to the same operator.}
%     \item Higgs portal dark matter. \todo{(1) Higgs invisible decay constraints on HPDM assuming direct $H\to$ 2 DM decay. (2) Rare Higgs decays, e.g. $H\to 4b$, in dark sector models connected to asymmetric DM. Need ILC studies of rare Higgs decays.} 
%    MN: $H\to 4b$ is also  studied in the context of EWPT context by Kozaczuk et al.,      1911.10210 .
  
%     \item Dark photons. \todo{Make ILC reach plots for visibly and invisibly decaying dark photons.}
    
%     {\bf MN: a lesson is that we may need to discuss particular process in next section because a single channel can have different implication.  }
    
%  \end{itemize}

    \section{What is the energy scale of new physics?}
   \label{sec:BSMscale}

The Higgs boson is an exquisitely sensitive barometer for new physics, with any deviation in its properties from the Standard Model prediction providing a smoking gun indication of new physics. If new physics enters at or above the weak scale, these deviations can be systematically captured in effective field theory extensions of the Standard Model that encode the energy scale of new physics. In this section we interpret the SMEFT projections of section \ref{chap:SMEFT} in terms of motivated scenarios for new physics, translating ILC precision into qualitative lessons about the nature of the Higgs boson, its potential, and its coupling to other Standard Model particles.

In the next chapter, we will describe the relation betwen the levels of precision that will be reached in the ILC experiments and the predictions of specific models of new physics.   We will demonstrate that the ILC is robustly sensitive to the predictions of these models, and that the pattern of 
deviations of the Higgs couplings from the SM predictions gives insight into the nature of new physics responsible for those deviation.   Here, we will discuss a higher-level issue:   What does 
the high-precision study of the Higgs boson tell us in general about the scale of new physics?
Can we use this information to make fundamental tests of the SMEFT framework and of the 
quantum field theory description of the Higgs boson more generally?

% We further explore the broad lessons that may be extracted from potential Higgs coupling deviations in different EFT frameworks, including tests of the bedrock principles of quantum field theory via positivity constraints and the implications of deviations that are poorly fit by SMEFT.   

\paragraph{The scale of new physics:} The observation of any deviation from Standard Model predictions would be an unambiguous indicator of new physics.  As we have discussed in the previous chapter, this can then be interpreted within the SMEFT framework, in  terms of nonzero Wilson coefficients $c_i / \Lambda^2$ for a set of irrelevant operators. If such deviations can be well-described by dimension-6 operators in SMEFT, their size would allow us to infer the ratio of the couplings and masses of new physics. At the ILC, the anticipated sensitivity to Wilson coefficients of dimension-6 operators ranges from the few percent to per-mil level, depending on both the nature and number of operators in question. If new particles interact with the Standard Model at tree level with generic $\mathcal{O}(1)$ couplings, this could provide indirect evidence for particles as heavy as tens of TeV.  If new particles instead interact only at loop level, the ILC remains sensitive to new particles between the weak scale and the TeV scale. Such particles need not carry Standard Model quantum numbers, in which case they would have remained undetected at the LHC. 

Constraints on Wilson coefficients coming from null results at the ILC would provide strong evidence for a mass gap between the weak scale and the TeV scale, though the strength of the  inferred bounds varies from model to model. It should be noted that constraints on dimension-6 operators do not generally provide an unambiguous exclusion of new physics, since contributions from different UV degrees of freedom to a given Wilson coefficient may partially or wholly cancel. As we will discuss shortly, constraints on dimension-8 operators can provide an unambiguous exclusion of new physics up to the corresponding scale due to positivity bounds that forbid cancellations among different UV contributions.

\paragraph{The ``size'' of the Higgs:} A key higher-dimension SMEFT operator of broad significance is 
\beq
\mathcal{O}_H = \frac{1}{2 \Lambda^2} \left(\partial_\mu |H|^2 \right)^2 \ , 
\eeqn 
the leading nontrivial form factor for the Higgs field. The scale $\Lambda$ associated with $\mathcal{O}_H$ encodes the effective ``size'' of the Higgs boson, which may arise due to quantum corrections from new particles or compositeness of the Higgs itself. The leading effect of $\mathcal{O}_H$ on Higgs properties is to generate a universal shift in Higgs couplings relative to their Standard Model values. This shift necessarily drops out of ratios of branching ratios typically measured at hadron colliders.   We can sensitive to this parameter only if we can measure the Higgs partial width in absolute terms.   Thus, the direct measurement of the $Zh$ cross section at the ILC
using the recoil $Z$ boson as a tag allows the first unambiguous probe of $\mathcal{O}_H$. 

Among other things, bounds on (or measurement of) $\mathcal{O}_H$ quantify the extent to which the observed Higgs boson is an elementary or composite scalar. A sharp target is provided by the ratio of the Higgs' size to its Compton wavelength.  This ratio is of order unity for fully composite scalars, while smaller values correspond to increasingly elementary scalars. To date the neutral pion is the most elementary-seeming (pseudo)scalar yet observed in nature, with a ratio of size to Compton wavelength on the order of $\sim 1/6$. LHC measurements of Higgs properties do not yet probe pion-like levels of compositeness, and retain some degree of model-dependence. At the ILC, observation of $\mathcal{O}_H$ would provide compelling evidence for the compositeness of the Higgs, while sufficiently stringent bounds would ultimately indicate that the Higgs is the most elementary scalar observed to date.

\paragraph{The Higgs self-coupling:} A second key operator at dimension 6 is $\mathcal{O}_6 =  |H|^6/\Lambda^2$, which gives the leading correction to the Higgs self-coupling in the SMEFT framework. The anticipated precision of the ILC's constraint on $\mathcal{O}_6$ is sufficient to unambiguously establish the non-zero self-interaction of the Higgs boson. This would, in turn, be the first observation of a self-interacting particle whose interaction preserves all of its internal quantum numbers. Conversely, if the ILC measures a nonzero value of the SMEFT coefficient of $\mathcal{O}_6$, this would immediately indicate new physics below the TeV scale.

\paragraph{Positivity tests of analyticity and unitarity:}  In local, unitary quantum field theories, the basic assumption of relativistic causality implies that amplitudes are analytic functions of their kinematic variables.   This analyticity in turn implies positivity bounds in the space of SMEFT couplings~\cite{Adams:2006sv}. On one hand, these may be viewed as theoretical constraints that sharpen the interpretation of experimental results by narrowing the space of allowed couplings and precluding cancellations between different UV contributions. On the other hand, they may be viewed as an opportunity for direct experimental tests of the axiomatic principles of quantum field theory such as analyticity, unitarity, and locality~\cite{Remmen:2019cyz}. Experimental probes of positivity bounds are challenging because the vast majority apply to operators at dimension eight and higher on account of the energy growth required to impose UV-insensitive bounds. The effects of dimension-8 operators are typically subleading to those of dimension-6 operators, which are generally not subject to generic positivity bounds. 

Nonetheless, there are a number of observables for which dimension-8 operators provide the leading contributions, enabling tests of positivity bounds at colliders. At the LHC, diboson production allows for sharp tests of positivity bounds on anomalous quartic 
gauge couplings~\cite{Bellazzini:2018paj, Bi:2019phv}. But the ILC is particularly well-positioned to test positivity bounds on account of its clean environment and the ability to make measurements at multiple well-defined center-of-mass energies, which can be used to disentangle contributions from operators with different scaling dimensions. 

Particularly interesting tests can be carried out in a process that is very straightforward to measure at the ILC, $\ee\to \gamma\gamma$.   In SMEFT, this process receives no corrections at dimension  6; the first higher-dimension corrections are of dimension 8.   It is shown in \cite{Gu:2020ldn},
the dimension 8 corrections necessarily increase the differential cross section; a correction that would contribute negatively is forbidden by positivity.   Because a dimension 8 operator is involved, the test is sensitive to new mass scales $M$ only in the few-TeV range, though the question might also be pursued at higher-energy $\ee$ colliders. 

Further channels in which it is possible to test positivity 
include $e^+ e^- \rightarrow e^+ e^-$ scattering~\cite{Fuks:2020ujk}, 
$\ee\to Z\gamma$, $\ee\to ZZ$ are presented in 
\cite{Bellazzini:2018paj}.  The last of these reactions gets no dimension 6 corrections; in the other cases, ILC sensitivity to dimension-8 operators is sufficient to give unambiguous tests of positivity bounds, even in the presence of dimension-6 operators.   In
other processes of $\ee$ annihilation to vector bosons, the assumption that deviations from the SM arise from dimension-6 operators leads to specific predictions, such as relations between the $\gamma WW$ and $ZWW$ trilinear couplings and the absence of corrections to $\ee\to ZZ$, that can be tested with detailed measurements of the differential cross sections.  Deviations from these predictions must be attributed to dimension-8 contributions.  Through these analyses,  the ILC can probe bedrock principles of quantum field theory, and,  in the event of null results, can unambiguously exclude new physics in the relevant channels.

\paragraph{The linear realization of electroweak symmetry:} Although the $SU(2)_L \times U(1)_Y$-symmetric Standard Model EFT (SMEFT) is currently the preferred effective field theory extension of the Standard Model, it is not the only possibility. As we have already described, it is an assumption in SMEFT that any additional sources of electroweak symmetry breaking beyond the observed Higgs boson are associated with large mass scales that can be cleanly integrated out. If there are additional sources of electroweak symmetry breaking below 1~TeV or if there are heavy particles that still acquire most of their mass from the Higgs field, this would require using a different, more inclusive effective field theory. In Sec.~\ref{sec:EFT}, we described an alternative HEFT in which the Higgs field belongs to a nonlinear realization of weak-interaction $SU(2)$.  At present, it is possible for either SMEFT or HEFT to describe deviations from the Standard Model, leaving unresolved whether electroweak symmetry is linearly or non-linearly realized by the known fundamental particles. This question is unlikely to be answered decisively at the LHC, leaving a compelling open question for the ILC.

If precision measurements of Higgs couplings at the ILC are not well-fit by SMEFT operators at dimension 6, HEFT may provide the more appropriate description. This would suggest that electroweak symmetry is not linearly realized by the particles of the Standard Model and signal the presence of non-decoupling new physics between the weak scale and a few TeV. On the other hand, consistency of ILC precision measurements with Standard Model predictions---and, in particular, verification of the Higgs coupling constant relations predicted by  SMEFT---would significantly narrow the types of UV physics associated with HEFT. Future energy upgrades of the ILC could decisively determine whether electroweak symmetry is linearly realized by the known fundamental particles by probing scattering processes at the $\sim$ few TeV scale.

     \section{Why is electroweak symmetry broken?}
       \label{sec:EW-weak}

Behind all of these questions, there is another very important one.
All of the questions that we have discussed in this section eventually
point back to mysteries about the Higgs boson.  

The
structure of the SM is such that the interactions of gauge bosons and
fermions are specified completely by their quantum numbers and the
values of the $SU(3)\times SU(2)\times U(1)$ gauge couplings. These 
couplings are dimensionless.  For
energies above a few GeV, all three of these couplings are weak. This 
part of the SM is easy to understand and has been tested in great
detail through precision electroweak measurements and measurements of
quark and gluon reactions at the LHC. 

Any property of the SM that goes beyond this---including the basic 
mass scale of the model,  the mass
spectrum of quarks and leptons, and  the origin of CP
violation---necessarily involves the Higgs boson. The explanation that
the SM gives for these aspects comes in the form of renormalizable
parameters, the Higgs field mass and quartic terms and the
Higgs-fermion Yukawa couplings.   These are adjustable inputs to the
quantum field theory.  These input parameters are subject to some
general phenomenological constraints, but attempt to compute these
parameters from first principles have always led to paradoxes (such as
the Gauge Hierarchy Problem).  This is why the SM is often described
as an effective theory that represents a more fundamental theory at
higher energies.  We are now at the point where we need to know how
that more fundamental theory is constructed.

A basic physics question that we can ask about that more fundamental theory
is, why is the $SU(2)\times U(1)$ symmetry of SM spontaneously broken?
Like the values of the fermion masses, spontaneous symmetry breaking
is an input to the SM.  It comes in the assignment of a negative value
to the Higgs field mass parameter $\mu^2$.  This value cannot be
determined from first principles.  The connection between the physical
and the ``bare'' value of $\mu^2$ is not well-defined and these quantities can
easily have different signs.  This is a symptom of the fact that the
SM is only a phenomenological theory.  It cannot answer the why
questions, not this one, not any of the others that we have listed
above.

This situation stands in sharp contrast to our knowledge about
spontaneous symmetry breaking acquired from the study of
superconductivity, magnetism, and other condensed
matter phenomena, pairing in nuclear physics, and even chiral symmetry breaking in
low-energy QCD.  In each case, there is a fascinating story that
explains the why
of the broken symmetry state.   Some theorists are dismissive of
similar explanations in ``fundamental'' physics.  We disagree.  It is
true that any explanation of EWSB requires new physics beyond the SM.
But, to us, this means that there is an opportunity to discover new
fundamental forces now unknown.  We ought to be grasping for it.

Models that explain the phenomenon of electroweak symmetry
breaking (EWSB), require structure beyond the SM, but this can come in
one of
many forms.   The theoretical literature
contains a large number of different  types of models that address this question.  It is
useful to divide these models roughly into categories. New theoretical
ideas can give rise to new categories, but always with the imperative
to explain the mass parameter of EWSB, the Higgs field vacuum
expectation value $v = 246$~GeV. In the
following, we will refer to physics at the ``TeV scale'', with new
particles of mass from 100~GeV to a few TeV, the ``10 TeV scale'',
with new particles in the range 5--50~TeV, and a ``very high scale'',
with new particles above $10^9$~GeV and possibly up to the Planck scale.

Here is a sampling of models found in the literature:
\begin{itemize}
\item {\bf Models with a fundamental scalar field  at the TeV scale}:
Here the Higgs field is a fundamental field.  To avoid the conceptual problems of the SM and to allow the Higgs potential to be computable, this the Higgs field must be supplemented by additional fields providing add structure.  An example is the Minimal Supersymmetric Standard Model.
Here, the Higgs potential can be computed in terms of the masses and couplings of supersymmetric particles, which in principle can be measured independently by experiments.
The negative value of $\mu^2$ can be generated by
a loop diagram involving $\s t_L$, $\s t_R$, and the Higgs field $\Phi_u$, and this mechanism is testable after observation of these particles.
\item {\bf Models with a scalar field composite at the TeV scale}:  Here
EWSB is due to new strong interactions at the TeV scale, as in the
original Technicolor models.  These models do not include a light
Higgs boson doublet, but they may include a Higgs ``imposter'', for
example, a light scalar dilaton.  These models are allowed by the
current LHC data only with considerable tuning~\cite{Low:2012rj}.
\item {\bf Models with a scalar field composite at the 10~TeV scale}:
Here EWSB is due to new strong interactions at a higher scale, with
the Higgs field mass term protected by symmetry.  For example, the
Higgs doublet field can appear as a set of Goldstone bosons of the
strong interaction theory.  Little Higgs models are examples of models
of this type.  In these models, additional new TeV-scale particles
such as vectorlike top quark partners are needed to build computable
models of EWSB.   These partners can be evade LHC constraints by being
heavier than the limits or by being color-singlet, a class of models
called ``neutral naturalness''. 
\item {\bf Models with extra dimensions}:  In such models, the Higgs
doublet field can arise as the 5th component of a 5- or
higher-dimensional gauge field.   Randall-Sundrum models fall into
this class.  The higher-dimensional field excitations (``Kaluza-Klein
excitations'') play an essential role in the computation of the Higgs
potential and EWSB.
\item {\bf Models with fundamental scalar fields from very high
scales}:  Here the Higgs doublet is a fundamental scalar field arising
at very high energy scales.  For example, in the Relaxion model, the
Higgs potential evolves on cosmological time scales along with the
early expansion of the universe.  Another example is Nnaturalness, in
which the fundamental theory at the Planck scale contains a large
number $N$ of  copies
of the Higgs doublet with random $\mu^2$ values, of which one has a
mass at the TeV scale~\cite{Arkani-Hamed:2016rle}.  In these models,
the presence of the fundamental scalar field is given and the
mechanism only serves to solve the Gauge Hierarchy Problem. Often,
extreme parameter values are needed.  For
example, in Nnaturalness, one requires $N\sim 10^{60}$. 
\end{itemize}

The type of model dictates whether the model has the power to solve
other questions about the SM such as the values of the fermion
masses.  In supersymmetry, these values are set by the values of
Yukawa couplings at the scale of Grand Unification.  In models in
which the Higgs field is a Goldstone boson or an extra-dimensional
vector field component, there is a possibility that the fermion Yukawa
couplings can be generated dynamically at the TeV or 10~TeV mass
scale.

Though some of these models, especially those of the last class, can
be 
very difficult to test with colliders, all of the classes contain
models with new particles at the TeV scale, plausibly within the reach
or just beyond the reach of the LHC.  These particles can also give
tree-level or 
radiative corrections to the properties of the Higgs boson at the 5\%
level that can be discovered in a program of precision Higgs
measurements. The very different physics origins of EWSB in these
classes of models implies that the predictions for new particles and 
 anomalous Higgs coupling are very different from one class of models
 to another.   This gives the possibility that both direct and
 indirect effects of new particles can distiguish the classes and set
 us on the road to understanding correctly the origin of EWSB

In the next chapter, we will see how the various issues described in
this chapter can be addressed by measurements that the ILC will make possible.

\chapter{ILC and Models of Physics Beyond the Standard Model}  \label{ILCBSM}

\def\gmt{\ensuremath{(g-2)_\mu}}
\def\amu{\ensuremath{a_\mu}}
\def\msl#1{\ensuremath{m_{\tilde{\ell}_{#1}}}}
\def\cha#1{\ensuremath{\tilde{\chi}^\pm_{#1}}}
\def\neu#1{\ensuremath{\tilde{\chi}^0_{#1}}}
\def\mcha#1{\ensuremath{m_{\cha{#1}}}}
\def\mneu#1{\ensuremath{m_{\neu{#1}}}}
\def\gev{\,{\mathrm{GeV}}}
\def\tev{\,{\mathrm{TeV}}}
\def\Slpm{\ensuremath{\tilde{\ell}}}

In the previous chapter, we discussed the major questions of particle
physics and explained in general terms the ability of the ILC to
address those questions.   In this chapter, we will continue that
discussion by reviewing the ILC capabilities in terms of specific
models of new physics.   This will make more concrete the relationship
between the big questions and the ILC capabilities  for
measurements and new particle searches that we have presented  in
Chapters 8--12. 

It is not clear to us that the case for the ILC needs to be tied to specific
model-dependent goals.   As we have explained in Sec.~\ref{sec:EFT},
under the general assumptions of SMEFT, the relative deviations of the Higgs
boson couplings from their SM expectations are expected to be less
than 10\%.   These levels  are not yet probed by LHC results and are
expected to be out of the reach of the HL-LHC for 5~$\sigma$
discovery.  The same is true of the top quark couplings to the SM
gauge sectors.   On the other hand, new physics models predict
deviations accessible to the ILC, and different models predict
different patterns.   Thus, the ILC will provide a window into
physics beyond the SM that is rich in character and, today, totally
unexplored.  Nevertheless, considerations of specific models can be
useful in illustrating the variety of insights that the ILC could
produce.

\section{ILC and dark matter}
\label{sec:ILCdark}

The ILC can give insight into the particle identity of dark matter by
discovering the dark matter particle or associated particles of a dark
sector.  The search for the dark matter particle is a broad program
that is being carried out using many different strategies, including
searches at accelerators both at GeV and TeV energies.   The ILC will
add a number of new capabilities that extend or complement these
programs.

The possible candidates for the dark matter particle span an enormous
range, from axions whose Compton wavelength is the size of a galaxy to
black holes of almost a solar mass.   But a particularly attractive
class of candidates lies in the center of this range.  This is the
``thermal WIMP'', a stable,  weakly-interacting neutral particle that interacts
strongly enough with SM particles to come to thermal equilibrium in
the early universe.   The number density of such particles is set by
their freeze-out density, the density at which, when the universe
becomes sufficiently cold, they cannot find partners with which to
annihilate and thus reach a constant density co-moving with the
expansion of the universe.   The simplest models of thermal freeze-out
predict that an annihilation cross section of roughly 1~pb leads to a
WIMP density comparable to the measured dark matter density in the
universe.   This value can be met in several ways:   (1) by a particle
with mass of order 100~GeV and couplings of electroweak strength,  (2)
by particles with weaker couplings and comparably smaller masses, and
(3) by particles with stronger couplings and comparably larger
masses.   

Models of supersymmetry and other models of electroweak
symmetry breaking often contain stable neutral particles of the first
type.  However, these models are now strongly challenged by the limits
on supersymmetric particles from the LHC and the limits on the WIMP
cross section on matter from direct detection experiments.   Specific
cases remain in play, as we will discuss below.  Models of the second
type can be achieved through the idea of a dark
sector, a new sector of particles with zero quantum numbers under that
SM gauge symmetries that are connected very weakly to SM particles through
specific operators  called ``portals''.   We have reviewed the
structure of such models in Sec.~\ref{sec:physfixedtarget}.  Models of
the third type require higher-energy accelerators.  They turn out to be
especially difficult to study with hadron colliders and  thus provide
a motivation for higher energy lepton colliders.   We will discuss
this point further in Secs.~\ref{sec:multiTeV} and \ref{sec:multi10TeV}.

We have explained in Secs.~\ref{sec:WIMPDM} and \ref{sec:dark500}
that the search for events of the type $\ee\to \gamma +$~missing
energy at the ILC can be a powerful probe both for WIMPs in the
100~GeV mass region 
and for dark sector particles.  For WIMPs of the first type, the
production cross sections are of electroweak strength and thus are
large enough to provide a substantial event sample.  The WIMPs would
not be observed in collider detectors, but their presence can be
infered by the observation of initial state radiation.   We have
emphasized in Sec.~\ref{sec:WIMPDM} that linear $\ee$ colliders have
many advantages for this search.   Initial-state photon radiation does
not depend on non-perturbative quantities such as parton distribution
functions but rather is given by QED theory at part-per-mil precision.
The ILC detectors operate without a trigger and  are capable of recognizing very small energy
distributions at angles close to the beam director.  The use of
electron and positron beam polarization allows backgrounds to be directly
measured.  All of these features allow sensitivity to WIMPs at masses
close to 1/2 of the CM energy.   For dark sector models, the
sensitivity of the ILC is limited by the strength of the mixing
through the gauge portal.  Still, the ILC can be sensitive to mixing
parameters as small as  $|\eps|^2 \sim 10^{-5}$ for mass regions
that extend beyond those of other colliders.

In supersymmetric models, and in other models of electroweak symmetry
breaking that predict dark matter candidates, there is another important
production mechanism for the dark matter particle.   A collider can
produce a heavier state that decays to the dark matter matter
particle, depositing in the process observable energy.  ILC examples in
supersymmetry are the reactions $\ee\to \XP{1}\XN{1}$ or $\ee\to
\stau^+ \stau^-$ with subsequent decays of the SUSY particles $\XN{1}$
plus SM particles.  This process can be difficult to observe at hadron
colliders if the
mass gap between the the heavier SUSY particle and the  $\XN{1}$ is
small, leading to a small energy deposition.   In practice, mass gaps
less than 10~GeV lead to difficult for  LHC experiments in triggering
and signal recognition~\cite{ATLAS:2019lng}.   The LHC experiments can
recover sensitivity for very small mass gaps that lead to observable
lifetimes.   However, there are physics reasons for models to fall
into this gap, first, because the Higgsino sector of SUSY naturally
has small mass splittings of just a few GeV or less, second, because
models of SUSY dark matter often require coannhilation of the dark
matter particle with some other species to produce a sufficiently
small dark matter density, and this requires a mass gap of order
5~GeV.   The ILC, using the advantages already presented for initial
state radiation, can fill this gap.   For example, simulations of SUSY
analyses at the ILC for models with sub-GeV mass gaps are presented
in~\cite{Berggren:2013vfa}.

ILC offers another method to search for dark sector particles using
its fixed-target capabilities, as discussed in
Sec.~\ref{sec:beamdump}.  The ILC will produce an electron beam that
combines high intensity and high energy.    This beam can be used
parasitically, with detectors mounted behind the ILC beam dumps, and
directly at dedicated fixed-target interaction halls.  Both types of
experiments extend the expected reach for dark sector particles both
in particle mass and in sensitivity to small couplings.

%%%%%%%%%%%%%%%%%%%%%%%%%%%%%%%%%%%%%%%%%%%%%%%%%%%%%%%%%%%%%%%%%%%%%%%%%

\section{ILC and supersymmetry}
\label{sec:ILCSUSY}

The ILC can give insight into supersymmetric models of new physics in
two different ways, first, through direct searches for supersymmetric
particles and, second, through observation of corrections to the Higgs
boson couplings induced by loop effects of supersymmetric particles.

\subsection{Direct SUSY particle production}
\label{sec:directSUSY}

In the previous section, we have already discussed the ability of the ILC to
discover new particles in models with stable neutral particles
separated from heavier partners by small mass gaps.   This is a
general issue, but it has taken on more importance in view of the
recent result of the Muon $g-2$ experiment, which gives a deviation of
4.2~$\sigma$ between the measured result and the theoretical consensus
value~\cite{Abi:2021gix}.

There are various more ad hoc models that can explain the muon $g-2$
deviation, but it is relevant that the anomaly can be 
explained within the Minimal Supersymmetric Standard Model (MSSM). This has been studied in 
\cite{Baum:2021qzx}, in the series of papers~\cite{Chakraborti:2020vjp,Chakraborti:2021kkr,Chakraborti:2021dli,Chakraborti:2021mbr}, and in \cite{Endo:2020mqz,Endo:2021zal}.
It is shown that there are MSSM parameter sets that can provide this
explanation consistently with LHC SUSY searches by making use of the
region of small mass gaps within the chargino-neutralino sector.  The
points found by this analysis are consistent with the observed
cosmic dark matter abundance~\cite{Planck:2018vyg} and constraints
from dark matter direct detection
experiments~\cite{XENON:2018voc,LUX:2016ggv,PandaX-II:2017hlx}. 

%%%%%%%%%%%%%%%%%%%%%%%%%%%%%%%%%%%%%%%%%%%%%%%%%%%%%
\begin{figure}
\begin{center}
\includegraphics[width=0.42\hsize]{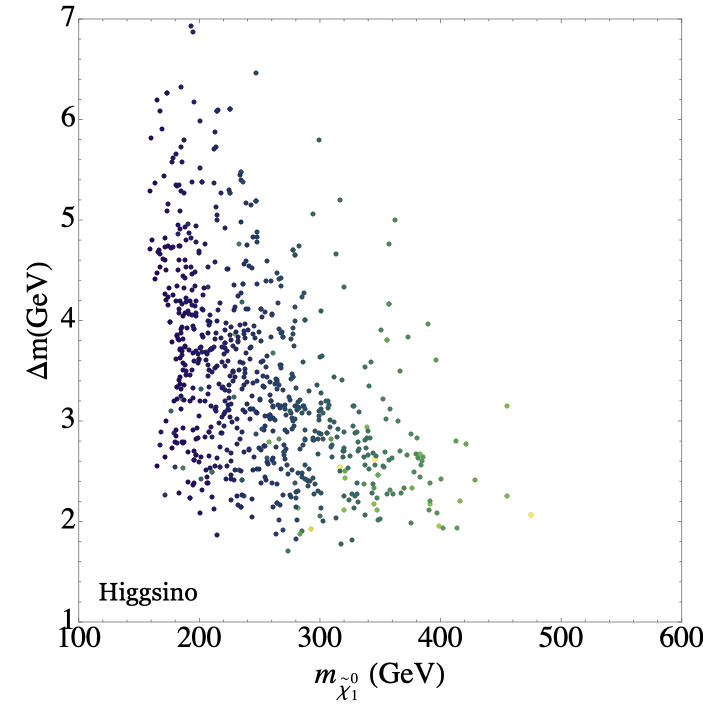}
\ \ 
\includegraphics[width=0.42\hsize]{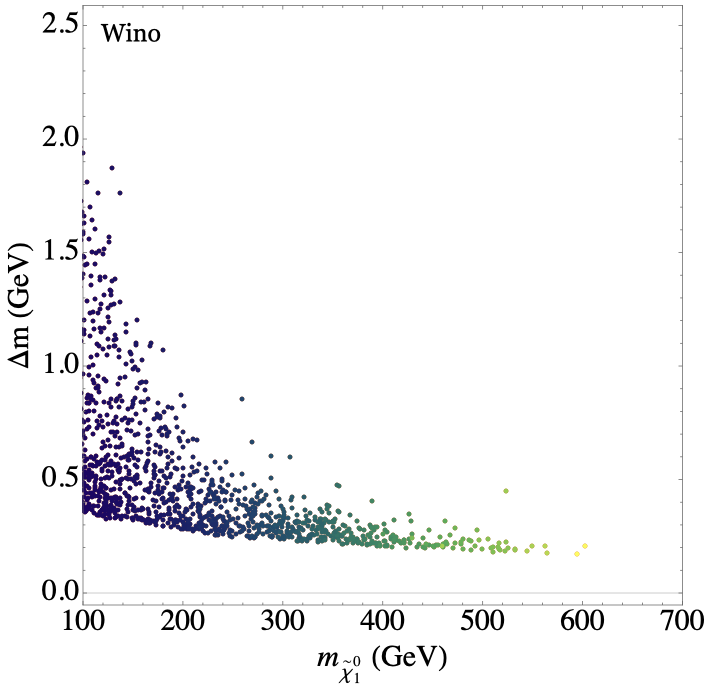}
\end{center}
\caption{Scans over the parameter set of the MSSM giving models that
  account for the current discrepancy between the observed value of
  the muon $g-2$ and the consensus SM prediction, from
  \cite{Chakraborti:2021mbr}.   Left: Higgsino LSP; Right: Wino LSP.
  Other scenarios are described in the text.  Color encodes the
  predicted value of the LSP dark matter density, with green
  indicating a higher value.}
\label{fig:gm2scans}
\end{figure}
%%%%%%%%%%%%%%%%%%%%%%%%%%%%%%%%%%%%%%%%%%%%%%%%%%

There are five different scenarios, all characterized by a mass region
for the lightest SUSY particle $M_{LSP}$ and the mass gap to its 
heavier partner $\Delta M$.   In the first two scenarios, this SUSY
sector can supplies only a fraction of the total dark matter density.
In all of the scenarios, the requirement to explain the $g-2$ result
leads to an upper bound on $M_{LSP}$.
\begin{itemize}
\item[(i)]
higgsino DM :
$M_{\rm LSP} \lsim 500 \gev$ with $\Delta M \sim 5 \gev$;
\item[(ii)]
wino DM : 
$m_{\rm LSP} \lsim 600 \gev$ with $\Delta M \sim 0.3 \gev$;
\item[(iii)]
mixed bino/wino DM with $\cha1$-coannihilation :
$m_{\rm LSP} \lsim 650  \gev$   with $15 \gev < \Delta M < 60 \gev$;
\item[(iv)]
bino DM with $\Slpm$-coannihilation with $\stau_L$:
$m_{\rm LSP} \lsim 650 \gev$  with $10 \gev < \Delta M < 80 \gev$;
\item[(v)]
bino DM with $\Slpm$-coannihilation with $\stau_R$ :
$m_{\rm LSP} \lsim 650 \gev$   with $10 \gev < \Delta M < 100 \gev$;
\end{itemize}
 The next round of direct detection
 experiments~\cite{Aprile:2020vtw,LUX-ZEPLIN:2018poe} will give us
 more information~\cite{Chakraborti:2021mbr}. In the case that no
 signal is observed, the upper limit on  the LSP goes down to $\sim
 500 \gev$ and the entire parameter region will be covered by the
 1~TeV ILC.  For certain choices of the signs of the MSSM parameters (in particular, 
$\mu \times M_1 < 0$)  the different contributions to the spin-independent direct 
detection cross section for a bino-like DM candidate interfere destructively, giving even more space for that solution within the current direct detection constraints~\cite{Baum:2021qzx}.

The parameter scans in the first two cases are shown in
Fig.~\ref{fig:gm2scans}~\cite{Chakraborti:2021mbr}.   These emphasize that the SUSY particles
solving the $g-2$ anomaly may be close at hand and uniquely accessible
to an $\ee$ collider.    The other cases give substantial opportunity
for first discovery of new physics at LHC.   However, even in those
cases, an $\ee$ collider will be needed to characterize  the actual
scenario by measuring the quantum numbers and mixing angles of the
observed SUSY particles, and to clarify the role of the light SUSY particle in the cosmic dark matter.   It has been shown that the $\ee$ studies can measure the masses and mixing angles needed to evaluate the contribution to $(g-2)$ from the new particles and verify that the anomaly indeed has an understood supersymmetric origin~\cite{Endo:2013xka,Endo:2022qnm}.

%%%%%%%%%%%%%%%%%%%%%%%%%%%%%%%%%%%%%%%%%%%%%%%%%%%%%
\begin{figure}[p]
\begin{center}
\includegraphics[width=0.50\hsize]{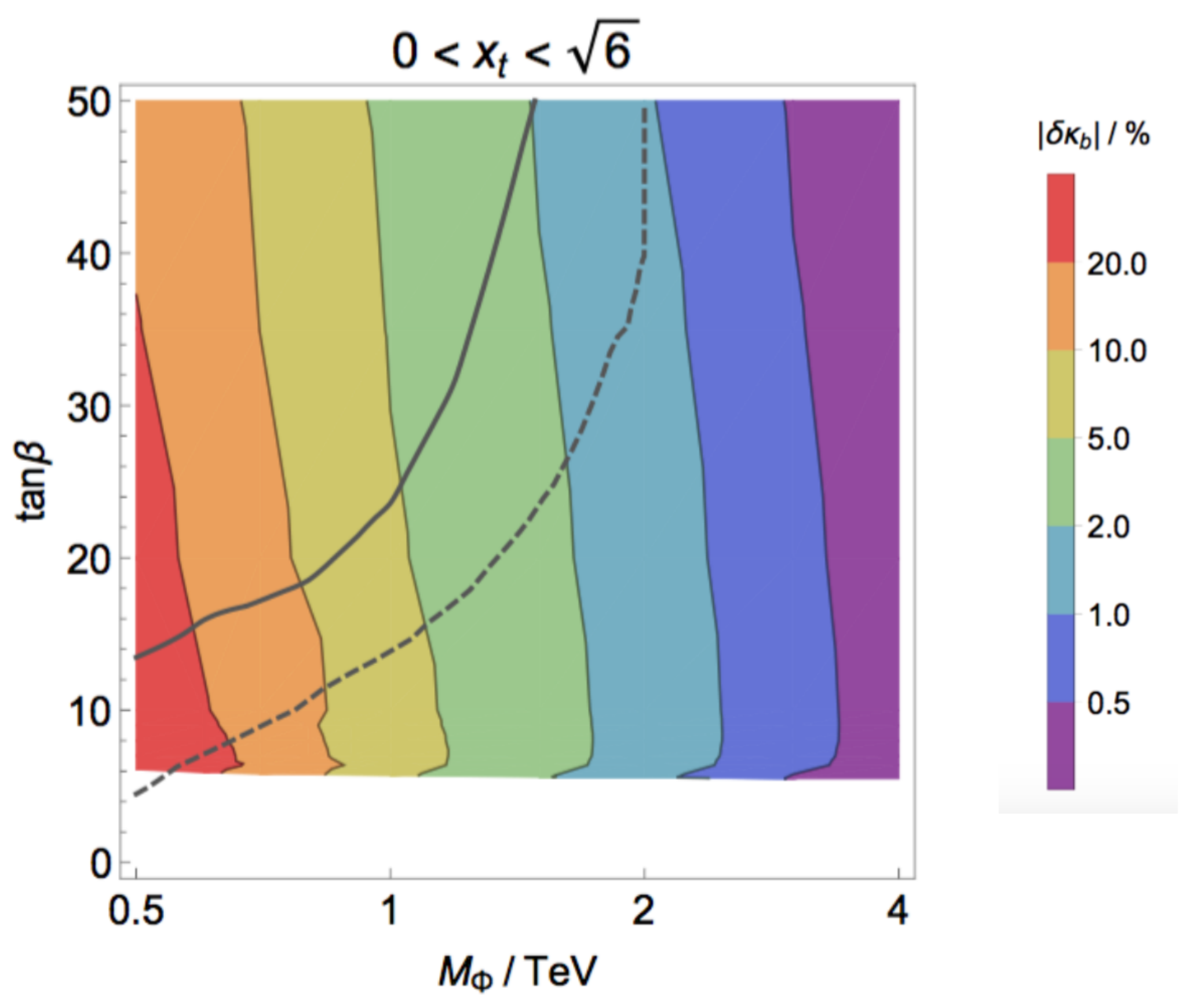}\\
\includegraphics[width=0.50\hsize]{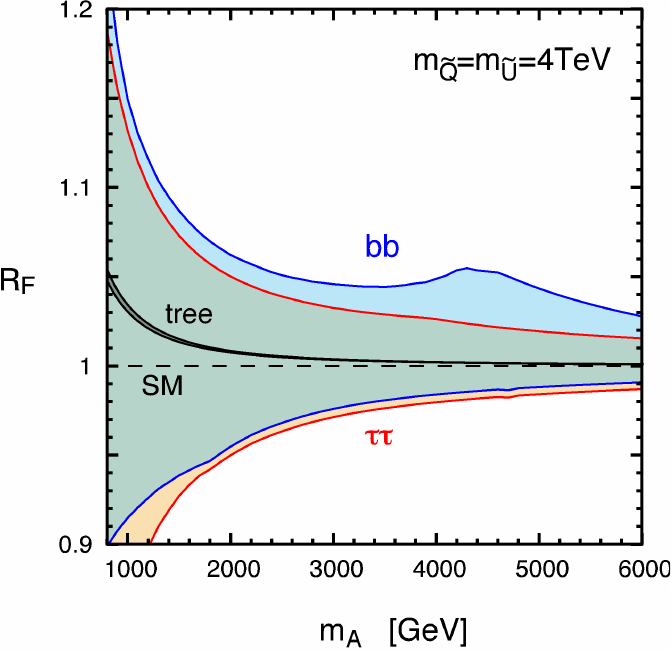}

\end{center}
\caption{Fractional deviations of the Higgs boson couplings from their SM values in 
supersymmetry models with very heavy SUSY partners:  Top: effect on the $Hbb$ couplings
  in a  class of supersymmetry models with $b$-$\tau$ Yukawa
  unification, from~\cite{Wells:2017vla}.  $M_\Phi$ is the mass of a heavy boson
  in the extended Higgs sector of the model.   The extended Higgs
  sector is excluded by LHC searches above the solid line.  The
  expected exclusion limit at the HL-LHC is shown by the dotted line.  Bottom: range of effects on the $Hbb$ and $H\tau\tau$ couplings induced by loop corrections involving heavy SUSY states, from~\cite{Endo:2015oia}. The black line shows the tree-level prediction due to the 2-Higgs doublet structure, for $\tan\beta = 5$~--50.}
 \label{fig:heavySUSY}
\end{figure}
%%%%%%%%%%%%%%%%%%%%%%%%%%%%%%%%%%%%%%%%%%%%%%%%%%

\subsection{Observation of SUSY effects on the Higgs boson}
\label{sec:HiggsandSUSY}

If the masses of SUSY particles are so large that they cannot be
discovered directly at the LHC, it is still possible that their
effects can be observed through their effect on  the Higgs boson
couplings.   There are many possible sources of such effects, but two
are especially important:   (1) the shift of the Higgs boson couplings
to $b\bar b$ and $\tau^+\tau^-$ coming from the extended Higgs sector
required in SUSY, and (2) the shift of the Higgs boson coupling to
$b\bar b$ due to $\s b$-gluino loop diagrams with large $\s b_L$-$\s
b_R$ mixing.  A broad survey of SUSY effects on the Higgs boson couplings in the parameter region in which the SUSY particles are very heavy is given in Ref.~\cite{Endo:2015oia}. Further enhancements are possible in the MSSM with general squark generation mixing, as studied in~\cite{Hidaka:2021cup}.  These can appear both in $\Gamma(H\to b\overline{b})$ and especially in  $\Gamma(H\to c\overline{c})$, for which the precision measurement is unique to $\ee$ colliders.

Illustrations of these effects are shown in  Fig.~\ref{fig:heavySUSY}. 
Figure~\ref{fig:heavySUSY}(a)  shows the fractional deviation in the
$Hbb$ coupling in a class of SUSY models with
$b$-$\tau$ Yukawa unification at the grand unification
scale~\cite{Wells:2017vla}.
   The models are chosen such that the gluino and stop masses are
above 5~TeV, well out of the reach of the LHC.   The heavy Higgs
sector is excluded by LHC searches in the region above the solid
line.   The HL-LHC is expected to improve this limit to the dotted
line.  This still leaves considerable parameter space that can be
accessed by precision Higgs boson measurements.
Figure~\ref{fig:heavySUSY}(b) shows the range of values allowed for relative 
enhancement or depression of the $Hbb$ and $H\tau\tau$ couplings in the MSSM with the squark masses $m_{{\s Q}} = m_{{\s U}}  = m_{{\s g}}=$~4 TeV
and other sfermion masses greater than 4 TeV~\cite{Endo:2015oia} .
  The enhanced range in the $Hbb$ case comes from solutions with large $\s b_L$-$\s
b_R$ mixing, respecting the condition of vacuum stability.\footnote{We
  caution that the results from public SUSY codes for models with SUSY
  masses of several TeV are quite unstable
  with respect to addition of small corrections. The analysis of
  \cite{Endo:2015oia} used FeynHiggs 2.10.2.  Using  FeynHiggs 2.18.1, the
  point previously giving the largest deviation in the $b\bar b$
  coupling for $m_A = 4600$~GeV (5\%) now gives a coupling deviation
  of 2\%.  However, it is still not difficult to find points with
  deviations of 3\% (6$\sigma$ for ILC).  It would be good to
  study this issue more systematically.}

These figures make a point that is more general.   The direct search for
new particles and the search for new physics through precision
measurement do not compete directly with one another.   Instead, they
generally access complementary regions of the parameter space.  We
should make use of both techniques to make the broadest search for new physics.

%%%%%%%%%%%%%%%%%%%%%%%%%%%%%%%%%%%%%%%%%%%%%%%%%%%%%%%%%%%%%%%%%%%%%%%%%
 
 \section{ILC and composite Higgs fields}
\label{sec:ILCcomposite}

The ILC can give insight into models in which the  Higgs field is
composite through a number of different  measurements.   These include
direct probes of the Higgs boson properties, as we have already
discussed in the section on the ``Higgs size'' in
Sec.~\ref{sec:BSMscale}.   Because composite Higgs bosons must communicate
strongly with the top quark, probes of the top quark also play an
important role here.

There is some subtlety to the construction of composite models of the
Higgs boson.  In such models, the composite scalar sector is
parametrized by a mass scale $F$ analogous to the pion decay constant
$f_\pi = 93$~MeV
in the familiar strong interactions.   The first composite models of
the Higgs boson, models of ``technicolor'', required  $F = v =
246$~GeV~\cite{Weinberg:1975gm,Susskind:1978ms}.   These models are now excluded, because such a light
compositeness scale generates large corrections to the precision
electroweak observables and strong interaction resonances light enough
to be observed at the LHC.   An attractive picture that move the new
binding interactions to higher energies is the idea of the Higgs boson
as a Goldstone boson.   A lucid review of this idea is given
in~\cite{Contino:2010rs}.
  Spontaneous symmetry
breaking in a new strong interaction theory at  a high energy scale generates the Higgs $SU(2)$ doublet as
a multiplet of Goldstone bosons, with the associated $F$ parameter
related to the high scale of symmetry-breaking.   Perturbations of
this theory, perhaps generated by coupling to SM particles, break the
symmetry weakly and lead to a potential for the Higgs multiplet.
This gives a model for the origin of electroweak symmetry breaking.

It is also possible to model the Higgs boson using models with extra
space dimensions.  Here the Higgs boson appears as the 5th component
of a gauge field in the higher-dimensional space.  This description
can be viewed as dual description of the previous one, with some
advantages for calculation~\cite{Arkani-Hamed:2000ijo,Contino:2003ve}.

The Goldstone boson origin of the Higgs boson leads to modification of
its kinetic energy term and, in particular, generation of a SMEFT
correction
\beq
     c_H =   a  v^2/F^2 \ ,
\eeqn
where $a$ is a number of order 1.  This leads to a uniform decrease of
all Higgs boson partial widths by the factor  $(1-c_H)$. This effect is most  visible as a
shift of the Higgs boson partial widths to $W$ and $Z$.   At the ILC, this effect can be visible for multi-TeV
values of $F$.

Very often in models, the largest effect that breaks the original
strong interaction symmetry is the influence of the top quark and
heavy vectorlike top quark
partners of that appear as part of the composite model.   The heavy
partners can also modify the Higgs
boson couplings through loop corrections, in particular, modifying the
partial width of the Higgs boson into
$gg$~\cite{Han:2003wu,Han:2003gf}.

The interplay of the top quark and its partners with the new strong
interactions also leads to effects that are visible directly in
precision top quark physics.  Often, the top quark must acquire some 
composite structure in order to receive its large mass.   This leads
to observable effects.  The couplings of the top quark to the
photon and gluon are fixed by Ward identities, but the couplings to
the $Z$ and $W$ bosons can be modified and obtain corrections of order
$v^2/F^2$.   Both positive and negative corrections are possible in
models, and the corrections are expected to affect differently the
left- and right-handed top quark
couplings~\cite{Berger:2005ht,Richard:2014upa,Funatsu:2020haj,Yoon:2018xud}.

The ILC has exceptional sensitivity to modifications of the top quark
coupling to the $Z$ boson~\cite{Amjad:2015mma}; see also the
discussion of precision top quark physics in Sec.~\ref{sec:top}.   At the ILC500, top quarks are  pair-produced
through $s$-channel photon and $Z$ exchange.   The compositeness
corrections appear as interference terms in the  production
amplitude.   Using beam polarization and measuring the polarization of
the final top quarks, it is possible separate the various chiral
contributions to the production amplitude and to measure these at the
parts-per-mil level.

Through these three effects and others, the ILC can give a detailed
characterization of the influence of possible composite structure on the Higgs
boson and the top quark.

%   \section{ILC and CP violation}
%   \label{sec:ILCCP}

% In this section, we will review the various probes of new CP violation
% mechanisms available at the ILC, including measurements of Higgs
% decays, $\ee\to W^+W^-$, and top quark production and decay.
%
%  (  We hope to add this later.)

\section{ILC and flavor}
\label{sec:ILCflavor}

The ILC can give insight into models of flavor-dependence of fermion
masses.  In the SM, the quark and lepton masses are proportional to
the fermion-Higgs Yukawa couplings, which are renormalizable---and
therefore freely adjustible---parameters of the model.  At this level,
there is no explanation for the fermion mass hierarchy.

It is possible that the explanation for the Yukawa couplings comes
from physics at a very high mass scale---the grand unification scale
or the scale of string compactification---and is inaccessible to
forseen colliders.   However, it is also possible that this hierarchy
could be generated at the TeV.   If the opportunity is there, we
should test it.

%%%%%%%%%%%%%%%%%%%%%%%%%%%%%%%%%%%%%%%%%%%%%%%%%%%%%
\begin{figure}
\begin{center}
\includegraphics[width=0.60\hsize]{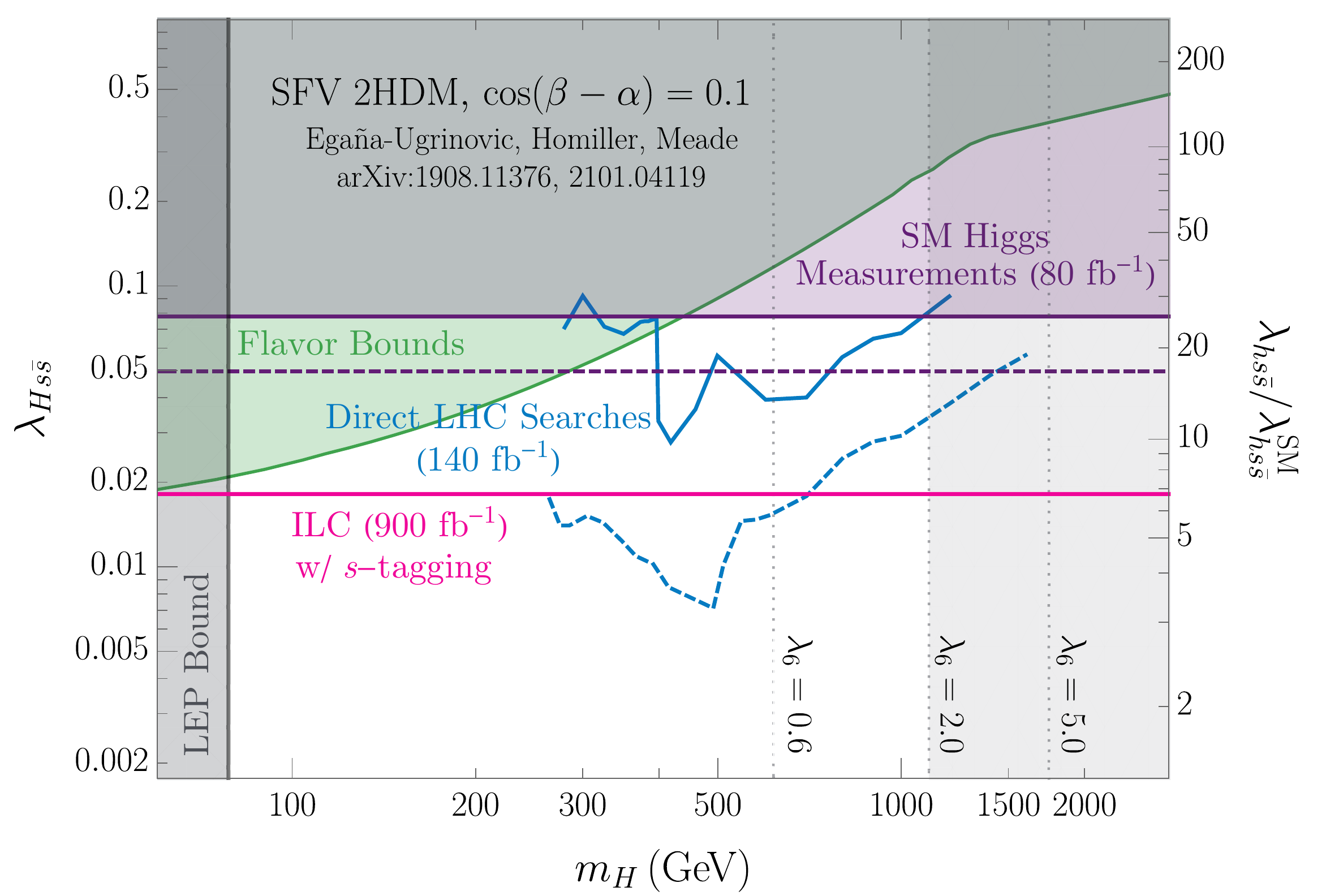}
\end{center}
\caption{LHC bounds on a 2-Higgs-doublet model with specific couplings
  to the strange quarks as a function of the heavy Higgs mass
  $m_H$. The scale on the right-hand side shows show the corresponding
  enhancement of the 125GeV Higgs Yukawa coupling to strange quarks.
  From~\cite{Egana-Ugrinovic:2021uew}, with the ILC expectation from
  \cite{Albert:2022mpk} added.}
 \label{fig:stag}
\end{figure}
%%%%%%%%%%%%%%%%%%%%%%%%%%%%%%%%%%%%%%%%%%%%%%%%%%

An approach to the problem of fermion mass generation comes from the
idea that there are multiple Higgs bosons, each with a fundamental
coupling to one particular generation of fermions.  Many models have
been proposed that have this general 
structure~\cite{Das:1995df,Blechman:2010cs,Altmannshofer:2015esa,Ghosh:2015gpa,Botella:2016krk}.
It is possible for additional Higgs bosons coupling differently to
fermion generations to be relatively light, with masses below
1~TeV~\cite{Altmannshofer:2016zrn,Egana-Ugrinovic:2019dqu,Egana-Ugrinovic:2021uew}.   It is
possible to investigate these models through searches at the LHC and,
for sufficiently light scalars, at the ILC.

An important feature of these models, offering an alternative search
strategy, is that they can lead to values of the Yukawa couplings of the
lighter generations that violate the usual SM relation between Yukawa
couplings and fermion mass.  Both the LHC and ILC can probe the muon
Yukawa coupling, but it is a unique feature of $\ee$ colliders that
they can probe the charm Yukawa coupling with high precision and also put
significant constraints on the strange Yukawa coupling.    We have
discussed these analyses at the ILC in Sec.~\ref{sec:Higgs250}.
Figure~\ref{fig:stag} shows the current constraints on the enhancement
of the strange quark Yukawa coupling, from
\cite{Egana-Ugrinovic:2021uew},  and the improvement currently
expected from ILC250~\cite{Albert:2022mpk}.

In addition to the possible enhancement of lighter-generation Yukawa
couplings, generation-dependent extended Higgs models predict
flavor-off-diagonal Higgs boson couplings.  In the SM, it is always
possible to redefine fields so that the Higgs boson couplings are
diagonal in flavor.  However, in SMEFT, when we reach dimension-6
operators, this freedom is already used up, and  so, generally, it
is not expected that these preserve flavor.   Naive  estimates of the
size of the coefficients would put flavor-violating Higgs decays at
the $10^{-3}$--$10^{-4}$ level.  This level of Higgs flavor violation  is quite compatible with the
current strong bounds on $\tau\to \mu\gamma$~\cite{Vicente:2019ykr}.   Within the 
MSSM with general squark generation mixings, branching ratios for $H\to bs$ of order $10^{-3}$  can be generated by 
loop effects~\cite{Gomez:2015duj,Hidaka:2021cup}.
 The LHC will be able to test for $H\to \tau\mu$, but tests  in the
quark sector are much more difficult at hadron colliders.  At the ILC,
though, searches for the exotic decay $H\to bs$ should reach this level.

\section{Mass Reach of Precision Higgs Measurements}

It is interesting to ask whether the precision measurement of Higgs
boson couplings gives access to new physics particles that are outside
the range of HL-LHC direct searches.  Actually, we have already
presented many examples of this in previous sections, but one might
ask whether these are special cases or represent more generic
situations.  In this section, we discuss this question from another
point of view.

When the new physics corrections to Higgs couplings can be described
by SMEFT, these corrections arise from dimension-6 operators and thus
are of the order of $v^2/M^2$, where $M$ is the mass scale of new
particles that have been integrated out.   Naively estimating the size
of these corrections by by setting $M = 2$~TeV and 
putting the coefficient 1 in front of
the dimensional estimate.  This gives the size of these effects
as
\beqa
\mbox{tree\ level\ effects:} &\quad &  v^2/M^2 \sim  1\%  \CR 
  \mbox{loop\ level\ effects:} &\quad & (g^2/4\pi)  v^2/M^2 \sim
  0.1\%
  \eeqan
The tree level estimate is roughly the same size as the uncertainties
estimated for the ILC.  One might conclude from this that the program
of precision Higgs measurements can access new particles of mass up to
2~TeV but cannot definitively prove their existence.

%%%%%%%%%%%%%%%%%%%%%%%%%%%%%%%%%%%%%%%%%%%%%%%%%%%%%
\begin{figure}[p]
\begin{center}
\includegraphics[width=0.42\hsize]{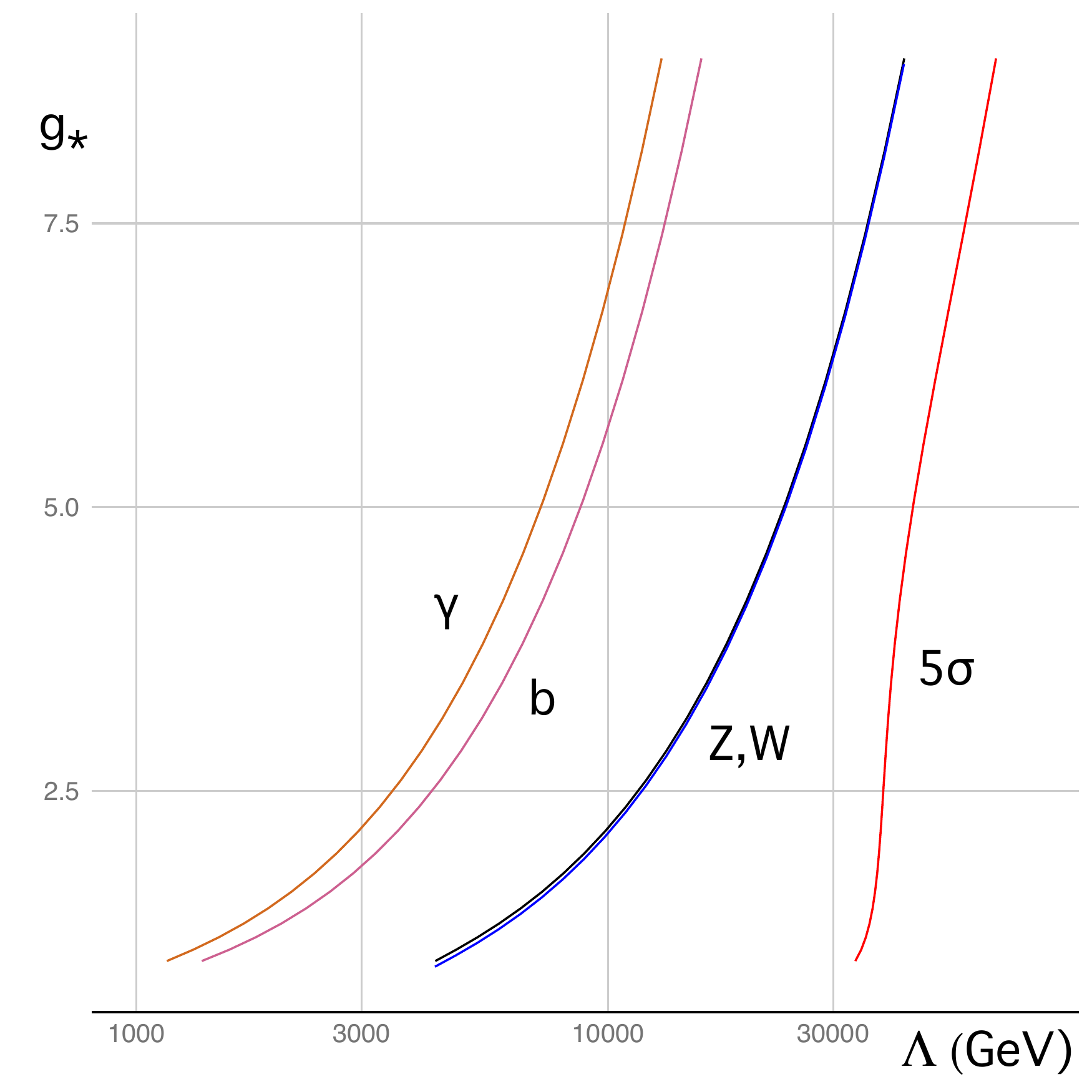}
\  
\ \includegraphics[width=0.42\hsize]{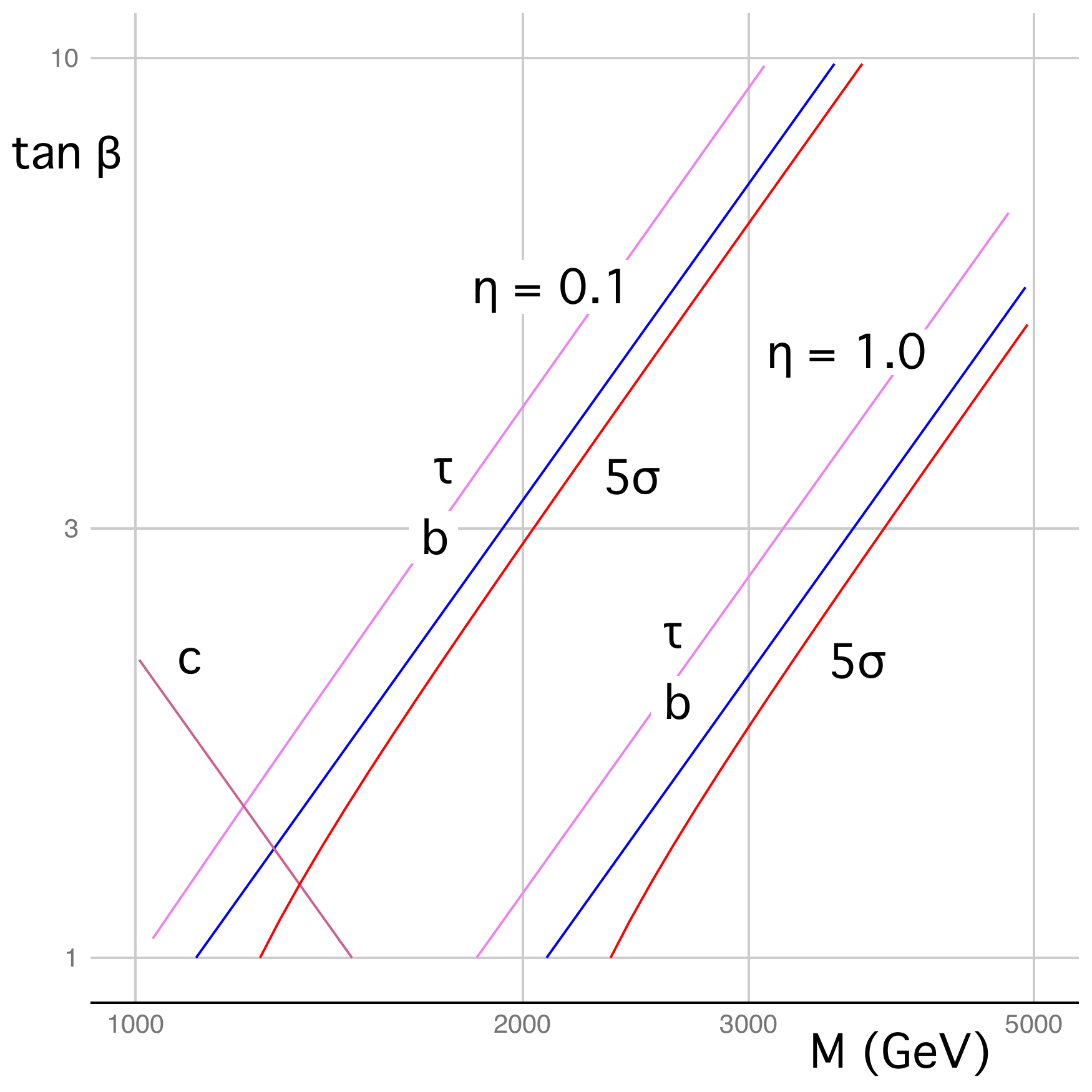}
\\
\includegraphics[width=0.42\hsize]{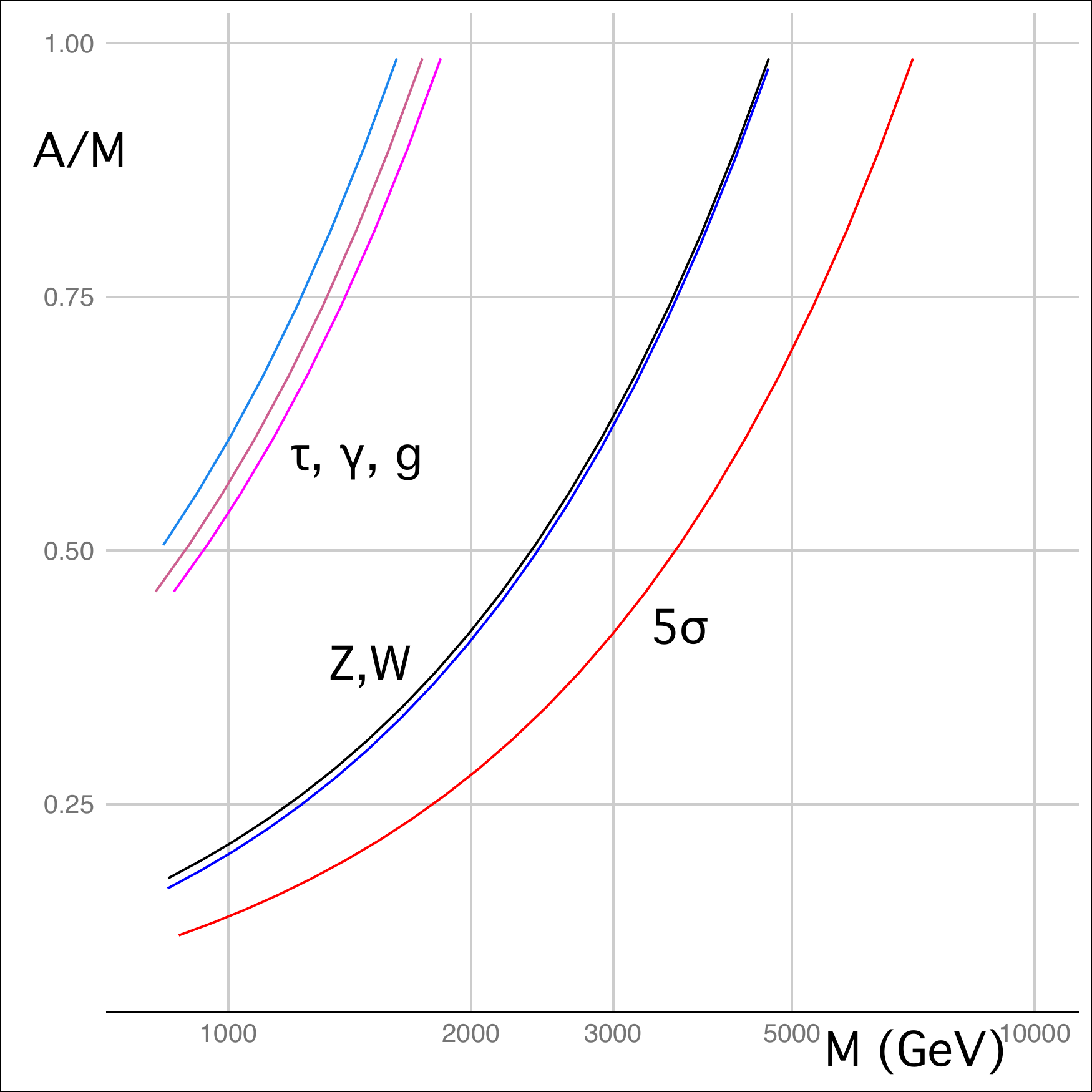}\
\  
\ \includegraphics[width=0.42\hsize]{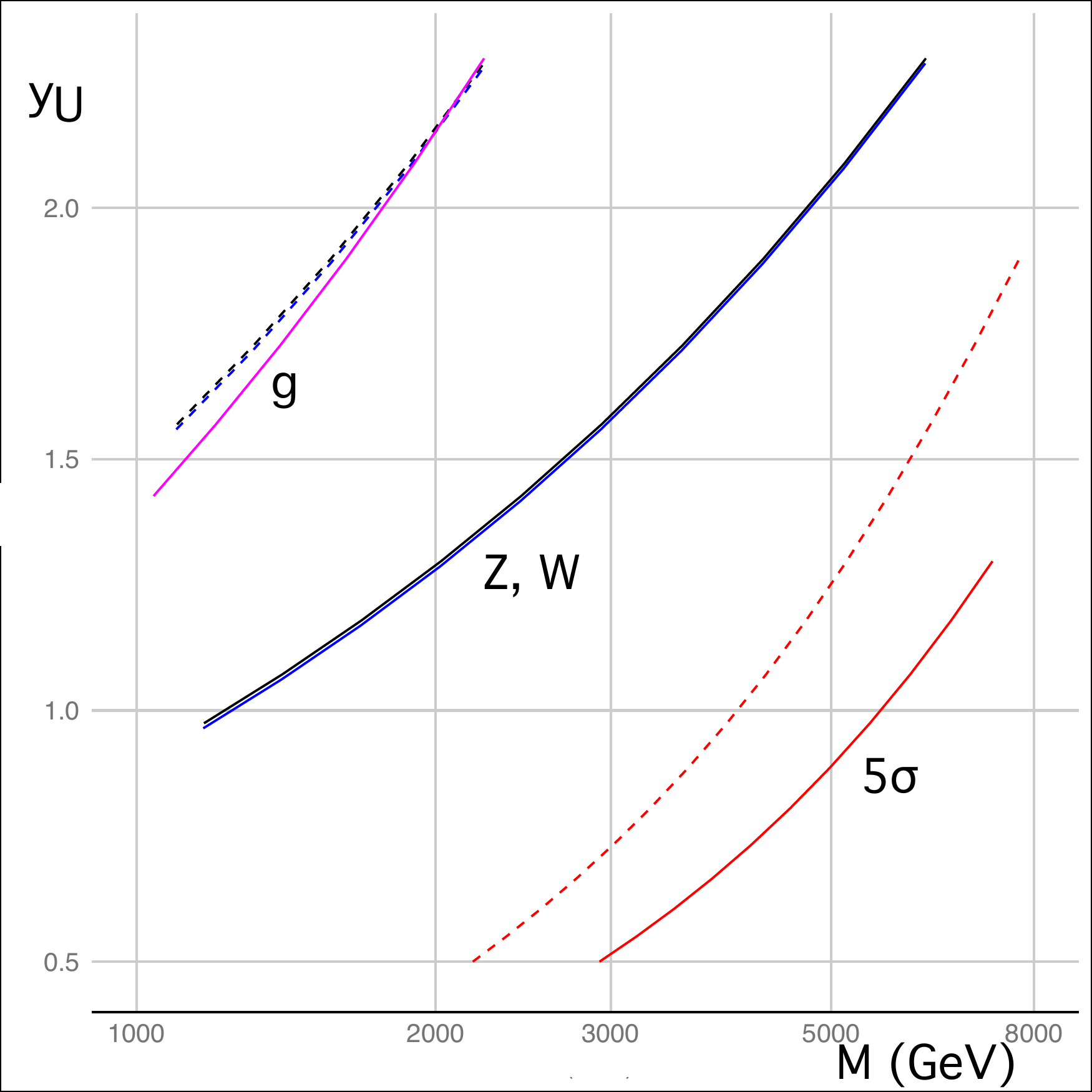}
\end{center}
\caption{Mass reach of Higgs boson coupling deviations in  a variety
  of scenarios reviewed in~\cite{Peskin:2022pfv}.   The curves give the 3$\sigma$
  sensitivity
for particular Higgs boson couplings as a function
of the new particle mass and a dimensionless parameter relevant to
each scenario.  The red curves on the right give the 5$\sigma$
discovery contour for a fit of the full set of Higgs boson coupling
measurements to ILC data.  The scenarios are:  (upper left) the
Strongly Interacting Light Higgs, (upper right) two-Higgs doublet
models, (lower left) models with a scalar singlet, (lower right)
integration out of a vectorlike quark doublet.}
 \label{fig:massreach}
\end{figure}
%%%%%%%%%%%%%%%%%%%%%%%%%%%%%%%%%%%%%%%%%%%%%%%%%%

However, it is quite possible that the coefficient of the
$v^2/M^2$ dependence is a large dimensionless number.  In scenarios
with this property, the mass reach of precision Higgs boson
measurements can be much larger.   The study \cite{Peskin:2022pfv} reviews a
number of 
scenarios that illustrate this using a wide variety of physical
mechanisms. The scenarios are presented, not as complete new physics
models, but in terms of the dimension-6 SMEFT coefficients that are
produced by integrating out specific sections.  These scenarios can
then appear as elements in a variety of complete models.    These
scenarios offer discovery sensitivity to new particle masses above
2~TeV and even in the multi-TeV range.  

We illustrate the mass reach for four of these scenarios in Fig.~\ref{fig:massreach}.
One way to generate such large coefficients is to consider models in
which the Higgs boson is composite, so that higher-dimension operators
contain a new strong coupling constant.   This is illustrated by the
example of the Strongly Interacting Light Higgs (SILH) in which some
SMEFT coefficients are multiplied by the strong couplings $g_*^2$~\cite{Giudice:2007fh}.   In
two-Higgs doublet models in the decoupling limit, the tree level
$v^2/M^2$ corrections are proportional to a Lagrangian term that mixes
the light ($H$) and heavy ($\Phi$ Higgs fields,
\beq
     \Delta\L =   \eta |H|^2 (H^\dagger \Phi + \Phi^\dagger H) \ .
\eeqn
Supersymmetric models tend to predict small values of $\eta$, but
still they lead to sensitivity to heavy Higgs bosons of mass about
1~TeV.  For $\tan\beta < 7$, this is beyond the search reach of
HL-LHC~~\cite{CMS:2019qzn}.  In non-supersymmetric models, $\eta$ can
be 1 or larger, leading to mass reach in the 3~TeV region.  Mixing of
the Higgs boson with a scalar singlet can lead to a significant change
in the overall normalization of Higgs couplings.   Though the HL-LHC
can discover the singlet Higgs up to 2.5~TeV when the mixing angle is of order
10\%~\cite{deBlas:2018mhx}, precision Higgs couplings measurements
  are sensitive to mixings of 1\% and below, with mass reach above
  3~TeV.  Integrating out a heavy vectorlike quark doublet leads to
  Wilson coefficients for SMEFT dimension-6 operators depending on 
  the new quark Yukawa couplings as $y_U^2$ and even $y_U^4$.  The
  Yukawa coupling of the top quark is $y_t\sim 1$ or $\alpha_t =
  0.08$, and larger Yukawa couplings appear often in models with
  vectorlike quarks.  Then the mass reach for precision measurement
  can extend above 3~TeV and thus well above HL-LHC search reach for
  vectorlike quark of about 1.5~TeV~\cite{CidVidal:2018eel}.
  Additional examples, including a specifically supersymmetric effect,
  are discussed in~\cite{Peskin:2022pfv}.

It is straightforward, then, to identify scenarios in which the mass
sensitivity of Higgs boson coupling measurements extends well above 2
and even 3~TeV.   It should be noted that the relevant scenarios are
distinct from those in which the lightest new particles are readily
discovered at the HL-LHC.  As we have emphasized already, precision
Higgs measurements do not explore the same windows accessed by the LHC
and HL-LHC but rather open new windows that have not been sufficiently
explored up to now.

  \section{The Higgs Inverse Problem}
  \label{sec:HiggsInverse}

In our discussion of the influence of different models on the Higgs
boson coupling, we have noted that the various classes of models lead
to different effects.   This suggests that, by measuring the pattern
of deviations of Higgs couplings from their SM values, we can infer
properties of the new physics that led to them.  These models 
 of new physics are accessed only indirectly, through their
 effects on precision measurements, but nevertheless we can gain
 concrete information about their nature.

%%%%%%%%%%%%%%%%%%%%%%%%%%%%%%%%%%%%%%%%%%%%%%%%%%%%%
\begin{figure}[p]
\begin{center}
\includegraphics[width=0.42\hsize]{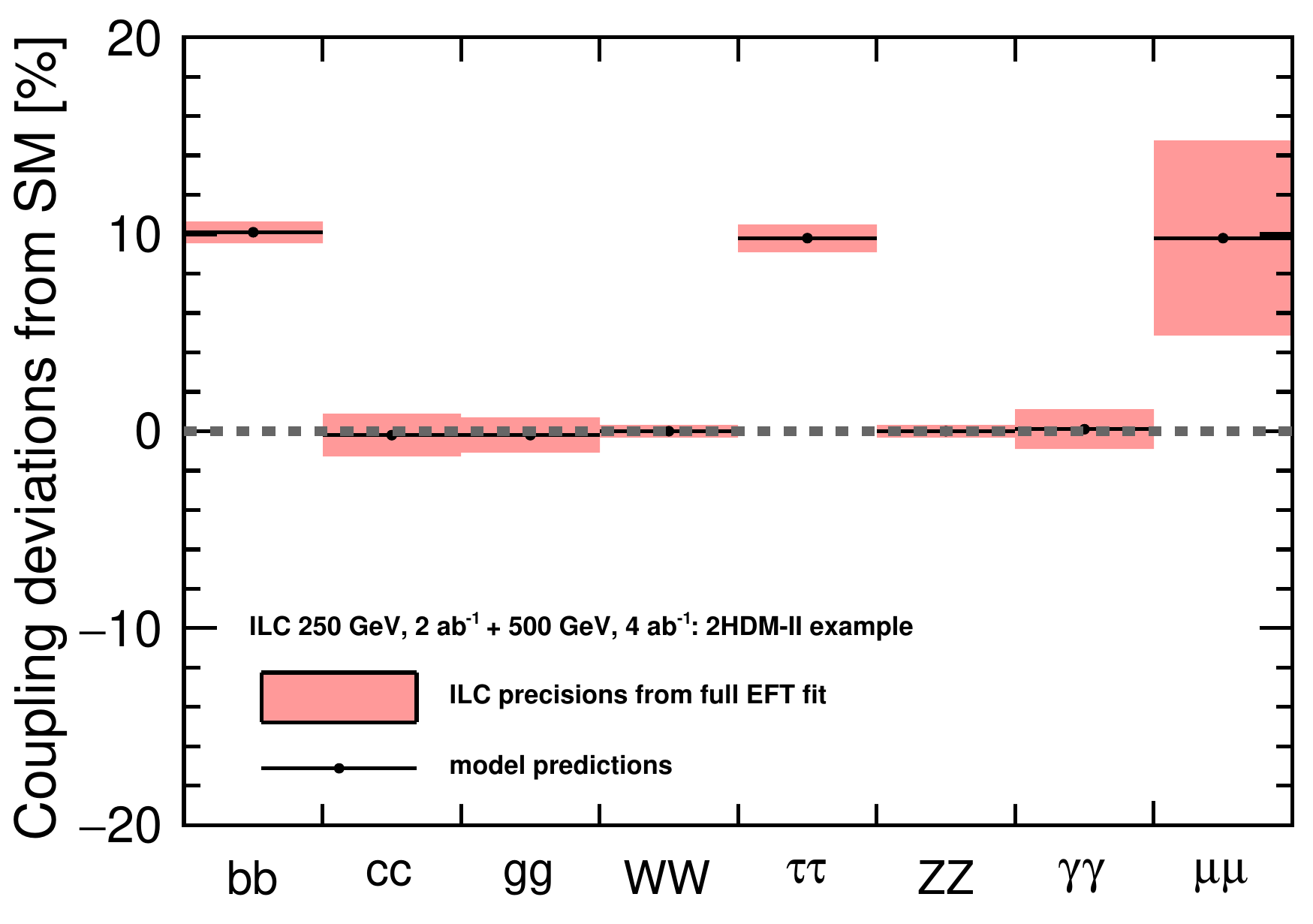}
\  
\ \includegraphics[width=0.42\hsize]{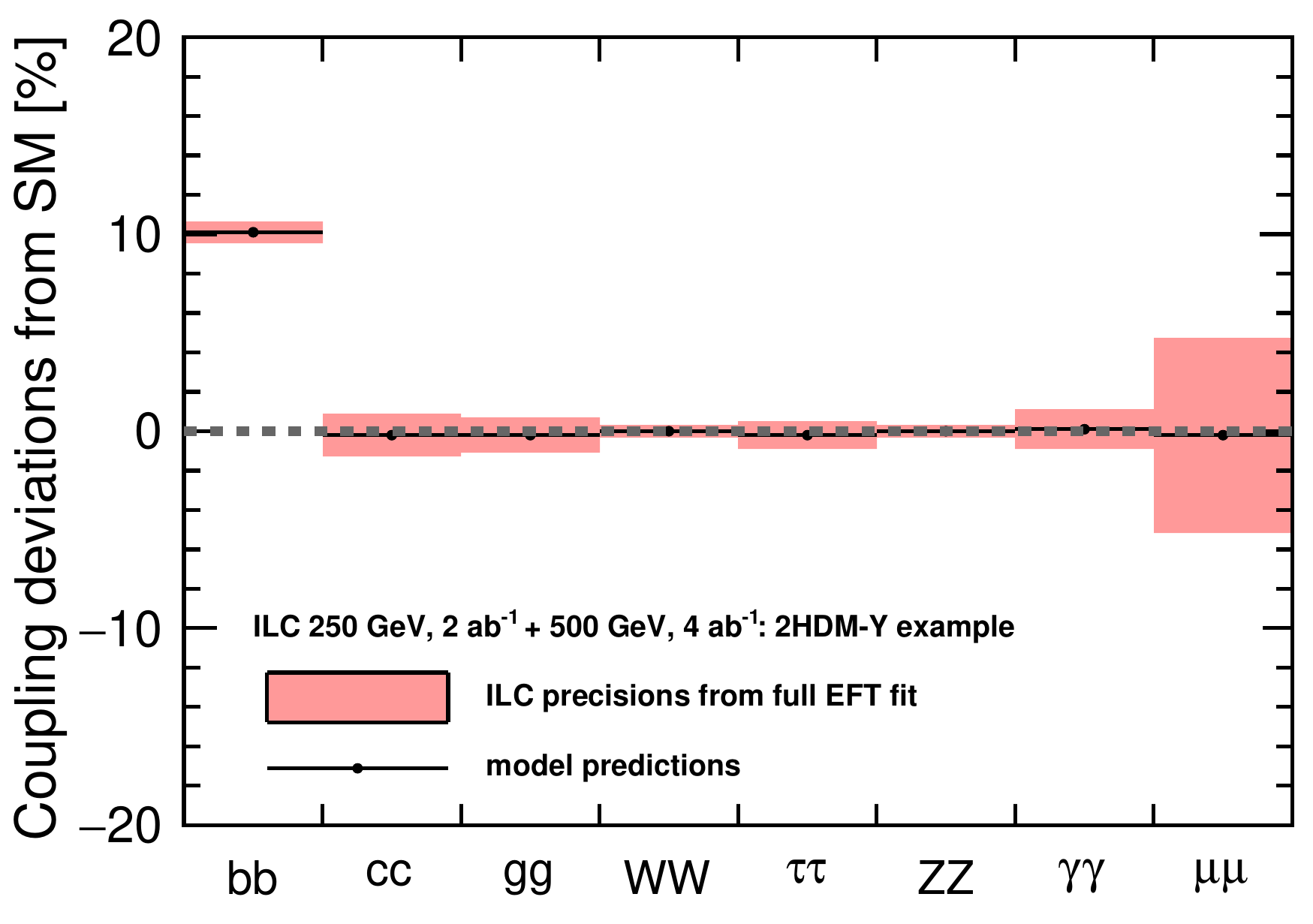}
\\
\includegraphics[width=0.42\hsize]{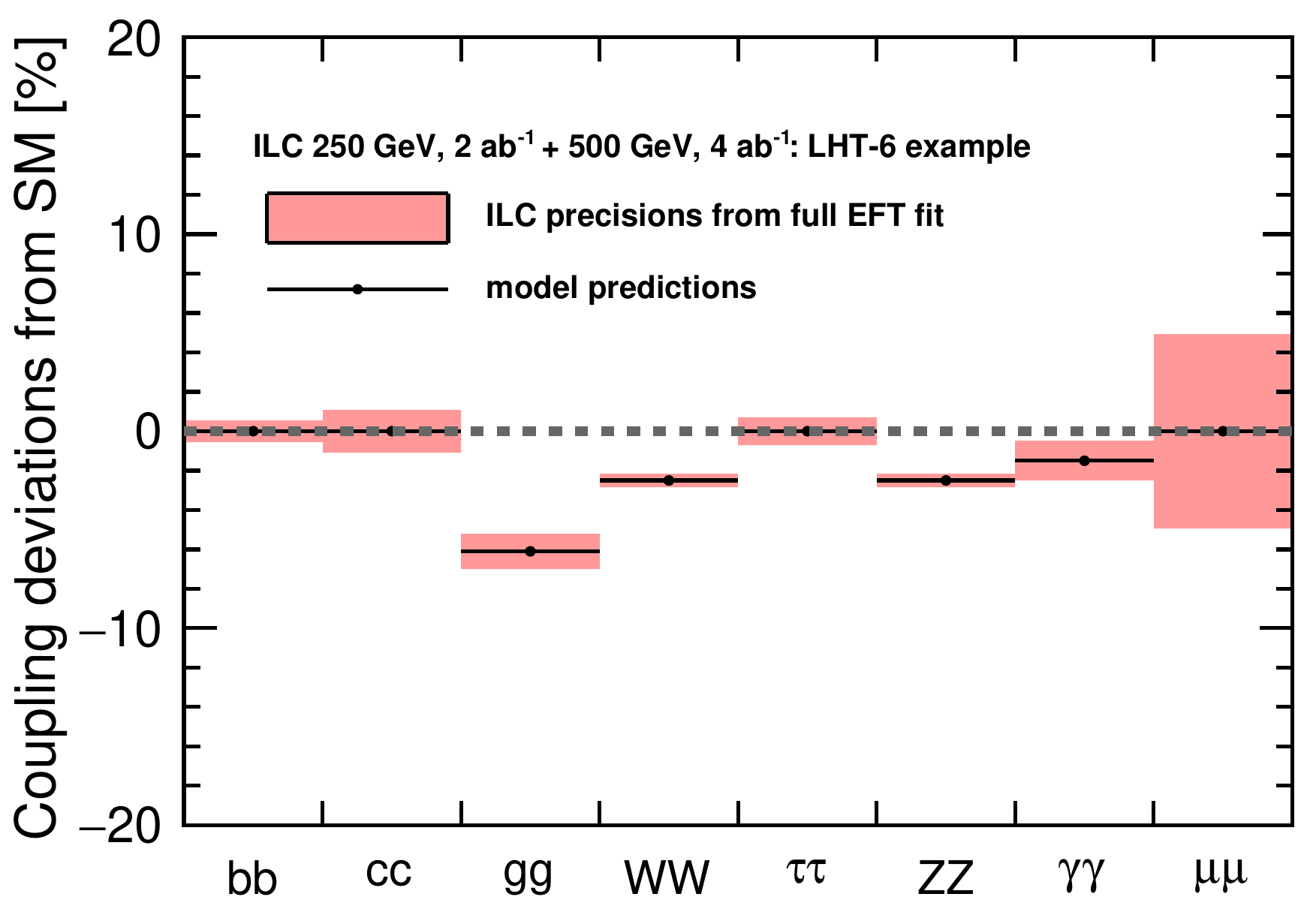}\
\  
\ \includegraphics[width=0.42\hsize]{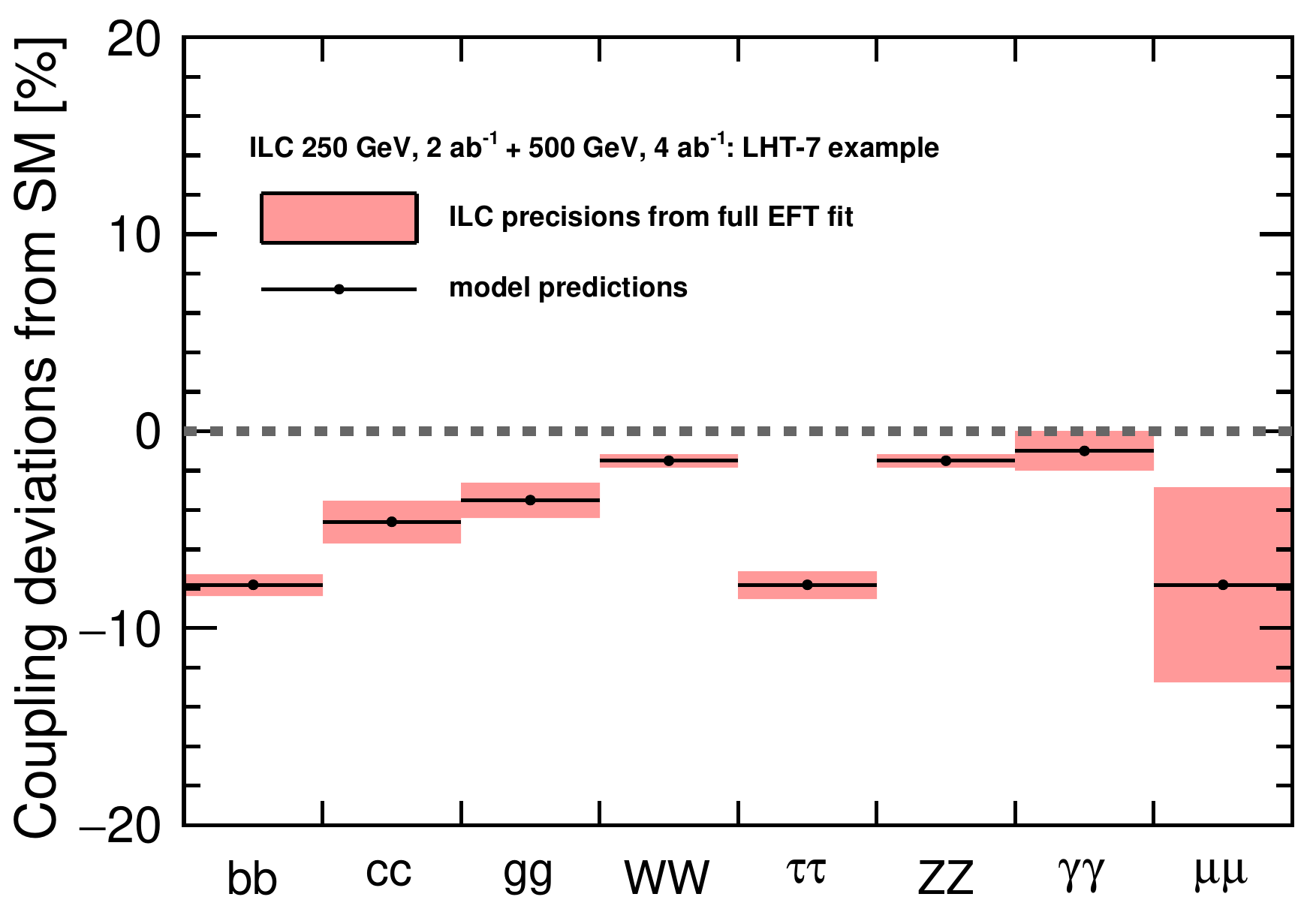}\\
\includegraphics[width=0.42\hsize]{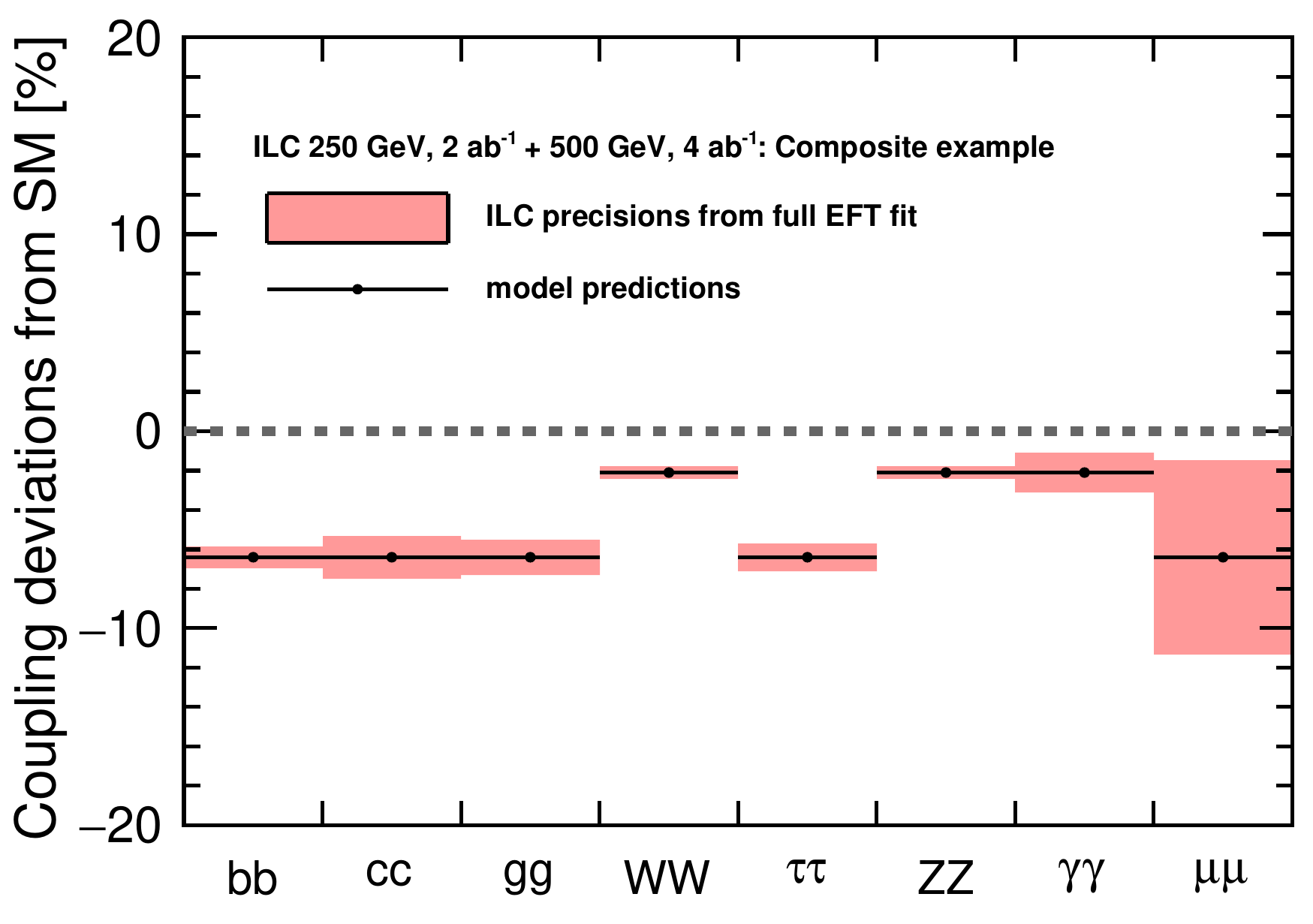}\ 
\  
\ \includegraphics[width=0.42\hsize]{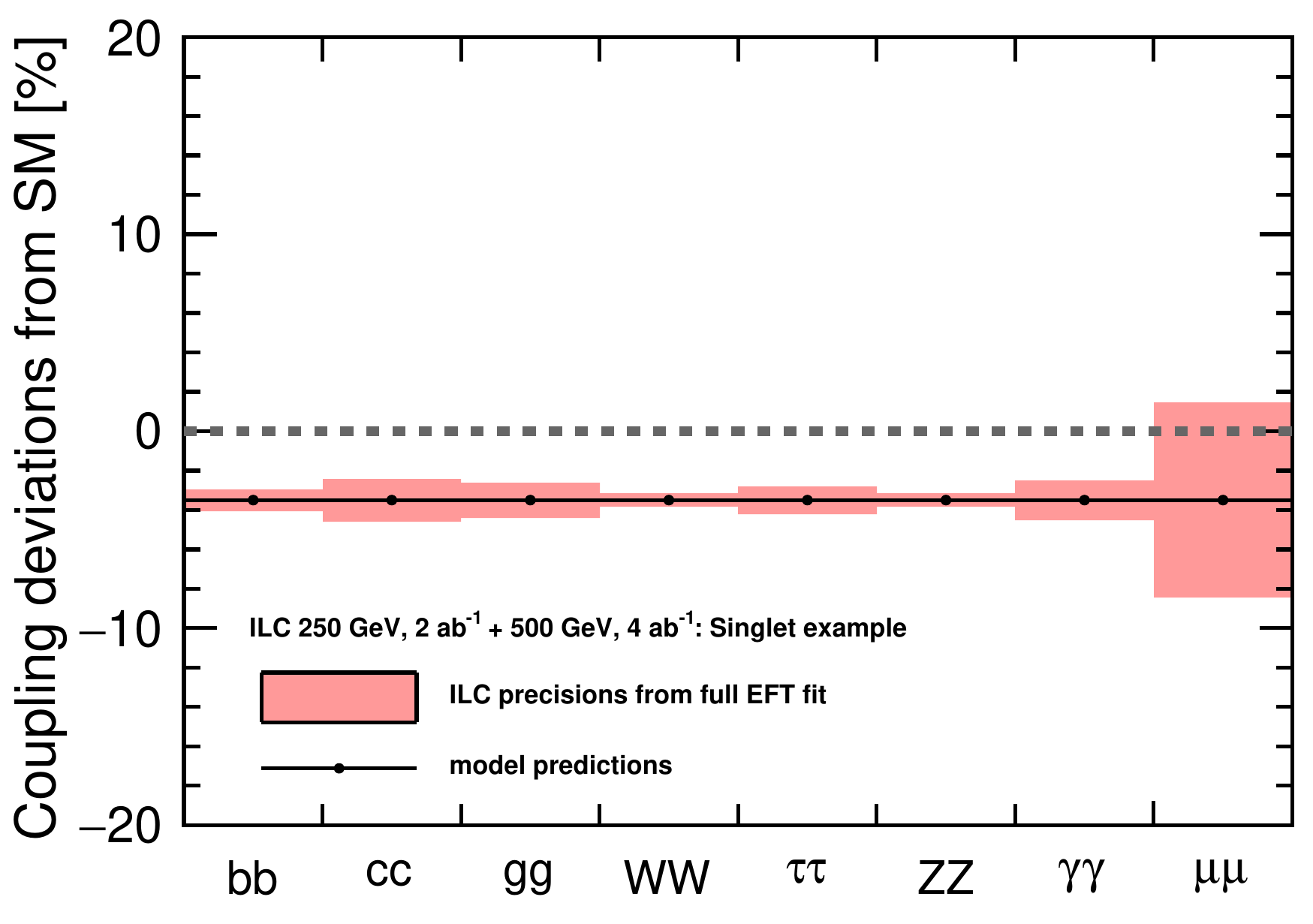}
\end{center}
\caption{Relative deviations of the Higgs boson couplings in six  diverse
  models of new physics.
First row: 2-Higgs doublet models; Second
  row: Little Higgs models; Third row: a Composite Higgs model and a
  Scalar Singlet model. Error intervals shown are those for the
  ILC500.  From~\cite{Barklow:2017suo}; see this
  reference for more details.}
 \label{fig:variousHiggs}
\end{figure}
%%%%%%%%%%%%%%%%%%%%%%%%%%%%%%%%%%%%%%%%%%%%%%%%%%

In the review above, we have pointed out that
\begin{itemize}
\item  In models with extended Higgs sections, including in particular
  supersymmetry, the major effects are on the Higgs boson couplings to
  $b$ and $\tau$
\item In models with new SM singlet scalars and mix with the Higgs
  boson, and in composite Higgs boson models, there is an overall
  shift of the scale of the Higgs boson couplings.   This shift is
  especially visible as shifts in the Higgs boson couplings to $W$ and
  $Z$, which are expected to be the best measured couplings at the
  ILC.  Models with scalar mixing can also lead to large changes in
  the Higgs boson self-coupling.
\item In models with top quark partners, the Higgs boson coupling to
  gluons may be shifted, either positively or negatively, depending on
  the model.   This same effect should be seen in models with extra
  space dimensions.   These models generally produce large deviations
  in the direct Higgs boson coupling to the top quark.
\item In models in which the quark and lepton flavor hierarchies
  originate from extended Higgs sector models with different Higgs
  bosons coupling to each generation, the Higgs boson couplings to
  first and second generation fermions can be enhanced.   At the ILC,
  this effect can be seen in the Higgs boson couplings to $c$, $s$,
  and $\mu$.
\end{itemize}

To illustrate the diversity of expectations for the Higgs boson
couplings, we show in Fig.~\ref{fig:variousHiggs} the predictions for
Higgs coupling deviations from the SM in six specific models of
different types, including extended Higgs models, models with
composite Higgs bosons, and models with scalar singlets.

The ILC, with its comprehensive, high-precision program of Higgs boson
measurements, may well be able to see the overall pattern of Higgs
boson couplings.  This will be vital information to plan the future
stages of exploration with higher energy accelerators.

\chapter{Long-Term Future of the ILC Laboratory}  
\label{chap:farfuture}

The story of the ILC does not end at 500 GeV, or even at 1~TeV.  For the studies that we have described in this report, the ILC will create a new international laboratory with substantial capabilities and infrastructure.  It will be a major world center for particle physics.  If the ILC is constructed in Japan, this laboratory will play a large role in furthering the current rapid growth of physics research in the Asia-Pacific region.

Although we are now proposing only the first stages of this laboratory, it is important that the ILC laboratory should have a longer-term vision that continues the study of particle physics.  Especially if the precision study of the Higgs boson and the top quark reveals the existence of new physics at higher energies, it will be imperative to use the resources of the ILC laboratory to go there and fully characterize the interactions that extend the current Standard Model.  Any electron accelerator at energies above 500~GeV must be a linear collider.  Thus, it is natural to consider extensions of the ILC to meet this goal.

The design of the stages of the ILC up to 500~GeV is mature, well-supported by concrete demonstrations, and ready for construction.  We have presented the relevant ideas and supporting R\&D in Chapter~\ref{chap:accelerator}.   In contrast, 
  the ideas presented in this section are 
frankly speculative.   However, enough is understood to claim that there is a path from the currently proposed ILC to much higher energies.  The routes presented here can be evaluated more concretely and the technologies brought to realization during the construction of the ILC and the course of the experimental program that we have described in the previous chapters.

The layout and geology of the currently favored ILC site allow the construction of linear accelerators as long as 50~km.   Still, to reach multi-TeV or higher energies, we will need to develop new accelerating technologies with much higher accelerating gradients.   These can be based on  superconducting RF, normal-conducting RF, or advanced concepts such as plasma wakefield acceleration. 

In this Section, we will describe visions for the long-term future of the ILC Laboratory.  We first discuss the physics case.  Up to 
3~TeV in the center of mass, the physics case for $\ee$ colliders has been studied in detail as part of the preparations for the CLIC 
program.   We will review the most important items in Sec.~\ref{sec:multiTeV}.  Less study has been done for $\ee$ above 10~TeV, but there are important reasons to study these energies with lepton collider.  We discuss these in Sec.~\ref{sec:multi10TeV}.  We then present
possible accelerator technologies to reach these energies.   Section~\ref{sec:futureSRF} will discuss designs for superconducting RF accelerators.  Section~\ref{sec:futurecopper} will discuss designs based on normal conducting accelerators, both with two-beam acceleration, as in CLIC, and using new advances in direct RF acceleration.  Section~\ref{sec:wakefield}  will discuss designs based on advanced acceleration ideas such as plasma wakefield acceleration.

All of these ideas point to a long-term future for the ILC Laboratory, in which this laboratory remains at the forefront of discovery in particle physics.

\section{Physics opportunities for a multi-TeV collider} 
\label{sec:multiTeV}

 In this section, we will describe the physics issues for $\ee$ colliders up to a CM energy of 3~TeV. Multi-TeV collider options open the gates to accessing TeV new physics directly and exploring new physics in a way that is 
complementary to the lower energy Higgs factory and high energy proton colliders. A multi-TeV lepton collider will shed light on many core puzzles of particle physics. We can probe the Higgs self-coupling, the top quark Yukawa coupling, and electroweak precision observables in di-boson production, adding precision measurements of the SM and testing electroweak symmetry breaking and universal theories.  This program  can also access flavor physics through measurements of flavor-changing neutral currents and  lepton flavor universality violation and through top quark and Higgs boson  exotic decays. A multi-TeV collider enables us to directly produce and probe new particles that might be within its energy range, such as top quark partners, dark matter particles, and hidden sector states.  The prospects for $\ee$ physics up to $E_{CM}$ of 3~TeV have been studied in detail in for the CLIC program at CERN~\cite{deBlas:2018mhx, Franceschini:2019zsg}.   In this section, we will select a few representative physics topics that are
targeted at these energies.

%%%%%%%%%%%%%%%%%%%%%%%%%    
    \begin{figure}
        \centering%
        \includegraphics[width=0.49\linewidth]{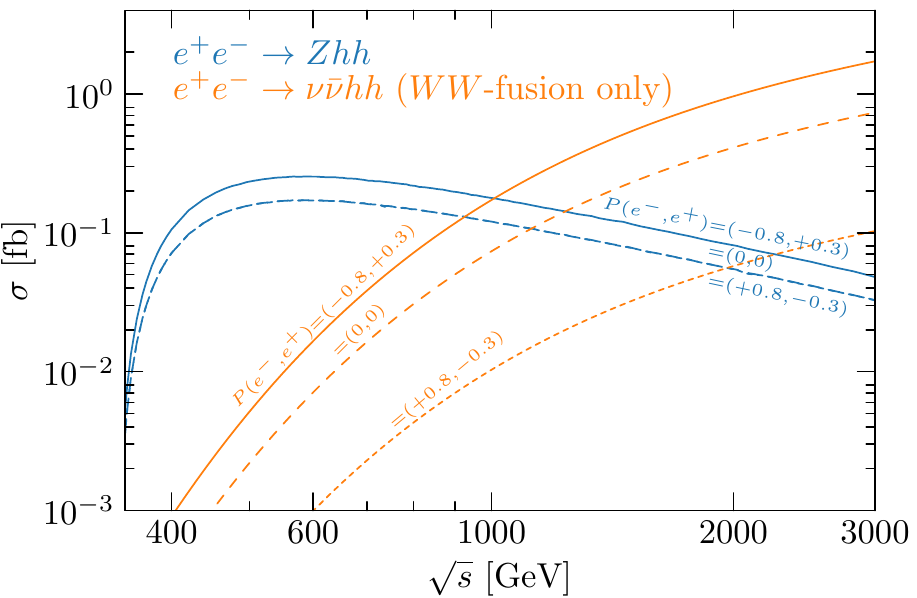}%
        \hfill%
    \includegraphics[width=0.49\linewidth]{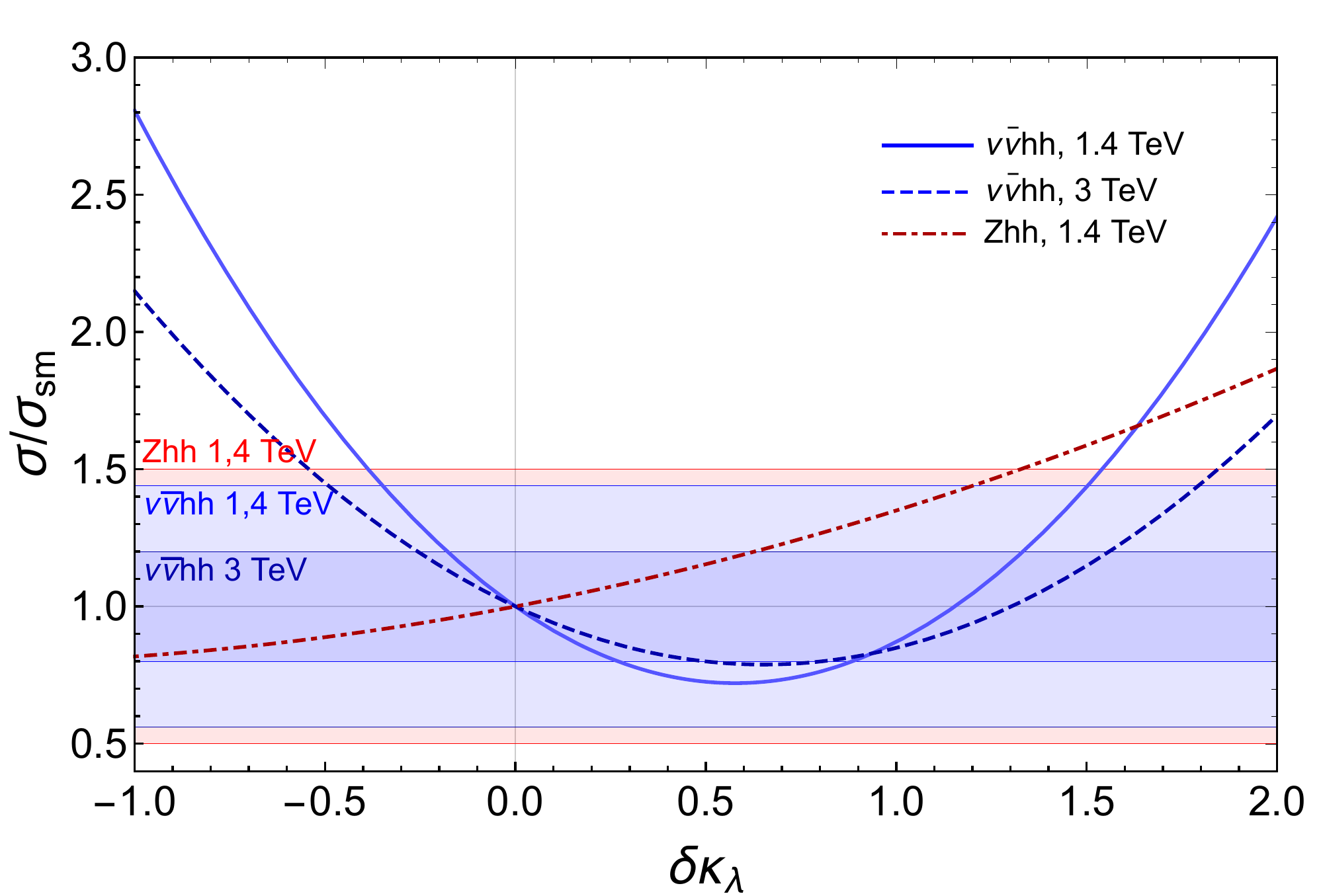}%
     \caption{Left: Cross section of the two leading diHiggs production modes in a lepton collider as a function of the center-of-mass energy. Right: Dependence of the signal strengths on the trilinear coupling of the Higgs with the horizontal bands showing the estimated sensitivities.}
        \label{fig:hhxsec}
    \end{figure}
%%%%%%%%%%%%%%%%%%%%%%%%%%%

\subsubsection{Higgs Self-Coupling}

As we have discussed in Sec.~\ref{sec:HiggsSelf}, the Higgs field self-coupling can be measured using the processes of di-Higgs production at $\ee$ colliders. 
A multi-TeV lepton collider offers two main di-Higgs production modes~\cite{Abramowicz:2016zbo}: 
double Higgs-strahlung ($e^+e^-\rightarrow Zhh$) and vector boson fusion ($e^+e^-\rightarrow\nu \overbar{\nu} hh$). 
The cross sections for the two channels have different dependences on the center of mass energy of the collider, as shown in the  left-hand panel of Fig.~\ref{fig:hhxsec}. Double Higgs-strahlung reaches a maximum not far above threshold (at $\sqrt{s} \sim 500$\,GeV) and then decreases due to the $s$-channel $Z$ boson propagator.
The vector boson fusion cross section receives a logarithmic enhancement at higher collider energies, giving an advantage for 
its study at multi-TeV energies.  The CLIC study considered measurements at $\ee$ colliders 
with $\sqrt{s}=1.4$ and $3\,$TeV,  with $1.5$ and 2~ab$^{-1}$ of integrated luminosity, respectively, with unpolarized beams~\cite{Abramowicz:2016zbo}.

In the right-hand panel of  Fig.~\ref{fig:hhxsec},
we show how the trilinear Higgs self-coupling changes the total cross-section in the two leading di-Higgs channels.
The result is shown as a function of $\delta\kappa_\lambda = \lambda/\lambda_{SM} - 1$, the correction to the Higgs self coupling normalized to its SM value. 
We can see an interesting complementarity between these two leading di-Higgs production channels. The $Zhh$ cross-section grows for $\delta \kappa_\lambda > 0$ through constructive interference, more sensitive to positive deviations in the trilinear Higgs self-coupling.
The $\nu \overbar{\nu} hh$ production, instead, is more sensitive to negative shifts of the trilinear coupling.   Note, though, that, at high energies, even if the total cross section is insensitive to the presence of $\delta\kappa_\lambda$, the di-Higgs mass distribution shifts toward values closer to $2m_H$ when $\delta\kappa_\lambda$ becomes large~\cite{Roloff:2019crr}.

After combining both vector boson fusion and double Higgsstrahlung channels, the two runs at $1.4\,$TeV and $3\,$TeV
are sufficient to exclude the second fit minimum at $\delta \kappa_\lambda \sim 1$ at $95\%$CL. We show the results in Table~\ref{tab:clicvvhh01}. 
 
        \begin{table}[t]
            \centering
            \begin{tabular}{c|c|c}
                & $\Delta\chi^2 =1$ & $\Delta\chi^2 =4$\\
                \hline
                \hline
                \rule[-.5em]{0pt}{1.6em} 1.4\,TeV   &  $[-0.22,~0.48]$    &  $[-0.40,~1.05]$    \\ \hline
                \rule[-.5em]{0pt}{1.6em} 3\,TeV   & $[-0.13,~0.16] \cup [1.13,~1.42]$     &  $[-0.24,~0.42] \cup [0.87,~1.53]$   \\ \hline
                \rule[-.5em]{0pt}{1.6em} combined   &  $[-0.12,~0.14] $     &  $[-0.21,~0.35]$    \\ \hline
                \rule[-.5em]{0pt}{1.6em} 5 bins $m_H$ for $\nu \bar{\nu} hh$   &  $[-0.11,~0.13]$   &  $[-0.21,~0.29]$
            \end{tabular}
            \caption{
              Single operator constraints on $\delta \kappa_\lambda$ deriving from the measurements of $Zhh$ and $\nu \overbar{\nu} hh$ cross sections, with all other parameters fixed to their standard-model values. In the fourth row, a differential $m_{hh}$ measurement in weak boson fusion di-Higgs production at $\sqrt{s}=3\,$TeV is further included.
            }
            \label{tab:clicvvhh01}

        \begin{tabular}{l|c|c}
            & 68 \%CL & 95\%CL\\
            \hline\hline
            \rule[-.5em]{0pt}{1.6em}1.4\,TeV, exclusive  &  $[-0.21,~0.34]$     &  $[-0.38,~0.89]$   \\ \hline
            \rule[-.5em]{0pt}{1.6em}1.4\,TeV, global  &  $[-0.22,~0.40]$     &  $[-0.39,~1.00]$    \\ \hline\hline
           \rule[-.5em]{0pt}{1.6em}1.4\,+\,3\,TeV, exclusive\hspace{.5cm}  &  $[-0.11,~0.12]$     &  $[-0.20,~0.27]$    \\ \hline
           \rule[-.5em]{0pt}{1.6em}1.4\,+\,3\,TeV, global  &  $[-0.11,~0.13]$     &  $[-0.21,~0.29]$    \\ \hline
        \end{tabular}
        \caption{
        Single and global constraints on $\delta \kappa_\lambda$ after the $1.4$ and $3\,$TeV runs of CLIC. We also show the combined results with the HL-LHC. }
        \label{tab:highE}
\end{table}

The di-Higgs production is also affected by modifications in other Higgs couplings. To consistently reduce the model dependence, we performed a study comparing single-operator constraints to that of a global fit, shown in Table~\ref{tab:highE}. The
$\nu \overbar{\nu} hh$ production with a differential analysis including $4$ bins in the $m_{hh}$ distribution, and 
the inclusive $Zhh$ cross-section and the $\delta \kappa_\lambda$ dependence of the single-Higgs processes are included in this fit.
The $3\,$TeV run will markedly increase the Higgs self-coupling sensitivity over that for $1.4\,$TeV due to the increase in statistics allowing  access to the differential distributions.

\subsubsection{Higgs and Top Compositeness}

\begin{figure}
\begin{centering}
\subfloat[\label{fig:ch}]{\centering{}\includegraphics[width=0.45\linewidth]{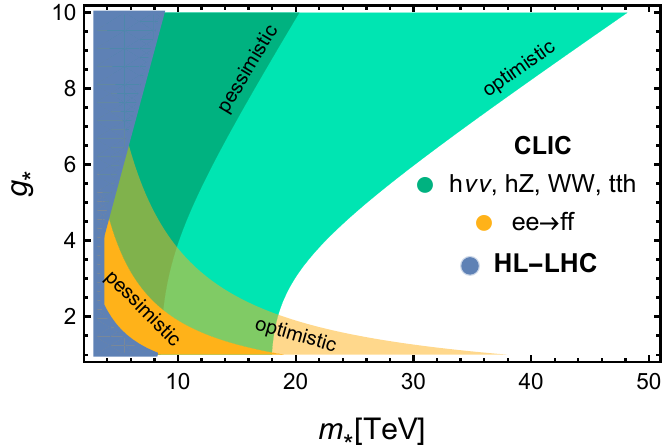}
}~~\subfloat[\label{fig:ch-1}]{\centering{}\includegraphics[width=0.45\linewidth]{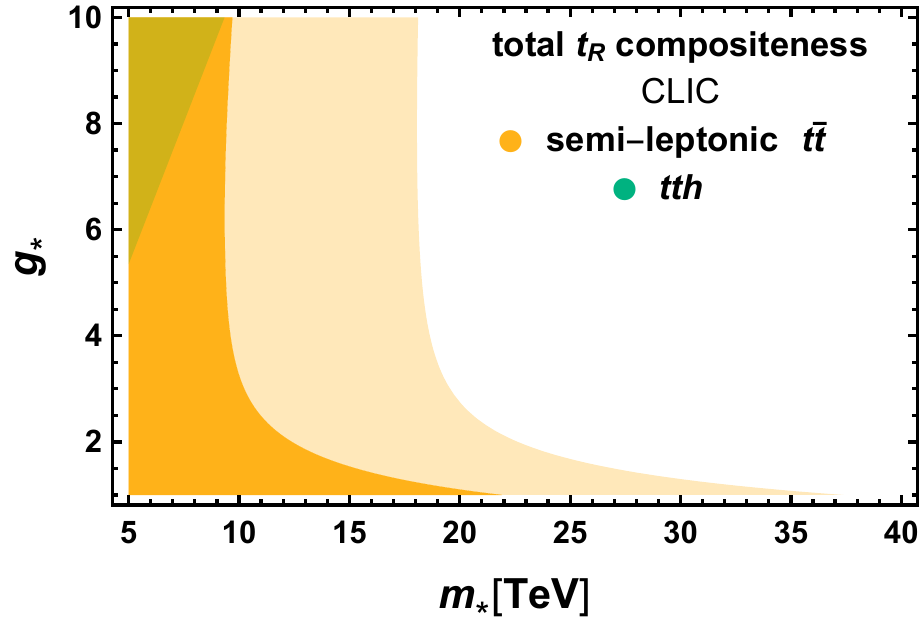}
}
\par\end{centering}
\caption{Composite Higgs reach from Higgs boson, top quark and Drell-Yan studies
taken from Refs.~\cite{deBlas:2018mhx} and \cite{CLICdp:2018esa}. Left panel: $5\sigma$ discovery contours for Higgs compositeness
in the $(m_{*},g_{*})$ plane, and as well the $2\sigma$ projected
exclusions from the HL-LHC. Right panel: The $5\sigma$ top quark compositeness discovery contours in the $(m_{*},g_{*})$ planes from studies of $t\overbar{t}$ and $t\overbar{t}h$ final states.
In both panels, darker and lighter shaded areas correspond to the variations of the size of the operators' coefficients by a factor of 2 or 1/2 on top of the baseline expectation from the values
of $m_{*}$ and $g_{*}$.}
\end{figure}

The Higgs precision program at a multi-TeV lepton collider not only reveals the Higgs trilinear coupling but, more importantly, provides a holistic understanding of the dynamics of the Higgs boson. We illustrate this through an analysis sensitive to the  geometric size $l_{H}$ of the composite Higgs boson.   As we have discussed in Sec.~\ref{sec:BSMscale}, the size of a composite Higgs boson is 
measured by the SMEFT coefficient  $c_H$.  This and other  operator coefficients are enhanced or suppressed by positive or negative powers of the composite coupling parameter $g_*$~\cite{Giudice:2007fh}.  Constraints on the SMEFT coefficients can be translated into constraints on $g_*$ and a SMEFT decoupling scale $m_*$ as illustrated in 
Figure~\ref{fig:ch} ~\cite{deBlas:2018mhx}.
The projected HL-LHC exclusion reach (as opposed to discovery lines shown for CLIC) is also shown in the figures.
The improvement achieved by CLIC at small and intermediate
$g_{*}$ is due to the high-energy stages that allow for a very precise
determination of the $c_{H}$, $c_{T}$, $c_{WW}$ and $c_{BB}$
SMEFT coefficients. 
Single Higgs boson couplings measurements provide the most stringent constraints at large $g_{*}$.

FIgure~\ref{fig:ch} clarifies the  complementarity between precision and high mass searches. Precision measurements of the Higgs boson couplings probe one combination of the two characteristic parameters of this scenario, while the other combination is probed with less copiously produced events at $E_{CM} = 3$~TeV and  high invariant mass. 
We can also consider top quark compositeness in connection with the naturalness puzzle. SMEFT operators in the top sector can be probed by measuring the top Yukawa coupling and as well as $t{\overline{t}}$ production at high-energy~\cite{AlAli:2021let}. The reach in the ``total $t_{R}$ compositeness'' scenario is displayed on Fig.~\ref{fig:ch-1}. For further details in the case of ``partial top compositeness'' see Sec.~2.1 of Ref.~\cite{deBlas:2018mhx} and Sec.~10.2 of Refs.~\cite{CLICdp:2018esa}.

\subsubsection{Dark Matter}

\begin{figure}
\begin{centering}
\subfloat[\label{fig:reach-DM}]{\begin{centering}
\includegraphics[width=0.48\linewidth]{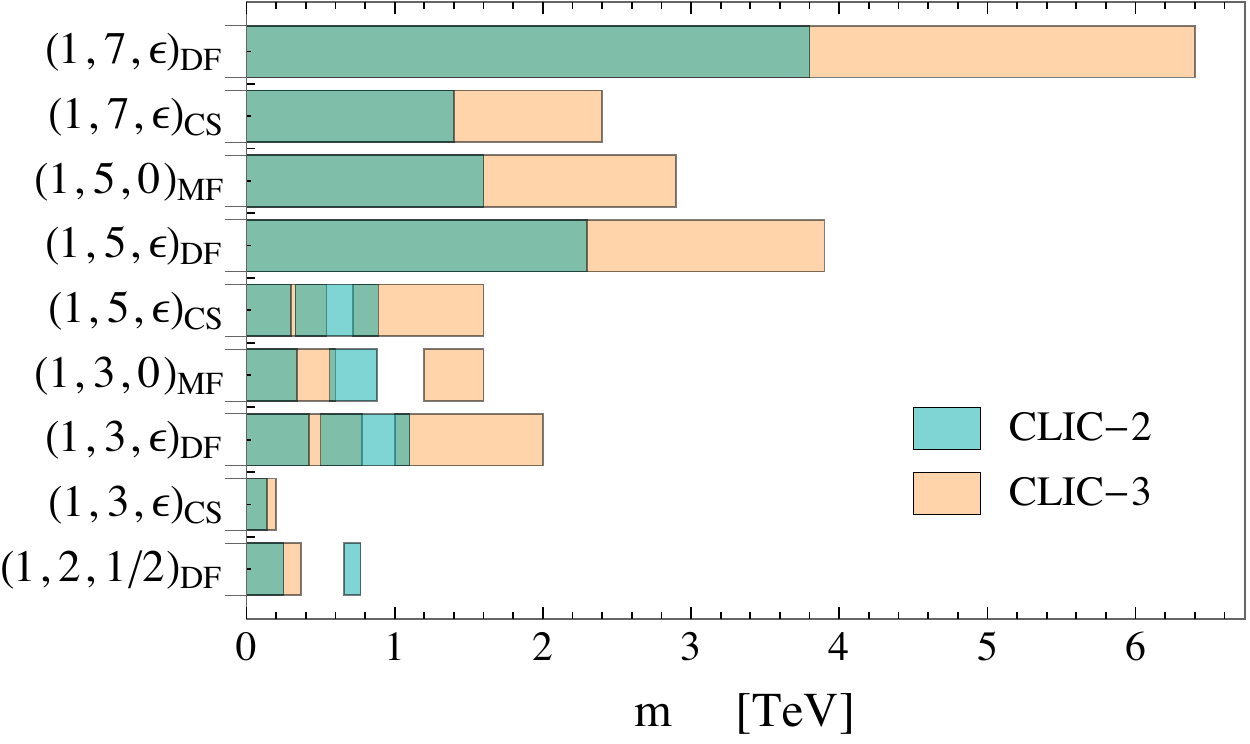}
\par\end{centering}
}~~~\subfloat[\label{fig:reach-higgsino}]{\begin{centering}
\includegraphics[width=0.44\linewidth]{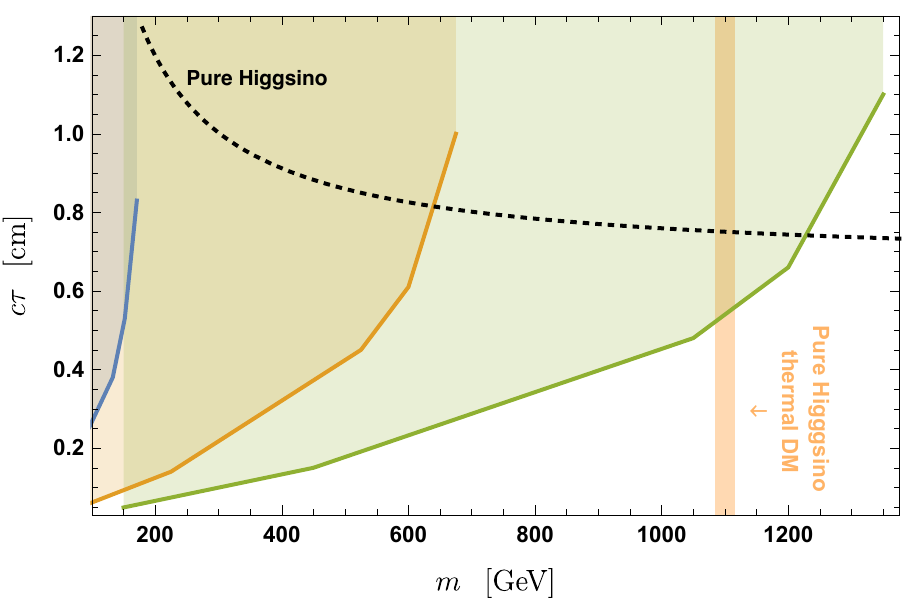}
\par\end{centering}
}
\par\end{centering}
\caption{Reach of direct searches for Dark Matter. Left panel: 95\% excluded masses for new electroweak $n$-plet states with hypercharge $Y$. The exclusion results for each state denoted by (1,n,$Y$) at CLIC Stage2 and Stage3 are presented in green and yellow bar \cite{DiLuzio:2018jwd}. Right panel: 95\% excluded region for pure Higgsino in the mass-lifetime plane. The black dashed line denotes the lifetime of a pure Higgsino. The green, yellow and blue areas correspond to 3 TeV, 1.5 TeV and 380 GeV CLIC expected exclusions, respectively.}
\end{figure}

As we have emphasized in Sec.~\ref{sec:ILCdark}, 
little is known about the particle identity of cosmic dark matter. A general and  compelling candidate is the thermal Weakly Interacting Massive Particle (WIMP). WIMPs are in thermal equilibrium in the early universe. From this boundary condition, it is possible to predict the current abundance of these particles.   This yields the observed abundance of dark matter for masses of order 
\beq
       M_{\text{WIMP}}\simeq\text{TeV} \bigl(g_{\text{DM}}/g_w\bigr)^2, 
\eeqn
where $g_{\text{DM}}$ gives the strength of the WIMP coupling in its annihilation reactions and $g_w$ is the 
SM weak interaction coupling.

In Refs.~\cite{deBlas:2018mhx,Han:2020uak,AlAli:2021let}
comprehensive studies are presented to test general WIMPs.
WIMPs specified by masses and quantum numbers of new states are dubbed ``Minimal DM''~\cite{Cirelli:2005uq}.
It is shown in these studies that WIMPs can come from many different representations of the SM $SU(2)$ gauge group, 
even as large as the $7$-plet.  This whole variety of candidate WIMPs are targets 
for future colliders.    TeV lepton colliders can probe them in different ways, including the ISR photon searches 
discussed in Sec.~\ref{sec:WIMPDM} but also in processes involving their charged $SU(2)$ partners . We show the $95\%$~CL sensitivities from of Ref.~\cite{DiLuzio:2018jwd} in Fig.~\ref{fig:reach-DM}. 
The sensitivity reaches the thermal targets in the case of the Dirac fermion triplet candidate $(1,3,\epsilon)_{{\rm {DF}}}$. 
Next, one can exploit the long-lived particle signatures from the charged state in the electroweak multiplet. Its distinctive signature is thus a disappearing
track. In figure~\ref{fig:reach-higgsino} we shows that a 3~TeV $\ee$ collider  can discover the thermal Higgsino at $1.1$~TeV.

Beyond the minimal dark matter scenarios, thermal WIMPs can show up in different ways at a TeV lepton collider. For instance, in co-annihilation
scenarios, two nearly degenerate states can scatter with a larger rate than the DM alone.  The Inert Doublet model, discussed earlier in Sec.~\ref{sec:newscalars},
can also be thoroughly explored at TeV lepton colliders, extending significantly the domain of the parameters space probed in comparison to the HL-LHC capabilities \cite{Kalinowski:2018kdn,Kalinowski:2018ylg}.  Details on these and other models are presented in~\cite{deBlas:2018mhx,Han:2020uak}.
Here we are content with stating that, in general, a TeV lepton collider can effectively probe DM models with a sufficient mass-splitting that the DM particles are produced promptly, filling the gap left by the LHC searches.

\subsubsection{Hidden Sector}

Hidden sector dynamics represents a large class of well-motivated BSM physics that is elusive at hadron colliders. Specialized search strategies are often needed. Here we choose two examples: an RPV electroweakino in connection to baryogenesis, and Higgs decaying into long-lived hadronic particles in connection to neutral naturalness. The first example represents the reach for heavy new states and the second example represents the reach for light states through Higgs decays.

\paragraph{R-parity Violating Long-Lived Wino and Higgsino:}

We consider a weak scale particle X that decays after thermal freeze-out and has an R-parity violating, baryon number violating decay. The particle freezes out when its annihilation rate falls below the Hubble expansion rate. The temperature at freeze-out, $T_{\rm fo}$, depends only logarithmically on the annihilation cross-section, such that $T_{\rm fo}\sim M_X/20$ for annihilation cross-sections $\sim$ fb.

The cosmological condition that $X$ decay out of equilibrium requires that
\beq
 c\tau_X \gsim 1~{\rm cm}\left(\frac{100~{\rm GeV}}{M_X}\right)^2.
\eeqn
Scattering with the SM may keep $X$ in thermal equilibrium down to $T_{\rm fo}$, in which case the decay length should be somewhat longer. If $X$ decays after the freeze-out, this leads to a final baryon asymmetry proportional the relic abundance that this particle would produce if it did not decay~\cite{Cui:2012jh}.   In any case, This model predicts new particles that can decay with a range of 
possible lifetimes, visible in various components of a detector at a collider,  typically in the displaced vertex regime 
(or out of the detector as missing energy).

If the decay temperature is less than the freeze-out temperature, $T_{\rm fo}>T_{\rm d}>T_{\rm BBN}$, and assuming that we can neglect washout processes, the baryon asymmetry is given by
\begin{align}
\Delta_B = \epsilon_{\rm CP} n_X(T_{\rm fo}),
\end{align}
where $\epsilon_{\rm CP}<1$ is a measure of $CP$ violation in the decays that can be generated by interference between tree-level and loop-level decay diagrams. Directly measuring such a CP violation effect tied to baryogenesis at collider experiments is exciting yet generally challenging. We will focus on displaced decay signals tied to the other Sakharov condition for baryogenesis. 

Note that the lifetime of the parent particle $X$ can be naturally very different from that expected from the couplings that lead to its production. For example, suppose that an approximately conserved $Z_2$ symmetry is responsible for the long lifetime. In that case, $X$ particles can still be produced in pairs via $Z_2$ conserving interactions but decay slowly through interactions that violate the symmetry. The TeV lepton collider could copiously produce these particles. An earlier study proposed simplified models for WIMP baryogenesis mechanisms and studies of sensitivity to these models in some searches at ATLAS and CMS  \cite{Cui:2014twa}. 

\begin{figure}[t]
	\centering
	\includegraphics[scale=0.3,clip]{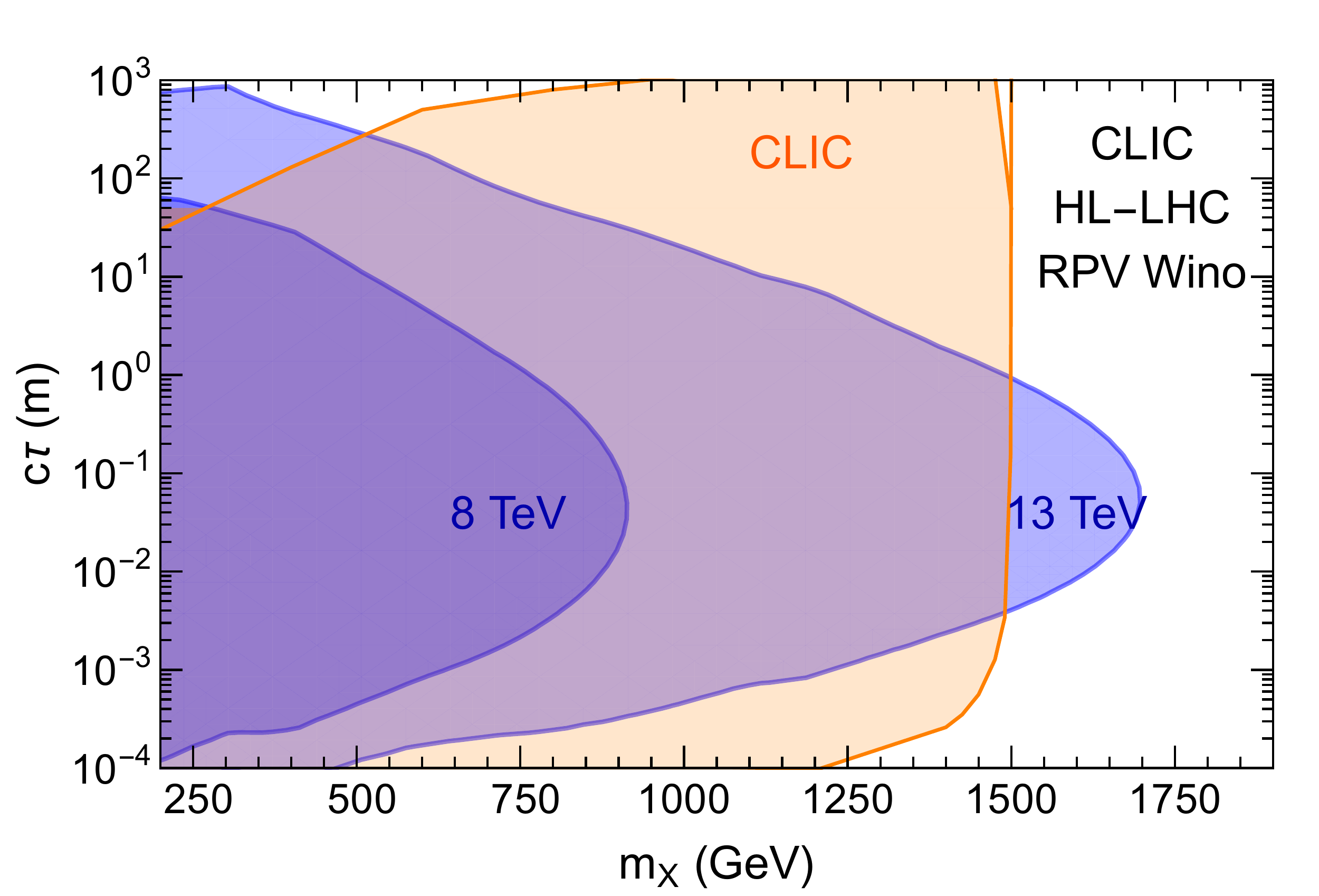}
	\includegraphics[scale=0.3,clip]{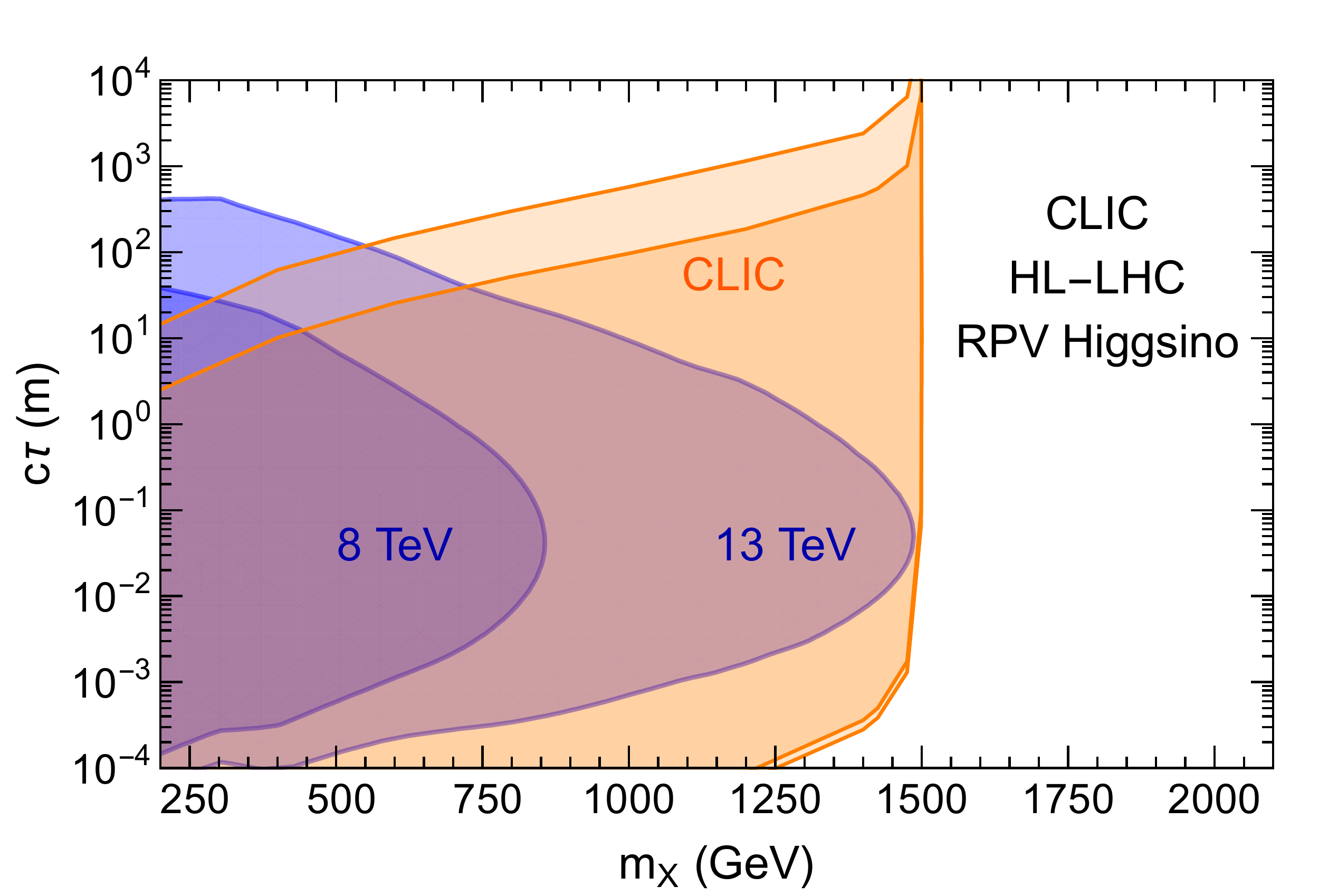}
	\caption{Event rates and exclusions for the wino and higgsino signal in the lifetime vs. mass plane.
Orange: darker region corresponds to $N > 30$ events in the CLIC acceptance, lighter orange regions
corresponds to $N > 3$ events and correspond to a projected 95\% C.L. exclusion limit for zero expected
background. The left (right) panel refers to the RPV wino (higgsino) signal. Blue region: the recasted current and HL-LHC (3 $\rm ab^{-1}$) projected 95\% C.L. exclusion limit as the function of Wino mass and its lifetime.}
	\label{fig:LLPCLIC}
\end{figure}

The coverage extends to long and short lifetimes, covering 0.1 millimeters to 500~meters for a 500 GeV wino.
These pair-produced winos have low boost factors and therefore move slowly. Further development in using the precision timing for LLPs at the LHC, similar to the GMSB Higgsino benchmark study in Ref.~\cite{Liu:2018wte}, could improve the HL-LHC sensitivity significantly, especially for the long lifetime regime.

The advantage of high collision energy enables the LHC to cover wino mass up to 1650 GeV in the most sensitive $c\tau$ range ($\sim 10$ cm). 3 TeV CLIC thus cannot compete with LHC in terms of the mass reach of the wino in general. But there is ample parameter space in $c\tau$ at masses below 1.5 TeV that HL-LHC is not sensitive to. This is due to the extensive QCD background at the LHC and the current limit in vertex reconstruction efficiency. In contrast, an $\ee$ collider provides a much cleaner environment for these searches, with almost complete coverage for electroweak states below 1.5 TeV mass. With a much lower background (in particular for hadronic channel) and improved vertex reconstruction techniques, a 3~TeV collider  has the potential to close up the region that HL-LHC is not capable of effectively probing, as illustrated in Fig.~\ref{fig:LLPCLIC}.

In Fig.~\ref{fig:LLPCLIC}, the projected exclusion limit for a 3 TeV$\ee$ collider  at $95\%$ C.L. for the luminosity of $3\,\text{ab}^{-1}$ is indicated by the orange region in the wino mass and $c\tau$ parameter space, overlaid on the blue regions showing the LHC sensitivity. Here we simulated pair production of wino-like charginos at 3~TeV. The charginos almost exclusively decay to wino-like neutralino, and a hefty $\mu$ term heavily suppresses a soft pion since the couplings to bino-like neutralino states. The wino-like neutralino decays hadronically via RPV couplings. We make a simplifying assumption for charginos: $c\tau_{\chi^\pm\rightarrow\chi^0}<<c\tau_{\chi^\pm({\rm RPV})}$, so that tracks contributing to DVs come entirely from wino-like neutralinos and the soft pion track is not associated with any vertex.
We assume a nearly perfect vertex reconstruction efficiency in the $c\tau$ range of $0.3-100$ mm for the analysis.
An $\ee$ collider at  3~TeV with $3\,\text{ab}^{-1}$ luminosity is sensitive to the large parts of parameter space that LHC is not, below the wino mass of $1500$~GeV. It almost entirely covers the open parameter space for $c\tau>1$~cm and $m_\chi<1500$~GeV. For lower $c\tau$, this lepton collider can offer up to an order of magnitude improvement in the reach in $c\tau$.
\begin{figure}[t]
	\centering
	\includegraphics[scale=0.4,clip]{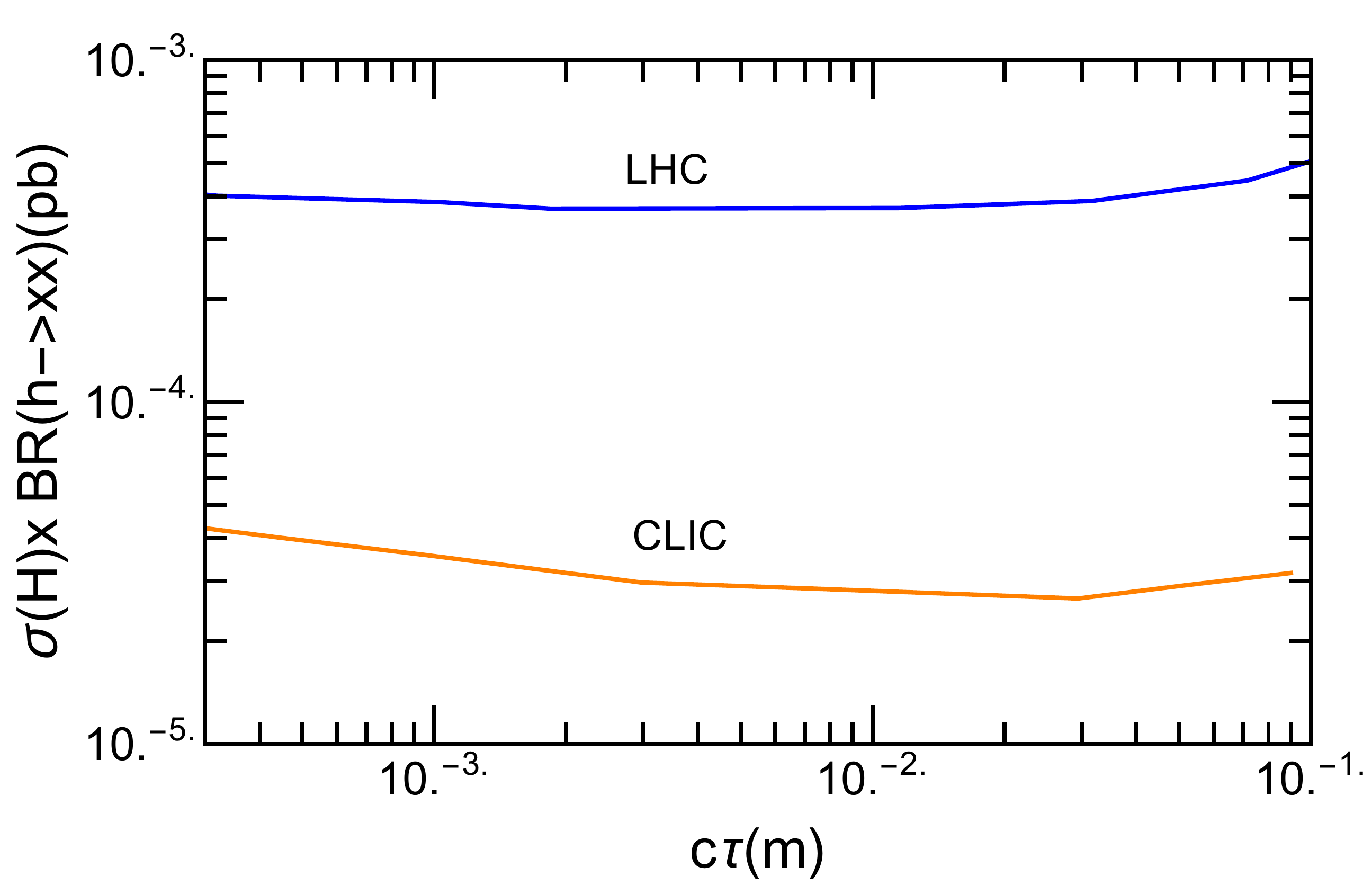}
	\caption{Blue line: HL-LHC projected 95\% C.L. exclusion limit for the Higgs portal singlet model as the function of $c\tau_\chi$ for $m_\chi=30$~GeV. Orange line: projection for CLIC with the same model.}
	\label{fig:hportal}
\end{figure}

\paragraph{Higgs-portal singlet model:}

Whether a secluded sector exists in a ``hidden world'' is an open question. Due to a tiny coupling to Standard Model states, these particles may be secluded to us. Such feeble interactions may be helpful in many contexts to address open issues of the Standard Model, see, e.g., Ref.~\cite{Alexander:2016aln} for a discussion. These searches are very challenging as the properties of the new physics states vary. Consequently, a broad program of searches needs to be put in place to explore this idea effectively. In this context, new physics may manifest itself with light new particles. A TeV lepton collider can make progress on the experimental exploration of this scenario. For example, the clean environment and the absence of triggers allow one to improve significantly over the HL-LHC in the search for Higgs or Higgs-like bosons decay to long-lived particles.

In Fig.~\ref{fig:hportal}, we compare the $95\%$ C.L. reach of HL-LHC with a 3~TeV $\ee$ collider  for this class of models. The LHC sensitivity of various Higgs portal models is studied in~\cite{Craig:2015pha}. These include Twin Higgs models, Folded SUSY models, quirky little Higgs models, etc. Similar LHC sensitivity is obtained for the Higgs portal singlet model embedded in RPV-NMSSM that decays to SM quarks via RPV couplings as shown by the blue line in Fig.~\ref{fig:hportal} for $m_\chi=30$~GeV.

At an $\ee$ collider,  the dominant mode of production is via $WW$ fusion, which has a cross-section an order of magnitude lower than that at the LHC. Since we are dealing with on-shell production of light states, the cleaner environments and vertex reconstruction efficiencies can enable an $\ee$ collider to have better $c\tau_\chi$ coverage than HL-LHC for a given mass of the exotic particle, similar to the heavier case of RPV wino discussed before. This can be observed in Fig.~\ref{fig:hportal}. The CLIC sensitivity to $h\rightarrow \chi\chi$ at 3 TeV for 95$\%$ C.L. is indicated by the orange line, using the sensitivity given for the Hidden Valley models. This is  an order of magnitude better reach in the $c\tau_\chi$ range favored by the WIMP baryogenesis models with light singlets $(<100\,\text{GeV})$.

A TeV lepton collider can also search for relatively heavy Axion-Like Particles motivated by various theoretical
contexts. As a high energy collider, it can probe
ALPs outside the reach of dedicated low-energy
experiments~\cite{Bauer:2018uxu}. We can see that for the photo-phobic ALP~\cite{Craig:2018kne} case, one can improve about one order of magnitude of improvement.

\section{Physics opportunities for a  multi-10 TeV collider} 
\label{sec:multi10TeV}

Beyond 3~TeV in the parton-parton CM energy,  there are important issues that particle physics must explore.   In this section, we will present arguments that call for an advanced collider with parton CM energies of order 10~TeV and discuss some aspects of the experimental program to reach these energies with an electron accelerator.

  There are several proposals for such a collider already on the table, including the FCC-hh~\cite{FCC:2018vvp} and a muon collider~\cite{Schulte:2019bdl}.  The report \cite{ALEGRO:2019alc} put forward preliminary designs for a 30~TeV $\ee$ collider (ALIC), based on technologies now under development for advanced electron acceleration.  This possibility has also been discussed earlier in this document, in Sec.~\ref{sec:wakefield}.   Given the uncertainties in technology and cost and the long time expected to realize such a collider, it is important to continue R\&D along all threee of these lines.  In particular, it would be wonderful if a true highest energy accelerator could also take advantage of the experimental features such as freedom from QCD backgrounds, full event reconstruction, and polarization, that we have stressed for ILC in this report.

We are high-energy physicists, so it is a part of our culture to demand to explore continually higher energies.  But it is important for us to enumerate issues that explicitly require colliders with higher CM energies, beyond simply extending search limits to higher mass or precision.  In fact, there are three issues that now call for colliders with parton CM energies of 10~TeV and above. 

The first motivation for 10 TeV scale experiments is the issue of electroweak symmetry breaking.   Before the start of the LHC, most theorists believed that the particles most directly responsible for creating the Higgs field potential had masses near 1~TeV and would be discovered early in the LHC program. This led to optimism that the entire spectrum of supersymmetry partners of SM particles might be accessible at the LHC.  Of course, this belief proved not to be correct.   New expectations were established, in particular, for models of supersymmetry, based on the idea of ``more minimal'' or ``natural'' supersymmetry~\cite{Cohen:1996vb}, in which only the Higgs boson partners and the top squarks are light.   The paper \cite{Papucci:2011wy} set out expectations for the masses of supersymmetric particles in this framework, but, today, these are also excluded by LHC searches.  Other mechanisms for generating the Higgs boson mass scale discussed in the 2000's have met similar problems.

There have been three responses to this situation.  The first is to dismiss a mechanical explanation for the Higgs boson mass scale, ascribing this to the  anthropic principle or other physics that originates from quantum gravity.   Unfortunately, there is no specific predictive model of this type.    The second is to predict  the Higgs mass scale as a consequence of new physics in the process of cosmic inflation, as,  for example, in the relaxion~\cite{Graham:2015cka} or self-organized localization~\cite{Giudice:2021viw} mechanisms.  

However, the most concrete models that address this question are those that embed the generation of the Higgs potential in a more general structure.   A particular example is given by Little Higgs models~\cite{Arkani-Hamed:2001nha,Arkani-Hamed:2002ikv}  where new strong interactions at a very high mass scale $\Lambda$ drive a spontaneous symmetry breaking that produces a multiplet of Goldstone bosons that includes the Higgs boson.  Coupling of this sector to the SM induces some light particles to acquire a masses $M^2 \sim (\alpha_c/4\pi)\Lambda^2$, but the Higgs multiplet escapes obtaining a mass at this level.   Then, finally, the Higgs multiplet acquires a potential at the level of $M^2 \sim (\alpha_c/4\pi)^2\Lambda^2$.  Putting $\alpha_c ~\sim \alpha_w \sim 1/30$, we need $M\sim$~ few TeV and $\Lambda~\sim$~tens of TeV. 
A similar pattern is found in Randall-Sundrum models with a 5th dimension of space~\cite{Randall:1999ee}. If the lowest-mass Kaluza-Klein extra-dimensional excitations have  masses of a few TeV, these can radiatively generate the Higgs 
potential~\cite{Agashe:2004rs}. However, much higher energies are needed to discover a series of recurrences of Kaluza-Klein that would give evidence for the 5-dimensional geometry.  Within models of supersymmetry, there are also ideas that can extend the 
range of superparticle masses, including Dirac gauginos  and ``supersoft'' supersymmetry breaking~\cite{Kribs:2012gx,Nelson:2015cea,Carpenter:2016lgo,Baer:2017yqq,Chakraborty:2018izc}. 

The second motivation for 10 TeV scale experiments is the flavor problem. In the SM, the quark and lepton masses and mixings are generated from the matrix of Yukawa couplings of the fermions to the Higgs boson. Taking the SM literally, the elements of this matrix are renormalized parameters that need to be specified externally---that is, they have no explanation. In models of  flavor dynamics that extend the SM, new particles are severely constrained if their masses are below 1 TeV.  But for particles with masses in the multi-TeV region, more possibilities open up.   In particular, models of multi-TeV mass leptoquarks can  explain the anomalies in $B$ meson decays suggested by results of the LHCb experiment~\cite{Aebischer:2019mlg}.  Vector leptoquarks have also been proposed to explain the apparent anomaly in the muon $(g-2)$~\cite{Ban:2021tos}. In these models, the lightest leptoquarks are those that involve particular flavor combinations, for example, coupling to $\mu$ and $b$.   But if leptoquarks exist, there should be bosons that couple to all combinations of quark and lepton generations.  The full expanantion of this structure will again take us into the multi-10-TeV regime.

The third motivation for 10 TeV scale experiments comes from the exploration of the space of dark matter candidates.   The most attractive candidates for particle dark matter are those whose cosmic density is generated in the early universe as particles in 
thermal equilibrium that freeze out as the universe cools.  For particles with electroweak couplings, the assumption of thermal freeze-out leads to particle masses near 100~GeV.  Such models are strongly challenged now by the limits on direct detection cross sections.  One way to address this problem is to consider more weakly coupled dark matter candidates such as appear in the dark sector models discussed in Secs.~\ref{sec:dark500} and \ref{sec:beamdump}.  But the opposite direction is also possible. For dark matter particle
sectors with strong interactions, the condition of thermal freeze-out  leads to the prediction of heavier masses.  The mass limit 
from unitarity bounds is in region of 100s of TeV~\cite{Griest:1989wd}.   But this indicates that there is a large range to explore. Given the difficulties discussed earlier  for finding even 1~TeV WIMPs at proton-proton colliders, this exploration cannot be done except at an electroweak collider with energies well above the TeV range.

For all of these motivations, evidence for physics beyond the SM can be found in current experiments.  The search for this evidence will be extended by particle searches at the HL-LHC and, as we have explained in this report, by precision measurements on the Higgs boson and the top quark at the ILC.  To fully explore the consequences of these discoveries, however, we will need accelerators in the 10s of TeV energy range.  Access to the electroweak sector at these energies will be essential, and thus a lepton collider will be the instrument of choice.

For an electron-based collider, such a high energy machine must be a  linear collider.  Then the facilities of a linear collider 
laboratory such as the ILC Laboratory and the experience of operating a linear collider for high luminosity will be essential to 
proceed down this road. We now turn to discussion of technologies that could take the ILC Laboratory to energies well above
those of the ILC program.

\section{Very high gradient superconducting RF} 
\label{sec:futureSRF}

In this section we will consider ILC upgrade paths beyond 1 TeV using Superconducting RF cavities with improved performance.  We will discuss extensions in energy  (1) to 2 TeV and (2) to 3 TeV,  depending on the needs of high energy physics.  

Cost are quoted in this section for the comparison of options.   These are not detailed bottom-up costs as have been presented in Sec.~\ref{sec:costandschedule} for the 250~GeV ILC design but rather are extrapolations based on the costing scheme used there. Absolute cost numbers should be used with caution, especially in comparison to other technologies.  As in Sec.~\ref{sec:costandschedule}, costs  are quoted in ILC currency units,  1~ILCU = \$ 1 US using 2012 prices and are capital costs that do not include manpower and detectors.

\begin{enumerate}
\item From 1 TeV to 2 TeV, the design will be  based one of the paths:
\begin{enumerate}
\item Gradient advances of Nb cavities to 55 MV/m anticipated from on-going SRF R\&D on Nb structures discussed in Sec.~\ref{sec:acc-beyond}.
\item Radically new travelling wave (TW) superconducting structures~\cite{Kanareykin:2005wn,Kostin:2015tws,Shemelin:2021} optimized for effective gradients 
of 70+~MV/m. 
\end{enumerate}
\item   From 1 TeV to 3 TeV based on  one of the paths:
\begin{enumerate}
\item Radically new travelling wave (TW) superconducting structures~\cite{Kanareykin:2005wn,Kostin:2015tws,Shemelin:2021} optimized for effective 
gradients of 70+~MV/m. 
\item 80 MV/m gradient potential for Nb$_3$Sn~\cite{Posen:2019kks} with a $Q$ of $1\times 10^{10}$. Further, the operating temperature is 4.2~K instead of 2~K. 
\end{enumerate}
\end{enumerate}
We will discuss each of these paths in turn and estimate for each the cost increment beyond the 1 TeV ILC and the power requirements.

\subsubsection{Cost and power estimates}

ILC Energy upgrades beyond 1 TeV  (except path 2b) require 300--400~MW AC power for operation.   We can expect further reductions in AC power from on-going developments under the Green-ILC program described in Sec.~\ref{sec:greenilc}. Efforts under this umbrella are preparing to explore multiple paths to make ILC and its upgrades environmentally sustainable. Wind power is one avenue following the example of ESS in Sweden~\cite{ESSAC}.  A 30--40 unit wind turbine farm is capable of providing 100~MW at a cost of 150~MEuro.  Combined heat and power production using bioenergy or solar photovoltaic cells integrated in the buildings are other examples.  New ways of recycling low heat water (below 50$^\circ$C) would also enable agricultural use of recycled heat, such as greenhouse heating.   

The 1 TeV upgrade discussion in the TDR does not apply any learning curve cost reduction to cavity, cryomodules or klystrons.  Between the baseline ILC at 0.25~TeV and the upgrade options to 2~TeV and 3~TeV the total number of cavities increases by a factor of 5 from 8000 to about 40,000, and the total number of klystrons increases by a factor of 5.6 from 250 to 1500.  Accordingly, we apply a 25\% cost reduction for cavities and klystrons for 2.5 doublings, using the 90\% learning curve in the TDR.  We further assume that due to RF power developments, the efficiency of klystrons will improve from 65\% (TDR) to 85\%.  Taking into account modulator and distribution efficiencies of 90\% each, we use 65\% efficiency for newly installed RF systems for 1~TeV, 2~TeV and 3~TeV upgrades but continue to use 50\% efficiency for RF systems installed for the first 0.5~TeV.  We expect further cost reductions from several areas of R\&D already started.  Among the areas under exploration are niobium material cost reduction (25\%) for sheet production directly from ingots (large grains), and/or from seamless cavity manufacturing from tubes with hydroforming or spinning to reduce the number of electron beam welds and weld preparations (15 - 20\%).   Based on the above ideas, we use an overall cost reduction of 50\% in the cost of large productions of SW cavities.  After including these reductions, we expect the cost of TW cavities will be 30\% higher, leading to 15\% increase in the cost of CM for TW structures.

Cost-reducing features for cryomodules~\cite{Peterson:2016teh}  are to connect cryomodules in continuous, long strings similar to cryostats for long strings of superconducting magnets, saving the cost for the expensive ends. The elimination of the external cryogenic transfer line by placing all cryogenic supply and return services in the cryomodule also reduce costs, not only directly for the cryogenic components, but also by reducing tunnel space required.  We estimate that by this method the filling factor from cavities to ``linac tunnel length'' will improve from 0.7 to 0.75.

\subsubsection{Path 1a: 2 TeV Upgrade with 55 MV/m Nb}

Scenario B of the ILC TDR~\cite{Adolphsen:2013jya} assumes a gradient/$Q$ of 45~MV/m/$2\times 10^{10}$
 for the upgrade from 500~GeV to 1~TeV.   This is based on the assumption that improvements in SRF technology discussed in Sec.~\ref{sec:acc-beyond} will already be implemented in the design of this upgrade.  Recall that the gradient/$Q$ for the first 
500~GeV is 31.5~MV/m/$1\times 10^{10}$.  We expect that R\&D in SRF technology will continue in parallel to both construction and operation of the earlier ILC stages to reach 45 MV/m/$2\times 10^{10}$.

For the 2 TeV upgrade Option 1a we consider advances in SRF performance (as discussed in  Sec.~\ref{sec:acc-beyond}) to gradients/$Q$ of 55 MV/m/$2\times 10^{10}$ based on the best new treatments applied to advanced shape structures such as the Re-entrant, Low-Loss, or Low-Surface-Field (LSF) candidates. Therefore, applying the best new treatments to the advanced shapes we can optimistically expect gradients from 56--59~MV/m with successful R\&D.   

The strategy adopted for path 1a is to replace the lowest gradient (31.5~MV/m) 0.5~TeV section of cavities/cryomodules, re-using the tunnel, RF and Refrigeration of this section, keep the 0.5~TeV section 
with 45~MV/m gradient (11,000~cavities), running with the slightly lower bunch charge (Table~\ref{tab:ILCSRFdetailed}), and add 1.5~TeV with 55 MV/m and $Q=2\times 10^{10}$.  With this approach it is possible to keep the total linac length  to 52~km well below the  currently expected 65~km site limit.

Table~\ref{tab:ILCSRFhighlevel} shows high level parameters for the 2~TeV upgrade as compared to 1~TeV in the ILC TDR.  
%The luminosity is $7.9\times 10^{34}$ which is higher than the $3.75\times 10^{34}$ for CLIC~1.5 TeV~\cite{Aicheler:2012bya}.  
Table~\ref{tab:ILCSRFdetailed} gives more detail parameters for beam and accelerator.  The number of particles per bunch is slightly lower than for the 1~TeV case, but the number of bunches and repetition rate are the same.  The peak beam current is therefore slightly lower.  The total beam power for two beams increases from 27~MW to 47~MW. Other beam parameters are adjusted so that the spot size at collision is reduced to 1.6~nm (from 2.7~nm).  

As shown in Table~\ref{tab:ILCSRFdetailed}, the total number of new cavities at 55~MV/m required for 1.5~TeV is 27,000, spanning a linac length of 36~km, of which 22~km can be installed into the empty tunnel (from the removed 0.5~TeV),
 leaving 14~km of new tunnel to be installed.  Adding in the length (16~km) of the 0.5~TeV section remaining with 45~MV/m cavities, the total linac length will be 52~km, below the expected site limit of 65~km. There are savings from cryomodule parts if the tear down and replacement are staged so that some of the removed cryomodules parts are re-used.   From 1600~CM removed from the 0.5~TeV section,  we estimate the parts savings to be in the range of 0.5~B provided the removal and production of CMs are properly staged.  For the new 1.5~TeV section, the cavity loaded $Q$ is $6.7\times 10^6$, the input power per cavity will be 365~kW, with RF pulse length 2.0~ms, similar to the RF pulse length for 1~TeV.  The total number of klystrons required is 1150 of which 360 klystrons are re-used from the 0.5~TeV removed section, and 65 klystrons are available from the 0.5~TeV remaining section (which operates with the new, lower bunch charge), leaving 725 new klystrons to be added.  We use 65\% efficiency for RF systems installed for 1~TeV and above, and 50\% efficiency for the RF system installed for the first 0.5~TeV, to give an average efficiency of 60\%.  The total 2~K refrigeration required will be 66~kW, of which 33~kW is re-used, leaving 33~kW new refrigeration to be installed.  We assume a cryoload safety factor and RF power overhead of 20\% each for the new installations.  The damping ring and positron source will be same as for 1~TeV, due to the same number of bunches, but the beam dump cost will increase.  Summing all the cost components outlined, the additional cost for the 2~TeV upgrade will be 6.0 B.  The AC power to operate 2~TeV will be 345~MW, making ILC with SRF attractive for 2~TeV. 

\subsubsection{Path 1b:  2 TeV Upgrade from 1 TeV with 70 MV/m TW Nb structures}

This is the more attractive option because of the lower AC power and lower capital cost   (see Table~\ref{tab:ILCSRFhighlevel}), but it depends strongly on the success of the development of the Travelling Wave Nb cavities with 70 MV/m. 

As discussed in Sec.~\ref{sec:acc-beyond}, TW structures offer several advantages compared to
 standing wave (SW) structures: substantially lower peak magnetic ($H_{pk}/E_{acc}$) and lower peak electric field ($E_{pk}/E_{acc}$) ratios, together with substantially higher $R$/Q (for lower cryogenic losses and lower AC power).  We expect that the cost of TW SRF cavities will be 30\% higher, leading to 15\% increase in the cost of CM for TW structures.  

The first strategy adopted in this option is again to remove the lowest gradient (31.5~MV/m) 0.5~TeV section, re-use the tunnel, RF and Refrigeration of this section, keep the 0.5~TeV section (11,000 cavities) with 45~MV/m gradient (running with the slightly lower bunch charge for 2~TeV), and add 1.5~TeV with TW SRF cavities at 70~MV/m/Q= $2\times 10^{10}$ and $R$/Q  2 times higher than SW Nb cavities.  With this approach it is possible to keep the total linac length to 44~km, well below the currently expected 65~km site limit.  

\begin{table}
\begin{center}
\begin{tabular}{lccccccc}
   &          &     ILC  &    ILC    &  ILC 2 TeV   &  ILC   & ILC   &  CLIC  \\
   &          &      1 TeV  &    2 TeV   &  2 TeV   &  3 TeV   & 3 TeV   &  3 TeV \\
  &  units &        TDR      &       path 1a    &   path 1b    &   path 2a   &   path 2b     &     \cite{Aicheler:2012bya} \\ 
   \hline
Energy &                TeV      &       1    &    2  &     2  &    3 &     3   &   3  \\
Luminosity &        $10^{34} $ &    4.9     &    7.9  &     7.9   &    6.1 &     6.1     &   5.9  \\
AC Power &            MW    &      $<$ 300  &   345  &    315  &   400 &    525    &   590  \\
Cap. Cost &      B ILCU  &      + 5.5  &  +11.5   &  +10.4   & +17.3   &  +16.5  &   \\
%\ \ (total)  &                 &    13.3    &    19.3   &    18.2   &   25.1  &   24.3    &       24.2 BCHF \\ 
Gradient&  MV/m  &  45   &   55  &    70   &   70 &    80     &   72 / 100\\
\ \ (new linac) &  \\
Q new linac &     $ 10^{10}$  &  2     &    2  &    3  &    3 &     2 (4.2 K)     &   \\
CM unit cost &    M ILCU  &  1.85    &   1.15  &   1.32   &    1.32 &    1.15   &     \\
\end{tabular}
\caption{High level parameters for ILC energy upgrades. Costs are quoted as estimated additions to the costs from the ILC TDR.   These are given in ILC currents units (as in Sec.~\ref{sec:costandschedule}) and are capital costs not including manpower and detectors.  Incremental costs are given relative to the ILC 250 GeV plus 500 GeV upgrade (7.8 B ILCU). } 
\label{tab:ILCSRFhighlevel}
\end{center}
\end{table}

\begin{sidewaystable*}[p] %thanks to Alex Aryshev 
\centering 
\begin{tabular}{lccccccc}
   &          &     ILC 1 TeV  &    ILC 2 TeV   &  ILC 2 TeV   &  ILC 3 TeV   & ILC 3 TeV   &  CLIC 3 TeV \\
   &  units &        TDR      &       path 1a    &   path 1b    &   path 2a   &   path 2b     &     \cite{Aicheler:2012bya} \\ 
   \hline
Energy &                TeV      &       1    &    2  &     2  &    3 &     3   &   3  \\
particles/bunch &          $10^{10}$    &     1.74   &    1.5 &  1.5 &  0.65  &   0.65   &  0.37  \\
bunches/train  &                &      2450   &   2450 &    2450&  4900 & 4900  &   312  \\
bunch spacing  &       nsec    &      366   &   366  &    366&   250  &     150   &   0.5  \\
pulse current  &               mA    &      7.6    &  6.6  &     6.6 &   4.16&     4.16   &   3  \\
rep. rate  &       Hz     &     4    &   4  &    4  &   4 &     4   &   50 \\
RF pulse length &         ms     &       1.94    &   2.0 &     1.76  &    2.6 &    2.6   &   0.00024  \\
  (added linac) &          \\  
Beam power  &        MW    &   27.2    &  47  &  47 &   61  &    61   &  28  \\
\ \ (2 beams)  & \\ 
$\epsilon_x/\epsilon_y$&           $10^{-8}$m  &     500/3    &   500/2  &    500/2 &   500/2 &   500/2   &   66/2 \\
$\beta_x/\beta_y$ &           $10^{-3}$m  &     22/0.23   &   22/0.23  &   22/0.23 & 16/0.15 &   16/0.15  &   \\
$\sigma_x/\sigma_y$  &           $10^{-9}$ m   &   335/2.7   &  237/1.6 &   237/1.6&  165/1.0 &  165/1.0   &   40/1 \\
$\sigma_z$ &          $10^{-3}$m      &      0.225    &   0.225  &    0.225 &  0.1 &   0.1  &  0.044 \\
$\Psi$ (beamstr.  &                &    0.21   &   0.5  &  0.5  &  1/045 &   1.045  &   5  \\
 \ \ parameter)   &   \\ 
$\delta$ (RMS  &   \%   &     10.5  &   20  &  20 &  16 &   16 &   35  \\
 \ \ energy spread)  & \\ 
\hline
Luminosity &        $10^{34} $ &    4.9     &    7.9  &     7.9   &    6.1 &     6.1     &   5.9  \\
photons/electron &           &    1.95  & 2.1  &   2.1  &   1.2 &   1.2   &   2.2 \\
coherent pairs  &                 &   0    &   $2\times 10^4$  &     $2\times 10^4$   &    $7.9\times 10^5$  &     $7.9\times 10^5$ &       $6.8 \times 10^8$\\ 
incoh. pairs  &    & 383   &  49 &   49  &  5&    5    & $3\times 10^5$\\
No. of klystrons &     $ 10^{3}$  &  460 + 320     &  820 + 460 &   755 +  425 &   690 + 820  &   1680 + 820    &   \\
(new + existing) &                   &    = 820        &     = 1280        &    = 1180     &    = 1500   &      = 2500            &      \\
No. of cavities &     $ 10^{3}$  &  11 + 16    &  27 + 11 &   21 + 11 &   43 + 0  &   37.5 + 0    &  160 (0.25 m) \\
(new + existing) &                   &    = 27        &     = 38        &    = 32     &    = 43   &      = 37.5 &      \\
$Q_L$ (new cavities)  &   $10^6$   &  5.6    &   8 &   5   &    8 &    10   &     \\
input power   &   kW  &  350   &   365 &   460  &   300  &   550  &     \\
\ \ (new cavities) & \\ 
linac length  &   km   &  16 + 22  &  14 + 38  &   6 + 38   &   19 + 38 &  12 + 38   &     \\
\ \ (new + existing) &                   &    = 38       &     = 52       &    = 44     &    = 57   &      = 50           & 42 \\ 
\end{tabular}
\caption{Detailed parameters for the proposed ILC energy upgrades compared with the CLIC 3~TeV design~\cite{Aicheler:2012bya}.}
\label{tab:ILCSRFdetailed}
\end{sidewaystable*}

As shown in Table~\ref{tab:ILCSRFdetailed}, the total number of  new TW cavities at 70~MV/m required is 21,000,  spanning a linac length of 28~km, of which 22~km can be installed into the empty tunnel (from the removed 0.5~TeV), requiring 
6~km of new tunnel to be installed.  For 1600 CMs removed from the 0.5~TeV section, we estimate the savings in re-used parts to be in the range of 0.5B, provided the removal and production of CMs are properly staged.   For the new 1.5 TeV section, the cavity loaded $Q$ is $5\times 10^6$, the input power per cavity will be 460~kW, with RF pulse length 1.76~ms.    The total number of klystrons required is 1180, of which 360 klystrons are re-used from the 0.5~TeV removed section, and 65~klystrons are available from the 0.5~TeV remaining section (because it operates with the lower bunch charge than for 1~TeV), leaving 755 new klystrons to be added.  The average RF power efficiency of new RF systems will be 65\% and the existing RF systems from the first 0.5 TeV installation will be 0.5, giving an overall RF efficiency of  61\%.  The total 2~K refrigeration required will be 37~kW, of which 33~kW is re-used, leaving 4~kW new refrigeration to be installed.  We assume a cryoload safety factor and RF power overhead of 20\% each for the new installations.  The damping ring and positron source will be same as for 1~TeV, due to the same number of bunches, but the beam dump cost will increase.  Summing all the cost components outlined, the additional cost for the 2~TeV upgrade will be 4.9~B.  The AC power to operate 2~TeV will be 315~MW, making this path attractive for the improved environmental impact.  Note the substantial benefit to the AC power due to the 2 times  higher $R/Q$ of the TW cavities. 
If we follow the alternative path of removing the entire 1 TeV linac, keeping the RF, tunnel and Refrigerator, to install a brand new linac using 70~MV/m TW cavities, we will need to populate the existing 38~km of tunnel with 28,000 TW cavities (no new tunnel needed), and use the existing Refrigeration (no new refrigeration needed), adding 755 klystrons.  Savings from re-using CM parts from $>$ 3000 CM from the 1~TeV section is estimated to be 1~B.  The additional capital cost for this path will be 5.2~B, comparable to the path above, and the AC power will be 240~MW, less than the path above.  The shorter tunnel and lower AC power may dominate the choice of this path. 

\subsubsection{Path 2a:  3~TeV Upgrade from 1 TeV with 70 MV/m TW Nb structures} 

The beam bunch charge for the 3~TeV upgrade is chosen to be 3 times lower than the bunch charge for 0.5~TeV stage to obtain a luminosity comparable to CLIC 3TeV~\cite{Aicheler:2012bya}.  The lower bunch charge helps with wakefields and with IP backgrounds.  The number of bunches per RF pulse is doubled to 4900, and the bunch spacing is lowered due to the lower bunch charge (see Table~\ref{tab:ILCSRFdetailed}). 

The option adopted here is to remove ALL  of the installed cryomodules for 1~TeV and replace them with new 70~MV/m TW cavities/cryomodules, plus add new linac sections to reach 3~TeV energy.  We would re-use the existing RF and Refrigeration and CM parts from the removed 1~TeV section.    As shown inTable~\ref{tab:ILCSRFdetailed}, a total of 43,000 TW cavities will be required, so that with the (cavity to linac tunnel) filling factor of 0.75, the total length of the 3 TeV linac will be 57~km, under the expected site limit of 65~km.  38~km of tunnel would already be present from the 1~TeV removed, requiring 19~km of new linac tunnel. The total number of klystrons required will be 1500, of which 820 are available from the 1~TeV installation.   The RF system cost will be higher due to the longer RF pulse length.  Also, the existing 820 klystrons and RF system will have to be upgraded to provide longer RF pulses, which will incur a cost of about 0.4~B.  The efficiency of the first RF system installed with 360 klystrons for 0.5~TeV is 50\%,  and for the later installed RF system for the next 0.5~TeV with 460 klystrons it is 65\%.  Hence the average RF system efficiency used is 61\%.    The input power per cavity will be 300~kW due to the high gradient. The loaded $Q$ will be $8\times 10^6$.   The total 2~K refrigeration requirement will be 95~kW of which 51~kW is already present, leaving a balance of 44~kW to be installed.  Add in the cost of needed damping rings, positron source and beam dump for increasing the number of bunches from 2450 to 4900.   The total additional capital cost for 3~TeV (from 1~TeV) will be 11.8~B, shown in Table~\ref{tab:ILCSRFhighlevel}  The total AC power to run 3~TeV will be 400~MW, with substantial benefit from the 100\%  higher $R/Q$  of TW structures.

Table~\ref{tab:ILCSRFdetailed} gives detailed parameters (for beam and accelerator) for ILC 3~TeV (Option 2a) with 70~MV/m TW structures as compared to CLIC 3~TeV.  Note that the backgrounds at the IP for the ILC 3~TeV are much lower than for CLIC, and final beamstrahlung energy spread is 16\% compared to 35\% for CLIC.  To reach the desired luminosity, the beam power is 61~MW with twice the number of bunches (4900) spaced closer together in the linac (250ns instead of 366 for 1~TeV) as allowed by the lower bunch charge.  The peak beam current is 4.16~mA.  The final vertical spot size is 1~nm, comparable to the CLIC case. %Figure~\ref{fig:ILCSRFupgrade} shows the rough breakdowns for the costs of the various systems.

\subsubsection{Path 2b: 3 TeV Upgrade with 80 MV/m Nb$_3$Sn structures at 4.2 K}

Option 2b for 3~TeV is to consider 80~MV/m Standing Wave Nb$_3$Sn TESLA-like structures at 4.2~K with $Q$ values of $1\times 10^{10}$.  In this case the challenge is to develop high performance Nb$_3$Sn.  Due to the combined improvement of Carnot and technical efficiency at 4.2~K over 2~K, the comparison of  AC power/cryo power improves from a ratio of 730 to 230.  We assume that the capital cost of 4.2~K refrigeration will be a factor 3 lower than for 2~K, and that the refrigerator units installed for 1~TeV are designed so that 1 watt of cooling at 2~K would be later equivalent to 3 watts of cooling at 4.2~K when the conversion is made for the 3~TeV upgrade at 4.2~K. 

Our plan would be to install Nb$_3$Sn cavities for 3~TeV,    removing  all of  the cryomodules for 1~TeV and replacing them with new 80~MV/m/$Q = 1\times 10^{10}$  cavities/cryomodules, plus install new linac sections to reach 3~TeV energy.  We will re-use the RF, Refrigeration and CM parts of the removed 1~TeV section, converting the 2~K refrigeration to remove heat load at 4.2~K.  A total of 37,500 Nb$_3$Sn cavities will be required, so that with the filling factor (cavity to tunnel length) of 0.75, and  the total length of the 3~TeV linac will be 50~km, well under the expected Japan site constraint of 65~km.  38~km of tunnel has already been installed for 1~TeV, so that 12~km of new linac will be required.  The total number of klystrons required will be 2500, of which 820 are available from the removed 1~TeV installation.  The existing klystrons and RF system will have to be upgraded to provide longer RF pulses (2.6~ms), which will incur a cost of about 0.4~B.   The number of new klystrons required is 1680.  The average efficiency of old and new RF systems will be 63\%.  The input power per cavity will be 550~kW, at a loaded $Q$ of $1\times 10^7$, so couplers will need to be improved.   The total 4.2~K refrigeration required will be 352~kW of which 51~kW (at 2~K) is already present for 1~TeV, equivalent to 150~kW at 4.2~K.   The balance of 200~kW at 4.2~K needs to be installed.  Add in the cost of needed damping rings, positron source and beam dump for increasing the number of bunches from 2450 to 4900.   The total additional capital cost for 3~TeV will be 11.0~B, as shown in Table~\ref{tab:ILCSRFhighlevel}.  The total AC power to run 3~TeV will 525~MW.

\section{Very high gradient normal conducting accelerators} 
\label{sec:futurecopper}

The infrastructure that will be put in place for the ILC at either the 250 or the 500~GeV center of mass energies provides a unique resource for future experiments after the currently proposed ILC program has been completed. This statement of course applies to superconducting accelerators, but is also true for normal conducting accelerators that could follow the ILC.  The ILC civil construction represents a significant investment with tunnels, electrical power distribution, cryogenic systems, \etc, that could be re-utilized with modifications for future experiments. The particle sources could also potentially be re-utilized depending on the details of the electron and positron bunch structures required.  An advanced accelerator technology could also  largely re-utilize parts of the accelerator infrastructure, including the damping rings and elements of the beam 
delivery system. Given the time required for construction and data collection for the ILC, novel technologies presently under investigation that could be developed into collider concepts will be able to make significant technical progress.

Having discussed in some detail the options for advanced superconducting RF acceleration, we now turn to options based on normal conducting acceleration.  In this section, we will discuss possible upgrades of the ILC infrastructure based on two-beam acceleration
as studied for CLIC and described in~\cite{CLICdp:2018cto, Aicheler:2018arh}, and also 
on conventional RF distribution in novel structures based either on copper cavities or more advanced materials. An example of the
latter  approach using copper is the Cool Copper Collider (C$^3$), for which the concept and the proposed R\&D program have been presented in~\cite{Bai:2021rdg}.  We will explain both of these concepts as potential examples for reusing the ILC infrastructure, before presenting ideas about more advanced possibilities with even higher gradients. 

\subsubsection{ILC Energy Upgrade with CLIC  Technology}

The Compact Linear Collider (CLIC) is a multi-TeV high-luminosity linear $\ee$ collider under development by the CLIC accelerator collaboration~\cite{clic-study}. As a standalone proposal the CLIC accelerator has been optimised for three energy stages at centre-of-mass energies \SI{380}{\GeV}, \SI{1.5}{\TeV} and \SI{3}{\TeV}~\cite{CLIC:2016zwp}.  A future re-use of ILC infrastructure could move directly to a multi-TeV stage.  Detailed studies of the CLIC accelerator, detector studies and physics potential are documented in detail at~\cite{CLICsite}. 

\textbf{CLIC layout}

\begin{figure}[!h]
\centering
\includegraphics[width=\columnwidth]{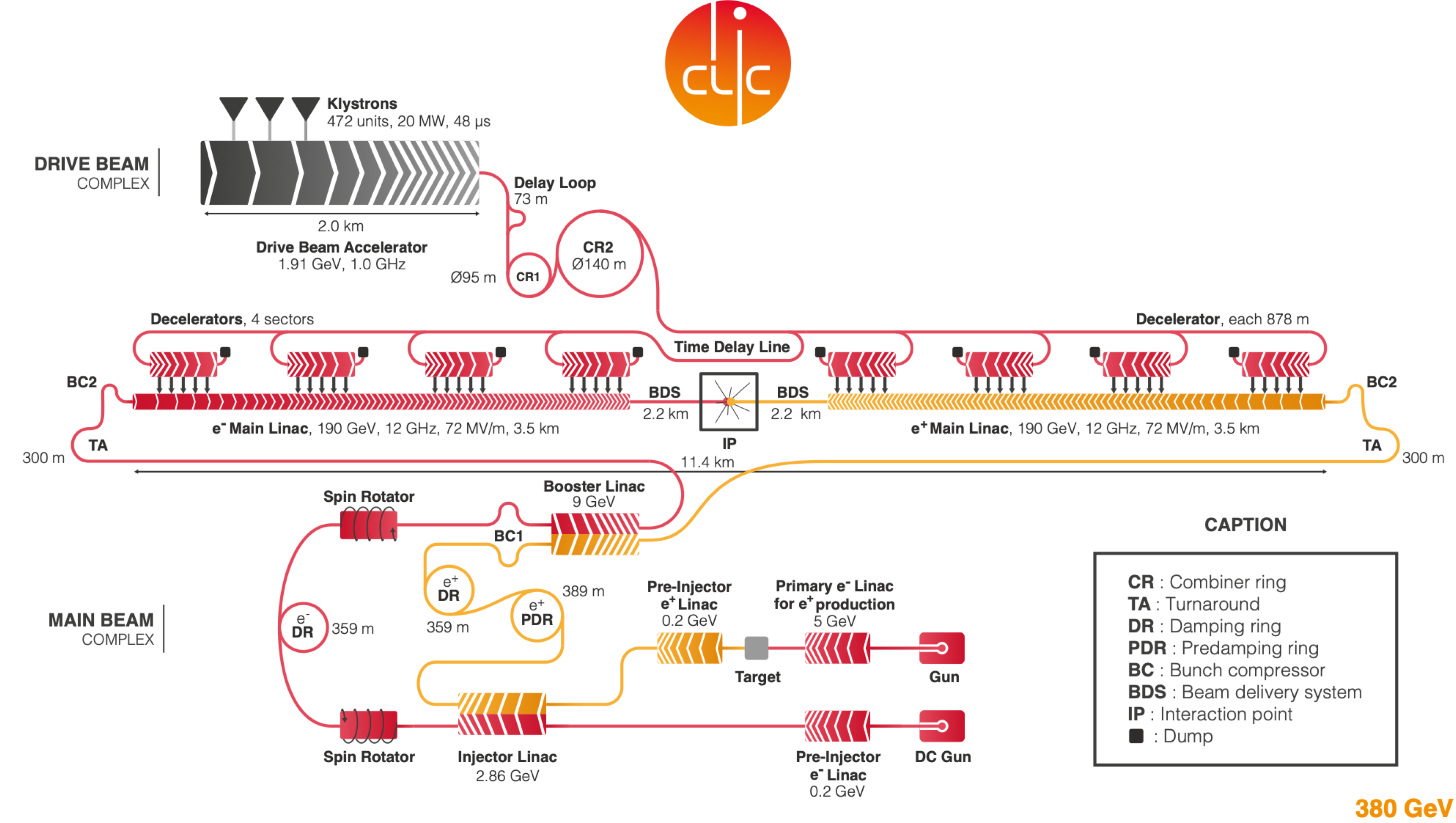}
\caption{Schematic layout of the CLIC complex at \SI{380}{\GeV}.}
\label{scd:clic_layout}
\end{figure}

A schematic overview of the accelerator configuration for the initially proposed \SI{380}{\GeV}, energy stage is shown in Figure~\ref{scd:clic_layout}. To reach multi-TeV collision energies in an acceptable site length and at affordable cost, the main linacs use normal conducting X-band accelerating structures.
These achieve a high accelerating gradient of \SI{100}{\mega\volt/\meter}.
For the first energy stage, a lower gradient of \SI{72}{\mega\volt/\meter} is the optimum to achieve the luminosity goal, which requires a larger beam current than at higher energies.

In order to provide the necessary high peak power, the novel drive-beam scheme uses low-frequency klystrons
to efficiently generate long RF pulses and to store their energy
in a long, high-current drive-beam pulse.
This beam pulse is used to generate many short, even higher intensity pulses that are distributed alongside the main linac,
where they release the stored energy in power extraction and transfer structures (PETS) 
in the form of short RF power pulses, transferred via waveguides into the accelerating structures. 
This concept strongly reduces the cost and power consumption compared with powering the structures directly by klystrons.

The upgrade to higher energies is done by lengthening the main linacs. While the upgrade to \SI{1.5}{\TeV} can be done by increasing the energy and pulse length of the primary drive-beam, a second drive-beam complex must be added for the upgrade to \SI{3}{\TeV}. 

\textbf{Parameter overview}

The parameters for the three energy stages of CLIC are given in Table~\ref{t:scdup1}.
The baseline plan for operating CLIC results in an integrated
luminosity per year equivalent to operating at full luminosity for \SI{1.2e7}{\second}~\cite{Bordry:2018gri}. 
With 8, 7 and 8 years of running at 380, 1500 and 3000 GeV respectively, and a luminosity ramp up for the first years at each stage, integrated luminosities of 1.0, 2.5 and 5.0 ab$^{-1}$ are reached for the three stages. 

CLIC provides $\pm 80$\% longitudinal electron polarization and proposes a sharing between the two polarization states at each energy stage for optimal physics reach~\cite{Roloff:2018dqu}. 

\textbf{Luminosity margins and performance}

In order to achieve high luminosity, CLIC requires very small beam sizes at the collision point, as listed in Table~\ref{t:scdup1}. Recent studies have explored the margins and possibilities for increasing the luminosity, operation at the $Z$-pole and gamma-gamma collisions~\cite{Latina:2687090}.

The vertical emittance and consequently the luminosity are to a large extent determined by
imperfections in the accelerator complex. Significant margin has been added to the known effects to enhance the robustness of the design; without imperfections 
a factor three higher luminosity would be reached at $\SI{380}{\GeV}$~\cite{Gohil:2020tzn}. At this energy also the repetition rate of the facility, and consequently luminosity, could be doubled from 50\,Hz to 100\,Hz without major changes and with relatively little increase in the overall power consumption and cost (at the $\sim 30\%$ and $\sim 5\%$ levels, respectively). This is because a large fraction of the power is used by systems in which the consumption is independent of the repetition rate.

\begin{table}[t!]
\small
\small \caption{Key parameters of the CLIC energy stages.}
\label{t:scdup1}
\centering
\footnotesize{
\begin{tabular}{l l l l l}
\toprule
Parameter                  &  Unit         &    Stage 1 &   Stage 2 &   Stage 3 \\
\midrule
Centre-of-mass energy               & \si{\GeV}                                     & 380     & 1500          & 3000\\
Repetition frequency                & \si{\Hz}                                     & 50      & 50            & 50\\
Nb. of bunches per train         &                                               & 352     & 312           & 312\\
Bunch separation                    & \si{\ns}                                      & 0.5     & 0.5           & 0.5\\
Pulse length                        & \si{\ns}                                      & 244     &244            & 244\\
\midrule
Accelerating gradient               & \si{\mega\volt/\meter}                        & 72      & 72/100        & 72/100\\
\midrule
Total luminosity                    & \SI{e34} & 1.5     & 3.7           & 5.9 \\
Lum. above \SI{99}{\percent} of $\sqrt{s}$ & \SI{e34} & 0.9     & 1.4           & 2\\
Total int. lum. per year            & fb$^{-1}$                                  & 180     & 444           & 708 \\ 
\midrule
Main linac tunnel length                  &  \si{\km}                                      & 11.4    & 29.0          & 50.1\\
Nb. of particles per bunch       & \num{e9}                                      & 5.2     & 3.7           & 3.7\\
Bunch length                        & \si{\um}                                      & 70      & 44            & 44\\
IP beam size                        & \si{\nm}                                      & 149/2.9 & $\sim$60/1.5 & $\sim$40/1\\
Norm. emitt. (end linac) & \si{\nm}                                      & 900/20  & 660/20        & 660/20\\
Final RMS energy spread & \si{\percent} & 0.35 & 0.35 & 0.35 \\
\midrule
Crossing angle (at IP)              & mrad                                    & 16.5    & 20            & 20 \\
\bottomrule
\end{tabular}}
\end{table}

\textbf{Technical maturity}

Accelerating gradients of up to \SI{145}{MV/m} have been reached with the two-beam concept at the CLIC Test Facility (CTF3).
Breakdown rates of the accelerating structures well below the limit of $3 \times 10^{-7}$/m  per beam pulse 
are being stably achieved at X-band test platforms.

Substantial progress has been made towards realising the nanometer-sized beams required by CLIC for high luminosities: 
the low emittances needed for the CLIC damping rings are achieved by modern synchrotron light sources;
special alignment procedures for the main linac are now available;
and sub-nanometer stabilisation of the final focus quadrupoles has been demonstrated.
The advanced beam-based alignment of the CLIC main linac has successfully been tested at FACET and FERMI~\cite{FACET, FERMI}.

Other technology developments include the main linac modules and their auxiliary sub-systems
such as vacuum, stable supports, and instrumentation. Beam instrumentation, including sub-micron level resolution beam-position monitors
with time accuracy better than \SI{20}{ns} and bunch-length monitors with resolution better than \SI{20}{fs},
have been developed and tested with beam in CTF3.  

Recent developments, among others, of high efficiency klystrons have resulted in an improved energy efficiency for the \SI{380}{\GeV} stage, as well as a lower estimated cost.

\textbf{Schedule, cost estimate, and power consumption}

\begin{figure}[h!]
\centering
\includegraphics[width=\columnwidth]{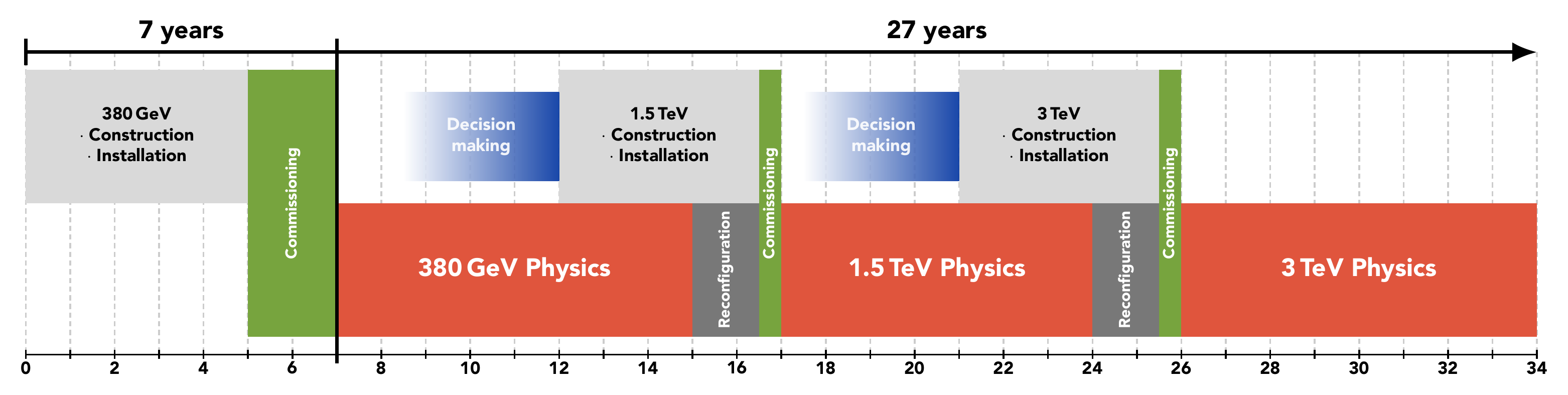}
\caption{\label{fig_IMP_9} \small Technology and construction-driven CLIC schedule.
The time needed for reconfiguration (connection, hardware commissioning) between the stages is also indicated.}
\end{figure}

The technology and construction-driven timeline for the CLIC programme as a standalone project is shown in Figure~\ref{fig_IMP_9}~\cite{Aicheler:2018arh}.
This schedule has seven years of initial construction and commissioning.  
The 27 years of CLIC data-taking include two intervals of two years between the stages.

%\vspace*{.2cm}

The cost estimate of the initial stage as a standalone project is approximately \num{5.9}~billion~\si{CHF}.  The energy upgrade
to \SI{1.5}{\TeV} has an estimated cost of approximately \num{5.1}~billion~\si{CHF}, including the upgrade of the
drive-beam RF power. The cost of the further energy upgrade to \SI{3}{\TeV}
has been estimated at approximately \num{7.3}~billion~\si{CHF}, including the construction of a second
drive-beam complex.

The nominal power consumption at the \SI{380}{\GeV} stage is approximately \SI{170}{MW}. Earlier estimates for the
\SI{1.5}{\TeV} and \SI{3}{\TeV} stages yield
approximately \SI{370}{MW} and \SI{590}{MW}, respectively~\cite{Aicheler:2012bya}, however recent power savings applied to the \SI{380}{\GeV} design have not yet been implemented for these higher energy stages.
The annual energy consumption for nominal running at the initial energy stage
is estimated to be \SI{0.8}{TWh}.  For comparison, CERN's current energy consumption is
approximately \SI{1.2}{TWh} per year, of which the accelerator complex uses approximately 90\%.

\textbf{Future Programme}

The design and implementation studies for the CLIC e$^+$e$^-$ multi-TeV linear collider are at an advanced stage. The main technical issues, cost and project timelines have been developed, demonstrated and documented.

During the coming years the focus will remain on core technology development and dissemination, which will capitalise on existing facilities (X-band test stands and the CLEAR beam facility at CERN), as well as optimising X-band components and RF-systems, involving extensive collaborations with laboratories and universities using the technology. 

The use of the CLIC technology -- primarily X-band RF, associated components and nano-beams -- in compact medical, industrial and research linacs has become an increasingly important development and test ground for CLIC, and is destined to grow further~\cite{DAuria:2019sjj}. The adoption of CLIC technology for these
applications is now providing a significant boost to CLIC,
especially through an enlarging commercial supplier base. 

On the design side the parameters for running at multi-\si{\TeV} energies, with X-band or other RF technologies, will be studied further, in particular with energy efficiency guiding the designs.

\subsubsection{ILC Energy Upgrade with C$^3$ Technology}

C$^3$ is a concept that is 
aimed at developing normal conducting RF (NCRF) accelerator technology to operates at high gradient with high RF-to-beam 
efficiency~\cite{Bai:2021rdg,bane2018advanced}. C$^3$ accelerators are bringing recent advances in the understanding of high-gradient operation \cite{grudiev2009new,simakov2018advances, othman2020experimental}, cavity design and RF power distribution \cite{tantawi2020design}, RF pulse compression \cite{wang2017development}, and cryogenic operation \cite{nasr2021experimental} to improve the performance of normal-conducting RF (NCRF) accelerators for high-gradient, high-brightness and high-luminosity applications. The two principal innovations for the C$^3$ concept are: (1) the use of highly-optimized reentrant cells with distributed coupling to power the linac without cell-to-cell RF coupling, and (2) the operation of the  copper accelerating structure at liquid nitrogen temperatures (77~K) to increase the RF efficiency of the structure by a factor of three, while also increasing the strength of the material.  This has been found to correlate with the achievable operating gradient. The  C$^3$ approach has the potential of operating at extremely high gradients. Prototype structures have been operated with beam up to 160~MeV/m \cite{bane2018advanced,nasr2021experimental} and single cell test cavities have exceeded 200~MeV/m gradients \cite{cahill2018high,Simakov:2021lcw}. The C$^3$ cryomodule that is under development for the main linac and the RF sources that accompany it fits comfortably within the existing diameter tunnel that is planned for the ILC main linac tunnel. The nominal operating parameters for C$^3$ technology are 120 MeV/m gradient with $\sim$90\% fill factor for the cryomodule that forms the basis of the main linac design. A higher gradient version at 155~MeV/m is also being explored. The proponents of C$^3$ are developing an R\&D plan to realize a fully engineered cryomodule and test it with full beam loading over the next decade \cite{Nanni:2022oha}. During the timescale of this demonstration R\&D plan, we will have the opportunity to push the accelerating gradient of the C$^3$ linac well beyond 120 MeV/m, measure the break down rate in realistic operating conditions and make a determination of the physical limit of the installed cryomodules capacity for higher gradient.

For the re-utilization of ILC infrastructure with a C$^3$ energy upgrade we will consider the case of a 30 (20) km tunnel constructed for ILC including the main linac and beam delivery system (BDS). For $\ee$ collisions, a center of mass energy upgrade to  3~(2)~TeV is presently considered the limit for a practical beam delivery system that accounts for beamsstrahlung effects. This BDS would need to be on the order of 2~km per side leaving 13 (7)~km of tunnel to reach the colliding beam energy of 1.5~(1)~TeV. This main linac length would be achievable with the nominal C$^3$ operating gradient of 120~(155) MeV/m  and a fill factor of $\sim$90\%. A summary of the parameters is shown in Table \ref{tab:TeVmainlinacparam}.

\begin{table}
\begin{center}
\begin{tabular}{|c | c | c | c |} 
 \hline
  Parameter  & Unit & Value  & Value \\
 \hline\hline
Center of Mass Energy   & GeV & 3000 & 2000 \\
 \hline\hline
Site Length & km & 30 & 20 \\
Main Linac Length (per side) & km & 13 & 7\\
Accel. Grad. & MeV/m & 120 & 155\\
RF Compression for Pre-Pulse &   & 3X & 3X \\
Flat-Top Pulse Length  & ns & 260 &  195 \\
Cryogenic Load at 77 K & MW & 36.6 &  24.7 \\
Est. AC Power for RF Sources  & MW  & 322 &  215 \\
Est. Electrical Power for Cryogenic Cooling  & MW  & 255 & 170 \\
RF Source efficiency (AC line to linac) & $\%$  & 80 & 80\\
Luminosity  & x$10^{34}$ cm$^{-2}$s$^{-1}$  & 6 & 4 \\ 
Single Beam Power  & MW & 13.5 & 9 \\ 
Injection Energy Main Linac & GeV  & 10 & 10  \\
Train Rep. Rate & Hz & 120 & 120  \\
Bunch Charge & nC & 1 & 1 \\
Flat-Top RF Pulse Length & ns & 260 &195 \\
Bunch Spacing & Periods (ns)  & 20 (3.5) &  15 (2.6) \\
Average Current  &$\mu$A  & 9 & 9 \\
Peak Current & A & 0.3 & 0.385 \\
RF Power for Structure Flat-Top & MW/m &  80 & 140\\
\hline
 \end{tabular}
\end{center}
\caption{Main Linac parameters for C$^3$ at 3~TeV and 2~TeV center of mass energy.}
 \label{tab:TeVmainlinacparam}
\end{table}

Beyond the main linac tunnel, the DC electron gun, polarized positron source and damping ring tunnels could also be re-utilized. The helium cryoplants for the ILC could also be repurposed in part if they use a liquid nitrogen pre-cooler stage for the cryoplant. For  C$^3$, only the nitrogen pre-cooler would need to be operated. It is likely that additional cryoplants would need to be installed for the required cooling capacity of the main linac.

\subsubsection{C$^3$ R\&D for Multi-TeV Operation}

Given the high center of mass energy for a C$^3$ energy upgrade after ILC operation, electrical power consumption will be a key area of concern. This will require RF pulse compression technology and RF source technology to also undergo R\&D in order to improve the electrical efficiency to a practical level at multi-TeV center of mass. These topics are considered for parallel R\&D in the C$^3$ development plan \cite{Nanni:2022oha} as they are essential for multi-TeV operation but not for its utilization as a Higgs factory \cite{Bai:2021rdg}.

Significant electrical power savings are possible with the inclusion of a pulse compressor. This is due to the reduced power loss in the cavity during the long fill time of the cavity, which reduces both the klystron electrical power requirements, as well as the cryogenic cooling requirements.  For C$^3$, the extremely high Q-factor of the cryogenic cavities results in a fill time that requires a significant amount of average RF power to be delivered to the accelerating structure when there is no electron beam. It is possible however to reduce the average power loss during the filling of the cavities with appropriate tailoring of the RF pulse to have a very high power pre-pulse. Recent advances in pulse compressors \cite{wang2017development} with super-compact spherical cavities has dramatically changed the performance potential for these systems. It has been shown that a chain of spherical cavities can produce tailored pulse formats with extremely high conversion efficiency if the power gain is kept low~\cite{wang2017rf}.  A pre-pulse power compression ratio of three can be achieved with $>$80$\%$ conversion efficiency. We have explored the impact of operating with and without a compressed pre-pulse on the thermal and electrical power requirements for the linac and observed a 35$\%$ reduction in the electrical power required for the same beam energy if pulse compression is included.

In addition, we have also considered operating with a compressed pulse during the flat-top portion of the RF pulse. This does not prove to be realistic for the power ratios required during the pre-pulse of the system to maintain reasonable power consumption. Increasing the
compression for the full pulse by an additional factor of 2 (6X for the pre-pulse) reduces the overall efficiency of the system by a factor of three. This is a startling result considering that pre-pulse compression actually increases the system efficiency by 35$\%$ for overall power consumption. In practice this is due to the rather modest requirements of pre-pulse compression becoming excessive in the presence of flat-top compression.

To realize this concept, extensive work on RF source cost and performance is also required. For a normal conducting linear accelerator the cost is dominated by the required RF sources. As we enter the multi-TeV range, RF source cost becomes the dominant factor for the entire normal conducting accelerator complex. While significant progress has been made over the past decade incorporating new concepts in RF sources, significant work remains with key challenges include production, capital cost, lifetime, operating voltage and efficiency all playing key roles. The present state of the art is capable of achieving 65$\%$ AC to RF efficiency in klystron RF sources of the power levels that are required. New concepts are being explored to push this up to 80$\%$ efficiency \cite{baikov2015toward,constable2017magic2,weatherford2018high,weatherford2020modular} which would dramatically reduce the number of sources needed and the electrical power consumed. The techniques that are being developed include core oscillation method, bunch align compress, use of permanent or superconducting solenoids, and implementation of depressed collectors that are externally or self-biased.

\subsubsection{Towards GeV/m Accelerating Gradients}

Space limitations, significant reductions in RF power cost, or improved accelerator structure shunt impedance (efficiency) may render operation at higher gradient more appealing. Operating beyond 250 MeV/m for an RF accelerator would require a new topology for the linac. One possibility that is actively being researched is the use of shorter RF pulses and higher repetition rates at significantly higher THz frequencies (100-300 GHz). Structures in this frequency range have exceeded GV/m surface fields \cite{dal2016rf, dal2016rf, forno2017high}. High frequency structures made significant progress in recent years and are now being utilized for beam acceleration and beam manipulation \cite{othman2020experimental,li2019terahertz,othman2021visualizing,snively2020femtosecond,tang2021stable,zhang2020cascaded} but still require significant R\&D before formulating a proposal for a high energy facility. Extensive investigations into the beam dynamics of such structures are required to confirm the viability for a high luminosity application such as a linear collider.

Studies of beam transport in these structures with high gradients (500 MeV/m) and pC bunches indicate it is possible to transport the beam while accounting for effects from short and long range wakes~\cite{nanniTHzSnowmass}.   Long range wakefields are particularly challenging at high frequency.   One approach to increasing the beam current is to operate with a bunch in every RF period and allow only for excitation of higher order modes that are  harmonics at integer multiples of the drive frequency.  This would greatly reduce the number of modes which must be damped. With a bunch charge of 1~pC in every cycle, operation at 300 GHz would provide the same peak current during the RF pulse, as shown in Table~\ref{tab:TeVmainlinacparam} for a C$^3$ accelerator. Due to the shortened RF pulse, recovering luminosity would require high repetition rates and reduced beam emittance. Powering of such structures would also require very different approach than presently envisioned for RF accelerator based linear colliders. High frequency RF sources can be extremely efficient with fast-wave cyclotron resonance masers and long pulse formats. Matching these RF sources to high luminosity applications requires the development of active quasi-optical pulse compression.   R\&D on this issue is in progress. Beam-driven (wakefield) RF sources, as envisioned for CLIC, may also be a possible option for such high frequency operation. Finally, it has been discussed since the 1990's that higher frequencies could be preferred for operation with smaller bunches to obtain greater power efficiency at higher gradients (see, for example, \cite{zimmermann1998final}). At that time, attempts to realize this approach stumbled on the issue of breakdown. It is now appealing to revisit this program in the light of the advances in THz accelerator technology, to see whether operation at X- or even W-band can give a practical route to very high gradients.

\section{Plasma, laser, and structure wakefield accelerators}
\label{sec:wakefield}

In this section, we will discuss possible upgrades of the ILC infrastructure based on wakefield acceleration.  This is an advanced technique capable of extremely high gradiant electron acceleration.  Today it still very much in the research stage,  with many important issues still unanswered.   However, this technology  has the potential not only to deliver extremely high-energy beams, but do so in a highly efficient manner and to
 achieve the high luminosities needed for physics at these energies~\cite{ALEGRO:2019alc}.   In this section, we discuss three options for wakefield acceleration; this is not meant to exhaust the possible options.

\begin{table}
\begin{center}
\begin{tabular}{lcccccc}
 & $e^-$  & Drive & Interstage & Plasma &  \\
Status &Source  & complex & coupling & medium & BDS \\
\hline
Conventional & Damping Ring &  Pulsed RF & Warm magnets & Laser-ionized cell & ILC-type \\
Upgraded & Photoinjector &  CW RF & Warm magnets  & Laser-ionized & CLIC-type \\
  &  &  &  and plasma lenses &  gas jet & \\
Advanced & Plasma injector &  CW High-Q & Combined function & Beam-ionized & plasma-based \\
  &  &  & plasma  &  gas jet &  \\
\end{tabular}
\caption{Envisioned evolution of a plasma-based linear collider.}
\label{tab:PWFAupgrade}
\end{center}
\end{table}

\subsubsection{Beam-Driven Plasma Wakefield Accelerators}

Research on beam-driven plasma wakefield acceleration (PWFA) is motivated by the ultimate goal of creating a linear collider that is affordable, highly-efficient, and operates at the highest possible energies. There are many challenges on the path to a plasma-based upgrade to the ILC, but the field has shown steady progress on multiple fronts since the previous Snowmass study  in 2013. Among many highlights are the first demonstration of highly-efficient plasma acceleration of electron beams~\cite{Litos:2014yqa}, acceleration of positron beams in the non-linear regime~\cite{Corde:2015zxa}, proton beam-driven acceleration~\cite{AWAKE:2018gdq}, staged laser-plasma acceleration~\cite{Steinke:2016cyx}, plasma photocathodes for generating  ultralow-emittance beams~\cite{Deng:2019gix}, and emittance preservation in an active plasma lens~\cite{Lindstrom:2018hqq}.

The remaining challenges associated with the development of a linear collider based on PWFA have been identified in a variety of papers, workshops, and strategy sessions~\cite{Rosenzweig:1998vh,Seryi:2009zza,Schroeder:2010zza,Adli:2013npa,Cros:2019tns}.
 We enumerate some of them here:
\begin{enumerate}
    \item High-efficiency, high-quality acceleration in a single plasma stage.
    \item Coupling between plasma stages.
    \item Positron acceleration in plasma.
    \item Preservation of beam polarization.
    \item High repetition-rate plasma acceleration and energy deposition in the plasma source.
    \item Final focusing and alignment of beams at the collision point.
\end{enumerate}
Experiments to demonstrate high-efficiency, high-quality electron acceleration in plasma are currently underway at FLASHForward at DESY and preparing to start at FACET-II at SLAC. These experiments will demonstrate the viability of PWFA technology and establish the tolerances for producing high-quality beams. Experiments at FLASHForward will also study high-repetition rate PWFA, while experiments at FACET-II will cover positron acceleration in plasma and beam focusing based on thin plasma lenses. Both FLASHForward and FACET-II need to be modified in order to demonstrate staged PWFA.  This is a high priority for the field. 

Because of the challenges to the PWFA concept from limitations on power and tolerances, it is important to take a long-term view, going beyond the ideas for the baseline PWFA accelerator to elements that use plasma-based concepts in a more sophisticated way.  In Table~\ref{tab:PWFAupgrade}, we sketch such upgrade paths for the various elements of a PWFA collider.

\subsubsection{Laser Wakefield Accelerators}

Laser wakefield accelerators (LWFAs)~\cite{Esarey:2009zz} rely on an intense, ultrashort laser pulses to resonantly excite large amplitude electron plasma waves with relativistic phase velocities.  The accelerating fields of the plasma wave, or wakefields, are 1-10 GV/m, orders of magnitude larger than conventional accelerating structures, enabling compact acceleration of charged particle beams.  LWFA technology provides an opportunity to upgrade the ILC to higher beam energy using the planned ILC main linac tunnel, site power, and infrastructure.  An LWFA-based linac arm would consist of multiple plasma stages, each stage yielding a few GeV/stage energy gain, driven by a multi-J, short pulse laser~\cite{Schroeder:2010zza,Schroeder:2016mrg}.  Laser drivers are highly flexible, and plasma mirror technology enables compact coupling of the laser driver into the plasma accelerating cells.  The multi-Joule-class laser systems, potentially based on fiber laser combination, occupy an area of a few $m^2$ and both the drive lasers and plasma accelerating stages may be placed in the ILC Main Linac tunnel.  LWFAs accelerate short bunches, of order 10~microns, and the resulting beamstrahulung reduction at the IP yields significant power savings for a given target luminosity~\cite{Yakimenko:2018kih}.  To reach $E_{CM}=1$~TeV, an LWFA-based linac requires potentially only 0.2~km in each linac arm, and 100~MW of power for both beams to reach a luminosity of 10$^{34}$ cm$^{-2}$s$^{-1}$.  This could be upgraded to $E_{CM} = 3$~TeV with luminosity of 10$^{35}$ cm$^{-2}$s$^{-1}$, requiring a 0.65 km LWFA linac in each linac arm and 300 MW of power for both beams.  The LWFA beam power for 1 TeV and 3 TeV would be 4 MW and 12 MW, respectively, and are within the power rating of the planned ILC beam dump.  The unused main linac tunnel length could be employed to extend the BDS system to accommodate $E_{CM} = 3$ TeV, as well as space for linear cooling sections to further reduce the beam emittance.  The bunch structure employed is one bunch each 20 $\mu$s, and additional bunch compressors would be required to achieve the short, 10-micron-scale, bunch length.  Furthermore, achieving high beam energies ($E_{CM} > $3~TeV) is straightforward by adding additional LWFA stages, although the required increased luminosity would require site power beyond the planned ILC design.  This provides a long-term upgrade path to continue realizing new physics reach in realistic stages using the infrastructure of a linear collider.  Significant R\&D is required to realize an LWFA-based linac, and, in particular, further development of high average power, short-pulse laser systems operating at tens of kHz repetition rates~\cite{kBella:2017}.
 
\bigskip

\subsubsection{Structure Wakefield Accelerators}

Structure Wakefield Acceleration (SWFA) has been proposed as the backbone for a high-gradient and high-efficiency accelerator for a multi-TeV linear collider~\cite{Gai:2012}.  Two separate SWFA schemes, two-beam acceleration (TBA) and collinear wakefield acceleration (CWA) are under consideration. This contribution will explore the application of the relatively mature SWFA schemes (both in the TBA and CWA implementations) as a possible upgrade path to the ILC.    The ILC beam format (a train of 3.2~nC single-bunch with an ${\cal O}$(MHz) micropulse repetition rate) is comparable  to the 182-GHz~CWA-based XFEL design that is being pursued at Argonne.   The challenge for the CWA based linear collider would be to raise the overall efficiency due to its single pulse nature.  Alternatively, the TBA technology currently under development at Argonne is a 26 GHz accelerator based on a high charge drive beam. Therefore, a TBA contribution to the ILC application would explore two avenues: either operating ILC with higher charge or raising the TBA operating frequency to operate at lower drive charge.  Critical to both the TBA and CWA approaches would be continued development of the SWFA bunch control R\&D program.  This program develops the bunch shaping technology critical for the main and drive beams.  For example, we will explore the possibility of shaping the ILC 3.2nC Gaussian bunch for the CWA scheme with a transformer ratio of 5 to produce a 5~TeV LC in the ILC tunnel at high efficiency.   Note that bunch control is critical to both beam-driven wakefield acceleration methods: SWFA and plasma wakefield acceleration (PWFA).

\chapter{Conclusions}  
\label{chap:conclusion}

In this report, we have surveyed all aspects of the International
Linear Collider.   We have, first of all, explained the importance of
this machine for physics.    The Higgs boson is at the center of the
Standard Model of particle physics, and almost all of the major
questions about this model go back to questions about its nature.
The ILC will give us a clear and complete view of the properties of
this particle, its couplings to all particles of the Standard Model,
its self-coupling, and its possible couplings to new particles not
yet discovered.   At its 500~GeV stage, the ILC will also give us a
detailed picture of the top quark that will illuminate its relation to
the Higgs boson and the electroweak sector.  The ILC will be able to
search for weakly coupled particles and the particle of dark matter in
a way that is almost free of model assumptions.  The ILC also has 
opportunities for discovery in studies of quark and lepton production,
$W$ and $Z$ properties, and high-precision tests of QCD.

We have explained that the ILC will provide an ideal environment for
precision studies in particle physics.  Electroweak processes at the
highest energies dominate the event samples.  Individual events can be
fully reconstructed, including identification of heavy flavors.  Beam
polarization can add to the variety of events sampled and offer
compelling clues to their interpretation.  We have described detector
designs that make use of this remarkable physics environment, and also
new technologies, some developed from the LHC experiments, that will
extend their capabilities further.

We have reviewed the accelerator design of the ILC and shown how it
gives a robust solution to deliver electron and positron beams in the
energy region of the Higgs boson.  This
design has been created and refined over more than twenty years.  It
addresses all of the major technical problems of such beams within
a well-understood budget and schedule.

We have discussed how the ILC can be the first step on a road to much higher center of mass energies.   We envision the ILC Laboratory as being a major center for particle physics long after the measurements we have presented in this report are completed.

We can look ahead to the future of particle physics and dream of
accelerator experiments at very high energies.   But the mysteries of the Standard Model are
with us today.  It is has become clear that we need a new approach
today to gain insight into the key problems of particle physics. The
ILC gives us a strategy to address these questions and a technology
that is ready for construction. It is time to 
make the ILC a reality.

\bigskip

\Acknowledgements

We are grateful to Marina Chadeeva and Alexey Drutskoy  (Lebedev Institute) and to Valery Telnov (Budker Institue) for their contributions to this report.

The work of the DESY group is supported by the Deutsche Forschungsgemeinschaft under Germany’s Excellence Strategy, EXC 2121 “Quantum Universe”, grant 390833306. 
The work of IFIC is supported by Projects No. PGC2018-094856-B-100
(MCIN/AEI), PROMETEO-2018/060 and CIDEGENT/2020/21 (Generalitat
Valenciana) and iLINK Grant No. LINKB20065 (CSIC).
The work of the KEK group is supported in part by JSPS KAKENHI Grant Numbers 16H02173 and 21H01077.
The work of the SLAC group is supported by the US Department of Energy, contract DE–AC02–76SF00515. 
The work of James Brau is supported by the US Department of Energy grant  DE-SC0017996. 
The work of Francesco Giovanni Celiberto is supported by the INFN/NINPHA project. 
 The work of Sven Heinemeyer is supported in part by the grant
PID2019-110058GB-C21 funded by MCIN/AEI/10.13039/501100011033 and by
"ERDF A way of making Europe", and in part by the grant CEX2020-001007-S
funded by MCIN/AEI/10.13039/501100011033.
The work of Sunghoon Jung is supported by the National Research Foundation of Korea under grant NRF-2017R1D1A1B03030820.
The work of Zhen Liu is supported in part by the U.S. Department of Energy (DOE) under grant No. DE-SC0022345. The work of Nathaniel Craig is supported in part by the U.S.~Department of Energy under the grant DE-SC0011702.
The work of Alessandro Papa is supported by the INFN/QFT\@COLLIDERS project.The work of Junping Tian is supported in part by the Japan Society for the Promotion of Science under the Grant-in-Aid for Science Research 15H02083.
The work of Graham Wilson is supported by the US National Science Foundation under award NSF 2013007.

\newpage

\bibliographystyle{JHEP}

\providecommand{\href}[2]{#2}\begingroup\raggedright\endgroup

% \bibliography{chapters/intro/intro,
% chapters/case/case,
% chapters/accelerator/accelerator,
% chapters/ILCorg/ILCorg,
% chapters/gen-phys/gen-phys,
% chapters/detectors/detectors,
% chapters/sim/sim,
% chapters/ILC250/ILC250,
% chapters/PEW/PEW,
% chapters/ILC500/ILC500,
% chapters/fixedtarget/fixedtarget,
% chapters/SMEFT/SMEFT,
% chapters/bigquestions/bigquestions,
% chapters/ILCBSM/ILCBSM,
% chapters/farfuture/farfuture,
% chapters/conclusion/conclusion}

\end{document}